\documentclass[12pt,a4paper]{article}

\usepackage{amsmath}
\usepackage{amssymb}
\usepackage[nottoc]{tocbibind}
\usepackage[usenames,dvipsnames,svgnames,table]{xcolor}
\usepackage[colorlinks,citecolor=DarkGreen,linkcolor=FireBrick,urlcolor=FireBrick,linktocpage]{hyperref}
\usepackage{graphicx}
\usepackage{float}
\usepackage{young-kugo}
\usepackage[height=8.8in,width=6.45in]{geometry}
\usepackage{tgtermes}
\usepackage{tgadventor}
\usepackage[T1]{fontenc}

\def\cF{\mathcal{F}}
\def\cG{\mathcal{G}}

\def\cM{\mathcal{M}}
\def\cN{\mathcal{N}}

\def\cQ{\mathcal{Q}}

\def\cT{\mathcal{T}}

\def\bC{\mathbb{C}}
\def\bH{\mathbb{H}}
\def\bR{\mathbb{R}}
\def\bZ{\mathbb{Z}}

\def\slash#1{\ooalign{{\text{$#1$}}\crcr \hss\big/\hss}}

\def\Arg{\mathop{\mathrm{Arg}}\nolimits}

\def\tr{\mathop{\mathrm{tr}}\nolimits}

\def\diag{\mathop{\mathrm{diag}}}

\def\rank{\mathop{\mathrm{rank}}}
\def\Im{\mathop{\mathrm{Im}}}

\def\Re{\mathop{\mathrm{Re}}}

\def\Res{\mathop{\mathrm{Res}}}

\def\SO{\mathop{\mathrm{SO}}}
\def\OO{\mathrm{O}}
\def\SU{\mathop{\mathrm{SU}}}
\def\SL{\mathop{\mathrm{SL}}}

\def\Sp{\mathop{\mathrm{Sp}}}

\def\U{{\mathrm{U}}}

\def\vev#1{\langle#1\rangle}
\def\ket#1{|#1\rangle}
\def\ii{\mathrm{i}}
\def\LambdaRG{E}
\usepackage{mathtools}
\def\primeM{\prescript{\prime}{}M}
\let\oldstar\star
\def\star{{\oldstar}}

\def\SeibergWitten{Seiberg-Witten}
\def\Gaiotto{ultraviolet}

\def\inc#1{\vcenter{\hbox{\includegraphics[scale=.2]{#1}}}}
\def\incc#1{\vcenter{\hbox{\includegraphics[scale=.6]{#1}}}}

\numberwithin{equation}{subsection}
\numberwithin{figure}{section}
\numberwithin{table}{section}

\usepackage{tikz}
\usetikzlibrary{backgrounds,fit,calc}

\begin{document}

\begin{titlepage}

\begin{flushright}
IPMU 13-0234 \\
UT-13-42\\
%preliminary version 0.3
version 3
\end{flushright}
\vskip 2cm

\begin{center}
{\Large \bfseries
$\cN{=}2$ supersymmetric dynamics for pedestrians
}

\vskip 1.2cm

Yuji Tachikawa$^{\sharp,\flat}$

\bigskip
\bigskip

\begin{tabular}{ll}
$^\flat$  & Department of Physics, Faculty of Science, \\
& University of Tokyo,  Bunkyo-ku, Tokyo 133-0022, Japan\\
$^\sharp$  & Kavli Institute for the Physics and Mathematics of the Universe (WPI), \\
& University of Tokyo,  Kashiwa, Chiba 277-8583, Japan
\end{tabular}

\vskip 1.5cm

\textbf{abstract}
\end{center}

\medskip
\noindent

We give a pedagogical introduction to the dynamics of $\cN{=}2$ supersymmetric systems in four dimensions.  
The topic ranges from the Lagrangian and the Seiberg-Witten solutions of $\SU(2)$ gauge theories to Argyres-Douglas CFTs and Gaiotto dualities. 

This lecture note is a write-up of the author's lectures at Tohoku University, Nagoya University and Rikkyo University. 

\bigskip

\bigskip

\vfill
\end{titlepage}

\setcounter{tocdepth}{2}
\tableofcontents

\newpage

\setcounter{section}{-1}
\section{Introduction}

The study of $\cN{=}2$ supersymmetric quantum field theories in four-dimensions has been a fertile field for theoretical physicists for quite some time.  These theories always have non-chiral matter representations, and therefore can never be directly relevant for describing the real world. That said, the existence of two sets of supersymmetries allows us to study their properties in much greater detail than both non-supersymmetric theories and $\cN{=}1$ supersymmetric theories. Being able to do so is quite fun in itself, and hopefully the general lessons thus learned concerning $\cN{=}2$ supersymmetric theories might be useful when we study the dynamics of theories with lower supersymmetry.  At least, the physical properties of $\cN{=}2$ theories have been successfully used to point mathematicians to a number of new mathematical phenomena unknown to them. 

These words would not probably be persuasive enough for non-motivated people to start studying $\cN{=}2$ dynamics. It is not, however, the author's plan to present here a convincing argument why you should want to study it anyway; the fact that you are reading this sentence should mean that you are already somewhat interested in this subject and are looking for a place to start. 

There have been many important contributions to the study of $\cN{=}2$ theories since its introduction \cite{Fayet:1975yi}. The four most significant ones in the author's very personal opinion are the following: 
\begin{itemize}
%\item[\textbf{1975}]   Fayet found in \cite{Fayet:1975yi}  four-dimensional theories with $\cN{=}2$ supersymmetry.
%	\begin{itemize}
%	\item[\textbf{1987}] Hyperk\"ahler quotient construction was found in \cite{Hitchin:1986ea}.
%	\end{itemize}
\item In 1994,  Seiberg and Witten found in \cite{Seiberg:1994rs,Seiberg:1994aj} exact low-energy solutions to $\cN{=}2$ supersymmetric $\SU(2)$ gauge theories  by using holomorphy and by introducing the concept of the \SeibergWitten\ curves. 
\item In 1996-7, the \SeibergWitten\ curves, which were so far mathematical auxiliary objects, were identified as  physical objects appearing in various string theory constructions of $\cN{=}2$ supersymmetric theories \cite{Klemm:1996bj,Banks:1996nj,Witten:1997sc}.
%	\begin{itemize}
%	\item[\textbf{1994}] The monopole equation was found to be very useful to study the geometry of four-dimensional manifolds \cite{Witten:1994cg}.
%	\end{itemize}
\item In 2002, Nekrasov found in \cite{Nekrasov:2002qd} a concise method to obtain the solutions of Seiberg and Witten via the instanton counting.
%	\begin{itemize}
%	\item[\textbf{2003}] His conjecture was rigorously proved in \cite{Nekrasov:2003rj,Nakajima:2003pg,Braverman:2004vv}.
%	\end{itemize}
\item In 2009, Gaiotto found in \cite{Gaiotto:2009we} a huge web of S-dualities acting upon $\cN{=}2$ supersymmetric systems.
%	\begin{itemize}
%	\item[\textbf{2012}] A mathematical conjecture \cite{Alday:2009aq} based on this work was again rigorously proved \cite{SchiffmannVasserot,MaulikOkounkov}.
%	\end{itemize}
\end{itemize}

The developments before 2002 have been described in  many nice introductory reviews and lecture notes, e.g.~\cite{Bilal:1995hc,Lerche:1996xu,AlvarezGaume:1996mv,Klemm:1997gg,Peskin:1997qi,DiVecchia:1998ky,Argyres1998Lecture,Giveon:1998sr,D'Hoker:1999ft}. Newer textbooks also have sections  on them, see e.g.~Chap.~29.5 of \cite{WeinbergIII} and Chap.~13 of \cite{Terning}.  A short review on the instanton counting is also available \cite{Tachikawa:2014dja}.  A comprehensive review on the newer developments since 2009 would then surely be useful to have, but this lecture note is not exactly that. 
Rather, the main aim of this lecture note is to present the same old results covered in the lectures and reviews listed above under a new light introduced in 2009 and developed in the last few years, so that readers would be naturally prepared to the study of recent works once they go through this note. A good review with an emphasis on more recent developments can be found in \cite{JaewonThesis,Giacomelli:2013tia}.

\begin{figure}[h]
\centering
\tikzset{sec/.style={circle,draw,inner sep=0pt,minimum size=.8cm,fill=white}}
\tikzset{>=stealth}
\begin{tikzpicture}

\tikzstyle{box} = [rounded corners, rectangle,minimum width=12cm]

\node[box,fill=black!10,minimum height=2.5cm] at (2,.5) {};
\node[box,fill=black!10,minimum height=3.5cm] at (2,-3) {};
\node[box,fill=black!10,minimum height=1.5cm] at (2,-6) {};
\node[box,fill=black!10,minimum height=1.5cm] at (2,-8) {};
\node[box,fill=black!10,minimum height=3.5cm] at (2,-11) {};

\node[sec] (1) at (0,0) {2};
\node[sec] (2) at (2,0) {1};
\node[sec] (3) at (4,0) {3};
\node[circle,inner sep=0pt,minimum size=.8cm] (4d) at (0,-2) {};
\node[sec] (4) at (4,-2) {4};
\node[sec] (A) at (2,-4) {A};
\node[sec] (5) at (4,-4) {5};
\node[sec] (6) at (4,-6) {6};
\node[sec] (7) at (0,-6) {7};
\node[sec] (8) at (4,-8) {8};
\node[sec] (9) at (0,-8) {9};
\node[sec] (10) at (4,-10) {10};
\node[sec] (11) at (2,-10) {11};
\node[sec] (12) at (0,-12) {12};
%\node[sec] (B) at (2,-14) {B};
\draw[->,thick] (1) to (4);
\draw[->,thick] (2) to (4);
\draw[->,thick] (3) to (4);
\draw[->,thick] (4) to (5);
\draw[->,thick] (5) to (6);
\draw[->,thick] (1) to (7);
\draw[->,thick] (7) to (8);
\draw[->,thick] (7) to (9);
\draw[->,thick] (6) to (8);
\draw[->,thick] (6) to (9);
\draw[->,thick] (6) to (11);
\draw[->,thick] (9) to (12);
\draw[->,thick] (11) to (12);
\draw[->,thick] (10) to (12);
\draw[->,thick] (8) to (10);
\draw[->,thick,dashed] (10) to (11);
\draw[->,thick,dashed] (4) to (A);
\node[left] (1t) at (1.west) {Lagrangians};
\node[left,align=right] (7t) at (7.west) {More on\\ hypermultiplets};
\node[left,align=right] (9t) at (9.west) {$\SU(2)$ $N_f=4$ \\ and S-duality};
\node[left,align=right] (12t) at (12.west) {Argyres-Seiberg-\\ Gaiotto duality};
%\node[right,align=left] (Bt) at (B.east) {Summary of \\ various theories};
\node[left,align=right] (11t) at (11.west) {$\SU(N)$};
\node[right,align=left] (3t) at (3.east) {Renormalization\\ and anomalies};
\node[right] (4t) at (4.east) {pure $\SU(2)$};
\node[above,align=center] (At) at (A.north) {Comparison  \\ with instantons};
\node[left] (4dt) at (4d.west) {};
\node[right] (5t)  at (5.east) {$\SU(2)$ $N_f=1$};
\node[right] (6t) at (6.east) {6d interpretation};
\node[right] (8t)  at (8.east) {$\SU(2)$ $N_f=2,3$};
\node[right,align=left] (10t) at (10.east) {Argyres-Douglas \\ CFTs};
\node[above,align=center] (2t) at (2.north) {Abelian duality\\ and monopoles};

\end{tikzpicture}
\caption{Interdependence of sections of this lecture note\label{fig:relation}}
\end{figure}
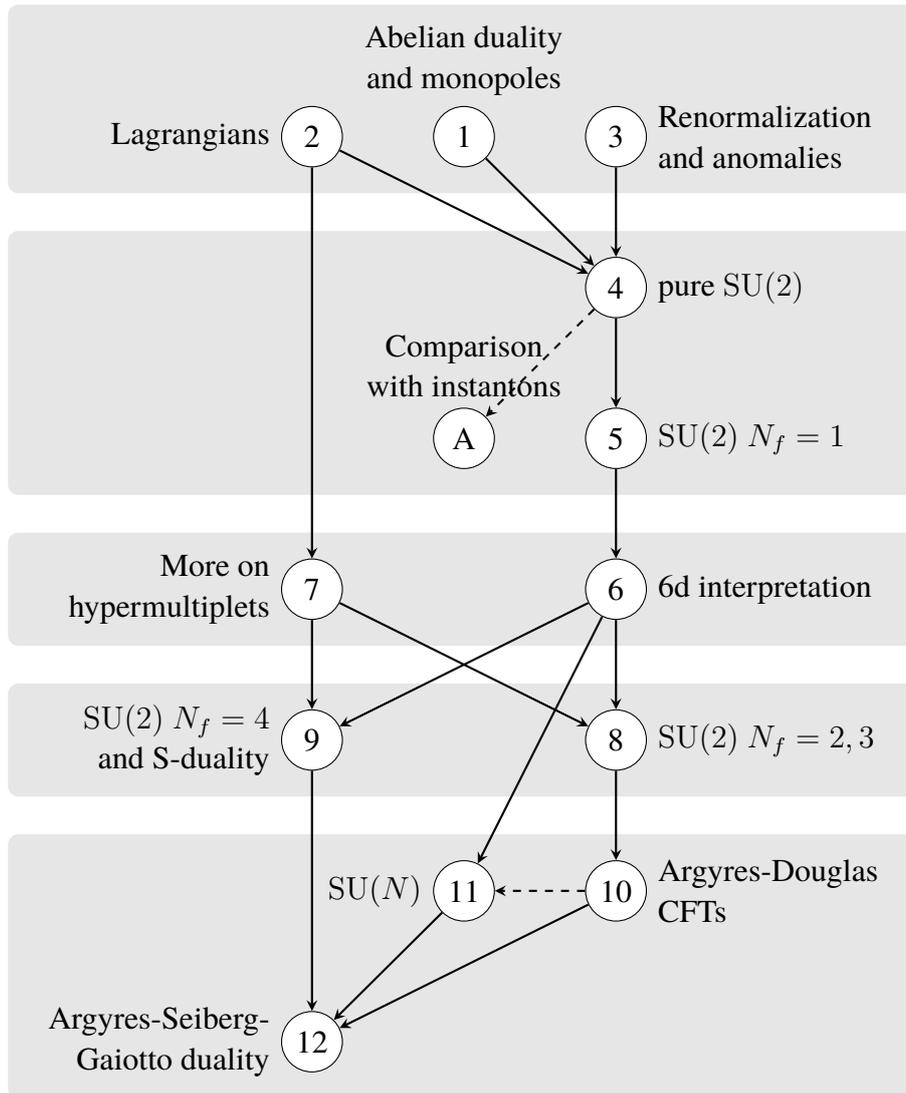

 %So, without further ado, here is how 
The rest  of the lecture note is organized as follows. First three sections are there to prepare ourselves to the study of $\cN{=}2$ dynamics. 
\begin{itemize}
\item We start in Sec.~\ref{sec:emdual} by introducing the electromagnetic dualities of $\U(1)$ gauge theories and recalling the basic semiclassical features of monopoles. 
\item In Sec.~\ref{sec:lagrangian}, we construct the $\cN{=}2$ supersymmetric Lagrangians and studying their classical features.  We introduce the concepts of the Coulomb branch and the Higgs branch.
\item In Sec.~\ref{sec:renormalization}, we will first see that the renormalization of $\cN{=}2$ gauge theories are one-loop exact perturbatively. We also study the anomalous R-symmetry of supersymmetric theories. As an application, we will quickly study the behavior of pure $\cN{=}1$ gauge theories.
\end{itemize}
\noindent The next two sections are devoted to the solutions of the two most basic cases. 
\begin{itemize}
\item In Sec.~\ref{sec:pureSU2}, we discuss the solution to the pure $\cN{=}2$ supersymmetric $\SU(2)$ gauge theory in great detail. Two important concepts, the \SeibergWitten\ curve and the \Gaiotto\ curve\footnote{The concept of the \SeibergWitten\ curve was introduced in \cite{Seiberg:1994rs}. The concept of the \Gaiotto\ curve, applicable in a  general setting, can be traced back to \cite{Klemm:1996bj}, see Fig.~1 there. As also stated there, it was already implicitly  used in \cite{Gorsky:1995zq,Martinec:1995by,Itoyama:1995nv}. In \cite{Gaiotto:2009we}, the \Gaiotto\ curve was used very effectively to uncover the duality of $\cN{=}2$ theories. Privately, the author often calls the \Gaiotto\ curve as the Gaiotto curve, but this usage  would not be quite fair to every party involved.  In view of \href{http://en.wikipedia.org/wiki/Stigler's_law_of_eponymy}{Stiegler's law}, the author could have used this terminology in this lecture note, but in the end he     opted for a more neutral term `the \Gaiotto\ curve', which contains more scientific information at the same time.  The author could have similarly used `the infrared curve' for the \SeibergWitten\ curve.  As there is no bibliographical issue in this case, however, the author decided to stick to the standard usage to call it the Seiberg-Witten curve.}, will be introduced.
\item In Sec.~\ref{sec:nf=1},  we solve the $\cN{=}2$ supersymmetric $\SU(2)$ gauge  theory with one hypermultiplet in the doublet representation. We will see again that the solution can be given in terms of the curves.
\end{itemize}
\noindent The sections~\ref{sec:6d} and \ref{sec:higgsbranch} are again preparatory.
\begin{itemize}
\item In Sec.~\ref{sec:6d}, we give a physical meaning to the \SeibergWitten\ curves and the \Gaiotto\ curves, in terms of six-dimensional theory. With this we will be able to guess the solutions to $\SU(2)$ gauge theory with arbitrary number of hypermultiplets in the doublet representations.  
This section will not be self-contained at all, but it should give the reader the minimum with which to work from this point on.
\item Up to the previous section, we will be mainly concerned with the Coulomb branch. As the analysis of the Higgs branch will become also useful and instructive later, we will study the features of the Higgs branch in slightly more detail in Sec.~\ref{sec:higgsbranch}.
\end{itemize}
 We resume the study of $\SU(2)$ gauge theories in the next two sections.
 \begin{itemize}
 \item In Sec.~\ref{sec:nf=2,3}, we will see that the solutions of $\SU(2)$ gauge theories with two or three hypermultiplets in the doublet representation, which we will have guessed in Sec.~\ref{sec:6d}, indeed pass all the checks to be the correct ones. 
 \item In Sec.~\ref{sec:gaiotto}, we first study the $\SU(2)$ gauge theory with four hypermultiplets in the doublet representation. We will see that it has an S-duality acting on the $\SO(8)$ flavor symmetry via its outer-automorphism. Then the analysis will be generalized, following Gaiotto, to arbitrary theories with gauge group of the form $\SU(2)^n$.
\end{itemize}
We will consider more diverse examples in the final three sections of the main part. 
\begin{itemize}
\item In Sec.~\ref{sec:AD}, we will study various superconformal field theories of the type first found by Argyres and Douglas, which arises when electrically and magnetically charged particles become simultaneously very light. 
\item In Sec.~\ref{sec:higherrank}, the solutions to $\SU(N)$ and $\SO(2N)$ gauge theories with and without hypermultiplets in the fundamental or vector representation will be quickly described. 
\item In Sec.~\ref{sec:duality}, we will analyze  the S-duality of the $\SU(N)$ gauge theory with $2N$ flavors and its generalization.  Important roles will be played by punctures on the \Gaiotto\ curve labeled by Young diagrams with $N$ boxes, whose relation to the Higgs branch will also be explained. As an application, we will construct superconformal field theories with exceptional flavor symmetries $E_{6,7,8}$.
\end{itemize}
We conclude the lecture note by a discussion of further directions of study in Sec.~\ref{sec:conclusions}. 
We have two appendices: \begin{itemize}
\item In Appendix~\ref{sec:instcount}, we explicitly compute the weak-coupling expansion of the prepotential of pure $\SU(2)$ theory obtained in Sec.~\ref{sec:pureSU2}, and see that it agrees with the prepotential as obtained by instanton computation, which we explain very briefly. 
\item In Appendix~\ref{sec:zoo}, we list various $\cN{=}2$ theories we encounter in this lecture note in one place, and summarize their constructions. 
\end{itemize}
The inter-relation of the sections within this lecture note is summarized in Fig.~\ref{fig:relation}.

\section*{Prerequisites, disclaimer, and acknowledgments}
A working knowledge of $\cN{=}1$ superfields is required; we set up our notation in Sec.~\ref{sec:lagrangian}.
Similarly, a reader should know one-loop renormalization and perturbative anomalies, and should have at least heard about instantons and monopoles, although we give a quick summary and references.
No prior knowledge of string theory or M-theory is assumed, but a reader should be open to the concept of theories defined in spacetime whose dimension is larger than four.
Unless otherwise stated, we use the common physics convention of calling $\SU(N)$ whatever gauge group whose gauge algebra is $\mathfrak{su}(N)$, etc.

Signs and powers of $i$ in the terms in the Lagrangian are not completely consistent or correct, but the overall ideas presented in the lecture note should be alright.  The author is sorry that he used the same letter $i$ for the imaginary unit and for the indices, and the same letter $\theta$ for the theta angle and for the supercoordinates. 
In general, readers are encouraged to  read not just what is written, but what should be written instead. Presumably there are many other typos, errors, and points to be improved.  The author would welcome whatever comments from you, so please do not hesitate to write an email to the author at \href{mailto:yuji.tachikawa@ipmu.jp}{yuji.tachikawa@ipmu.jp}.

The deficiencies concerning citations are most obvious, as the number of relevant papers is immense. The author is quite sure that he cited too much of his own papers. Other than that, the author at least tried to give a few pointers to recent papers from whose references the interested readers should be able to start exploring the literature. The author is open to add more references in this lecture note itself, and any reader is again encouraged to send emails. 

This lecture note is based on the author's lectures at Nagoya University and Tohoku University on 2011 and at Rikkyo University on 2013. The author thanks the hosts in these three universities for giving him the opportunity to present a review of the $\cN{=}2$ supersymmetric dynamics using new techniques. He also thanks the participants of these lectures for giving him many useful comments along the course of the lectures. 
The author's approach to this topic has been formed and heavily influenced by the discussions with various colleagues, and most of the new arguments in this note, except for those which are wrong, should not be credited to the author.  
Ofer Aharony, Lakshya Bhardwaj, Chi-Ming Chang, Jacques Distler,  Sheng-Lan Ko, John Mangual,  Satoshi Nawata,  Vasily Pestun, Futoshi Yagi and an anonymous referee  gave helpful comments on the draft version of this lecture note. 
Simone Giacomelli, Brian Henning, Greg Moore,  Tatsuma Nishioka, Jun'ya Yagi and Kazuya Yonekura in particular read  the draft in detail and suggested many points to be improved to the author. It is a pleasure and indeed a privilege that the author can thank them. 
Finally, the author would like to thank Teppei Kitahara, who  helped the author preparing the figures, without which this lecture note would lose much of its value.

The author also thanks the right amount of duties associated to his position, with which he cannot concentrate any longer on cutting-edge researches but still has some time to summarize what he already knows.
In particular, he thanks various stupid faculty meetings he needs to participate, during which time he drew most of the figures on his laptop. The readers should therefore thank the overly bureaucratic system prevalent in University of Tokyo, which made this lecture note materialize.  
This work  is supported  in part by JSPS Grant-in-Aid for Scientific Research No.~25870159 and in part by WPI Initiative, MEXT, Japan at IPMU, the University of Tokyo.

Acknowledgements added in v3: The author is grateful that, since the publication of lecture notes,
many people took the time to read it very carefully, and kindly reported back many errors to be corrected to me.  
These kind people include T. Burton, . W. Kim, S. Nawata, J, Y. Pang and T. Proch\'azka (in alphabetic order), and the author would like to thank all of them.
Among them, the author would like to thank T. Proch\'azka in particular,
since it was him who provided the bulk of corrections, 
which allowed this set of lecture notes to be improved greatly.
The author still welcomes whatever new comments (errors or otherwise) to this set of lecture notes,
and the known errors will be added to the following webpage: 
\begin{quotation}
\url{https://member.ipmu.jp/yuji.tachikawa/errata/pedestrians.html} .
\end{quotation}

\section{Electromagnetic duality and monopoles}\label{sec:emdual}
The electromagnetic duality of the Maxwell theory, exchanging electric and magnetic fields, plays a central role in this lecture note. It is therefore convenient to review it here, without the extra complication of supersymmetry.  The basic features of magnetic monopoles will also be recalled. 
\subsection{Electric and magnetic charges}\label{charges}
Consider a $\U(1)$ gauge field, described by the gauge potential $A=A_\mu dx^\mu$ and the field strength $F=\frac12F_{\mu\nu} dx^\mu \wedge dx^\nu$, where $F_{\mu\nu}=\partial_{[\mu} A_{\nu]}$.
This is invariant under the gauge transformation \begin{equation}
A\to A+i g^{-1} dg
\end{equation} where $g$ is a map from the spacetime to complex numbers with absolute value one, $|g|=1$. 
We can write $g=e^{i\chi}$ with a real function $\chi$, and we then have a more familiar \begin{equation}
A\to A-d\chi,\label{agauge}
\end{equation} but it will be important for us that $\chi$ can be multi-valued, so that we identify 
\begin{equation}
\chi\sim \chi+2\pi. \label{chiperiod}
\end{equation}

Consider a field $\phi$, with the gauge transformation given by \begin{equation}
\phi \to g^n \phi
\end{equation} 
We require here that $g$ specifies the transformations of all fields in the system uniquely.
Then  $n$ needs to be an integer; fractional powers are not uniquely defined.

The covariant derivative given by \begin{equation}
D_\mu\phi = \partial_\mu\phi + inA_\mu \phi,
\end{equation} and the kinetic term $|D_\mu\phi|^2$ is gauge-invariant. 
We write the action of the gauge field as \begin{equation}
S_\text{Maxwell}=\int d^4 x \frac1{2e^2} F_{\mu\nu} F_{\mu\nu}.\label{maxwell}
\end{equation} The coefficient $2$ in the denominator is slightly unconventional, but this choice removes various annoying factors later. 
Then the force between two particles obtained by quantizing the field $\phi$ is proportional to $e^2n^2$.
In phenomenological literature the combination $en$ is often called the electric charge, but in this lecture note we call the integer $n$ the electric charge. 
It might  also be tempting to rescale $F$ to eliminate the factor of $e^2$ from the denominator above.
But we stick to the convention that the periodicity of $\chi$  is $2\pi$, see \eqref{chiperiod}.

An electric particle with charge $n$ in the first quantized setup, Wick-rotated to the Euclidean signature, couples to the gauge field via \begin{equation}
S= in \int_L dx^\mu A_\mu\label{particle}
\end{equation} where $L$ is the worldline. The integrality of $n$ in this approach can be seen as follows. Due to the periodicity of $\chi$ \eqref{chiperiod}, the line integral $\int dx^\mu A_\mu$ is determined only up to an addition of an integral multiple of $2\pi$. Inside the path integral, $e^{iS}$ needs to be well defined. Then $n$ needs to be an integer. 

Adding \eqref{maxwell} and \eqref{particle} and writing down the equation of motion for $A_\mu$, we see that \begin{equation}
\int_{S^2}  \frac{4\pi}{e^2}\vec E \cdot d\vec n=
\int_{S^2}  \frac{4\pi}{e^2} \star F=2\pi n,\label{equantize}
\end{equation}  where $E_i=F_{0i}$ are the electric field components, $S^2$ is the sphere at infinity, \begin{equation}
\star F=\frac12 (\star F)_{\mu\nu} dx^\mu\wedge dx^\nu
\end{equation} where \begin{equation}
(\star F)_{\mu\nu}=\frac12\epsilon_{\mu\nu\rho\sigma} F^{\rho\sigma}
\end{equation} is the dual field strength. We also use the notation $\tilde F=\star F$ interchangeably.

Next, consider a space with the origin removed.
Surround the origin by a sphere. The gauge fields $A_{N,S}$ on the northern and  the southern hemispheres are related  by gauge transformation: \begin{equation}
A_N = A_S+ig^{-1} dg
\end{equation} on the equator. Then we have \begin{equation}
\int_{S^2} F = \int_{N} F + \int_{S} F = \int_\text{equator}(A_N-A_S) =\int_{\theta=0}^{\theta =2\pi} \frac{d\chi}{d\theta} d\theta=2\pi m,\label{mquantize}
\end{equation} where $m$ is an integer. We call $m$ the magnetic charge of the configuration. 
The energy contained in the Coulombic magnetic field diverges at the origin; but you should not worry too much about it, as the quantized electric particle also has a Coulombic electric field whose energy diverges. They are both  rendered finite by renormalization. 
When $m$ is nonzero, the configuration is called a magnetic monopole. Usually we simply call it a monopole. 

\begin{figure}[h]
\[
\inc{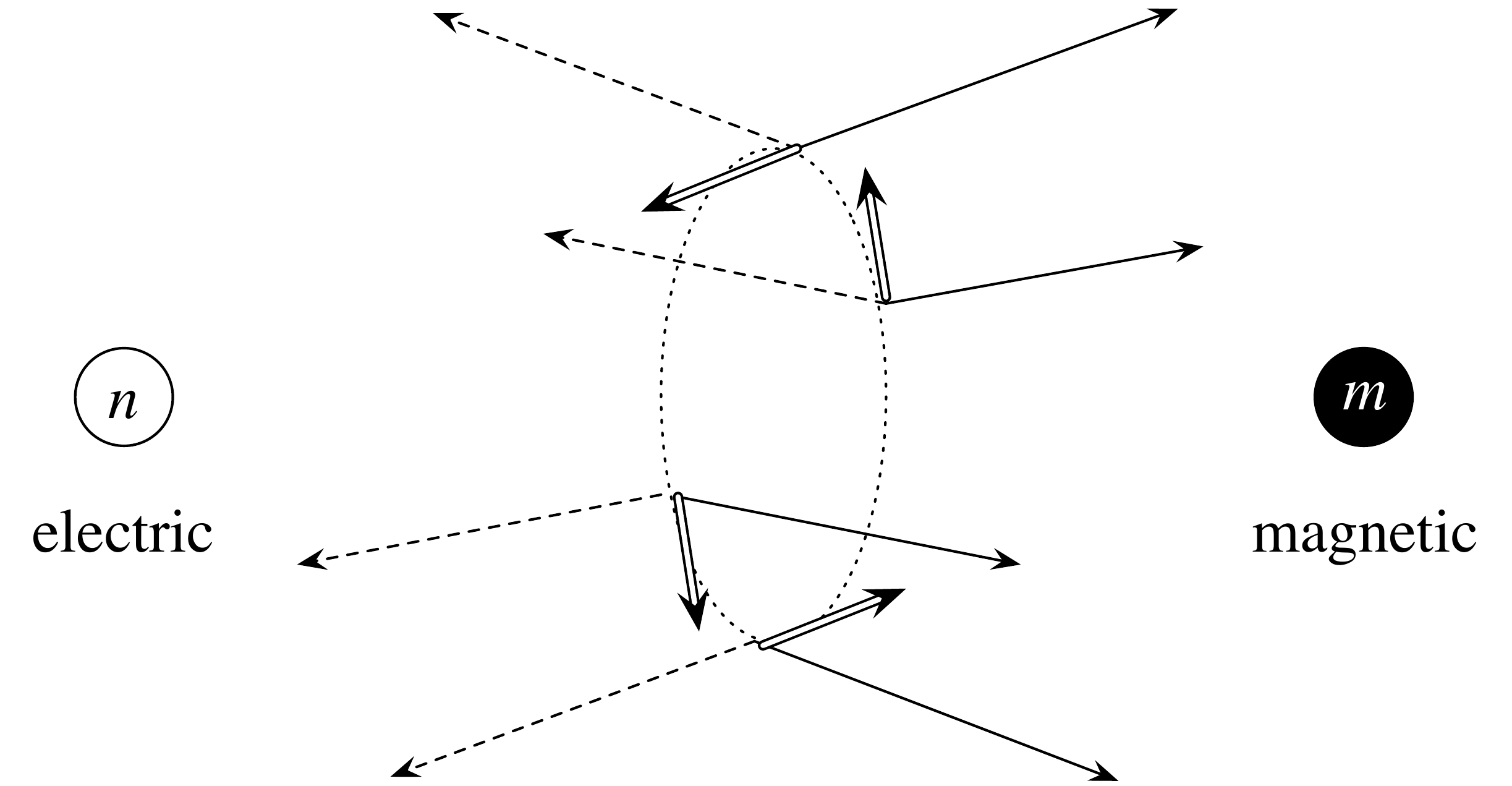}
\]
\caption{Angular momentum generated in the presence of both electric and magnetic particles. The straight, dashed and double arrows are for electric fields, magnetic fields and Poynting vectors, respectively.\label{fig:angular}}
\end{figure}

Put a particle with electric charge $n$, and another particle with magnetic charge $m$ on two separate points. 
The combined electric and magnetic field generate an angular momentum around the axis connecting two points via their Poynting vector, see Fig.~\ref{fig:angular}.
A careful computation shows that the total angular momentum contained in the electromagnetic field is $\hbar nm/2$,
which is consistent with the quantum-mechanical quantization of the angular momentum.

More generally, we can consider dyons, which are particles with both electric and magnetic charges. If we have a particle with electric charge $n$ and magnetic charge $m$,
and another particle with electric charge $n'$ and magnetic charge $m'$, the total angular momentum is $\hbar/2$ times \begin{equation}
 nm'-mn'.
\end{equation} We call this combination  the Dirac pairing of two sets of charges $(n,m)$ and $(n',m')$. 

\subsection{The S and the T transformations}\label{ST}
The Maxwell equation is given by \begin{equation}
\partial_{[\mu}F_{\nu\rho]}=0,\qquad 
\partial_{\mu}F_{\mu\nu}=0
\end{equation} or equivalently in the differential form notation by \begin{equation}
dF=d\star F=0.
\end{equation}
This set of equations is invariant under the exchange \begin{equation}
F\leftrightarrow \star F.
\end{equation} 
In terms of the electric field $\vec E$ and the magnetic field $\vec B$, which we schematically denote by  $F=(\vec E,\vec B)$,
the transformation does \begin{equation}
F=(\vec E,\vec B) \longrightarrow
\star F=(\vec B,-\vec E) \longrightarrow
\star^2 F=(-\vec E,-\vec B).
\end{equation} 
This operation is often called the $S$ transformation.

To preserve the quantization of the electric and magnetic charges \eqref{equantize}, \eqref{mquantize}, the dual field strength $F_D$ and the dual coupling $e_D$ need to be defined so that \begin{equation}
F_D=\frac{4\pi}{e^2}{}\star F,\qquad
\frac{4\pi}{e^2}\frac{4\pi}{e_D^2}=1.
\end{equation}

 Under this transformation, the charge $(n,m)$ is transformed as 
\begin{equation}
\begin{array}{r|ccc}
\text{particle 1} & (n,m) & \stackrel{S}{\longrightarrow} & (-m,n), \\
\text{particle 2} &(n',m') & \stackrel{S}{\longrightarrow} & (-m',n'), \\
\text{Dirac pairing} & nm'-n'm & = &  -mn'-(-m')n.
\end{array}
\end{equation}
Note that the Dirac pairing is preserved under the operation. 

%\subsection{The Witten effect and the T transformation}
Let us suppose that we have a neutral real scalar field $\phi$ and the action of the $\U(1)$ gauge field is given by \begin{equation}
\frac{1}{2e(\phi)^2} F_{\mu\nu}F_{\mu\nu} + \frac{\theta(\phi)}{16\pi^2} F_{\mu\nu} \tilde F_{\mu\nu}. 
\label{u1coupling}
\end{equation}
The Maxwell equation is now \begin{align}
\partial_{[\mu}F_{\nu\rho]}&=0,\label{MaxwellEq1}\\
\partial_{\mu}\left[ \frac{4\pi}{e(\phi)^2} F_{\mu\nu}+\frac{\theta(\phi)}{2\pi}\tilde F_{\mu\nu}\right]&=0.
\label{MaxwellEq2}
\end{align}
Decompose $F=(\vec E,\vec B)$ as before. The equations above show that the magnetic field satisfying the Gauss law is still $\vec B$, but the electric field satisfying the Gauss law is now the combination \begin{equation}
\vec E_\text{conserved} = \frac{4\pi }{e(\phi)^2} \vec E+\frac{\theta(\phi)}{2\pi} \vec B.
\end{equation} Therefore , we have  \begin{equation}
\int_{S^2} \vec B \cdot d\vec n= 2\pi m,\quad
\int_{S^2} \vec E_\text{conserved} \cdot d\vec n = 2\pi n
\end{equation}  where $m$ and $n$ are the integers introduced in Sec.~\ref{charges}.
This shows an interesting fact: let us change $\phi$ adiabatically to change $\theta(\phi)$. 
As $n$ is an integer, it cannot change. Therefore, $\vec E$ gets a contribution proportional to $\theta(\phi)\vec B$ to keep $\vec E_\text{conserved}$ fixed. This is called the Witten effect \cite{Witten:1979ey}. 

The S transformation, then, exchanges $(\vec E)_\text{conserved}$ and $\vec B$. 
The dual gauge field strength $F_D$ should then be \begin{equation}
F_D=\frac{4\pi}{e(\phi)^2} {}\star F - \frac{\theta}{2\pi}F
\end{equation}  so that the equation \eqref{MaxwellEq2} should just be
\begin{equation}
\partial_{[\mu}F_{D,\nu\rho]}=0.
\end{equation} We find that the equation \eqref{MaxwellEq1} becomes \begin{equation}
\partial_{\mu}\left[ \frac{4\pi}{e_D(\phi)^2} F_{D,\mu\nu}+\frac{\theta_D(\phi)}{2\pi}\tilde F_{D,\mu\nu}\right]=0
\end{equation}
where  $e_D(\phi)$, $\theta_D(\phi)$ are given by  \begin{equation}
\tau_D(\phi)  = -\frac{1}{\tau(\phi)}
\end{equation} where \begin{equation}
\tau(\phi)= \frac{4\pi i}{e(\phi)^2} + \frac{\theta(\phi)}{2\pi},\qquad
\tau_D(\phi)= \frac{4\pi i}{e_D(\phi)^2} + \frac{\theta_D(\phi)}{2\pi}. \label{u1tau}
\end{equation} This combination $\tau(\phi)$ is called the complexified coupling. 

We also know that, quantum mechanically, $\theta(\phi)$ and $\theta(\phi)+2\pi$ cannot be distinguished, since the change in the integrand of the Euclidean path integral is \begin{equation}
\exp\left[i\int d^4x \frac{1}{8\pi}  F_{\mu\nu} \tilde F_{\mu\nu}\right]
\end{equation} which is always one\footnote{Strictly speaking this is only true on a spin manifold. 
Note that $\int d^4x (8\pi)^{-1}  F_{\mu\nu} \tilde F_{\mu\nu}= \pi \int (F/2\pi)^2=\pi \int c_1(F)^2$.
On a spin manifold, the intersection form is even, and the last expression is an integral multiple of $2\pi$. For the subtlety on non-spin manifolds, see \cite{Witten:1995gf}.}.  We call it the $T$ transformation.
This does change $\vec E_\text{conserved}$ by adding $\vec B$, however. Equivalently, it changes the set of charges $(n,m)$ as follows: \begin{equation}
\begin{array}{r|ccc}
\text{particle 1} & (n,m) & \stackrel{T}{\longrightarrow} & (n+m,m), \\
\text{particle 2} &(n',m') & \stackrel{T}{\longrightarrow} & (n'+m',m'), \\
\text{Dirac pairing} & nm'-n'm & = &  (n+m)m'-(n'+m')m.
\end{array}
\end{equation} We see that the Dirac pairing of two particles remain unchanged. 
On the complexified coupling $\tau(\phi)$, it operates as \begin{equation}
\tau(\phi)_\text{old}  \stackrel{T}{\longrightarrow} \tau(\phi)_\text{new} = \tau(\phi)_\text{old}+1.
\end{equation}
The transformations $S$ and $T$ generates the action of $\SL(2,\bZ)$ on the set of charge $(n,m)$:
\begin{align}
S&=\begin{pmatrix}
0&-1\\
1&0
\end{pmatrix}, & 
S \begin{pmatrix}
n \\
m
\end{pmatrix} &=
\begin{pmatrix}
-m \\
n
\end{pmatrix},&
S\tau &= -\frac1\tau \label{S}\\
T&=\begin{pmatrix}
1&1\\
0&1
\end{pmatrix}, & 
T \begin{pmatrix}
n \\
m
\end{pmatrix} &=
\begin{pmatrix}
n+m \\
m
\end{pmatrix},&
T\tau &= \tau+1. \label{T}
\end{align}
In general the action on $\tau$ is the fractional linear transformation \begin{equation}
\begin{pmatrix}
a&b \\
c&d
\end{pmatrix} \in \SL(2,\bZ) : 
\tau \to \frac{d\tau+b}{c\tau+a}.
\end{equation}
%%% comment to self 
% this looks a bit funny, but the monodromy matrix acts on \tau "from the right",
% e.g. the potential energy between two particles of the same (n,m) is given by 
% (n,m) f(\tau) (n,m)^T
%%%

\subsection{'t Hooft-Polyakov monopoles}\label{sec:monopole}
Here we summarize the features of magnetic monopoles which we will repeatedly quote in the rest of the lecture note. For a detailed exposition of topics discussed in this subsection, the readers should consult the reviews such as \cite{Harvey:1996ur,Weinberg:2006rq}, or the textbook \cite{Shifman}. The review by Coleman \cite{Coleman:1982cx} is also very instructive.\footnote{Unfortunately this review is not in the compilation ``Aspects of Symmetry''.  A french translation by R.~Stora is also available as  \cite{Coleman1982cxF}, which was typeset much more beautifully than the version in \cite{Coleman:1982cx}.}
\subsubsection{Classical features}
Consider an $\SU(2)$ gauge theory with a scalar in the adjoint representation, with the action \begin{equation}
S=\int d^4 x\frac{1}{g^2}\left[\frac{1}{2}\tr F_{\mu\nu} F_{\mu\nu} + \tr D_\mu\Phi D_\mu\Phi \right].
\end{equation} The field $\Phi$ is a traceless Hermitean $2\times 2$ matrix.

Consider the vacuum where \begin{equation}
\Phi= \begin{pmatrix}
a & 0 \\
0 & -a
\end{pmatrix}.\label{phivev}
\end{equation} When $a\neq 0$, the $\SU(2)$ gauge symmetry is broken to $\U(1)$. Indeed, the vev \eqref{phivev} commutes with a gauge field strength of the form \begin{equation}
F_{\mu\nu}=\begin{pmatrix}
F_{\mu\nu}^{\U(1)} & 0\\
0 & -F_{\mu\nu}^{\U(1)} 
\end{pmatrix}\label{u1embed}
\end{equation} where $F_{\mu\nu}^{\U(1)}$ is a $\U(1)$ gauge field strength normalized as in Sec.~\ref{charges}.
Note that the quanta of off-diagonal components of the scalar field $\Phi$ have electric charge $\pm 2$ under this $\U(1)$ field, as  can be found by expanding the covariant derivative. 

 We are considering a gauge theory; therefore the field $\Phi$ does not have to be given exactly as in the right hand side of \eqref{phivev}. Rather, we just need that $\Phi$ has eigenvalues $\pm a$. 
  Then we can consider a configuration of the form  \begin{equation}
\Phi(x)=\frac{x_i \sigma^i}{|x|} f(|x|) a
%F_{ij}(x)=\epsilon_{ija}\sigma^a h(|x|) \frac{1}{|x|^2},
\label{tHP}
\end{equation} where $i=1,2,3$ and $f(r)$ is a dimensionless function such that \begin{equation}
\lim_{r\to 0} f(r)=0,\qquad
\lim_{r\to \infty} f(r)=1.
\end{equation} At the spatial infinity, the vev of $\Phi$ is conjugate to \eqref{phivev}, and therefore this configuration can be thought of as an excitation of the vacuum given by \eqref{phivev}. 

The unbroken $\U(1)$ within $\SU(2)$ is along $\Phi$. A more general definition of the $\U(1)$ field strength $F_{\mu\nu}^{\U(1)}$, at least when $r\gg 0$, is then the combination \begin{equation}
F_{\mu\nu}^{\U(1)}:=\frac{1}{2a}\tr F_{\mu\nu} \Phi .\label{gauge-inv-def-of-u1}
\end{equation} 
In the region $r \gg 0$, let us try to bring the configuration \eqref{tHP} to \eqref{phivev} by a gauge transformation. This can be done smoothly except at the south pole, by using the gauge transformation \begin{equation}
\exp[ i\frac{\varphi}{2} (-\sigma^1 \sin\theta + \sigma^2 \cos\theta )],\quad
\text{where}\quad \frac{\vec x}{|x|}=(\cos\varphi,\sin\varphi\cos\theta,\sin\varphi\sin\theta).
\end{equation} 
This gives a gauge transformation around the south pole given by \begin{equation}
i (-\sigma^1 \sin\theta + \sigma^2\cos\theta) =  \exp[ -i\theta \sigma^3 ] \cdot (i\sigma^2).
\end{equation} 
As $\theta$ goes from $0$ to $2\pi$, we see that the $\U(1)$ field $F^{\U(1)}_{\mu\nu}$ has  the magnetic charge $m=1$, and therefore is a monopole. This was originally found by 't Hooft and Polyakov. 
Note that its Dirac pairing with the particle of the field $\Phi$ is 2, which is twice the minimum allowed value. 

Let us evaluate the energy contained in the field configuration. The kinetic energy is $1/g^2$ times \begin{align}
\int d^3x \left[\tr B_{i} B_i + \tr D_i\Phi D_i\Phi \right]
&=\int d^3 x \left[ \tr(B_i\mp D_i\Phi)^2 \pm 2 \tr B_i D_i\Phi \right]\\
&\ge \pm 2 \int d^3 x  \tr B_i D_i\Phi = \pm 2 \int d^3 x \partial_i \tr B_i\Phi \\
&= \pm 2 \int_{S^2}  d\vec n \cdot \tr \vec B\Phi 
\end{align}  where the final integral is over the sphere at the spatial infinity,
which according to \eqref{gauge-inv-def-of-u1} evaluates to $\pm 2(2a)(2\pi m)$, where $m$ is the magnetic charge.
 Therefore we have the  bound \begin{equation}
\text{(energy of the monopole)}\ge \frac{4\pi }{g^2} (2a) |m|  \label{monopole-energy}
\end{equation} This is called the Bogomolnyi-Prasad-Sommerfield (BPS) bound. The inequality is saturated if and only if \begin{equation}
B_i=\pm D_i\Phi,
\end{equation} which is called the BPS equation. This fixes the form of the function $f(r)$ in \eqref{tHP}.

\subsubsection{Semiclassical features}
Given such an explicit monopole solution, there is a way to construct other solutions related by the symmetry. 
First, the configuration \eqref{tHP} has a center at the origin of the coordinate system.
We can shift the center of the monopole at an arbitrary point $\vec y$ of the spatial $\bR^3$. 
These give three zero-modes. 

Another zero mode is obtained by the gauge transformation:
 \begin{equation}
e^{ i \alpha \Phi/a }\label{gauge-zm}
\end{equation}
Note that a gauge transformation which vanishes at infinity is a redundancy of the physical system, but a gauge transformation which does not vanish at infinity is considered to change the classical configuration. 
For general $\alpha$,  this transformation \eqref{gauge-zm} changes the asymptotic behavior of $F_{ij}(x)$,
but for $\alpha=\pi$, the transformation \eqref{gauge-zm} trivially acts on the fields in the adjoint representation.
Therefore $\alpha$ is an angular variable $0\le \alpha < \pi$. 

The semiclassical quantization of the monopole involves the Fock space of non-zero modes, together with a wavefunction $\psi(\vec y,\alpha)$ depending on the zero modes $\vec y$ and $\alpha$.
The wavefunction along $\vec y$ represents the spatial motion of the center of mass of the monopole.
The wavefunction along $\alpha$ represents the electric charge of the monopole, which can be seen as follows.

By comparing \eqref{gauge-zm} with \eqref{u1embed} and \eqref{gauge-inv-def-of-u1},
we see that the unbroken $\U(1)$ global gauge transformation by $e^{i\varphi}$
shifts $\alpha$ by \begin{equation}
\alpha\to \alpha+\varphi.\label{shift-el}
\end{equation}
Recall that a state $\ket\psi$ with electric charge $n$ behaves under the $\U(1)$ global transformation by $e^{i\varphi}$ by \begin{equation}
\ket\psi \to e^{in\varphi} \ket\psi.\label{www}
\end{equation}
Now, as $\alpha$ is a variable with period $\pi$, $\psi(\alpha)$ can be expanded as a linear combination of $e^{i2d\alpha}$ where $d$ is an integer.  Under \eqref{shift-el}
the wavefunction changes as in \eqref{www} with $n=2d$, therefore we see that the monopole state with this zero-mode wave function has the electric charge $2d$.

Summarizing, the combination of the electric charge and the magnetic charge $(n,m)$ we obtain from the semi-classical quantization has the form $(n,m)=(2d,1)$ where $d$ is an integer. 
This was found originally by Julia and Zee: once we quantize the 't Hooft-Polyakov monopole, we not only have a purely-magnetic monopole but a whole tower of dyon states, with $d=-\infty$ to $+\infty$.

Finally let us consider the effect of the fermionic zero modes in the 't Hooft-Polyakov monopole \eqref{tHP}.
First let us consider two Weyl fermions $\lambda$, $\tilde\lambda$ in the adjoint representation, with the Lagrangian \begin{equation}
\tr \bar\lambda \slash D \lambda + \tr\bar{\tilde\lambda} \slash D \tilde\lambda 
+ c(\tr  \lambda [\Phi,  \tilde\lambda ] + \tr \bar \lambda [\Phi,\bar{\tilde\lambda}]).
\end{equation}
We regard both the gauge potential in the covariant derivative $D$ and the scalar field as backgrounds, and  decompose $\lambda$, $\tilde \lambda$ into eigenstates of the angular momentum. 
The lower bound of the orbital angular momentum is given by the Dirac pairing, which is $\hbar$ here. 
The spinor fields have spin $\hbar/2$. Therefore the state with lowest angular momenta has spin $\hbar/2$.
When the coefficient $c$ takes a value in a certain range, it is known that there is a pair of zero modes 
$b_\alpha$ where $\alpha=1,2$ the spinor index of the $\SO(3)$ spatial rotation. The semiclassical quantization promotes them into a pair of fermionic oscillators \begin{equation}
\{b_\alpha,b_\beta^\dagger\}=\delta_{\alpha\beta}.
\end{equation}
This creates four states starting from one state $\ket\psi$ from the semiclassical quantization of the bosonic part: \begin{equation}
\begin{array}{ccccc}
&\raisebox{-.8ex}{\rotatebox{30}{$\leftrightarrow$}}& b^\dagger_1 \ket\psi & \leftrightarrow & b^\dagger_1 b^\dagger_2 \ket\psi \\
\ket\psi & \leftrightarrow & b^\dagger_2\ket\psi & \rotatebox{30}{$\leftrightarrow$}  
\end{array}.
\end{equation} This counts as one complex boson and one fermion. 

Suppose we introduce another pair $\lambda'$, $\tilde\lambda'$ of  the adjoint Weyl fermions. Then we will have another pair of fermionic oscillators $b_\alpha'$. Together, they generate $2^4=16$ states,
consisting of one massive vector (with 3 states), four massive spinors (with 8 states) and five massive scalars. 

Next, consider having $2N$ Weyl fermions $\psi_i^a$ in the doublet representation where $a=1,2$ and $i=1,\ldots,2N$, with the Lagrangian \begin{equation}
 \bar\psi_i \slash D \psi_i 
+ c'  (\psi^a_i \Phi_{(ab)}  \psi^b_i +\bar\psi^a_i \Phi_{(ab)}  \bar\psi^b_i ).
\end{equation}  Note that the Lagrangian has an $\SO(2N)$ flavor symmetry acting on the index $i$.

The electric charge of the quanta of $\psi$, $\tilde \psi$ with respect to the unbroken $\U(1)$ is now $1$. 
Then the Dirac pairing is $\hbar/2$. Tensoring with the intrinsic spin $\hbar/2$, we find that the minimal orbital angular momentum is $0$.  
It is known that for a suitable choice of $c'$, this fermion system has zero modes $\gamma_i$, $i=1,\ldots,2N$.
After semiclassical quantization, it becomes a set of fermionic operators with the commutation relation \begin{equation}
\{\gamma_i,\gamma_j\}=\delta_{ij}.
\end{equation}
This is the commutation relation of the gamma matrices of $\SO(2N)$.
Monopole states are representations of $\gamma_i$'s, meaning that they transform as a spinor representation of the flavor symmetry $\SO(2N)$.  

Fields in a doublet representation of the $\SU(2)$ gauge symmetry has an another effect. Namely, in the gauge zero mode \eqref{gauge-zm}, $\alpha=\pi$ gives the matrix \begin{equation}
\begin{pmatrix}
-1 & 0 \\
0 & -1
\end{pmatrix} 
\end{equation} which acts nontrivially on the fields in the doublet representation. 
Then the periodicity of the gauge zero mode $\alpha$ is now $2\pi$, and the wavefunction along the $\alpha$ direction can now be $e^{in\alpha}$ for arbitrary integer $n$.
Therefore, the electric charge $n$ can either be even or odd. 
The operators $\gamma_i$ come from the modes of the fields in the doublet representation, and therefore it changes the electric charge by $\pm 1$. 

We can define the flavor spinor chirality $\Gamma$ by \begin{equation}
\Gamma=\gamma_1\gamma_2 \cdots \gamma_{2N},
\end{equation} by which the spinor of $\SO(2N)$ can be split into positive-chirality and negative-chirality spinors.  The action of the operators $\gamma_i$ changes the chirality of the flavor spinors. 
Combined with the behavior of the $\U(1)$ electric charge we saw in the previous paragraph, we conclude that the parity of the $\U(1)$ electric charges of the monopole states 
is correlated with the chirality of the flavor spinor representation. 

\section{$\cN{=}2$ multiplets and Lagrangians}\label{sec:lagrangian}

\subsection{Microscopic Lagrangian}
\subsubsection{$\cN{=}1$ superfields}
Let us now move on to the  construction of the Lagrangian with $\cN{=}2$ supersymmetry.
An $\cN{=}2$ supersymmetric theory is in particular an $\cN{=}1$ supersymmetric theory.
Therefore it is convenient to use $\cN{=}1$ superfields to describe $\cN{=}2$ systems. 
For this purpose let us quickly recall the $\cN{=}1$ formalism.  
In this section only, we distinguish the imaginary unit by writing it as $\ii$.

An $\cN{=}1$ vector multiplet consists of a Weyl fermion $\lambda_\alpha$ and a vector field $A_\mu$,
both in the adjoint representation of the gauge group $G$. We combine them into the superfield $W_\alpha$ with the expansion \begin{equation}
W_\alpha=\lambda_\alpha+F_{(\alpha\beta)}\theta^\beta + D\theta_\alpha+\cdots
\end{equation} where $D$ is an auxiliary field, again in the adjoint of the gauge group. 
$F_{\alpha\beta}=\frac \ii2 \sigma^\mu{}^\beta_{\dot\gamma} \bar\sigma^{\nu}{}^{\dot \gamma}_\alpha F_{\mu\nu}$ is the anti-self-dual part of the field strength $F_{\mu\nu}$.

The kinetic term for a vector multiplet is given by \begin{equation}
\int d^2\theta \frac{-\ii}{8\pi} \tau \tr W_\alpha W^\alpha + cc. 
\end{equation} where \begin{equation}
\tau=\frac{4\pi \ii}{g^2}+\frac{\theta}{2\pi}
\end{equation} is a complex number combining the inverse of the coupling constant and the theta angle. We call it the complexified coupling of the gauge multiplet.
Expanding in components, we have \begin{equation}
\frac{1}{2g^2} \tr F_{\mu\nu}F^{\mu\nu} + \frac{\theta}{16\pi^2} \tr F_{\mu\nu} \tilde F^{\mu\nu} + \frac{1}{g^2}\tr D^2 -\frac{2\ii}{g^2}\tr \bar\lambda\slash\partial \lambda.
\end{equation} We use the convention that $\tr T^a T^b = \frac12 \delta^{ab}$ for the standard generators of gauge algebras, which explain why we have the factors $1/(2g^2)$ in front of the gauge kinetic term. 
The $\theta$ term is a total derivative of a gauge-dependent term. Therefore, it does not affect to perturbative computations. It does affect non-perturbative computations, to which we will come back later. 

An $\cN{=}1$ chiral multiplet $Q$ consists of a complex scalar $Q$ and a Weyl fermion $\psi_{\alpha}$, both in the same representation of the gauge group. 
In terms of a superfield we have \begin{equation}
Q=Q\big|_{\theta=0} + 2\psi_\alpha \theta^\alpha + F \theta_\alpha \theta^\alpha
\end{equation} where $F$ is auxiliary. The coefficient 2 in front of the middle component is unconventional, but this choice allows us to remove various annoying factors of $\sqrt{2}$ appearing in the formulas later. 
The chiral multiplet $Q_{1,\ldots}$ can be in an arbitrary complex representation $R$ of the gauge group $G$. The Lagrangian density is then \begin{equation}
\int d^4\theta Q^\dagger{}^{j} e^{V^a \rho_a{}^{i}_{j}} Q_i + \int d^2\theta W(Q)+ cc.
\end{equation}  where $V$ is the vector superfield, $\rho_a{}^i_j$ is the matrix representation of the gauge algebra, and $W(Q)$ is a gauge invariant holomorphic function of $Q_{1,\ldots}$.

The supersymmetric vacua is obtained by demanding that the supersymmetry transformation of various fields are zero. The nontrivial conditions come from\begin{equation}
\delta \lambda_\alpha =0, \qquad \delta\psi_\alpha=0
\end{equation} which give \begin{equation}
D_a=0, \quad F_i=0.
\end{equation} By solving the algebraic equations of motion of the auxiliary fields, we find \begin{equation}
Q^\dagger_{\bar j} \rho_a^{\bar ji}Q_i =0,\quad
\frac{\partial W}{\partial Q_i}=0.
\end{equation}

\subsubsection{Vector multiplets and hypermultiplets}\label{sec:hyperdef}
An $\cN{=}2$ vector multiplet consists of the following $\cN{=}1$ multiplets, both in the adjoint of the gauge group $G$: \begin{equation}
\begin{array}{ccccc@{\qquad}c}
&\raisebox{-.8ex}{\rotatebox{30}{$\leftrightarrow$}}& \lambda_\alpha & \leftrightarrow & A_\mu & \text{$\cN{=}1$ vector multiplet},\\
\Phi & \leftrightarrow & \tilde\lambda _\alpha &  \rotatebox{30}{$\leftrightarrow$} &  & \text{$\cN{=}1$ chiral multiplet}.
\end{array}
\end{equation} Here, the horizontal arrows signify the $\cN{=}1$ sub-supersymmetry generator manifest in the $\cN{=}1$ superfield formalism, and the slanted arrows are for the second $\cN{=}1$ sub-supersymmetry.

One easy way to construct the second supersymmetry action is to demand that the theory is symmetric under the $\SU(2)$ rotation acting on $\lambda_\alpha$ and $\tilde\lambda _\alpha$. 
A symmetry which does not commute with the supersymmetry generators is called an R-symmetry in general.
Therefore this $\SU(2)$ symmetry is often called the $\SU(2)_R$ symmetry.
It is by now a standard technique to combine  the supersymmetry manifest in a superfield formalism and an R-symmetry to construct a theory with more supersymmetries, see e.g.~\cite{Aharony:2008ug} for an application. 
It is also to be kept in mind that there can be and indeed are $\cN{=}2$ supersymmetric theories without $\SU(2)_R$ symmetry: there can just be two sets of supersymmetry generators without $\SU(2)$ symmetry relating them, see e.g.~\cite{Antoniadis:1995vb,FujiwaraItoyamaSakaguchi}.
That said, for simplicity, we only deal with $\cN{=}2$ supersymmetric systems with $\SU(2)_R$ symmetry in this lecture note. 

The Lagrangian is then \begin{equation}
\frac{\Im\tau}{4\pi} \int d^4\theta \tr \Phi^\dagger e^{[V,\cdot]} \Phi +\int d^2\theta \frac{-\ii}{8\pi}\tau \tr W_\alpha W^\alpha + cc.\label{vectorlag}
\end{equation} The ratio between the prefactors of the K\"ahler potential and of the gauge kinetic term is fixed by demanding $\SU(2)_R$ symmetry.

An $\cN{=}2$  hypermultiplet\footnote{There is a stupid convention that we use a space between `vector' and `multiplets' to spell ``vector multiplets'', but not for ``hypermultiplets''.  Colloquially, hypermultiplets are often just called hypers.} consists of the following fields:
\begin{equation}
\begin{array}{ccccc@{\qquad}c}
&\raisebox{-.8ex}{\rotatebox{30}{$\leftrightarrow$}}& Q & \leftrightarrow & \psi & \text{$\cN{=}1$ chiral multiplet}\\
\tilde\psi^\dagger & \leftrightarrow &\tilde Q^\dagger &   \rotatebox{30}{$\leftrightarrow$}  &  & \text{$\cN{=}1$ antichiral multiplet}
\end{array} 
\end{equation} They are both in the same representation $R$ of the gauge group.
Therefore, the $\cN{=}1$ chiral multiplets $Q$ and $\tilde Q$ are in the conjugate representations of the gauge group.
We demand again that the theory is symmetric under the $\SU(2)$ rotation acting on $Q$ and $\tilde Q^\dagger$, to have $\cN{=}2$ supersymmetry.

For definiteness, let us consider $G=\SU(N)$ and $N_f$ hypermultiplets $Q_i^a$, $\tilde Q^i_a$ in the fundamental $N$-dimensional representation, where $a=1,\ldots,N$ and $i=1,\ldots,N_f$. 
This set of fields is often called $N_f$ flavors of fundamentals of $\SU(N)$.  The gauge transformation acts on them as \begin{equation}
Q_i \to e^{\Lambda} Q_i,\qquad
\tilde Q^i \to  \tilde Q^i e^{-\Lambda}
\end{equation} where $\Lambda$ is a traceless $N\times N$ matrix of chiral superfields; the gauge indices are suppressed. 

The Lagrangian for the  hypermultiplets is  \begin{equation}
c\int d^4\theta (Q^\dagger{}^i e^V Q_i+ \tilde Q^i e^{-V} \tilde Q^\dagger{}_i)
+c'(\int d^2 \theta \tilde Q^i \Phi Q_i + cc.)
+(\int d^2\theta \mu^i_j \tilde Q^jQ_i + cc.)\label{hyperlag}
\end{equation} where the gauge index $a$ is suppressed again.
The existence of $\SU(2)_R$ symmetry fixes the ratio of $c$ and $c'$: it can be done e.g.~by comparing the coefficients of $ Q^i \lambda \psi$ from the first term and of $ \tilde Q^i  \tilde\lambda \psi$ from the second term.  We find the choice $c=c'$ does the job. In the following we take $c=c'=1$ unless otherwise mentioned. 
The $\SU(2)_R$ symmetry also demands that the mass term $\mu^i_j$ satisfies $[\mu,\mu^\dagger]=0$.
Then $\mu$ can be diagonalized, and consequently the mass term is often written as \begin{equation}
\sum_i \int d^2\theta \mu_i \tilde Q^i Q_i + cc.
\end{equation}

As another example, let us consider the case when we have a hypermultiplet $(Z,\tilde Z)$ in the adjoint representation, i.e.~they are both $N\times N$ traceless matrices. 
The following discussion can easily be generalized to arbitrary gauge group too. 
When the hypermultiplet is massless, the total Lagrangian has the form \begin{multline}\int d^2\theta \frac{-\ii}{8\pi}\tau \tr W_\alpha W^\alpha + cc. 
+\frac{\Im\tau}{4\pi} \int d^4\theta \tr \Phi^\dagger e^{[V,\cdot]} \Phi \\
+\frac{\Im\tau}{4\pi} \int d^4\theta (Z^\dagger{} e^{[V,\cdot]} Z+ \tilde Z e^{-[V,\cdot]} \tilde Z^\dagger{})
+\frac{\Im\tau}{4\pi} \int d^2 \theta \tilde Z [\Phi, Z] + cc.\label{n4lag}
\end{multline} where we made a different choice of $c=c'$ in \eqref{hyperlag}.
This Lagrangian clearly has $\SU(3)_F$ flavor symmetry rotating $\Phi$, $Z$ and $\tilde Z$. This commutes with the $\cN{=}1$ supersymmetry manifest in the superfield formalism.  We also know that this theory has an $\SU(2)_R$ symmetry rotating $Z$ and $\tilde Z^\dagger$. These two symmetries $\SU(3)_F$ and $\SU(2)_R$ does not commute: we find that there is an $\SO(6)_R$ symmetry, acting on \begin{equation}
\Re \Phi, \Im \Phi, \Re Z, \Im Z, \Re \tilde Z, \Im \tilde Z.
\end{equation} 
Note that $\SO(6)_R$ can also be regarded as $\SU(4)_R$, as $\SO(6)$ and $\SU(4)$ have the same Lie algebra. 
Then the $\SU(4)_R$ symmetry acts on the four Weyl fermions \begin{equation}
\lambda, \tilde\lambda, \psi, \tilde \psi
\end{equation} in the system, where $\lambda$ and $\tilde\lambda$  are in the $\cN{=}2$ vector multiplet, and $\psi$, $\tilde \psi$ are in the $\cN{=}2$ hypermultiplet.  
We conclude that this system has in fact $\cN{=}4$ supersymmetry, whose four supersymmetry generators are acted on by $\SU(4)_R\simeq \SO(6)_R$.  The argument here is another application of the combination of the manifest and non-manifest symmetries in the superfield formalism. 

We can add the mass term $\int d^2\theta \mu Z\tilde Z + cc.$ to \eqref{n4lag}. This preserves the $\cN{=}2$ supersymmetry but it breaks $\cN{=}4$ supersymmetry. The resulting theory is sometimes called the $\cN{=}2^*$ theory.

Before closing this section, we should mention the concept of half-hypermultiplet.  Let us start from a full hypermultiplet $(Q_a,\tilde Q^a)$ so that $Q_a$ and $\tilde Q^a$ are in the representations $R$, $\bar R$, respectively. When $R$ is pseudo-real, or equivalently when there is an antisymmetric invariant tensor $\epsilon_{ab}$, we can impose the constraint \begin{equation}
Q_a=\epsilon_{ab} \tilde Q^b
\end{equation} compatible with $\cN{=}2$ supersymmetry, which halves the number of degrees of freedom in the multiplet. The resulting multiplet is called a half-hypermultiplet in the representation $R$. We will come back to this in Sec.~\ref{sec:hyperrevisited}.

\subsection{Vacua}\label{sec:classicalvac}

The combined system of the vector multiplet and the hypermultiplets has the Lagrangian which is the sum of \eqref{vectorlag} and \eqref{hyperlag}. The supersymmetric vacua are given by the following conditions. 

First, the variation of the $D$ auxiliary fields gives \begin{equation}
\frac{1}{g^2}[\Phi^\dagger,\Phi]+ (Q_i Q^\dagger{}^i-\tilde Q^\dagger{}_i \tilde Q^i)\big|_\text{traceless}=0,\label{Deq}
\end{equation}  where $X|_\text{traceless}$ for an $N\times N$ matrix is defined by \begin{equation}
X|_\text{traceless}=X-\frac1N\tr X.
\end{equation} We use the convention that a scalar is multiplied by a unit matrix when necessary. 

Second, the variation of the $F$ auxiliary field of $\Phi$ gives \begin{equation}
Q_i \tilde Q^i \big|_\text{traceless}=0\label{Feq}
\end{equation} and the $F$ auxiliary fields of $Q_i$, $\tilde Q^i$ give \begin{equation}
\Phi Q_i +\mu^j_i Q_j =0, \qquad \tilde Q^i \Phi + \mu^i_j \tilde Q^j=0 \label{Qeq}
\end{equation} for all $i$.
The total scalar potential is a weighted sum of absolute values squared of \eqref{Deq}, \eqref{Feq} and \eqref{Qeq}. 

So far we only used the supersymmetry condition with respect to the $\cN{=}1$ supersymmetry manifest in the superfield notation. By massaging the cross terms between the first term and the second term of \eqref{Deq}  and combining them with the squares of \eqref{Qeq}, 
 we can re-write the total scalar potential as a weighted sum of the following objects.
First, we have one term purely of $\Phi$: \begin{equation}
[\Phi^\dagger,\Phi]=0.\label{phieq}
\end{equation} Second, we have terms purely of $Q$ and $\tilde Q$: one is \begin{equation}
(Q_i Q^\dagger{}^i-\tilde Q^\dagger{}_i \tilde Q^i)\big|_\text{traceless}=0\label{tripleteq}
\end{equation} and another is \eqref{Feq}. 
Finally, we have terms mixing $\Phi$ and $Q$, which are \eqref{Qeq} together with \begin{equation}
\Phi^\dagger Q_i +\mu^\dagger{}^j_i Q_j =0, \qquad \tilde Q^i \Phi^\dagger + \mu^\dagger{}^i_j \tilde Q^j=0 \label{QQeq}.
\end{equation}

Note that \eqref{phieq} and \eqref{tripleteq} are the $\SU(2)_R$ singlet and triplet parts of the equation \eqref{Deq}, respectively. Furthermore, the equation \eqref{tripleteq} together with the real and the imaginary parts of the equation \eqref{Feq} form  the triplet of $\SU(2)_R$. 
Finally, the equations \eqref{Qeq} and \eqref{QQeq} transform as a doublet of $\SU(2)_R$. 

Let us summarize. We first  demanded that one $\cN{=}1$ sub-supersymmetry is unbroken in \eqref{Deq}, \eqref{Feq} and \eqref{Qeq}. We found the equations satisfied are automatically $\SU(2)_R$ invariant, and therefore we see that all the $\cN{=}2$ supersymmetry is automatically unbroken.

One easy way to have supersymmetry is to demand \eqref{phieq} and set $Q=\tilde Q=0$. This subspace of the supersymmetric vacuum moduli is called the Coulomb branch, since there usually remain a number of Abelian gauge fields in the infrared. 
 
When the mass terms $\mu^i_j$ are nonzero, it is not straightforward to discuss other vacuum configurations in general. When $\mu^i_j=0$, there is another class of vacuum configurations, given by  
 just  demanding \eqref{tripleteq} and \eqref{Feq}, and setting $\Phi=0$. This is called the Higgs branch. Some people in the field reserve the word the Higgs branch for the branch where the gauge group is completely broken, but theoretically the Higgs branch as defined here behaves more uniformly under various operations. 

The branches with when both the hypermultiplet scalars $Q$, $\tilde Q$ and the vector multiplet scalars $\Phi$ are nonzero are called the mixed branches. 

From \eqref{phieq} we see that $\Phi$ can be diagonalized in the supersymmetric vacua. 
For definiteness let $G=\SU(2)$. Then $\Phi=\diag(a,-a)$. When $a\neq 0$ this breaks the gauge group to $\U(1)$.
As there is a Coulomb field remaining in the infrared, these vacua are called the Coulomb branch. 
Let us compute the mass of the resulting W-bosons. From \begin{equation}
\frac1{g^2}\tr|D_\mu \Phi|^2 = \frac1{g^2}\tr (\partial_\mu \Phi+[A_\mu,\Phi])^2
\end{equation} we have a term \begin{equation}
\frac1{g^2}\tr [A_\mu,\vev{\Phi}]^2
\end{equation} in the Lagrangian, which gives a mass to the vector field. Writing \begin{equation}
A_\mu=\begin{pmatrix}
A^0 & W^+ \\
W^- & -A^0
\end{pmatrix}_\mu,
\end{equation} we find \begin{equation}
\left[
\begin{pmatrix}
0 & W^+_\mu \\
0 & 0 
\end{pmatrix},
\begin{pmatrix}
a & 0\\
0 & -a
\end{pmatrix}
\right]=-2a \begin{pmatrix}
0 & W^+_\mu \\
0 & 0 
\end{pmatrix}.
\end{equation} The kinetic term in our convention is $\tr F_{\mu\nu}F_{\mu\nu}/(2g^2)$, and therefore this gives the mass \begin{equation}
M_W=|2a|.
\end{equation}  The mass terms of the fields $Q_i$, $\tilde Q^i$ for fixed $i$ are \begin{equation}
\tilde Q^i \Phi Q_i+ \mu_i \tilde Q^i Q_i =
(\tilde Q_1^i, \tilde Q_2^i)
\begin{pmatrix}
a+\mu_i & 0 \\
0 & -a+\mu_i
\end{pmatrix}
\begin{pmatrix}
Q_1^i \\
Q_2^i
\end{pmatrix}.
\end{equation} Therefore we have \begin{equation}
M_{Q_{i,1}}=|a+\mu|,\qquad
M_{Q_{i,2}}=|{-a}+\mu|.
\end{equation}

We studied the classical mass of the monopole in this model in \eqref{monopole-energy} when $\theta=0$. In general, this is given by  \begin{equation}
M_\text{monopole}=|2\tau a|.
\end{equation} Classically, there is a general inequality for the mass of a particle \begin{equation}
M \geq |na+m(2\tau a) + \sum_i f_i \mu_i| \label{classicalBPS}
\end{equation}
where $n$, $m$, $f_i$ are the electric, magnetic and flavor charges of the particle.
Here the $i$-th flavor charges are associated to the symmetry \begin{equation}
Q_i \to e^{\ii\varphi_i} Q_i,\quad
\tilde Q^i \to e^{-\ii\varphi_i} \tilde Q^i.
\end{equation}
This inequality, called the Bogomolnyi-Prasad-Sommerfield (BPS) bound, persists in the quantum system, once quantum corrections are taken into account to $a$ and $2\tau a$. Let us study this point next.

\subsection{BPS bound}
The general $\cN{=}2$ supersymmetry algebra has the following form \begin{align}
\{Q^I_\alpha,Q^\dagger{}^{\bar J}_{\dot \beta}\} &= \delta^{I\bar J} P_{\mu} \sigma^\mu_{\alpha\dot\beta},\\
\{Q^I_\alpha , Q^J_\beta\}&= \epsilon^{IJ} \epsilon_{\alpha\beta}Z.\label{Zcomm}
\end{align} Here $I=1,2$ are the index distinguishing two supersymmetry generators,
and  $Z$ is a complex quantity which commutes with everything. 
Let us take the coordinate system where \begin{equation}
P_\mu = (M,0,0,0).
\end{equation} This choice breaks the Lorentz symmetry $\SO(3,1)$ to the spatial rotation $\SO(3)$, which allows us to identify the undotted and the dotted spinor indices. Let us then define \begin{equation}
{}^{(\varphi)}Q_\alpha=\frac{1}{\sqrt2}(Q^1_\alpha + e^{-\ii\varphi} \sigma^0{}_{\alpha}{}^{\dot\beta} Q^\dagger{}^2_{\dot\beta})
\end{equation} for which we have \begin{equation}
\{
{}^{(\varphi)}Q_\alpha,
{}^{(\varphi)}Q^\dagger_\beta
\}=\delta_{\alpha\beta} (M-\Re (e^{-\ii\varphi} Z)).\label{varphi}
\end{equation}

In general, if there is an operator $a$ satisfying $\{a,a^\dagger\}=c$ with a constant $c$, $c$ is necessarily non-negative.
Indeed, take a ket vector $\ket{\psi}$ then \begin{equation}
\left|a^\dagger \ket{\psi}\right|^2
+\left|a \ket{\psi}\right|^2
=\vev{\psi|a a^\dagger|\psi}+\vev{\psi|a^\dagger a|\psi} = c\vev{\psi|\psi},\label{tmp}
\end{equation} meaning that $c\ge 0$. From \eqref{varphi}, then, we see \begin{equation}
M\ge \Re (e^{-\ii\varphi} Z)
\end{equation} for all $\varphi$. Choosing $\varphi=\Arg Z$, 
we find  the inequality \begin{equation}
M\geq |Z|. \label{BPS}
\end{equation}

In general, the multiplet of the supertranslations $Q^I_\alpha$ and $Q^J{}^\dagger_\alpha$ 
generates $2^4=16$ states in the supermultiplet. 
When the  inequality \eqref{BPS} is saturated, $c$ in the equation \eqref{tmp} for $a_\alpha={}^{(\Arg Z)}Q_\alpha$  is zero, forcing the operators ${}^{(\Arg Z)}Q_\alpha$ themselves to vanish.
Then the supertranslations only generate $2^2=4$ states.   Such multiplets are called BPS,
and those multiplets with 16 states under the action of supertranslations are called non-BPS. 
A BPS state is rather robust: under a generic perturbation, the number of states in a multiplet can not jump. Therefore the BPS state will generically stay BPS. 

What is this quantity $Z$, which commutes with everything? 
A quantity commuting with everything is by definition a conserved charge.
When the low-energy theory is  a weakly-coupled $\U(1)$ gauge theory, 
$Z$ is a linear combination of the electric charge $n$, the magnetic charge $m$, 
and the flavor charges $f_i$. We define the coefficients appearing in the linear combination to be $a$, $a_D$ and $\mu_i$ in the quantum theory: \begin{equation}
Z=na+ma_D+\sum_i \mu_i f_i.\label{Zlin}
\end{equation} When the theory is weakly-coupled, we can identify $a$ to be the diagonal entry of the field $\Phi$,
$a_D$ to be $2\tau a$, and $\mu_i$ to be the coefficients of the mass terms in the Lagrangian, by comparing the quantum BPS mass formula \eqref{BPS}  and its classical counterpart \eqref{classicalBPS}.
In the strongly-coupled regime, there is no meaning in saying that $a$ is the diagonal entry of a gauge-dependent field $\Phi$. Rather, we should think of \eqref{Zlin} as the definition of the quantity $a$. 

\subsection{Low energy Lagrangian}\label{sec:lel}
Let us consider a general effective Lagrangian which describes $\U(1)^n$ gauge fields in the infrared. Let us denote $n$ $\U(1)$ vector multiplets by 
\begin{equation}
\begin{array}{ccccc@{\qquad}c}
&\raisebox{-.8ex}{\rotatebox{30}{$\leftrightarrow$}}& \lambda_\alpha & \leftrightarrow & A_\mu & \text{$\cN{=}1$ vector multiplet}\\
a & \leftrightarrow & \tilde\lambda _\alpha & \rotatebox{30}{$\leftrightarrow$}  &  & \text{$\cN{=}1$ chiral multiplet}
\end{array}\label{u1mult}
\end{equation} with additional scripts $i=1,\ldots,n$. A general $\cN{=}1$  supersymmetric Lagrangian is given by \begin{equation}
\frac{1}{8\pi}\int d^4\theta K(\bar a_i,a_j)+\int d^2\theta \frac{-\ii}{8\pi}\tau^{ij}(a) W_{\alpha,i} W^\alpha{}_j + cc.\label{u1kin}
\end{equation} 
Note that we allowed the K\"ahler potential and the gauge coupling matrix to have nontrivial dependence on $a_i$.

We  demand the existence of the $\SU(2)_R$ symmetry rotating $\lambda_\alpha$ and $\tilde\lambda _\alpha$ to guarantee the existence of $\cN{=}2$ supersymmetry.
The kinetic matrix of $\tilde\lambda _\alpha$ is \begin{equation}
\frac{1}{4\pi}\frac{\partial^2 K}{\partial a_i \partial\bar a_j}
\end{equation} and that of $\lambda$ is \begin{equation}
\frac{\Im \tau^{ij}}{2\pi}=\frac{\tau^{ij}-\bar \tau^{ij}}{4\pi \ii}.
\end{equation} Equating them, we have \begin{equation}
\frac{\tau^{ij}-\bar \tau^{ij}}\ii =\frac{\partial^2 K}{\partial a_i \partial\bar a_j}. 
\end{equation} Taking the derivative of both sides by $a_k$, we have \begin{equation}
\frac{\partial}{\partial a_k} \frac{\tau^{ij}}\ii=\frac{\partial^3 K}{\partial a_k\partial a_i \partial\bar a_j}. 
\end{equation} The left hand side is symmetric under $i\leftrightarrow j$,
and the right hand side is symmetric under $k\leftrightarrow i$. 
Therefore, at least locally, $\tau^{ij}$ can be integrated twice: \begin{equation}
\tau^{ij}=\frac{\partial^2 F}{\partial a_i\partial a_j}\label{pretau}
\end{equation} for a locally holomorphic function $F(a)$. 
We define \begin{equation}
a_D^i=\frac{\partial F}{\partial a_i},\label{adi}
\end{equation}then we have \begin{equation}
K=\ii(\bar a_D^i a_i-\bar a_i a_D^i ).\label{Keq}
\end{equation}

A K\"ahler manifold with this additional structure is often called a special K\"ahler manifold. With supergravity, a slightly different structure appears. To distinguish from it, it is also called a rigid special K\"ahler manifold. The same geometry is also called a Seiberg-Witten integrable system, or a Donagi-Witten integrable system.  See e.g.~\cite{Donagi:1997sr,Craps:1997gp,Freed:1997dp} for a review. 
In this context, the fields $a_i$ and $a_D^i$ are called the special coordinates. 

The notations $a_i$ and $a_D^i$ can be justified as follows. 
Suppose we have a hypermultiplet $Q$, $\tilde Q$ charged under the $i$-th vector multiplet only. It has the  superpotential \begin{equation}
W=Qa_i \tilde Q,
\end{equation} which gives the mass \begin{equation}
M_Q=|a_i|.
\end{equation} Therefore, $a_i$ is indeed the coefficient appearing in \eqref{Zlin}. 
To justify the notation $a_D^i$, write down the Lagrangian for the bosons in components: \begin{equation}
\frac{\Im \tau^{ij}}{4\pi} \partial_\mu\bar a_i \partial^\mu a_j
+\frac{\Im \tau^{ij}}{8\pi} F_{\mu\nu\ i} F^{\mu\nu}_j
+\frac{\Re \tau^{ij}}{8\pi} F_{\mu\nu\ i} \tilde F^{\mu\nu}_j.
\end{equation}
Generalizing the argument in Sec.~\ref{ST}, the dual electromagnetic field $F_D$ is given by \begin{equation}
F_D{}_{\mu\nu}^i = \Im \tau^{ij} F_{\mu\nu\ j} + \Re \tau^{ij} \tilde F_{\mu\nu\ j},
\end{equation} in terms of which the kinetic term of the gauge fields is \begin{equation}
\frac{1}{8\pi} \left(\Im \tau_D{}_{ij} F_D{}_{\mu\nu}^{i} F_D{}^{\mu\nu\ j}
+\Re \tau_D{}_{ij} F_D{}_{\mu\nu\ i} \tilde F_D{}^{\mu\nu\ j} \right)
\end{equation}where \begin{equation}
\tau_D{}_{ij} = (-\tau^{-1})_{ij}.
\end{equation} Then we find \begin{equation}
\frac{1}{4\pi}\Im \tau^{ij} \partial_\mu\bar a_i \partial^\mu a_j
=\frac{1}{4\pi}\Im \tau_D{}_{ij} \partial_\mu\bar a_D^i \partial^\mu a_D^j
\end{equation} where $a_D$ is as defined in \eqref{adi}. This means that we have the dual $\cN{=}2$ multiplets
\begin{equation}
\begin{array}{ccccc@{\qquad}c}
&\raisebox{-.8ex}{\rotatebox{30}{$\leftrightarrow$}}& \lambda_D{}_\alpha & \leftrightarrow & A_D{}_\mu & \text{$\cN{=}1$ vector multiplet}\\
a_D & \leftrightarrow & \tilde\lambda _D{}_\alpha & \rotatebox{30}{$\leftrightarrow$}   &  & \text{$\cN{=}1$ chiral multiplet}
\end{array}
\end{equation} where $A_D{}_\mu$ is the gauge potential of $F_D{}_{\mu\nu}$ introduced above, with additional superscripts $i$. 

We introduced the prepotential $F$  in a rather indirect manner in this section, by saying that the kinetic term of the $\U(1)$ vector multiplets \eqref{u1kin} should be given by \eqref{pretau} and \eqref{Keq}.
This can be better understood using $\cN{=}2$ superspace, % the point of view of $\cN{=}2$ vector superfield.
since it is known that the prepotential is the Lagrangian density in the $\cN{=}2$ superspace.
This is similar to the situation where  the K\"ahler potential gives the Lagrangian density in the $\cN{=}1$ superspace. 

Recall that the multiplets \eqref{u1mult} can be summarized in $\cN{=}1$ superfields \begin{equation}
\Phi_i=a_i+2\tilde\lambda _i{}^\alpha \theta_\alpha+\cdots,\qquad
W_i=\lambda_{\alpha\ i} +F_{\alpha\beta}\theta^\beta+\cdots. 
\end{equation} We can introduce another set of supercoordinates $\tilde \theta_\alpha$ to combine them: \begin{equation}
\mathbf{\Phi}_i=\Phi_i+2W_{\alpha\ i} \tilde\theta^\alpha
=a_i + 2\tilde\lambda_{\alpha\ i} \theta^\alpha + 2\lambda_{\alpha\ i}\tilde \theta^\alpha +2 F_{\alpha\beta\ i}\theta^{(\alpha} \tilde\theta^{\beta)}+\cdots.
\end{equation} Then the $\SU(2)$ R-symmetry rotating $\lambda$ and $\tilde \lambda$ 
acts on the two sets of supercoordinates $\theta_\alpha$ and $\tilde \theta_\alpha$.

Now, take an arbitrary holomorphic function of $n$ variables $F(a_1,\ldots,a_n)$, and consider its integral over the chiral  $\cN{=}2$ superspace: \begin{equation}
\int d^2\theta d^2\tilde \theta F(\mathbf{\Phi}_1,\ldots,\mathbf{\Phi}_n) + cc.
\end{equation} It is clear that this gives rise to the structure \eqref{pretau} for the gauge kinetic matrix. 
To obtain the K\"ahler potential \eqref{Keq} one needs to study the structure of the constraints and the auxiliary fields of the $\cN{=}2$ superfields, see e.g.~Sec. 2.10 of \cite{D'Hoker:1999ft}.  
The non-Abelian microscopic action \eqref{vectorlag} has the prepotential $F(\mathbf{\Phi})=\frac12\tau \tr\mathbf{\Phi}^2$.

\section{Renormalization and anomaly}\label{sec:renormalization}
In the last section we constructed the Lagrangian of $\cN{=}2$ supersymmetric field theories. Before going into the analysis of their dynamics, we would like to recall a few basic methods here, namely one-loop renormalization and anomalies. 

\subsection{Renormalization}
Recall the one-loop renormalization of the gauge coupling in a general Lagrangian field theory: \begin{equation}
\LambdaRG\frac{d}{d\LambdaRG}g=-\frac{g^3}{(4\pi)^2}\left[
\frac{11}{3}C(\text{adj})-\frac23  C(R_f)-\frac13   C(R_s)
\right].
\end{equation}Here, $\LambdaRG$ is the energy scale at which $g$ is measured, and we use the convention that all fermions are written in terms of left-handed Weyl fermions.
Then $R_f$ and $R_s$ are the representations of the gauge group to which the Weyl fermions and the complex scalars belong, respectively.
The quantity $C(\rho)$ is defined so that \begin{equation}
\tr \rho(T^a) \rho(T^b) =C(\rho) \delta^{ab}
\end{equation} where $T^a$ are the  generators of the gauge algebra and $\rho(T^a)$ is the matrix in the representation $\rho$, normalized so that $C(\text{adj})$ is equal to the dual Coxeter number. 
For $\SU(N)$, we have \begin{equation}
C(\text{adj})=N,\qquad C(\text{fund})=\frac12.
\end{equation}
In an $\cN{=}1$ gauge theory, the equation simplifies to \begin{equation}
\LambdaRG\frac{d}{d\LambdaRG}g=-\frac{g^3}{(4\pi)^2} \left[
3C(\text{adj})-C(R)
\right] 
\end{equation} or equivalently \begin{equation}
\LambdaRG\frac{d}{d\LambdaRG}\frac{8\pi^2 }{g^2}=
3C(\text{adj})-C(R),\label{n1running}
\end{equation}where $R$ is the representation of the chiral multiplet.
In an $\cN{=}2$ gauge theory, one adjoint chiral multiplet $\Phi$ is considered to be a part of the vector multiplet. Then we have \begin{equation}
\LambdaRG\frac{d}{d\LambdaRG}\frac{8\pi^2 }{g^2}=
2C(\text{adj})-C(R),\label{n2running}
\end{equation} where $R$ is now the representation of the $\cN{=}1$ chiral multiplets describing the hypermultiplets of the system. 
If one has one adjoint hypermultiplet, consisting of two $\cN{=}1$ chiral multiplets $A$ and $B$, we have zero one-loop beta function. 
When the mass terms for $A$, $B$ are zero, the system in fact has a further enlarged supersymmetry, 
 and is the $\cN{=}4$ super Yang-Mills. When the mass term is nonzero, it is called the $\cN{=}2^*$ theory. 

In a supersymmetric theory, the coupling $g$ is combined with the theta angle $\theta$ and enters in the Lagrangian as\begin{equation}
\ \int d^2\theta \frac{-i}{8\pi}\tau \tr W_\alpha W^\alpha + cc. 
\end{equation}  where $\tau$ is given by \begin{equation}
\tau=\frac{4\pi i}{g^2} +\frac{\theta}{2\pi}.
\end{equation} 
We call this $\tau$ the complexified gauge coupling.

We can consider $\tau$ to be an expectation value of a background chiral superfield. 
There is a renormalization scheme where the superpotential remains a holomorphic function of the chiral superfields, including background fields whose vevs are the gauge and superpotential couplings \cite{Intriligator:1995au}. We call it Seiberg's holomorphy principle. 

In this scheme, the one-loop running coupling at the energy scale $\LambdaRG$ can be expressed as \begin{equation}
\tau(\LambdaRG )=\tau_{UV} - \frac{b}{2\pi i} \log\frac{\LambdaRG }{\Lambda_{UV}} + \cdots
\label{1looprunning}
\end{equation} where $b$ is the rational number appearing on the right hand side of  \eqref{n1running} or \eqref{n2running}.
Note that the coupling $\tau$ starts from  $1/g^2$, and therefore the $n$ loop diagram would have the dependence $g^{2(n-1)}$.  The constant shift as in the imaginary part in \eqref{1looprunning} is then a one-loop effect.

Perturbation theory is independent of the $\theta$ angle, since $F_{\mu\nu}\tilde F_{\mu\nu}$ is a total derivative, although of a gauge-dependent quantity.
Therefore the $n$ loop effect is a function of $(\Im \tau)^{1-n}$, which is not holomorphic unless $n=1$.
We conclude that the running \eqref{1looprunning} is one-loop exact in the holomorphic scheme. 
We find that the combination \begin{equation}
\Lambda^b = \LambdaRG ^b e^{2\pi i  \tau(\LambdaRG )}
\end{equation} is invariant to all orders in perturbation theory. We call this $\Lambda$ the complexified  dynamical scale of the theory.\footnote{A redefinition of the form $\Lambda\to c\Lambda$ by a real constant $c$  corresponds to a redefinition of the coupling of the form $1/g^2 \to 1/g^2 - c'$ where $c'$ is another constant, or equivalently $g^2 \to g^2 + c' g^4 + \cdots$. Therefore this is a redefinition starting at the one-loop order, keeping the leading order definition of $g^2$ fixed. In this lecture note, we do not track such finite renormalization of the coupling very carefully. } Note that $\Lambda$ is a complex quantity, and can be considered as a vev of a background chiral superfield. 

This one-loop exactness  does not necessarily mean that the physical gauge coupling, which controls the scattering process for example, is one-loop exact. In the holomorphic scheme in generic $\cN{=}1$ supersymmetric theories, we have nontrivial wave-function renormalization factors $Z_{ij}$ \begin{equation}
\int d^4\theta Z^{\bar ij}(\LambdaRG ) Q^\dagger_{\bar i}  e^{V} Q_j \label{wr}
\end{equation} which need to be taken into account by a further field redefinition to compute physical scattering amplitudes.
This is known to produce further perturbative contributions to the physical running of the gauge coupling. For more on this point, see e.g.~\cite{ArkaniHamed:1997mj}.

For $\cN{=}2$ supersymmetric theories, however, one can make a stronger statement. We assume that there is a holomorphic renormalization scheme which is compatible with the existence of $\SU(2)_R$ symmetry. Then, the structure of the Lagrangian is restricted to be of the form \eqref{vectorlag} for the vector multiplets and of the form \eqref{hyperlag} for the hypermultiplets.  We consider $\tau$ as the vev of a background field.
Then, on the vector multiplet side, one finds that we cannot have nontrivial wavefunction renormalization factors $Z_{\bar ij}$ as in \eqref{wr} in the vector multiplet Lagrangian \eqref{vectorlag}. 
On the hypermultiplet side, the coefficient $c'$ in \eqref{hyperlag} is not renormalized in the holomorphic scheme.  Since $c=c'$, the K\"ahler potential is not renormalized. Therefore, there is no renormalization in the hypermultiplet Lagrangian \eqref{hyperlag}.

Then, in particular when $b=0$, the beta function is zero to all orders in perturbation theory.
This makes the system conformal, and the value of $\tau$ becomes an exactly marginal coupling parameter. 
The non-perturbative corrections will induce finite renormalization, but are not thought to introduce any additional infinite renormalization. 

For example, the $\cN{=}4$ super Yang-Mills is automatically superconformal, with one exactly marginal coupling. Another example with $b=0$ is $\cN{=}2$ supersymmetric $\SU(N)$ gauge theory with $2N$ hypermultiplets in the fundamental representation. Indeed, in \eqref{n2running}, we have $C(\text{adj})=N$ and $C(R)=2\cdot 2N\cdot 1/2$. 

\subsection{Anomalies}
\subsubsection{Anomalies of global symmetry}\label{sec:anomaly}
Non-abelian gauge theories have an important source of non-perturbative effects, called instantons. This is a nontrivial classical field configuration in the Euclidean $\mathbb{R}^4$ with nonzero integral of \begin{equation}
16\pi^2 k :=\int_{\mathbb{R}^4} \tr F_{\mu\nu}\tilde F^{\mu\nu}.
\end{equation} In the standard normalization of the trace for $\SU(N)$, $k$ is automatically an integer, and is called the instanton number.
%For other gauge groups, we define the trace symbol so that we have this normalization. 
The theta term in the Euclidean path integral appears as \begin{equation}
\exp\left[i\frac{\theta}{16\pi^2} \tr F_{\mu\nu} \tilde F^{\mu\nu}\right].
\end{equation} Therefore, a configuration with the instanton number $k$ has a nontrivial phase $e^{i\theta k}$.
Note that a shift of $\theta$ by $2\pi$ does not change this phase at all. Therefore, even in a quantum theory,
the shift $\theta\to\theta+2\pi$ is a symmetry.

Using \begin{equation}
\tr F_{\mu\nu}F_{\mu\nu} = \frac12 \tr(F_{\mu\nu}\pm \tilde F_{\mu\nu})^2 \mp  \tr F_{\mu\nu} \tilde F_{\mu\nu} 
\ge \mp \tr F_{\mu\nu}\tilde F_{\mu\nu}, 
\end{equation} we find that \begin{equation}
\int d^4 x \tr F_{\mu\nu} F_{\mu\nu} \ge 16\pi^2 |k| \label{bound}
\end{equation}
 which is saturated only when \begin{equation}
F_{\mu\nu} + \tilde F_{\mu\nu}\propto F_{\alpha\beta}=0 \quad
\text{or}
\quad F_{\mu\nu} - \tilde F_{\mu\nu}\propto F_{\dot\alpha\dot\beta}=0 
\label{SD}
\end{equation} depending on the sign of $k$.
Therefore, within configurations of fixed $k$, those satisfying relations \eqref{SD} give the dominant contributions to the path integral.  The solutions to \eqref{SD} are called instantons or anti-instantons, depending on the sign of $k$. 

In an instanton background, the weight in the path integral coming from the gauge kinetic term is \begin{equation}
\exp\left[ 
-\frac{1}{2g^2}\int \tr F_{\mu\nu}F^{\mu\nu}
+
i\frac{\theta}{16\pi^2}\int \tr F_{\mu\nu}\tilde F^{\mu\nu}
\right]
=e^{2\pi i \tau k}.\label{instanton_contri}
\end{equation} We similarly have the contribution $e^{2\pi i\bar\tau |k|}$ in an anti-instanton background. 
The fact that we have just $\tau$ or $\bar \tau$, instead of more complicated combinations, is related to the fact that 
in the instanton background in a supersymmetric theory, $\delta\lambda_{\dot\alpha}=F_{\dot\alpha\dot\beta} \epsilon^{\dot\beta}=0$ assuming the D-term is also zero, and thus the dotted supertranslation is preserved. 
Similarly, the undotted supersymmetry is unbroken in the anti-instanton background.

Now, consider charged Weyl fermions $\psi_\alpha$ coupled to the gauge field, with the kinetic term \begin{equation}
\bar\psi_{\dot\alpha} {D}{}_\mu \sigma^{\mu \dot\alpha\alpha} \psi_\alpha.
\end{equation} Let us say $\psi_\alpha$ is in the representation $R$ of the gauge group. 
It is known that the number of zero modes in $\psi_\alpha$ minus the number of zero modes in $\bar\psi_{\dot\alpha}$ 
is $2C(R)k$. 
In particular, the path integral restricted to the $k$-instanton configuration with positive $k$ is vanishing unless we insert $k$ additional $\psi$'s in the integrand. More explicitly, \begin{equation}
\vev{O_1O_2\cdots} = \int [D\psi] [D\bar\psi] O_1O_2 \cdots e^{-S} =0    
\end{equation} 
unless the product of the operators $O_1 O_2\cdots$ contains $2C(R)k$ more $\psi$'s than $\bar\psi$'s.
This is interpreted as follows: the path integral measures $[D\psi]$ and  $[D\bar\psi]$ contain both infinite number of integrations. However, there is a finite number, $2C(R)k$, of difference in the number of integration variables. 
Equivalently, under the constant rotation \begin{equation}
\psi \to e^{i\varphi} \psi,\quad
\bar\psi \to e^{-i\varphi} \bar\psi,
\end{equation} the fermionic path integration measure rotates as \begin{equation}
\begin{aligned}\relax
[D\psi] & \to [D\psi]e^{+\infty i\varphi + 2C(R)k i\varphi  }, \\
[D\bar\psi] & \to [D\bar\psi]e^{-\infty i\varphi }. 
\end{aligned}
\end{equation} 
When combined, we have  \begin{equation}
[D\psi][D\bar\psi] \to [D\psi][D\bar\psi] e^{2C(R)k i\varphi} = 
[D\psi][D\bar\psi] \exp\left[
2C(R)\varphi \frac{i}{16\pi^2}\int \tr F_{\mu\nu}\tilde F^{\mu\nu}
\right].
\end{equation}
This can be compensated by a shift of the $\theta$ angle, $\theta\to \theta+2C(R)\varphi$. 
As we recalled before, the shift $\theta\to\theta+2\pi$ is a symmetry.
Therefore, the rotation of the field $\psi$ by $\exp(\frac{2\pi i}{2C(R)})$ is a genuine, unbroken symmetry.

\subsubsection{Anomalies of gauge symmetry}
In $\cN{=}2$ gauge theories, fermions always come in non-chiral representations. Indeed, the fermions in the vector multiplets are always in the adjoint,  the $\cN{=}1$ chiral superfields in a full hypermultiplet is a sum of a representation $R$ and its conjugate $\bar R$, and a half-hypermultiplet counts as an $\cN{=}1$ chiral superfield in a pseudo-real representation $R$. 
Therefore there are no perturbative gauge anomalies. 

One needs to be careful about Witten's global anomaly \cite{Witten:1982fp}, though, as this can arise even for real representations. 
It is known that a Weyl fermion in the doublet of gauge $\SU(2)$  is anomalous, due to the following fact.
When we perform the path integral of this system, we first need to consider \begin{equation}
Z[A_\mu] = \int [D\psi_{\alpha i}][D\bar\psi_{\dot\alpha i}] e^{-\int \bar\psi D_\mu \sigma^\mu \psi}
\end{equation} where $i=1,2$ is the $\SU(2)$ doublet index. 
To perform a further integration over $A_\mu$ consistently, we need \begin{equation}
Z[A_\mu]= Z[A^g_\mu], \quad A^g_\mu = g^{-1} A_\mu g  + g^{-1} \partial_\mu g.
\end{equation} for any gauge transformation $g:\mathbb{R}^4\to \SU(2)$.
These maps are characterized by $\pi_4(\SU(2))$. It is known that \begin{equation}
\pi_4(\SU(2))=\pi_4(S^3)=\mathbb{Z}_2.
\end{equation} Let $g_0:\mathbb{R}^4\to \SU(2)$ be the one corresponding to the nontrivial element in this $\bZ_2$. Then it is known that \begin{equation}
[D\psi_{\alpha i}][D\bar\psi_{\dot\alpha i}] \stackrel{g_0}{\longrightarrow}
-[D\psi_{\alpha i}][D\bar\psi_{\dot\alpha i}] 
\end{equation} resulting in \begin{equation}
Z[A^{g_0}_\mu] = -Z[A_\mu],
\end{equation} thus making the path integral over $A_\mu$ inconsistent.

In general $\pi_4(G)=\mathbb{Z}_2$ if $G=\Sp(n)$, and $\pi_4(G)=1$ otherwise. 
Therefore Witten's global anomaly can be there only for Weyl fermions in a representation $R$ under gauge $\Sp(n)$. A short computation reveals that there is an anomaly only when $C(R)$ is half-integral. 

Witten's anomaly is always $\bZ_2$ valued in four dimensions. Therefore full hypermultiplets are always free of Witten's global anomaly. The danger only  exists  for half-hypermultiplets of gauge $\Sp(n)$. 
For example, one cannot have odd number of half-hypermultiplets in the doublet representation of gauge $\SU(2)$,
or more generally, one cannot have half-hypermultiplets in a pseudo-real representation $R$ of gauge $\Sp(n)$ such that $C(R)$ is half-integral.

\subsection{$\cN{=}1$ pure Yang-Mills}\label{sec:n=1pure}
\subsubsection{Confinement and gaugino condensate}
As an example of the application of what we learned  in this section, let us consider the $\cN{=}1$ pure supersymmetric Yang-Mills theory with gauge group $\SU(N)$. The content of this section will not be used much in the rest of the lecture note.

This theory has just the vector multiplet, with the Lagrangian \begin{equation}
L=\int d^2\theta \frac{-i}{8\pi}\tau \tr W_\alpha W^\alpha+ cc., \qquad
W_\alpha=\lambda_\alpha + F_{\alpha\beta}\theta^\beta + \cdots
\end{equation} The one-loop running of the coupling is given by \begin{equation}
E\frac{\partial}{\partial E}\tau(E) = \frac{i}{2\pi} 3N,
\end{equation} and therefore we define the dynamical scale $\Lambda$ by the relation \begin{equation}
\Lambda^{3N}= e^{2\pi i\tau_{UV} }\Lambda_{UV}^{3N}.
\end{equation}

We assign R-charge zero to the gauge field, and R-charge 1 to the gaugino $\lambda_\alpha$. 
The phase rotation $\lambda_\alpha\to e^{i\varphi} \lambda_\alpha$ is anomalous, and needs to be compensated by $\theta\to \theta + 2N \varphi$. The shift of $\theta$ by $2\pi$ is still a symmetry, therefore the discrete rotation \begin{equation}
\lambda_\alpha \to e^{\pi i/N} \lambda_\alpha, \qquad \theta\to \theta+2\pi
\end{equation} is a symmetry generating $\bZ_{2N}$. 
Note that under this symmetry, $\Lambda$ defined above has the transformation \begin{equation}
\Lambda \to e^{2\pi i/(3N)} \Lambda.\label{lambdatrans}
\end{equation}

This theory is believed to confine, with nonzero gaugino condensate $\vev{\lambda_\alpha\lambda^\alpha}$.
What would be the value of this condensate?  This should be of mass dimension 3 and of R-charge 2. The only candidate is 
\begin{equation}
\vev{\lambda_\alpha\lambda^\alpha}=c \Lambda^3
\end{equation}  for some constant $c$.
The  symmetry \eqref{lambdatrans} acts in the same way on both sides by the multiplication by $e^{2\pi i/N}$.
Assuming that the numerical constant $c$ is non-zero, this $\bZ_{2N}$ is further spontaneously broken to $\bZ_2$,
 generating $N$ distinct solutions \begin{equation}
\vev{\lambda_\alpha\lambda^\alpha}=c e^{2\pi i \ell/N } \Lambda^3
\end{equation} where $\ell=0,1,\ldots, N-1$.  Unbroken $\bZ_2$ acts on the fermions by $\lambda_\alpha \to -\lambda_\alpha$, which is a $360^\circ$ rotation. This $\bZ_2$ symmetry is hard to break.

It is now generally believed that this theory has these $N$ supersymmetric vacua and not more. For other gauge groups, the analysis proceeds in the same manner, by replacing $N$ by the dual Coxeter number $C(\text{adj})$ of the gauge group under consideration. For example, we have $N-2$ vacua for the pure $\cN{=}1$ $\SO(N)$ gauge theory.

\subsubsection{The theory in a box}\label{sec:box}
It is instructive to recall another way to compute the number of vacua in the $\cN{=}1$ pure  Yang-Mills theory with gauge group $G$, originally discussed in \cite{Witten:1982df}.  We put the system in a spatial box of size $L\times L\times L$  with the periodic boundary condition in each direction. We keep the time direction as $\bR$. 
By performing the Kaluza-Klein reduction along the three spatial directions, the system becomes supersymmetric quantum mechanics with infinite number of degrees of freedom. 

The box still preserves the translation generators $P^\mu$ and the supertranslations $Q_\alpha$ unbroken. 
We just use a  linear combination $\cQ$  of $Q_\alpha$ and $Q_\alpha^\dagger$, satisfying \begin{equation}
H=P^0=  \{\cQ,\cQ^\dagger\}.
\end{equation} 
We also have the fermion number operator $(-1)^F$ such that \begin{equation}
\{(-1)^F,\cQ\}=0.
\end{equation}
Consider eigenstates of the Hamiltonian $H$, given by \begin{equation}
H\ket{E}=E\ket{E}.
\end{equation} 
In general, the multiplet structure under the algebra of $\cQ$, $\cQ^\dagger$, $H$ and $(-1)^F$ is of the form 
\begin{equation}
\begin{array}{ccccc@{\qquad}c}
&\raisebox{-.8ex}{\rotatebox{30}{$\leftrightarrow$}}& \cQ^\dagger \ket{E} & \leftrightarrow & (\cQ^\dagger\cQ-\cQ\cQ^\dagger)\ket{E} \\
\ket{E} & \leftrightarrow & \phantom{^\dagger}\cQ\ket{E} &  \rotatebox{30}{$\leftrightarrow$} 
\end{array}
\end{equation}
involving four states.  When $\cQ\ket{E}=0$ or $\cQ^\dagger\ket{E}=0$, the multiplet only has two states.
If $\cQ\ket{E}=\cQ^\dagger\ket{E}=0$, the multiplet has only one state, and $E$ is automatically zero
due to the equality
\begin{equation}
E\vev{EE}
=\vev{E|H|E}
=\vev{E|(\cQ \cQ^\dagger+\cQ^\dagger \cQ)|E}=
|\cQ\ket{E}|^2+
|\cQ^\dagger\ket{E}|^2 .
\end{equation} 
We see that a bosonic state is always paired with a fermionic state unless $E=0$.

This guarantees that  the Witten index \begin{equation}
\tr e^{-\beta H} (-1)^F =\tr\big|_{E=0} (-1)^F
\end{equation} is a robust quantity independent of the change in the size $L$ of the box: when a perturbation makes  a number of zero-energy states to non-zero energy $E\neq 0$, the states involved are necessarily  composed of  pairs of a fermionic state and a bosonic state. Thus it cannot change $\tr (-1)^F$. 

Therefore, we can compute the Witten index in the limit where the box size $L$ is far smaller than the scale $\Lambda^{-1}$ set by the dynamics. Then the system is weakly coupled, and we can use perturbative analysis. 
To have almost zero energy, we need to have $F_{\mu\nu}=0$ in the spatial directions, since   magnetic fields contribute to the energy.  Then the only low-energy degrees of freedom in the system are the holonomies \begin{equation}
U_x, U_y, U_z \in \SU(N),
\end{equation} which commute with each other. Assuming that they can be simultaneously diagonalized, we have \begin{align}
U_x&=\diag(e^{i\theta^x_1},\ldots,e^{i\theta^x_N}),\\
U_y&=\diag(e^{i\theta^y_1},\ldots,e^{i\theta^y_N}),\\
U_z&=\diag(e^{i\theta^z_1},\ldots,e^{i\theta^z_N}).\label{Uxyz}
\end{align} together with gaugino zero modes \begin{equation}
\lambda^{\alpha=1}_1,\ldots,\lambda^{\alpha=1}_N,\quad
\lambda^{\alpha=2}_1,\ldots,\lambda^{\alpha=2}_N
\end{equation} with the condition that \begin{equation}
\sum_i \theta^x_i=\sum_i \theta^y_i=\sum_i \theta^z_i=0,\quad
\sum_i\lambda^{\alpha=1}_i=
\sum_i\lambda^{\alpha=2}_i=0.
\end{equation}
The wavefunction of this truncated quantum system is given by a linear combination of states of the form \begin{equation}
\lambda^{\alpha_1}_{i_1}\lambda^{\alpha_2}_{i_2}\cdots \lambda^{\alpha_\ell}_{i_\ell}\psi(\theta^x_i;\theta^y_i;\theta^z_i)
\end{equation} which is invariant under the permutation acting on the index $i=1,\ldots N$.
To have zero energy, the wavefunction cannot have dependence on $\theta^{x,y,z}_i$ anyway, since the derivatives with respect to them are the components of the electric field, and they contribute to the energy.
Thus the only possible zero energy states are just invariant polynomials of $\lambda$s.
We find $N$ states with the wavefunctions given by \begin{equation}
1,\ S,\ S^2,\ \ldots,\ S^{N-1}\label{tow}
\end{equation} where $S=\sum_i\lambda^{\alpha=1}_i\lambda^{\alpha=2}_i$. They all have the same Grassmann parity, and contribute to the Witten index with the same sign. Thus we found $N$ states in the limit of small box, too. 

The construction so far, when applied to other groups, only gives $1+\rank G$ states.
For example, let us consider for $G=\SO(N)$ for $N>4$. Then the method explained so far  only gives  $\lfloor N/2\rfloor+1$ states \begin{equation}
1,\ S,\ S^2,\ \ldots,\ S^{\lfloor N/2\rfloor},
\end{equation}  and does not agree with $C(\text{adj})=N-2$ when $N\ge 7$. This conundrum  was already pointed out in \cite{Witten:1982df} and resolved later in the Appendix I of \cite{Witten:1997bs} by the same author.\footnote{It is a sad state of affairs that a problem reported in such an important paper as \cite{Witten:1982df} was not resolved for 15 years by any other physicist. It seems that people in our field rely too much on the author of \cite{Witten:1982df,Witten:1997bs}.}  What was wrong was the assumption that three commuting matrices $U_{x,y,z}$ can be simultaneously diagonalized as in \eqref{Uxyz}. It is known that there is another component where they cannot be simultaneously diagonalized into the Cartan torus. For $\SO(7)$, an example is given by the triple \begin{align}
U_x^{(7)}&=\diag(++---+-),\\
U_y^{(7)}&=\diag(+-+-+--),\\
U_z^{(7)}&=\diag(-+++---).
\end{align} 
These three matrices might look diagonal, but not in the same Cartan subgroup.
This component adds one supersymmetric state.
Then, in total, we have $(\lfloor7/2\rfloor+1)+1=5=7-2$, reproducing $C(\text{adj})$.  

For larger $N$, one can consider $U_{x,y,z}$ given by 
the form \begin{equation}
U_x=U_x^{(7)} \oplus U_x',\quad
U_y=U_y^{(7)} \oplus U_y',\quad
U_z=U_z^{(7)} \oplus U_z',\label{N-7}
\end{equation} where $U_{x,y,z}'$ are in the Cartan subgroup of $\SO(N-7)$.
 Applying the analysis leading to \eqref{tow} in both components, i.e.~in the component where $U_{x,y,z}$ are in the Cartan subgroup of $\SO(N)$, and in the component where $U_{x,y,z}$ has the form \eqref{N-7},
 we find in total \begin{equation}
(\lfloor N/2\rfloor+1)+(\lfloor (N-7)/2\rfloor+1) = N-2
\end{equation} zero-energy states, thus reproducing $C(\text{adj})$ states. This analysis has been extended to arbitrary gauge groups \cite{Kac:1999gw,BorelFreedmanMorgan}.

\section{Seiberg-Witten solution to pure $\SU(2)$ theory}\label{sec:pureSU2}
We are finally prepared enough to start the analysis of the simplest of non-Abelian $\cN{=}2$ supersymmetric  theory, namely the pure $\SU(2)$ gauge theory. We mainly follow the presentation of the original paper \cite{Seiberg:1994rs}, except that we use the \SeibergWitten\ curve in the form first found in \cite{Martinec:1995by}, which is more suited to the generalization later.  

\subsection{One-loop running and the monodromy at infinity}

 The pure $\SU(2)$ theory contains only an $\cN{=}2$  vector multiplet for the $\SU(2)$ gauge group, with its Lagrangian given by \eqref{vectorlag}. For reference we reproduce it here: 
 \begin{equation}
L=\frac{\Im\tau}{4\pi} \int d^4\theta \tr \Phi^\dagger e^{[V,\cdot]} \Phi +\int d^2\theta \frac{-\ii}{8\pi}\tau \tr W_\alpha W^\alpha + cc. \label{high}
\end{equation}

A supersymmetric vacuum is classically characterized by the solution to the D-term constraint \begin{equation}
[\Phi^\dagger,\Phi]=0.
\end{equation} This means that $\Phi$ can be diagonalized by a gauge rotation. Let \begin{equation}
\Phi=\diag(a,-a).\label{dv}
\end{equation} 

\begin{figure}[h]
\[
\inc{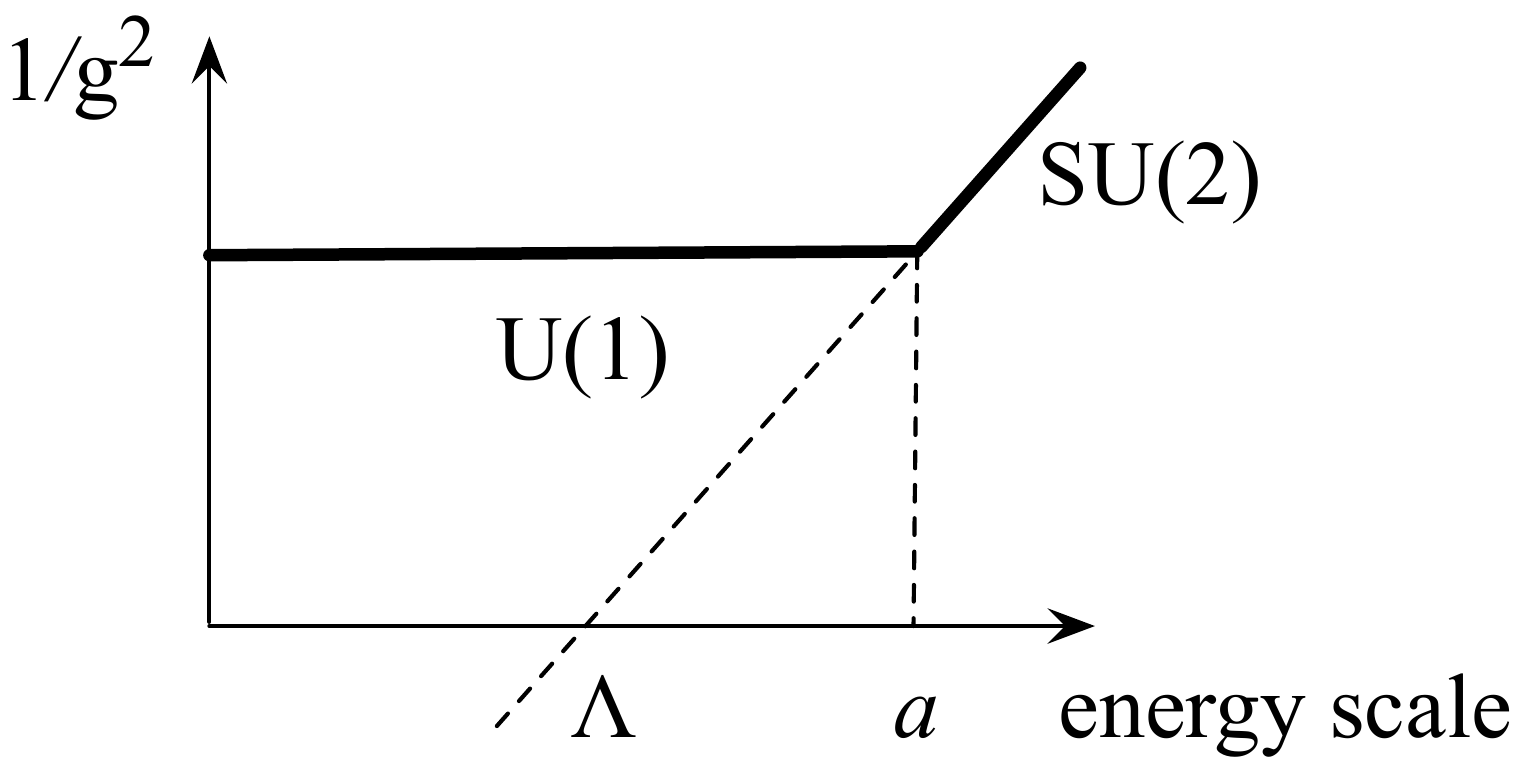}
\]
\caption{Schematic drawing of the running of the coupling. \label{fig:roughrunning}}
\end{figure}

Roughly speaking, the gauge coupling $\tau$ runs from a very high energy scale down to the energy scale $a$ according to the one-loop renormalization of the $\SU(2)$ theory. Then the vev $a$ breaks the gauge group $\SU(2)$ to $\U(1)$. There are massive excitations charged under the unbroken $\U(1)$, but they will soon decouple, and the coupling remains almost constant below the energy scale $a$. 
This evolution is shown in Fig.~\ref{fig:roughrunning}.

Let us describe it slightly more quantitatively.  Our normalization of the $\U(1)$ Lagrangian and the gauge coupling was given in \eqref{u1coupling} and \eqref{u1tau}, which we reproduce here: \begin{equation}
\frac{1}{2e^2} F_{\mu\nu}^{\U(1)}F_{\mu\nu}^{\U(1)} + \frac{\theta}{16\pi^2} F_{\mu\nu}^{\U(1)} \tilde F_{\mu\nu}^{\U(1)},
\qquad \text{and}\qquad
\tau^{\U(1)}= \frac{4\pi i}{e^2} + \frac{\theta}{2\pi}.
\end{equation} In the broken vacuum, the low-energy $\U(1)$ and the high-energy $\SU(2)$ are related as in \eqref{u1embed}, which we also reproduce here \begin{equation}
F_{\mu\nu}^{\SU(2)}=\diag(F_{\mu\nu}^{\U(1)},-F_{\mu\nu}^{\U(1)}).
\end{equation} Plugging this in to the high-energy Lagrangian \eqref{high} and comparing the definitions of $\tau$s, we find \begin{equation}
\tau^{\U(1)} = 2\tau^{\SU(2)}.
\end{equation} This relation gets modified by the quantum corrections.

Let us denote by $\tau(a)$ the low-energy coupling of the $\U(1)$ gauge field when the vev is given by \eqref{dv}, and by $\tau_{UV}$ the high-energy coupling of the $\SU(2)$ gauge field at the high-energy renormalization point $\Lambda_{UV}$. 
The one-loop running \eqref{n2running} then gives  \begin{align}
\tau(a)&=2\tau_{UV}-\frac{8}{2\pi i}\log \frac{a}{\Lambda_{UV}}+\cdots \\
&=-\frac{8}{2\pi i}\log \frac{a}{\Lambda} + \cdots\label{puretau}
\end{align} where we defined \begin{equation}
\Lambda^4 = \Lambda_{UV}^4 e^{2\pi i \tau_{UV}}.\label{lambdaSU2pure}
\end{equation}
The dual variable $a_D$ can be obtained by integrating \eqref{puretau} once, and we find \begin{equation}
a_D=-\frac{8a}{2\pi i}\log\frac{a}{\Lambda} + \cdots.
\end{equation}
As long as we keep $|a|\gg |\Lambda|$, the coupling $\tau(a)$ remains weak, and the computation above gives a reliable approximation. 

A gauge-invariant way to label the supersymmetric vacua is to use \begin{equation}
u=\frac12\vev{\tr\phi^2}=a^2+\cdots
\end{equation} where $\cdots$ are quantum corrections.  
Let us consider adiabatically rotating the phase of $u$ by $2\pi$: \begin{equation}
u=e^{i\theta}|u|, \qquad \theta=0 \sim 2\pi
\end{equation}
We have $a\mapsto -a$. From the explicit form of $a_D$ we find $a_D\to -a_D+4a$. 
We denote it as \begin{equation}
(a,a_D)\to (a,a_D) \begin{pmatrix} 
-1 & 4 \\
0 & -1
\end{pmatrix}.\label{amat}
\end{equation}
The mass formula of BPS particles is \begin{equation}
M=|na+ma_D|=\left|
(a,a_D) 
\begin{pmatrix}
n \\ m
\end{pmatrix}
\right|.
\end{equation}  Therefore, the transformation \eqref{m_inf_nf_0} can also be ascribed to the transformation of the charges: \begin{equation}
\begin{pmatrix}
n \\ m
\end{pmatrix}
\to 
 \begin{pmatrix} 
-1 & 4 \\
0 & -1
\end{pmatrix}
\begin{pmatrix}
n \\ m
\end{pmatrix}.\label{chargemat}
\end{equation}

\begin{figure}[h]
\[
\inc{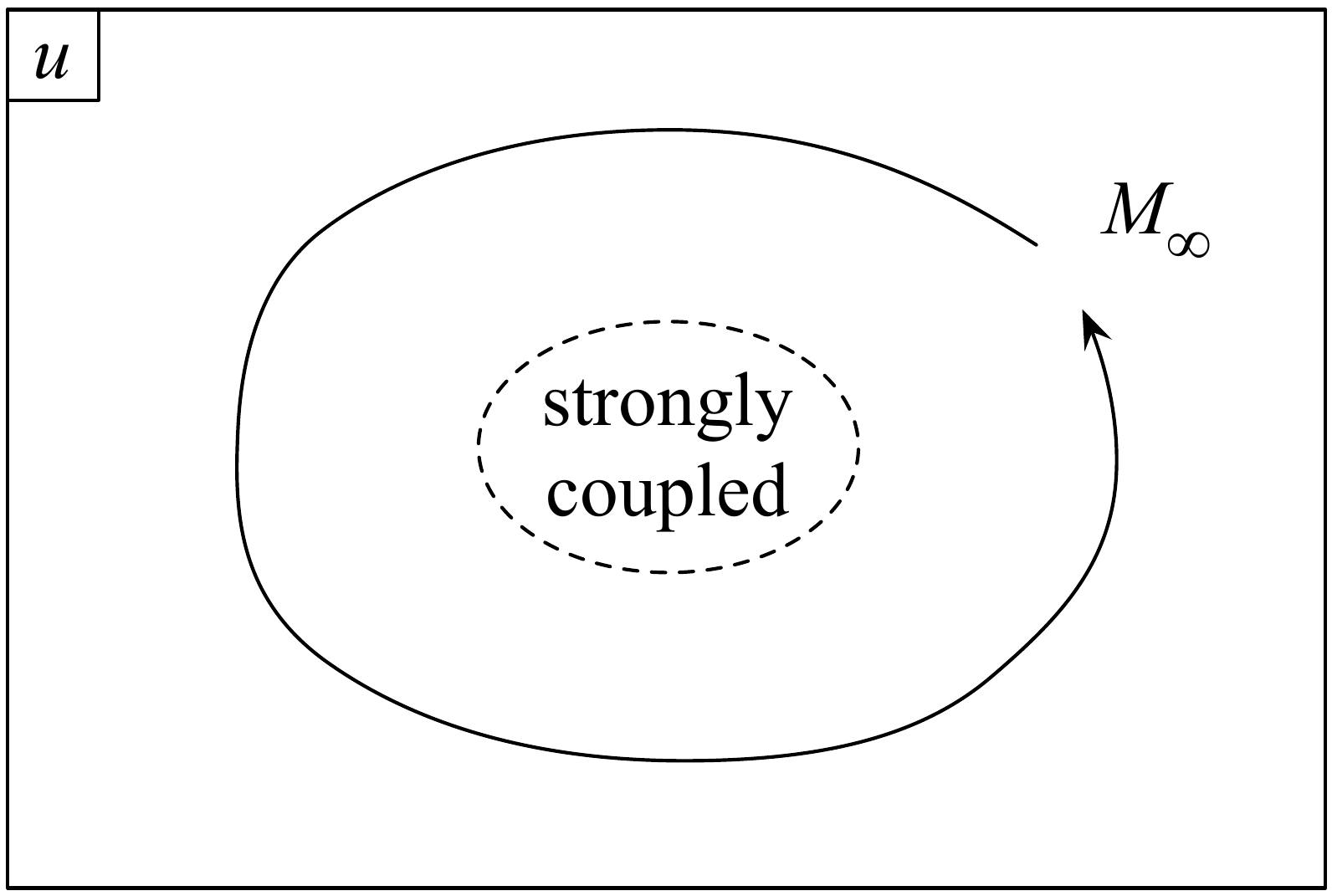}
\]
\caption{Monodromy at infinity. \label{fig:m_inf_nf_0}}
\end{figure}

We call this matrix \begin{equation}
M_\infty= \begin{pmatrix} 
-1 & 4 \\
0 & -1
\end{pmatrix}\label{m_inf_nf_0}
\end{equation} the monodromy at infinity. The situation is schematically shown in Fig~\ref{fig:m_inf_nf_0}.
The space of the supersymmetric vacua, parametrized by $u$, is often called the $u$-plane. 

In our argument, the matrix \eqref{amat} could have had non-integral entries, as we read the matrix elements off from an approximate formula of $a$ and $a_D$. However, the transformation \eqref{chargemat} should necessarily map integral vectors to integral vectors, which guarantees that the matrix \eqref{chargemat} is integral. Not only that, this transformation is just a relabeling of the charges and should not change the Dirac pairing \begin{equation}
nm'-mn'=\det \begin{pmatrix}
n & n' \\
m & m'
\end{pmatrix}
\end{equation}
 which measure the angular momentum carried in the space when we have two particles with charges $(n,m)$ and $(n',m')$, respectively. A transformation given by \begin{equation}
 \begin{pmatrix}
n \\ m
\end{pmatrix}
\to 
M
\begin{pmatrix}
n \\ m
\end{pmatrix}
\end{equation} affects the Dirac pairing as \begin{equation}
\det \begin{pmatrix}
n & n' \\
m & m'
\end{pmatrix}
\to 
\det M
\det
\begin{pmatrix}
n & n' \\
m & m'
\end{pmatrix}.
\end{equation} Therefore, $M$ should necessarily has unit determinant.  
A $2\times 2$ integral matrix with unit determinant is called an element of $\SL(2,\bZ)$.
It is reassuring  that the matrix \eqref{m_inf_nf_0} satisfies this condition.

\subsection{Behavior in the strongly-coupled region}

\begin{figure}[h]
\[
\inc{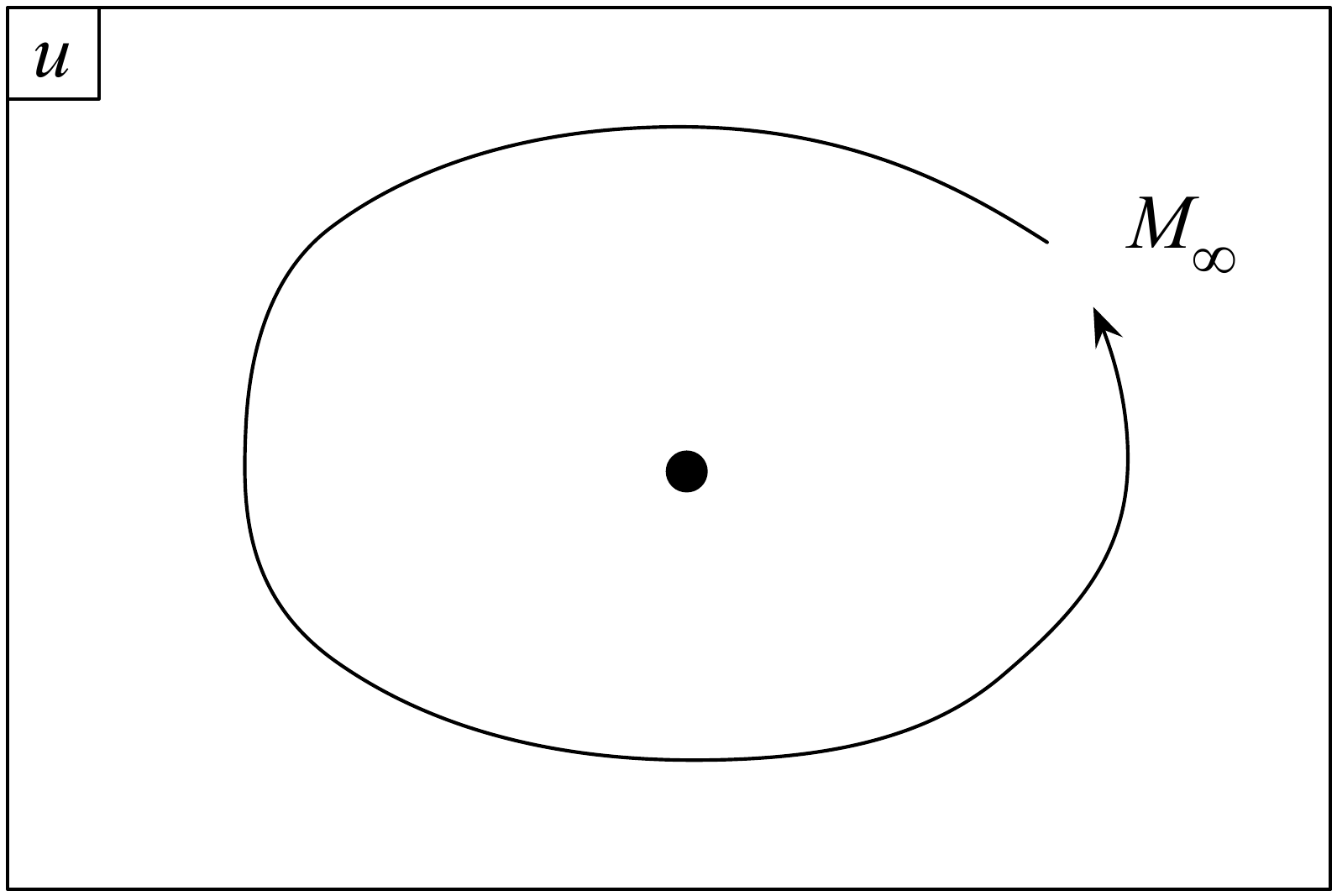}
\]
\caption{Naive guess which does not work \label{fig:m_0_nf_0}}
\end{figure}

Let us study what is going on in the strongly coupled region which is the interior of the $u$-plane.
There needs to be at least one singularity in this interior region to realize the monodromy $M_\infty$ of holomorphic functions $a$ and $a_D$. 
So, most naively, we would expect the structure as in Fig.~\ref{fig:m_0_nf_0}.
Where will the singularity be? 
Here the discrete unbroken $\U(1)$ R-symmetry of the system is useful. 
Recall our $\cN{=}2$ theory has an $\SU(2)_R$ symmetry. Classically,
we can also consider a $\U(1)_R$ symmetry with the 
 standard R-charge assignment given as follows: \begin{equation}
\begin{array}{r|ccc}
R=0 & & A\\
1& \lambda & & \lambda \\
2 & & \Phi
\end{array}.
\end{equation} 
Different components in the same supersymmetry multiplet have different charges, and therefore this is an R-symmetry. 

Quantum mechanically, the rotation \begin{equation}
\lambda\to e^{i\varphi} \lambda,
\end{equation}  is anomalous, but can be compensated by \begin{equation}
\theta_{UV}\to \theta_{UV}+8\varphi,
\end{equation}  as we learned in Sec.~\ref{sec:anomaly}.

Therefore $\varphi=\pi/4$ is a genuine symmetry, which does \begin{equation}
\theta\to \theta+2\pi,\quad
\Phi\to e^{\pi i/2}\Phi.
\end{equation} This generates a $\bZ_4$ discrete R-symmetry of the system. 
In the low-energy variables, it acts as \begin{equation}
\theta_{IR}\to \theta_{IR}+4\pi,\qquad
u\to -u.\label{discrete_r_nf_0}
\end{equation}
Then, if there is a singularity at $u=u_0$, there should be another at $u=-u_0.$
Therefore, if there is only one singularity, it is at $u=0$. 

If this were really the case, we would find that $\tau(a)$ is given  by \begin{equation}
\tau(a)=-\frac{8}{2\pi i}\log \frac{a}{\Lambda} + f(a).\label{wrongt}
\end{equation}   where $f(a)$  is a meromorphic function whose only singularity is at $a=0$. 
This does not sound right, however.  The coupling is given by $\Im \tau(a)$, which is the imaginary part of a holomorphic function. Then, it has no lower bound, and therefore it becomes negative for some value of $a$.
This means that the coupling $g^2$ is negative there, and the system becomes unstable. 
 For example, supposing $f(a)=0$,  the imaginary part is negative when $|a|$ is small enough. 
We conclude that our assumption of having just one singularity at $u=0$ was too naive.

\begin{figure}[h]
\[
\inc{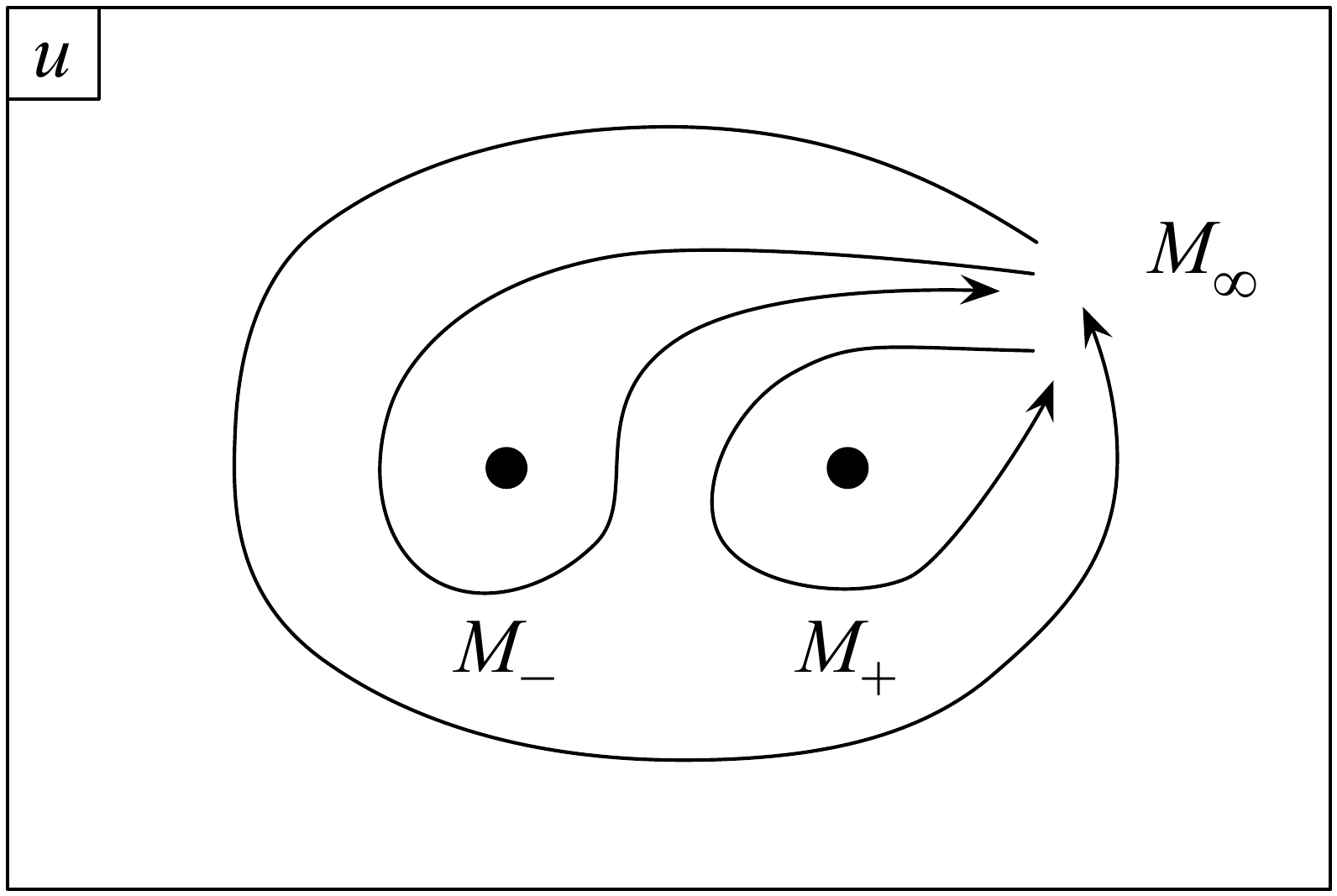}
\]
\caption{Next guess which turns out to be correct\label{fig:monodromy_nf_0_correct}}
\end{figure}

\begin{figure}[h]
\[
\inc{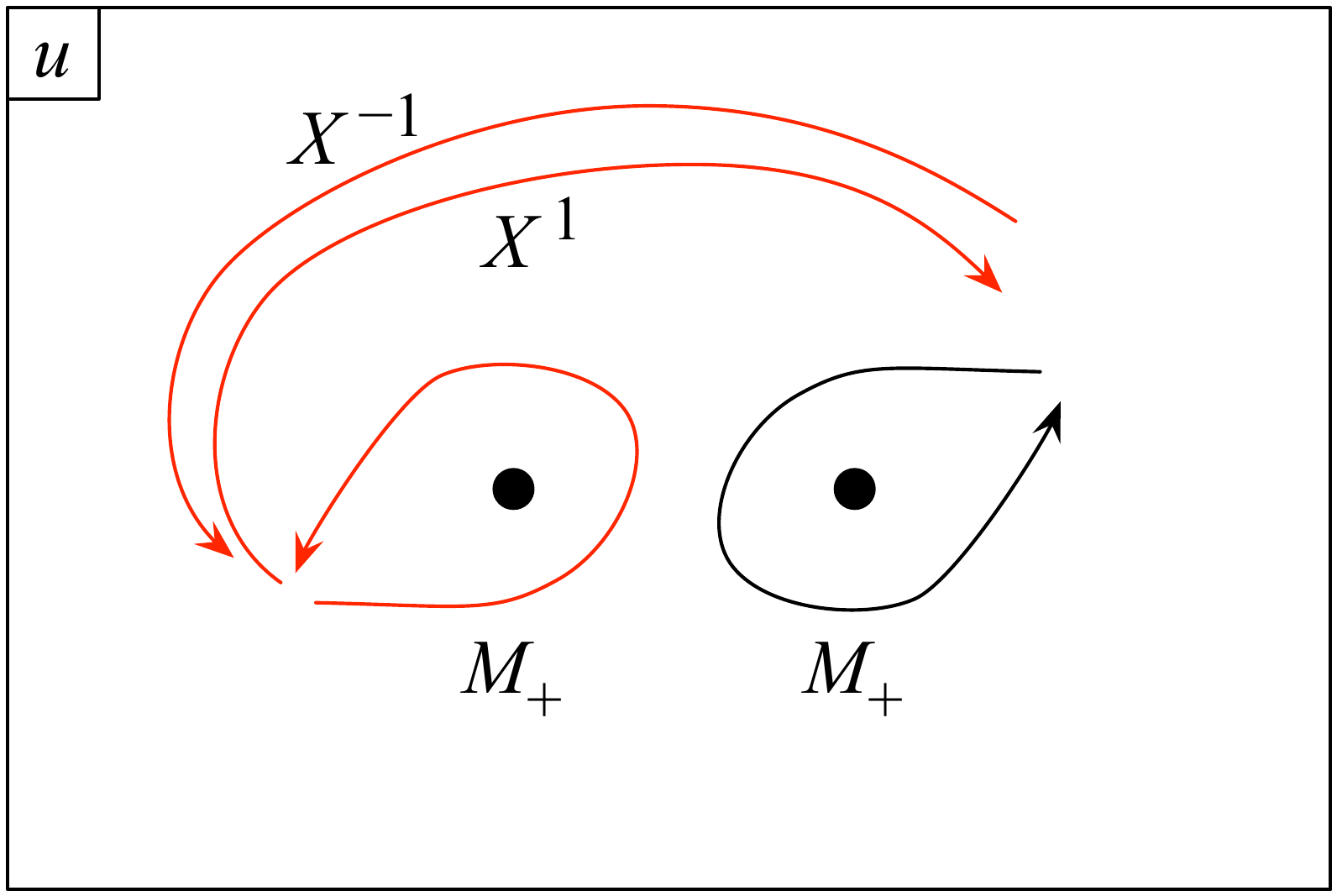}
\]
\caption{The relation between $M_+$ and $M_-$\label{fig:discrete_r_nf_0}}
\end{figure}

The next simplest possibility is then to suppose that there are two singularities at $u=\pm u_0$, see Fig.~\ref{fig:monodromy_nf_0_correct}.
The only scale in the system is the dynamical scale $\Lambda$, therefore $u_0$ should be given by $c\Lambda$ where $c$ is a number.
Denoting the monodromies around two singularities $M_\pm$, we should have \begin{equation}
M_\infty=M_+ M_-,
\end{equation} since the path going around the infinity of the $u$-plane is topologically the same as the path which first goes around $u=-u_0$ and then around $u=u_0$.
As two singularities are exchanged by a symmetry, the monodromies around them should be essentially the same, except for the relabeling of the charges. Or equivalently, they should be conjugate \begin{equation}
M_-=X M_+ X^{-1}
\end{equation} by an $\SL(2,\bZ)$ matrix $X$. Note that this matrix $X$ can be thought of a half-monodromy associated to the symmetry operation \eqref{discrete_r_nf_0}, see Fig.~\ref{fig:discrete_r_nf_0}

A solution to these equations is given by \begin{equation}
M_+=STS^{-1}=\begin{pmatrix}
1 & 0\\
-1 & 1
\end{pmatrix}, \quad
M_-=T^2 S T S^{-1} T^{-2}=\begin{pmatrix}
-1 & 4 \\
-1 & 3
\end{pmatrix}\label{mpm}
\end{equation} 
where $S$ and $T$ were given in \eqref{S}, \eqref{T}.
Note that we have \begin{equation}
X=T^2,
\end{equation} which is roughly compatible with the fact that the discrete R-symmetry \eqref{discrete_r_nf_0} shifts $\theta_{IR}$ by $4\pi$. 

\subsection{The Seiberg-Witten solution}\label{sec:purecurve}

Let us construct holomorphic functions $a$ and $a_D$ satisfying these monodromies explicitly. 
Note that a holomorphic function is uniquely determined by its singularities. Therefore, if we find a candidate with the correct properties around the singularities and at infinity of its domain of definition, it is necessarily the correct answer itself, assuming that we identified the singularities correctly. 
Therefore, it suffices to construct a candidate and then check that it satisfies the conditions. 

\subsubsection{The curve}
We first introduce two auxiliary complex variables $x$ and $z$, and 
then we consider an equation \begin{equation}
\Sigma:\quad
\Lambda^2 z + \frac{\Lambda^2}z=x^2-u.\label{curve_pure}
\end{equation} 
We consider this equation as defining a complex one-dimensional subspace of a complex two-dimensional space of $x$ and $z$.\footnote{Our usage of $(z,x)$ for the coordinates follows the convention of  \cite{Martinec:1995by}.  Using $(t,v)$ for what we call $(z,x)$ is also common, which comes from \cite{Witten:1997sc}. } As the equation changes as we change $u$, the shape of this subspace also changes. 
This complex one-dimensional object is called the \SeibergWitten\ curve.\footnote{It is real two-dimensional, and therefore it is a surface from a usual point of view. Mathematicians are strange and they consider one-dimensional objects curves, whether it is complex one-dimensional or real one-dimensional.} 
A  differential \begin{equation}
\lambda=x\frac{dz}z,
\end{equation} called the Seiberg-Witten differential, plays an important role later.\footnote{The symbol $\lambda$ were for adjoint fermions up to this point, but  we use $\lambda$ mainly for the differential from now on, unless otherwise noted.}

\begin{figure}[h]
\[
\inc{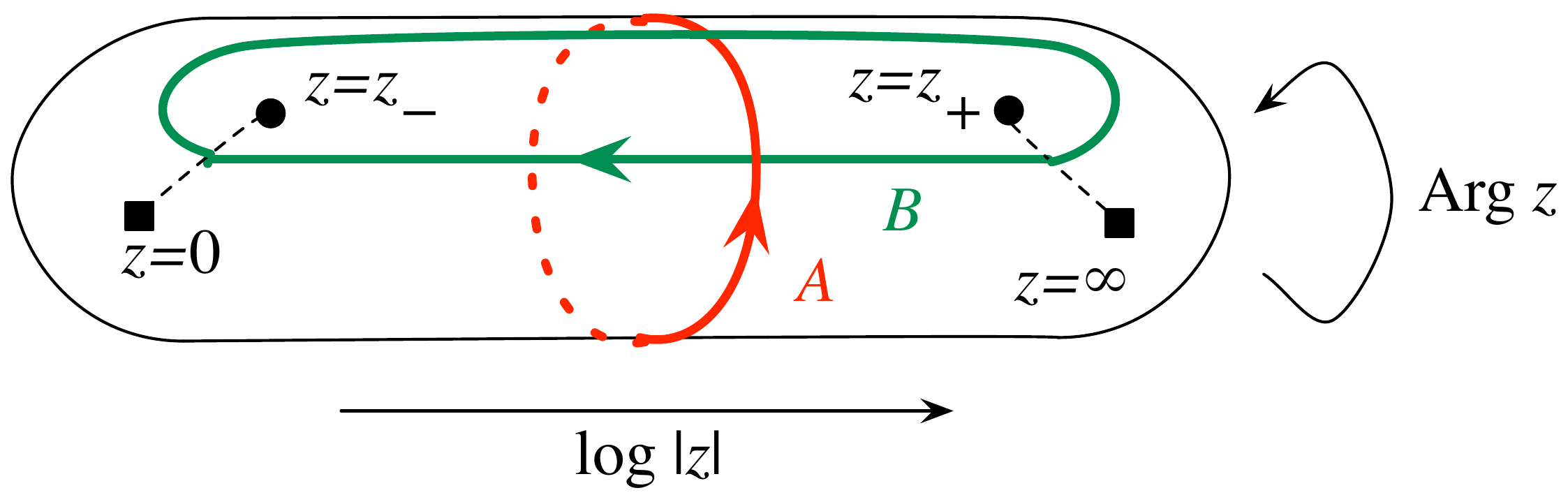}
\]
\caption{The \Gaiotto\ curve $C$ of pure $\SU(2)$ theory.\label{fig:gaiotto_curve_nf_0}}
\end{figure}

The space parametrized by $z$ is important in itself. 
We add the point at $z=\infty$ to the complex plane of $z$, or equivalently, we regard $z$ to be the complex coordinate of a sphere. We denote this sphere by $C$, and call it the \Gaiotto\ curve of this system.  
The variable $x$ as a function of $z$ has four square-root branch points, see Fig.~\ref{fig:gaiotto_curve_nf_0}.

\begin{figure}[h]
\[
\inc{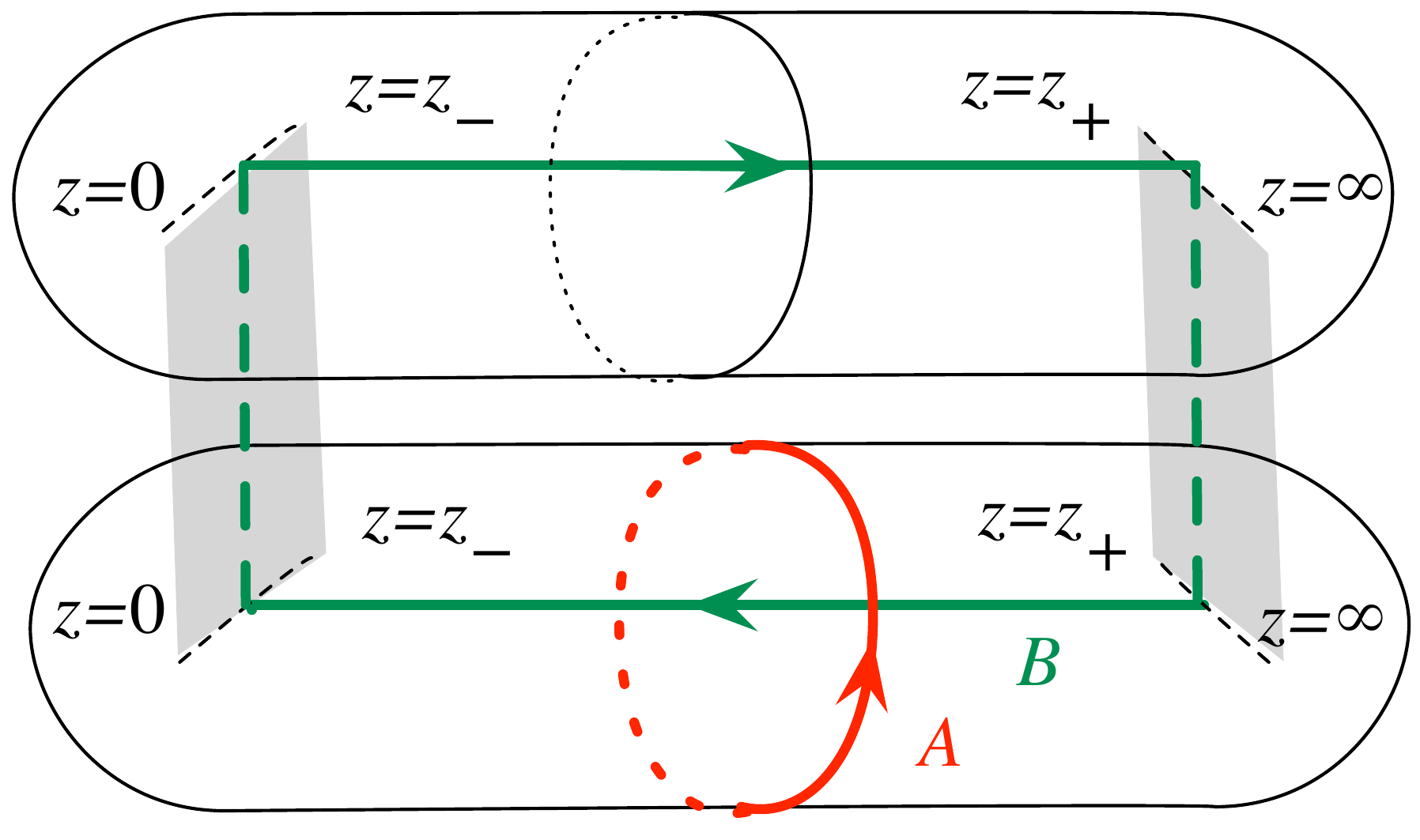}
\]
\caption{The sheets of the \SeibergWitten\ curve $\Sigma$ of pure $\SU(2)$ theory.\label{fig:curve_nf_0}}
\end{figure}

Then the curve $\Sigma$ is a two-sheeted cover of $C$, \begin{equation}
\Sigma\stackrel{2:1}{\longrightarrow} C,
\end{equation}
 see Fig.~\ref{fig:curve_nf_0}.
We then draw two one-dimensional cycles $A$, $B$ on the curves as shown in the figures, and we declare that \begin{equation}
a=\frac{1}{2\pi i}\oint_A \lambda,\qquad
a_D=\frac{1}{2\pi i}\oint_B \lambda.
\end{equation}

Let us check that the functions $a(u)$ and $a_D(u)$ thus defined satisfy physically expected properties. First, let us compute $\tau(a)$: \begin{equation}
\tau(a)=\frac{\partial a_D}{\partial a}
=\frac{\partial a_D/\partial u}{\partial a/\partial u}.
\end{equation}
The $u$ derivatives can be computed in the following way: \begin{align}
\frac{\partial a}{\partial u}&=\int_A \frac{\partial}{\partial u} \lambda 
=\int_A \frac{dz}{2xz}\\
\frac{\partial a_D}{\partial u}&=\int_B \frac{\partial}{\partial u} \lambda 
=\int_B \frac{dz}{2xz}
\end{align} where the $u$ derivative within the integral is taken at fixed $z$. 
The differential $\omega=dz/(xz)$ is finite on $\Sigma$, even at apparently dangerous points $z=0$, $z=\infty$ or at $x=0$. For example, when $x=0$, $z\sim c+c'x^2$ for some constants $c$ and $c'$. Then $dz/(xz)\sim (2c'/c) dx$.

\begin{figure}[h]
\[
\inc{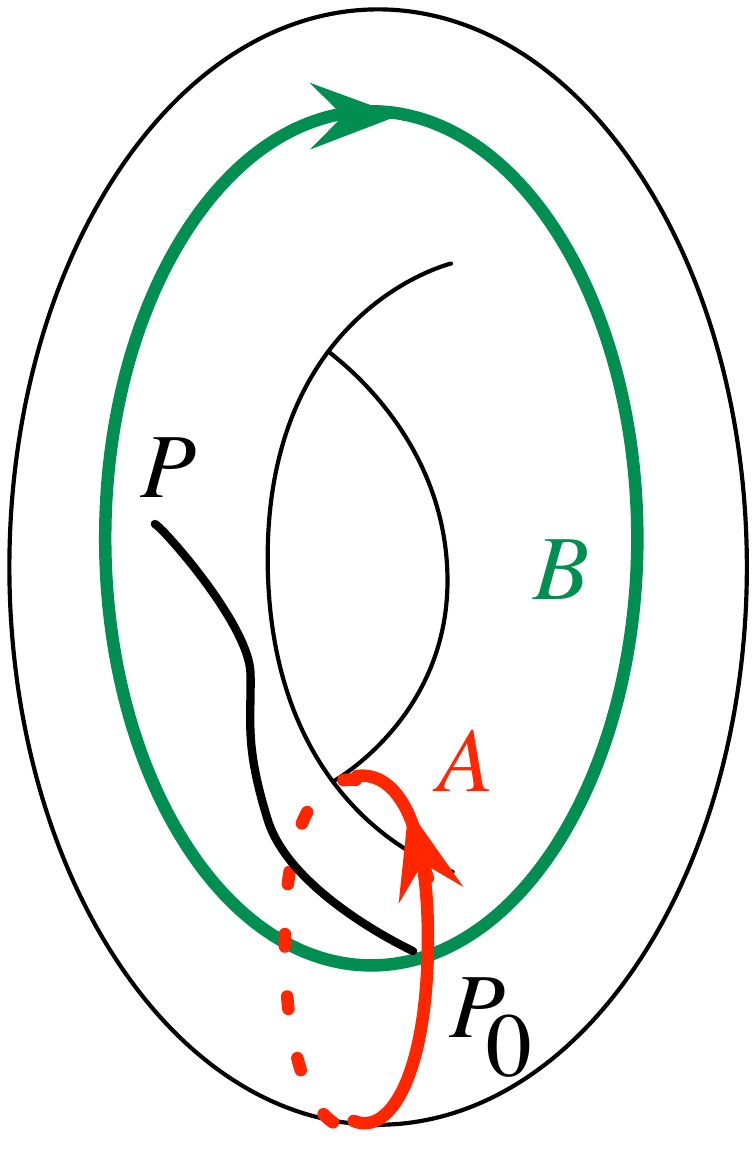}\qquad \Longrightarrow \qquad
\inc{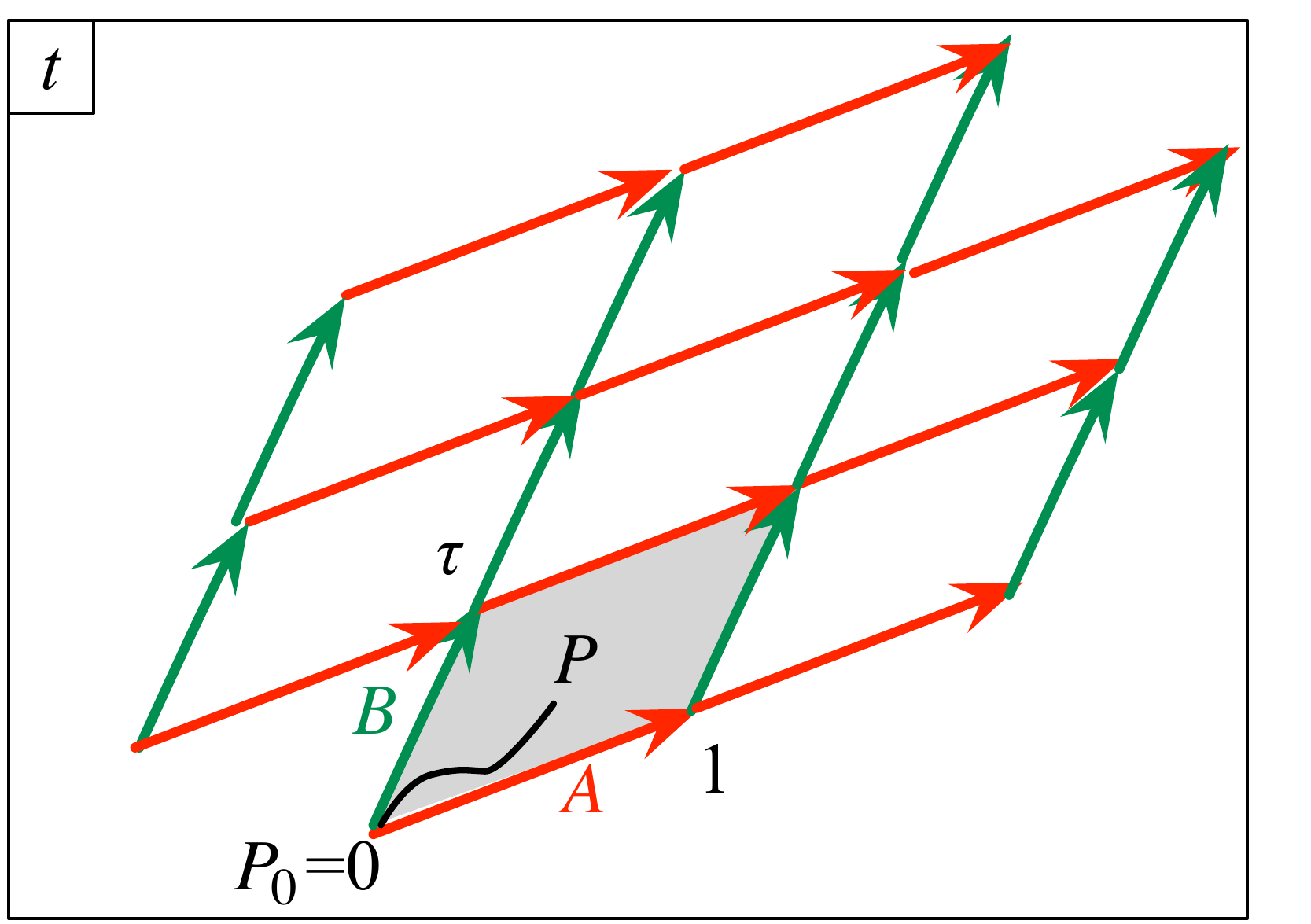}
\]
\caption{The \SeibergWitten\ curve $\Sigma$ of the pure $\SU(2)$ theory, when smoothed out, is a torus.\label{fig:para_nf_0}}
\end{figure}

Given an open path on the curve $\Sigma$ from a fixed point $P_0$, 
we find a map from the endpoint of the path to another complex plane \begin{equation}
t=\int_{P_0}^P \omega.\label{mapping_to_t}
\end{equation} 
As shown in Fig.~\ref{fig:para_nf_0}, the curve $\Sigma$ is mapped to a parallelogram in the complex plane, bounded by the lines which are the images of the cycles $A$ and $B$. 
Now, any holomorphic mapping such as \eqref{mapping_to_t} preserves the angles. Therefore, the image of the cycle $B$ is always to the left of the image of the cycle $A$. 
Then \begin{equation}
\tau(a)=\frac{\partial a_D/\partial u}{\partial a/\partial u}=
\frac{\int_B {dz}/(xz)}{\int_A {dz}/(xz)}
\end{equation} takes the values to the left of the real axis, and therefore \begin{equation}
\Im \tau(a)>0,
\end{equation} which guarantees that the coupling squared $g^2(a)$ is always positive. 
This complex number $\tau(a)$ is called the period or the complex structure of the torus. 

\subsubsection{The monodromy $M_\infty$}
Let us check the curve \eqref{curve_pure} reproduces the monodromy we determined from physical considerations. Write the curve $\Sigma$ as \begin{equation}
z+\frac1z=\frac{x^2}{\Lambda^2}-\frac{u}{\Lambda^2}.\label{simplified_eq}
\end{equation}
From this we see that when $|u|\gg \Lambda^2$, we find two  branch points $z_\pm$ of the function $x(z)$ around \begin{equation}
z_+\sim -u/\Lambda^2,\qquad z_-\sim -\Lambda^2/u.
\end{equation} 
We also have branch points at $z=0$ and $z=\infty$, and we take the branch cuts to run from $z=0$ to $z=z_-$, and from $z=z_+$ to $z=\infty$.

We put the $A$-cycle  at $|z|=1$.
Then the integral over it is very easy: $x\simeq \sqrt{u}$ around $|z|=1$, and therefore \begin{equation}
a=\frac{1}{2\pi i}\oint x\frac{dz}z\simeq \sqrt{u}.
\end{equation}
As for the $B$-cycle integral, the dominant contribution comes when the variable $z$ is not very close to the branch points. The variable $x$ can be again approximated by $\sqrt{u}\simeq a$, and therefore \begin{equation}
a_D=\frac{2}{2\pi i} \int^{z_-}_{z_+} x\frac{dz}z\simeq \frac{2\cdot 2}{2\pi i}  \int^1_{u/\Lambda^2} a\frac{dz}z \simeq -\frac{8a}{2\pi i}\log \frac{a}{\Lambda}.\label{aDpure}
\end{equation}
From these two equations we find that $a$ and $a_D$ defined via the curve $\Sigma$ have the correct monodromy around $u\sim \infty$, \begin{equation}
M_\infty=\begin{pmatrix}
-1 & 4 \\
0  & -1
\end{pmatrix}.
\end{equation}

By a more careful computation, we can explicitly find corrections to \eqref{aDpure}, or to its derivative $\tau(a)$. 
From the form of the curve \eqref{curve_pure}, it is clear that the corrections can be expanded in powers of $\Lambda^2$, but in fact they are given by powers of $\Lambda^4$. We find \begin{equation}
\tau(a)= -\frac{8}{2\pi i}\log\frac{a}{\Lambda} + \sum_{k=0}^\infty c_k \left(\frac{\Lambda}{a}\right)^{4k}
\label{tauexpansion}
\end{equation} where $c_k$ are dimensionless rational numbers.  We now know the terms hidden as $\cdots$ in \eqref{puretau}.   This expansion can be understood for example by introducing $\tilde z=\Lambda^2 z$. Then the curve is $\tilde z+\Lambda^4/\tilde z = x^2-u$, and we can compute $a$, $\tilde a_D$ by considering $\Lambda^4/\tilde z$ as a perturbation to the leading-order form of the curve $\tilde z=x^2-u$. 

Let us interpret these corrections in the powers of $\Lambda^{4}$. From \eqref{lambdaSU2pure}, we know that the term $\Lambda^{4k}$ carries the phase $e^{ik\theta_{UV}}$ where $\theta_{UV}$ is the theta angle. 
It corresponds to a configuration with instanton number $k$, as we learned in \eqref{instanton_contri}. 
This expansion explicitly demonstrates that the only perturbative correction to the low-energy coupling $\tau(a)$ is from the one-loop level, and there are non-perturbative corrections from the instantons. 
An honest path-integral computation in the instanton background should reproduce the coefficients $c_k$. For the one-instanton contribution $c_1$ this was done in \cite{Finnell:1995dr}. It was later extended to all $k$ in \cite{Nekrasov:2002qd,Nekrasov:2003rj}.
Summarizing, we see that various quantities are given by a combination of a one-loop logarithmic contribution plus instanton corrections. It is now known that they agree to all orders in the instanton expansion, thanks to the developments starting from \cite{Nekrasov:2002qd}.
In the Appendix \ref{sec:instcount}, we compute the coefficients $c_k$ directly from the curve, and  see that they agree with the results from microscopic instanton computation.

\subsubsection{The monodromies $M_\pm$}
Let us next study the monodromy around the strongly-coupled singularities.
Taking a look at \eqref{simplified_eq} again, it is clear that when $z+1/z=\pm 2$  we have a  rather special situation.
When $u=2\Lambda^2$, the two branch points collide at $z_\pm=-1$,
and when $u=-2\Lambda^2$, they collide at $z_\pm=+1$.
These are the singularities $u=\pm u_0$ introduced in Fig.~\ref{fig:monodromy_nf_0_correct}.

\begin{figure}[h]
\[
\inc{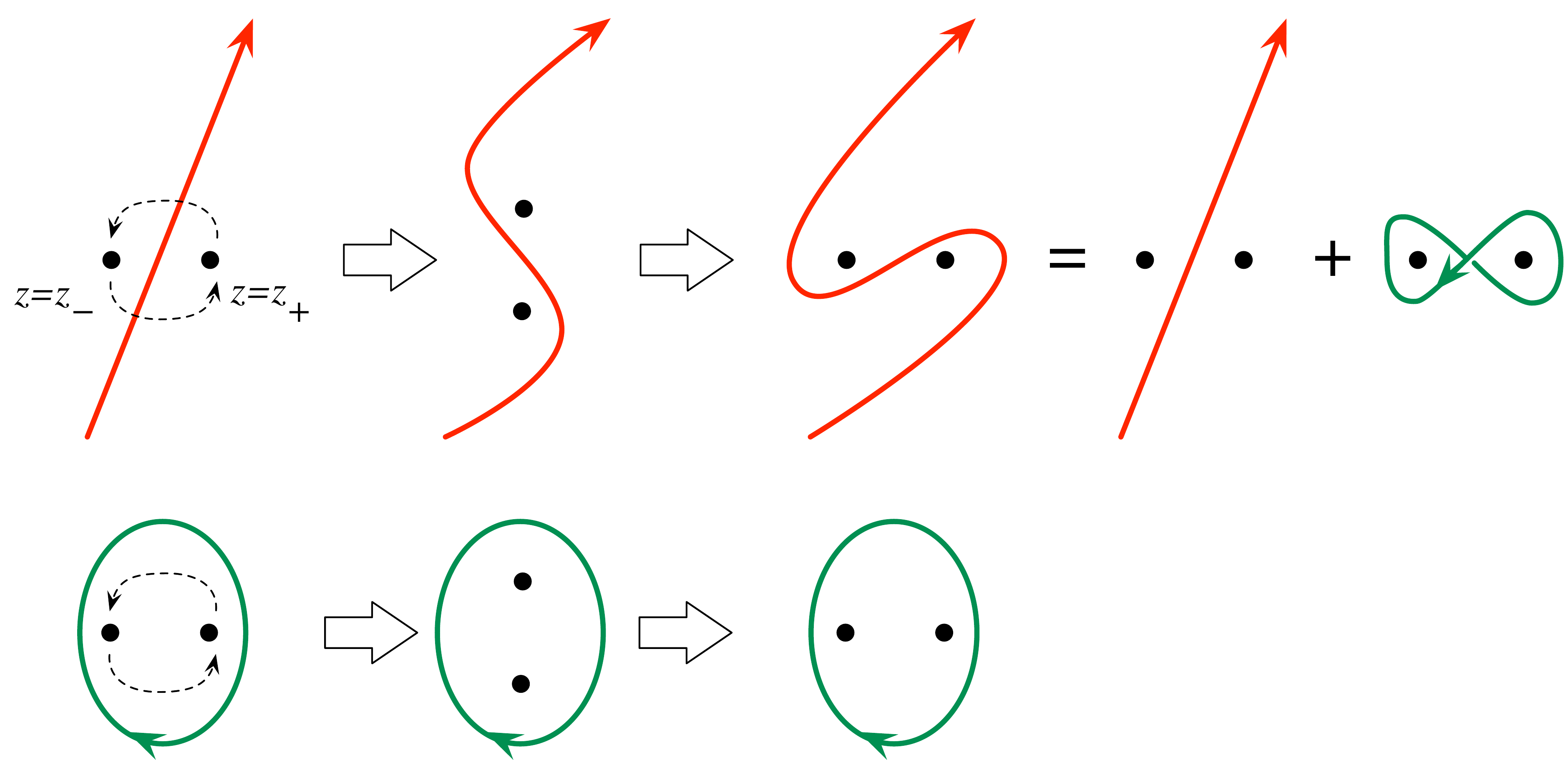}
\]
\caption{Monodromy action on cycles around the monopole point. \label{fig:closeup_nf_0}}
\end{figure}

Let us study the behavior close to $u=u_0=2\Lambda^2$ as an example. 
We let $u=2\Lambda^2+\delta u$. Then the branch points are at \begin{equation}
z_\pm-1\propto \pm \sqrt{\delta u}.
\end{equation}
The close up of the branch points $z_\pm$ and the cycles $A$, $B$ are shown in Fig.~\ref{fig:closeup_nf_0}.
When we slowly change the value of $u$ around $u=2\Lambda^2$, two branch points 
$z=z_\pm$ are exchanged. This modifies the cycle $A$ as shown in the figure,
which is equivalent to the original cycle $A$ minus the cycle $B$. 
The cycle $B$ is clearly unchanged. 
Therefore we have \begin{equation}
a \to a-a_D,\qquad
a_D\to a_D
\end{equation} or equivalently, the monodromy is \begin{equation}
M_+=\begin{pmatrix}
1 & 0 \\
-1 & 1
\end{pmatrix}, 
\end{equation}reproducing \eqref{mpm}.

Let us study the physics at $u=u_0=2\Lambda^2$. 
We perform the $S$ transformation \eqref{S} \begin{equation}
a'=-a_D,\quad
a_D'=a
\end{equation} 
exchanging the electric and magnetic charges. 
These are given as functions of $u$ by  \begin{equation}
a'=c(u-u_0),\quad
a_D'=\frac{a'}{2\pi i}\log c'(u-u_0)\label{cc}
\end{equation} where  $c$ and $c'$ are two constants, from which we find \begin{equation}
\tau_D(a')=\frac{\partial a_D'}{\partial a'}\sim +\frac{\log a'}{2\pi i}.\label{dual_run}
\end{equation} 
Note that $a'$ sets the energy scale of the system.
The result shows the same behavior as the running of the coupling of an $\cN{=}2$ supersymmetric $\U(1)$ gauge theory with one charged hypermultiplet, consisting of $\cN{=}1$ chiral multiplets $(Q,\tilde Q)$. The superpotential coupling is then \begin{equation}
\int d^2\theta Qa'\tilde Q.
\end{equation}
Writing \begin{equation}
\tau_D=\frac{4\pi i}{g_D^2} + \frac{\theta_D}{2\pi},
\end{equation} we find \begin{equation}
g_D\to 0
\end{equation} as we approach $u\to u_0$. 
The mass of the quantum of $Q$ is given by the BPS mass formula  to be \begin{equation}
\text{mass of quantum of $Q$}=|a'|=|a_D|.
\end{equation} Therefore, we identify  the charged chiral multiplet $Q$ as the second quantized version of the monopole in the original theory. The monopoles, which were very heavy in the weakly coupled region, are now very light. 

The behavior at $u=-u_0$ is easily given by applying the discrete R-symmetry \eqref{discrete_r_nf_0}. As we map by $T^2$, we find that the very light particles now have 
 electric charge $n=2$ and magnetic charge $m=1$, i.e.~they are dyons.   From these reasons, the point $u=u_0=2\Lambda$ is often called the monopole point, and the point $u=-u_0=-2\Lambda$ the dyon point. 

\subsection{Less supersymmetric cases}

Before continuing the study of $\cN{=}2$ systems, let us pause here and  see what we can learn about  less supersymmetric theories from the solution of the pure $\cN{=}2$ $\SU(2)$ theory. A general Lagrangian we consider in this section is given by \begin{equation}
(\int d^2\theta \frac{-i}{8\pi}\tau \tr W_\alpha W^\alpha + cc.)
+\frac{\Im \tau}{4\pi}\int d^4\theta \Phi^\dagger \Phi 
+(\int d^2\theta \frac{m}2\tr\Phi^2 + cc.)
+(\mu \lambda_\alpha\lambda^\alpha + cc.)\label{441}
\end{equation}
The setup is $\cN{=}2$ supersymmetric when $m=\mu=0$. 
When we let $|m|\to \infty$, we decouple the chiral superfield $\Phi$, and we end up with $\cN{=}1$ pure $\SU(2)$ theory which we discussed in Sec.~\ref{sec:n=1pure}.
Next, by letting $|\mu| \to \infty$, we decouple the gaugino $\lambda$ and recover pure bosonic Yang-Mills.
\subsubsection{$\cN{=}1$ system}
First let us consider the $\cN{=}1$ system. When $m$ is very small, the term $m\tr\Phi^2$ can be considered as a perturbation to the $\cN{=}2$ solution we just obtained. In terms of the variable $u$, the term $\int d^2\theta m\tr\Phi^2$ is $\sim \int d^2\theta m u$, and therefore the F-term equation with respect to $u$ cannot be satisfied unless $u$ is at the singularity. There is no supersymmetric vacuum at generic value of $u$. 

When $u$ is close to $u_0=2\Lambda^2$, there are additional terms in the superpotential given by \begin{equation}
\int d^2\theta  Qa'\tilde Q = \int d^2\theta c(u-u_0) Q\tilde Q
\end{equation} where the constant $c$ was introduced in \eqref{cc}.
Together with the term $-\int d^2\theta m u$, the F-term equations with respect to $u$, $Q$ and $\tilde Q$ are given respectively by \begin{equation}
m=cQ\tilde Q,\quad
(u-u_0)\tilde Q=0,\quad
(u-u_0)Q=0.
\end{equation} Then we find a solution at \begin{equation}
u=u_0, \quad Q\tilde Q =m/c.
\end{equation} The vacuum is pinned at $u=u_0$, and there is a nonzero condensate of the monopole $Q\tilde Q=m/c$. 
A similar argument at $u=-u_0$ says that there is another supersymmetric vacuum given by \begin{equation}
u=-u_0, \quad Q'\tilde Q'=m/c
\end{equation} where $Q'$, $\tilde Q'$ are the dyon fields.  

\begin{figure}[h]
\[
\inc{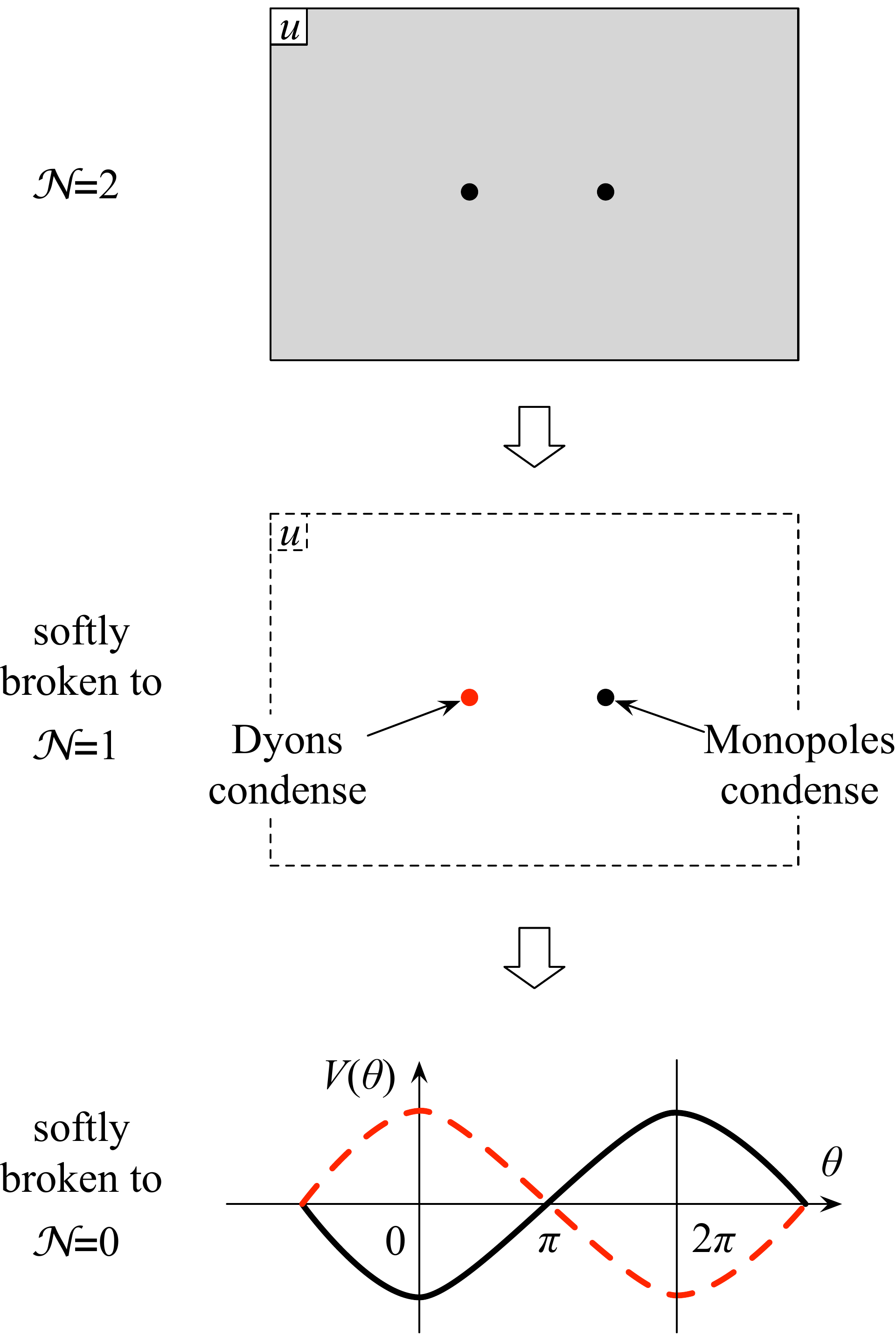}
\]
\caption{Vacua for the softly broken $\cN{=}1$, $\cN{=}0$ theories\label{fig:less_supersymmetric}}
\end{figure}

Summarizing, we found two supersymmetric vacua at $u=\pm u_0$, where monopoles or dyons condense, concretely realizing the idea that the confinement is given by condensation of magnetically-charged objects, see Fig.~\ref{fig:less_supersymmetric}.

Recall that the anomalously broken continuous R-symmetry \begin{equation}
\Phi \to e^{i\varphi} \Phi,
\end{equation}
can be compensated by the 
\begin{equation}
\theta_{UV} \to \theta_{UV}+4\varphi.
\end{equation} 
Applying it to the Lagrangian \eqref{441},
we see that \begin{equation}
m\vev{\tr \Phi^2} = \frac{-i}{2\pi} \vev{\tr W^\alpha W_\alpha} 
\end{equation} with which we find \begin{equation}
\vev{\lambda_\alpha \lambda^\alpha} \propto \pm 2\pi i m \Lambda^2 =: \pm \Lambda_{\cN{=}1}^3.
\end{equation} It is important to keep in mind that the right hand side contains
$e^{i\theta_{UV}/2}$ as the phase.

We now take the limit $m\to \infty$ keeping $\Lambda_{\cN{=}1}$ fixed. 
This should give the pure $\cN{=}1$ $\SU(2)$ Yang-Mills theory. 
It is reassuring to find that we also see two vacua here, as in Sec.~\ref{sec:n=1pure}.

\subsubsection{Pure bosonic system}
Let us now make $\mu\neq 0$, keeping $|\mu| \ll |\Lambda_{\cN{=}1}|$. 
In this limit, the effect of the gaugino mass term $\mu \lambda_\alpha\lambda^\alpha$ is given by the first order perturbation theory, and the vacuum energy is given by \begin{equation}
V\propto \Re (\pm \mu  \Lambda^3_{\cN{=}1})
\propto \Lambda^4_{\cN{=}0} \Re (\pm e^{i\theta_{UV}/2}).
\end{equation} This was first pointed out in \cite{Konishi:1996iz}.

We see that two degenerate vacua of the $\cN{=}1$ supersymmetric theory are split into two levels with different energy density, corresponding to monopole condensation and dyon condensation, respectively.
A slow change of $\theta_{UV}$ from $0$ to $2\pi$ exchanges the two levels,
which cross at $\theta_{UV}=\pi$. So there is a first-order phase transition at $\theta_{UV}=\pi$, at least when $|\mu|$ is sufficiently small.

It is an interesting question to ask if this first order phase transition persists in the limit $|\mu|\to \infty$, i.e.~in the pure bosonic Yang-Mills theory. Let us give an argument for the persistence. The idea is to use the behavior of the potential between two external particles which are magnetically  or dyonically charged as the order parameter \cite{Aharony:2013hda}. 

First let us consider the dynamics more carefully. Two branches differ in the types of particles which condense: we can call the branches the monopole branch and the dyon branch, accordingly.  
In our convention, the charges of the particles are $(n,m)=(0,1)$ and $(2,1)$, respectively.
The charge of the $\SU(2)$ adjoint fields, under the unbroken $\U(1)$ symmetry, is $(2,0)$ in our normalization. As there are no dynamical particles of charge $(1,0)$, 
the charge $(0,1)$ of the monopole is twice that of a minimally allowed one. The charge of this external monopole can then be written as $(n,m)=(0,1/2)$. 

Consider first introducing two external electric particles with charge $(n,m)=(1,0)$.
In both branches, the electric field is made into a flux tube by the condensed monopoles or dyons. The flux tube has constant tension, and cannot pair-create dynamical particles, since all the dynamical particles have charge $(\pm 2,0)$. Therefore the flux tube does not break, and the potential is linear. The electric particles with charge $(1,0)$ are confined. 

Instead, let us consider introducing external monopoles into the system, and measure the potential between the two. 
At $\theta=0$, we can assume, without loss of generality, that the monopole branch has lower energy. 
There are dynamical monopole particles with charge $(n,m)=(0,1)$ condensing in the background. 
Let us introduce two external monopoles of charge $(n,m)=(0,1/2)$.
The magnetic field produced by the external particles with charges $(n,m)=(0,1)$ is screened and damped exponentially. The potential between them is then basically constant.

Instead, consider introducing two external particles with charge $(1,1/2)$ into the monopole branch. The dynamical monopole cannot screen the electric charge, which is then confined into a flux tube. The potential between them is linear and they are confined. 

We can repeat the analysis in the dyon branch. The behavior of the potential between external particles can be summarized as follows: 
\begin{center}
\begin{tabular}{r|cc}
& $(0,1/2)$ & $(1,1/2)$ \\
\hline
monopole branch & screened & confined \\
dyon branch & confined & screened
\end{tabular}
\end{center}
These two behaviors are exchanged under a slow continuous change of $\theta$ from $0$ to $2\pi$. Therefore, there should be at least one phase transition. It would be interesting to confirm this analysis by a lattice strong-coupling expansion, or by a computer simulation. 

%\eject

\subsection{$\SU(2)$ vs $\SO(3)$}
At this point, it might be useful to discuss a rather sutble point concerning the \SeibergWitten\ curve of the theory which depends on the precise choice of the gauge group to be $\SU(2)$ or $\SO(3)$. 
This subsection can be skipped on a first reading. 

In this section, our choice of the charges has been that \begin{equation}
(n,m)=(1,0)
\end{equation} represents an electric doublet in $\SU(2)$,  \begin{equation}
(n,m)=(0,1)
\end{equation} represents a 't Hooft-Polyakov monopole associated to the breaking $\SU(2)\to \U(1)$.

That said, the  dynamical particles in the theory all has the charge of the form \begin{equation}
(n,m)=(2k,m)\label{dync}
\end{equation} for some integers $k$ and $m$. Furthermore, as we do not have any dynamical fields in the doublet of $\SU(2)$, we can consider an external monopole with charge \begin{equation}
(n,m)=(0,1/2).
\end{equation} This still satisfies the Dirac quantization condition with respect to any of the dynamical particles in the theory, whose chages are given by \eqref{dync}.

 Correspondingly, the monodromy matrices \begin{equation}
M_\infty=\begin{pmatrix}
-1 & 4 \\
0 & -1
\end{pmatrix},\qquad
M_+=\begin{pmatrix}
1 & 0 \\
-1 & 1
\end{pmatrix},\qquad
M_-=\begin{pmatrix}
-1 & 4 \\
-1 & 3
\end{pmatrix}
\end{equation} all had an integral multiple of 4 in the upper right corner. 

\begin{figure}[h]
\[
\inc{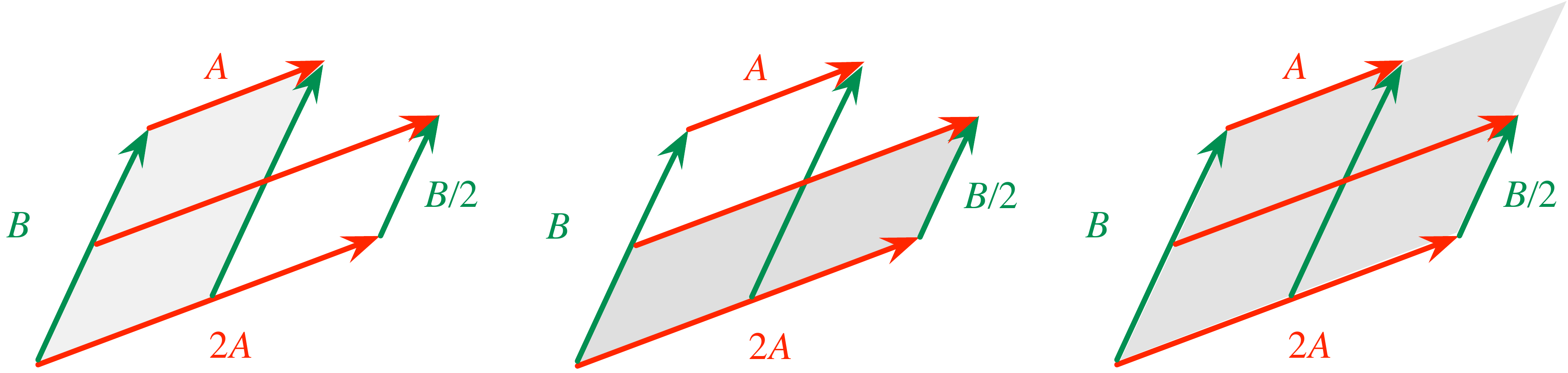}
\]
\caption{Various choices of the \SeibergWitten curves. From the left to right: the pure $\SU(2)$ theory, the pure $\SO(3)$ theory, and the choice in \cite{Seiberg:1994rs}\label{fig:so3}}
\end{figure}

Therefore, we can do the following. We define rescaled electric and magnetic charges via\begin{equation}
(n',m')=(n/2,2m)
\end{equation} and still the monodromy are still integer valued: \begin{equation}
M_\infty'=\begin{pmatrix}
-1 & 1 \\
0 & -1
\end{pmatrix},\qquad
M_+'=\begin{pmatrix}
1 & 0 \\
-4 & 1
\end{pmatrix},\qquad
M_-'=\begin{pmatrix}
-1 & 1 \\
-4 & 3
\end{pmatrix}.
\end{equation}  
The BPS mass formula is now \begin{equation}
|na+ma_D|=|n'a'+m'a'_D| \qquad \text{where}\quad (a',a'_D)=(2a,a_D/2)
\end{equation}  and correspondingly, the new $A$ and $B$ cycles are related to the old ones via \begin{equation}
A'=2A,\qquad B'=B/2.
\end{equation} The respective \SeibergWitten\ curves, as quotients of the complex plane as in Fig.~\ref{fig:para_nf_0}, are given in Fig.~\ref{fig:so3}.

The standard interpretation is that  the \SeibergWitten\ curve with cycles $A$ and $B$ as the curve for the pure $\SU(2)$ theory, and that the \SeibergWitten\ curve with cycles $A'=2A$ and $B'=B/2$ as the curve for the pure $\SO(3)$ theory. The difference manifests in a rather subtle manner. 

At the monopole point, with $\SO(3)$ gauge group, the monodromy is $M_+'$ given above. This means that the charge of the light particle there is 2 with respect to the low-energy $\U(1)$ field: the entry $-4$ in the lower left corner is given by the square of the charges.  This is due to the fact that the periodicity of low-energy $\U(1)$ is reduced by a factor of two, as it is embedded in $\SO(3)$ rather than $\SU(2)$.  
The monopole has $(n',m')=(0,2)$ and the g.c.d.~of $n'$ and $m'$ is two.

At the dyon point, the monodromy $M_-'$ is still conjugate to $\begin{pmatrix}
1 &1 \\
0 & 1
\end{pmatrix}$, that is, the light dyon has charge $1$ with respect to the $\U(1)$. 
The dyon has $(n',m')=(1,2)$ and the g.c.d.~of $n'$ and $m'$ is one. 
Therefore one loses the physical equivalence of the monopole point and the dyon point.
These subtle differences affect the system more drastically when the system is put on $\bR^3\times S^1$ or more complicated manifolds. 

Another interesting fact is that this combination $M_\infty'$, $M_+'$, and $M_-'$ is exactly the same as the monodromy matrices of that of the $\SU(2)$ theory with 3 massless flavors, which we discuss in Sec.~\ref{sec:nf3}.
Still, the physics of the pure $\SO(3)$ theory and the $\SU(2)$ theory with 3 massless flavors are drastically different, as we will learn later. 
This shows an obvious point that the structure of the Coulomb branch alone does not fix the entire theory. 

Finally, we should mention that in the very orignal paper on the pure $\SU(2)$ theory \cite{Seiberg:1994rs} another curve is used, which had $2A$ and $B$ as two cycles, as shown in Fig.~\ref{fig:so3}. 
This choice is adapted to the spectrum of the dynamical particles \eqref{dync}, but it is now known to be a not very well motivated when we consider the theory on nontrivial manifolds and the properties of line operators. 
Therefore, it is advisable to stick to either the pure $\SU(2)$ curve or the pure $\SO(3)$ curve, given as the first two entries in Fig.~\ref{fig:so3}.

\section{$\SU(2)$ theory with one flavor}\label{sec:nf=1}
Our next task is to study $\cN{=}2$ supersymmetric $\SU(2)$ gauge theory with one hypermultiplet in the doublet representation. This is often called the $\SU(2)$ theory with one flavor, or more simply $N_f=1$. 
We will see that all the methods employed in the last section are readily adapted to this theory, too.
We again follow the presentation of the original paper \cite{Seiberg:1994aj}, but  we use the \SeibergWitten\ curve in a form more suitable for the generalization later. Appendix C of \cite{Hollands:2010xa} is a good source where many different forms of the \SeibergWitten\ curves of $\SU(2)$ theories are summarized. 

\subsection{Structure of the $u$-plane}
\subsubsection{Schematic running of the coupling}
In terms of $\cN{=}1$ chiral multiplets, the hypermultiplet consists of two $\SU(2)$ doublets $Q^a$ and $\tilde Q_a$ where $a=1,2$ is the $\SU(2)$ index.
There is an $\cN{=}1$ superpotential \begin{equation}
W=Q\Phi \tilde Q + \mu Q\tilde Q
\end{equation} where $\mu$ is the bare mass of the hypermultiplet. 
Classically, $\Phi=\diag(a,-a)$ together with $Q=\tilde Q=0$ still gives supersymmetric vacua.
With nonzero $a$, the gauge group is broken to $\U(1)$, and the physical mass of the fields $Q$ and $\tilde Q$ can be found by explicitly expanding the superpotential above:
\begin{equation}
W=(Q_1,Q_2)\begin{pmatrix}
a & 0\\
0 & -a 
\end{pmatrix}
\begin{pmatrix}
\tilde Q_1\\
\tilde Q_2
\end{pmatrix}
+\mu (Q_1,Q_2)\begin{pmatrix}
\tilde Q_1\\
\tilde Q_2
\end{pmatrix}.
\end{equation} We see that the masses are \begin{equation}
|{\pm a\pm\mu}|\label{qmass}
\end{equation} where we allow all four choices of signs. 

In general, the BPS mass formula is \begin{equation}
\text{mass}\ge |na+ma_D+f\mu|\label{bps_nf_1}
\end{equation} where $f$ is the charge under the $\U(1)$ flavor symmetry, under which $Q$ has charge $1$ and $\tilde Q$ has charge $-1$.

 From the one-loop running of the coupling constant, we find \begin{align}
\tau(a)&=2\tau_{UV}-\frac{6}{2\pi i}\log \frac{a}{\Lambda_{UV}}+\cdots \\
&=-\frac{6}{2\pi i}\log \frac{a}{\Lambda_1} + \cdots \label{running_nf_1}
\end{align} in the ultraviolet region. Here we defined \begin{equation}
\Lambda_1^6 = \Lambda_{UV}^6 e^{4\pi i \tau_{UV}}
\end{equation} where the subscript $1$ is a reminder that we are dealing with the $N_f=1$ theory.
From this, we can determine the monodromy $M_\infty$ at infinity acting on $(a,a_D)$: \begin{equation}
M_\infty=\begin{pmatrix}
-1 & 3 \\
0 & -1
\end{pmatrix}\label{Minf_nf_1}
\end{equation} exactly as in the pure $\SU(2)$ case.

\begin{figure}[h]
\[
\inc{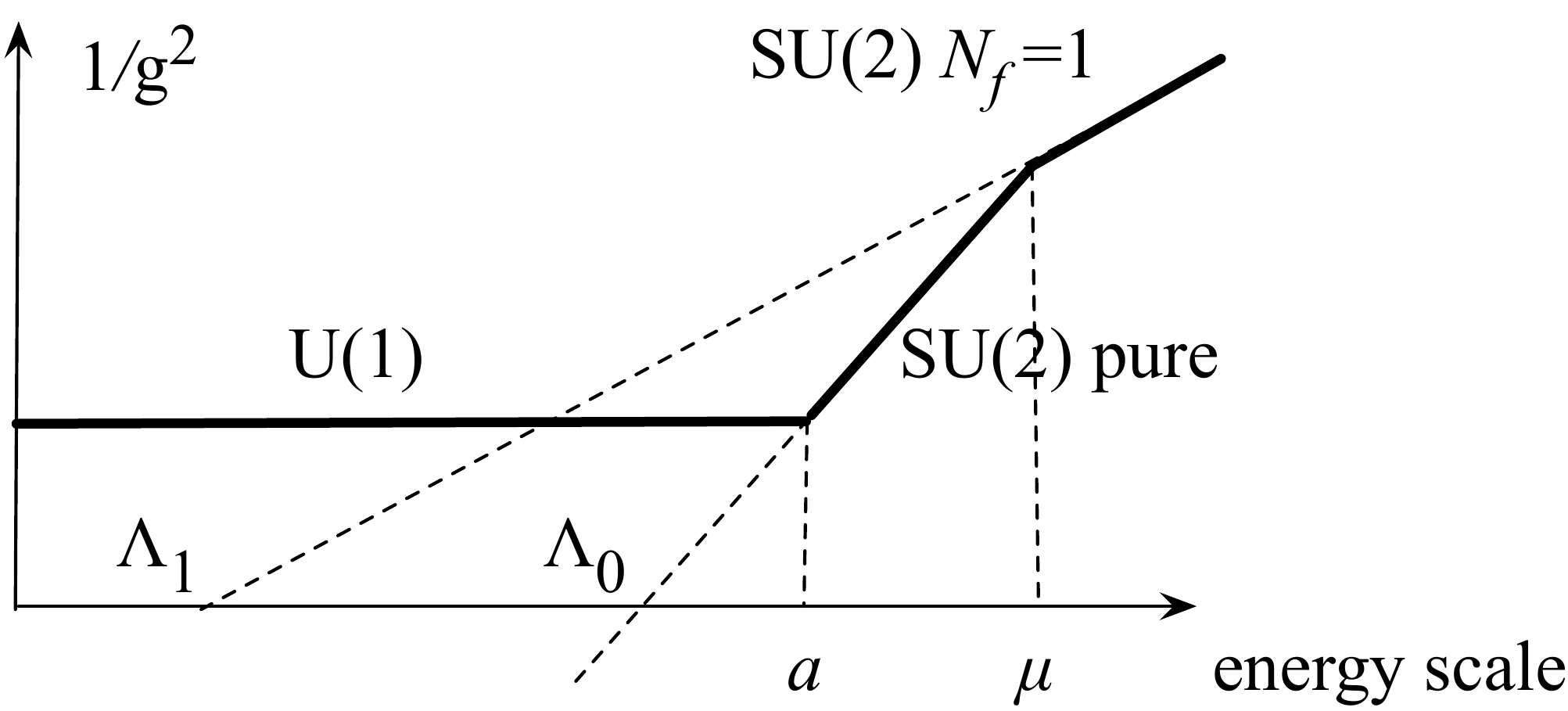}
\]
\caption{Schematic running of the coupling of $N_f=1$ theory, when $|\Lambda|\ll |a|\ll |\mu|$\label{fig:running_nf_1}}
\end{figure}

To study the strong coupling region of the system, let us first consider two extreme cases. When $|\mu|$ is very big, we expect the running of the coupling to be given roughly as in Fig.~\ref{fig:running_nf_1}. Namely,
at around the scale $|\mu|$, the fields $Q$ and $\tilde Q$ decouple, and the system effectively becomes the pure $\SU(2)$ gauge theory, which we studied in the last section. Correspondingly, the structure of the  $u$-plane in the region $|u|\ll |\mu^2|$ should be effectively the same with that of the pure $\SU(2)$ theory,
with two singularities at $u=\pm 2\Lambda_0^2$. 

A rough relation between $\Lambda_0$ and $\Lambda_1$ 
  can be read off from the schematic graph of the running coupling shown in Fig.~\ref{fig:running_nf_1}. The rightmost segment in the graph is given by \begin{equation}
\tau(\LambdaRG )=-\frac{6}{2\pi i}\log \frac{\LambdaRG }{\Lambda_1}
\end{equation} and the middle segment in the graph, representing the  effectively pure $\SU(2)$ theory, is \begin{equation}
\tau(\LambdaRG )=-\frac{8}{2\pi i}\log \frac{\LambdaRG }{\Lambda_0}.
\end{equation} Equating these two values at $\LambdaRG =\mu$, we obtain 
\begin{equation}
\Lambda_0^4 = \mu \Lambda_1^3.
\label{l1vsl0}
\end{equation}

\begin{figure}[h]
\[
\inc{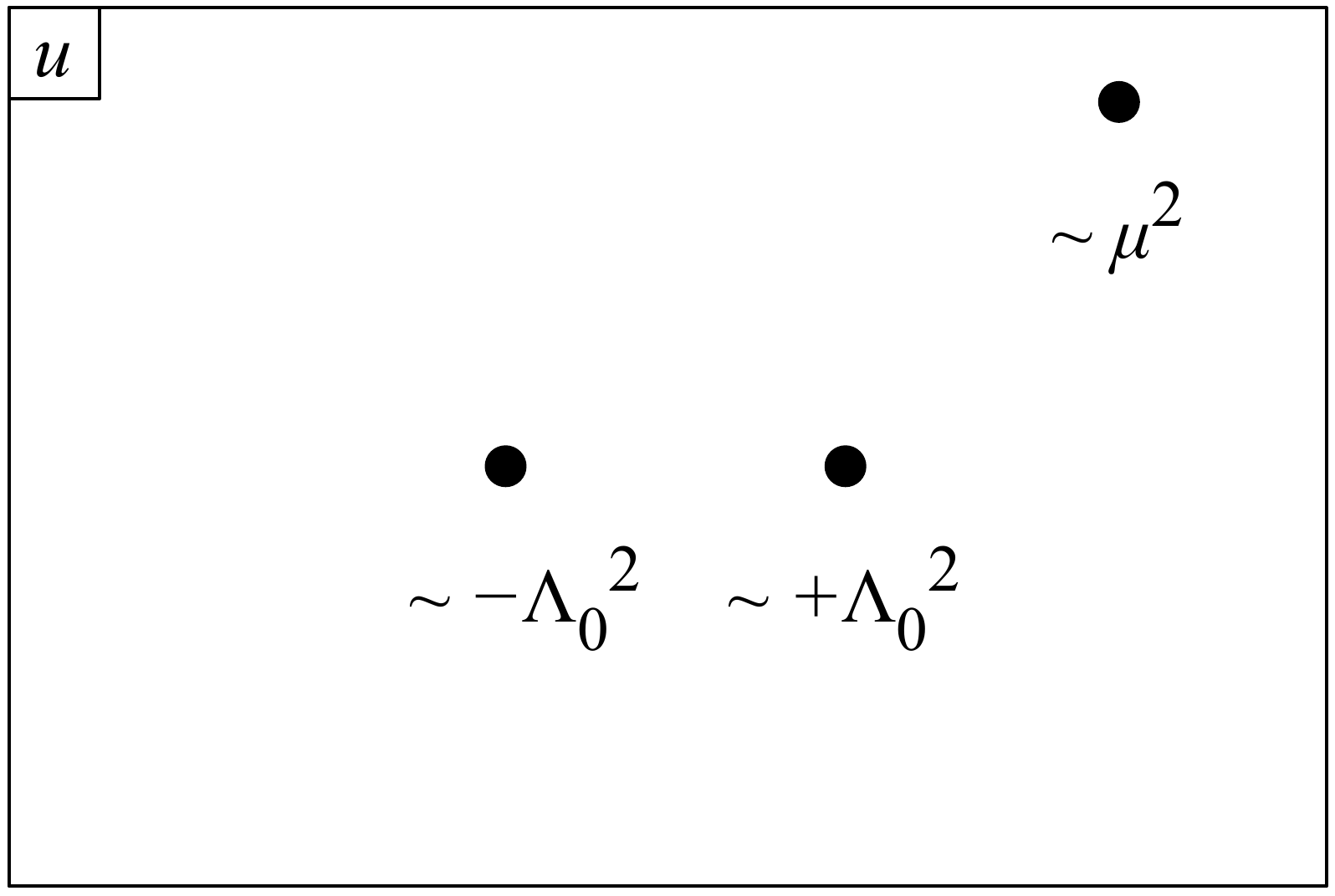}
\]
\caption{Singularities on the $u$-plane when $m\gg \Lambda$\label{fig:uplane_nf_1_m_large}}
\end{figure}

In addition, we know from \eqref{qmass} that 
the quanta of one component of $Q$ and $\tilde Q$ become very light  when $\pm a\sim \mu$. This should produce a singularity in the $u$-plane at around $u\simeq \mu^2$. We therefore expect that the $u$-plane to have three singularities, as shown in Fig.~\ref{fig:uplane_nf_1_m_large}.
Note that local physics at the three singularities, at $u\simeq \mu^2$ and at $u\simeq \pm 2\Lambda_0^2$, is always just $\U(1)$ gauge theory with one charged hypermultiplet. 

\begin{figure}[h]
\[
\inc{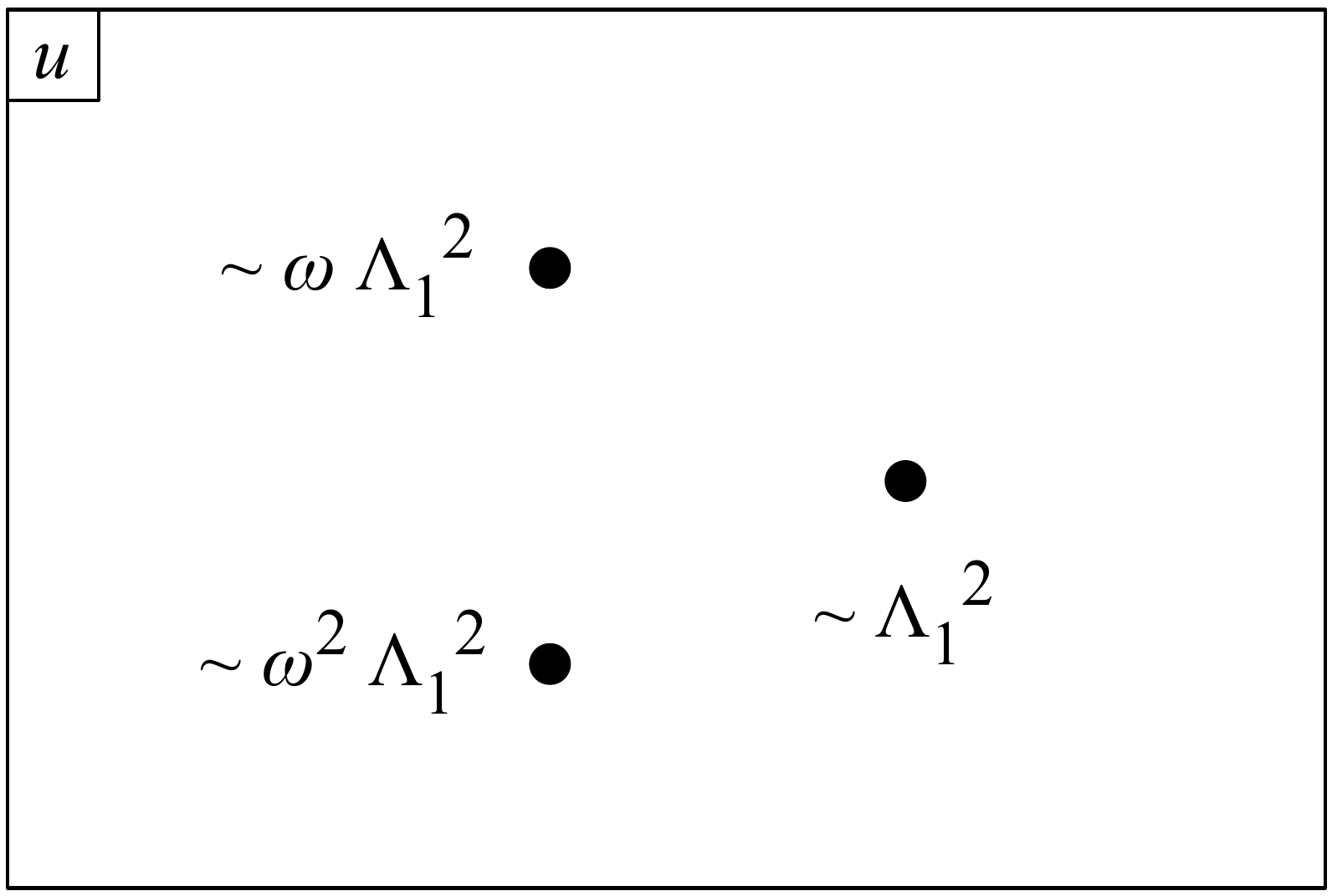}
\]
\caption{Singularities on the $u$-plane when $m=0$\label{fig:uplane_nf_1_mless}}
\end{figure}

In the other extreme when $\mu=0$, we can make use of the discrete R symmetry. 
The standard R-charge assignment is as follows: \begin{equation}
\begin{array}{r|ccc}
R=0 & & A\\
1& \lambda & & \lambda \\
2 & & \Phi
\end{array}, \qquad
\begin{array}{r|ccc}
R=-1 & & \psi_Q\\
0& Q & & \tilde Q^\dagger \\
1 & & \psi_{\tilde Q}^\dagger
\end{array}.
\end{equation}
The rotation \begin{equation}
\lambda\to e^{i\varphi} \lambda,\qquad
\psi_{Q,\tilde Q}\to e^{-i\varphi} \psi_{Q,\tilde Q}
\end{equation}  is anomalous, but can be compensated by \begin{equation}
\theta_{UV}\to \theta_{UV}+6\varphi.
\end{equation} Therefore $\varphi=2\pi/6$ is a genuine symmetry, which does \begin{equation}
\theta\to \theta+2\pi,\quad
\Phi\to e^{2\pi i/3}\Phi,\quad
u\to e^{4\pi i/3} u.
\end{equation}
This guarantees that singularities in the $u$-plane should appear in triples, related by $120^\circ$ rotation. 
A minimal assumption is then to have exactly three singularities, as shown in Fig.~\ref{fig:uplane_nf_1_mless}.
Having three singularities is consistent with our previous analysis when $|\mu|$ was very big.
We expect that the situation in Fig.~\ref{fig:uplane_nf_1_m_large} will smoothly change into
the one in Fig.~\ref{fig:uplane_nf_1_mless} when $\mu$ is adiabatically changed. 

\subsubsection{Monodromies}

\begin{figure}[h]
\[
\inc{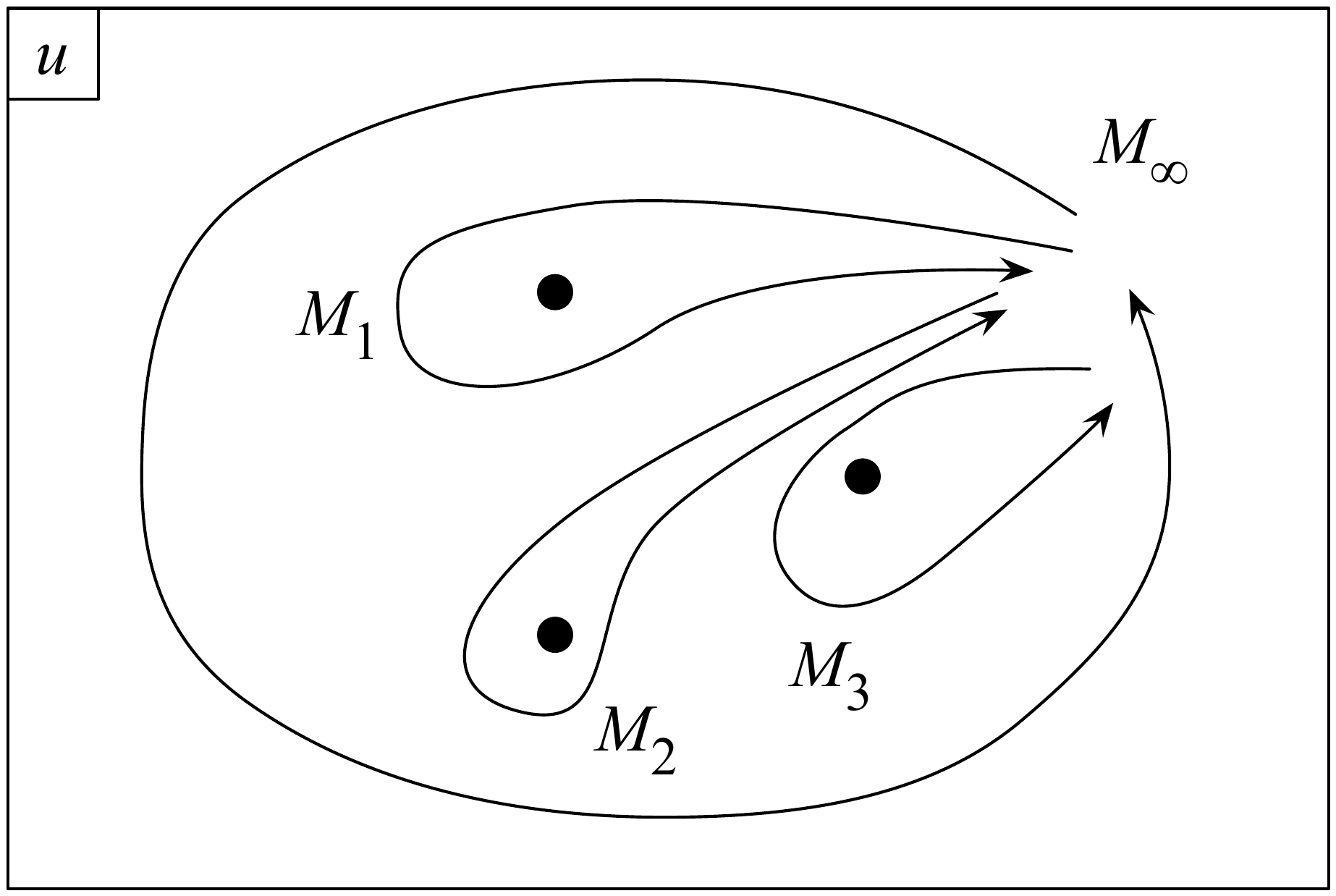}
\]
\caption{Monodromy of $N_f=1$\label{fig:monodromy_nf_1}}
\end{figure}
\begin{figure}[h]
\[
\inc{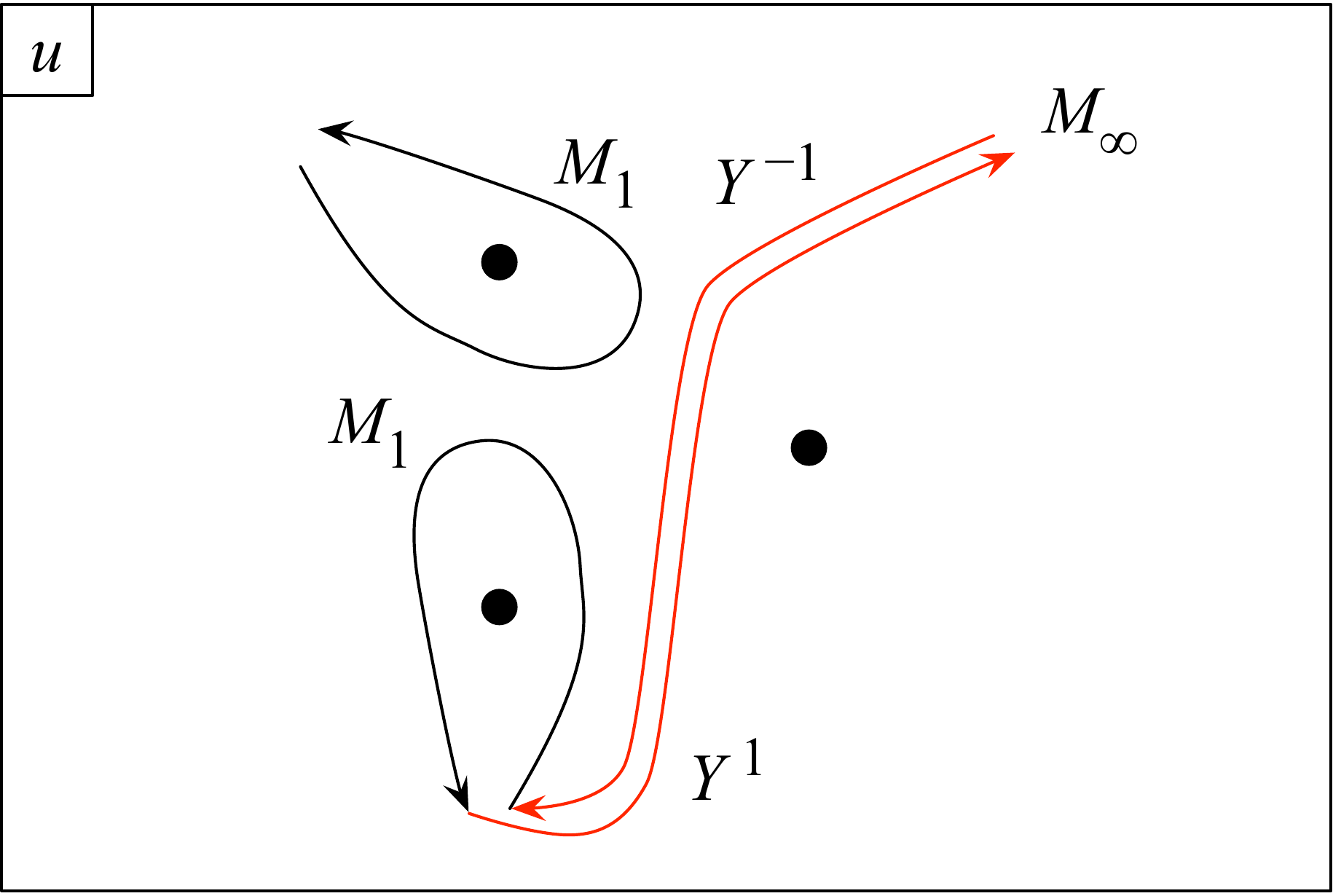}
\]
\caption{Relation of $M_1$, $M_2$\label{fig:conj_nf_1}}
\end{figure}

Let us denote the monodromies around each of the three singularities by $M_{1,2,3}$, see Fig.~\ref{fig:monodromy_nf_1}.
Clearly, we should have \begin{equation}
M_\infty = M_3 M_2 M_1
\end{equation} where $M_\infty$ was given in \eqref{Minf_nf_1}. As the three singularities are related by discrete R-symmetry, they should be conjugate. For example, as shown in Fig.~\ref{fig:conj_nf_1}, we expect $M_2=YM_1Y^{-1}$.
A solution is given by \begin{equation}
M_2 =T^{-1} M_1 T^{1},\quad
M_3=T^{-2}M_1 T^{2},
\end{equation} together with \begin{equation}
M_1=STS^{-1}=\begin{pmatrix}
1 & 0 \\
-1 & 1
\end{pmatrix}.
\end{equation}
As $M_1$ found here is the same as $M_+$ found in the pure case \eqref{mpm}, the local physics close to the singularity is also the same, i.e.~it is described by an $\cN{=}2$ $\U(1)$ gauge theory coupled to one charged hypermultiplet.
The same can be said for $M_2$ and $M_3$. 

For the pure case, we saw that the light charged hypermultiplet in this low energy $\U(1)$ description was a monopole in the original description. Is the same true in this case? 
It is easier to give a definitive answer when $|\mu|$ is very big.
Then, the two singularities in the strong coupled region have the same physics as that of the pure $\SU(2)$ theory, and thus we should have light monopoles and dyons there. 
At the third singularity $u\simeq \mu^2$, one component of the doublet hypermultiplet $(Q,\tilde Q)$ becomes very light. 
For all three singularities, the low-energy description is that of a $\U(1)$ gauge theory coupled to one charged hypermultiplet. 

By gradually decreasing $\mu$ to be zero, these three singularities become the three singularities related by the discrete R symmetry.  At this stage, it is not possible to say which of the three was originally the one whose light particle came from the doublet hypermultiplet and which two of the three were the ones with monopoles and dyons. 
This loss of the distinction between the  hypermultiplets which were elementary fields
and the hypermultiplets which came from solitons such as monopoles or dyons is somewhat  surprising to an eye trained in the classical field theory.  We will see this more explicitly below, in Fig.~\ref{fig:motion}.

\subsection{The curve}\label{sec:nf_1_curve}
Let us now construct the holomorphic functions $a(u)$, $a_D(u)$ satisfying the monodromies determined above.
It is again done by using the \SeibergWitten\ curve, which is given in this case by \begin{equation}
\Sigma:\quad
  \frac{2\Lambda(x-\mu)}z +\Lambda^2 z = x^2 -u\label{curve_nf_1}
\end{equation} with auxiliary complex variables $z$ and $x$, together with the Seiberg-Witten differential \begin{equation}
\lambda=x\frac{dz}z.
\end{equation}  We dropped the subscript $1$ from $\Lambda$ to lighten the notation. 

\begin{figure}[h]
\[
\inc{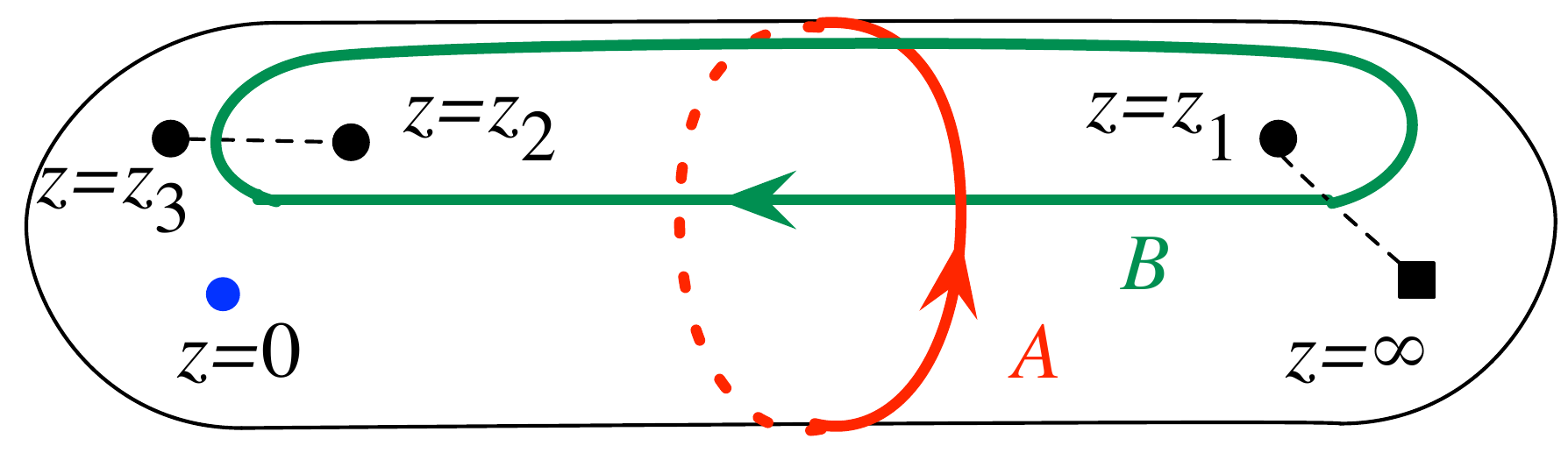}
\]
\caption{The \Gaiotto\ curve of $\SU(2)$ $N_f=1$ theory.\label{fig:gaiotto_nf_1}}
\end{figure}
\begin{figure}[h]
\[
\inc{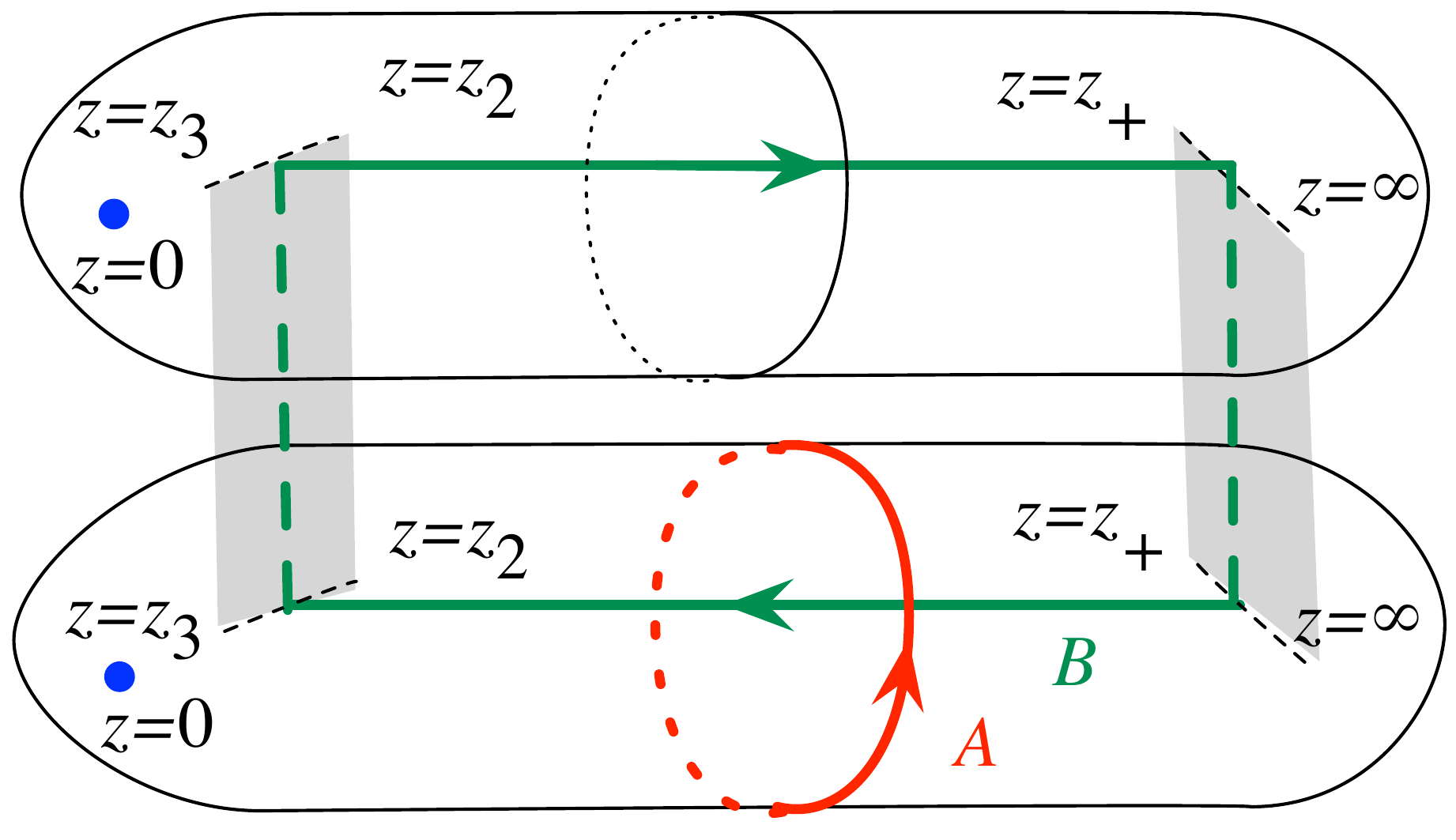}
\]
\caption{The sheets of the \SeibergWitten\ curve of $\SU(2)$ $N_f=1$ theory.\label{fig:swcurve_nf_1}}
\end{figure}

Again, we add a point $z=\infty$ and regard $z$ as a complex coordinate on the sphere $C$. This is the \Gaiotto\ curve.
The variable $x$ is now a function on it, see Fig.~\ref{fig:gaiotto_nf_1}.
Note that $z=0$ is no longer a branch point; indeed, the local behavior of $x$ there is now \begin{align}
x_+ &\sim \frac{2\Lambda}z - \mu + O(z),\\
x_- & \sim \phantom{\frac{2\Lambda}z } + \mu +O(z).
\end{align} Note also that $\lambda$ has a residue $\mp \mu$ at $z=0$. 
The curve $\Sigma$ is a two-sheeted cover of $C$ shown in Fig.~\ref{fig:swcurve_nf_1}.
We define cycles $A$ and $B$ as shown, and then the functions $a(u)$ and $a_D(u)$ are given by \begin{equation}
a=\frac{1}{2\pi i} \oint_A \lambda,\qquad
a_D=\frac{1}{2\pi i} \oint_B \lambda.
\end{equation}

\begin{figure}[h]
\[
\inc{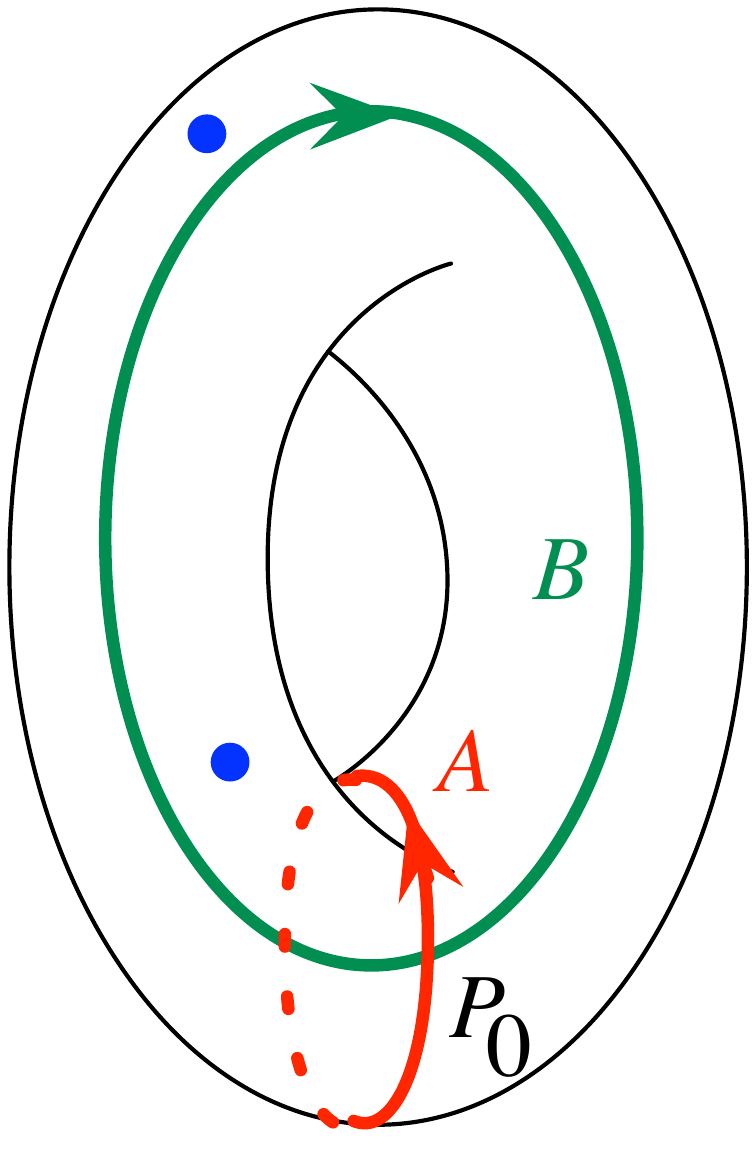}\qquad \Longrightarrow \qquad
\inc{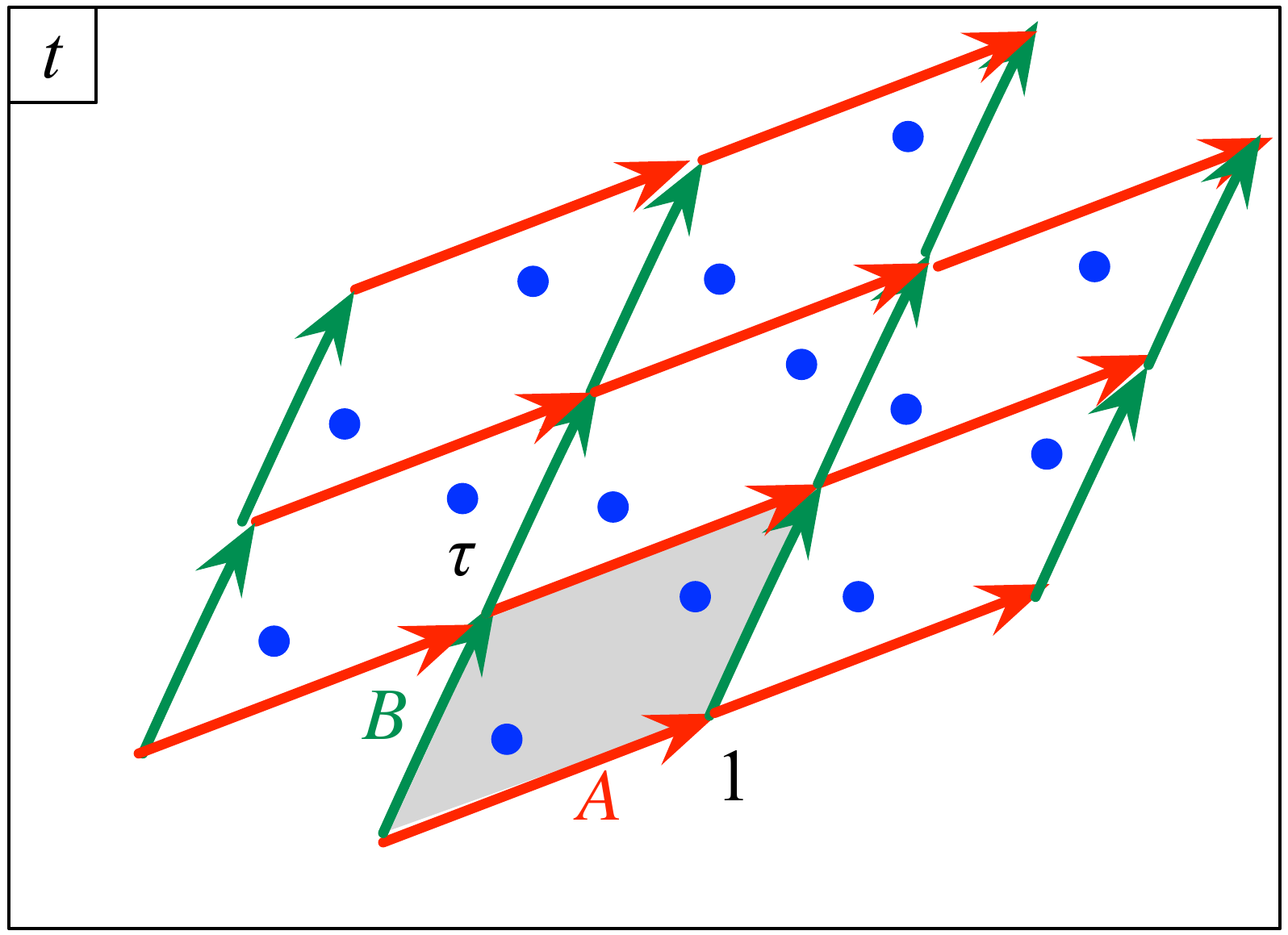}
\]
\caption{The smoothed-out torus of the curve of the  $\SU(2)$ $N_f=1$ theory.\label{fig:torus_nf_1}}
\end{figure}
The proof $\Im \tau(a)>0$ goes exactly as in the pure case. The curve $\Sigma$ can be mapped to a parallelogram within a complex $t$ plane by $\int_{P_0}^P \partial\lambda/\partial u=\int_{P_0}^P dz/(2(xz-\Lambda))$, see Fig.~\ref{fig:torus_nf_1}.
The poles with residues $\pm \mu$ of  $\lambda$ are denoted explicitly in the figure. 
When a closed cycle $L$ on the torus winds the $A$ cycles $n$ times, $B$ cycles $m$ times, and the poles $f$ times, 
the integral of $\lambda$ is then \begin{equation}
\frac{1}{2\pi i}\oint_L \lambda=na+ma_D + f\mu,
\end{equation} just as in the BPS mass formula \eqref{bps_nf_1}.

Let us check that the curve correctly reproduces the running of the coupling in the weakly-coupled region. 
For simplicity, set $\mu=0$, and assume $|u|\gg |\Lambda|$. 
We put the $A$ cycle at  $|z|=1$.
 We easily find \begin{equation}
\frac{1}{2\pi i}\oint_A x\frac{dz}z\sim \sqrt{u}
\end{equation} as before. 
As for the $B$ integral, two branch points are around $z\sim \Lambda/\sqrt{u}$
and one branch point is around $z\sim u/\Lambda^2$. The dominant contribution to the integral is then \begin{equation}
\frac{1}{2\pi i}\oint_B x\frac{dz}z\sim 
\frac{2}{2\pi i}  \int^{\Lambda/\sqrt{u}}_{u/\Lambda^2} a\frac{dz}z
=-\frac{6}{2\pi i} a\log \frac{a}{\Lambda}.
\end{equation} Then we find \begin{equation}
\tau(a)=\frac{\partial a_D}{\partial a}=-\frac{6}{2\pi i}\log \frac{a}{\Lambda},
\end{equation} reproducing the running \eqref{running_nf_1}. 

Let us next check that the curve correctly reproduces the singularity structure on the $u$-plane. 
The branch points of the function $x(z)$ can be determined by studying when the equation of $\Sigma$, given in \eqref{curve_nf_1}, has double roots. 
The equation for the branch points  is given by \begin{equation}
z^3+\frac{uz^2}{\Lambda^2}-\frac{2\mu z}{\Lambda} + 1=0.
\end{equation} 
The singularity in the $u$-plane is caused by two of the branch points of $x(z)$ colliding in the \Gaiotto\ curve $C$ with the coordinate  $z$.
This condition can be found by taking the discriminant of the equation of $z$ above, giving \begin{equation}
u^3-\mu^2u^2+9\Lambda^3\mu u+\frac{27}{4}\Lambda^6-8\Lambda^3\mu^3=0.\label{u-eq}
\end{equation}

When $\mu=0$, this equation simplifies to $u^3+\frac{27}4 \Lambda^6=0$, giving the solutions \begin{equation}
u=c\Lambda^2,\ e^{2\pi i/3} c\Lambda^2,\ e^{4\pi i/3} c\Lambda^2
\end{equation} for a constant $c$, reproducing Fig.~\ref{fig:uplane_nf_1_mless}.

When $|\mu|\gg |\Lambda|$, the equation \eqref{u-eq} can be solved by making two separate approximations.
Assuming $u$ is rather big, we can truncate the equation to just $u^3-\mu^2u^2\sim 0$, finding a singularity at \begin{equation}
u\sim \mu^2.
\end{equation} Next, assuming $u$ is rather small, we find $-\mu^2 u^2 -8\Lambda^3\mu^3\sim 0$ giving \begin{equation}
u\sim \pm\sqrt{-8\Lambda^3\mu}. \label{asd}
\end{equation} Together, they reproduce Fig.~\ref{fig:uplane_nf_1_m_large}.
From this, we find that the effective pure $\SU(2)$ theory in the region $|u|\ll |\mu|$ has the dynamical scale \begin{equation}
\Lambda_0^2 \sim \sqrt{\Lambda^3 \mu}.\label{relation01}
\end{equation} This agrees with what we saw in \eqref{l1vsl0}.

\begin{figure}[h]
\[
\begin{array}{c@{\qquad}c}
\inc{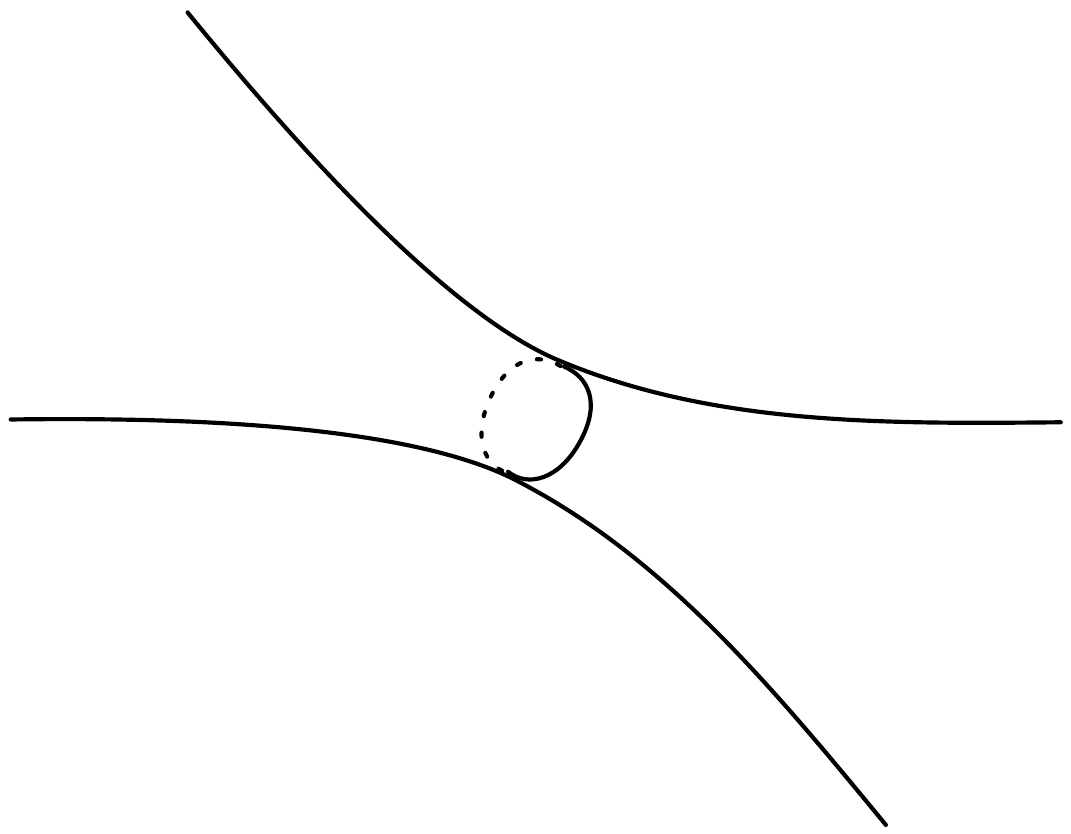} & 
\inc{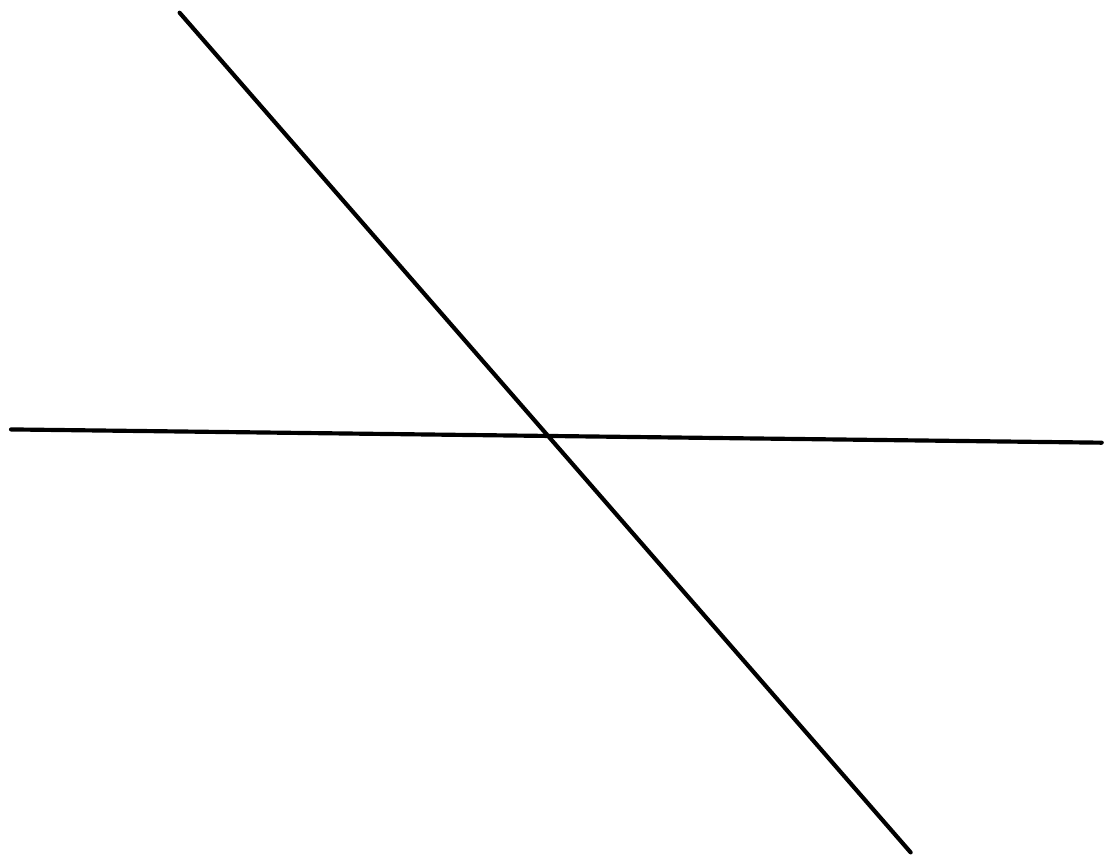}\\
u\neq \mu^2 & u=\mu^2
\end{array}
\]
\caption{The schematic change in the \SeibergWitten\ curve when $u\to \mu^2$.\label{fig:connection}}
\end{figure}

It is instructive to study another way to derive the singularity at $u\sim \mu^2$ from the curve. 
We would like to take the approximation $|\Lambda| \ll |\mu|$. To facilitate to take the limit, we introduce $\tilde z=z/\Lambda$ in \eqref{curve_nf_1} and find \begin{equation}
 \frac{2(x-\mu)}{\tilde z} +\Lambda^3 \tilde z = x^2 -u.
\end{equation} Now the limit is easy to take: we just find \begin{equation}
 \frac{2(x-\mu)}{\tilde z} = x^2 -u.
\end{equation} Then it is clear that when $u=\mu^2$, the equation can be factorized to \begin{equation}
(x-\mu)(x+\mu - \frac{2}{\tilde z})=0,
\end{equation} therefore it represents two sheets intersecting at a point.
When $u\neq \mu^2$, two sheets are connected smoothly. The change is schematically shown in Fig.~\ref{fig:connection}. 
We learned that the singularity at $u\sim \mu^2$ arises essentially from the structure $2\Lambda(x-\mu)/z$ in the curve. 

\subsection{Some notable features}

\begin{figure}[h]
\[
\incc{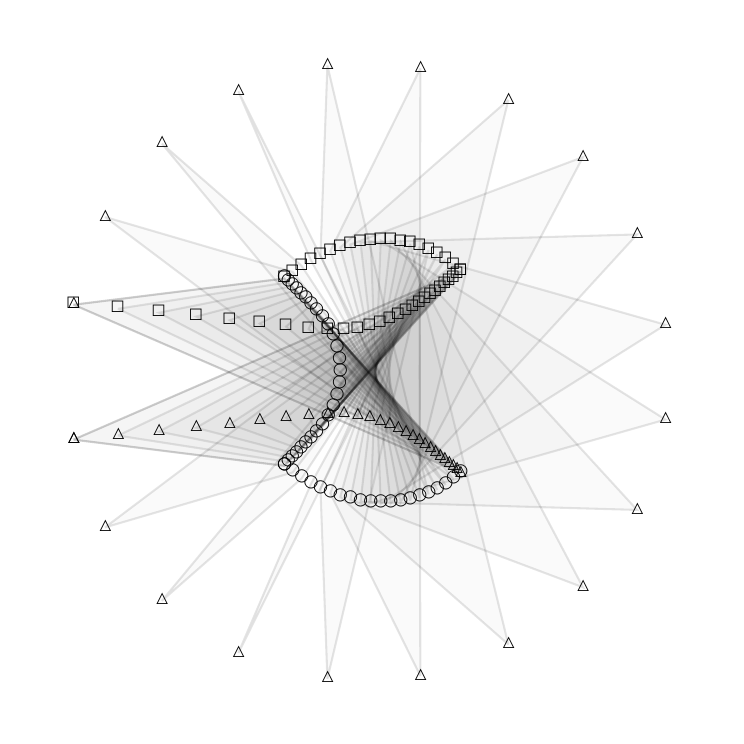}\qquad
\inc{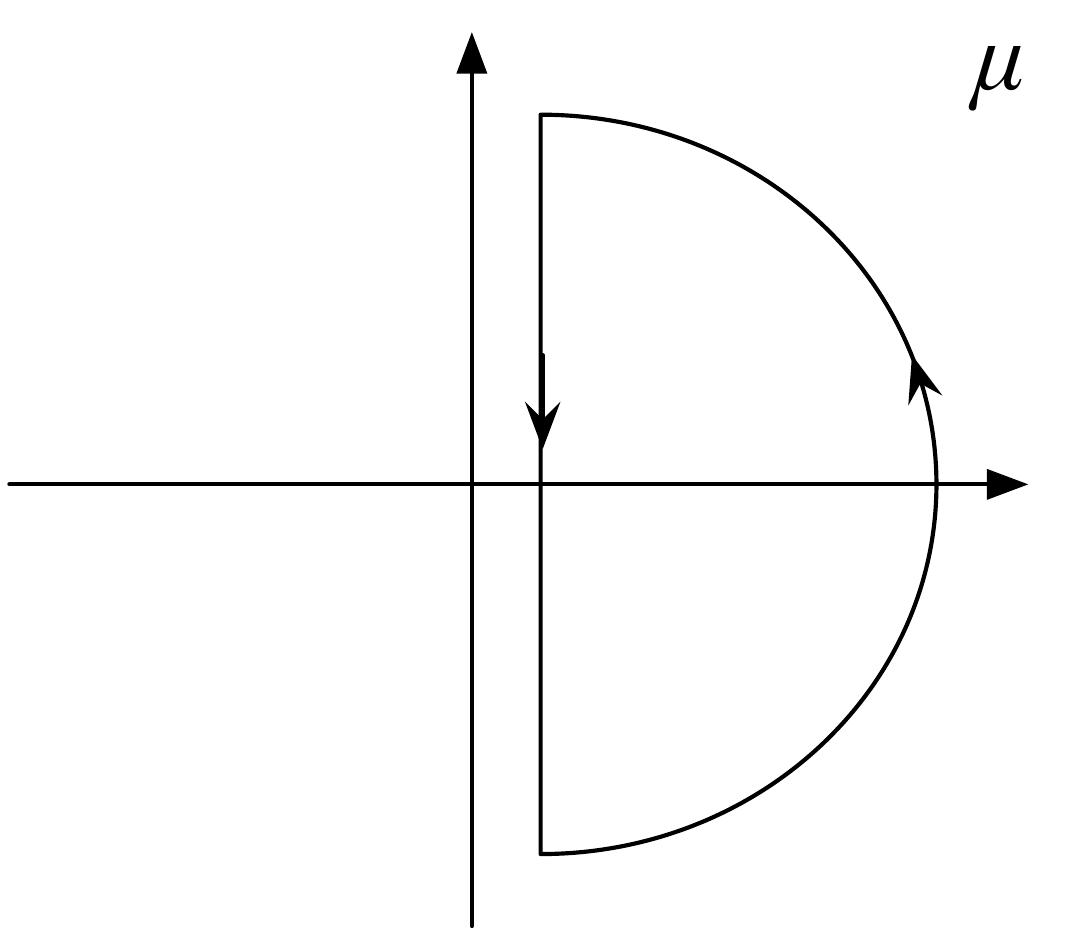}
\]
\caption{Motion of the singularities on the $u$-plane.\label{fig:motion}}
\end{figure}

Let us see how  three singularities on the $u$-plane move as we change $\mu$, by solving \eqref{u-eq}.  An example is shown in Fig.~\ref{fig:motion}.
On the right, the path in the $\mu$ space is given. 
On the left, the three singularities for a given $\mu$ is shown with three dots marked by $\triangle$, $\square$ and \raisebox{.1ex}{$\bigcirc$}, connected by a light-gray triangle. 
As $\mu$ moves along a semicircle with constant, large $|\mu|$, the quark point $u\sim \mu^2$ rotates the $u$-plane once.
At the same time, the monopole point and the dyon point of the effective pure $\SU(2)$ theory rotates by $90$ degrees, as we see from \eqref{asd}. 
Now we make $|\mu|$ decrease first; all three singularities come close to the origin of the $u$-plane. 
Finally, we make $|\mu|$ come back to the same semicircle again.
As can be seen in the figure, this process exchanges the quark point and the monopole point. 
We learned that, using  the strongly-coupled region, we can continuously change a quark into a monopole.

\begin{figure}[h]
\[
\inc{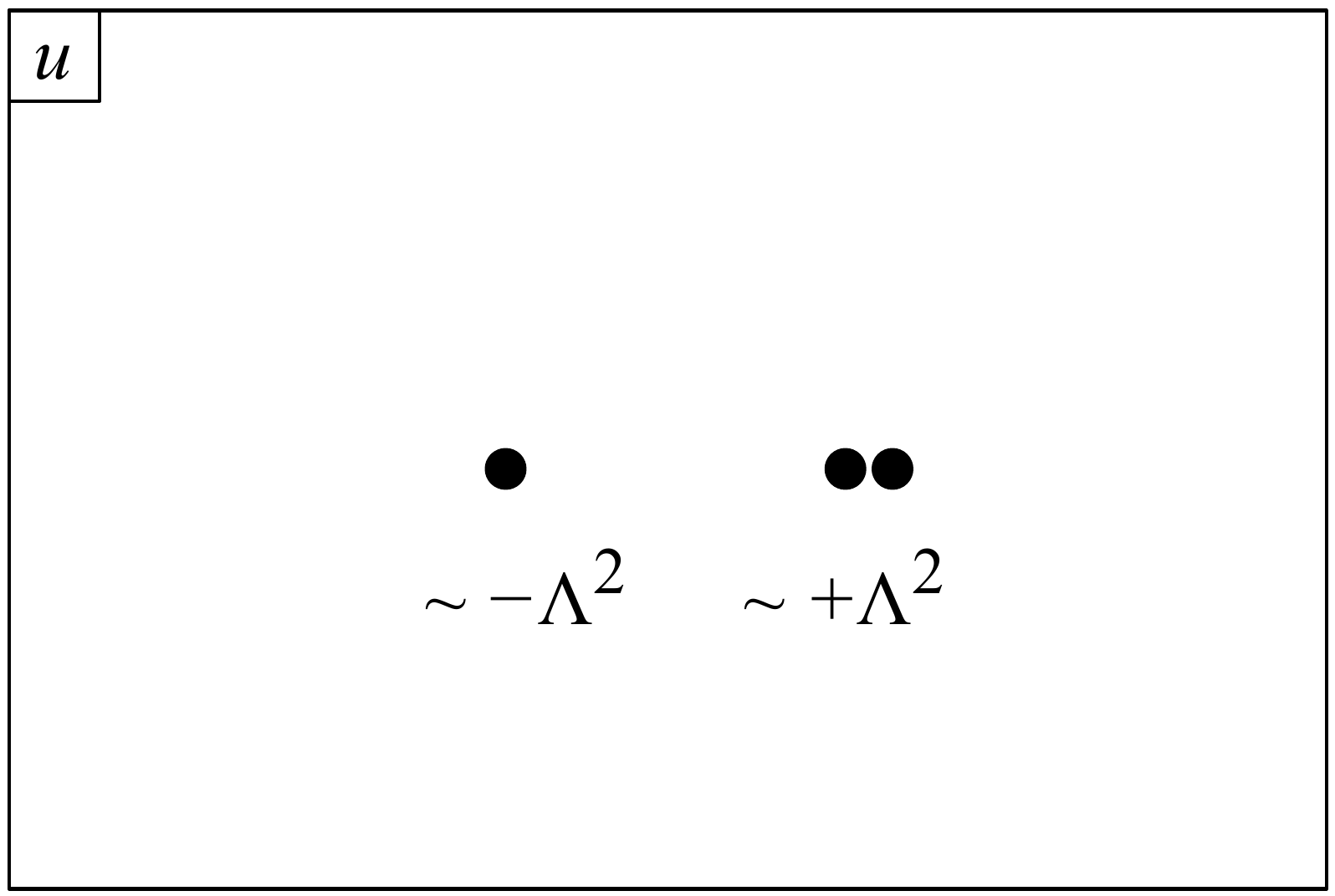}
\qquad
\inc{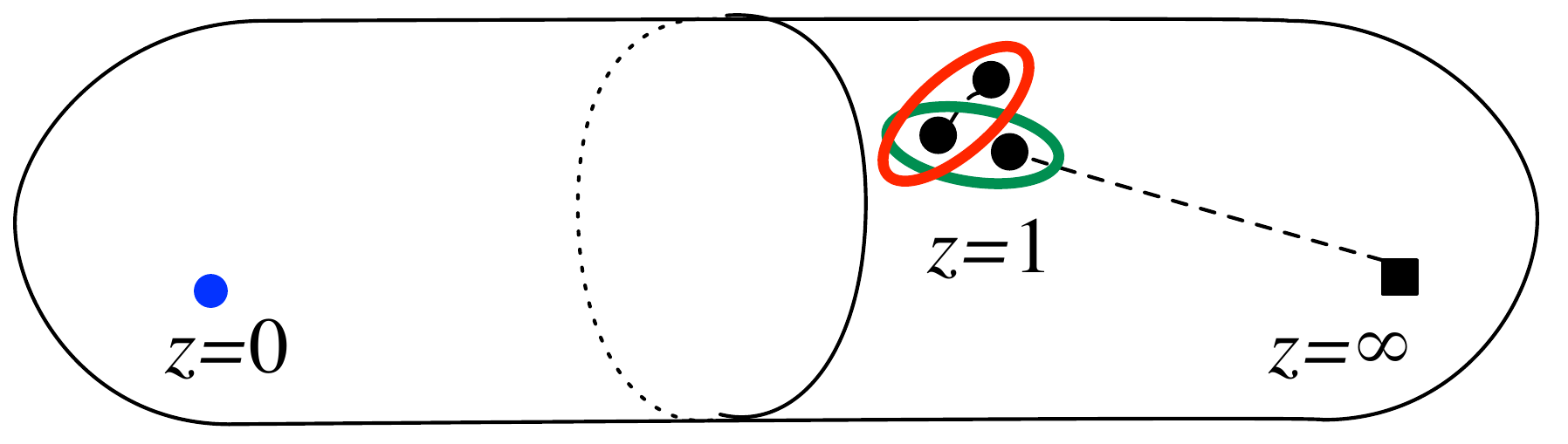}
\]
\caption{Two out of three singularities can collide on the $u$-plane.
Then three branch points collide on the \Gaiotto\ curve.\label{fig:uplane_nf1_ad}}
\end{figure}
Finally, let us study the discriminant of the equation \eqref{u-eq} itself, which is given by \begin{equation}
(\mu^3+\frac{27}{8}\Lambda^3)^3=0.
\end{equation} Take $\mu=-3\Lambda/2$ as an explicit choice. 
Then there is one singularity in the $u$-plane at $u=-15\Lambda^2/4$, and two singularities collide at $u=3\Lambda^2$.
In the curve, we find that the branch points of $x(z)$ consist of one at $z=\infty$ and three colliding at $z=-1$. See Fig.~\ref{fig:uplane_nf1_ad}.
From the curve, we immediately see that $a=a_D=0$, since the integration cycles shrink.  Using the BPS mass formula, we see that both electrically charged particles and magnetically charged particles are simultaneously becoming very light. 
This is a rather unusual situation for an eye trained in the classical field theory. 
Semiclassically, the magnetically charged particles come from solitons, which are always parametrically heavier than the electrically charged particles which are quanta of elementary fields in the theory. 
We will study this system in more details Sec.~\ref{sec:AD}.

%\eject

\section{Curves and 6d $\cN{=}(2,0)$ theory}\label{sec:6d}
We have seen that the low energy dynamics of the $\SU(2)$ pure gauge theory and  the $\SU(2)$ gauge theory with one flavor can both be expressed in terms of the complex curves \eqref{curve_pure}, \eqref{curve_nf_1}. 
The aim of this section is to explain that these two-dimensional spaces can be given a physical interpretation. 

The ideas which will be presented in this section were originally obtained by exploiting various deep properties of string theory and M-theory, namely Calabi-Yau compactifications, brane constructions, and string dualities. 
The approach using Calabi-Yau compactifications goes back to \cite{Kachru:1995wm,Kachru:1995fv,Klemm:1996bj} and the brane construction approach was introduced in \cite{Witten:1997sc}.
 Learning these constructions definitely helps in understanding $\cN{=}2$ supersymmetric dynamics, and vice versa.  This lecture note is not, however, the place where you can learn them. 
 
The presentation here is analogical rather than being logical, and the author intentionally tried to phrase it in such a way that the knowledge of string theory and M-theory  required to read it is kept to the minimum. 
Anyone interested in more details should refer to the original articles, or the reviews such as ~\cite{Lerche:1996xu,Giveon:1998sr} and Sec.~3 of \cite{Gaiotto:2009hg}.

\subsection{Strings with variable tension}

Recall the BPS mass formula of the pure theory of a particle with electric charge $n$ and magnetic charge $m$, \begin{equation}
M\ge |na+ma_D|=| \int_L \lambda|\label{bpsagain}
\end{equation} where $L$ is a cycle on the curve which goes around $n$ times along the $A$ direction and $m$ times along the $B$ direction. The basic idea we employ is to take this equation seriously: we regard the four-dimensional particle as arising from a string wrapped on the cycle $L$. Then $\lambda$ is something like the tension of the string. In this section we introduce a factor of $2\pi i$ in the definition of $\lambda$, to lighten the equations. 

\begin{figure}[h]
\[
\inc{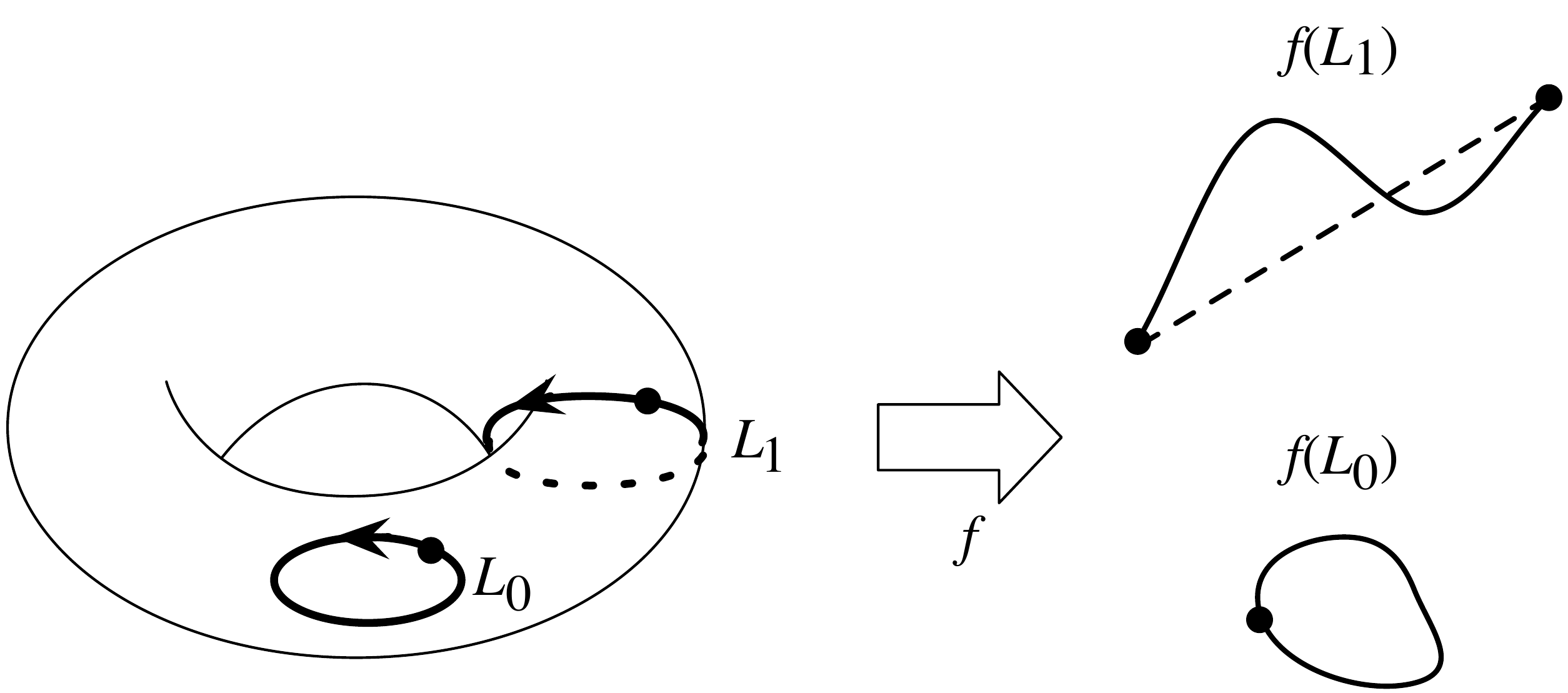}
\]
\caption{Variable-tension strings. Its tension is controlled by $f(s)=\int^s_{P_0} \lambda$.  \label{fig:var_tension_string} }
\end{figure}

To make this idea more concrete, suppose  a six-dimensional theory which has strings as its excitation\footnote{Some of the young readers who just started learning string theory might wonder at this point: aren't relativistic Lorentz-invariant string theories only possible in 26 dimensions if bosonic, and in 10 dimensions if supersymmetric? The catch is that the standard arguments in the textbooks assume that the interaction among strings is perturbative. %\emph{That said, can't we just quantize free 6d strings?} 
}, and assume this theory is  on a two-dimensional space $C$ times the four-dimensional Minkowski space $\bR^{1,3}$.  Further assume that the tension of the string depends on these extra-dimensional directions. Namely, let us assume that there is a locally-holomorphic one-form $\lambda$ such that the tension of an infinitesimal segment of a string, parameterized by $s$, is given by \begin{equation}
|\lambda| := |\frac{\lambda}{ds}| ds,
\end{equation} see the left hand side of Fig.~\ref{fig:var_tension_string}. 
There, the two-dimensional space is taken to be a torus for definiteness.

A string looks like a particle from the point of view of the uncompactified four dimensions, and its mass is given by the integral of its variable tension: \begin{equation}
M=\int_L |\lambda| \label{m-l}
\end{equation} The right hand side can be bounded below using an integral version of  the triangle inequality: \begin{equation}
\int_L |\lambda| \ge |\int_L \lambda|.\label{tri-ineq}
\end{equation} The inequality can be visualized by considering the curve in the complex plane defined by \begin{equation}
f(s)=\int^s_{P_0} \lambda
\end{equation} parameterized by $s$, where $P_0$ is a fixed point on the cycle $L$. Then the left hand side of \eqref{tri-ineq} is the length of the parameterized curve $f(s)$, while the right hand side is the distance between the end-points of $f(s)$, see the right hand side of Fig.~\ref{fig:var_tension_string}.
Then clearly the former is longer than the latter, and the equality is attained only when the line $f(s)$ itself is a straight line. Or equivalently \begin{equation}
\Arg \frac{\lambda}{ds} = \text{constant}.\label{curveBPSeq}
\end{equation}

When the cycle $L$ is topologically trivial, the image of the function $f(s)$ is itself a loop, and the right hand side of \eqref{tri-ineq} is zero.  When the cycle $L$ is nontrivial, the image of the function $f(s)$ can be an open segment.  As $\lambda$ is holomorphic, the difference between the two ends of the segment only depends on the topology of the cycle $L$. Say $L$ wraps the $A$-cycle $n$ times and the $B$-cycle $m$ times.
Combining \eqref{m-l} and \eqref{tri-ineq}, we find \begin{equation}
M\ge |na+ma_D|
\end{equation} where $a$, $a_D$ are defined by the relations \begin{equation}
a=\int_A\lambda,\qquad
a_D=\int_B\lambda.
\end{equation} 
This reproduces the BPS mass formula \eqref{bpsagain}. 
We learned furthermore that the inequality is saturated only when \eqref{curveBPSeq} is satisfied.  Therefore we regard \eqref{curveBPSeq} as the BPS equation for the string excitation. 

\subsection{Strings with variable tension from membranes}
\subsubsection{General idea}
\begin{figure}[h]
\[
\inc{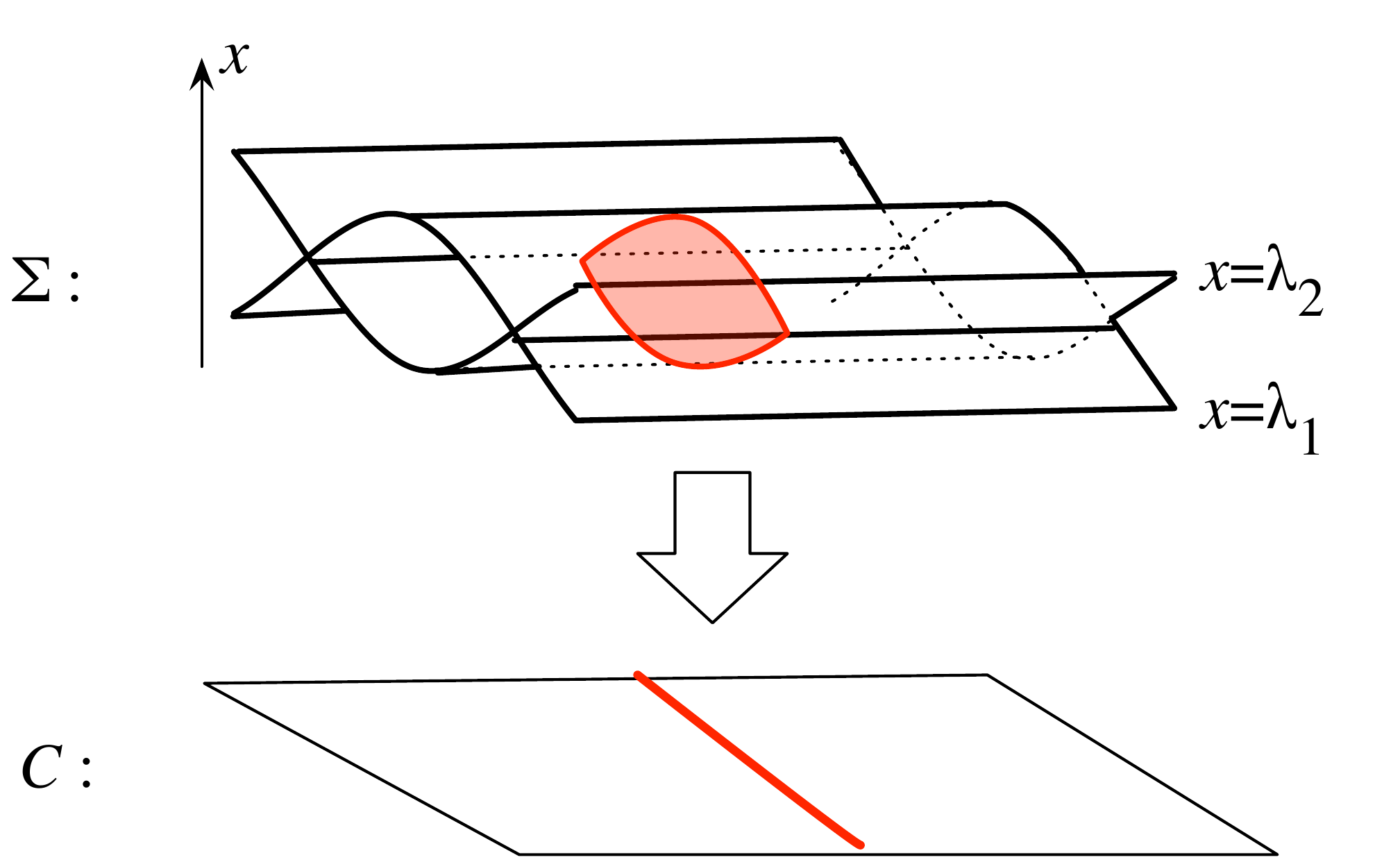} 
\]
\caption{How the variable-tension string arises from higher dimensions.\label{fig:membrane-to-string}}
\end{figure}
One might say strings with variable tension is slightly weird.  One way to realize this variation of the tension in a natural manner is to consider that the extra-dimensional space $C$ which have two dimensions is further embedded in a four-dimensional ambient space $X$, and there are two sheets of $\Sigma$  covering $C$ separated in the additional directions of $X$.
We then furthermore suppose that there is a membrane extending along two spatial directions plus one temporal direction, which can have ends on the sheets of $\Sigma$. 
The situation is depicted in Fig.~\ref{fig:membrane-to-string}. 
Let $z$ be the coordinate of $C$, and $X$ has complex coordinates $(z,x)$. Then two sheets of $\Sigma$ define two functions $x_1(z)$ and $x_2(z)$. Then, a membrane with constant tension $|dx|\wedge |d\log z|$ , suspended between two sheets, can be regarded as a string with variable string whose tension at a given value of $z$ is given by \begin{equation}
\text{(tension at $z$)}\ge \left|\int_{x_2(z)}^{x_1(z)} dx\wedge d\log z\right| =
\left| x_1\frac{dz}z - x_2\frac{dz}z\right|.\label{integ}
\end{equation} Denoting $\lambda_i(z)=x_i dz/z$, we find that
 \begin{equation}
\text{(tension at $z$)}\ge |\lambda(z)| \qquad \text{where}\ \lambda(z)=\lambda_1(z)-\lambda_2(z).
\end{equation}

In M-theory, there are indeed higher-dimensional objects with these properties. 
We consider an eleven dimensional spacetime of the form \begin{equation}
\mathbb{R}^{3,1}\times X\times \mathbb{R}^3. 
\end{equation} M-theory has six-dimensional objects called M5-branes.
We put one M5-brane on \begin{equation}
\mathbb{R}^{3,1}\times \Sigma \times \{0\}
\end{equation} where $\Sigma\subset X$ is the curve, and $0$ is the origin of the additional $\bR^3$. This gives a four-dimensional theory. 
M-theory also has three-dimensional objects called M2-branes, which can have ends on M5-branes. We can take one M2-brane on \begin{equation}
\mathbb{R}^{0,1} \times \text{disc}\times \{0\}
\end{equation} where $\mathbb{R}^{0,1}\subset \mathbb{R}^{3,1}$ is the worldline of a particle in the four-dimensional spacetime, and the $\text{disc}\subset X$ has its boundary on $\Sigma$ as depicted in Fig.~\ref{fig:membrane-to-string}. For more details on this point, the reader should start from the original paper \cite{Mikhailov:1997jv}.

It is also useful to regard the intermediate situation when we regard the system as a six-dimensional one on $\mathbb{R}^{3,1}\times C$. This six-dimensional theory is known as the 6d $\cN{=}(2,0)$ theory of type $\SU(2)$. 

\subsubsection{Example: pure $\SU(2)$ theory}
\begin{figure}[h]
\[
\inc{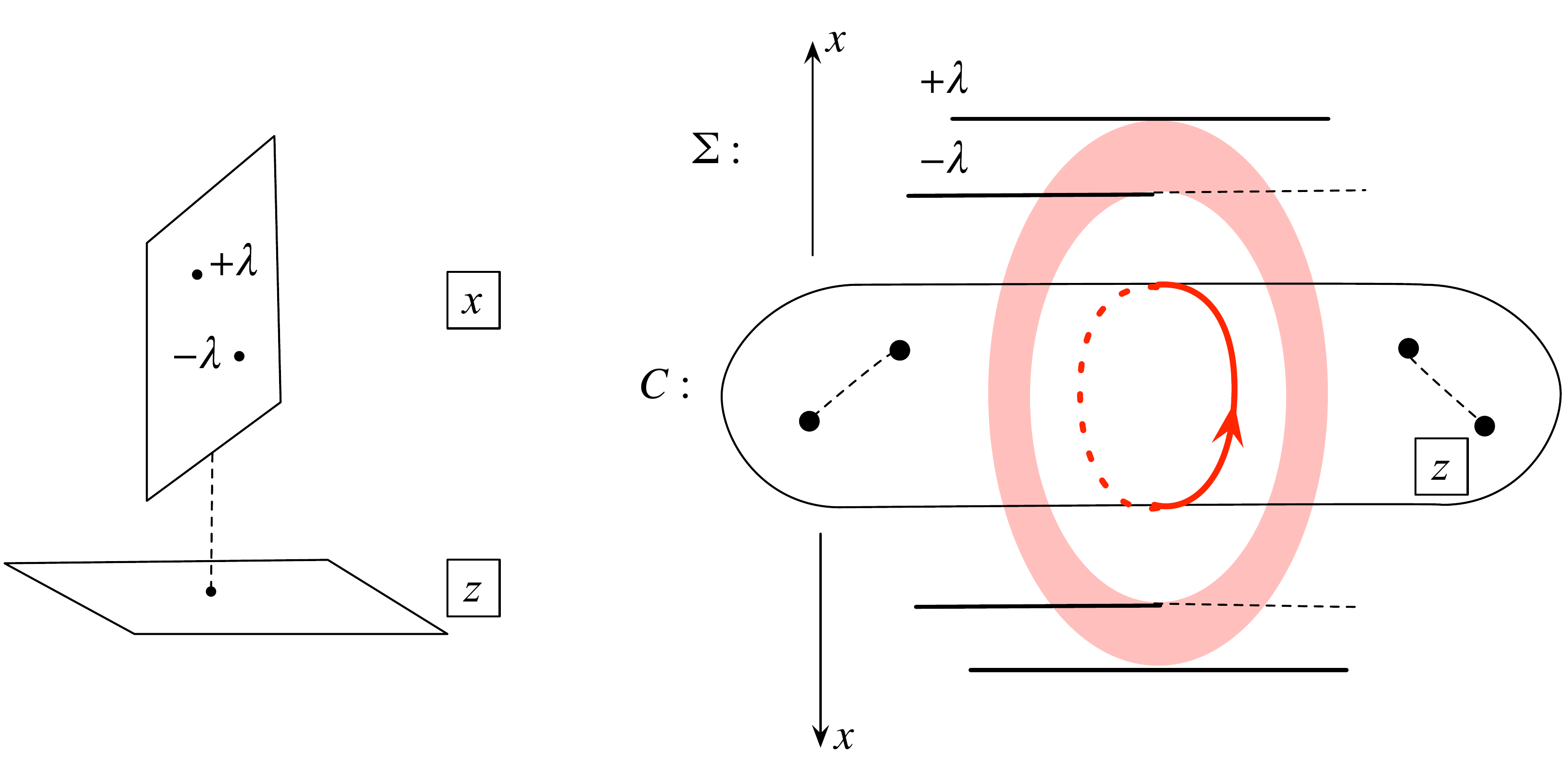}
\]
\caption{W-boson as a string and as a suspended membrane\label{fig:wboson_from_membrane}}
\end{figure}

Let us apply this higher-dimensional idea to the curve \eqref{curve_pure} of the pure $\SU(2)$ theory concretely. For easy reference we reproduce the curve here: \begin{equation}
\Sigma:\qquad \Lambda^2 z+\frac{\Lambda^2}z = x^2-u. 
\end{equation}
We consider $\Sigma$ to be embedded in a four-dimensional space $X$. 
Given a point $z$ on $C$, we find two  $x$ coordinates by solving the quadratic equation above,
as depicted on the left hand side of Fig.~\ref{fig:wboson_from_membrane}.
Let the solutions be $\pm x(z)$.
As the point $z$ moves on $C$, they form two sheets of the curve $\Sigma$, see the right hand side of Fig.~\ref{fig:wboson_from_membrane}.
 The coordinate $x$ always appears as a way to describe the one-form on $C$ giving the tension, so it is convenient to multiply them always by $dz/z$, and say that two sheets have coordinates $\pm\lambda=x(z)dz/z$. We use  this convention from now on. 

We can now consider a ring-shaped membrane suspended between the two sheets over the $A$ cycle, see Fig.~\ref{fig:wboson_from_membrane}. Note that the tension as a string on $C$ is $2\lambda$, and the mass is given by \begin{equation}
M\ge |2\int_A \lambda|=|2a|.
\end{equation} We can minimize the tension by solving \eqref{curveBPSeq}, which give rise to a configuration with the mass \begin{equation}
M=|2a|.
\end{equation} Note that this has the correct mass to be a W-boson, which has electric charge $n=2$ in our normalization, which is for the triplets of $\SU(2)$.
It is also to be noted that there is no way to have a membrane whose mass is given by \begin{equation}
M'=|a|,
\end{equation} because there is simply no way to suspend the membrane to have just one ends over the $A$-cycle. 
Therefore, this higher-dimensional reasoning has more explanatory power than just regarding the curve $\Sigma$ as an auxiliary object producing the holomorphic functions $a(u)$ and $a_D(u)$ with the correct monodromy properties. This procedure knows that there is no dynamical  particle with electric  charge $n=1$ in this system.

\begin{figure}[h]
\[
\inc{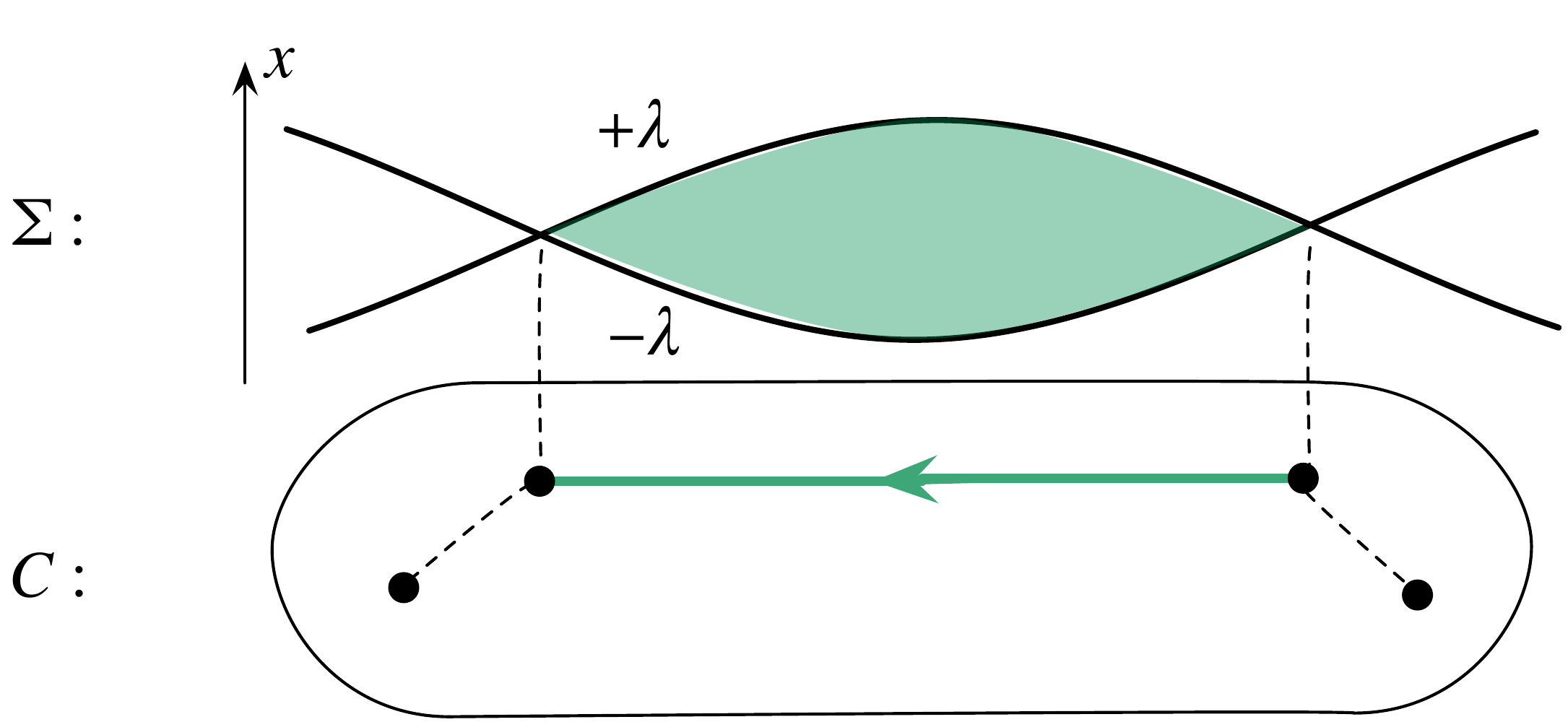}
\]
\caption{Monopole as a string and as a suspended membrane\label{fig:monopole_as_membrane}}
\end{figure}
Next, we can consider a disc-shaped membrane suspended between the sheets of $\Sigma$ so that they have endpoints over the branch points $z_+$, $z_-$ of $C$, see Fig.~\ref{fig:monopole_as_membrane}. By a similar reasoning as above, the mass of this membrane is \begin{equation}
M=2|\int_{z_-}^{z_+}\lambda|=|\int_B\lambda|=|a_D|.
\end{equation} This is a correct mass formula for the monopole, whose  magnetic charge is  $m=1$.  In terms of a variable-tension string on $C$, it is to be noted that this corresponds to an open string, ending at the points where the tension $2\lambda$ becomes zero. 

\begin{figure}[h]
\[
\inc{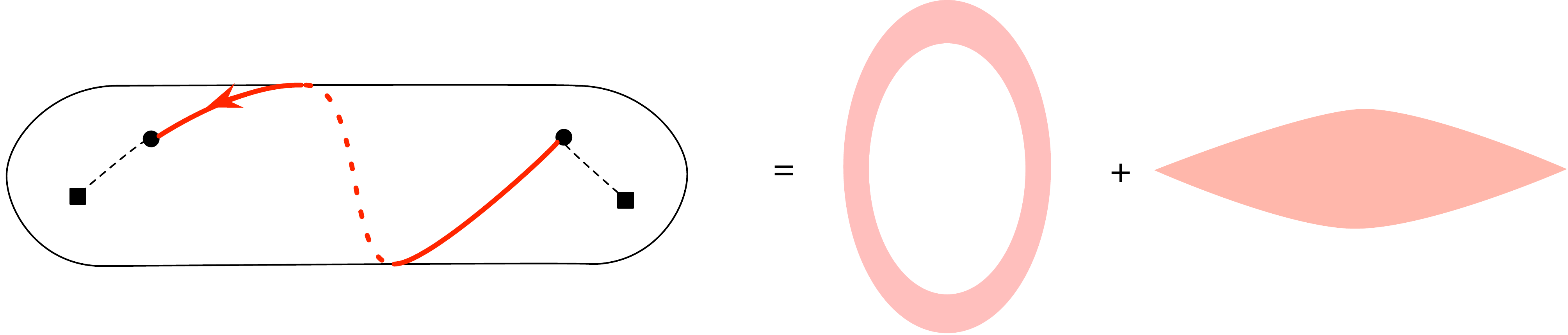}
\]
\caption{Dyon as a string and as a suspended membrane. Note that it automatically has the charge $a_D+2a$, \emph{not} $a_D+a$.\label{fig:dyon_as_membrane}}
\end{figure}
We can also connect the two branch points $z_\pm$ by going around the phase direction of $z$, as shown in Fig.~\ref{fig:dyon_as_membrane}.
As shown there, the membrane is topologically the sum of the two configurations considered so far, and we find that the mass of this configuration  is \begin{equation}
M=|2a+a_D|.
\end{equation} This is the correct mass formula for the dyon, with the electric charge $n$ and the magnetic $m$ given by $(n,m)=(2,1)$.  
By going around $n$ times when we connect the branch points, we see that there are dyons with mass $|2na+a_D|$ for integral $n$. We also see there is no way to connect the branch points to have dyons with mass $|(2n+1)a+a_D|$, which is compatible with the field theory analysis in Sec.~\ref{sec:monopole}.

\subsection{Self-duality of the 6d theory}

\begin{figure}[h]
\[
\inc{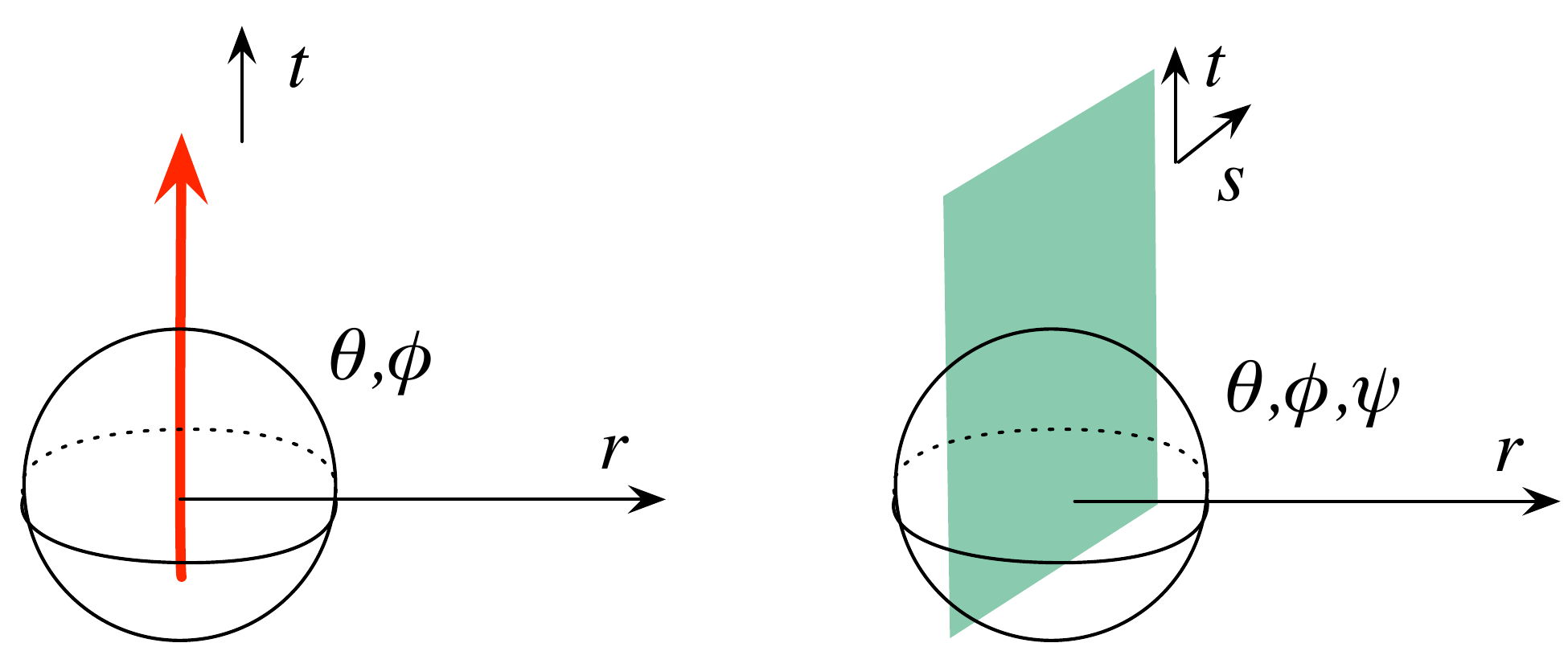}
\qquad
\]
\caption{Charged things in 4d and 6d.\label{fig:charged_objs}}
\end{figure}

Now we found that a single type of objects, the membrane of M-theory or equivalently the string of 6d $\cN{=}(2,0)$ theory, gives rise to both electrically charged objects such as W-bosons and magnetically charged objects such as monopoles, see Fig.~\ref{fig:wboson_from_membrane} and Fig.~\ref{fig:monopole_as_membrane}.
To get a handle of this property, let us first recall basic features of charged particles in four dimensions, see Fig.~\ref{fig:charged_objs}. 

In a first-quantized framework, an electric particle sitting at the origin of the space, extending along the time direction $t$, couples to the electromagnetic potential via \begin{equation}
\int_\text{worldline} A
\end{equation} which creates a nonzero electric field $F_{tr}\neq 0$ where \begin{equation}
F=dA
\end{equation} and $r$ is the radial direction.
The equations of motion are \begin{equation}
dF=d{}\star F=0
\end{equation} outside of the worldline.  
Note that in four dimensional Lorentzian space, we have $\star^2=-1$ acting on two-forms. 
Therefore we cannot impose the condition $\star F=F$.

Let us  consider a theory described by a two-form $B$ in six dimensions, to which a string couples via the term \begin{equation}
\int_\text{worldsheet} B.
\end{equation} Let us say that the string extends along the spatial direction $s$ and the time direction $t$. This configuration creates a nonzero electric field $G_{tsr}$, where $r$ is again the radial direction. 
The equations of motion are \begin{equation}
dG=d{}\star G=0
\end{equation} outside of the worldsheet.  Here $\star$ is the six-dimensional Hodge star operation, given by \begin{equation}
(\star G)_{\mu\nu\rho}=\frac 1{3!}
\epsilon_{\mu\nu\rho\alpha\beta\gamma} G^{\alpha\beta\gamma}.
\end{equation}

In six dimensions with Lorentizan signature, $\star^2=1$ acting on three-forms, so we can demand the equations of motion of the form \begin{equation}
dG=0, \quad G={}\star G.
\end{equation}  Then a worldsheet extending along the directions $t$ and $s$ has both nonzero electric field $G_{tsr}$ and nonzero magnetic field $G_{\theta\phi\psi}$ at the same time.

\begin{figure}[h]
\[
\inc{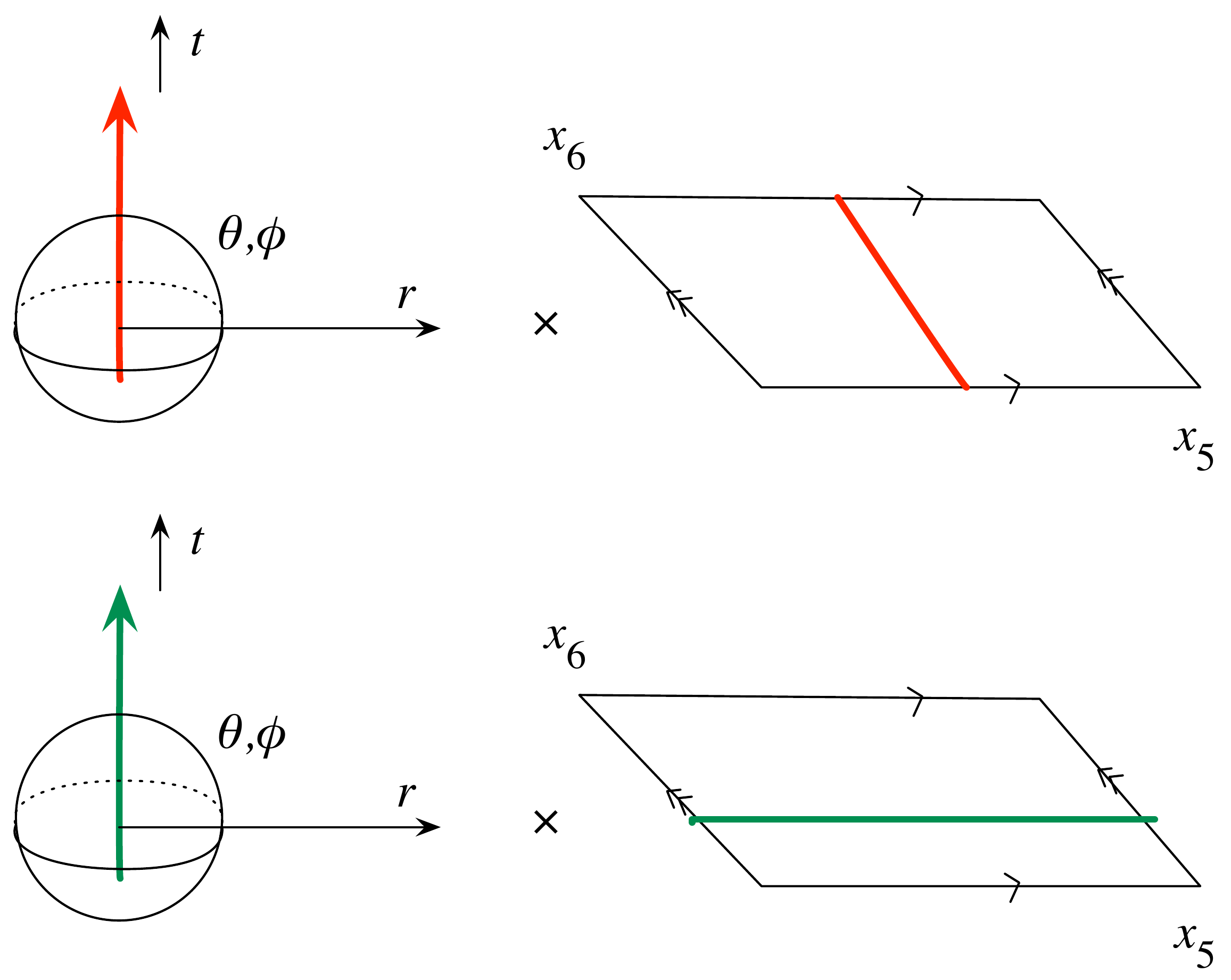}
\]
\caption{Electric and magnetic particles from a single type of objects in 6d \label{fig:em-from-6d}}
\end{figure}
Now, let us put this theory on a two-torus with coordinates $x_{5,6}$, and consider strings wrapped along each of the directions, as shown in Fig.~\ref{fig:em-from-6d}.
Denote the 6d three-form field-strength by $G_{ABC}$, where $A,\ldots$ are indices for six-dimensional spacetime. We can extract four-dimensional two-forms by considering 
\begin{equation}
F_{\mu\nu}:=G_{6\mu\nu},\qquad
F_{D\ \mu\nu}:=G_{5\mu\nu}.
\end{equation}
The 6d self-duality $G={}\star_6G$ translates to the equality \begin{equation}
F_D={}\star_4 F. 
\end{equation} Therefore, the single self-dual two-form field in 6d gives rise to a single $\U(1)$ field strength. 

Now, the string wrapped around $x_6$ has nonzero $G_{6tr}$ and $G_{5\theta\phi}$,
and therefore it has nonzero $F_{tr}$. Therefore this becomes an electric particle in four dimensions. 
Similarly, the string wrapped around $x_5$ has nonzero $G_{5tr}$ and $G_{6\theta\phi}$.
Therefore it has nonzero $F_{\theta\phi}$, meaning that  it is a magnetic particle in four dimensions. 

\begin{figure}[h]
\[
\inc{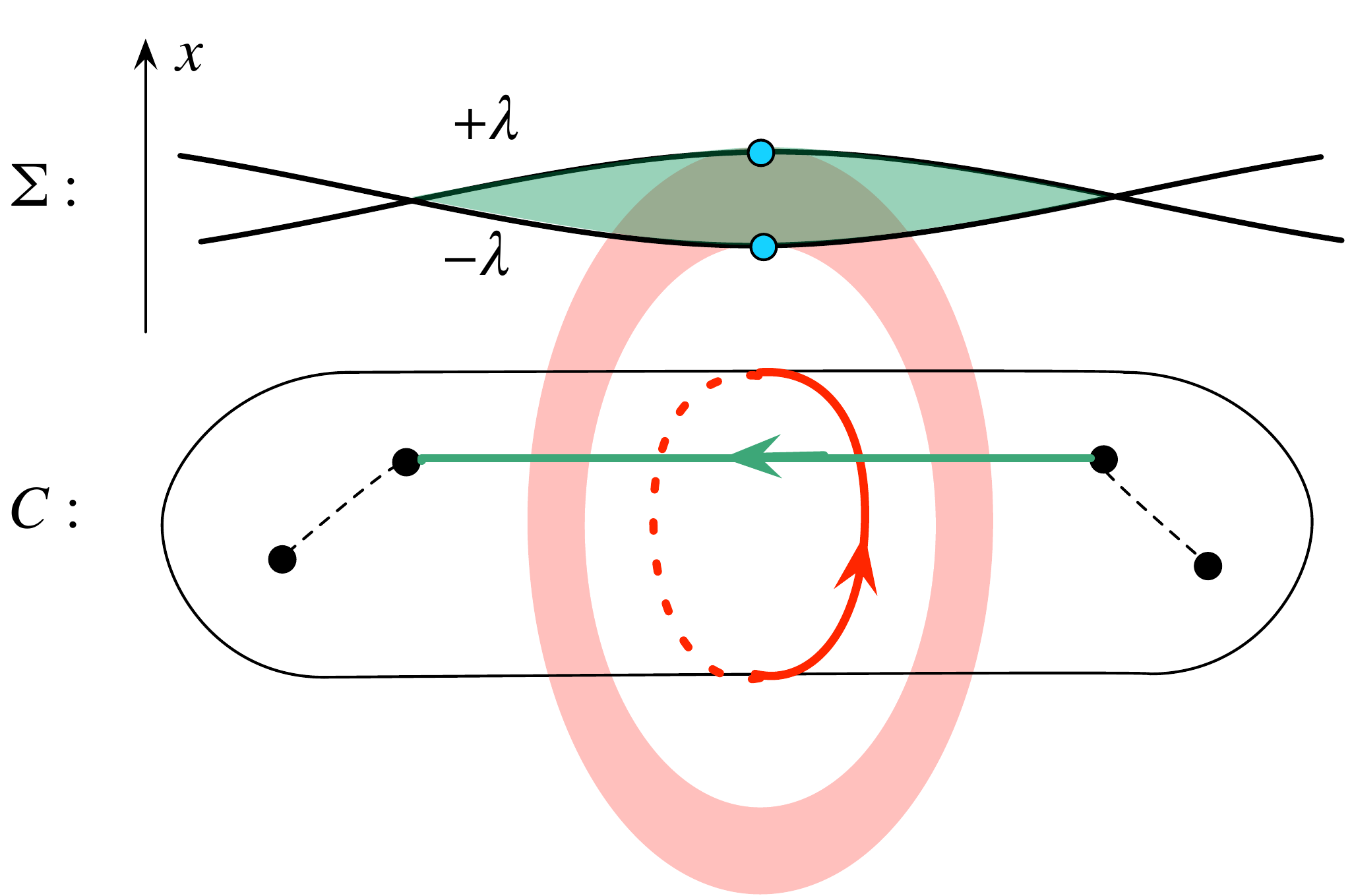}
\]
\caption{The boundaries of the membranes for a W-boson and a monopole intersect at two points. \label{fig:intersection}}
\end{figure}
In the concrete situation of the pure $\SU(2)$ theory, W-bosons and monopoles arise from the membranes as shown in Fig.~\ref{fig:intersection}. We see that the boundaries of the membrane for a W-boson 
and the boundary of the membrane for a monopole intersect at two points. In general, the Dirac pairing as particles in the four-dimensional spacetime  can be found in this way by counting the number of intersections, once  signs given by the orientation are included. 

\subsection{Intermediate 5d Yang-Mills theory and its boundary conditions}
\subsubsection{Five-dimensional maximally-supersymmetric Yang-Mills}
\begin{figure}[h]
\[
\inc{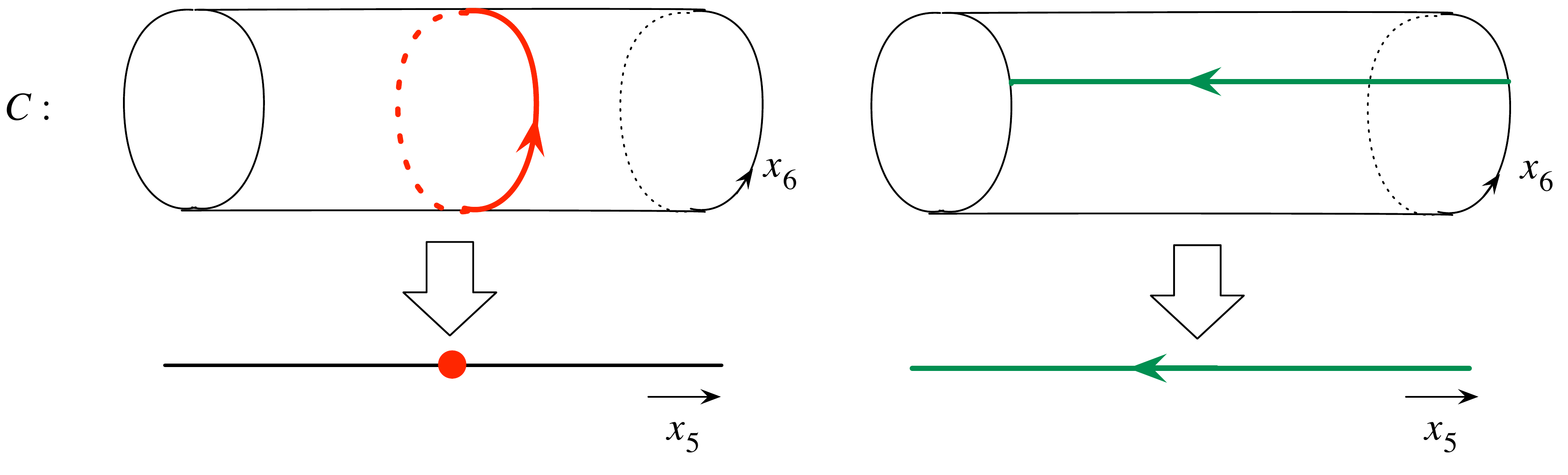}
\]
\caption{5d maximally supersymmetric Yang-Mills from 6d. A W-boson and a monopole-string are depicted there.\label{fig:5d-from-6d}}
\end{figure}
We have so far considered the situations where we put the six-dimensional theory on a two-dimensional space, with coordinates $x_5$ and $x_6$, say.
We can take a limit where the $x_5$ direction is much larger than the $x_6$ direction. Then we can first compactify along the $x_6$ direction and consider an intermediate five-dimensional theory, see Fig.~\ref{fig:5d-from-6d}. 
This is believed to give the maximally supersymmetric 5d Yang-Mills theory with gauge group $\SU(2)$. 

A string wrapped around the $x_6$ direction gives rise to a massive electric particle,
and a string not wrapped around the $x_6$ direction becomes a massive magnetic string. 
This agrees with a basic feature of the 5d Yang-Mills theory where $\SU(2)$ is broken to $\U(1)$:
First, we have massive W-bosons which are electric.
Second, the standard monopole solutions of 4d gauge theory can be regarded as a solution in 5d gauge theory, by declaring that there is no dependence of the fields on the additional fifth direction. 
Then the solutions should be now regarded as representing a magnetic-monopole-string.

\subsubsection{$\cN{=}4$ super Yang-Mills}
\begin{figure}[h]
\[
\inc{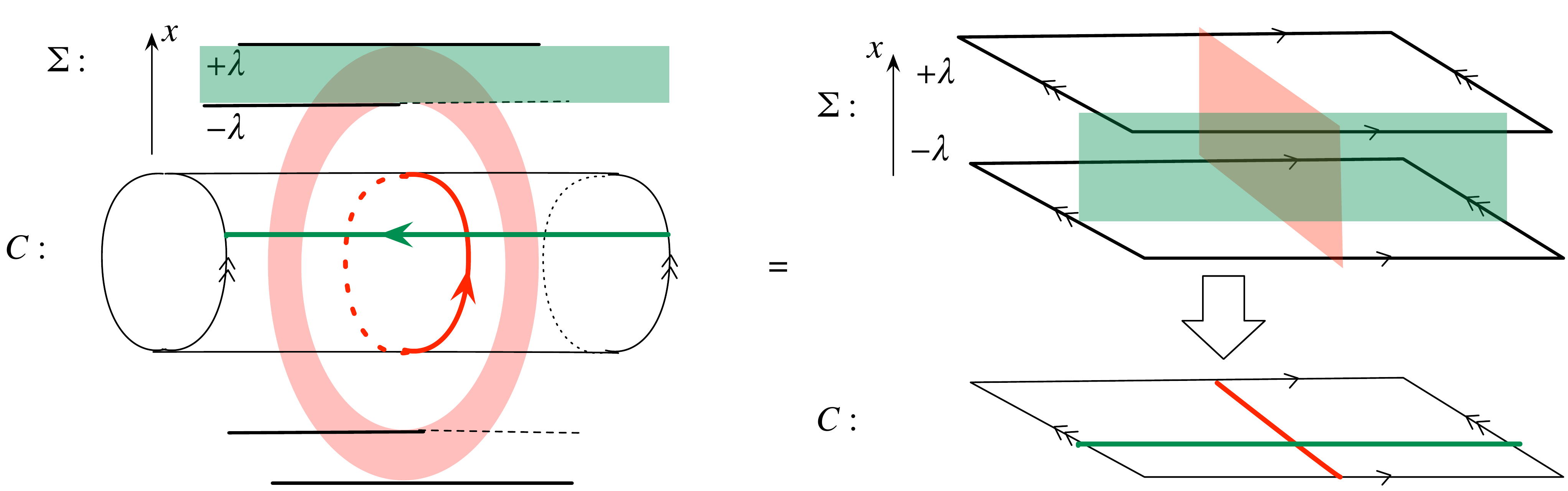}
\]
\caption{$\cN{=}4$ SYM from 6d. A W-boson and a monopole are depicted there.\label{fig:n4sym-from-6d}}
\end{figure}

By imposing periodic boundary condition in the $x^5$ direction, we have the situation of Fig.~\ref{fig:n4sym-from-6d}. 
We are compactifying  the maximally supersymmetric Yang-Mills in five dimensions on $S^1$. 
We therefore should obtain 4d $\cN{=}4$  super Yang-Mills theory.
%So far, we always took the two-dimensional space $C$ on which we put the 6d theory to be a sphere $S^2$, and its two-sheeted cover $\Sigma$ was a torus. 
%Here, let us take $C$ itself to be just $T^2$, as shown in Fig.~\ref{fig:n4sym-from-6d}. 
The \Gaiotto\ curve $C$ itself is now a torus $T^2$.  Let the complex structure of this $T^2$ be $\tau$.
The \SeibergWitten\ curve $\Sigma$ consists of two parallel copies of this torus embedded in $X$, separated by $2\lambda$ in the $x$ direction,
where $\lambda$ is now a constant. 
%This is known to give 4d $\cN{=}4$ supersymmetric $\SU(2)$ Yang-Mills theory. 

We can consider a cycle $L_{n,m}$ in $C$, wrapping $n$ times in the $x_6$ direction and $m$ times in the $x_5$ directions.
Then we can consider a ring-shaped membrane over this cycle, which gives rise to particles of masses \begin{equation}
M_{n,m}=2|na+ma_D|
\end{equation} where \begin{equation}
a=\int_A \lambda,\quad
a_D=\int_B\lambda=\tau a.
\end{equation}
The particles with $(n,m)=(1,0)$ are W-bosons, 
and the particles with $(n,m)=(0,1)$ are monopoles. 
The peculiar feature of this theory is that the monopoles and the W-bosons both come from ring-shaped membranes.
In fact, from the 6d point of view, the distinction of the two directions of the torus is completely arbitrary. 
Then this theory with a given value of $\tau=\tau_0$,
and the theory with another value of $\tau=-1/\tau_0$ are the same after the exchange of the W-bosons and the monopoles. 

Indeed they match the property of the $\cN{=}4$  supersymmetric $\SU(2)$ Yang-Mills.
This theory is conformal and has an exactly marginal coupling $\tau$. In the semi-classical region, the ratio of the mass of the monopole to that of the W-boson is $|\tau|$. 
The $\cN{=}4$ supersymmetric $\SU(2)$ Yang-Mills has four Weyl fermions in the adjoint representation.
The semiclassical quantization of the monopole solution in this situation, as was recalled briefly in Sec.~\ref{sec:monopole}, makes the monopole states into a massive $\cN{=}4$ vector multiplet.
This makes it possible to exchange it with the W-boson, which is also in a massive $\cN{=}4$ vector multiplet.  In general we expect that there is a massive $\cN{=}4$ vector multiplet with mass $|na+ma_D|$, for  any coprime pair of integers $(n,m)$. This should arise from a semi-classical quantization of charge-$m$ monopole background. This is the celebrated conjecture of Sen \cite{Sen:1994yi}. 

\subsubsection{$\cN{=}2$ pure $\SU(2)$ theory and the $N_f=1$ theory}

\begin{figure}[h]
\[
\inc{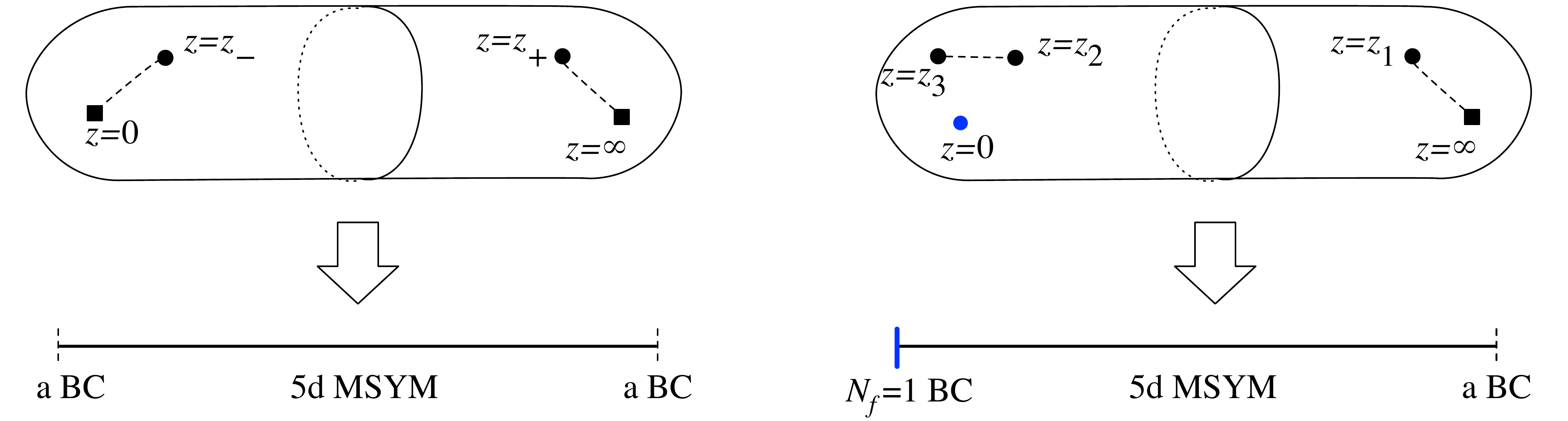}
\]
\caption{Pure and $N_f=1$ $\SU(2)$ theories via 5d construction.\label{fig:pure_nf_1_from_5d}}
\end{figure}
The curve of the pure $\cN{=}2$ $\SU(2)$ theory \begin{equation}
\frac{\Lambda^2}z + \Lambda^2 z = x^2-u 
\end{equation} and the curve of the $\cN{=}2$ $\SU(2)$ theory with one flavor \begin{equation}
\frac{2\Lambda(x-\mu)}z + \Lambda^2 z = x^2-u 
\end{equation} can be given a similar interpretation. 
The point is to take $x_5=\log |z|$ and $x_6=\Arg z$, and compactify along the $x_6$ direction first, see Fig.~\ref{fig:pure_nf_1_from_5d}.

Let us first consider the pure theory. 
The term on the left hand side, $\Lambda^2/z$, should be regarded as a boundary condition `terminating' the fifth direction $x_5$, although $x_5=\log |z|$ formally extends to $-\infty$. 
The bulk of the five dimensional theory is maximally supersymmetric. The resulting four-dimensional theory is $\cN{=}2$, and therefore the boundary   breaks half of the supersymmetry, without doing much other than that. A boundary condition which preserves half of the original supersymmetry is called a half-BPS boundary condition.  Then we see that the term $\Lambda^2/ z$ represents a half-BPS boundary condition of the 5d theory.

The term $\Lambda^2 z$ is obtained by the flip $x_5\leftrightarrow -x_5$, and therefore should represent the same boundary condition. 
In the end, we see that the system is a compactification of the maximally supersymmetric $\SU(2)$ Yang-Mills on a segment, terminated by two boundary conditions breaking half of the supersymmetry, realizing 4d pure $\SU(2)$ Yang-Mills. 

Next, let us consider the one-flavor theory. 
The term $\Lambda^2z$ is the same as the pure case, so it should give the same half-BPS boundary condition.
The boundary condition at $z\sim 0$ is different: now we have a term of the form $\Lambda(x-\mu)/z$.
This should mean that one hypermultiplet with the mass $\mu$ in the doublet of $\SU(2)$ lives on this boundary, coupling to the bulk five-dimensional gauge multiplets. 

\subsubsection{The $\SU(2)$ theories with $N_f=2,3,4$}\label{sec:su2nfpre}
From this interpretation, it is easy to get the 6d realization of $\SU(2)$ theory with $N_f=2,3,4$ flavors, namely the theory with $N_f=2,3,4$ hypermultiplets in the doublet representation.
In terms of $\cN{=}1$ chiral multiplets, we have $(Q_i^a,\tilde Q^i_a)$ for $a=1,2$ and  $i=1,\ldots,N_f$, with the superpotential \begin{equation}
\sum_i \left(Q_i \Phi \tilde Q^i + \mu_i Q_i\tilde Q^i\right)
\end{equation} where $\mu_{i}$ are mass terms. 

\begin{figure}[h]
\[
\inc{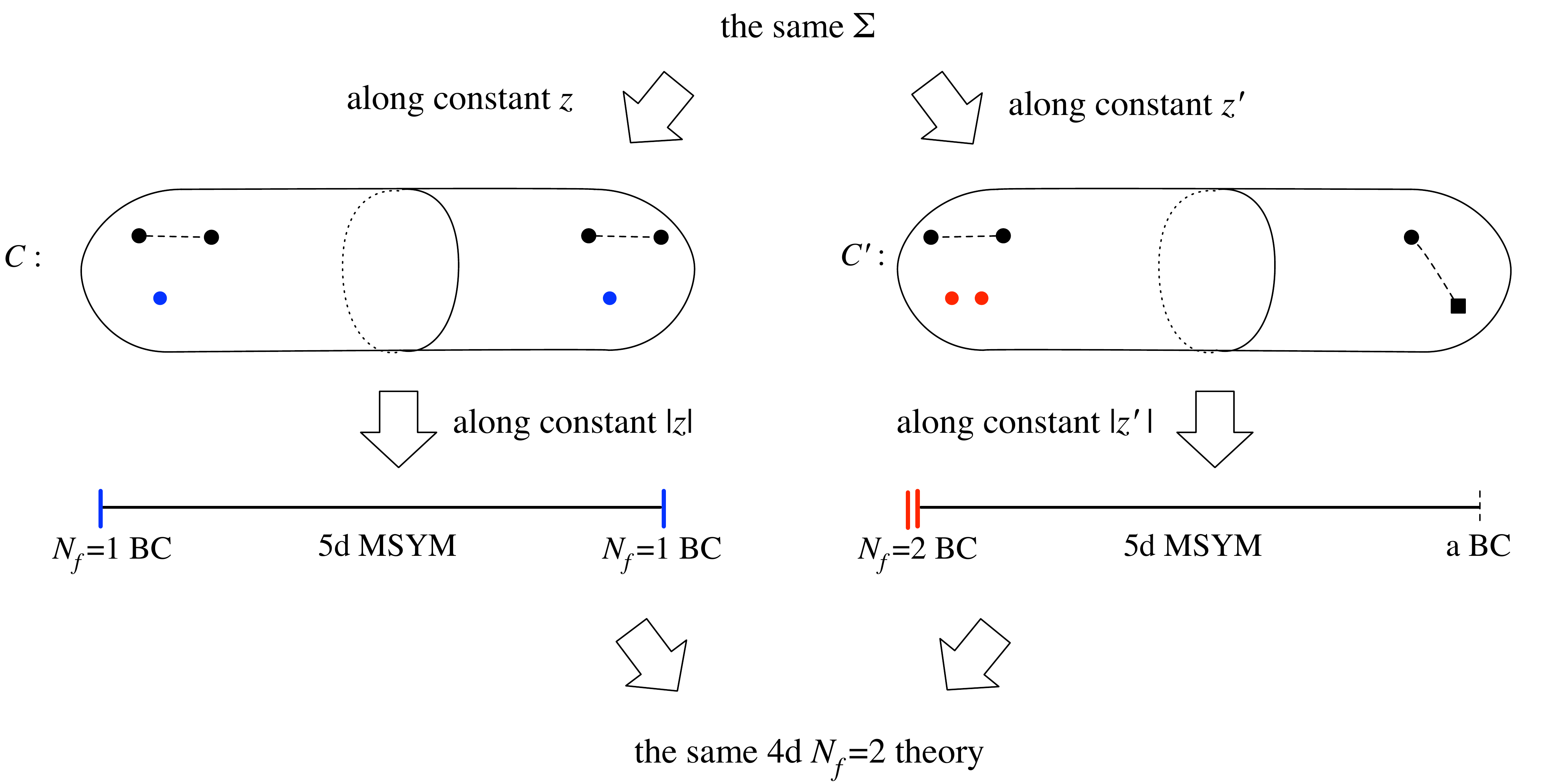}
\]
\caption{$N_f=2$ theory.\label{fig:nf_2_from_6d}}
\end{figure}

Let us start with the $N_f=2$ theory.
We know how to introduce one hypermultiplet in the doublet at the boundary on the side $z=0$.
To do the same on the side $z=\infty$, we just a  change of variables $z\leftrightarrow 1/z$. We end up with the setup shown on the left-hand side of Fig.~\ref{fig:nf_2_from_6d}, with the curve given by \begin{equation}
\frac{2\Lambda(x-\mu_1)}z + 2\Lambda(x-\mu_2) z = x^2-u \label{nf_2_curve_pre}
\end{equation} with the Seiberg-Witten differential $\lambda=xdz/z$.

The same curve can be rewritten using another variable $z'=(x-\mu_2)z/(2\Lambda)$: \begin{equation}
\frac{(x-\mu_1)(x-\mu_2)}{z'} + 4\Lambda^2 z' = x^2-u.\label{nf_2_curve_pre2}
\end{equation} But now we can consider $x_5'=\log |z'|$, $x_6'=\Arg z'$  to reduce first to a theory on $C'$ parameterized by $z'$, and then to a five-dimensional theory on a segment parameterized by $|z'|$. 
In this interpretation, the boundary condition on the $z'=\infty$ side is the same one in the pure $\SU(2)$ case.
Therefore, the boundary condition on the $z'=0$ side given by the term ${(x-\mu_1)(x-\mu_2)}/z$ should be the half-BPS condition such that two hypermultiplets in the doublet of $\SU(2)$ live on the boundary. 

The description of the system is not complete until we give the one-form $\lambda$ describing the variable tension. In \eqref{nf_2_curve_pre} it is $2\pi i\lambda=xdz/z$ and in \eqref{nf_2_curve_pre2} it is $2\pi i\lambda'=xdz'/z'$.  Both are obtained by integrating $dx\wedge d\log z= dx\wedge d\log z'$, see \eqref{integ}.
The two differentials are not quite equal, however: \begin{equation}
\lambda'-\lambda=\frac{1}{2\pi i}x d\log \frac{z'}{z}= \frac{1}{2\pi i}x d\log (x-\mu_2). \label{diff}
\end{equation} The difference is independent of $u$, and its non-zero residue is at $\mu_2$ at $x=\mu_2$. 
This means that, given a cycle $L$ on the \SeibergWitten\ curve $\Sigma$, we have \begin{equation}
\oint_L \lambda'-\oint_L \lambda= k\mu_2
\end{equation} where $k$ is an integer. 
Recall that the BPS mass formula is governed by the expansion \begin{equation}
\oint_L \lambda = na+ma_D + f_1\mu_1  + f_2\mu_2
\end{equation} where $f_{1,2}$ are flavor charges, see \eqref{Zlin}.
Therefore, the choice between the two Seiberg-Witten differentials $\lambda$ and $\lambda'$ affects the mapping of the flavor charge $f_2$ and the cycle $L$, but not much else. 
In general, a change in the Seiberg-Witten differential by a form which is independent of $u$ and whose residues are integral linear combinations of the hypermultiplet masses are safe. We will encounter them repeatedly later.  

\begin{figure}[h]
\[
\inc{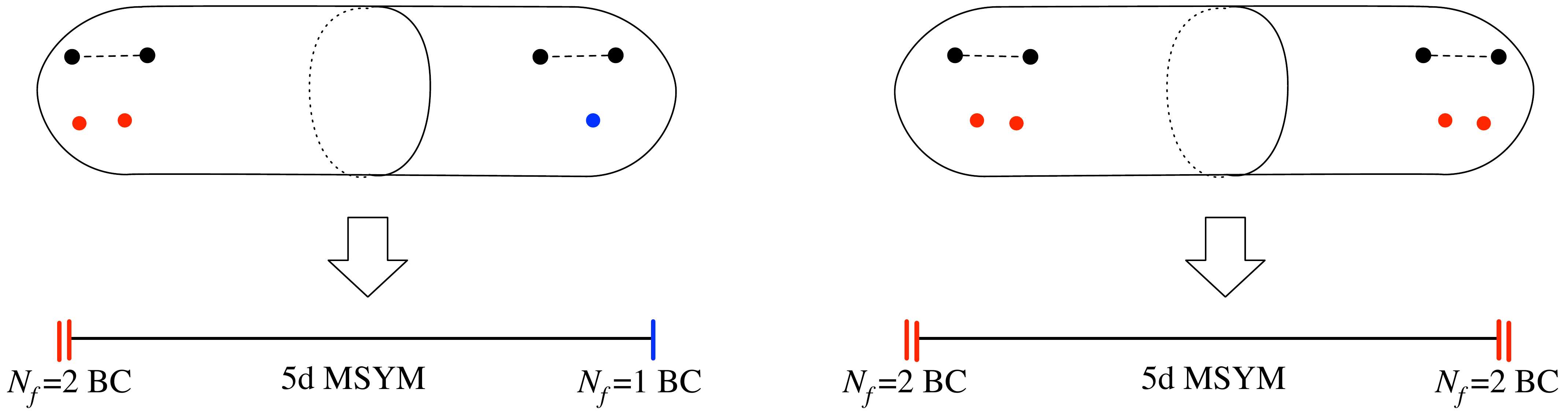}
\]
\caption{$N_f=3$ theory and $N_f=4$ theory.\label{fig:nf_3_4_from_6d}}
\end{figure}
Now that we have a boundary condition representing the existence of two doublet hypermultiplets, it is easy to guess the curve of the $N_f=3$ theory and $N_f=4$ theory. We just have to combine various boundary conditions which we already found, as in Fig.~\ref{fig:nf_3_4_from_6d}. For the $N_f=3$ theory we find \begin{equation}
\frac{(x-\mu_1)(x-\mu_2)}z + 2\Lambda(x-\mu_3) z = x^2-u,\label{nf_3_curve_pre}
\end{equation} and for the $N_f=4$ theory we find \begin{equation}
f\cdot \frac{(x-\mu_1)(x-\mu_2)}z + f'\cdot (x-\mu_3)(x-\mu_4) z = x^2-u\label{nf_4_curve_pre}
\end{equation} where we put complex numbers $f$ and $f'$. One of them can be eliminated by a rescaling of $z$. 

Our next task is to check that the curves thus obtained via the 6d construction have the correct properties to describe the respective four-dimensional theories. Before proceeding, we need to learn more about the Higgs branch of $\cN{=}2$ theories in general. 

\section{Higgs branches and hyperk\"ahler manifolds}\label{sec:higgsbranch}
So far we only considered the branch of the moduli space of the supersymmetric vacua where the scalar $\Phi$ in the vector multiplet is nonzero, and all the hypermultiplets are zero. Instead let us consider a branch where $\Phi=0$, but the hypermultiplet scalars are nonzero.   This branch is called the Higgs branch.  

\subsection{General structures of the Higgs branch Lagrangian}\label{sec:hypergeneral}
First, recall a general $\cN{=}1$ theory containing only scalars and fermions. 
Such a theory can be described by the Lagrangian \begin{equation}
\int d^4 \theta K(\bar\Phi_{\bar j},\Phi_i)=g_{i\bar j}\partial_\mu \phi^i \partial_\nu \bar \phi^{\bar j} + \cdots
\end{equation}
where \begin{equation}
g_{i\bar j}=\frac{\partial^2 K}{\partial \phi_i \partial \bar\phi_{\bar j}}.
\end{equation} This defines a K\"ahler manifold. In particular, the manifold is naturally a complex manifold.
This fact is almost implicit in our formalism, since the chiral multiplets are by definition complex valued. 
It is instructive to recall why this was so: we have the basic supersymmetry transformation \begin{equation}
\delta_\alpha \phi=\psi_\alpha, \quad
\delta^\dagger_{\dot\alpha} \psi_\alpha=i\sigma^\mu_{\alpha\dot\alpha} \partial_\mu \phi
\end{equation} A convention independent fact is that $\delta_\alpha\delta_{\dot\alpha}$ acting on a complex scalar involves a multiplication by $i$. In terms of the real and imaginary parts of $\phi$, we can schematically write this fact as \begin{equation}
\delta^\dagger_{\dot\alpha} \delta_\alpha 
\begin{pmatrix}
\Re \phi \\
\Im \phi
\end{pmatrix}
=
\sigma^\mu_{\dot\alpha\alpha}\partial_\mu  I\begin{pmatrix}
\Re \phi \\
\Im \phi
\end{pmatrix}
\end{equation}
where the matrix
\begin{equation}
I=\begin{pmatrix}
0 & -1 \\
1 & 0
\end{pmatrix}
\end{equation}  has the property $I^2=-1$. This is the crucial matrix defining the complex structure of the scalar manifold of an $\cN{=}1$ theory.

Now, let us consider an $\cN{=}2$ theory consisting of scalars and fermions only. 
Note that this means that there are no $\cN{=}2$  vector multiplets.
This theory has two sets of $\cN{=}1$ supersymmetries $\delta^{i=1,2}_{\alpha}$. In addition, \begin{equation}
\delta^{(c)}_\alpha := c_1 \delta^1_\alpha + c_2\delta^2_\alpha
\end{equation} also generates an $\cN{=}1$ sub-supersymmetry when $|c_1|^2+|c_2|^2=1$.
Applying the argument in the last paragraph for this $\cN{=}1$ subalgebra, we find that there are matrices \begin{equation}
I^{(c)}=I_a n^a, \qquad n^a=(\bar c_1,\bar c_2) \sigma^a \begin{pmatrix}
c_1\\
c_2
\end{pmatrix}\label{cn}
\end{equation} which always satisfy \begin{equation}
(I^{(c)}){}^2=-1.\label{Icsq}
\end{equation} Note that $n^a$ are real and $|n^1|^2+|n^2|^2+|n^3|^2=1$, i.e.~they are on $S^2$. 
Denoting $(I,J,K):=(I_1,I_2,I_3)$ for simplicity and expanding \eqref{Icsq}, one finds the relations \begin{equation}
I^2=J^2=K^2=-1,\qquad
IJ=K=-JI,\ 
JK=I=-KJ,\ 
KI=J=-IK. \label{qcom}
\end{equation} This commutation relation of $I$, $J$ and $K$ is that of a quaternion.
A manifold with an action of quaternion algebra on its tangent space is called a hyperk\"ahler manifold.
Therefore we found that the scalar manifold of an $\cN{=}2$ theory without massless vector multiplets is hyperk\"ahler. 

Note that the $\SU(2)_R$ symmetry acts on the doublet $(c_1,c_2)$, which is restricted to live on the three-sphere $|c_1|^2+|c_2|^2=1$.
The map \eqref{cn} from this $(c_1,c_2)$ to $n^a$ is the standard Hopf fibration $S^3\to S^2$, and the index $a$ transforms as the triplet of $\SU(2)_R$.

Combining with the analysis in Sec.~\ref{sec:lel}, we see that general low-energy $\cN{=}2$ theory has an action of the form \begin{equation}
\int d^2\theta \frac{-\ii}{8\pi}\tau^{ij} W_{\alpha i} W^{\alpha}_j + cc. + \int d^4\theta K_v(\bar a_{\bar j},a _i) + \int d^4\theta K_h(\bar q_{\bar t},q_s)
\end{equation} such that $K_h(\bar q_{\bar t}, q_s)$  gives a hyperk\"ahler manifold and that
there is a prepotential $F(a_i)$ giving $\tau^{ij}$ and $K_v$ via the standard formulas \eqref{pretau}, \eqref{adi} and \eqref{Keq}.

Note that the hypermultiplet side and the vector multiplet side are completely decoupled. The  dependence on the UV gauge coupling is implicitly there in the vector multiplet side. This means that the hypermultiplet side cannot receive quantum corrections depending on the gauge coupling. 

\subsection{Hypermultiplets revisited}\label{sec:hyperrevisited}
Let us revisit the structure of the full and half hypermultiplets introduced in Sec.~\ref{sec:hyperdef} from the viewpoints here.  First, let us recall the types of irreducible representations of compact groups: \begin{equation}
\begin{array}{rll}
\text{complex}& \text{if $R\not\simeq\bar R$ ;} & \\
\text{real} &\text{if $R\simeq \bar R$ :} & \begin{cases}
	\text{strictly real} & \text{if the invariant tensor $\delta_{ij}$ is symmetric,}\\
	\text{pseudo-real} & \text{if the invariant tensor $\epsilon_{ij}$ is antisymmetric.}
	\end{cases}
\end{array}
\end{equation}

In a non-supersymmetric theory with a number of real scalars $\phi^{i}$, $i=1,\ldots,n$,
they can have an action of the flavor symmetry $F$ or the gauge symmetry group $G$ if there are real $n\times n$ matrices $T^a$, $a=1,\ldots, \dim G$ representing the Lie algebra of $G$: \begin{equation}
F,G \curvearrowright \bR^n.
\end{equation}
This representation clearly has an invariant symmetric tensor $\delta_{ij}$ as it acts on $n$ real scalars with a kinetic term $\delta_{ij}\partial_\mu \phi^i \partial_\mu \phi^j$. The representation is therefore strictly real. 

In an $\cN{=}1$ supersymmetric theory with a number of real scalars $\phi^{i}$, $i=1,\ldots,n$ together with $n/2$ Weyl fermions,
the supersymmetry requires existence of a matrix $I$ with $I^2=-1$. The actions of the flavor symmetry $F$ and the gauge symmetry $G$ need to commute with this matrix $I$: \begin{equation}
F,G\curvearrowright \bR^n \curvearrowleft I.
\end{equation}
We can declare that a complex number $a+bi$ acts on the real scalars by the matrix $a+ bI$.
Then the space of scalars becomes a complex vector space,
and the symmetries act on them preserving the complex structure. 
So there are $m=n/2$ chiral multiplets $\Phi^s$, $s=1,\ldots,m$, and both $F$ and $G$ are represented on them in terms of $m\times m$ complex matrices representing their Lie algebras. We can summarize the situation in the following way: \begin{equation}
F,G\curvearrowright \bC^m .
\end{equation}

In an $\cN{=}2$ supersymmetric theory with a number of real scalars $\phi^{i}$, $i=1,\ldots,n$ together with $n/2$ Weyl fermions,
the supersymmetry requires existence of matrices $I$, $J$, $K$ with the commutation relations \eqref{qcom}. 
The actions of the flavor symmetry $F$ and the gauge symmetry $G$ need to commute with $I$, $J$, $K$:
\begin{equation}
F,G\curvearrowright \bR^n \curvearrowleft I, J, K.
\end{equation}
We can declare that a quaternion $a+bi+cj+dk$ acts on the real scalars by the matrix $a+bI+cJ+dK$.
Then the space of scalars becomes a quaternionic vector space,
and the symmetries act on them preserving the quaternion structure.
This requires $n$ to be automatically a multiple of four, $n=4\ell$
Both $F$ and $G$ are represented on them in terms of $\ell\times \ell$ quaternion matrices representing their Lie algebras. Summarizing, we have \begin{equation}
F,G\curvearrowright \bH^{\ell}
\end{equation} where $\bH$ is the skew-field of quaternions. 

As quaternions are not quite common among physicists, we usually just use $a+bI$ to think of the real scalars as complex scalars. Then we have a complex vector space of dimension $2\ell$, and we have $2\ell$ complex scalars $\Phi^s$, $s=1,\ldots,2\ell$, acted on by the flavor symmetry $F$ and the gauge symmetry $G$ in a complex representation $\tilde R$.
The matrix $J+iK$ then  determines an $2\ell \times 2\ell$ antisymmetric matrix $\epsilon_{st}$, which is invariant under the action of $F$ and $G$. 
This means that $\tilde R$ is a pseudoreal representation: \begin{equation}
F,G\curvearrowright \bC^{2\ell} \curvearrowleft \epsilon_{st}
\end{equation} This is the half-hypermultiplet in representation $\tilde R$, introduced briefly at the end of  Sec.~\ref{sec:hyperdef}.

From this point of view, a half-hypermultiplet  is more elementary than a full hypermultiplet, which is given as follows. Take an arbitrary complex representation $R$ of $F\times G$ of dimension $m$. 
Let $i=1,\ldots, m$ be its index. We have an invariant tensor $\delta_{i\bar j}$. 
Let  $\tilde R=R\oplus \bar R$. 
It has an index $s=1,\ldots,n,\bar 1\ldots, \bar n$, and automatically has an antisymmetric invariant tensor \begin{equation}
\epsilon_{st},\qquad \epsilon_{ij}=\epsilon_{\bar i\bar j}=0, \ \epsilon_{i\bar j}=\delta_{i\bar j}=-\epsilon_{\bar ji}.
\end{equation} Then the half-hypermultiplet based on this representation $\tilde R$ is the full hypermultiplet in the representation $R$. 

Concretely, consider four real scalars. This system has a natural symmetry $\SO(4)\simeq \SU(2)_l\times \SU(2)_r$. 
Add two Weyl fermions, with a natural symmetry $\SU(2)_{l}$. 
Then the total system has an $\cN{=}2$ supersymmetry where the $\SU(2)_R$ symmetry  of the $\cN{=}2$ algebra is the $\SU(2)_r$ acting on the scalars.
The symmetry $\SU(2)_l$ can be used as either a flavor or a gauge symmetry.
This whole system consists of just one full hypermultiplet, or one half-hypermultiplet in the $\SU(2)_l$ doublet. 

Next, let $i=1,\ldots, n$ and  $a=1,\ldots,m$ the indices for $\U(n)$ and $\U(m)$ symmetries, respectively.
Then, $\cN{=}1$ chiral multiplets $Q_{i\bar a}$, $\tilde Q_{\bar i a}$ form an $\cN{=}2$ hypermultiplet, in the bifundamental representation of $\U(n)\times \U(m)$. When $\U(n)$ is regarded as a gauge symmetry, $\U(m)$ becomes the flavor symmetry.

Another typical construction is to take $i=1,\ldots, 2n$ to be an index for $\Sp(n)$ symmetry 
and $a=1,\ldots,m$ to be that for $\SO(m)$ symmetry. 
Consider $\cN{=}1$ chiral multiplets $Q_{ia}$. Regard the pair of indices $ia$ as a single index $s=(ia)$, running from $1$ to $2nm$. This system has an antisymmetric invariant tensor $\epsilon_{st}=\epsilon_{(ia)(jb)}=J_{ij}\delta_{ab}$, thus they make up a hypermultiplet with the symmetry $\Sp(n)\times \SO(m)$, commuting with the superalgebra. 
When $\Sp(n)$ is made into a gauge symmetry, $\SO(m)$ becomes the flavor symmetry, and vice versa. 
This explains the fact that when there are $n$ hypermultiplets in the vector representation of gauge $\SO(m)$, we have $\Sp(n)$ flavor symmetry, and when there are $m$ half-hypermultiplets in the fundamental representation of gauge $\Sp(n)$, we have $\SO(m)$ flavor symmetry.

\subsection{The hyperk\"ahler quotient}\label{sec:hkq}
Let us come back to the study of the Higgs branch. 
The equations defining it were given in Sec.~\ref{sec:classicalvac} for the case of $\SU(N)$ gauge theory with $N_f$ flavors, see \eqref{Feq} and \eqref{tripleteq}. 
Let us write them down for the general case.
In this subsection we set all the mass parameters of the hypermultiplets to zero.

Consider an $\cN{=}2$ gauge theory with gauge group $G$ and a hypermultiplet $(Q^i,\tilde Q_i)$ in the representation $R$. Here the index $i=1,\ldots,\dim R$ is for the hypermultiplet
and we use the index $a=1,\ldots,\dim G$ for the adjoint representation.
The Higgs branch is given by \begin{equation}
\left\{
\begin{array}{r@{=0}}
(Q^i Q^\dagger{}_j-\tilde Q^\dagger{}^i \tilde Q_j)T^a{}_i^j \\
\Re Q^i \tilde Q_j T^a{}_i^j \\
\Im Q^i \tilde Q_j T^a{}_i^j
\end{array}
\right\} \Bigm/ (\text{identification by the gauge group})\label{hkq}
\end{equation} where $T^a{}_i^j$ is the matrix of the algebra of $G$ in the representation $R$.

There is no massless vector multiplet remaining in the generic point of the Higgs branch. From the general analysis in the preceding sections, we know that they form a hyperk\"aher manifold. The construction \eqref{hkq} is known as the hyperk\"ahler quotient construction in the literature both in mathematics and in physics, and found originally in \cite{Hitchin:1986ea}. 
The real dimension of any hyperk\"ahler manifold is always a multiple of four. 
Let us check this in this situation. 
Suppose the original hypermultiplets consist of $4m$ real scalars. 
The D-term condition imposes $\dim G$ real constraints, for each $I$, $J$ and $K$. Then we make the identification by the action of $G$. Therefore we have \begin{equation}
4m-3\dim G-\dim G=4(m-\dim G)\label{realdim}
\end{equation} real dimensions after the quotient. 

If we are only interested in the holomorphic  structure, we can drop the $D$-term equation
and instead perform the identification by the complexified gauge group \begin{equation}
\left\{
Q^i \tilde Q_j T^a{}_i^j   = 0
\right\} \Bigm/ (\text{identification by the \emph{complexified} gauge group}).\label{hsq}
\end{equation} 
Note that this is a more natural form in the $\cN{=}1$ superfield formulation, if we do not put the vector superfield into the Wess-Zumino gauge. 
The basic idea to show the equality of \eqref{hsq} with \eqref{hkq} is to minimize $|D|^2$ within each of the orbit of the complexfied gauge group. The minimization condition then gives $D=0$, recovering \eqref{hkq}. 
This rough analysis also shows that, more precisely speaking, we need to remove the so-called unstable orbits in \eqref{hsq}, in which there is no point where $|D|^2$ is minimized. 

In this approach, we start from $2m$ complex scalars. We then imposes $\dim G$ complex constraints and then perform the identification by the action of $G_\bC$, the complexified gauge group, removing $\dim G$ complex dimensions. We end up with \begin{equation}
2m-\dim G-\dim G = 2(m-\dim G )\label{complexdim}
\end{equation} complex scalars in the quotient. This is compatible with what we just found in \eqref{realdim}.  If we count the quaternionic dimension, we just have the formula \begin{equation}
m-\dim G.
\end{equation}

\subsubsection{$\U(1)$ gauge theory with one charged hypermultiplet}
\begin{figure}[h]
\[
\inc{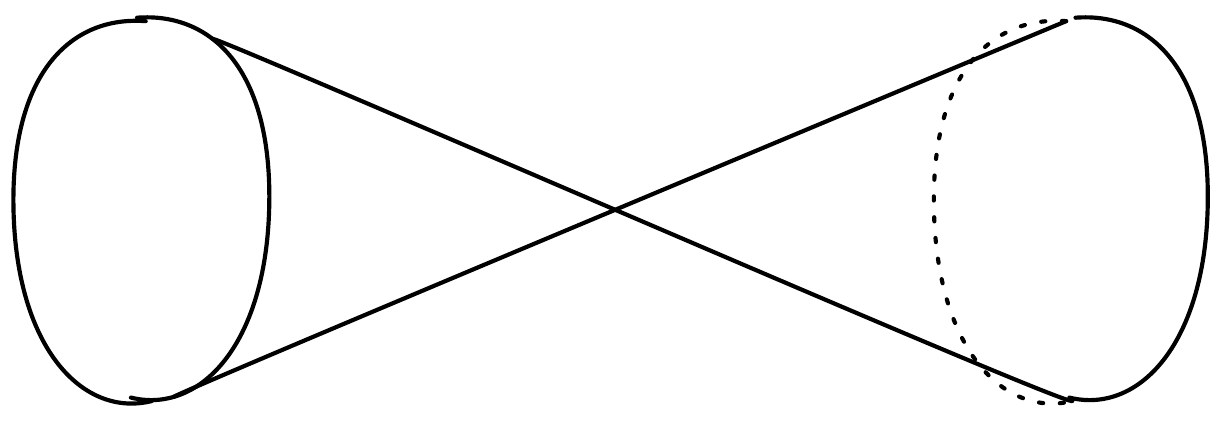}
\]
\caption{Not a very accurate depiction of $\bC^2/\bZ_2$\label{fig:c2z2}}
\end{figure}

Let us consider two examples. First, take an $\cN{=}2$ $\U(1)$ gauge theory with two hypermultiplets $(Q_i,\tilde Q^i)$ with charge $\pm1$. Here $i=1,2$.
We have $m=2$ and $\dim G=1$ in the expressions above, so we expect a complex two-dimensional Higgs branch.
First, let us determine the Higgs branch explicitly.  The F-term equation is \begin{equation}
Q_1 \tilde Q^1 + Q_2\tilde Q^2 =0.
\end{equation} Then we have \begin{equation}
(Q_1,\tilde Q^1)=(z,\tilde z t),\quad
(Q_2,\tilde Q^2)=(\tilde z,-zt)
\end{equation} 
for some complex numbers $z$, $\tilde z$ and $t$.
Then the D-term equation $|Q_1|^2+|Q_2|^2-|\tilde Q^1|^2-|\tilde Q_2|^2=0$ says \begin{equation}
|z|^2+|\tilde z|^2=|t|^2(|z|^2+|\tilde z|^2).
\end{equation} Therefore we see $|t|=1$. We can use the $\U(1)$ gauge rotation to eliminate $t$ almost completely, by demanding \begin{equation}
\Arg z=\Arg (\tilde z t). 
\end{equation} This still does not fix the $\U(1)$ gauge transformation given by the multiplication by $-1$ on $Q_i$, $\tilde Q^i$, sending the pair $(z,\tilde z)$ to $(-z,-\tilde z)$.
 We conclude that the Higgs branch is given by \begin{equation}
 \bC^2/\bZ_2  = \{(z,\tilde z)\in \bC^2\} /  (z,\tilde z)\leftrightarrow (-z,-\tilde z). \label{c2z2}
\end{equation}
 A not-quite-accurate schematic description is given in Fig.~\ref{fig:c2z2}.

Let us use the complex description \eqref{hsq} to obtain the same Higgs branch in a different way. 
Instead of identifying points connected by the complexified gauge group, we can just consider combinations of coordinates which are invariant under it. In this case, $Q_i$ has charge $+1$ and $\tilde Q^i$ has charge $-1$. Then, the gauge invariants are 
$Q_i \tilde Q^j$, for arbitrary choices of $i$ and $j$. 
We need to impose \begin{equation}
Q_1 \tilde Q^1 + Q_2 \tilde Q^2=0,
\end{equation} too. In total, we find three combinations \begin{equation}
A=Q_1 \tilde Q^2,\quad
B=Q_2\tilde Q^1,\quad
C=iQ_1\tilde Q^1 = -iQ_2 \tilde Q^2.
\end{equation} They satisfy one obvious relation \begin{equation}
AB=C^2.\label{abc}
\end{equation} With three variables $A$, $B$, $C$ and one relation above, we have complex two-dimensional space. This is the Higgs branch.

This description can also be found starting from the definition of $\bC^2/\bZ_2$ in \eqref{c2z2}. 
Combinations of $z$, $\tilde z$ invariant under $ (z,\tilde z)\leftrightarrow (-z,-\tilde z)$  are \begin{equation}
A=z^2,\quad B=\tilde z^2, \quad C=z\tilde z
\end{equation} which satisfy the same relation \eqref{abc}. Therefore they are the same spaces as complex manifolds. 

\subsubsection{$\SU(2)$ gauge theory with two  hypermultiplets in the doublet}
As the second example, consider $\cN{=}2$ $\SU(2)$ gauge theory with $N_f$ full hypermultiplets in the doublet representations. In terms of $\cN{=}1$ chiral multiplets, we have \begin{equation}
Q_i^a, \quad \tilde Q^i_a \qquad (a=1,2;\quad i=1,\ldots,N_f)
\end{equation} As the doublet and the anti-doublet representations are the same for $\SU(2)$ gauge theory, we can denote them also as \begin{equation}
Q_I^a,\qquad (a=1,2;\quad I=1,\ldots,2N_f)
\end{equation} which makes $\SO(2N_f)$ flavor symmetry more manifest.

We have $4N_f$ complex scalars and $\dim \SU(2)=3$. Then the complex dimension of the Higgs branch is \begin{equation}
4N_f -2\cdot 3.\label{dimform}
\end{equation} So we do not have the Higgs branch for $N_f=1$,
and expect a  Higgs branch with complex dimensions $2$, $6$, $10$ for $N_f=2,3,4$, respectively.  Let us study the case $N_f=2$ in more detail. 

Gauge-invariant combinations of $Q_I^a$ are \begin{equation}
M_{IJ}=Q_I^a Q_J^b \epsilon_{ab}.\label{su2mesons}
\end{equation} The left hand side is automatically  anti-symmetric under the exchange of $I$ and $J$. 
The F-term equation is \begin{equation}
Q_I^{(a} Q_J^{b)}\delta^{IJ}=0.\label{su2Dterm}
\end{equation}

For $\SO(2N_f)=\SO(4)$, we can split an antisymmetric matrix $M_{IJ}$ of $\SO(4)$ into the self-dual and the anti-self-dual parts. Equivalently, using $\SO(4)\simeq \SU(2)_u \times \SU(2)_v$, $M_{IJ}$ splits into the triplet $M_{(\alpha\beta)}$ of $\SU(2)_u$ and 
the triplet $M_{(\dot\alpha\dot\beta)}$ of $\SU(2)_v$, where $\alpha,\beta=1,2$ and $\dot\alpha,\dot\beta=1,2$ are doublet indices of $\SU(2)_{u,v}$ respectively. 
The index $I$ itself can be thought of a pair of indices: $I=(\alpha\dot\alpha)$.
Then the hypermultiplets we are dealing with can be written as \begin{equation}
Q_{a\alpha\dot\alpha},  \qquad a,\alpha,\dot\alpha=1,2
\end{equation} which makes the existence of $\SU(2)^3$ symmetry manifest. Then \begin{align}
M_{\alpha\beta}&=Q_{a\alpha\dot\alpha}Q_{b\beta\dot\beta}\epsilon^{ab}\epsilon^{\dot\alpha\dot\beta},\\
M_{\dot\alpha\dot\beta}&=Q_{a\alpha\dot\alpha}Q_{b\beta\dot\beta}\epsilon^{ab}\epsilon^{\alpha\beta}
\end{align}
are the self-dual and the anti-self-dual parts of $M_{IJ}$, respectively. 
 The F-term equation can be written as \begin{equation}
Q_{a\alpha\dot\alpha}Q_{b\beta\dot\beta}\epsilon^{\alpha\beta}\epsilon^{\dot\alpha\dot\beta}=0.
\end{equation}

Using this description, it is not very hard to check that \begin{align}
M_{\alpha\beta}M_{\gamma\delta}\epsilon^{\alpha\gamma}\epsilon^{\beta\delta}&=0,
\label{c2z2ABC}\\
M_{\dot\alpha\dot\beta}M_{\dot\gamma\dot\delta}\epsilon^{\dot\alpha\dot\gamma}\epsilon^{\dot\beta\dot\delta}&=0,\label{c2z2XYZ}\\
M_{\alpha\beta}M_{\dot\alpha\dot\beta}&=0.\label{ABCperpXYZ}
\end{align}
The structure becomes clearer by defining \begin{align}
A&=M_{11}, & B&=M_{22}, &  C&=M_{12}=M_{21};\\
X&=M_{\dot 1\dot1}, & Y&=M_{\dot2\dot2}, &  Z&=M_{\dot1\dot2}=M_{\dot2\dot1}.
\end{align}
The relations \eqref{c2z2ABC} and \eqref{c2z2XYZ} give \begin{equation}
AB=C^2,\qquad XY=Z^2,
\end{equation} whereas the relation \eqref{ABCperpXYZ} mean that 
 two vectors $(A,B,C)$ and $(X,Y,Z)$ cannot be both nonzero at the same time.

\begin{figure}[h]
\[
\inc{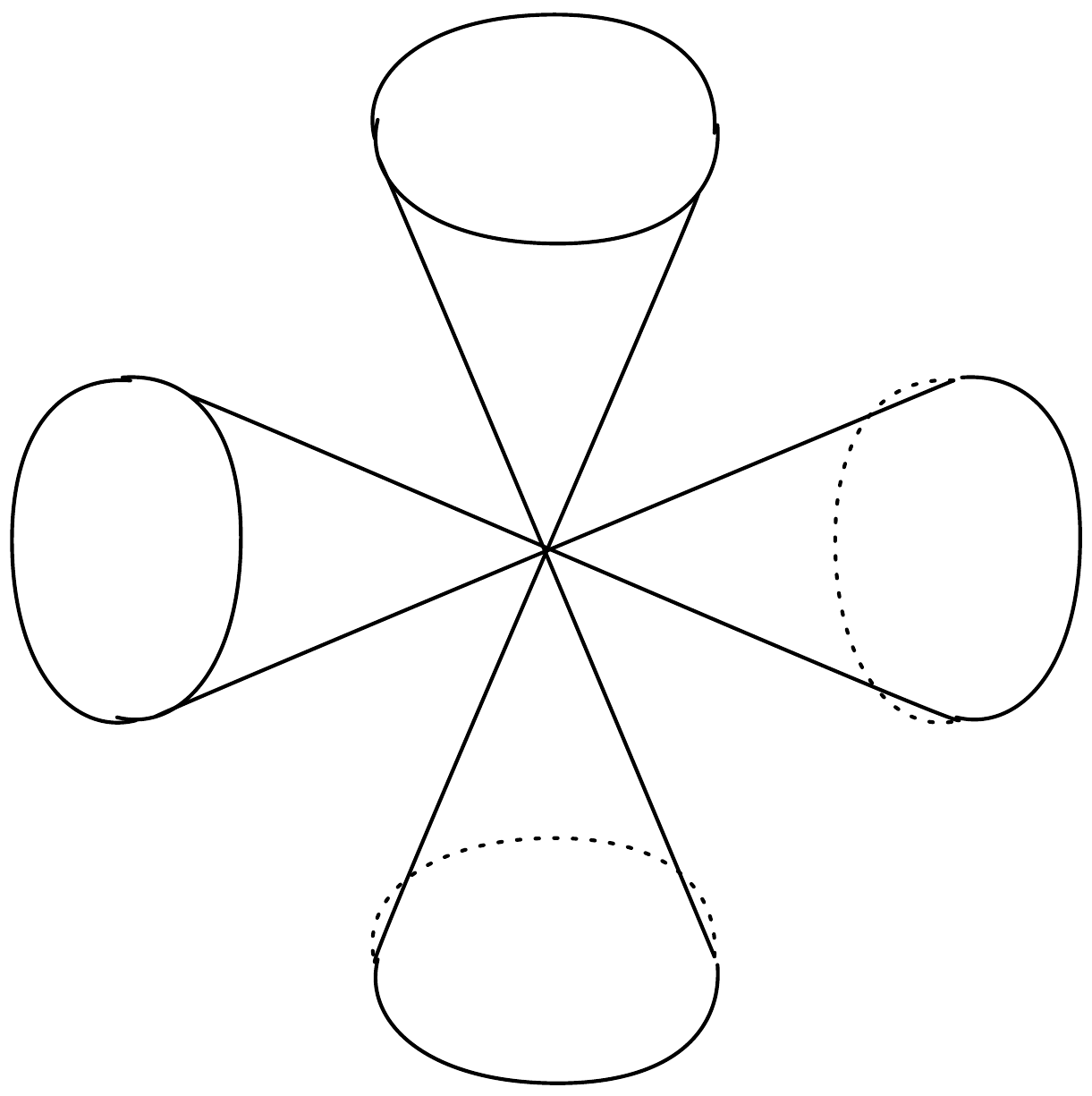}
\]
\caption{Not a very accurate depiction of $\bC^2/\bZ_2\wedge \bC^2/\bZ_2$\label{fig:c2z2copies}}
\end{figure}

Therefore we see that the Higgs branch has the structure schematically described in Fig.~\ref{fig:c2z2copies}: there are two copies of $\bC^2/\bZ_2$, described respectively by the variables $A,B,C$ and $X,Y,Z$. When the vacuum is on one of the $\bC^2/\bZ_2$ described by one set of variables $(A,B,C)$, the other variables are forced to be zero, and vice versa. 
Therefore two copies of $\bC^2/\bZ_2$ can be said to share the origin, where all of $A,B,C$ and $X,Y,Z$ are zero. The Higgs branch has complex dimension two, as expected. 

Recall we decomposed the flavor symmetry $\SO(4)$ into $\SU(2)_u\times \SU(2)_v$. 
The vectors $(A,B,C)$ and $(X,Y,Z)$ are triplets under $\SU(2)_u$ and $\SU(2)_v$, respectively.
Therefore, the flavor parity of $\OO(4)\supset \SO(4)$ exchanges the two copies of  $\bC^2/\bZ_2$ composing the Higgs branch.

\section{$\SU(2)$ theory with 2 and 3 flavors}\label{sec:nf=2,3}
\subsection{Generalities}
In this section and next, we  consider $\SU(2)$ gauge theory with $N_f$ flavors, with $N_f=2,3,4$.
In terms of $\cN{=}1$ chiral multiplets,
we have $(Q_i,\tilde Q^i)$ for $i=1,\ldots,N_f$ with the superpotential \begin{equation}
\sum_i \left(Q_i \Phi \tilde Q^i + \mu_i Q_i\tilde Q^i\right)
\end{equation} where $\mu_{i}$ are bare mass terms. 
With all $\mu_i$ are the same, there is a $\U(N_f)$ symmetry acting on the indices $i$ of $Q_i$ and $\tilde Q^i$.
On the Coulomb branch with $\Phi=\diag(a,-a)$, the physical masses of the hypermultiplets are given by \begin{equation}
|{}\pm a \pm \mu_i|.
\end{equation}

With $\mu_i=0$, we can combine  $Q_i$ and $\tilde Q^i$ into \begin{equation}
(q_I^a)_{I=1,2,\ldots,2N_f}=(Q_1^a,\ldots,Q_{N_f}^a,\epsilon^{ab} \tilde Q^1_b,\ldots,\epsilon^{ab} \tilde Q^{N_f}_b)
\end{equation} with $\SO(2N_f)$ symmetry. In this notation the superpotential is \begin{equation}
\propto \eta^{IJ} q_I^a \Phi_{ab} q_J^b,\qquad
\text{where}\ \eta=\left(
\begin{array}{c|c}
0 & \mathbf{1}_{N_f} \\
\hline
\mathbf{1}_{N_f} & 0
\end{array}
\right).
\end{equation} 
Since $\eta^{IJ}$ is a symmetric matrix, the flavor symmetry acting on the indices $I,$J is  $\SO(2N_f)$. 
Equivalently, we have $2N_f$ half-hypermultiplets in the doublet representation of $\SU(2)$. 

Classically, introducing an odd number of half-hypermultiplets in the doublet of $\SU(2)$ is all right, with $\SO(\text{odd})$ flavor symmetry. However, such a theory would have odd number of Weyl fermions in the doublet, and is plagued quantum mechanically by Witten's global anomaly, as reviewed in Sec.~\ref{sec:anomaly}. Therefore, for $\SU(2)$ gauge group, we can only consider an even number of half-hypermultiplets in the doublet, or equivalently, an integral number of full-hypermultiplets in the doublet. 

The one-loop running of this theory in the ultraviolet region $|a| \gg |\mu_i|$ is \begin{equation}
\tau(a)=2\tau_{UV}-\frac{2(4-N_f)}{2\pi i}\log \frac{a}{\Lambda_{UV}} +\cdots\label{su2running}
\end{equation}
which can further be rewritten as, when $N_f\neq 4$,
\begin{equation}
=-\frac{2(4-N_f)}{2\pi i}\log \frac{a}{\Lambda}\label{general_1_loop}
\quad\text{where}\quad
\Lambda^{4-N_f}=\Lambda_{UV}^{4-N_f} e^{2\pi i \tau_{UV}}.
\end{equation}

We guessed the form of the curves of these theories in Sec.~\ref{sec:su2nfpre}. The results were given in \eqref{nf_2_curve_pre}, \eqref{nf_3_curve_pre}, \eqref{nf_4_curve_pre}
for $N_f=2,3,4$ respectively. The aim of this section and the next section is to perform various checks that they do reproduce expected properties, and to study strong coupling dynamics using them. In this section we deal with $N_f=2$ and $N_f=3$.
The case $N_f=4$ opens up a whole new field, to which Sec.~\ref{sec:gaiotto} is dedicated. 

\subsection{$N_f=2$: the curve and the monodromies}

\begin{figure}[h]
\[
\inc{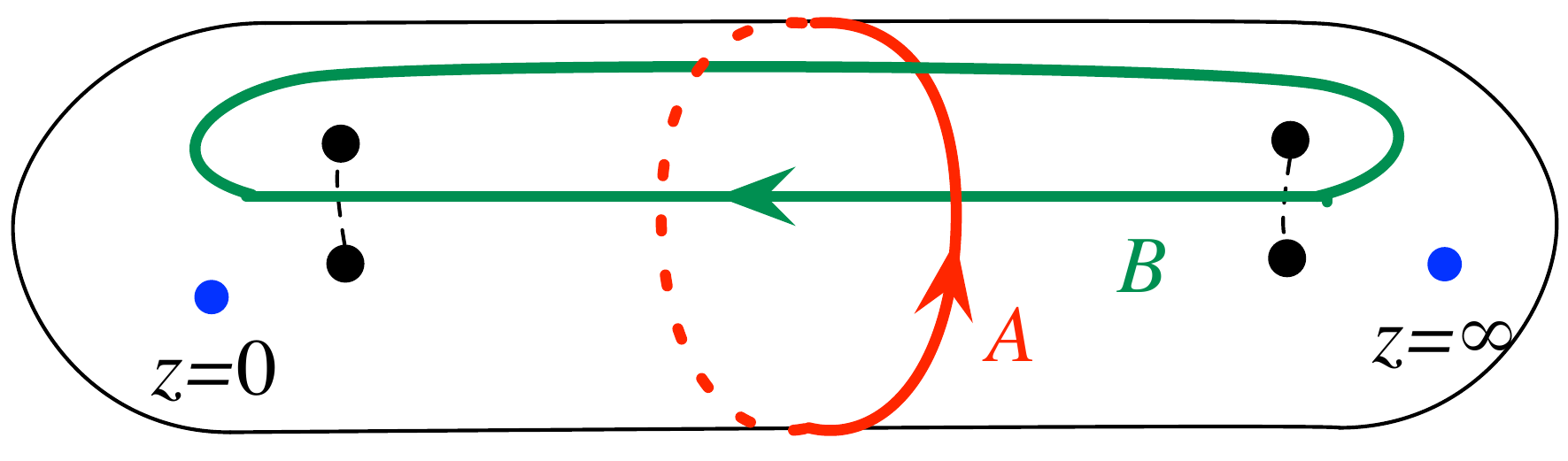}
\]
\caption{The curve of $N_f=2$ theory.\label{fig:nf_2_curve}}
\end{figure}

Let us start with the $\SU(2)$ with $N_f=2$ flavors. 
The \SeibergWitten\ curve was guessed in \eqref{nf_2_curve_pre}, which we repeat here: \begin{equation}
\Sigma:\qquad \frac{2\Lambda(x-\mu_1)}z + 2\Lambda(x-\mu_2) z = x^2-u\label{nf_2_curve}
\end{equation} with the Seiberg-Witten differential $\lambda=xdz/z$.
The \Gaiotto\ curve $C$ is still just an $S^2$, shown in Fig.~\ref{fig:nf_2_curve}.

When $|u|\gg |\Lambda^2|, |\mu_i|^2$, we can estimate the line integrals easily. First, 
we put the $A$-cycle at $|z|=1$. 
Using $x\simeq \sqrt{u}$ around there, we have \begin{equation}
a=\frac{1}{2\pi i}\oint_Ax\frac{dz}z  \simeq \sqrt{u}.\label{nf_2_a}
\end{equation} 
The positions of the branch points of $x(z)$ on the curve $C$ can also be easily estimated: there are two around $z\simeq \sqrt{u}/\Lambda$ and two more around $z\simeq \Lambda/\sqrt{u}$. Then we see \begin{equation}
a_D=\frac{1}{2\pi i}\oint_B x\frac{dz}z \simeq \frac{2\cdot 2}{2\pi i}\int^1_{\sqrt{u}/\Lambda} a\frac{dz}z
\simeq -\frac{4}{2\pi i}a \log \frac{a}{\Lambda}.\label{nf_2_ad}
\end{equation} 
From this we can compute $\tau(a)=\partial a_D/\partial a$, which reproduces the one-loop running \eqref{general_1_loop}.

\begin{figure}[h]
\[
\inc{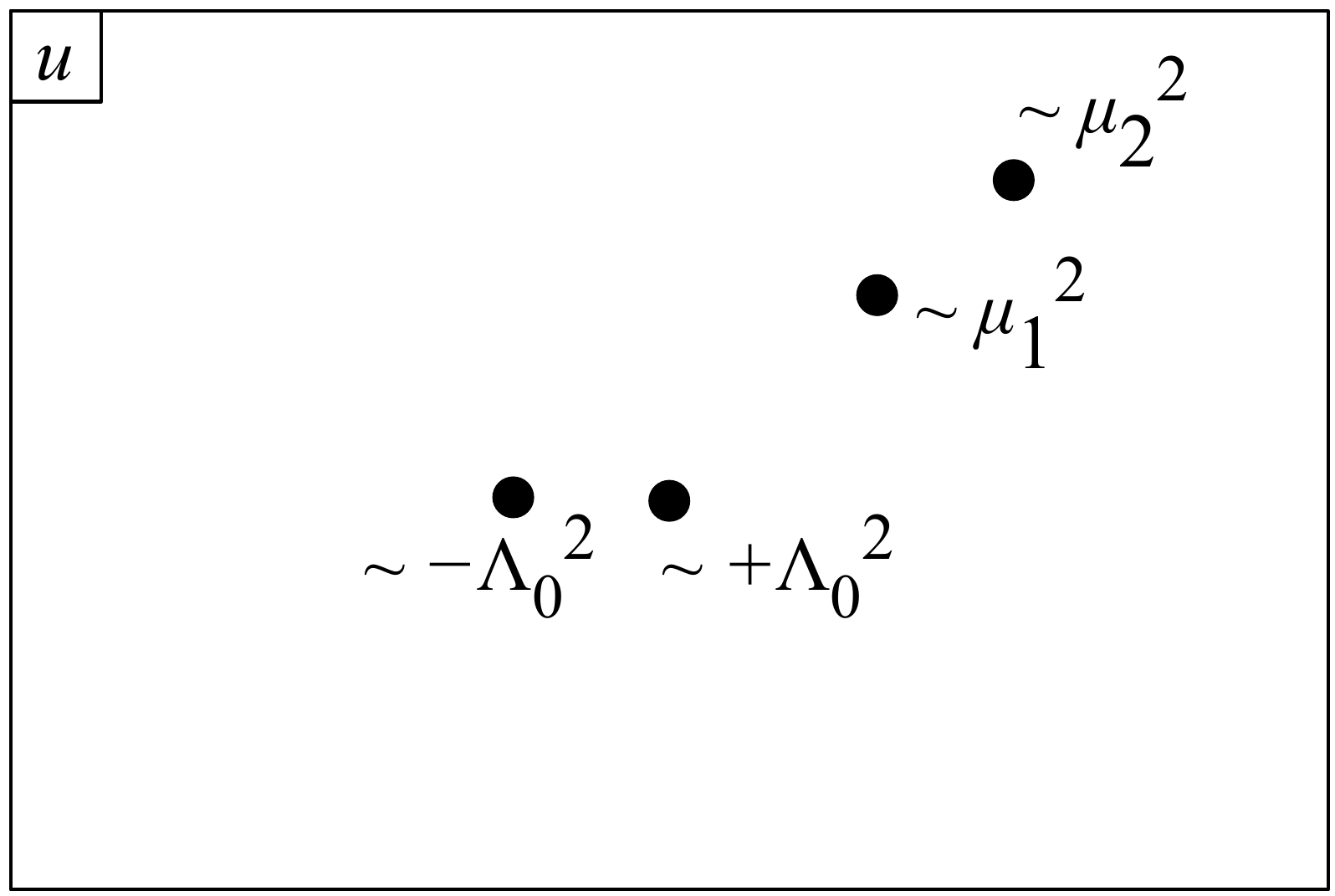}
\]
\caption{The $u$-plane for $N_f=2$.\label{fig:nf_2_u_plane_generic}}
\end{figure}

Let us next study the structure of the singularities on the $u$-plane. When $\mu_{1,2}\gg \Lambda$,  the gauge coupling is rather small around the energy scale $\mu_{1,2}$. Then we expect that when \begin{equation}
u\simeq \mu_i^2 \quad \text{for $i=1,2$}
\end{equation} one component of $(Q_i, \tilde Q^i)$ become very light, producing a singularity. 
Below the scale of $\mu_{1,2}$, the theory is effectively equivalent to pure $\SU(2)$ theory, which should have two singularities where either monopoles or dyons are very light. In total we expect that there are four singularities on the $u$-plane, see Fig.~\ref{fig:nf_2_u_plane_generic}.

\begin{figure}[h]
\[
\inc{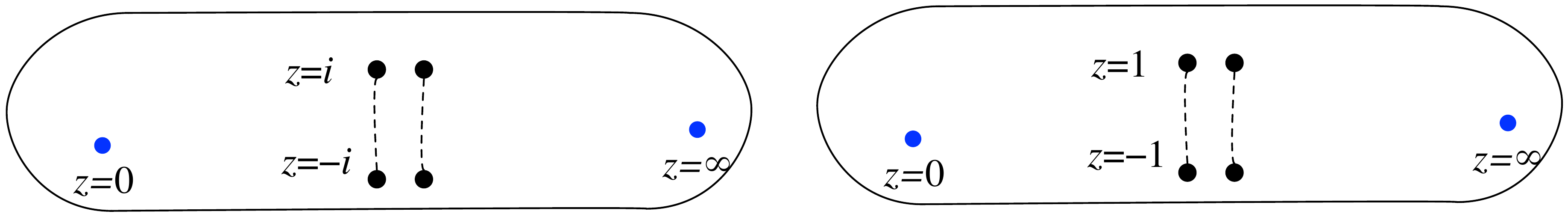}
\]
\caption{The curve of $N_f=2$ theory degenerates when $u=0$ or $u=-4\Lambda^2$. \label{fig:collision_nf_2}}
\end{figure}

\begin{figure}[h]
\[
\inc{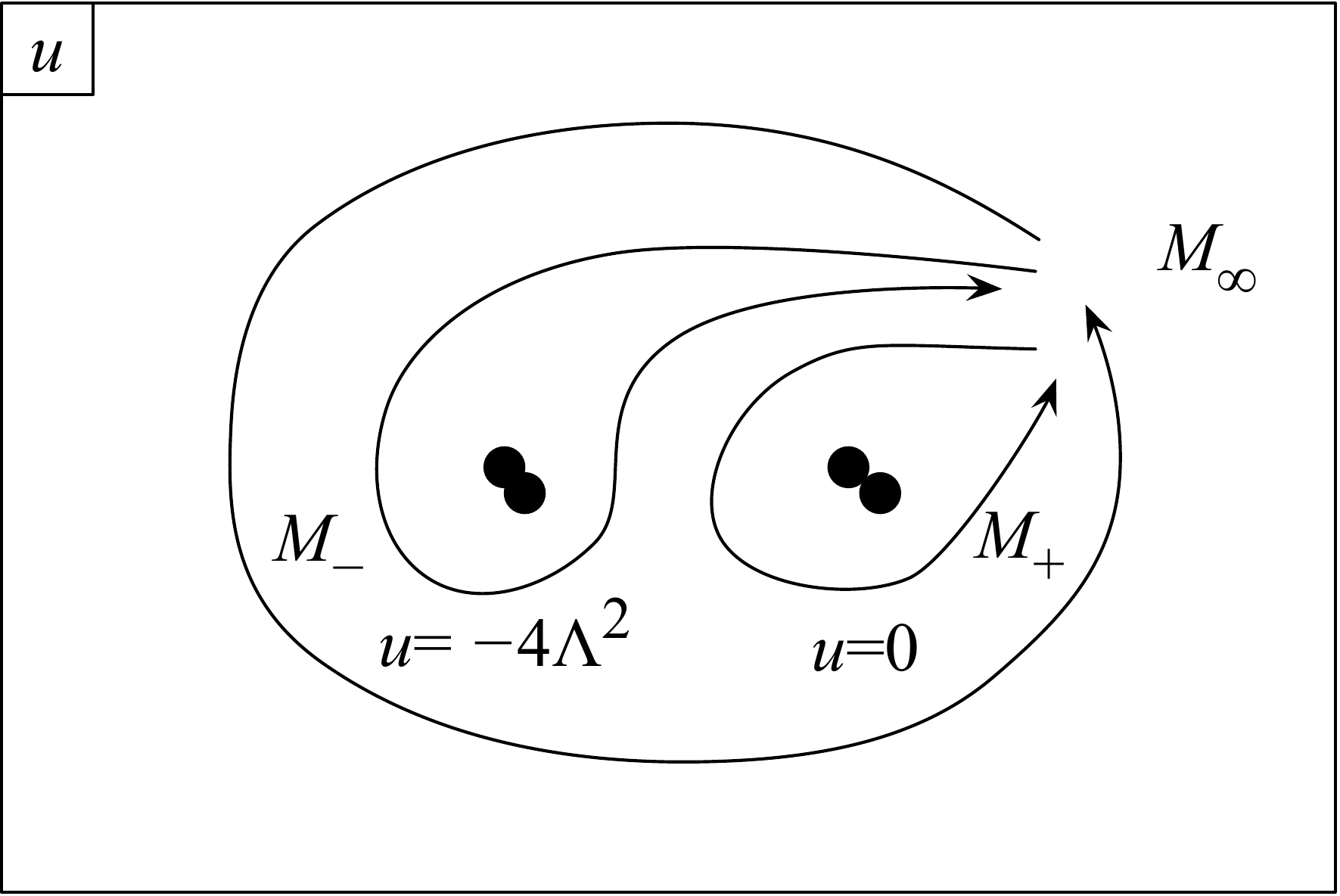}
\]
\caption{The $u$-plane for massless $N_f=2$.\label{fig:u_plane_nf_2_massless}}
\end{figure}

This structure can be checked starting from the curve \eqref{nf_2_curve} by studying its discriminant, which is left as an exercise to the reader. Here we study the massless case $\mu_1=\mu_2=0$ in detail. 
The \SeibergWitten\ curve for the massless case is simply \begin{equation}
x^2-2\Lambda(z+\frac1z)x-u=0.
\end{equation} Then we see that four branch points of $x(z)$ meet in pairs when $u=0$ or $u=-4\Lambda^2$ as depicted in Fig.~\ref{fig:collision_nf_2}. 
Explicitly, when $u=0$ they meet at $z=\pm i$ and when $u=-4\Lambda^2$ they meet at $z=\pm1$.
There are no other singularities on the $u$-plane, so we see that when $\mu_1=\mu_2=0$ the $u$-plane has the structure shown in Fig.~\ref{fig:u_plane_nf_2_massless}.

At each of $u=0$, $u=-4\Lambda^2$, two pairs of branch points of $x(z)$ collide. This means that each of $u=0$, $u=-4\Lambda^2$ should be considered as two singularities on the $u$-plane.
This situation was shown in Fig.~\ref{fig:u_plane_nf_2_massless} by putting almost overlapping two blobs at $u=0,$ $-4\Lambda^2$.
 In total there are four singularities, matching what we found above for $\mu_{1,2}\gg \Lambda$. 
Let us denote the monodromies around various closed paths as shown in Fig.~\ref{fig:u_plane_nf_2_massless}.

The monodromy $M_\infty$ at infinity can be found from the explicit form of $a$, $a_D$ found in \eqref{nf_2_a}, \eqref{nf_2_ad} to be \begin{equation}
M_\infty = \begin{pmatrix}
-1 & 2 \\
0 & -1
\end{pmatrix}.
\end{equation} The monodromy $M_+$ around $u=0$ can be found by following the motion of the branch points when we make a slow change along the path $u=\epsilon e^{i\theta}$ for a very small $\epsilon$ from $\theta=0$ to $\theta=2\pi$.
\begin{figure}[h]
\[
\inc{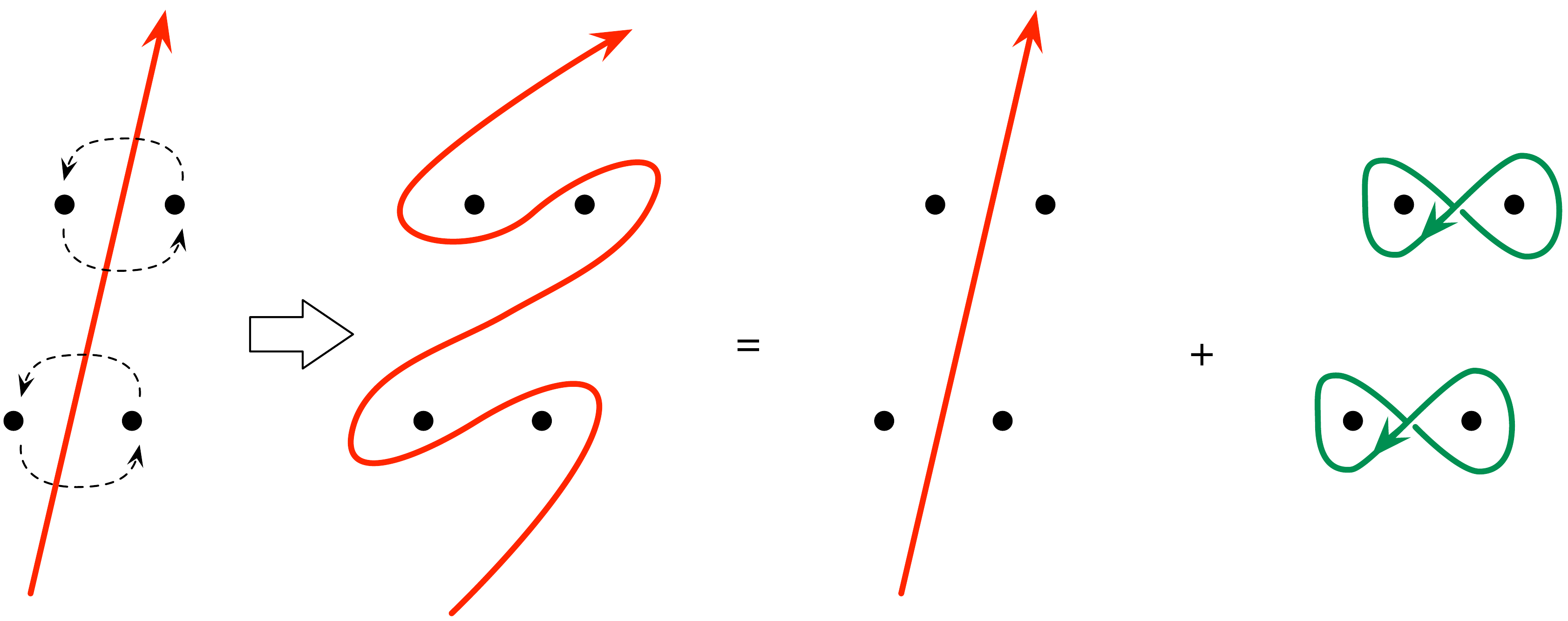}
\]
\caption{Monodromy action on cycles for $N_f=2$.\label{fig:nf_2_mon}}
\end{figure}
The pair of branch points exchanges positions as shown in Fig.~\ref{fig:nf_2_mon}. We see that the $B$ cycle remains the same, while $A$ is sent to $A-2B$, thus generating \begin{equation}
M_+ = \begin{pmatrix}
1 & 0 \\
-2 & 1
\end{pmatrix} = ST^2 S^{-1}.\label{nf_2_m+}
\end{equation} We see that \begin{equation}
M_\infty = M_+ M_-
\end{equation} with \begin{equation}
M_- = TM_+ T^{-1}.
\end{equation}

\begin{figure}[h]
\[
\inc{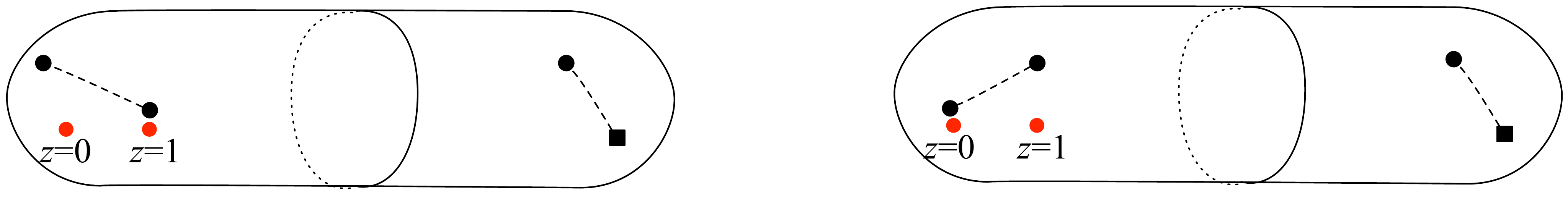}
\]
\caption{The curve of $N_f=2$ theory degenerates when $u=0$ or $u=-4\Lambda^2$, the second description.\label{fig:nf_2_second}}
\end{figure}

Before proceeding, it is instructive to use another description of the curve to find the same $u$-plane structure. 
The curve was given in \eqref{nf_2_curve_pre2}. When massless, this just becomes \begin{equation}
\frac{x^2}{z} + 4\Lambda^2 z = x^2- u. 
\end{equation} 
The branch points collide when $u=0$ or $u=-4\Lambda^2$ as before, but it looks rather different on the \Gaiotto\ curve, as shown in Fig.~\ref{fig:nf_2_second}. 

Note that $\lambda$ diverges at $z=0$ and $z=1$ independent of $u$.
One branch point of $x(z)$ on the \Gaiotto\ curve  moves as $u$ changes,
and this point hits either $z=0$ or $z=1$ at $u=0$ or $u=-4\Lambda^2$ respectively. 
It is left to the reader to recover the monodromies $M_\pm$ from this latter view point. 

\subsection{$N_f=2$: the discrete R-symmetry}

Let us now study the discrete R-symmetry. We assign the charges under continuous  R-symmetry to be given by
 \begin{equation}
\begin{array}{r|ccc}
R=0 & & A\\
1& \lambda & & \lambda \\
2 & & \Phi
\end{array}, \qquad
\begin{array}{r|ccc}
R=-1 & & \psi_I\\
0& q_I & 
\end{array}.
\end{equation}
The rotation \begin{equation}
\lambda\to e^{i\varphi} \lambda,\qquad
\psi_{I}\to e^{-i\varphi} \psi_{I}  \label{nf_2_shiftx}
\end{equation}  is anomalous, but can be compensated by \begin{equation}
\theta_{UV}\to \theta_{UV}+4\varphi.\label{nf_2_shift}
\end{equation} Equivalently, the dynamical scale $\Lambda$ transforms as \begin{equation}
\Lambda^2 \to e^{4i\varphi} \Lambda^2.
\end{equation}
Therefore $\varphi=\pi/2$ is a genuine symmetry, which does \begin{equation}
\theta_{UV}\to \theta_{UV}+2\pi,\quad
\Phi\to -\Phi,\quad
u'\to u'
\end{equation} 
where $u'=\vev{\tr\Phi^2/2}$. The reason why we put a prime to the symbol $u$ here will be explained shortly.
Unfortunately this does not tell us much about the structure on the $u'$-plane, as it acts trivially on it.

We can perform a slightly subtler operation. Consider the action on the hypermultiplets given by \begin{align}
(q_{I=1},q_{2},q_{3},q_{4}) & \mapsto (-q_{I=1},q_{2},q_{3},q_{4}), \\
(\psi_{I=1},\psi_{2},\psi_{3},\psi_{4}) & \mapsto (-\psi_{I=1},\psi_{2},\psi_{3},\psi_{4})  .\label{flavor_parity}
\end{align}
So far we always said that the flavor symmetry is $\SO(2N_f)=\SO(4)$. This operation is a flavor parity action \begin{equation}
\diag(-1,+1,+1,+1)\in \OO(4) \supset \SO(4).
\end{equation}
Recall that in an $\SU(2)$ $k$-instanton background, the number of zero-modes of $\psi_{I=1}$ is just $k$.
Then the operation \eqref{flavor_parity} multiplies the path integral measure by $(-1)^k$. 
This means that the parity part of the classical flavor symmetry $\mathrm{O}(4)$ is anomalous.
That said, as we have a term $e^{i\theta k}$ in the integrand of the path integral, we can compensate it  by the shift $\theta\to \theta+\pi$. 

Then, we can combine phase rotations \eqref{nf_2_shiftx}, \eqref{nf_2_shift} with $\varphi=\pi/4$ and the flavor parity \eqref{flavor_parity} to have a genuine unbroken symmetry. Summarizing, this is a combination of two actions: the first one is \begin{equation}
\theta\to \theta+\pi,\quad \Phi\to i\Phi, \quad u'\to -u'
\end{equation} and the second one is \begin{equation}
\theta+\pi \to \theta+2\pi, \quad q_{I=1}\to -q_{I=1}, \quad \psi_{I=1}\to -\psi_{I=1}.
\end{equation}
In total this is a $\bZ_4$ symmetry acting on the $u$-plane by $\bZ_2$. 

At the first sight this looks contradictory with the structure of the $u$-plane found in Fig.~\ref{fig:u_plane_nf_2_massless}: the two singularities are at $u=0$ and $u=-4\Lambda^2$.  The way out is to set \begin{equation}
u=u' -2\Lambda^2.\label{ushift}
\end{equation} This illustrates a subtlety which is often there in the non-perturbative analysis of field theories. 
Naively, $u$ is defined to be $\vev{\tr\Phi^2/2}$. But a composite operator needs to be defined with care, by carefully performing the regularization and the renormalization. As there are almost no divergence between two chiral operators in a supersymmetric theory, it is relatively safe to do this for chiral composite operators, although one still needs to take care of the point splitting between two gauge-dependent chiral operators,  which is known as a source of Konishi's anomaly \cite{Konishi:1983hf}, for example.
At least perturbatively, we can take the holomorphic scheme and that uniquely fixes the regularization and the renormalization of chiral composite operators to all orders in perturbation theory. There still is, however, non-perturbative ambiguity in the definition of the scheme. 
In our present case, $u$ and $\Lambda^2$ both have mass dimension two and has charge $2$ under the continuous broken R-symmetry, therefore they tend to mix.  When we guessed the curve in Sec.~\ref{sec:su2nfpre}, we did not take the discrete unbroken R-symmetry into account, thus there was a discrepancy between the $u$ appearing in the curve and the $u'$ which was constructed by definition to transform nicely under the discrete R-symmetry. 

We learned that the low energy behavior at $u=0$ and $u=-4\Lambda^2$, or equivalently at ${u}'=\pm 2\Lambda^2$  is related by the discrete R-symmetry combined with the flavor parity. Let us study them in more detail. 
We know that the monodromy at $u=0$ is given by \eqref{nf_2_m+}. 
Let us say $a_D\sim cu$ close to $u=0$, where $c$ is a constant.
Applying the $S$ transformation once, we see that the running of the dual coupling is \begin{equation}
\tau_D(E) \simeq + \frac{2}{2\pi i} \log \LambdaRG
\end{equation} where  $E\sim cu$ sets the energy scale.
 Compare this with the running of the dual coupling \eqref{dual_run} at the monopole point of the pure $\SU(2)$ theory. The factor $2$ in the numerator comes from the lower-left entry of $M_+$ in \eqref{nf_2_m+}, or more physically from the fact that two pairs of the branch points simultaneously collide as shown in Fig.~\ref{fig:collision_nf_2}. In general, when a $\U(1)$ gauge theory is coupled to several hypermultiplets with charges given by $q_i$, the running is given by \begin{equation}
\tau \simeq + \frac{\sum_i q_i^2}{2\pi i} \log \LambdaRG \label{general_u1_running}
\end{equation} Then we can conclude uniquely that there are two hypermultiplets with charge $1$. 
This can be seen from the higher-dimensional perspective: there are disk-shaped membranes as in Fig.~\ref{fig:monopole_as_membrane} for each pair of colliding branch points. They become massless when the branch points do collide, thus providing two charged hypermultiplets. 

\subsection{$N_f=2$: the moduli space}

\begin{figure}[h]
\[
\inc{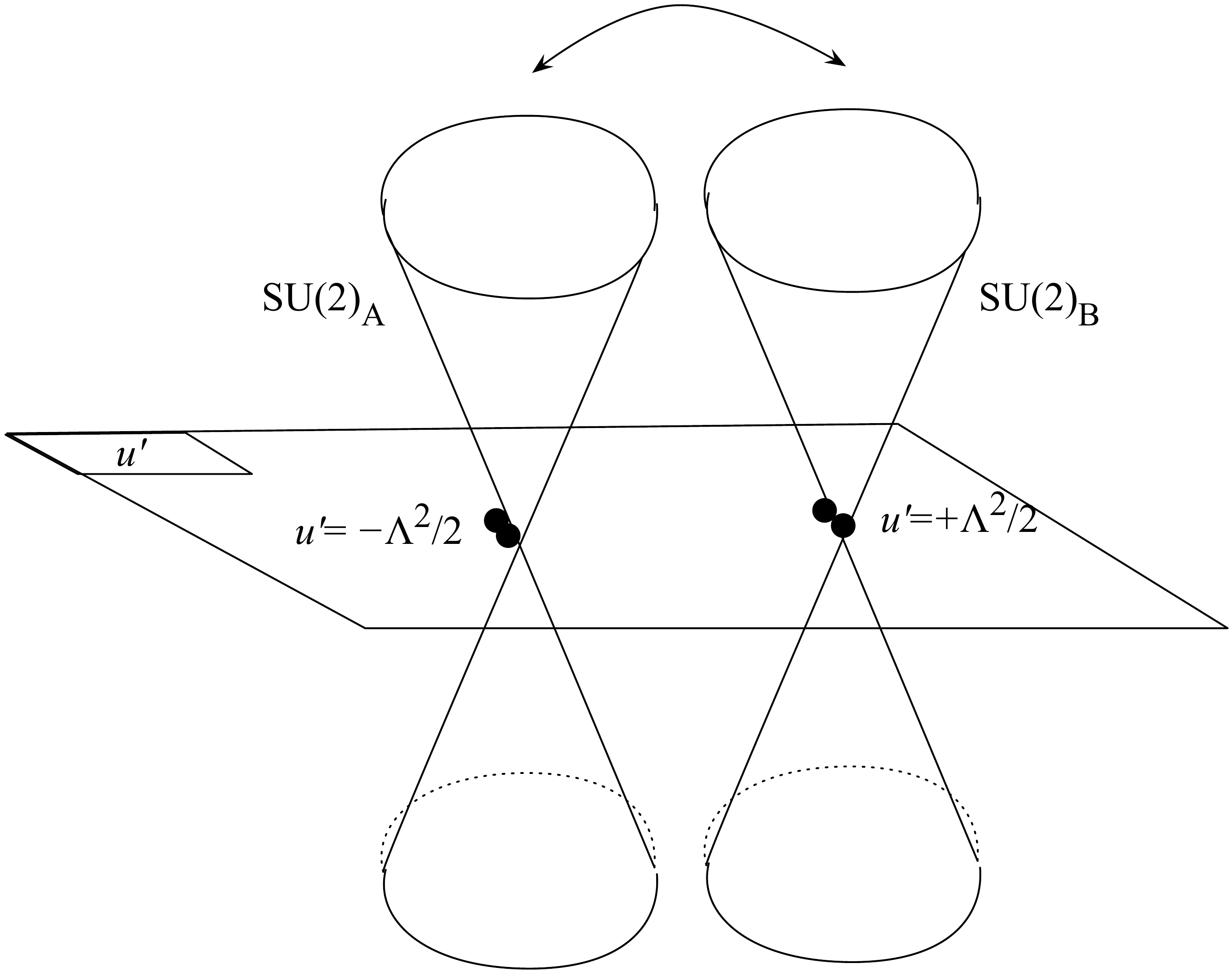}
\]
\caption{Quantum moduli space of the $N_f=2$ theory.\label{fig:nf_2_quantum_moduli}}
\end{figure}

We studied in Sec.~\ref{sec:hkq} that $\U(1)$ gauge theory with two charge-1 hypermultiplets has a Higgs branch of the form $\bC^2/\bZ_2$.  Together with the $u$-plane describing the Coulomb branch, we can visualize the totality of the supersymmetric vacuum moduli space as shown in Fig.~\ref{fig:nf_2_quantum_moduli}.
Note that two singular points on the $u$-plane where Higgs branches meet are exchanged by the discrete R-symmetry and the flavor parity. 

\begin{figure}[h]
\[
\inc{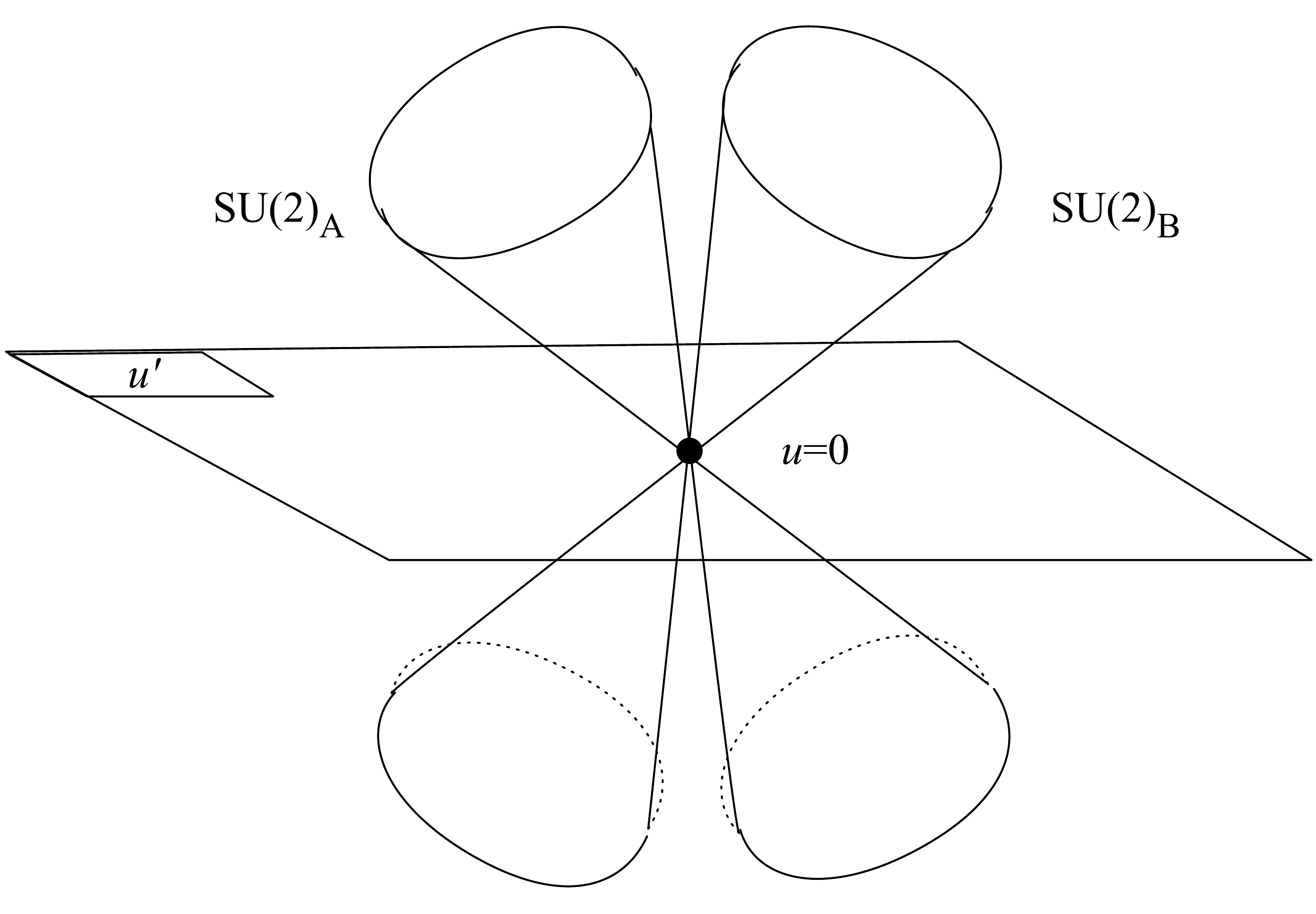}
\]
\caption{Classical moduli space of the $N_f=2$ theory.\label{fig:nf_2_classical_moduli}}
\end{figure}

Compare this with the classical moduli space of $\SU(2)$ theory with $N_f=2$ flavors. 
The Coulomb branch is still described by $u=\tr \Phi^2/2$. When $\Phi=0$, we can go to the Higgs branch;
we studied this system in Sec.~\ref{sec:hkq} too, where we saw that it is given by $\bC^2/\bZ_2\wedge \bC^2/\bZ_2$.
We can visualize them as in Fig.~\ref{fig:nf_2_classical_moduli}.

In Sec.~\ref{sec:hypergeneral}, we argued that the local metric of the Higgs branch cannot be corrected by the gauge dynamics. 
We see here that the quantum dynamics can still split the point where two copies of $\bC^2/\bZ_2$ meet the $u$-plane; the argument in that section is not applicable at the points where the metric is singular. 

Recall that there is a flavor symmetry $\SO(4)\simeq \SU(2)_A\times \SU(2)_B$, 
so that $\SU(2)_{A,B}$ acts separately on the two copies of $\bC^2/\bZ_2$. 
Then, after non-perturbative correction, $\SU(2)_A$ acts on the hypermultiplets at $u'=\Lambda^2/2$ 
and $\SU(2)_B$ at $u'=-\Lambda^2/2$. This is consistent with the action of the flavor parity exchanging $u'=\pm \Lambda^2/2$, recall \eqref{flavor_parity}.

We learned in Sec.~\ref{sec:monopole} that the monopole in this type of  theories transforms as the spinor representation of the $\SO(2N_f)$ flavor symmetry.
Here the spinor of $\SO(4)$ is the fundamental doublet of $\SU(2)_A$ or $\SU(2)_B$, and they are indeed interchanged by the flavor parity.  This is consistent with what we have found so far.

\begin{figure}[h]
\[
\inc{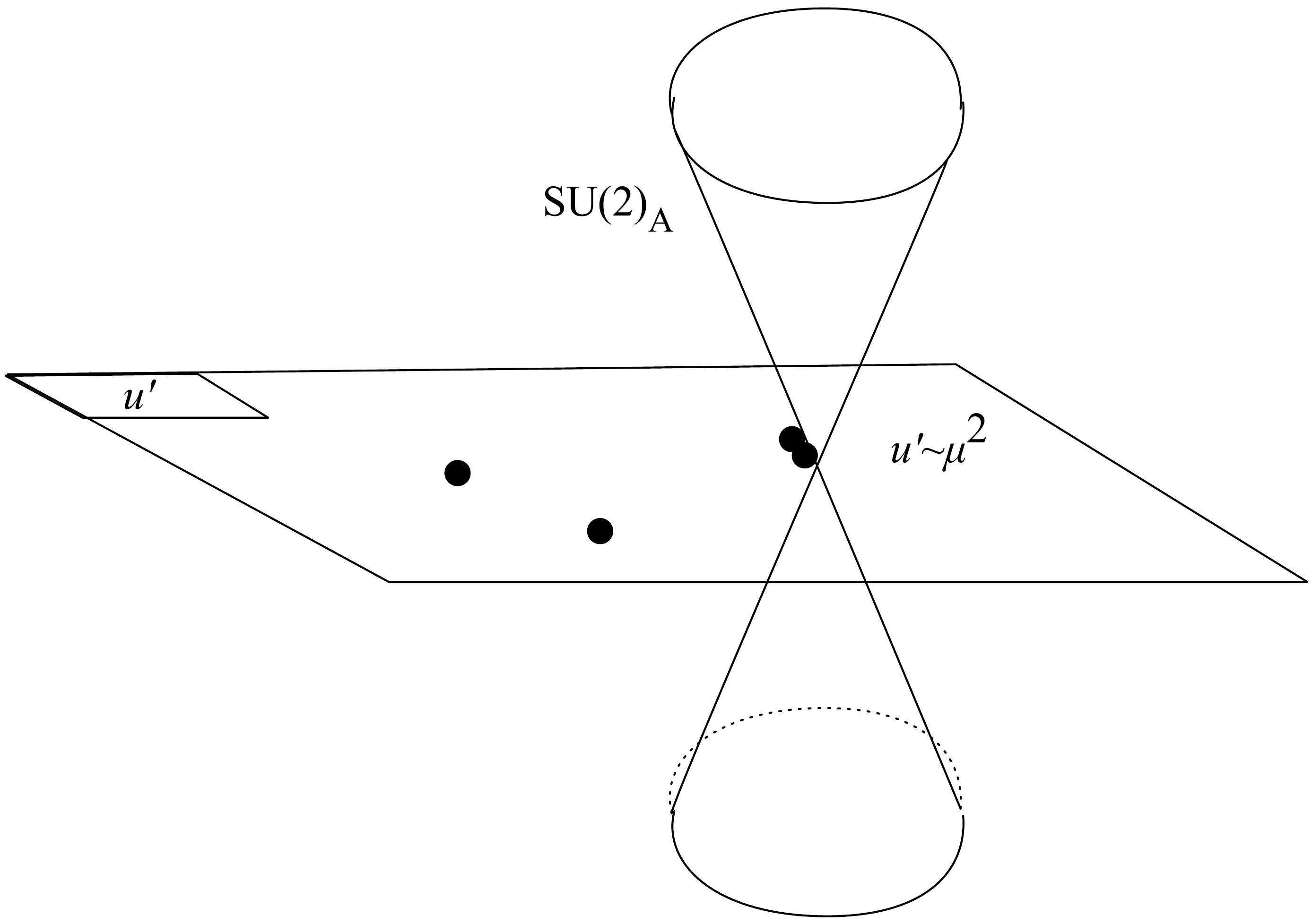}
\]
\caption{The moduli space of $N_f=2$ theory when $\mu_1=\mu_2$.\label{fig:nf_2_moduli}}
\end{figure}
Before closing this section, let us discuss what happens when we turn on a small but nonzero $\mu=\mu_1=\mu_2$.
This breaks the $\SO(4)=\SU(2)_A\times \SU(2)_B$ flavor symmetry to $\SU(2)_A$, say. 
Correspondingly, we can check that the two singularities sitting at the same point $u=-\Lambda^2$ splits into two,
by directly performing the analysis of the discriminant of the curve. 
We are still left with one point on the $u$-plane where two singularities still collide, and the local monodromy around it is unchanged from $M_+$. There, we have a Higgs branch of the form $\bC^2/\bZ_2$. The resulting structure is shown in Fig.~\ref{fig:nf_2_moduli}.
When $\mu$ is continuously made large, eventually the situation is better described as a special case of Fig.~\ref{fig:nf_2_u_plane_generic} with $\mu=\mu_1=\mu_2$. Namely, the gauge coupling at the scale $\mu$ is still very weak, and the classical Lagrangian analysis is valid. The superpotential is \begin{equation}
(Q_i^1, Q_i^2) \begin{pmatrix}
\mu+a & 0 \\
0 & -a+\mu
\end{pmatrix} \begin{pmatrix}
\tilde Q^i_1, \tilde Q^i_2
\end{pmatrix}\label{supernf}
\end{equation} and therefore when $a=\mu$, the components $(Q_i^2,\tilde Q^i_2)$ for $i=1,2$ remain massless. The gauge group is broken from $\SU(2)$ to $\U(1)$, and we have two charge-1 hypermultiplets, producing a Higgs branch of the form $\bC^2/\bZ_2$.

\subsection{$N_f=3$}\label{sec:nf3}
\begin{figure}[h]
\[
\inc{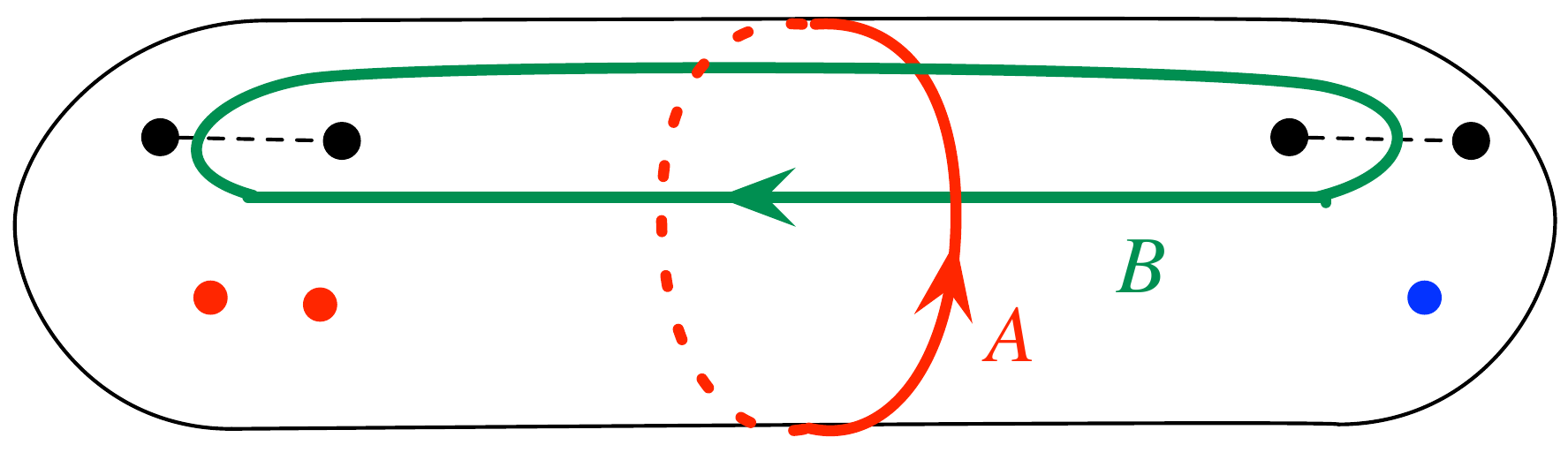}
\]
\caption{The curve of $N_f=3$ theory.\label{fig:nf_3_curve}}
\end{figure}

The curve for $\SU(2)$ theory with $N_f=3$ flavors was guessed in \eqref{nf_3_curve_pre}: \begin{equation}
\Sigma:\qquad \frac{(x-\tilde\mu_1)(x-\tilde\mu_2)}z + {2\Lambda(x-\tilde\mu_3)}z = x^2-u.\label{nf3curve!}
\end{equation}  We see that $\lambda$ diverges at $z=0,1,\infty$ independent of $u$, 
and there are four branch points which move as $u$ changes, see Fig.~\ref{fig:nf_3_curve}. The reason we put tildes above the mass parameters will become clear soon.

Let us check the behavior when $|u|\gg |\mu_i|^2, |\Lambda|^2$. 
Two branch points are at $z\sim O(1)$ and another branch point is at $\sqrt{u}/\Lambda$. 
 We now put the $A$-cycle around $|z|=c$, where $1 \ll c \ll \sqrt{u}/\Lambda $.
Then we see that the  integral is given as before by  \begin{equation}
a= \frac{1}{2\pi i}\oint_A x \frac{dz}z \simeq \sqrt{u}.
\end{equation}
The $B$-cycle integral can be approximated by \begin{equation}
a_D\sim \frac{2}{2\pi i}\int^1_{\sqrt{u}/\Lambda} a\frac{dz}z \simeq -\frac{2}{2\pi i}a \log \frac{a}{\Lambda}.
\end{equation} From this we find \begin{equation}
\tau(a)=\frac{\partial a_D}{\partial a}=-\frac{2}{2\pi i}a \log \frac{a}{\Lambda},
\end{equation} thus reproducing the field-theoretical one-loop computation \eqref{general_1_loop}.

\begin{figure}[h]
\[
\inc{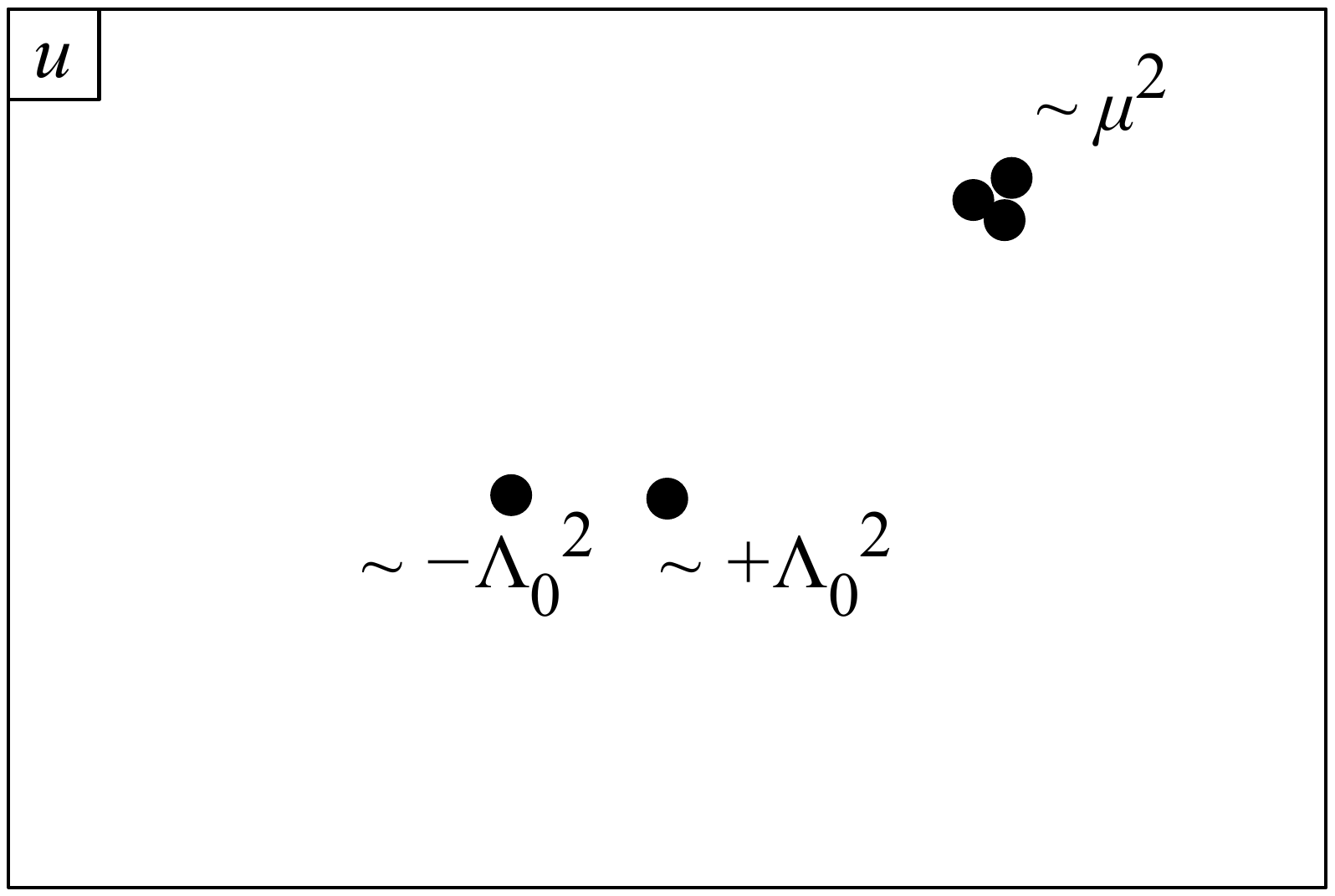}
\]
\caption{The $u$-plane of $N_f=3$ theory, for equal masses\label{fig:nf_3_u_plane_generic}}
\end{figure}
When $\mu_1=\mu_2=\mu_3=\mu$  and $|\mu| \gg |\Lambda|$, the coupling at the scale $\mu$ is still small,
and the classical analysis using the superpotential \eqref{supernf} is almost valid. We expect that around $a\simeq \mu$, 
i.e.~when $u\simeq \mu^2$,   the gauge group $\SU(2)$  is broken to $\U(1)$ with three charge-1 hypermultiplets. 
This point on the $u$-plane counts as three singularities, since when $\mu_{1,2,3}$ are 
slightly different, they should be at three slightly different points $u\simeq \mu^2_i$.
  When $|u|\ll |\mu|^2$, the theory can be effectively described by pure $\SU(2)$ gauge theory, which have the monopole point and the dyon point. In total we expect five singularities on the $u$-plane, see Fig.~\ref{fig:nf_3_u_plane_generic}.

We would like to study the massless case, $\mu=0$.  Here, we cannot just set $\tilde \mu_i=0$ in the curve \eqref{nf3curve!}.\footnote{The author thanks Kazuya Yonekura for  pointing this out.} We already saw that, when $N_f=2$, the vev $u$ can mix with the one-instanton factor $\Lambda^2$. Here, with $N_f=3$, the one-instanton factor is $\Lambda$ and it can mix with any neutral chiral dimension-1 operator. The curve makes only $\U(3)$ flavor symmetry manifest. The mass parameter corresponding to the $\U(1)$ flavor symmetry is neutral, chiral, and of dimension 1. Therefore there can be a mixing of the form \begin{equation}
\tilde\mu_i = \mu_i + c\Lambda
\end{equation} where $c$ is a constant.  Here, we fix the untilded mass parameter $\mu_i$ to transform linearly under the Weyl symmetry $\mu_i \to \pm \mu_i$ of the $\SO(2N_f)=\SO(6)$ flavor symmetry. 

To determine $c$, we set \begin{equation}
(\tilde \mu_1,\tilde \mu_2,\tilde \mu_3)=(-\mu + c\Lambda, \mu + c\Lambda, \mu + c\Lambda)
\end{equation} and study the singularities in the $u$-plane. This is just the $\SO(6)$ flavor Weyl transform of the $\SU(3)$ flavor symmetric choice of masses, therefore three out of five singularities on the $u$-plane should still collide as in \eqref{fig:nf_3_u_plane_generic}.  By an explicit computation, one finds that this happens only when $c=1$.

\begin{figure}[h]
\[
\inc{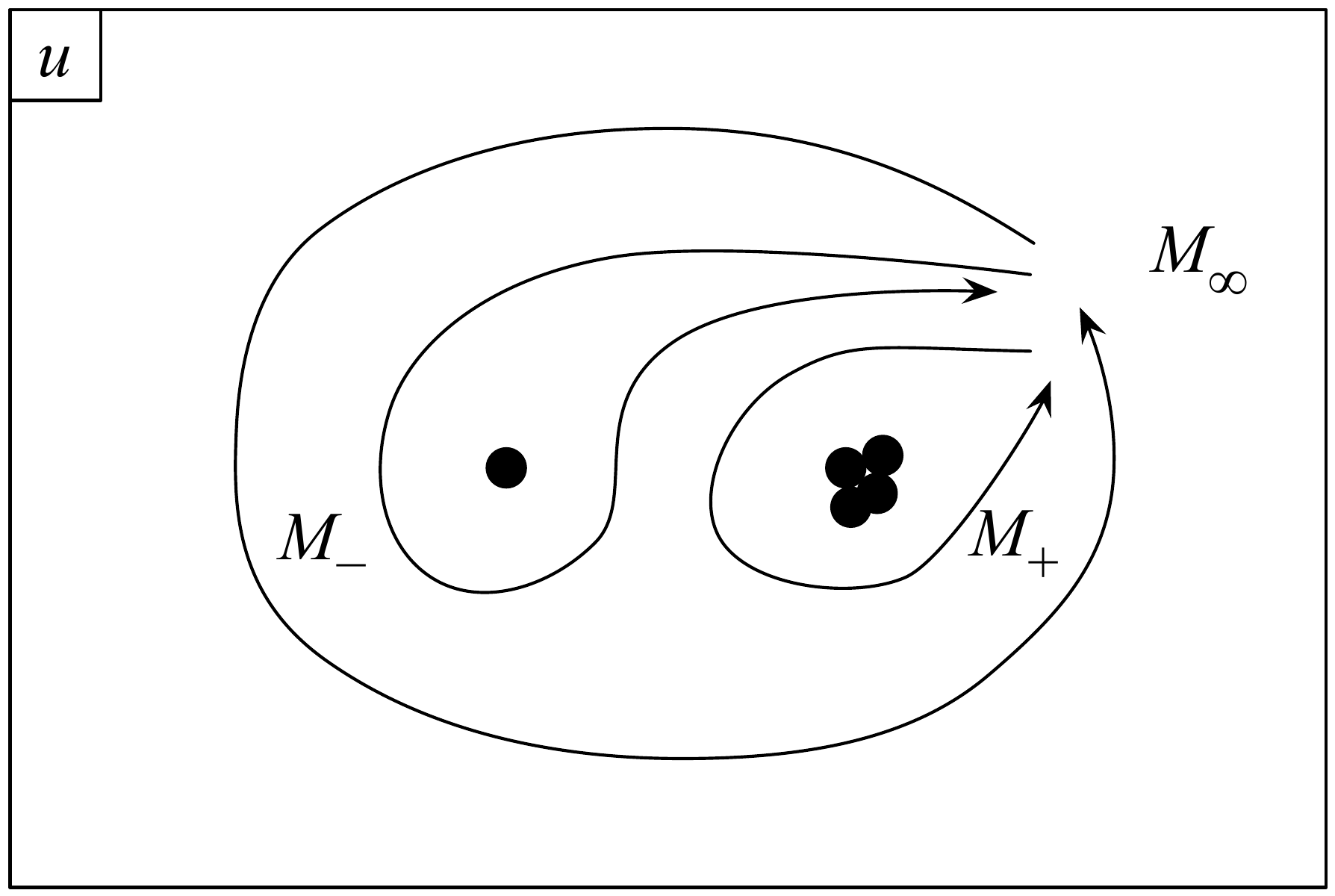}
\]
\caption{The $u$-plane of massless $N_f=3$ theory\label{fig:nf_3_uplane_massless}}
\end{figure}

Finally we can  set $\mu=0$. The curve is now \begin{equation}
\frac{(x-\Lambda)^2}z+ 2\Lambda (x-\Lambda)z=x^2-u. 
\end{equation} 
There is an $u$-independent branch point of $x(z)$ at $z=1$.
Two other branch points move with $z$, and are at  the solutions of \begin{equation}
\Lambda^2 z^2 -\Lambda^2 z + u -\Lambda^2=0.
\end{equation} 
The branch points collide when $u=\Lambda^2$ or $u=(5/4)\Lambda^2$:
\begin{itemize}
\item When $u=(5/4)\Lambda^2$, two $u$-dependent branch points meet at $z=1/2$.
The local physics there is just $\U(1)$ gauge theory with one charged hypermultiplet.
\item When $u=\Lambda^2$, one branch point moves to $z=0$ and the other branch point collides with the $u$-independent branch point at $z=1$.
From the general analysis we know that there are five singularities on the $u$-plane, therefore this point should count as four colliding singularities, see Fig.~\ref{fig:nf_3_uplane_massless}.
\end{itemize} 
The monodromy at infinity is \begin{equation}
M_\infty = \begin{pmatrix}
-1 & 1 \\
0 & -1
\end{pmatrix}.
\end{equation} Denoting the monodromies around $u=\Lambda^2$, $u=(5/4)\Lambda^2$ by $M_+$ and $M_-$, we have \begin{equation}
M_\infty=M_+ M_- ,\qquad M_+=\begin{pmatrix}
1 & 0 \\
-4 & 1
\end{pmatrix}, \quad
M_- = \begin{pmatrix}
-1 & 1 \\
-4 & 3
\end{pmatrix} \sim \begin{pmatrix}
1 & -1 \\ 
0 & 1
\end{pmatrix}.
\end{equation} 

By going to the S-dual frame at $u=\Lambda^2$, we find that the running of the dual coupling is \begin{equation}
\tau(\LambdaRG) = +\frac{4}{2\pi i} \log \LambdaRG
\end{equation} where the scale is set by $\LambdaRG\sim (u-\Lambda^2)$.
 Comparing with \eqref{general_u1_running}, the low energy physics can be guessed to be a $\U(1)$ gauge theory, coupled  either (i) to just one charge-2 hypermultiplets or (ii) to four charge-1 hypermultiplets. 

Recall that the classical theory has a Higgs branch.
The choice (i) does not have a Higgs branch at $u=\Lambda^2$. It does not have one at $u=(5/4)\Lambda^2$ either. The Higgs branch should be preserved by the quantum correction, and thus this choice is ruled out. 

The choice (ii) does have a Higgs branch at $u=\Lambda^2$. We have four charge-1 hypermultiplets coupled to the $\U(1)$ gauge multiplet.
Then the complex dimension of the Higgs branch is $2\cdot 4-2\cdot 1=6$. 
This is acted on by the $\SU(4)$ flavor symmetry rotating four hypermultiplets.

Classically, we have three hypermultiplets in the doublet of $\SU(2)$. Then the complex dimension of the Higgs branch is $4\cdot 3-2\cdot 3=6$. This agrees with the computation above. 
Recall that three hypermultiplets in the doublet of $\SU(2)$ count as six half-hypermultiplets of $\SU(2)$ doublet, with $\SO(6)$ flavor symmetry.  As $\SO(6)\simeq \SU(4)$, we see that the symmetry of the Higgs branch also agrees. 
We should recall that the monopole in this theory transforms as the spinor representation of the $\SO(2N_f)$ flavor symmetry.
In our case the spinor of $\SO(6)$ is the fundamental four-dimensional representation of $\SU(4)$. This is also consistent with our choice that at $u=\Lambda^2$ there are four charged hypermultiplets electrically coupled to the dual $\U(1)$.  By a more detailed analysis we can check that the Higgs branches agree as hyperk\"ahler manifolds.

\section{$\SU(2)$ theory with 4 flavors and Gaiotto's duality}\label{sec:gaiotto}
In this section we start with the analysis of $\SU(2)$ gauge theory with $N_f=4$ flavors. We will see that it can naturally generalized to the analysis of a whole zoo of theories with the gauge group of the form $\SU(2)^n$.  The discussions basically follow the first half of the seminal paper \cite{Gaiotto:2009we}.

\subsection{The curve as $\lambda^2=\phi_2(z)$}
Let us consider $\SU(2)$ gauge theory with four doublet hypermultiplets with masses $\mu_{1,2,3,4}$. 
In the very high energy region, the one-loop running is given by \eqref{general_1_loop}, which is just \begin{equation}
\tau(a) = 2\tau_{UV}.
\end{equation} From this we learn a distinguishing feature of the $N_f=4$ theory compared to the theories with less flavors: it has a dimensionless parameter $\tau_{UV}$. When $N_f<4$,  the bare coupling $\tau_{UV}$ was combined with the scale $\Lambda_{UV}$ to form the dynamical scale $\Lambda$, which just set the overall scale of the theory. 

\begin{figure}[h]
\[
\inc{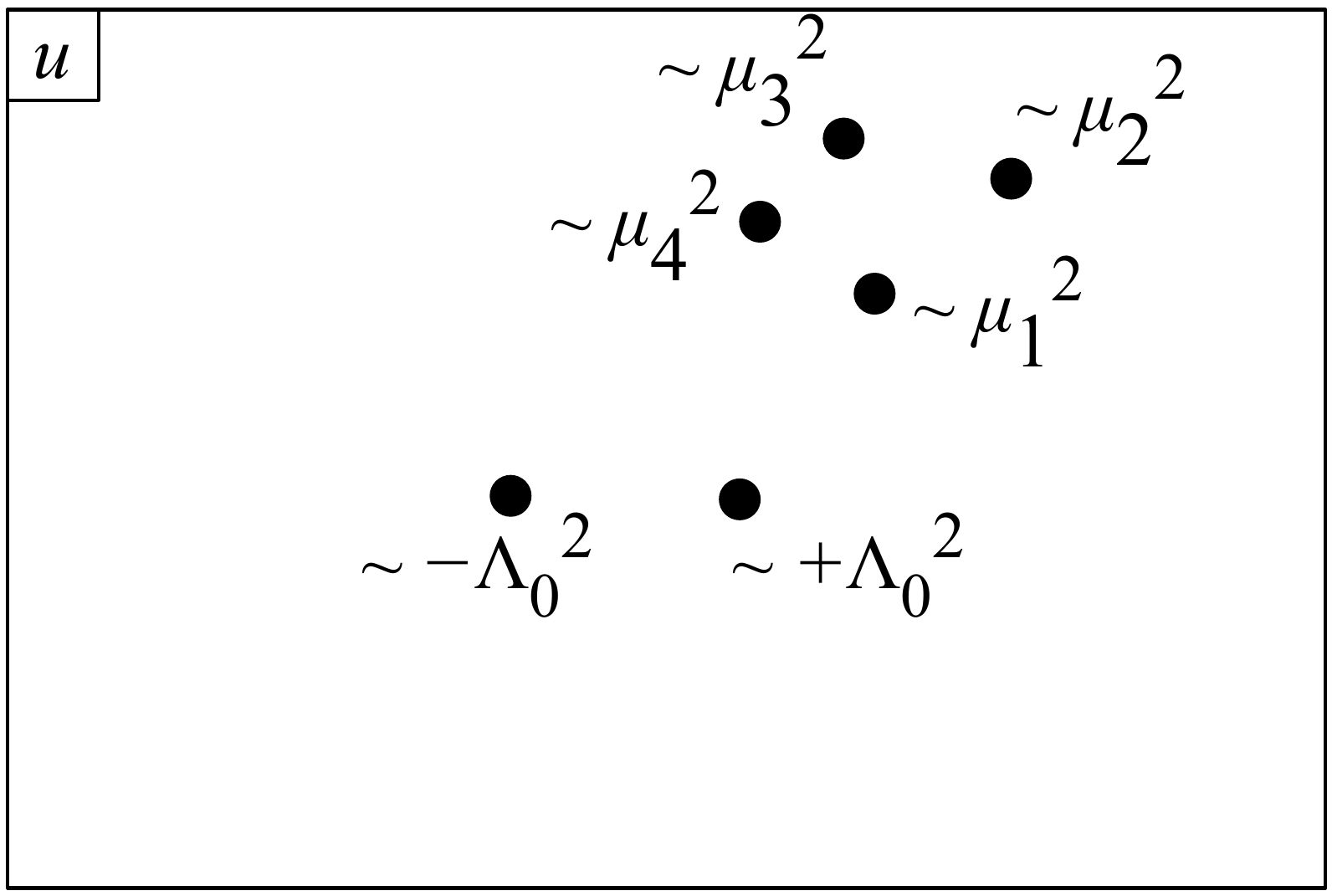}
\]
\caption{The $u$-plane of  $N_f=4$ theory\label{fig:nf_4_uplane_generic}}
\end{figure}

Now suppose the gauge coupling is small at the ultraviolet. Equivalently, suppose $\tau_{UV}$ has a large positive imaginary part. Further suppose $\mu_{1,2,3,4}$ are all of the same order, $\sim\mu$. Then the coupling at the energy scale $\sim \mu$ is small, and the semiclassical analysis is OK. We see that when $a\sim \pm \mu_i$, or equivalently when $u\sim \mu_i{}^2$, the low-energy limit is described  by $\U(1)$ gauge theory with one charged hypermultiplet.  Far below this scale, the theory is effectively the pure $\SU(2)$ theory, which has the monopole point and the dyon point.  Then the $u$-plane schematically has the structure shown in Fig.~\ref{fig:nf_4_uplane_generic}.

\begin{figure}[h]
\[
\inc{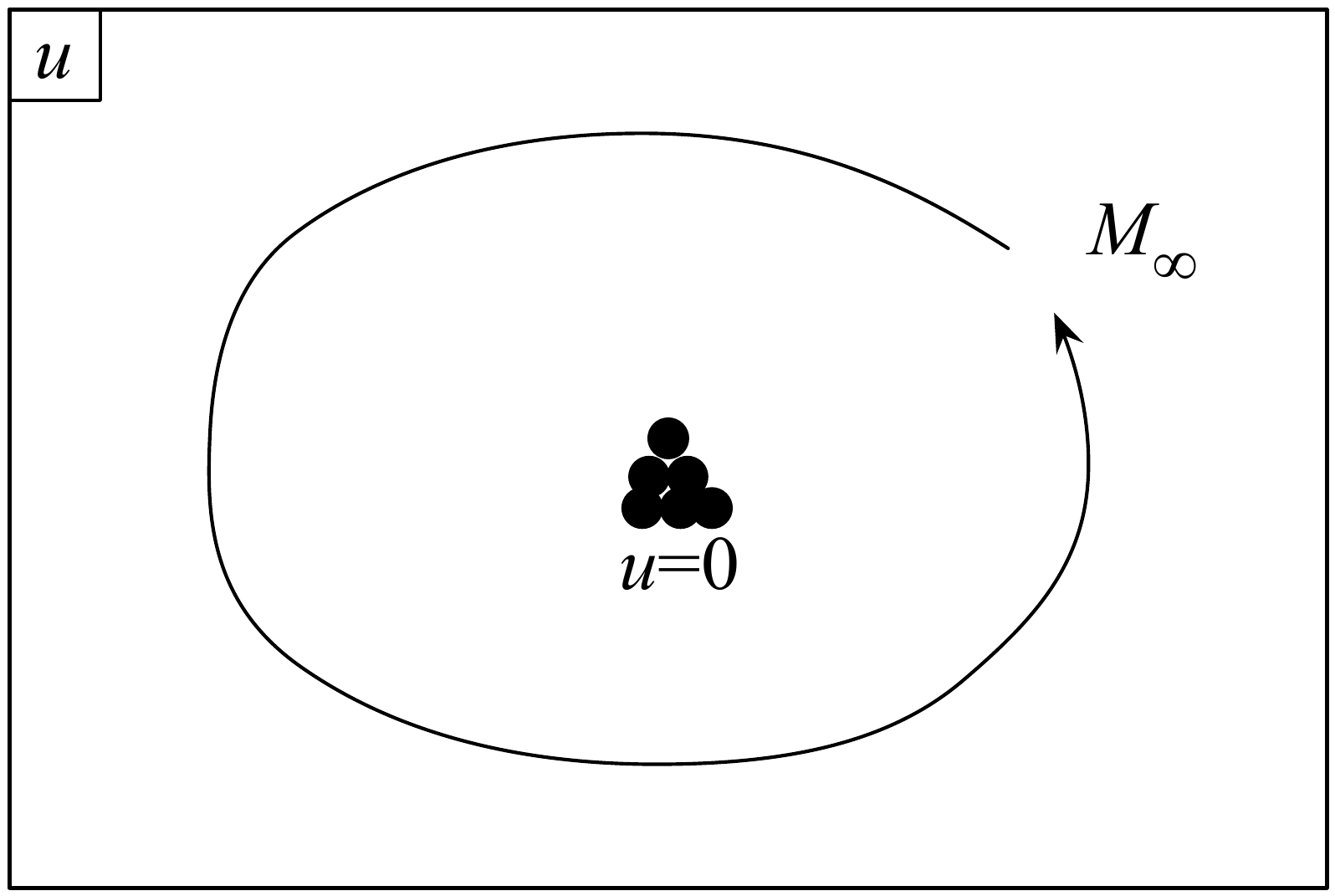}
\]
\caption{The $u$-plane of  massless $N_f=4$ theory\label{fig:nf_4_uplane_massless}}
\end{figure}

When $\mu_{1,2,3,4}=0$, we can consider the R-symmetry  with the charge assignment\begin{equation}
\begin{array}{r|ccc}
R=0 & & A\\
1& \lambda & & \lambda \\
2 & & \Phi
\end{array}, \qquad
\begin{array}{r|ccc}
R=-1 & & \psi_I\\
0& q_I & 
\end{array}.
\end{equation} This $\U(1)_R$ symmetry is not anomalous.  
The only way to put six singularities in a $\U(1)_R$ invariant way is to put all of them at the origin, where six singularities in the generic case collide, see Fig.~\ref{fig:nf_4_uplane_massless}.
The coupling is given by $\tau_{UV}$ everywhere, \begin{equation}
a=\sqrt{u}, \qquad a_D=2\tau_{UV} a_D.
\end{equation} Therefore the monodromy $M_\infty$ at infinity is just \begin{equation}
M_\infty = \begin{pmatrix}
-1 & 0\\
0 & -1
\end{pmatrix}.
\end{equation} 
It looks relatively uninteresting. We will see however that there is a lot of interesting physics going on when we study the dependence on $\tau_{UV}$.

\begin{figure}[h]
\[
\inc{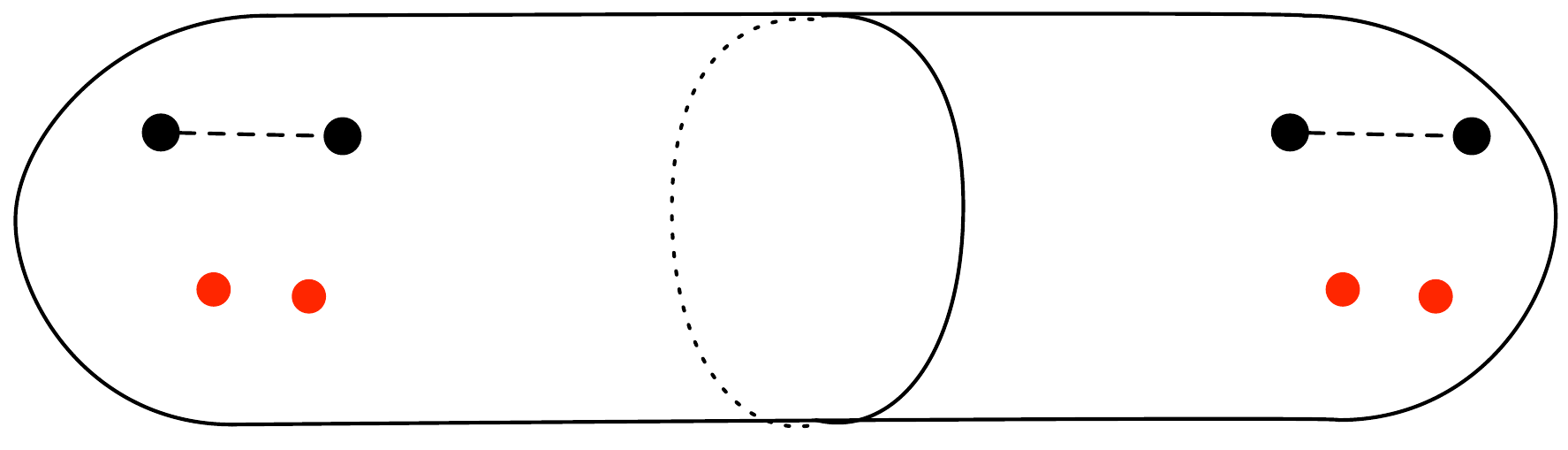}
\]
\caption{The curve of  %massless
 $N_f=4$ theory\label{fig:nf_4_curve}}
\end{figure}
The curve of the $N_f=4$ theory is given by \begin{equation}
\Sigma:\qquad  f\frac{(\tilde x-\tilde{\mu}_1)(\tilde x-\tilde{\mu}_2)}{\tilde z }
+ f'\cdot (\tilde x-{\tilde\mu}_3)(\tilde x-{\tilde\mu}_4)\tilde z =\tilde  x^2-u \label{naive_nf_4}
\end{equation}where $f$ and $f'$ are complex numbers,
 whose ratio will eventually be related to $\tau_{UV}$. The differential is $\tilde\lambda=\tilde x d\tilde z/\tilde z$.
 The reason for additional tildes will become clear later. 
 
The structure of the function $\tilde x(\tilde z)$ over the \Gaiotto\ curve $C$ which is a sphere with coordinate $z$ 
is shown in Fig.~\ref{fig:nf_4_curve}. 
The differential $\tilde \lambda$ always diverges at $\tilde z=0$, $\infty$, and at the two solutions $\tilde z=c_{1,2}(f)$ of $f/\tilde z+f'\tilde z=1$. 
These points do not move when $u$ is changed.
There are four additional branch points where $\tilde x(z)$ is finite, connected by dots in the figure to make branch cuts explicit. 

Let us rewrite the curve in a more illuminating way. 
We first rescale the coordinate $z$ to set $f'=1$.
We then collect terms with the same power of $\tilde x$: \begin{equation}
(1-\tilde z-\frac f{\tilde z})\tilde x^2 - \heartsuit \tilde x -\heartsuit' =0 
\end{equation} where $\heartsuit,\heartsuit'$ are some complicated expressions, which readers should fill in. We divide the whole expression by $(1-\tilde z-f/\tilde z)$, and find \begin{equation}
\tilde x^2 - \clubsuit \tilde x -\clubsuit' =0.
\end{equation} We note that $\clubsuit$ and $\clubsuit'$ have poles at the two solutions $c_{1,2}(f)$ of $1-\tilde z-f/\tilde z=0$. Here it is instructive to spell out $\clubsuit$, which is given by \begin{equation}
\clubsuit=-\frac{f\cdot (\tilde \mu_1+\tilde \mu_2)/\tilde z + (\tilde \mu_3+\tilde \mu_4)\tilde z}{1-\tilde z- f/\tilde z}.\label{clubsuit}
\end{equation}
Defining $x=\tilde x-\clubsuit/2$, we have \begin{equation}
x^2 - \diamondsuit=0
\end{equation} where $\diamondsuit$ now has double poles at $c_{1,2}(f)$ due to the completion of the square.
Instead of $\tilde \lambda =\tilde xd\tilde z/\tilde z$ we will use $\lambda = x d\tilde z/\tilde z$ henceforth. Note that \begin{equation}
\tilde\lambda-\lambda=\frac{\clubsuit}{2} \frac{d\tilde z}{\tilde z} =
-\frac12\frac{f\cdot (\tilde \mu_1+\tilde \mu_2)/\tilde z + (\tilde \mu_3+\tilde \mu_4)\tilde z}{1-\tilde z- f/\tilde z}
\frac{d\tilde z}{\tilde z}.
\end{equation}
This is independent of the Coulomb branch modulus $u$, and its residues are all linear combinations of $\tilde \mu_i$.  We encountered in \eqref{diff} a similar shift of $\lambda$ by a one-form which is independent of $u$ and whose residues are given by the mass terms only. Such shift only amounts to a re-definition of the flavor charge and the mass terms, and does not affect the physics, as discussed there. 
 
\begin{figure}[h]
\[
\inc{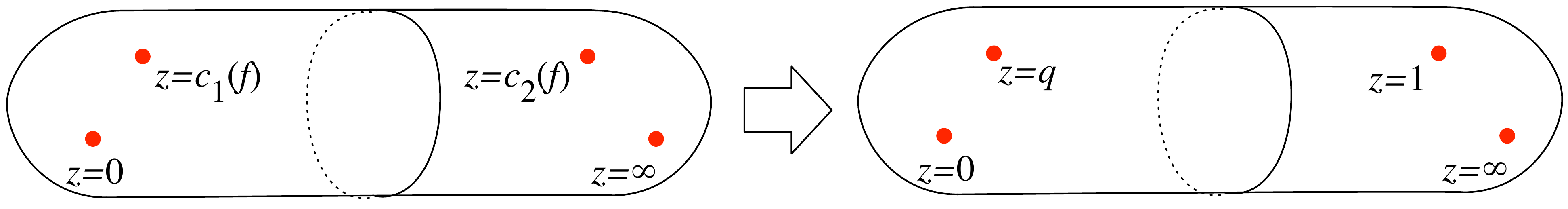}
\]
\caption{A step in the derivation of the curve in the Gaiotto form \label{fig:z_redef}}
\end{figure}
Now we define the coordinate $z=\tilde z/c_1(f)$ so that the double poles are at $z=q$ and $1$ for $|q|<1$, see Fig.~\ref{fig:z_redef}.
The final form of the curve is then: \begin{equation}
\lambda^2 -\phi_2(z)=0, \qquad \phi_2(z)=\frac{P(z)}{(z-1)^2(z-q)^2}\frac{dz^2}{z^2}
\end{equation} where $P(z)$ is a quartic polynomial, as can be seen by re-following the change of variables starting from \eqref{naive_nf_4}. The explicit expression of $P(z)$ in terms of $u$, $f$ and $\tilde \mu_i$ is not very important, however. 

The quadratic differential $\phi_2(z)$ has double poles at $z=0,q,1,\infty$. To see this for $z=\infty$, set $w=1/z$. Then
$dz^2/z^2=dw^2/w^2$. This has poles of order two when $w=0$, i.e.~when $z=\infty$.
We identify this dimensionless parameter $q$ as a function of the UV coupling $\tau_{UV}$. 

\subsection{Identification of parameters}\label{sec:ident}
\subsubsection{Coupling constant}
\begin{figure}[h]
\[
\inc{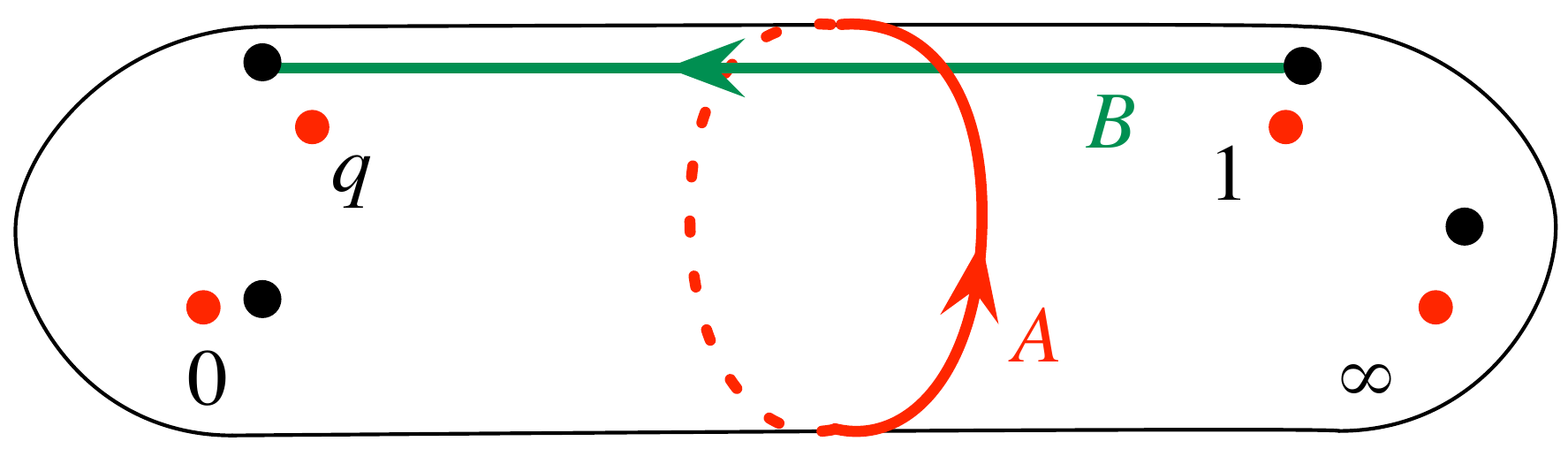}
\]
\caption{In $\SU(2)$ $N_f=4$, $q$ is related to the UV coupling\label{fig:q_vs_tau}}
\end{figure}
Let us first see concretely how $q$ and $\tau_{UV}$ are related.
This can be done by   computing $a$ and $a_D$ assuming $|q|\ll 1$.
As always, we put the $A$-cycle at $|z|=c$, where $|q|\ll c \ll 1$. Then we easily have \begin{equation}
a=\frac{1}{2\pi i}\oint_A x\frac{dz}z \sim \sqrt{u}.
\end{equation} The branch points of $\lambda$ are near $1$ and $q$ anyway, and therefore \begin{equation}
a_D=\frac{1}{2\pi i}\oint_B x\frac{dz}z \sim \frac{1}{2\pi i} 2 \int_1^q a \frac{dz}z =\frac{2a}{2\pi i}\log q.
\end{equation}
Then \begin{equation}
\tau_{\U(1)}=\frac{\partial a_D}{\partial a}\simeq \frac{2}{2\pi i}\log q,
\end{equation} or equivalently \begin{equation}
q\sim e^{2\pi i\tau_{UV}} 
\end{equation} in the limit of weak coupling; note our convention that $\tau_{\U(1)}\sim 2\tau_{UV}$.
This relation is often written as \begin{equation}
q_{C}= e^{2\pi i\tau_{UV,C}}\label{tauUV}
\end{equation} with  an equality.  This should be regarded as a nonperturbative definition of the renormalization and regularization scheme of $\tau_{UV}$. Here we added a subscript $C$ to both $q$ and $\tau_{UV}$, in order to emphasize that the coupling $q_C$ is given by the data on the \Gaiotto\ curve $C$. 

Another common nonperturbative definition of the UV coupling constant is to use the  low-energy $\U(1)$ coupling $\tau_{\U(1)}$ in the limit when the Coulomb vev is very large $|u| \gg |\tilde\mu_i|$: \begin{equation}
\tau_{UV,\Sigma} := \frac12\lim_{|u|\to \infty} \tau_{\U(1)}\label{tauIR}
\end{equation} 
This should isolate the $\SU(2)$ coupling whose running is stopped at a very large scale given by the Coulomb vev, and can be read off from the complex structure of the \SeibergWitten\ curve $\Sigma$.
That is why we used the subscript $\Sigma$ here. Let us also define 
\begin{equation}
q_\Sigma=e^{2\pi i \tau_{UV,\Sigma}}.
\end{equation}
This is also a perfectly good scheme, related to the one in \eqref{tauUV} via a finite renormalization.  

To explicitly determine the finite renormalization, we note that the \SeibergWitten\ curve $\Sigma$ in the $|u|\to\infty$ limit is just the torus which is a double-cover of $C$ branched at $z=0,1,q_C,\infty$.
Then $\tau_{\U(1)}$ in \eqref{tauIR} is given by the complex structure of this $\Sigma$. 
From a basic result in the theory of elliptic functions, we find \begin{equation}
q_C=\lambda(\tau_{\U(1)}) = \frac{\theta_2(q_{\Sigma}^2)^4}{\theta_3(q_{\Sigma}^2)^4}
=16q_{\Sigma} -128 q_{\Sigma}^2 + 704 q_{\Sigma}^{3} - 3072 q_{\Sigma}^4 +\cdots. 
\end{equation} This means that $\tau_{UV,C}$ and $\tau_{UV,\Sigma}$ are related by a constant shift of its imaginary part plus instanton corrections. 

For more extensive discussions on the non-perturbative finite renormalization, see e.g. Sec. 3.4 and Sec. 3.5 of ~\cite{JaewonThesis}.

\subsubsection{Mass parameters}
Next, let us study mass parameters. Recall that $\lambda$ has mass dimension 1, as its integral give the mass of BPS particles. 
This means that the five coefficients of the quartic polynomial $P(z)$ are of mass dimension two. 
We can identify these five coefficients with some combinations of five parameters $\mu_{i=1,2,3,4}$ and $u$.
The physical mass parameters are the residues at the poles of $\lambda$. 
Fixing $\mu_i$ fixes four linear combinations of the coefficients of $P(z)$.
The sole linear combination which does not change the coefficients of the double poles at $z=0,q,1,\infty$ can be identified with the parameter $u$, up to a certain coefficient.
Explicitly, we can write  \begin{equation}
\phi_2(z)=\frac{P_0(z)}{(z-q)^2(z-1)^2} \frac{dz^2}{z^2} + \frac{u}{(z-1)(z-q)}\frac{dz^2}z\label{++++}
\end{equation} where $P_0(z)$ is independent of $u$.

Let us now go back to the original curve and study the poles of $\lambda$. We can compute them  from \eqref{naive_nf_4} rather easily when the system is weakly coupled, $|q|\ll 1$. The residues are $\sim\tilde\mu_{1,2}$ at $z=0$
and $\sim\tilde\mu_{3,4}$ at $z=\infty$. 
When we went from $\tilde x$ to $x$, we subtracted $\clubsuit/2$ from $x$. 
We see that the residues are given by 
\begin{equation}
\begin{aligned}
\pm\frac{\mu_1-\mu_2}2 &\quad\text{at}\ z=0, & 
\pm\frac{\mu_3-\mu_4}2 &\quad\text{at}\ z=\infty, \\
\pm\frac{\mu_1+\mu_2}2 &\quad\text{at}\ z=q, & 
\pm\frac{\mu_3+\mu_4}2 &\quad\text{at}\ z=1
\end{aligned} \label{so8masses}
\end{equation}
where \begin{equation}
\mu_i =  \tilde \mu_i + O(q).
\end{equation}  The variables $\tilde \mu_i$ enter rather naturally  the \SeibergWitten\ curve we guessed in Sec.~\ref{sec:su2nfpre}, whereas the variables $\mu_i$ enter  the BPS mass formula. We see that they are related by a finite renormalization.

To understand the combinations in \eqref{so8masses} better, it is helpful to consider the $\cN{=}1$ superpotential.
With four doublet hypermultiplets, we have \begin{equation}
W=\sum_i( Q_i \Phi \tilde Q^i + \mu_i Q_i \tilde Q^i).
\end{equation} We combine $(Q_i,\tilde Q^i)$  for $i=1,2,3,4$ to $q_I$ with $I=1,\ldots,8$. Then the same term becomes \begin{equation}
W\propto q^a_I q^b_J \Phi_{ab} \delta^{IJ} + q^a_I q^b_J \epsilon_{ab} \mu^{IJ}
\end{equation} where $\mu^{IJ}$ is a constant matrix with $\SO(8)$ antisymmetric index: \begin{equation}
\mu^{IJ}=
\begin{pmatrix}
& -\mu_1 \\
\mu_1 
\end{pmatrix}
\oplus
\begin{pmatrix}
& -\mu_2 \\
\mu_2 
\end{pmatrix}
\oplus
\begin{pmatrix}
& -\mu_3 \\
\mu_3 
\end{pmatrix}
\oplus
\begin{pmatrix}
& -\mu_4 \\
\mu_4 
\end{pmatrix}\label{so8matrix}
\end{equation}

Under the decomposition \begin{equation}
\SO(8)\supset \SO(4)\times \SO(4) \simeq \SU(2)_A\times \SU(2)_B\times \SU(2)_C\times \SU(2)_D,
\end{equation} the entries of $\SO(8)$ antisymmetric matrix \eqref{so8matrix} decomposes to \begin{equation}
\begin{array}{c@{\quad}c@{\quad}c@{\quad}c}
\SU(2)_A & \SU(2)_B & \SU(2)_C & \SU(2)_D \\
\diag(\pm\frac{\mu_1-\mu_2}{2}) & \diag(\pm \frac{\mu_1+\mu_2}2) & \diag(\pm\frac{\mu_3+\mu_4}2) & \diag(\pm\frac{\mu_3-\mu_4}2)
\end{array}
\end{equation} which are exactly the residues we found in \eqref{so8masses} at $z=0,q,1$ and $=\infty$, respectively.
We can regard then that the singularity at $z=0$ carries the $\SU(2)_A$ symmetry,
and that the residue of $\lambda$ there is the mass parameter associated to this $\SU(2)_A$ symmetry;
similarly for $\SU(2)_B$ at $z=q$, $\SU(2)_C$ at $z=1$, and $\SU(2)_D$ at $z=\infty$.

We call these structures the punctures. From the 6d point of view, we consider a puncture at $z=0$ as a four-dimensional object extending along the Minkowski space $\bR^{3,1}$, which somehow carries an $\SU(2)$ flavor symmetry on it. We will see various other types of punctures below. To distinguish this one from them, we will call this a regular $\SU(2)$ puncture.

%Before getting further, let us pause for a moment and write the original form of the curve \eqref{naive_nf_4} in terms of the untilded mass terms $\mu_i$ and the coupling $q_C$ defined in \eqref{tauUV}. The result is \begin{equation}
%q_C\frac{(x-\mu_1)(x-\mu_2)}z +  (x-\mu_3)(x-\mu_4) z = (1+q_C) x^2 -u.
%\end{equation}

\subsection{Weak-coupling limit and trifundamentals}

\begin{figure}[h]
\[
\inc{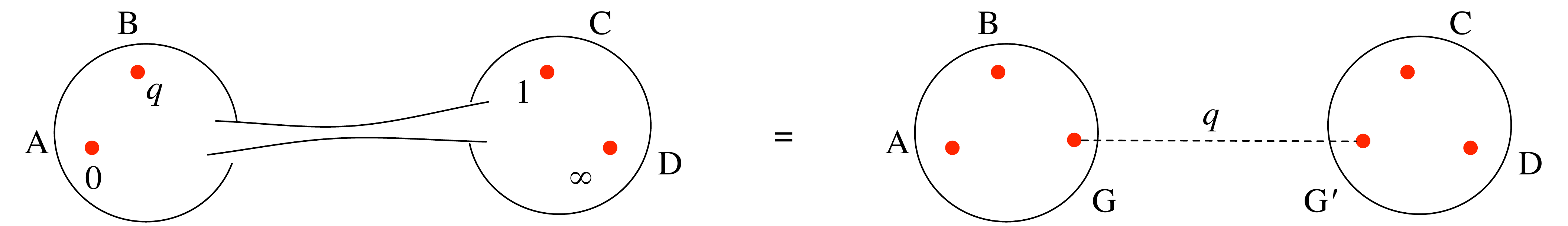}
\]
\caption{Weakly-coupled limit of $\SU(2)$ $N_f=4$\label{fig:nf_4_weaklimit}}
\end{figure}
Let us now take the limit $q\to 0$ to decouple the gauge $\SU(2)$, see Fig.~\ref{fig:nf_4_weaklimit}.
On the left hand side, we have a sphere parameterized by $z$, with four points at $z=0,q,1$ and $\infty$. 
On the right hand side, we have two spheres, parameterized by $z'$ and $z''$.
We put the points $A$, $B$, $G$ on the first sphere, at $z'=\infty$, $1$ and $0$, and then
the points $G'$, $C$, $D$ on the second sphere, at $z''=0$, $1$ and $\infty$. 
Then we glue the neighborhoods of $G$ and $G'$ by declaring \begin{equation}
z' z''=q.
\end{equation} Defining $z=z''$, we see that four points $A,B,C,D$ are exactly as in the first description. 
In this limit, around the tube connecting $G$ and $G'$, $\lambda \simeq \pm a dz/z$.
Then in the sphere containing $A$, $B$ and $G$, we have three singularities, with residues of $\lambda$ given by \begin{equation}
\pm\frac{\mu_1+\mu_2}2, \quad
\pm\frac{\mu_1-\mu_2}2, \quad
\pm a,
\end{equation} each corresponding to the symmetry $\SU(2)_A$, $\SU(2)_B$ and $\SU(2)_G$, respectively. 
Here $\SU(2)_G$ was originally the gauge symmetry. 

\begin{figure}[h]
\[
\inc{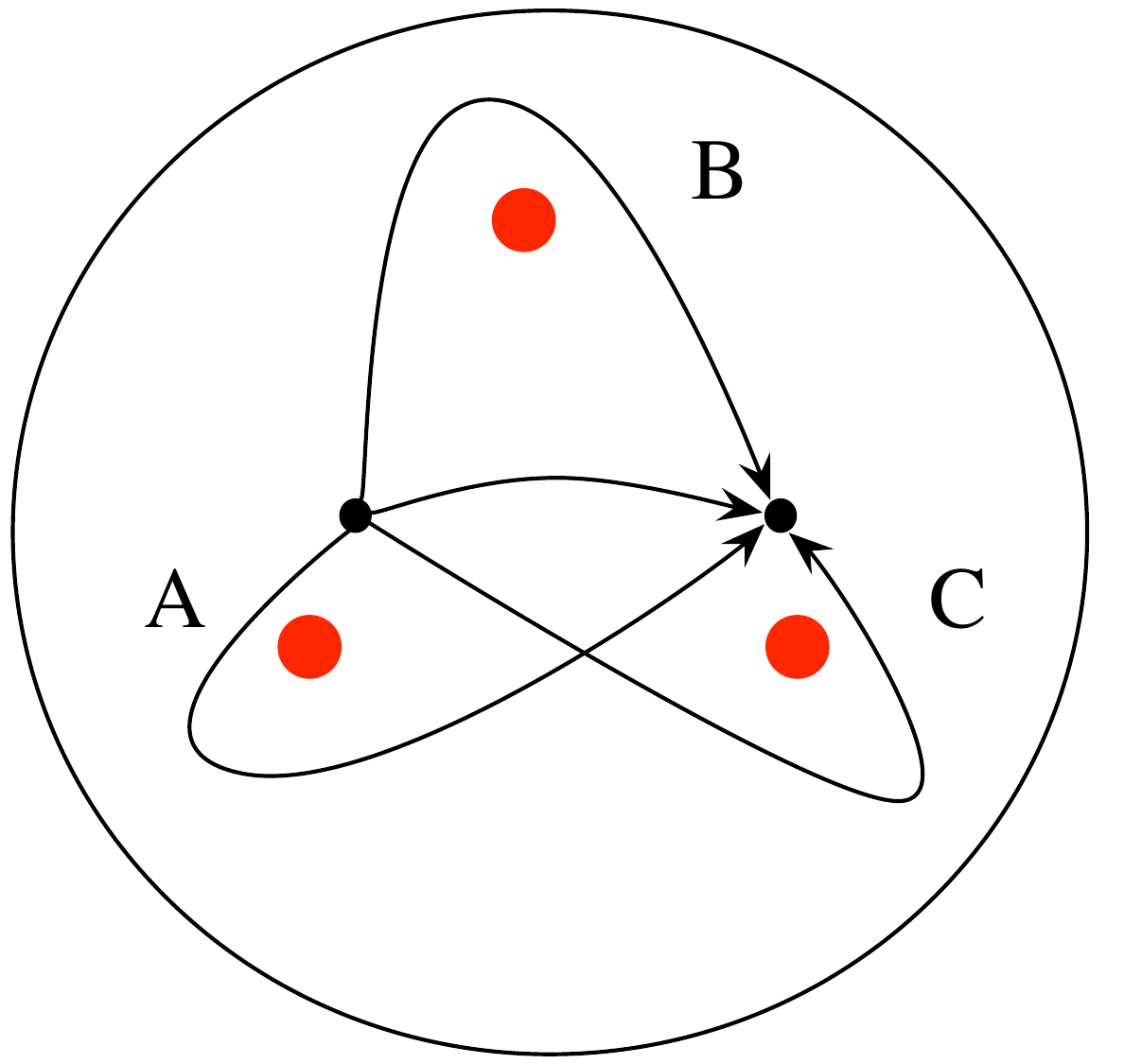}
\]
\caption{The \Gaiotto\ curve of the trifundamental, together with the BPS paths representing hypermultiplets \label{fig:curve_trifund}}
\end{figure}
We were talking about the $N_f=4$ theory. Then each of the sphere with three punctures should be associated to the $N_f=2$ hypermultiplet system, see Fig.~\ref{fig:curve_trifund}; note that this is not coupled  to any gauge group. 
Let us recall the structure of the hypermultiplets again. We start from two hypermultiplets $(Q_i^a, \tilde Q^i_a)$ in the doublet of $\SU(2)$, $i=1,2$ and $a=1,2$. We combine them to $q_I^a$, $a=1,2$ and $I=1,\ldots,4$, making $\SU(2)\times \SO(4)$ symmetry manifest. We then decompose the $\SO(4)$ index $I$ into the pair $(\alpha,u)$ where $\alpha=1,2$ and $u=1,2$: we have the trifundamental $q_{a\alpha u}$. The mass term for this hypermultiplet is \begin{equation}
\mu^{ab} q_{a\alpha u} q_{b\beta v} \epsilon^{\alpha\beta}\epsilon^{uv}
+\tilde\mu^{\alpha\beta} q_{a\alpha u} q_{b\beta v} \epsilon^{a b}\epsilon^{uv}
+\hat\mu^{uv} q_{a\alpha u} q_{b\beta v} \epsilon^{ab}\epsilon^{\alpha\beta},
\end{equation} where \begin{equation}
\mu^a_b=a \diag(1,-1),\quad
\tilde\mu^\alpha_\beta=\frac{\mu_1-\mu_2}2 \diag(1,-1),\quad
\hat\mu^\alpha_\beta=\frac{\mu_1+\mu_2}2 \diag(1,-1).
\end{equation} Then $(a,b)$ are the indices for $\SU(2)_G$, $(\alpha,\beta)$  for $\SU(2)_A$, 
and $(u,v)$ for $\SU(2)_B$. The physical masses of these fields are given by \begin{equation}
\pm a \pm \frac{\mu_1-\mu_2}2  \pm \frac{\mu_1+\mu_2}2  = 
%\left
\{
%\begin{array}{l}
\pm a \pm\mu_1,
\pm a \pm\mu_2
%\end{array}\right
\}.\label{trihyp}
\end{equation} which are the masses for the two doublets of $\SU(2)$ with bare masses $\mu_{1,2}$.

The curve of the system, shown in Fig.~\ref{fig:curve_trifund} is given by \begin{equation}
\lambda^2-\phi(z)=0,
\end{equation} where $\phi(z)$ has the asymptotic behavior \begin{equation}
\phi(z)\sim \frac{\tilde\mu^2}{z^2} dz^2, \qquad
\sim \frac{\hat \mu^2}{(z-1)^2} dz^2,\quad
\sim \frac{\mu^2}{w^2}dw^2 \label{phicondition_trifund}
\end{equation} at $z=0$, $z=1$, $z=\infty$ respectively. Here $w=1/z$ as always, and we set $\mu=a$, $\tilde \mu=(\mu_1-\mu_2)/2$ and $\hat \mu=(\mu_1+\mu_2)/2$.
Note that these asymptotic conditions uniquely fix the quadratic differential $\phi(z)$ to be \begin{equation}
\phi(z)=\frac{\mu^2 z^2 +(\hat \mu^2-\tilde\mu^2-\mu^2) z + \tilde \mu^2}{z^2(z-1)^2}dz^2\label{phi-trifund}
\end{equation}
As was discussed before, the BPS particles of this system can be found by solving the BPS equation \eqref{curveBPSeq} \begin{equation}
\Arg \frac{\lambda}{ds}=e^{i\theta} 
\end{equation} for a given $\theta$.  As $\phi(z)$ given above has two branch points only, the solution to the BPS equation should start from one and end on the other. A computer simulation shows that there are always four and only four such solutions, corresponding to the hypermultiplets with masses given in \eqref{trihyp}.

\subsection{Strong-coupling limit}\label{sec:su2-4-strong}

\begin{figure}[h]
\[
\inc{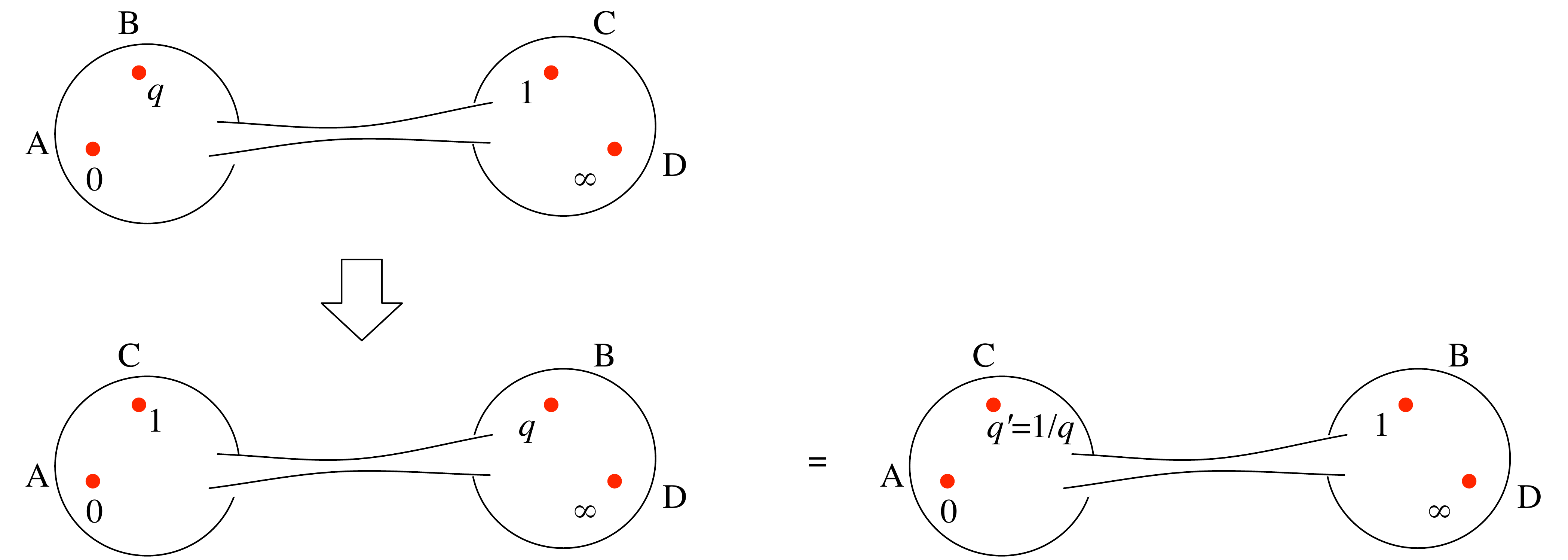}
\]
\caption{A strongly-coupled limit of $\SU(2)$, $N_f=4$\label{fig:nf_4_strong}}
\end{figure}
So far we mainly considered the weak coupling limit $q\to 0$.
Instead, consider sending $q\to \infty$, as shown in Fig.~\ref{fig:nf_4_strong}.
We immediately find that the strong coupling limit $q\to\infty$ is 
the weak coupling limit $q'=1/q\to 0$ of a similarly-looking $\SU(2)$ gauge theory with four flavors. 
Note however that the role of  the singularities $B$ and $C$ are exchanged. 

Originally, we had four flavors with masses \begin{equation}
\pm\mu_1,\quad \pm\mu_2, \quad \pm\mu_3, \quad \pm\mu_4.\label{so8original}
\end{equation}  The residues of $\lambda$ at the punctures were then \begin{equation}
\pm\mu_A,\quad \pm\mu_B, \quad \pm\mu_C, \quad \pm\mu_D
\end{equation} with \begin{equation}
\mu_A=\frac{\mu_1-\mu_2}2,\quad
\mu_B=\frac{\mu_1+\mu_2}2,\quad
\mu_C=\frac{\mu_3+\mu_4}2,\quad
\mu_D=\frac{\mu_3-\mu_4}2.
\end{equation} The original masses $\mu_i$ are \begin{equation}
\mu_1=\mu_A+\mu_B,\quad
\mu_2=-\mu_A+\mu_B,\quad
\mu_3=\mu_C+\mu_D,\quad
\mu_4=\mu_C-\mu_D.
\end{equation}
Now the singularities $B$ and $C$ are exchanged. Then, the masses $\mu'_{i}$ of the four hypermultiplets of the theory with the coupling $q'=1/q$ are instead  given by \begin{align}
\mu'_1&=\phantom{-}\mu_A+\mu_C =\phantom{-}\frac{\mu_1-\mu_2}2+\frac{\mu_3+\mu_4}2,\\
\mu'_2&=-\mu_A+\mu_C =-\frac{\mu_1-\mu_2}2+\frac{\mu_3+\mu_4}2,\\
\mu'_3&=\phantom{-}\mu_B+\mu_D=\phantom{-}\frac{\mu_1+\mu_2}2+\frac{\mu_3-\mu_4}2,\\
\mu'_4&=\phantom{-}\mu_B-\mu_D=\phantom{-}\frac{\mu_1+\mu_2}2-\frac{\mu_3-\mu_4}2.
\end{align}
 The original masses \eqref{so8original} can be thought of as the weights of the vector representation of $\SO(8)$.
The dual masses $\pm\mu'_i$ are then the weights of the spinor representation of $\SO(8)$.

\begin{figure}[h]
\[
\inc{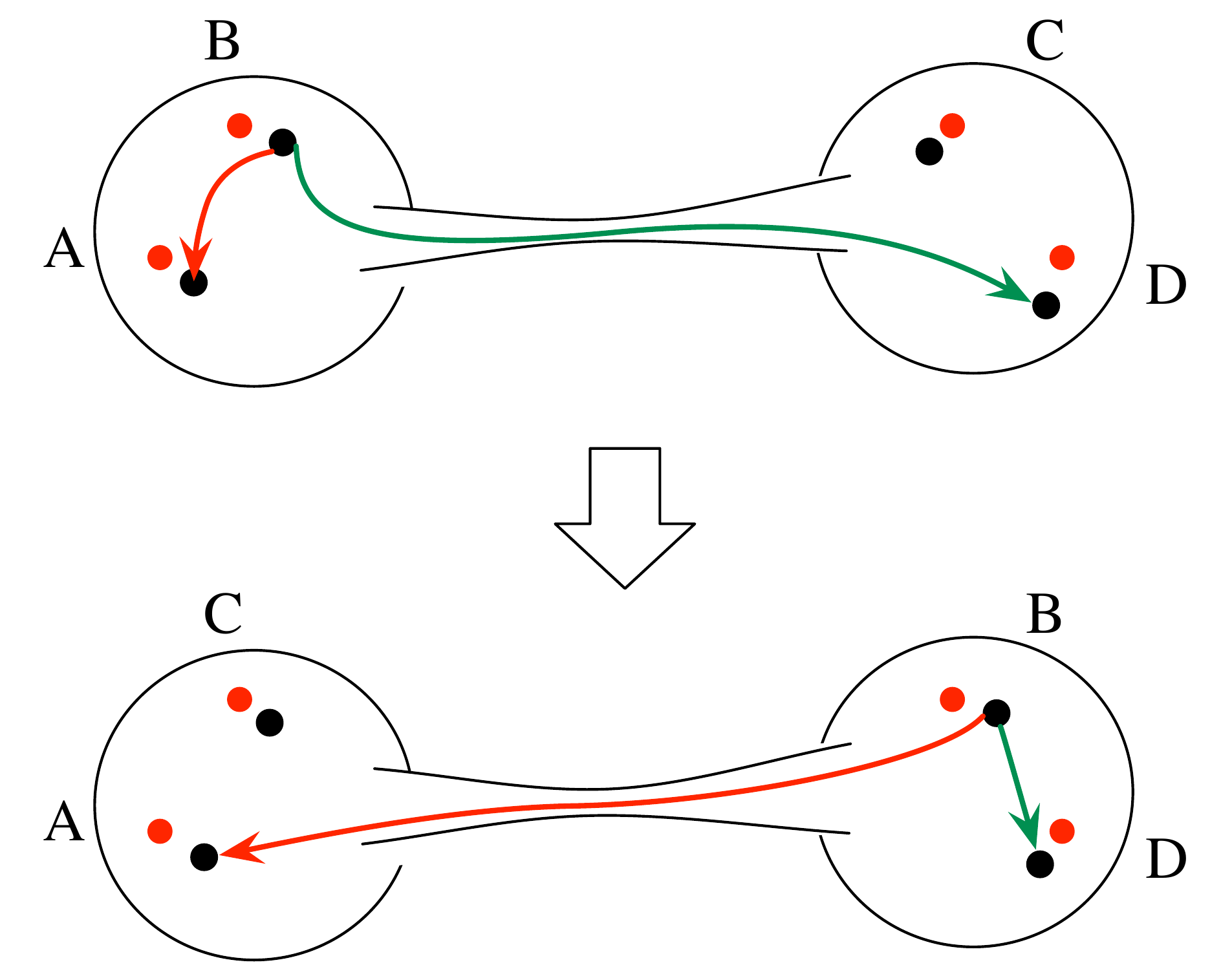}
\]
\caption{Monopoles and quarks are exchanged\label{fig:exchange_of_monopole_and_quark}}
\end{figure}
The dual quarks, therefore, transform in the spinor representation of the flavor $\SO(8)$ symmetry.
We can identify these dual quarks as the monopoles in the original description. This can be seen by slowly changing the value of $q$, following how various paths on the sphere change, see Fig.~\ref{fig:exchange_of_monopole_and_quark}.
Originally, the path connecting branch points close to the singularity $B$ and $D$ was a monopole.
Recall also that the semiclassical quantization of the monopole gave us a multiplet in the spinor representation of the flavor symmetry $\SO(2N_f)$ as we saw in Sec.~\ref{sec:monopole}.
In the limit $q\to \infty$, these monopoles become excitations whose  paths are totally contained in the sphere on the right. They are now the quark hypermultiplets in the trifundamental, as shown in Fig.~\ref{fig:curve_trifund}.

\begin{figure}[h]
\[
\inc{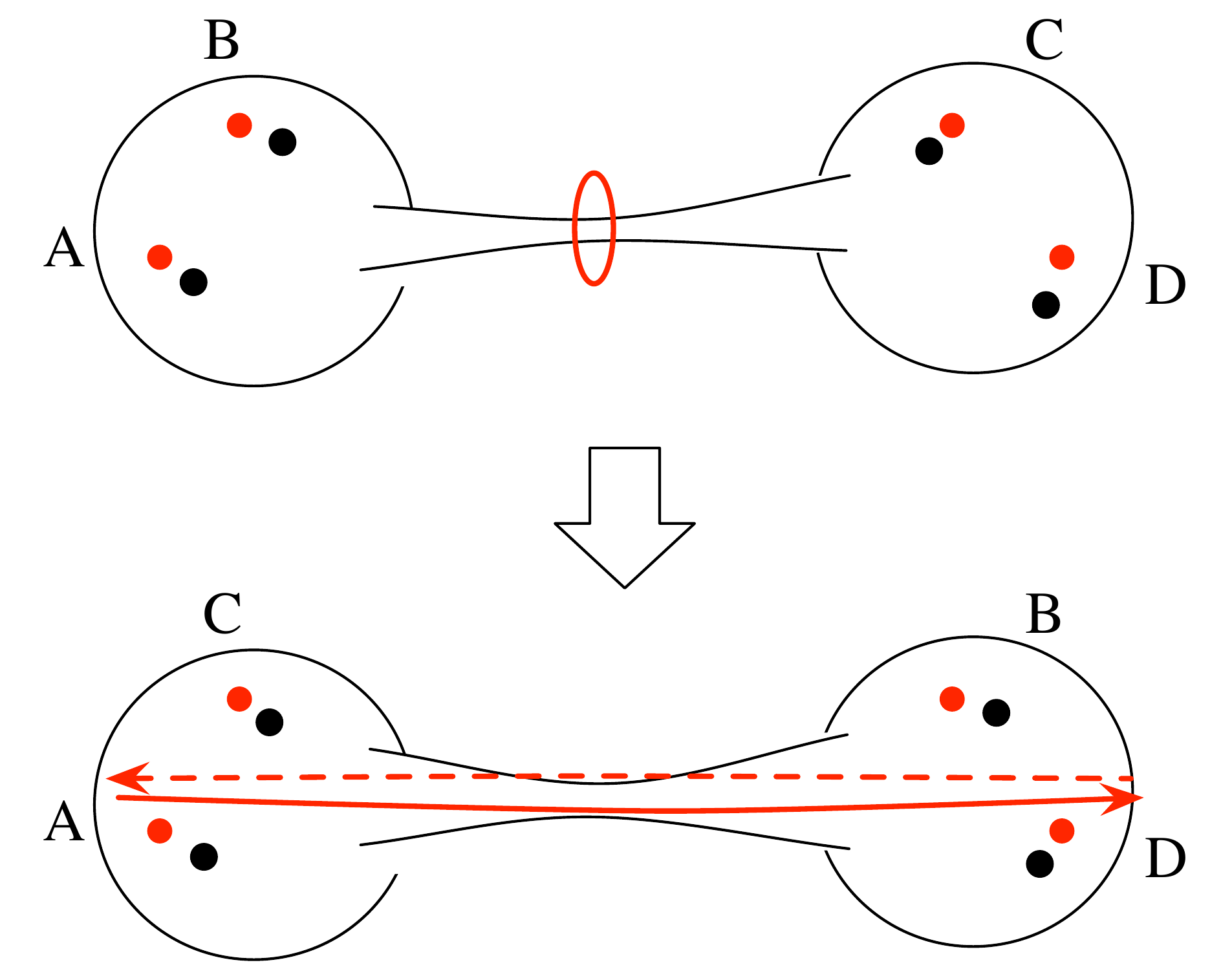}
\]
\caption{W bosons also come from  monopoles.\label{fig:su2nf4-exchange-wboson-monopole}}
\end{figure}
The same manipulation also shows that the $\SU(2)$ W-bosons in the dual description came from monopoles in the original description, see Fig.~\ref{fig:su2nf4-exchange-wboson-monopole}. 
Therefore it is important to keep in mind that the dual $\SU(2)$ gauge multiplet is not the same physical excitation as the original $\SU(2)$ gauge multiplet.
Note also that this monopole has twice the magnetic charge of the monopole which became the dual quarks.

\begin{figure}[h]
\[
\inc{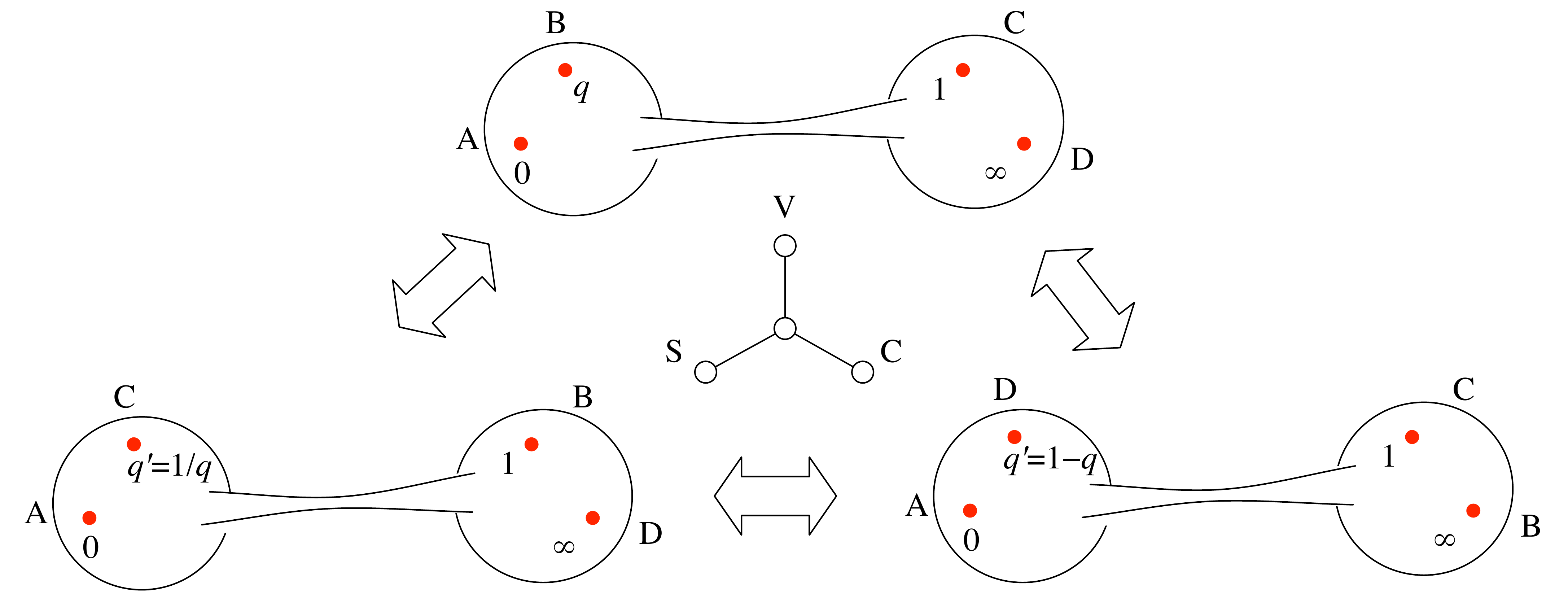}
\]
\caption{Triality\label{fig:triality}}
\end{figure}
There is also a limit where the singularity $B$ approaches the singularity $C$, $q\to 1$. This is again equivalent to a weakly-coupled $\SU(2)$ gauge theory with four flavors, but with the role of the singularities are permuted, see Fig.~\ref{fig:triality}.
The four mass parameters of the hypermultiplets are now given by 
\begin{align}
\mu''_1&=\phantom{-}\mu_A+\mu_D =\phantom{-}\frac{\mu_1-\mu_2}2+\frac{\mu_3-\mu_4}2,\\
\mu''_2&=-\mu_A+\mu_D =-\frac{\mu_1-\mu_2}2+\frac{\mu_3-\mu_4}2,\\
\mu''_3&=\phantom{-}\mu_C+\mu_B=\phantom{-}\frac{\mu_3+\mu_4}2+\frac{\mu_1+\mu_2}2,\\
\mu''_4&=\phantom{-}\mu_C-\mu_B=\phantom{-}\frac{\mu_3+\mu_4}2-\frac{\mu_1+\mu_2}2.
\end{align} These are the weights of the conjugate spinor representation of $\SO(8)$. 

Therefore, we learned that the strong-weak duality of the $\SU(2)$ gauge theory with four flavors, \begin{equation}
q \leftrightarrow q'=1/q \leftrightarrow q''=1-q
\end{equation} are accompanied by the exchange of the representations of the $\SO(8)$ flavor symmetry, \begin{equation}
\vcenter{\hbox{\begin{tikzpicture}
\node (V) at (0,1) {$V$};
\node (S) at (-1,0) {$S$};
\node (C) at (1,0) {$C$};
\draw[<->] (V) to (S);
\draw[<->] (S) to (C);
\draw[<->] (C) to (V);
\end{tikzpicture}}}
\end{equation} where $V$, $S$, $C$ are eight dimensional irreducible representations (vector, spinor, conjugate spinor) of $\SO(8)$. These exchanges of three irreducible eight dimensional representations are induced by the outer automorphism, and are known as the triality of the group $\SO(8)$, whose Dynkin diagram is also shown in Fig.~\ref{fig:triality}.  This triality was originally found in \cite{Seiberg:1994aj}. The exposition in this subsection followed the one given in \cite{Gaiotto:2009we}.

The Higgs branch of this system can be studied in any of these descriptions. 
Originally we have hypermultiplets $q^a_I$, where $a=1,2$ and $I=1,\ldots,8$. The gauge invariant combination is 
\begin{equation}
M_{[IJ]}=q^a_I q^b_J \epsilon_{ab}.
\end{equation}
In the dual, we have hypermultiplets $\tilde q^{\tilde a}_{\tilde I}$,
where $\tilde a=1,2$ are for dual $\SU(2)$ and $\tilde I=1,\ldots,8$  are for the spinor representation of the $\SO(8)$ flavor symmetry. The basic gauge invariant is then \begin{equation}
\tilde M_{[\tilde I\tilde J]}=\tilde q^{\tilde a}_{\tilde I}\tilde q^{\tilde b}_{\tilde J}\epsilon_{\tilde a\tilde b}.
\end{equation} Both $M_{[IJ]}$ and $\tilde M_{[\tilde I\tilde J]}$ are in the adjoint representation of $\SO(8)$, and can be naturally identified using the outer automorphism of $\SO(8)$. We can check that the constraints satisfied by $M_{[IJ]}$ and $\tilde M_{[\tilde I\tilde J]}$ are invariant under the outer automorphism.  This shows that the Higgs branch are the same as complex spaces. 

\subsection{Generalization}\label{sec:trivalent}
\subsubsection{Trivalent diagrams}
\begin{figure}[h]
\[
\inc{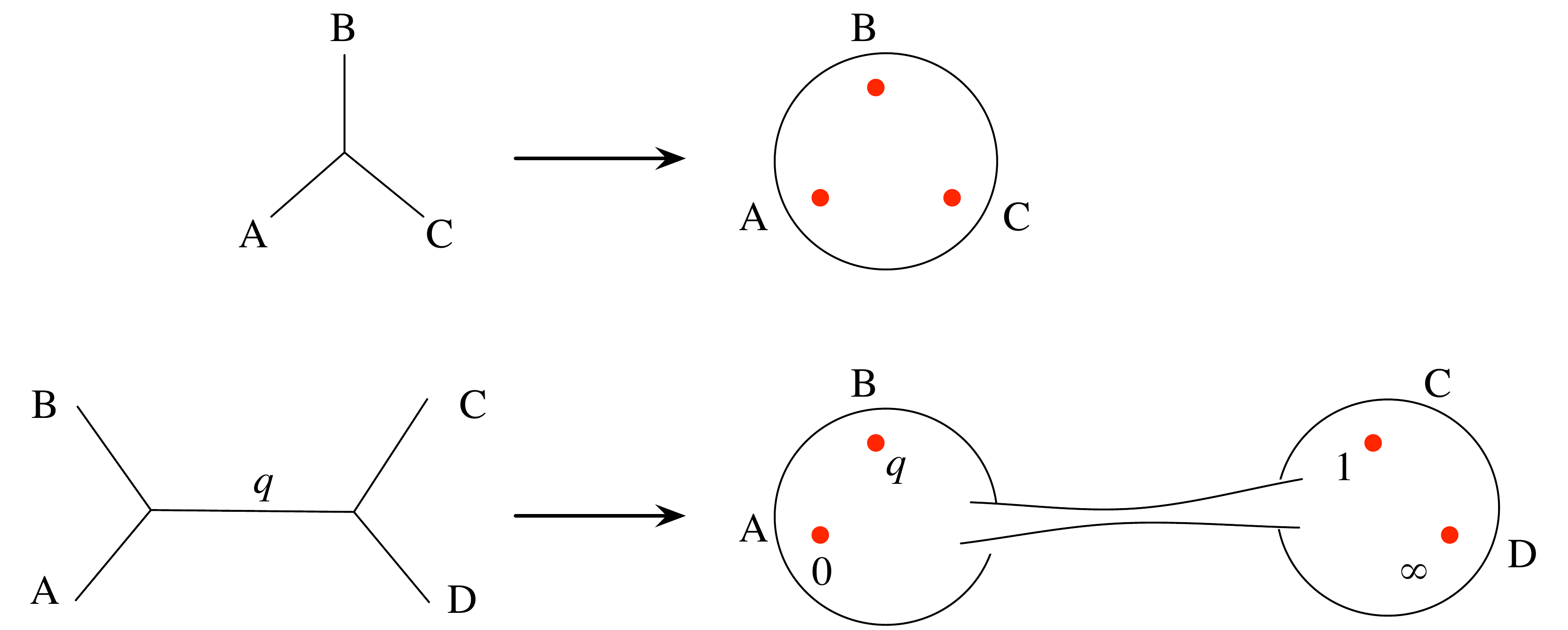}
\]
\caption{Trivalent diagrams and corresponding \Gaiotto\ curves\label{fig:trivalent}}
\end{figure}
The trifundamental hypermultiplet 
consisting of $\cN{=}1$ chiral multiplets $q_{a\alpha u}$ for $a,\alpha,u=1,2$ played the central role in the analysis so far. 
Let us introduce a shorthand notation for it, by representing it by a trivalent vertex with labels $A$, $B$, $C$ as in Fig.~\ref{fig:trivalent}, signifying the symmetries $\SU(2)_A$, $\SU(2)_B$, $\SU(2)_C$, acting on the indices $a$, $\alpha$, $u$ respectively.  %We assume that the corresponding mass parameters $\pm\mu_{A,B,C}$ are given.

The \Gaiotto\ curve for this system is given by a sphere with three punctures $A$, $B$, $C$, and the \SeibergWitten\ curve is given by \begin{equation}
\Sigma:\qquad\lambda^2-\phi(z)=0
\end{equation} where $\phi(z)$ is given by the condition that the coefficients of the double poles are given by $\mu_A^2$, $\mu_B^2$, $\mu_C^2$
at each of the punctures $A,B,C$, as in \eqref{phicondition_trifund}. 

Now, the $\SU(2)$ theory with four flavors can be obtained by taking two copies of trifundamentals,
and coupling an $\SU(2)$ gauge multiplet to them. We denote it by taking two trivalent vertices, and connecting them by a line, as shown in Fig.~\ref{fig:trivalent}. We put the exponentiated coupling $q$ on the connecting line. 
%Again, we assume that the mass parameters $\mu_{A,B,C,D}$ are implicitly given in the diagram.

\begin{figure}[h]
\[
\inc{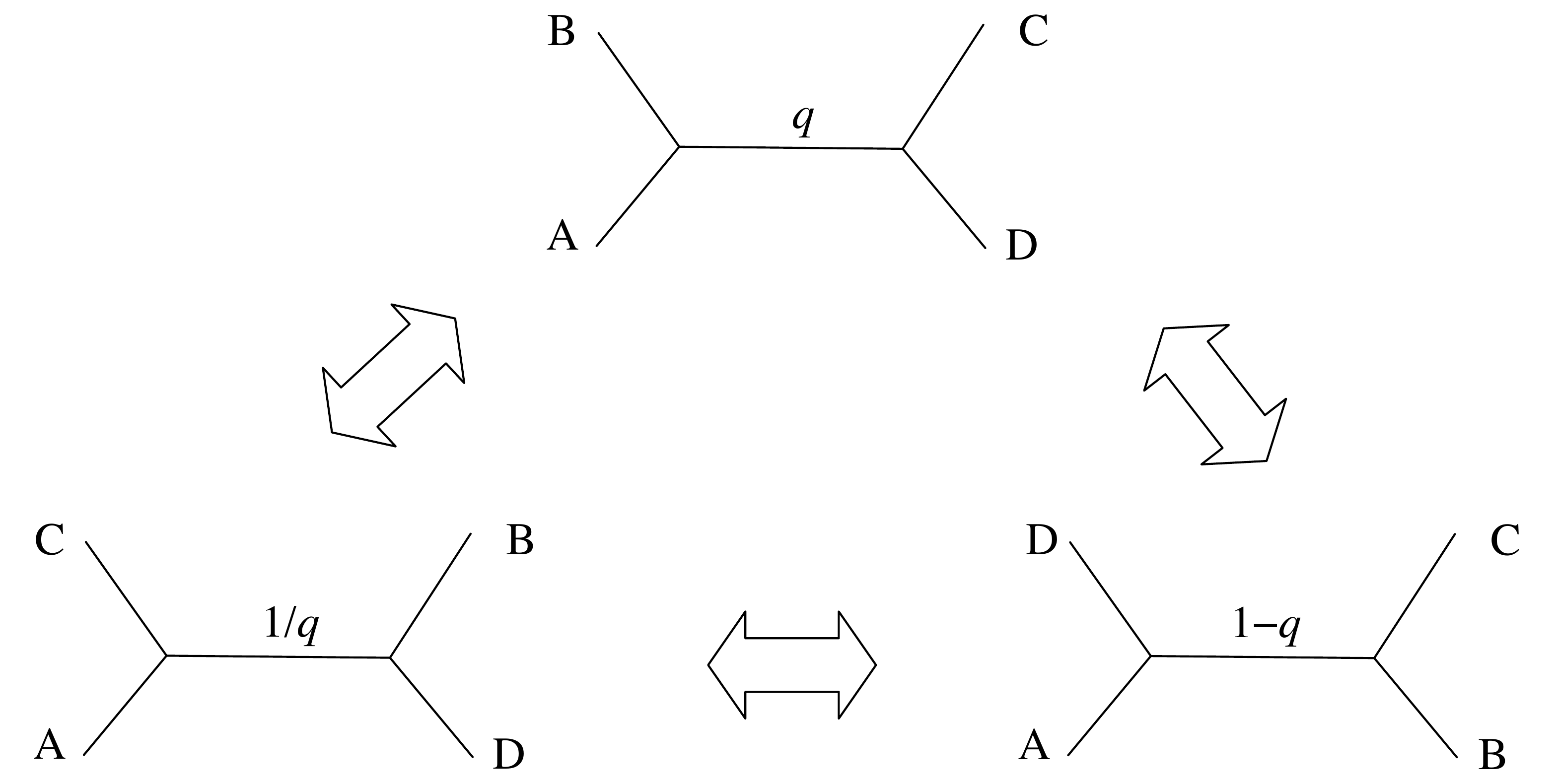}
\]
\caption{Triality, using trivalent diagrams.\label{fig:triality-trivalent}}
\end{figure}

Starting with this trivalent diagram, we can easily write down the Lagrangian of the theory.
The free parameters in the theory are the mass parameters $\mu_{A,B,C,D}$ and the UV coupling $q$. 
The \Gaiotto\ curve of this system is given by a sphere with four punctures $A$, $B$, $C$, $D$,
 and the \SeibergWitten\ curve is given by \begin{equation}
\Sigma:\qquad\lambda^2-\phi(z)=0
\end{equation} where $\phi(z)$ is given by the condition that its residues are given by $\mu_{X}^2$ at each of the punctures $X=A, B,C,D$ as in \eqref{so8masses}.
The triality of the $\SU(2)$ theory with four flavors, shown already in Fig.~\ref{fig:triality}, can be depicted in terms of the trivalent diagrams as in Fig.~\ref{fig:triality-trivalent}.

This way, we can regard  the trivalent diagram as a shorthand to represent the UV Lagrangian. The Seiberg-Witten solution to this given UV Lagrangian theory is given just by replacing each trivalent vertex with a three punctured sphere, and a connecting line with a connecting tube. This is a surprisingly concise method to obtain the Seiberg-Witten solutions to $\cN{=}2$ gauge theories.

\subsubsection{Example: torus with one puncture}

\begin{figure}[h]
\[
\inc{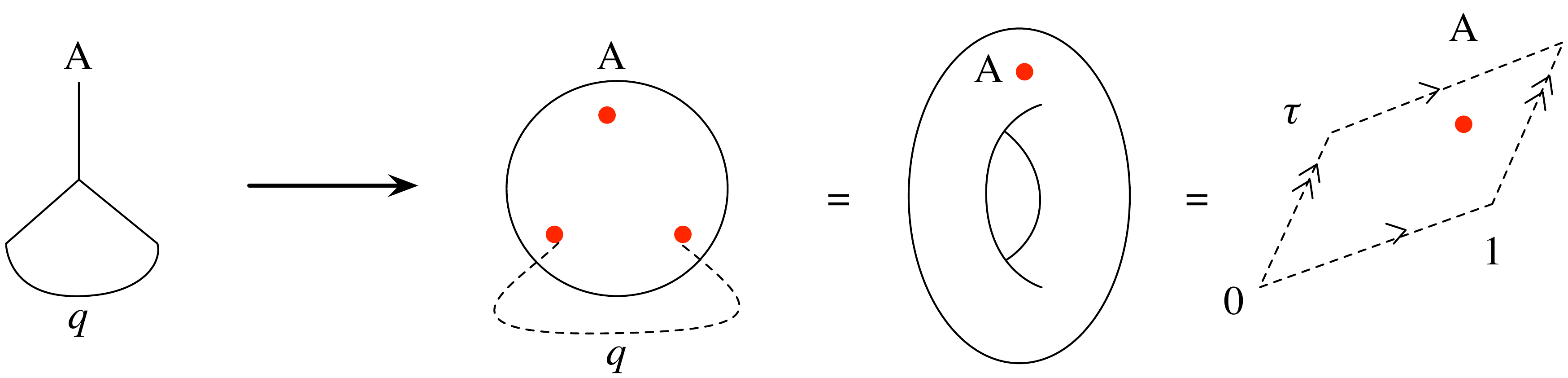}
\]
\caption{$\SU(2)$ with adjoint plus one free hyper\label{fig:n2star}}
\end{figure}
Let us see a few examples. First, take one trivalent vertex, and connect two out of the three lines by an edge, see Fig.~\ref{fig:n2star}.
We start from a trifundamental described by $\cN{=}1$ chiral multiplets $q_{a\alpha u}$, but we couple the same $\SU(2)$ gauge multiplet to the index $a$ and $\alpha$. Then the combination $(a,\alpha)$ is in the tensor product of two spin-$1/2$ representations. 
Therefore we can split it into a triplet and a singlet, with additional index $u=1,2$: \begin{equation}
q_{a\alpha u} \to q_{iu}', q_u'',
\end{equation}  where $i=1,2,3$ is the index for the triplet of $\SU(2)$. 

 In total, we just have one full hypermultiplet in the triplet, and another full hypermultiplet in the singlet which is completely decoupled. 
Therefore this is essentially the $\cN{=}2^*$ $\SU(2)$ theory, or equivalently the $\cN{=}4$ $\SU(2)$ theory with mass deformation to the hypermultiplet in the adjoint representation.  The adjoint mass $\mu$ is associated to the remaining one $\SU(2)$ flavor symmetry.

Its Seiberg-Witten solution is given by connecting two punctures of a three-punctured sphere by a tube. As shown in Fig.~\ref{fig:n2star}, the \Gaiotto\ curve is a torus with one puncture. The \SeibergWitten\ curve is then \begin{equation}
\lambda^2 - \phi(z)=0
\end{equation} where $z$ is now a coordinate of the torus, which we take to be the complex plane with the identification $z\sim z+1\sim z+\tau$. As the origin of the coordinate is arbitrary, we put the puncture at the origin. The $\phi(z)$ is given by the condition that it has a double pole with a given strength at $z=0$. This uniquely fixes the form of $\phi(z)$ to be \begin{equation}
\phi(z)=(\mu^2\wp(z;\tau)+u)dz^2
\end{equation} where $\wp$ is the Weierstra\ss\  function, and $u$ is the Coulomb branch vev $u=\vev{\tr\Phi^2}/2$.

Now it is clear that the theory at the coupling given by  $\tau$ and the same theory at the coupling given by $\tau'=-1/\tau$ 
are equivalent after exchanging the monopoles and the adjoint quarks.
The  space of the coupling can be identified with the moduli space $\cM_1$ of the tori, i.e~genus-1 Riemann surfaces,
which is given by \begin{equation}
\cM_1= \bH / \SL(2,\bZ)  \label{SL2Zfund}
\end{equation} where $\bH$ is the upper half plane where $\tau$ takes the value in, and $\SL(2,\bZ)$ is the modular group exchanging the edges of the torus. 
%The fundamental region is shown in Fig.~\ref{fig:sl2zfund}; 
The duality group can be identified with the modular group.

\subsubsection{Example: sphere with five punctures}

\begin{figure}[h]
\[
\inc{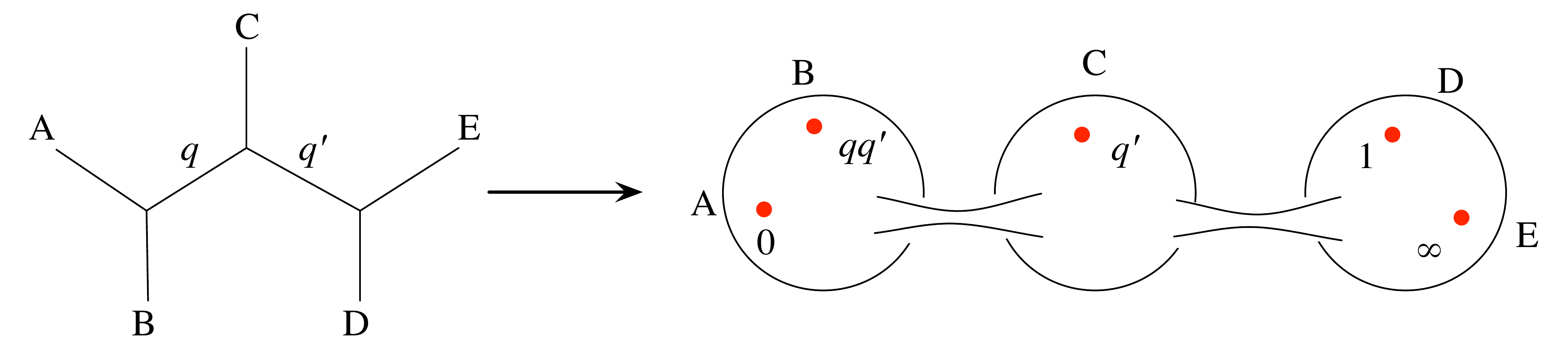}
\]
\caption{An $\SU(2)^2$ theory and its curve\label{fig:su2-quiver}}
\end{figure}
As another example, take three trivalent vertices and connect them as shown in Fig.~\ref{fig:su2-quiver}.
The leftmost trivalent vertex counts as $N_f=2$ flavors for $\SU(2)_1$,
and
the rightmost one counts as $N_f=2$ flavors for $\SU(2)_2$.
In addition, we have a hypermultiplet coming from the central trivalent vertex, $q_{ai u}$
where $a$ is for $\SU(2)_1$ and $i$ is for $\SU(2)_2$. The remaining index $u=1,2$ is an index for the flavor symmetry. 
In older literature it was more customary to denote this hypermultiplet charged under $\SU(2)_1\times \SU(2)_2$ using
$\cN{=}1$ chiral multiplets $(Q^i_a,\tilde Q_i^a)$ which are
\begin{equation}
Q^i_a=q_{aju=1}\epsilon^{ij},\qquad
\tilde Q_i^a =q_{biu=2} \epsilon^{ab}.
\end{equation} This is usually called the bifundamental multiplet charged under $\SU(2)_1\times \SU(2)_2$. 

The Seiberg-Witten solution to this theory is easily found, as shown in Fig.~\ref{fig:su2-quiver}. 
We start from three spheres, described by complex coordinates $z_1$, $z_2$, and $z_3$. 
The punctures $A,B$ are at $z_1=0,1$;
the puncture $C$ is at $z_2=1$;
the punctures $D,E$ are at $z_3=1,\infty$, respectively. 
To connect $z_1=\infty$ and $z_2=0$,
we introduce $w_1=1/z_1$ and require the relation $w_1z_2=q'$. 
This simply means that we have  via $z_1=q'z_2$. 
Similarly, by connecting  $z_2=\infty$ and $z_3=0$, we have $z_2=qz_3$.  
Now we introduce $z=z_3$ to describe the coordinate on the resulting sphere with five punctures.
Then the  punctures are at $z=0$, $qq'$, $q$, $1$ and $\infty$, each representing an $\SU(2)$ flavor symmetry which we call $\SU(2)_{A,B,C,D,E}$ respectively. 
The gauge couplings of $\SU(2)_1\times \SU(2)_2$ can be identified with $q$ and $q'$. 

Let us denote the  mass parameters associated to the flavor symmetries by $\mu_{A,B,C,D,E}$. 
The \SeibergWitten\ curve is \begin{equation}
\lambda^2-\phi(z)=0
\end{equation} where $\phi(z)$ needs to satisfy the asymptotic behavior \begin{equation}
\phi(z)\sim \frac{\mu_X^2}{z_X^2}dz_{X}^2\label{massive-full-su2}
\end{equation} where $z_X$ for $X=A,B,C,D,E$ is a local coordinate on the \Gaiotto\ curve such that the puncture $X$ is at $z_X=0$. From the conditions at $A$, $B$, $C$, $D$, we find that $\phi$ is given by \begin{equation}
\phi(z)=\frac{P(z)}{z^2(z-1)^2(z-q)^2(z-qq')^2} dz^2
\end{equation} where $P(z)$ is a polynomial. To impose the condition at $E$, we go to the coordinate $w=1/z$.
For $\phi(z)$ to behave as $\sim dw^2/w^2$, $P(z)$ can have terms of up to $z^6$. 
We see that $\phi(z)$ has seven coefficients. Five combinations are mass parameters, and two linear combinations that do not shift the coefficients of the double poles are the Coulomb branch parameters $u=\vev{\tr\Phi^2}/2$ and $u'=\vev{\tr\Phi'{}^2}/2$ of two gauge multiplets $\SU(2)_{1,2}$. 
From the Seiberg-Witten solution, we see that this theory has strong-weak coupling dualities where five flavor symmetry groups $\SU(2)_{A,B,C,D,E}$ can be arbitrarily permuted, with an appropriate change of the couplings $(q,q')$ of the two gauge groups. 
This extended duality was first found in \cite{Argyres:1999fc}.

\subsubsection{Example: a genus-two surface}

\begin{figure}[h]
\[
\inc{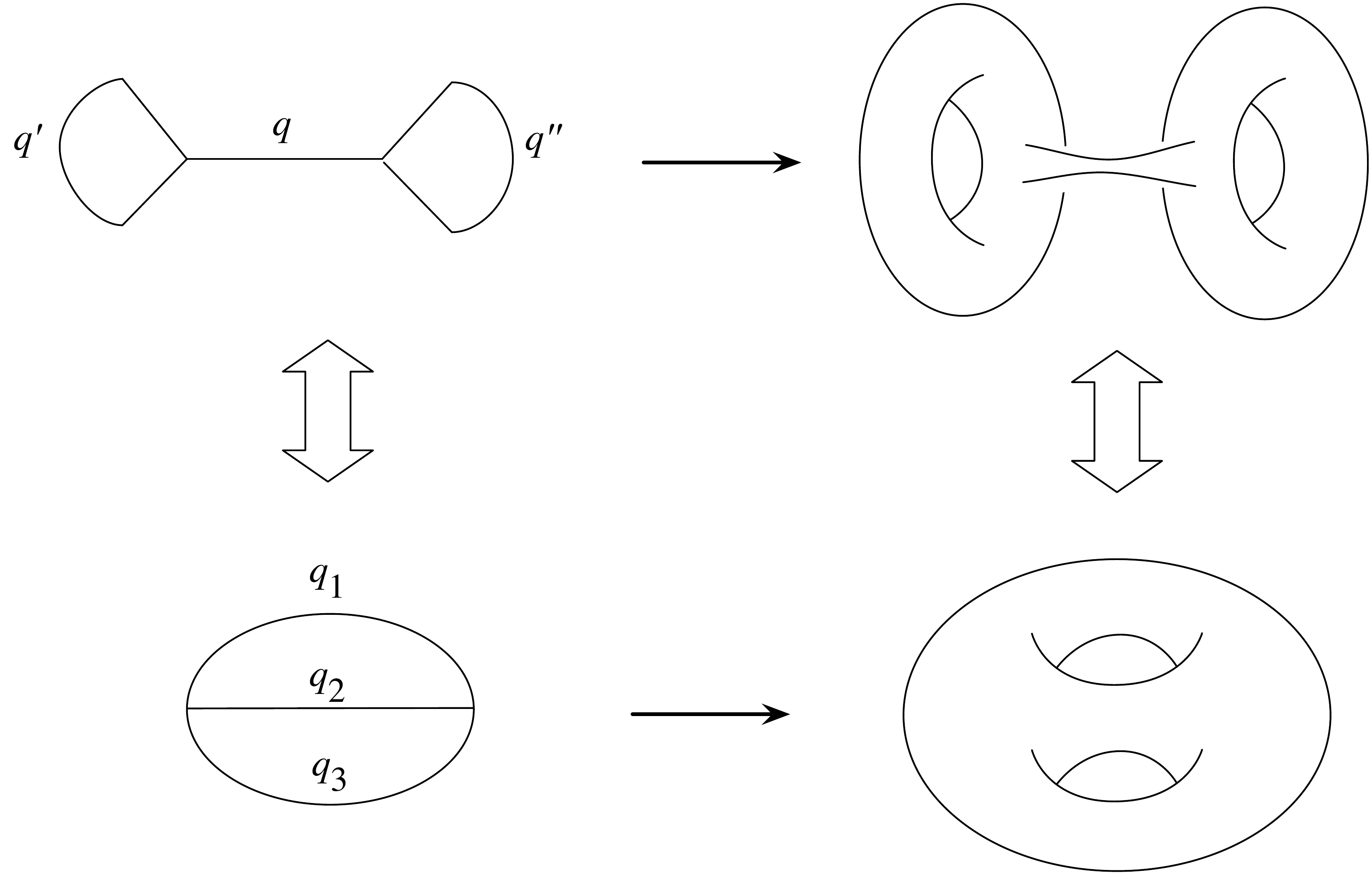}
\]
\caption{Two $\SU(2)^3$ theories and their curves\label{fig:genus-2}}
\end{figure}
As the third example, let us take two trivalent vertices and connect them with three edges. 
There are two topologically distinct ways to do this, as shown on the left hand side of Fig.~\ref{fig:genus-2}.

The upper theory is an $\SU(2)_l\times \SU(2)_m\times \SU(2)_r$ gauge theory.
There are half-hypermultiplets which are in \begin{equation}
\mathbf{3}\otimes \mathbf{2}\otimes \mathbf{1},\qquad
\mathbf{1}\otimes \mathbf{2}\otimes \mathbf{3}
\end{equation} and one full hypermultiplet charged under $\SU(2)_m$. 
Note that the trivalent-graph construction does not allow us to consider theories with non-zero mass term for this last full hypermultiplet.
The lower theory is an $\SU(2)_1\times \SU(2)_2\times \SU(2)_3$ theory with two half-hypermultiplets in the trifundamental representation. Again, the trivalent-graph construction does not allow us to introduce non-zero mass term for this full hypermultiplet in the trifundamental. 

The Seiberg-Witten solution is again easily obtained. To obtain the \Gaiotto\ curve, we replace two trivalent vertices with three-punctured spheres, and connect pairs of punctures with tubes. 
We see that both are given by a smooth genus-2 surface. The \SeibergWitten\ curve is a further double cover given by \begin{equation}
\lambda^2-\phi(z)=0
\end{equation} where $z$ is a complex coordinate of the genus-2 surface, and $\phi(z)$ is a smooth quadratic differential on it. The space of quadratic differentials on a fixed genus-2 surface is complex three dimensional, which we identify with the Coulomb branch vevs $u_i=\vev{\tr\Phi^2_i}/2$.

Now it is clear that we can continuously deform the upper theory to the lower theory by tuning the gauge couplings. The non-perturbative space of couplings can be identified with the moduli space $\cM_2$ of genus-2 Riemann surfaces, which is complex three dimensional. The duality group is identified with the mapping class group $\cG_2$ of the genus-2 surface, and $\cM_2=\cT_2/\cG_2$ where $\cT_2$ is the Teichm\"uller space of the genus-2 Riemann surface, compare the genus-1 case \eqref{SL2Zfund}. 

Now a somewhat surprising mathematical fact is that $\cT_2$ 
is equivalent to three copies of the upper half plane $\bH^3$ in the smooth sense, 
but not in the holomorphic sense\footnote{The author thanks Jacques Distler for very illuminating discussions on this point.}:  \begin{equation}
\begin{aligned}
\cT_2 \simeq \bH^3&\quad \text{in the smooth sense},\\
\cT_2 \not\simeq \bH^3  & \quad\text{in the holomorphic sense}.
\end{aligned}
\end{equation} Naive perturbative analysis tells us that the space of the couplings of $\SU(2)^3$ is just three copies of the upper half plane: \begin{equation}
 (\tau_1,\tau_2,\tau_3) \in \bH^3
\end{equation} including the complex structure. 
Therefore, we find that the non-perturbative corrections can make a rather drastic change in the complex structure of the parameter space of supersymmetric theories. 

\subsubsection{The curve and the Hitchin field}\label{sec:hitchin}
We learned that writing the \SeibergWitten\ curve in the form \begin{equation}
\Sigma:\qquad \lambda^2-\phi(z)=0\label{prehitchin}
\end{equation} is very useful for the understanding of the system.  This way of presenting the curve is closely related to the so-called Hitchin system on the \Gaiotto\ curve $C$. Explaining this technique would make another lecture note. A starting point for the reader is  e.g.~Sec.~3 of \cite{Gaiotto:2009hg}. 

Let us at least present a very crude aspect of it.  
We consider a meromoprhic one-form $\varphi(z)$ which is a traceless $2\times 2$ matrix. Then, we can form an equation of the form \begin{equation}
\det(\lambda-\varphi(z))=0\label{hitchin}
\end{equation} for a one-form $\lambda$.  Identifying \eqref{prehitchin} and \eqref{hitchin}, we find \begin{equation}
\frac12\tr\varphi(z)^2=\phi(z).\label{p-p}
\end{equation}  This matrix field $\varphi(z)$ is called the Hitchin field.\footnote{Hitchin himself called $\varphi(z)$ the Higgs field, but the author thinks this terminology is rather confusing in the $\cN{=}2$ supersymmetric context, as $\varphi(z)$ controls the physics of the Coulomb branch, not of the Higgs branch.}  It is possible to understand the existence of such a complex adjoint field on the \Gaiotto\ curve using string duality, but explaining it is outside the aim of this lecture note. 

The condition on the field $\phi(z)$ at a puncture at $z_X=0$ associated to a mass term $\mu_X$  was given in \eqref{massive-full-su2}: \begin{equation}
\phi(z)= \mu_X^2 \frac{dz_X^2}{z_X^2} + \text{(less singular terms)}.\label{phi-full}
\end{equation} This translates to the condition for the Hitchin field $\varphi(z)$ given by \begin{equation}
\varphi(z) \sim \begin{pmatrix}
\mu_X & 0 \\
0 & -\mu_X 
\end{pmatrix} \frac{dz_X}{z_X} + \text{(less singular terms)}.\label{varphi-full}
\end{equation} Here, the symbol $A\sim B$ means that $A$ and $B$ are conjugate, in the sense that there is an invertible $N\times N$ complex matrix $g$ such that $A=g B g^{-1}$. 

Now, let us consider what happens when we turn off $\mu_X$ to zero.  In the description \eqref{phi-full}, we find that the boundary condition becomes \begin{equation}
\phi(z)= c \frac{dz_X^2}{z_X} +  \text{(less singular terms)}\label{q-q}
\end{equation} for some constant $c$. In the description \eqref{varphi-full}, it is not that the residue just becomes a zero matrix. We note that \begin{equation}
\begin{pmatrix}
\mu_X & 0 \\
0 & -\mu_X 
\end{pmatrix} 
\sim
\begin{pmatrix}
\mu_X & 1 \\
0 & -\mu_X 
\end{pmatrix} 
\end{equation} as long as $\mu_X\neq 0$, and we can take the limit $\mu_X\to 0$ on the right hand side. 
This means that the massless limit results in the boundary condition of the form \begin{equation}
\varphi(z) \sim \begin{pmatrix}
0 & 1 \\
0 & 0 
\end{pmatrix} \frac{dz_X}{z_X} + \text{(less singular terms)}.
\end{equation} 
Using \eqref{p-p}, one finds that this reproduces the condition \eqref{q-q}.   

At this stage, one might not find the advantange of using $\varphi(z)$ instead of $\phi(z)$ very much. It turns out, however, that when we discuss a generalization to $\SU(N)$ gauge theories or gauge theories with more complicated gauge groups, it turns out to be crucial.

\subsection{Theories with less flavors revisited}\label{sec:less}
We found that writing the \SeibergWitten\ curve of the $N_f=4$ theory in the form \begin{equation}
\lambda^2-\phi(z)=0
\end{equation} helps greatly in understanding the structure of the duality. 
Let us apply this idea to the curves for theories with less number of flavors, $N_f<4$. 

\subsubsection{Rewriting of the curves}

\begin{figure}[h]
\[
\inc{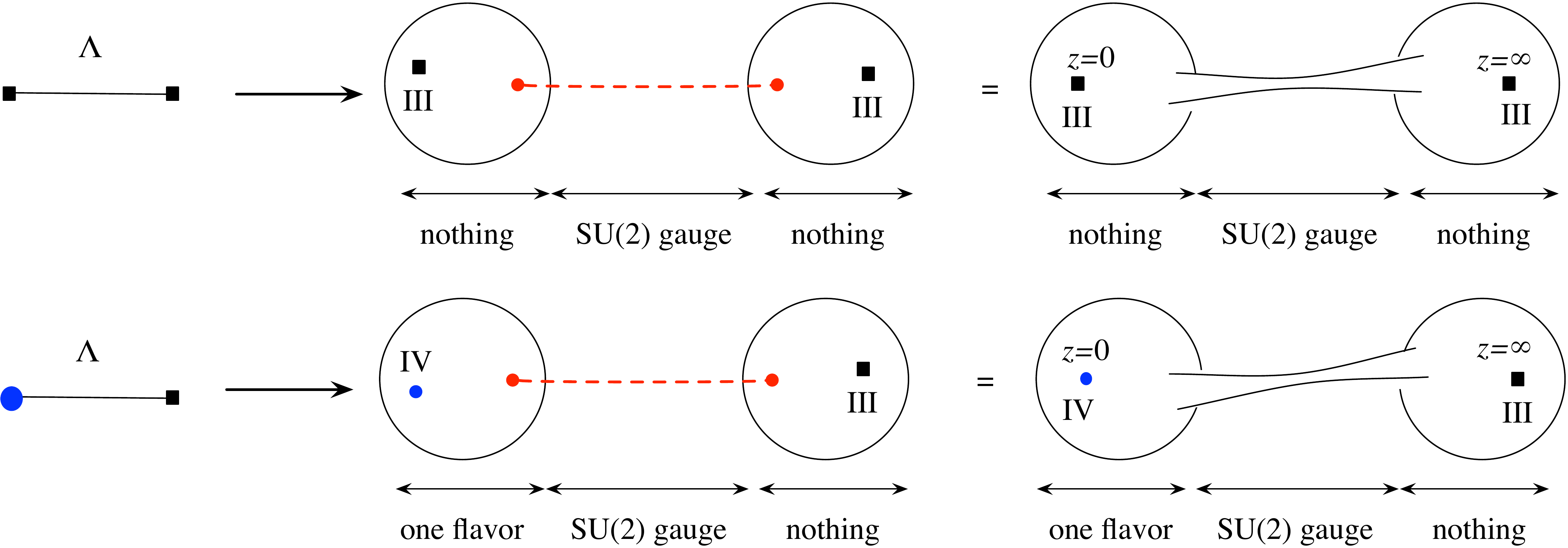}
\]
\caption{$\SU(2)$ theories with less flavors\label{fig:lessG}}
\end{figure}

First, consider the curve of the pure theory, \begin{equation}
\frac{\Lambda^2}z+\Lambda^2 z=x^2-u,\qquad \text{with}\ \lambda=x\frac{dz}z.
\end{equation} In terms of $\lambda$, this can be written as \begin{equation}
\lambda^2-\phi(z)=0,\qquad \phi(z)=(\frac{\Lambda^2}z+u+\Lambda^2 z)\frac{dz^2}{z^2}.
\end{equation} We see that the quadratic differential $\phi(z)$ has singularities worse than those in the $N_f=4$ theory: they now have order three poles at $z=0$ and $=\infty$. We can depict the situation of the curve as in  the upper row of Fig.~\ref{fig:lessG}. There, the roman numeral III shows that $\phi(z)$ has a third order pole at the puncture.  The singularities of $\phi(z)$ with higher poles form a new class of punctures, which we call wild $\SU(2)$  punctures.

As an extension of the trivalent diagram encoding the UV Lagrangian, let us introduce the notation that an edge stands for an $\cN{=}2$ $\SU(2)$ vector multiplet, and the black square at one end means that we do not introduce any hypermultiplet. 
Then the translation from the diagram representing the UV Lagrangian to the \Gaiotto\ curve can be simply seen, as also shown in Fig.~\ref{fig:lessG}.
Second, consider the curve of the $N_f=1$ theory, \begin{equation}
\frac{2\Lambda(\tilde x-\mu)}z+\Lambda^2 z=\tilde x^2-u,\qquad \text{with}\ \tilde\lambda=\tilde x\frac{dz}z.
\end{equation} This can be written as \begin{equation}
\lambda^2 -\phi(z)=0, \qquad \phi(z)=(\frac{\Lambda^2}{z^2}-\frac{2\Lambda\mu}z + u + \Lambda^2z )\frac{dz^2}{z^2}
\end{equation} where $\lambda= x dz/z$ is shifted from $\tilde\lambda$.
We find that the singularity at $z=0$ changes to a pole of order 4. The Lagrangian and its Seiberg-Witten solution can be concisely summarized as in the lower row of Fig.~\ref{fig:lessG}. The edge stands for an $\SU(2)$ gauge group.
A black square on one side means that we do not have any hypermultiplet there.
A  pale blob on another side means that we introduce one hypermultiplet in the doublet. 
The solution is obtained by associating to a black square by a sphere with a third order pole, denoted by III,
and by similarly associating to a pale blob a sphere with a fourth order pole, denoted by IV,
and finally connecting them by a tube. Note that a fourth order pole has its own $\SU(2)$ flavor symmetry and an associated mass parameter.

Summarizing, we consider a sphere with a regular puncture and a wild puncture of pole order III as an empty theory, and 
a sphere with a regular puncture and a wild puncture of pole order IV as a theory of decoupled doublet hypermultiplet, as shown in Fig.~\ref{fig:lessG}. Connecting the regular punctures with a tube, we find the \Gaiotto\ curves of less flavors. 

\subsubsection{Generalization}
\begin{figure}[h]
\[
\inc{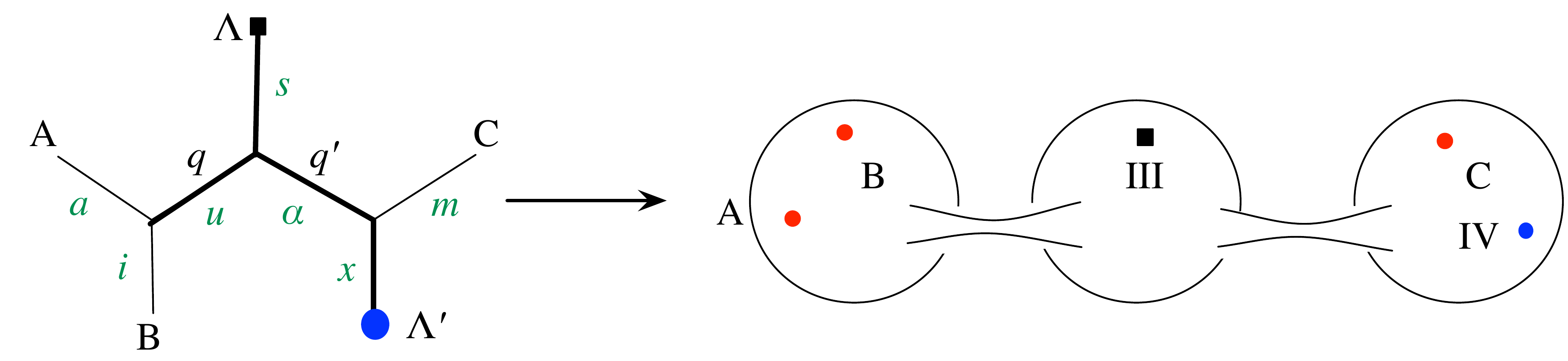}
\]
\caption{An $\SU(2)^4$ theory and its curve\label{fig:complicated-example}}
\end{figure}
This generalization allows us to find the Seiberg-Witten solutions to a huge class of $\cN{=}2$ theories whose gauge group is a product of  copies of $\SU(2)$. 
For example, consider a UV Lagrangian theory with gauge group $\SU(2)^4$ described by the left hand side of Fig.~\ref{fig:complicated-example}.
In words, we first take three copies of bifundamental hypermultiplets, \begin{equation}
Q_{aiu},\quad
Q'_{us\alpha},\quad
Q''_{\alpha x m}.
\end{equation} We showed in the figure how the indices are assigned to the edges of the trivalent diagram. 
We emphasized the edges corresponding to the dynamical gauge groups by making them thicker. 
In words, the indices $a,i,m$ are for $\SU(2)_{A,B,C}$ flavor symmetries. 
An $\SU(2)$ gauge multiplet couples to the index $u$, with exponentiated coupling constant $q$,
another $\SU(2)$ gauge multiplet to the index $\alpha$, with exponentiated coupling constant $q'$.
We introduce another $\SU(2)_1$ gauge multiplet which couples to the index $s$ corresponding to the black square,
and finally another $\SU(2)_2$ gauge multiplet which couples to the index $x$, with additional $N_f=1$ hypermultiplet $(Q''''{}_x,\tilde Q''''{}^x)$. 
We can write down the Lagrangian if required, but now we see how concise the trivalent diagram summarizes its structure.

Its Seiberg-Witten solution can be immediately obtained. It is given by \begin{equation}
\lambda^2-\phi(z)=0,
\end{equation}where $\phi(z)$ has three order two poles, one order three pole and finally an order four pole. 
Putting them at $z=0,1,z_0$, $z_1$ and at $\infty$ respectively, we see that $\phi(z)$ has the form \begin{equation}
\phi(z)=\frac{P(z)}{z^2(z-1)^2(z-z_0)^2(z-z_1)^3}dz^2
\end{equation} where $P(z)$ is a polynomial. To have an order four pole at $w=1/z=0$, $P(z)$ is seen to be a degree-9 polynomial. Among the ten coefficients, three are mass parameters for $\SU(2)_{A,B,C}$ flavor symmetry,
one is the scale of $\SU(2)_1$ for the black blob, another is the scale of $\SU(2)_2$, and another for the mass parameter of the additional $N_f=1$ flavor for $\SU(2)_2$.  The four remaining linear combinations can be identified with the four Coulomb branch parameters $u_i=\vev{\tr{\Phi_i^2}/2}$. 
This is not a conformal theory: there are two dynamical scales $\Lambda$ and $\Lambda'$. Still, we immediately see from the structure of the \Gaiotto\ curve that there are S-dualities exchanging the three regular punctures at $A$, $B$ and $C$.

\section{Argyres-Douglas CFTs}\label{sec:AD}
In this section, we come back to the observation made at the end of Sec.~\ref{sec:nf_1_curve} that there is a very singular point of the Coulomb branch of the $N_f=1$ theory. We study the physics at that point and its generalizations. 
\subsection{$N_f=1$ theory and the simplest Argyres-Douglas CFT}
\begin{figure}[h]
\[
\inc{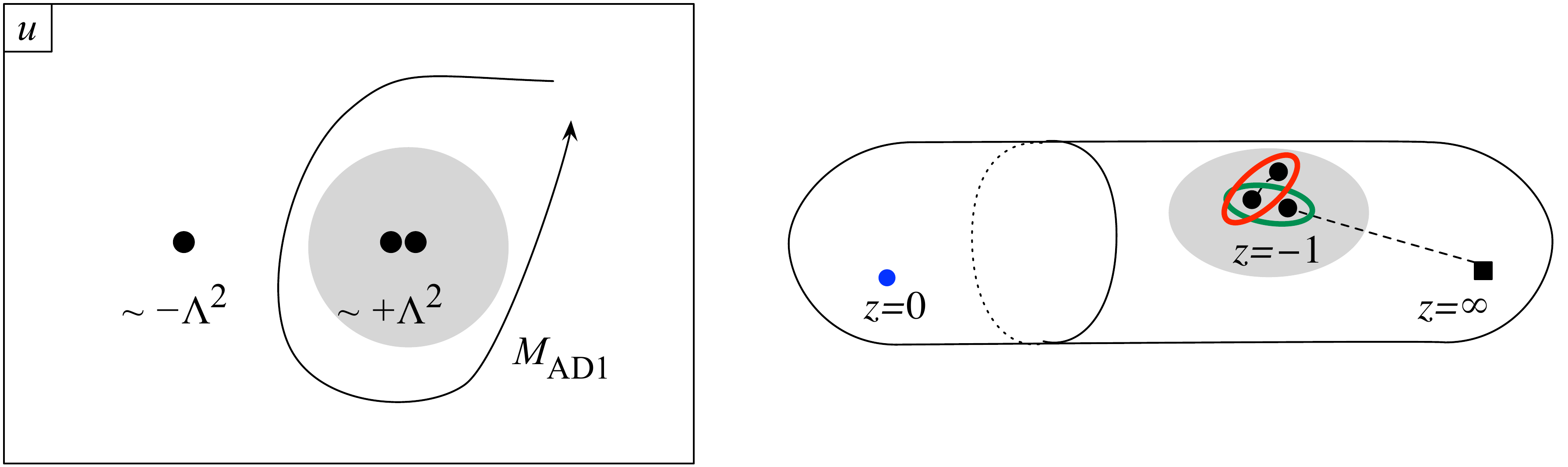}
\]
\caption{Argyres-Douglas point of $N_f=1$ theory\label{fig:nf_1_adpoint}}
\end{figure}

Let us come back to  the curve of $\SU(2)$ gauge theory with $N_f=1$ flavor again: \begin{equation}
\Sigma:\quad \frac{2\Lambda(x-\mu)}z + \Lambda^2z=x^2-u.\label{nf1curveagain}
\end{equation}
With a generic choice of $\Lambda$ and $\mu$, there are three singularities on the $u$-plane.
As we saw at the end of Sec.~\ref{sec:nf_1_curve}, 
two singularities collide at $u=3\Lambda^2$ when we set $\mu=-\frac32\Lambda$, see Fig.~\ref{fig:nf_1_adpoint}.
When $u\sim 3\Lambda^2$, three branch points of $x(z)$ collide at $z=-1$. 
Then, both the A-cycle and the B-cycle defining $a$ and $a_D$ can be taken to be small loops around $z=-1$.
This guarantees that both $a$ and $a_D$ are small. 
Therefore we simultaneously have very light electric and magnetic particles. 
Such a point on the Coulomb branch is called the Argyres-Douglas point. This was first identified in the case of pure $\SU(3)$ theory in \cite{Argyres:1995jj}, and extended to $\SU(2)$ theories with flavors in \cite{Argyres:1995xn}.

The monodromy $M_{AD1}$ around $u=3\Lambda^2$ can be found in various ways. One is to multiply the  monodromies of the two colliding singularities of the $N_f=1$ theory. Another is to follow how the three branch points move. Setting $u=3\Lambda^2+\delta u$, we find that the three branch points are at $z+1\propto \delta u^{1/3}$. This determines how the cycles are mapped, resulting in the monodromy.  In either method, we find \begin{equation}
M_{AD1} \sim \begin{pmatrix}
1 & 1 \\
-1 & 0
\end{pmatrix}.
\end{equation}
The transformation on the low energy coupling by $M_{AD1}$ is \begin{equation}
\tau \mapsto \tau'=\frac{1}{1-\tau}.
\end{equation}
Note that $\tau=e^{\pi i/3}$ is a fixed point of this transformation; by an explicit computation, we can check that 
$\tau-e^{\pi i/3}\propto \delta u^{1/3}$. 
We find that the coupling is pinned at this strongly-coupled value.

The low energy limit is believed to be conformal. To isolate the physics in this limit,
let us take \begin{equation}
\begin{aligned}
z&=-1+\tilde\delta z, &
x&=-\Lambda + \Lambda \delta x, &
u&=3\Lambda^2+ \Lambda^2 \tilde\delta u  , &
\mu&=-\frac 32\Lambda+\Lambda \tilde\delta\mu.
\end{aligned}
\end{equation}
and assume all variables prefixed with $\delta$ to be very small. 
It turns out that it is more useful to introduce further redefinitions \begin{equation}
\tilde \delta z = \delta z - (\delta z)^2,\quad
\tilde \delta u = \delta u - 2\delta \mu-2\delta x + (\delta z)^2,\quad
\tilde \delta \mu = \delta \mu + \delta x.\label{ADredefinition1}
\end{equation} 

Let us now plug  the relation \eqref{ADredefinition1} into the $N_f=1$ curve \eqref{nf1curveagain}. 
Expanding it to the third order in $\delta z$, we find the curve of the form \begin{equation}
(\delta x)^2-\delta u = (\delta z)^3 + 2\delta\mu \delta z.\label{nf1ADcurve}
\end{equation}
The differential is \begin{equation}
\lambda=x\frac{dz}{z}\sim \delta x d \delta z.
\end{equation}
As the integral $\int \lambda$ gives the mass of BPS particles, $\lambda$ itself should have the scaling dimension 1.
The relation \eqref{nf1ADcurve} means that the scaling dimensions $[\delta x]$ and $[\delta z]$ should satisfy \begin{equation}
[\delta x]:[\delta z]=3:2.
\end{equation} 
This fixes the scaling dimensions of all the  variables involved: \begin{equation}
[\delta x]=\frac35,\qquad [\delta z]=\frac25, \qquad
[\delta u]=\frac65,\qquad [\delta\mu]=\frac45.
\end{equation}
Note that the mass dimension, or equivalently the scaling dimension at the ultraviolet of the operator $u=\tr\Phi^2/2$ was 2.
We find that the anomalous dimension is of order one, reducing $[\delta u]$ significantly.

\begin{figure}[h]
\[
\inc{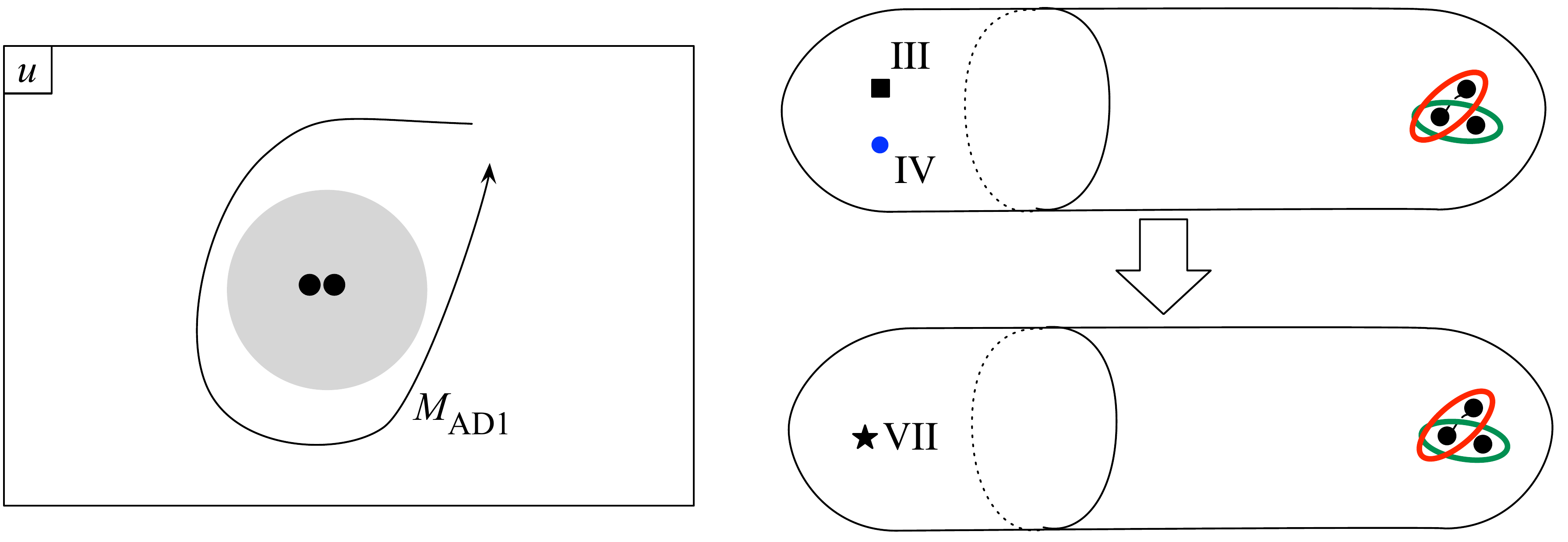}
\]
\caption{Argyres-Douglas theory $AD_{N_f=1}(\SU(2))$\label{fig:H1}}
\end{figure}
As we are taking the limit $\delta u\to 0$, we are zooming into the neighborhood of the u-plane around $u=3\Lambda^2$.
In the limit, we can think of the low energy theory to be described by a theory with only a singularity at $\delta u=0$,
as shown on the left hand side of Fig.~\ref{fig:H1}. We call the resulting theory the Argyres-Douglas CFT $AD_{N_f=1}(\SU(2))$.\footnote{Unfortunately, there is no accepted universal naming system for Argyres-Douglas theories in the literature. In this lecture note the author tries to provide one which might be more cumbersome than the ones in the literature but more explicit in distinguishing various constructions.}

Let us revisit this limiting procedure from the 6d point of view. We first write the original $N_f=1$ curve in the form $\lambda^2-\phi(z)=0$. Recall that $\phi(z)$ has one order-3 pole and one order-4 pole, 
as was studied in Sec.~\ref{sec:less} and shown in Fig.~\ref{fig:lessG}.
We also have three branch points of $\phi(z)$ on generic points.
Suppose that we tune the parameters carefully so that two poles of $\phi(z)$ collide: \begin{equation}
\phi(z)\sim (\frac{P_3(z)}{(z-\epsilon)^3} + \frac{P_4(z)}{z^4}){dz^2}
=\frac{P_7(z)}{(z-\epsilon)^3 z^4} {dz^2}
\to \frac{Q_7(z)}{ z^7} {dz^2} 
\end{equation} where $P_d$, $Q_d$ are generic polynomials of degree $d$ at this stage.
We end up having just one singularity with an order-7 pole, as shown on the right hand side of Fig.~\ref{fig:H1}.
To have no singularity at $z=\infty$, we see that $Q_7(z)$ should be in fact of degree 3: \begin{equation}
\lambda^2 = \frac{c+c'z+\mu z^2 + u z^3}{z^7} dz^2.
\end{equation} By the coordinate transformation $z\to z/(az-b)$, we can set $c=1$ and $c'=0$. We then have \begin{equation}
\lambda^2 = \frac{1+\mu z^2 + u z^3}{z^7} dz^2.
\end{equation}
As the left hand side is of scaling dimension $2$, we see that $[z]=-2/5$, and we conclude \begin{equation}
[\mu]=\frac45,\qquad [u]=\frac65\label{muu}
\end{equation} which agree with what we found above. 
Note that the variable $z$ is auxiliary, and therefore there is no reason for its dimension to match. 

In  general for any conformal field theory, any dynamical scalar operator $O$ should have scaling dimension larger  than or equal to one: \begin{equation}
[O]\ge 1,\label{generalCFTfeatures}
\end{equation} and the equality is only attained when $O$ describes a free decoupled scalar boson.
Then the operator $u$ with $[u]=6/5$ is a genuine operator in the theory $AD_{N_f=1}(\SU(2))$. The object $\mu$ is regarded as a parameter conjugate to $u$ in the following sense. In an $\cN{=}2$ theory, we can consider a deformation of the prepotential \begin{equation}
\int d^4\theta F \to\int d^4\theta (F   + mO).
\end{equation}  Here, $d^4\theta$ is the chiral $\cN{=}2$ superspace integral we briefly mentioned at the end of Sec.~\ref{sec:lel}, $O$ is an operator and $m$ is a parameter multiplying it.  In an $\cN{=}2$ superconformal theory, the combination $mO$ therefore needs to have a scaling dimension $2$. Then we should have \begin{equation}
[m]+[O]=2.
\end{equation} 

We see that the pair $\mu$ and $u$ satisfies this condition, see \eqref{muu}. We therefore regard $\mu$ as the deformation parameter corresponding to the operator $u$.

\subsection{Argyres-Douglas CFT from the $N_f=2$ theory  }

Consider the curve of the $N_f=2$ theory, \begin{equation}
\frac{2\Lambda(x-\mu)}{z}
+{2\Lambda(x-\mu)}{z}=x^2-u\label{firstform}
\end{equation} where we set the masses of the two flavors the same. 
We can also use the curve of the alternative form \begin{equation}
\frac{(x-\mu)^2}{z}
+{4\Lambda^2}{z}=x^2-u.\label{secondform}
\end{equation}
Its moduli space for generic $\mu$ was shown in Fig.~\ref{fig:nf_2_moduli}. 

\begin{figure}[h]
\[
\inc{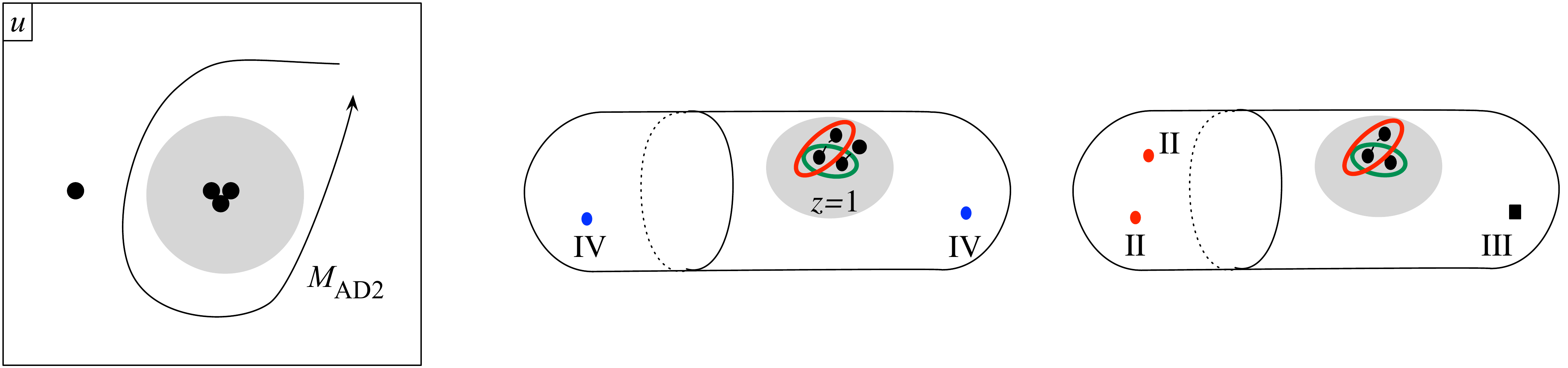}
\]
\caption{Argyres-Douglas point of $N_f=2$ theory\label{fig:nf_2_AD_plane}}
\end{figure}

When $\mu=0$, two singularities without the Higgs branch attached collide. Instead, let us tune the parameter $\mu$ so that the singularity with the Higgs branch collides with a singularity without, see Fig.~\ref{fig:nf_2_AD_plane}.
For definiteness let us use the first form of the curve.
Then this collision happens when $\mu=2\Lambda$, at $u=4\Lambda^2$. The four branch points then collide at $z=1$.
We find that the monodromy around the resulting singularity is \begin{equation}
M_{AD2}\sim \begin{pmatrix}
0 & 1\\
-1 & 0
\end{pmatrix},
\end{equation} acting on the coupling as \begin{equation}
\tau \to \tau'=-\frac{1}\tau.
\end{equation} The strong coupling value $\tau=i$ is the fixed point of this transformation, and the low-energy coupling approaches this value when we let $u\to 4\Lambda^2$. 

Expanding the variables as before, \begin{equation}
\begin{aligned}
z&=1+\delta z, &
x&=2\Lambda + \delta x, &
u&=4\Lambda^2+\delta u, &
\mu&=2\Lambda+\delta\mu,
\end{aligned}
\end{equation} we find that the curve  in the limit is 
\begin{equation}
(\delta x)^2+\delta u = (\delta z)^4 + \delta\mu \delta z^2+\Delta\mu \delta z \label{nf2ADcurve}
\end{equation} with the differential $\lambda\sim \delta x d\delta z$. Here we reinstated a small difference $\Delta\mu=\mu_1-\mu_2$ between the bare masses $\mu_1$, $\mu_2$ of two doublet hypermultiplets.
Demanding $\lambda$ to have scaling dimension 1, we see that \begin{equation}
[\delta x]=\frac23,\quad
[\delta z]=\frac13.
\end{equation} Then we find \begin{equation}
[\delta u]=\frac43,\quad
[\delta \mu]=\frac23,\quad
[\Delta\mu]=1.\label{scalingdimH2}
\end{equation}
We see again that $[\delta u]+[\delta \mu]=2$, and therefore $\delta\mu$ is a deformation parameter corresponding to the operator $\delta u$. 
$\Delta\mu$ is a mass parameter for the non-Abelian flavor symmetry $\SU(2)_F$. In general, in a conformal theory,  a non-Abelian flavor symmetry current $J^a$ should have scaling dimension 3.
The $\cN{=}2$ supersymmetry relates it to the mass term, which is given for a Lagrangian theory by the familiar term $Q\tilde Q$ and has scaling dimension 2. Therefore, the non-Abelian mass parameter of $\cN{=}2$ superconformal theory should always have scaling dimension 1. Our computation of $[\Delta\mu]$ is consistent with this general argument. We call this resulting theory $AD_{N_f=2}(\SU(2))$. 

\begin{figure}[h]
\[
\inc{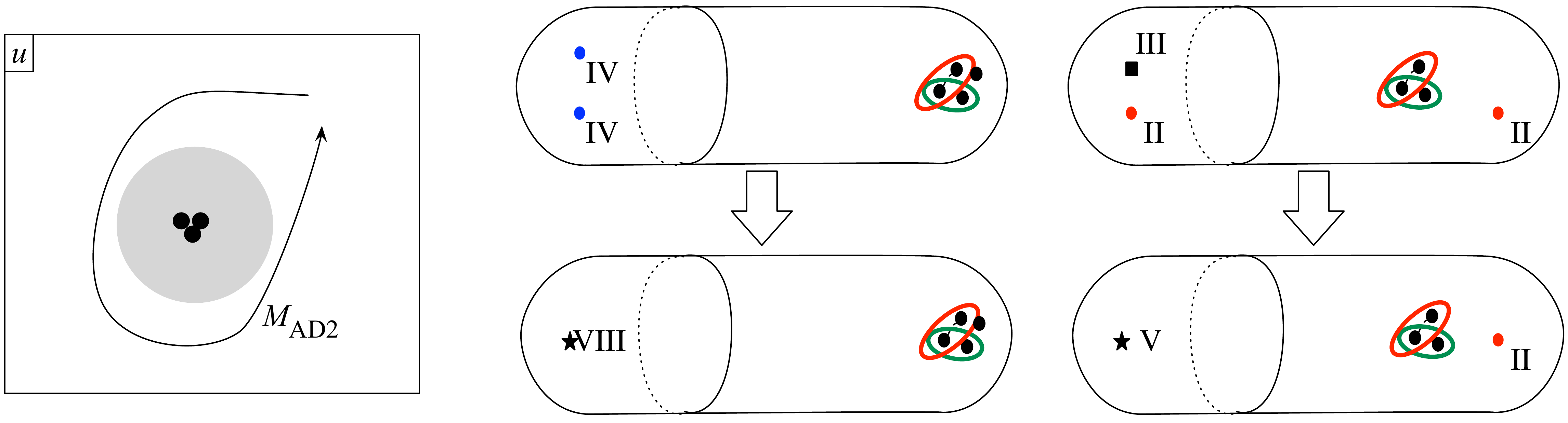}
\]
\caption{Argyres-Douglas theory $AD_{N_f=2}(\SU(2))$\label{fig:H2}}
\end{figure}
Let us study the limiting procedure of the $N_f=2$ theory from the 6d point of view. Before taking the limit, the curve
in the first form \eqref{firstform} was $\lambda^2=\phi(z)$
with two order-4 poles of $\phi(z)$.
We collide them, and we end up with a singularity of order 8. Just as in the analysis before, we conclude that the curve in the limit is given by \begin{equation}
\lambda^2=\frac{1+\mu z^2+\Delta\mu z^3+uz^4}{z^8}dz^2.
\end{equation} We easily see that $[z]=-1/3$. Then we find the same scaling dimensions as in \eqref{scalingdimH2}.

The curve in the second form \eqref{secondform}, when written as $\lambda^2=\phi(z)$, had two poles of order 2, and another of order 3. 
At the two order-two poles, the residues of $xdz/z$ are $\pm(\mu_1+\mu_2)/2$ and $\pm(\mu_1-\mu_2)/2$, respectively.
Let us collide an order-2 pole with the residue $\pm(\mu_1+\mu_2)/2$ and an order-3 pole to form a pole of order 5.
We end up having a $\phi(z)$ with one pole of order 5, say at $z=0$,
and another pole of order 2, with the residue $\pm(\mu_1-\mu_2)/2$, see Fig.~\ref{fig:H2}.
The curve in the limit can also be easily found: \begin{equation}
\lambda^2=\frac{1+\delta\mu z+\delta u z^2+ (\frac{\mu_1-\mu_2}{2})^2z^3 }{z^5}dz^2
\end{equation} The last coefficient was fixed by the condition at $z=\infty$. 
Demanding $\lambda$ to have scaling dimension 1, we see that $[z]=-2/3$, and \begin{equation}
[\delta \mu]=\frac23,\quad
[\delta u]=\frac43.
\end{equation} It is reassuring to find the same answer. 

\subsection{Argyres-Douglas CFT from the $N_f=3$ theory  }

\begin{figure}[h]
\[
\inc{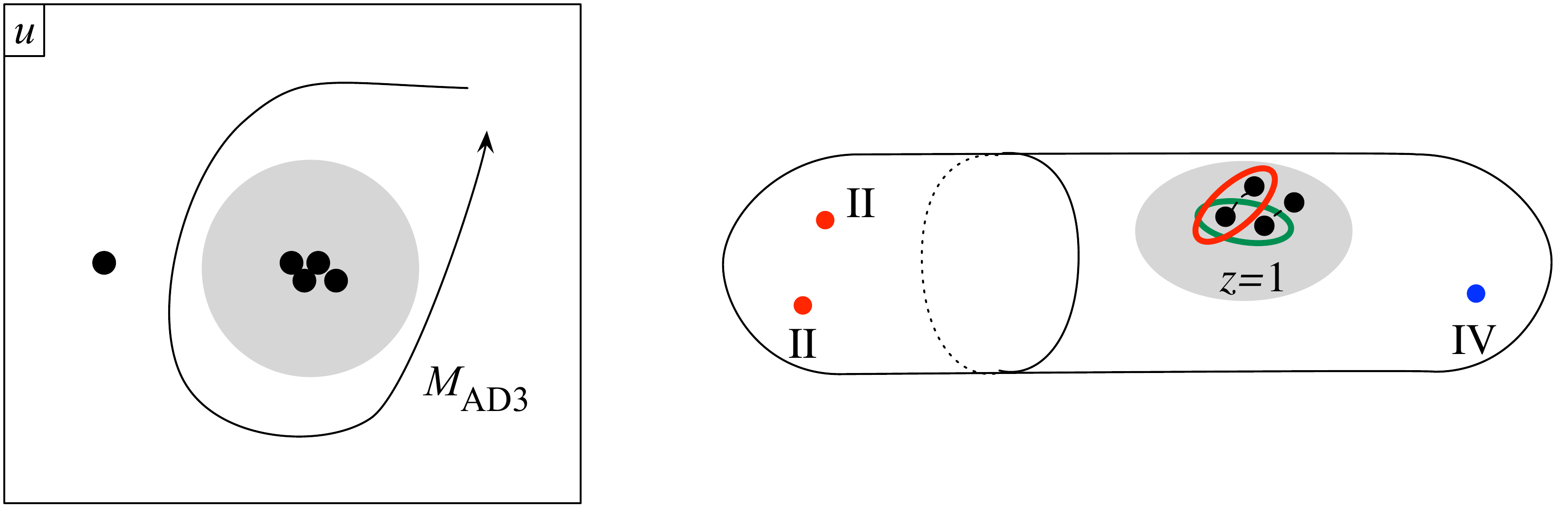}
\]
\caption{Argyres-Douglas point of $N_f=3$ theory\label{fig:nf_3_AD_uplane}}
\end{figure}

The special limit of $N_f=3$ theory can be found in exactly the same way. 
We start from the curve \eqref{nf3curve!} \begin{equation}
\frac{(x-\mu-\Lambda)^2}z + 2\Lambda(x-\mu-\Lambda)z = x^2-u
\end{equation} with the same mass for three flavors. 
On the $u$-plane, we have one singularity with the Higgs branch, and two singularities without. 
We tune $\mu$ so that singularity with the Higgs branch collides with another without, in a way that their monodromies do not commute. See Fig.~\ref{fig:nf_3_AD_uplane}.

The monodromy around the resulting singularities is \begin{equation}
M_{AD3}=\begin{pmatrix}
0& 1 \\
-1 & -1
\end{pmatrix}
\end{equation} with the action on the coupling given by \begin{equation}
\tau \mapsto \tau'=\frac{-\tau+1}{-\tau}.
\end{equation} The fixed point is at $\tau=e^{\pi i/3}$.

\begin{figure}[h]
\[
\inc{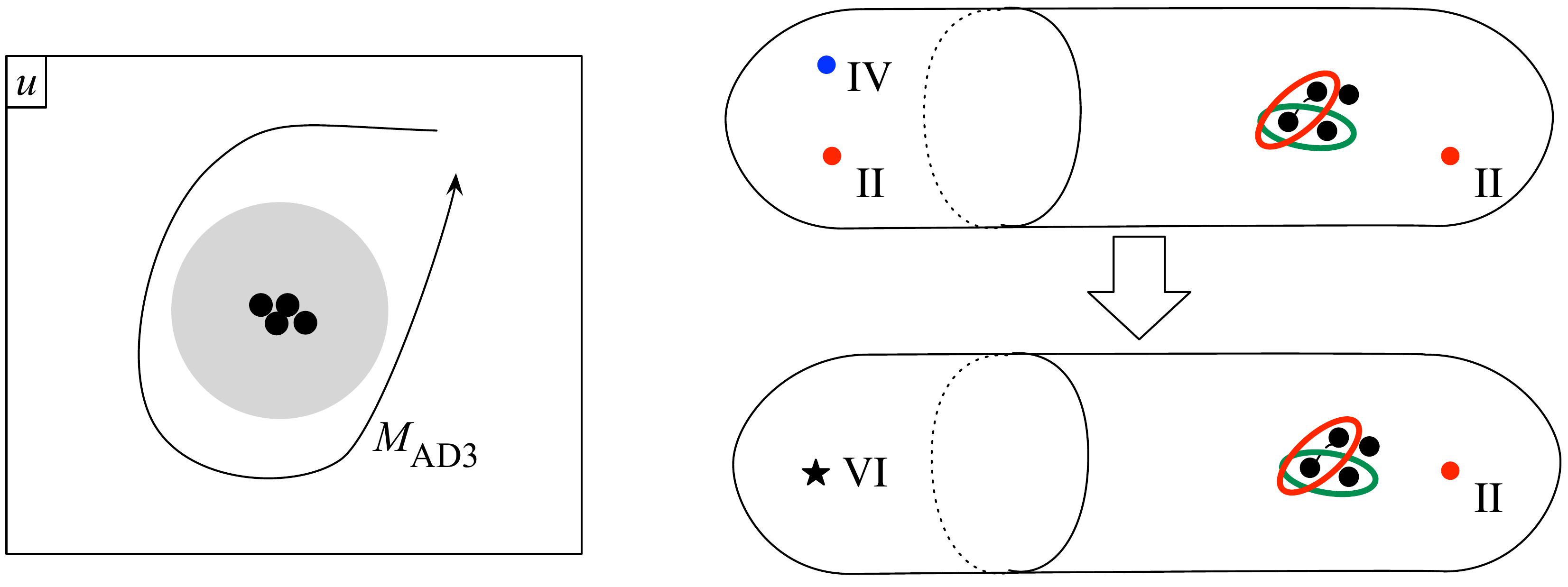}
\]
\caption{Argyres-Douglas theory $AD_{N_f=3}(\SU(2))$\label{fig:H3}}
\end{figure}
In the 6d description, we had two poles of order two and one pole of order four. We collide an order-2 pole and an order-4 pole, ending up with a pole of order six. The curve is then \begin{equation}
\lambda^2=\frac{1+\delta\mu z+\mu'z^2 + \delta u z^3+ (\frac{\mu_1-\mu_2}{2})^2z^4 }{z^6}dz^2
\end{equation} The differential $\lambda$ has scaling dimension 1. 
Then $[z]=-1/2$, and we find \begin{equation}
[\delta \mu]=\frac12,\quad
[\delta u]=\frac32,\quad
[\mu']=1,\quad
[\Delta\mu]=1,
\end{equation} where we defined $\Delta\mu=\mu_1-\mu_2$. 
Two parameters $\mu'$ and $\Delta\mu$ are of scaling dimension 1, and we 
identify them with  the mass parameters associated to the $\SU(3)$ flavor symmetry.
We also see $[\delta\mu]+[\delta u]=2$ again. We call this resulting theory $AD_{N_f=3}(\SU(2))$. 

\subsection{Summary of rank-1 theories}\label{sec:rank-1SCFT}

\subsubsection{Argyres-Douglas CFTs from $\SU(2)$ with flavors}
So far we studied the Argyres-Douglas CFTs which were obtained by special limits of $\SU(2)$ gauge theories with $N_f=1,2,3$ flavors. The data of these and other related CFTs are summarized in Table~\ref{tab:rank1sing}.
The Argyres-Douglas CFTs are the first three rows of the table. 
The fourth row is for the $\SU(2)$ theory with $N_f=4$ massless flavors.
The next two rows are for slightly different classes of theories.
Namely, if we consider $\SU(2)$ theory with more than $4$ flavors 
or $\U(1)$ theory with nonzero charged hypermultiplets, 
they are infrared free, see \eqref{su2running}  and \eqref{general_u1_running}. 
They  appeared repeatedly as a local behavior close to a singularity on the $u$-plane. 

In Table~\ref{tab:rank1sing} we also tabulated the dimension of the Higgs branch. 
Let us quickly recall how they are obtained. 
We know $AD_{N_f=1}(\SU(2))$ does not have one, since its parent theory $\SU(2)$ with $N_f=1$ does not have one either. 
For $AD_{N_f=2}(\SU(2))$, we consider $\SU(2)$ with $N_f=2$ with a $\U(1)$ mass term. Then the Higgs branch is $\bC^2/\bZ_2$, whose quaternionic dimension is 1. 
For $AD_{N_f=3}(\SU(2))$, we consider $\SU(2)$ with $N_f=3$. With a $\U(1)$ mass term, its Higgs branch can be found by studying a point $u=\mu_1^2=\mu_2^2=\mu_3^2$ in a weakly-coupled theory. The physics there is just $\U(1)$ with three charged hypermultiplets, with the Higgs branch of quaternionic dimension 2. 
For free $\SU(2)$ theory with $N_f\ge 4$ flavors, the quaternionic dimension is just $2N_f-\dim \SU(2)$, and similarly for $\U(1)$ theory with $N$ flavors it is given just by $N-\dim\U(1)$. 
One funny feature is that we see \begin{equation}
\dim_\bH (\text{Higgs branch}) = h^\vee(\text{flavor symmetry})-1\label{hf}
\end{equation} for the first six rows, where $h^\vee(G)$ is the dual Coxeter number, which is also a contribution to the one-loop running $C(\text{adj})$ from the adjoint representation of $G$. 
These theories have just one Coulomb branch modulus, and the low-energy theory on a generic point on the Coulomb branch is just a free $\U(1)$ theory. Such theories are called rank-1.

\begin{table}
\[
\begin{array}{c|cccccccccccc}
\text{name} & \text{monodromy} & \text{flavor} & [u] &  \# & \dim_{\bH}(\text{Higgs})\\
\hline
AD_{N_f=1}(\SU(2)) & \begin{pmatrix}
1 & 1 \\
-1 & 0
\end{pmatrix} & & 6/5 & 2 & \\
AD_{N_f=2}(\SU(2)) & \begin{pmatrix}
0 & 1 \\
-1 & 0
\end{pmatrix} & \SU(2) & 4/3 & 3 & 1\\
AD_{N_f=3}(\SU(2)) & \begin{pmatrix}
0 & 1 \\
-1 & -1
\end{pmatrix} & \SU(3) & 3/2 & 4 &  2\\
\SU(2)\ N_f=4 & \begin{pmatrix} 
-1 & 0 \\
0 & -1
\end{pmatrix} & \SO(8) & 2 & 6 &  5\\
\hline
\SU(2)\ N_f> 4 & \begin{pmatrix}
-1 & 4-N_f \\
0 & -1
\end{pmatrix} & \SO(2N_f) &    & N_f+2  & 2N_f-3\\
\U(1)\ \text{with}\ N\ \text{flavors} & \begin{pmatrix}
1 & N \\
0 & 1
\end{pmatrix} & \SU(N) &   & N  & N-1\\
\hline
MN(E_6) & \begin{pmatrix}
-1 & -1 \\
1 & 0
\end{pmatrix}& E_6 & 3 & 8 &  11\\
MN(E_7) & \begin{pmatrix}
0 & -1 \\
1 & 0
\end{pmatrix}& E_7 & 4 & 9  &  17\\
MN(E_8) & \begin{pmatrix}
0 & -1 \\
1 & 1
\end{pmatrix} & E_8 & 6 & 10 &  29
\end{array} 
\]
\caption{Data of various rank-1 CFTs.  
$\#$ is the number of singularities colliding at $u=0$,
and $\dim_\bH \text{Higgs}$ is the quaternionic dimension of the Higgs branch, i.e.~the real dimension $/4$. 
\label{tab:rank1sing}}
\end{table}

\begin{table}
\[
\begin{array}{c||c|c|c|c|c}
\Gamma & \bZ_n & \widehat {\mathcal{D}}_{n-2} & \widehat{\mathcal{T}} & \widehat{\mathcal{O}} & \widehat{\mathcal{I}}\\
\hline
G_\Gamma & \SU(n) & \SO(2n) & E_6 & E_7 & E_8
\end{array}
\]
\caption{Finite subgroups $\Gamma$ of $\SU(2)$ and simply-laced Lie groups $G_\Gamma$. Here, $\mathcal{D}_{n}$ is the dihedral group acting on the regular $n$-gon,
$\mathcal{T}$, $\mathcal{O}$, $\mathcal{I}$, are the tetra-, octa-, and icosahedral groups,
and the hat above them are the lift from $\SO(3)$ to $\SU(2)$. The resulting group $\widehat{\mathcal{T}}$ is called the binary tetrahedral group, for example. 
 \label{tab:ADE}}
\end{table}

\subsubsection{Exceptional theories of Minahan-Nemeschansky}\label{sec:F}
We have not discussed the theories listed in  the remaining three rows. 
One way to motivate them is to refer to a classical mathematical result of Kodaira. 
At a given point on the $u$-plane, we have the \Gaiotto\ curve $C$ and the \SeibergWitten\ curve $\Sigma$. 
The curve $\Sigma$ is a torus, whose shape is parameterized by its complex structure $\tau$, which depend
holomophically on $u$. 
Therefore we have a fibration of torus over the complex plane with the coordinate $u$.
The $u$-plane together with the fiber $\Sigma$ forms a complex two-dimensional space $X$.

Kodaira classified the possible types of singularities of such fibrations, and the first six rows of Table~\ref{tab:rank1sing} is an exact copy of part of that classification. The terminologies are of course different,  since he was a mathematician and we are studying $\cN{=}2$ gauge theories. Kodaira's classification had three more rows in addition to the first six rows, which motivated people that there should be three additional theories corresponding to them. The \SeibergWitten\ curves for these were constructed first by Minahan and Nemeschansky in \cite{Minahan:1996fg,Minahan:1996cj}.

In the mathematical language, a singularity in the torus fibration creates a singularity in the total space $X$ of complex dimension two. It is locally of the form $\bC^2/\Gamma$ where $\Gamma$ is a finite subgroup of $\SU(2)$. 
They have a natural ADE classification, and we can associate a Lie group $G_\Gamma$, see Table~\ref{tab:ADE}.

Mathematicians associate this group $G_\Gamma$ purely mathematically to a torus fibration,
and we see that they are   exactly the flavor symmetries of the gauge theories, at least to the first six. 
Mathematicians have associated exceptional groups $E_{6,7,8}$ to the last three cases. 
It was thus quite tempting that the putative theories which correspond to  the last three rows have these exceptional groups as the flavor symmetries. 
From the feature \eqref{hf} relating the flavor symmetry and the dimension of the Higgs branch, it is also tempting to guess the dimension of the Higgs branch of these theories. 
We call these CFTs $MN(E_6)$, $MN(E_7)$ and $MN(E_8)$, respectively. 

Note that it is rather hard to have an exceptional flavor symmetry in a classical Lagrangian $\cN{=}2$ theory.
We already know a general form of the Lagrangian: the superpotential as an $\cN{=}1$ theory is forced to be \begin{equation}
\sum_i \int Q_i\Phi \tilde Q_i,
\end{equation} and it is possible to check explicitly that the flavor symmetry visible in the ultraviolet is a product of $\SU$, $\SO$ and $\Sp$ groups \cite{McOrist:2013bga}.  Therefore, if the exceptional symmetries are to appear, they need to arise via strong-coupling effects. 
Once the reader comes to Sec.~\ref{sec:application} of this note, s/he will find exactly how this happens in the field theory setting.

Another way to construct the theories listed in the table uniformly is to use Type IIB string theory and its non-perturbative version F-theory. This approach originates in \cite{Banks:1996nj} for $\SU(2)$ with four flavors. For the general case, see e.g.~\cite{DeWolfe:1998bi}.  The \SeibergWitten\ curves of these rank-1 theories can be constructed most uniformly in this approach, see e.g.~\cite{Noguchi:1999xq}.

The type IIB string theory is ten-dimensional, and it has objects called 7-branes and 3-branes, where a $p$-brane extends along $p$ spatial direction and one time direction. 
Let us say the spacetime is of the form \begin{equation}
\bR^{1,3}\times \bR^2 \times \bR^4.
\end{equation} Put a 7-brane in the subspace \begin{equation}
\bR^{1,3}\times \{0\} \times \bR^4\label{7b}
\end{equation} and a D3-brane in the subspace \begin{equation}
\bR^{1,3}\times \{u\} \times \{0\}.
\end{equation}
There are various types of 7-branes in F-theory, corresponding to Kodaira's classification. 
They can all be obtained by taking a number of the simplest of the 7-branes, called $(p,q)$ 7-branes, separated along the $\bR^2$ direction and collapsing them at one point.  
Then the low-energy theory on the D3-brane gives the corresponding $\cN{=}2$ theories.

Due to its tension, one $(p,q)$ 7-brane creates deficit angles $\pi/6$. With $n$ $(p,q)$ 7-branes collapsed to a point, the remaining angle is $1-n/12$ of $2\pi$.  From this the scaling dimension of $u$ can be computed to be \begin{equation}
u=\frac{12}{12-n}, 
\end{equation}which explains an interesting pattern in Table~\ref{tab:rank1sing}.
These 7-branes obtained by collapsing a number of $(p,q)$ 7-branes has a gauge symmetry $F$ living on its eight-dimensional worldvolume. 
From the point of view, this gauge symmetry $F$ on the 7-brane becomes a flavor symmetry. 
The D3-brane can be absorbed into this 7-brane as an instanton in the internal $\bR^4$ direction of \eqref{7b}. Then, the Higgs branch should be given by the one instanton moduli space of the group $F$. The $k$-instanton moduli space of a group $F$ has quaternionic dimension $kh^\vee(F)-1$, explaining the relation \eqref{hf}.

\subsubsection{Newer rank-1 theories}
So far we saw that the structure of rank-1 theories closely follows that of the Kodaira classification, listed in Table~\ref{tab:rank1sing}. Before going further, it should be mentioned that there are even more rank-1 theories, first found through the analysis of S-dualities of various gauge theories in \cite{Argyres:2010py}. 
Their properties are reviewed from the point of view of the 6d construction in Sec.~7 of \cite{Chacaltana:2012ch}.

\subsection{More general Argyres-Douglas CFTs: $X_N$ and $Y_N$}\label{sec:generalAD}

\begin{figure}[h]
\[
\inc{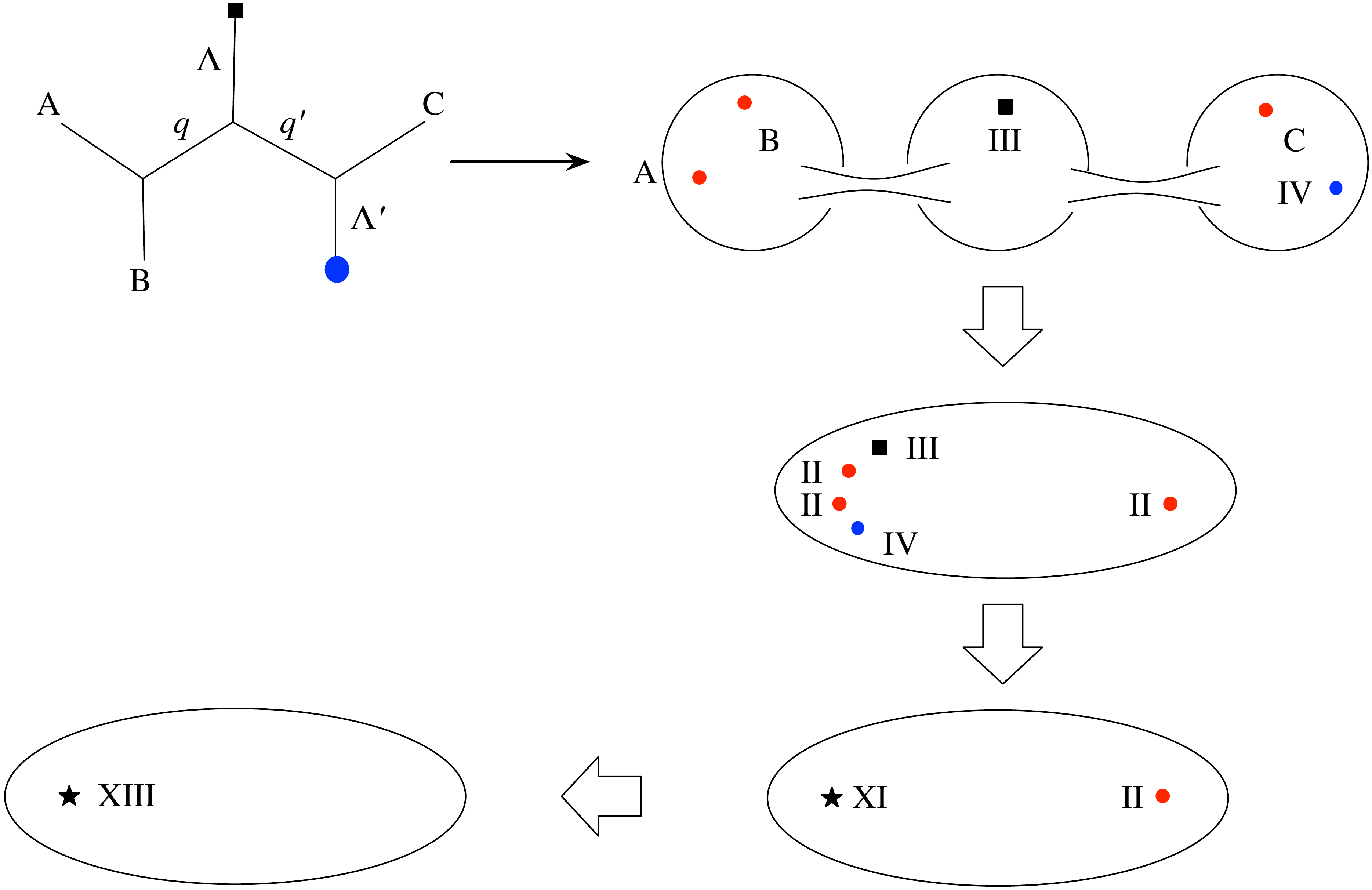}
\]
\caption{A higher Argyres-Douglas theory\label{fig:higherAD}}
\end{figure}

Let us switch gears and consider other Argyres-Douglas CFTs obtained from more complicated gauge theories with gauge group of the form $\SU(2)^n$. As an example, consider  a rather complicated theory with gauge group $\SU(2)^4$ studied at the end of Sec.~\ref{sec:less}. By performing the same limiting procedure we did in the $\SU(2)$ theory with $N_f=1,2,3$, we can consider the theory described by $\lambda^2-\phi(z)=0$
where $\phi(z)$ can have poles of very high order. The examples shown in Fig.~\ref{fig:higherAD} have
either just one pole of order 13 or one order-9 pole and an order-11 pole. 
They describe complicated 4d $\cN{=}2$ supersymmetric conformal field theories.

\begin{figure}[h]
\[
\begin{array}{cc}
\inc{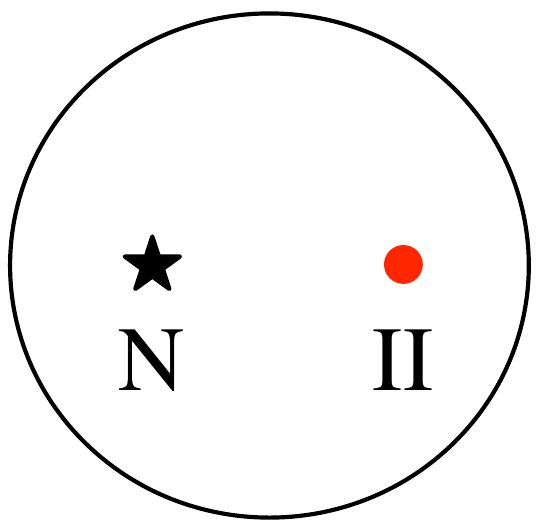} & \inc{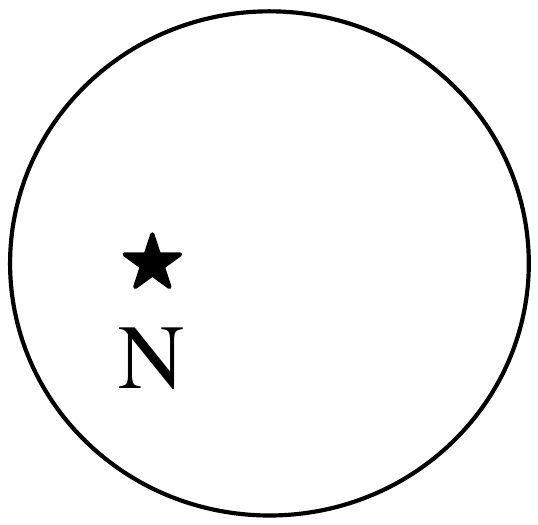} \\
X_N & Y_N
\end{array}
\]
\caption{The theory $X_N$ and the theory $Y_N$. \label{fig:XNYN}}
\end{figure}

Let us introduce names to these theories.  The $X_N$ theory is  the superconformal field theory corresponding to a sphere with one regular puncture and a puncture with an order-$N$ pole, 
and the $Y_N$ theory is  the superconformal field theory corresponding to a sphere with just a puncture with an order-$N$ pole, see Fig.~\ref{fig:XNYN}.
As can be seen from Fig.~\ref{fig:H1}, Fig.~\ref{fig:H2} and Fig.~\ref{fig:H3}, we know \begin{equation}
Y_7=AD_{N_f=1}(\SU(2)),\qquad
Y_8=AD_{N_f=2}(\SU(2))=X_5,\quad
AD_{N_f=3}(\SU(2))=X_6.\label{accidents}
\end{equation}

\begin{figure}[h]
\[
\inc{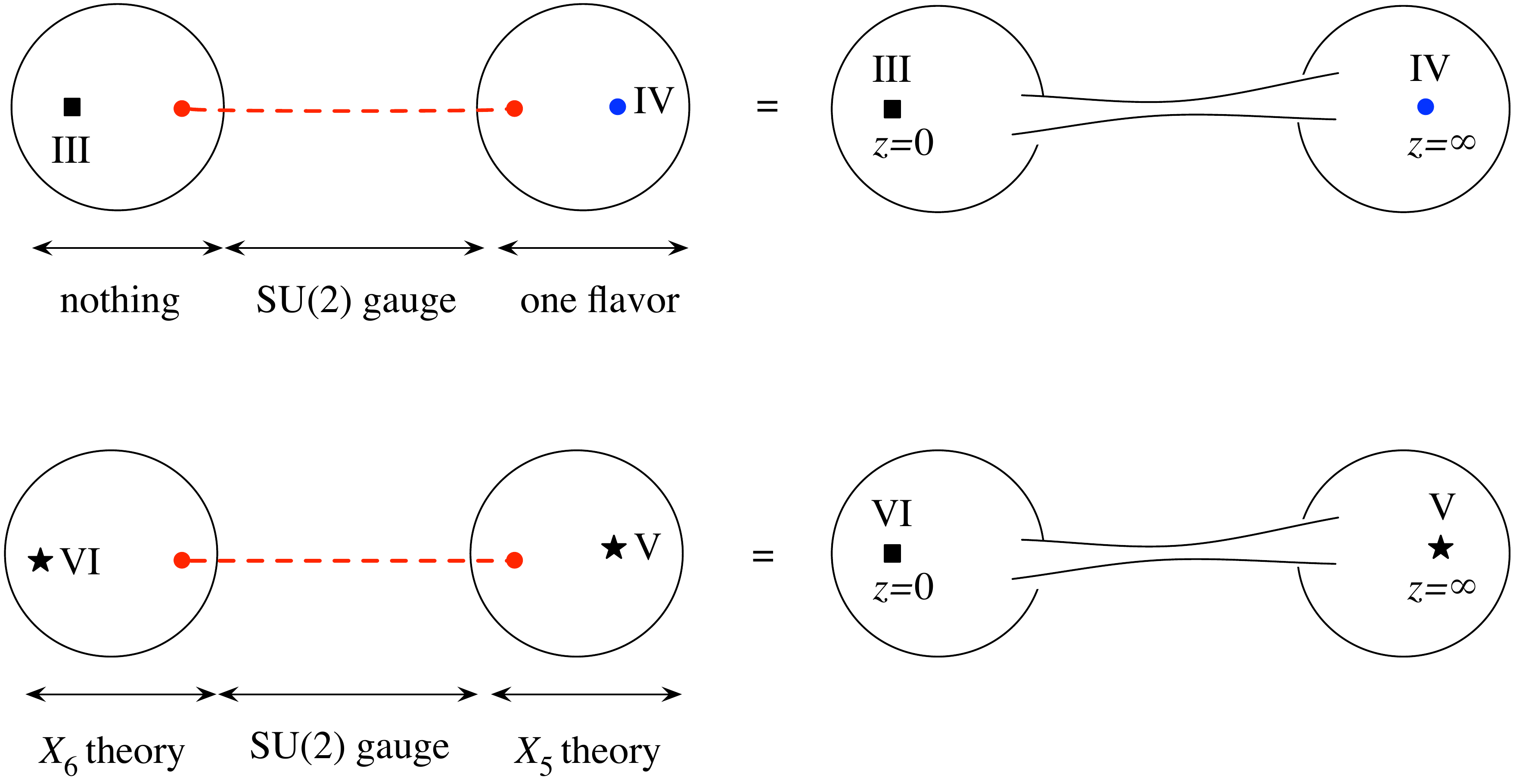}
\]
\caption{Ordinary and wild gauge theories\label{fig:wild}}
\end{figure}

Also, recall the construction of the $\SU(2)$ theory with one flavor given in Fig.~\ref{fig:lessG}.
There, a sphere with a regular puncture and a puncture of pole order 3 served as an empty boundary condition, and 
a sphere with a regular puncture and a puncture of pole order 4 behaves as a free hypermultiplet in the doublet of $\SU(2)$. Equivalently, we see \begin{equation}
X_3=\text{an empty theory},\qquad
X_4=\text{free hypermultiplet in the doublet of $\SU(2)$}.
\end{equation}
We depicted them in the first row of Fig.~\ref{fig:wild}. 

More generally,  we can have a two-punctured sphere with poles of arbitrary order $N$ and $N'$.
One example with $N=6$ and $N'=5$ is shown in the second row of Fig.~\ref{fig:wild}.
It can be understood as an $\SU(2)$ gauge theory with somewhat unusual matter contents, described by two strongly-interacting CFTs $X_N$ and $X_{N'}$.
Note that an order-2 pole always carries an $\SU(2)$ flavor symmetry, 
and therefore the $X_N$ theory always has an $\SU(2)$ flavor symmetry. 
The $\SU(2)$ gauge symmetry coming from the tube couples these two theories. 
This type of  gauge theory with $X_N$ as part of its matter contents is often called a wild gauge theory.

It is straightforward to find the running of the coupling of this theory. Assume  $a$ is very big, as always. The branch points of $\lambda^2=\phi(z)$ 
is around where \begin{equation}
\frac{\Lambda^2}{z^N} dz^2 \sim \frac{udz^2}{z^2} \quad\text{or}\quad
\Lambda^2z^{N'} dz^2 \sim \frac{udz^2}{z^2}.
\end{equation} Then they are around \begin{equation}
z_-\sim \left(\frac{\Lambda}{a}\right)^{2/(N-2)},\qquad
z_+\sim \left(\frac{a}{\Lambda}\right)^{2/(N'-2)}.
\end{equation} We find \begin{equation}
a_D\sim \frac{2}{2\pi i}\int_{z_+}^{z_-} x\frac{dz}z\sim
- \frac{2}{2\pi i} (\frac{2}{N-2}+\frac{2}{N'-2})a\log \frac{a}{\Lambda}.
\end{equation}
This means that the one-loop running is given by \begin{equation}
\Lambda\frac{d}{d\Lambda} \tau =\frac{2}{2\pi i}( b_N+b_{N'}-4)
\end{equation} where \begin{equation}
b_N=2-\frac{2}{N-2}.
\end{equation} 

\begin{figure}[h]
\[
\inc{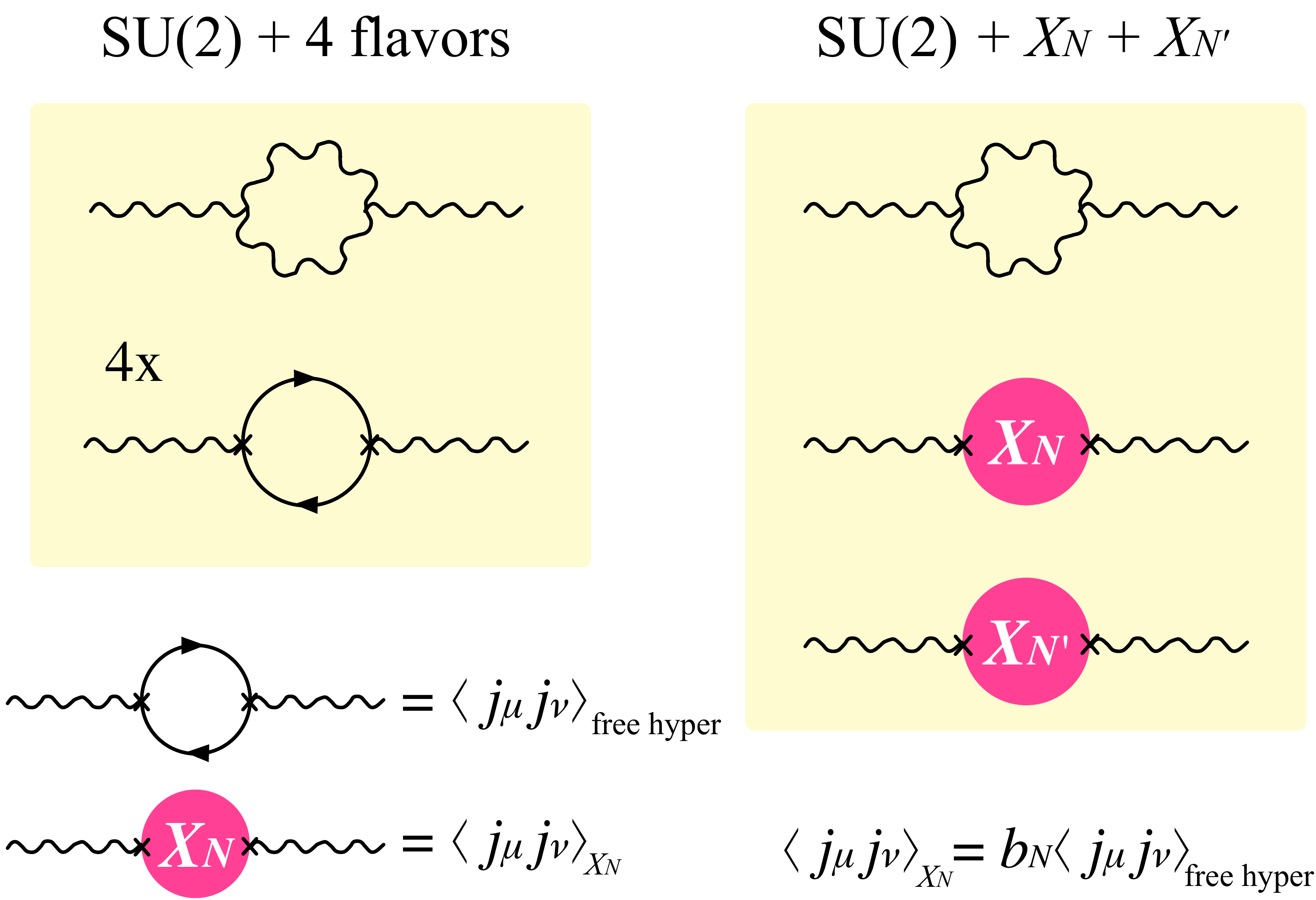}
\]
\caption{The running of the coupling measures the two-point correlator of the currents.\label{fig:cc}}
\end{figure}

The contribution to the one-loop running from one doublet hypermultiplet is given by $b=1$. Then
this $b_N$ can be roughly thought of as an effective number of doublet hypermultiplets,
carried by the theory $X_N$. More precisely, $b_N$ measures the coefficient of the correlator of the symmetry current $j_\mu$ of the $\SU(2)$ flavor symmetry, see Fig~\ref{fig:cc}. As shown there, for $\SU(2)$ with flavors, the running of the gauge coupling is caused by the loop of gauge multiplets (shown as wavy lines) or of hypermultiplets (shown as straight lines) coupled to the gauge fields via the $\SU(2)$ current operator $j_\mu$. Then the contribution to the one-loop running measures $\vev{j_\mu j_\nu}$. The fact that the $X_N$ theory contributes $b_N$ times a free flavor does means that \begin{equation}
\vev{j_\mu j_\nu}_{X_N}=b_N \vev{j_\mu j_\nu}_\text{free hyper in a doublet of $\SU(2)$}.\label{GOO}
\end{equation}
Recall that $X_3$ is just empty and $X_4$ is one free hypermultiplet in the doublet of flavor $\SU(2)$.
Our general formula correctly reproduces $b_3=0$ and $b_4=1$.

In the next section we will see that a singular limit of the pure $\SU(N)$ gauge theories 
becomes the theory $Y_{N+4}$,  whereas 
a singular limit of the pure $\SO(2N)$ gauge theories 
becomes the theory $X_{N+2}$. 
We will also see that $\SU(N)$ gauge theories with two flavors have a singular limit given by $X_{N+3}$.

Let us denote the Argyres-Douglas CFTs obtained from the pure $G$ gauge theory as $AD_{N_f=0}(G)$,
and the Argyres-Douglas CFTs obtained from the $\SU(N)$ with two flavors as $AD_{N_f=2}(\SU(N))$.
Then we can succinctly express these equivalences as \begin{equation}
\begin{array}{r@{}ll}
AD_{N_f=0}&(\SU(N))&=Y_{N+4},\\
AD_{N_f=0}&(\SO(2N))&=X_{N+2},\\
AD_{N_f=2}&(\SU(N))&=X_{N+3}.
\end{array}\label{ppp}
\end{equation}

We have already seen in \eqref{accidents}
that $AD_{N_f=2}(\SU(2))$, the Argyres-Douglas theories arising from $\SU(2)$ with $N_f=2$ flavors,
is equivalent to both $X_5$ and $Y_8$.
This coincidence is  a manifestation of the equivalence $\SU(4)\simeq \SO(6)$ from the point of view of \eqref{ppp}.
Also, consider the pure $\SO(4)$ theory, which is two copies of the pure $\SU(2)$ theory.
Its most singular point is where both copies are at the monopole point, thus realizing two free hypermultiplets.
Indeed, this has an $\SU(2)$ flavor symmetry, and is a doublet under it, realizing the fact \begin{equation}
AD_{N_f=0}(\SO(4))=X_{4}=\text{a free hypermultiplet in the doublet of $\SU(2)$}.
\end{equation}
These equations are rather interesting to the author, in the sense that they are equalities among the quantum field theories, not among the physical quantities in a single quantum field theory. 

\section{Theories with other simple gauge groups }\label{sec:higherrank}
We have spent so many pages to study $\cN{=}2$ gauge theories with gauge group $\SU(2)$. In this section we move on to the analysis of larger gauge groups. We will first study $\SU(N)$ gauge theories in some detail, and then go on to  the case $\SO(2N)$.  We also analyze the Argyres-Douglas CFTs obtained from these gauge theories, and show that they are given by the theories $X_N$ and $Y_N$ introduced in Sec.~\ref{sec:generalAD}.  
We close the section by briefly mentioning the Seiberg-Witten solutions to theories with other gauge groups in Sec.~\ref{sec:others}.

The curves that will be presented in this section might not be in the form most commonly found in the older literature. The relation between them would also be explained  in Sec.~\ref{sec:others}.

\subsection{Semiclassical analysis}
Let us consider $\SU(N)$ gauge theory with $N_f$ hypermultiplets in the fundamental $N$-dimensional representation. 
The $\cN{=}2$ vector multiplet consists of the $\cN{=}1$ adjoint chiral multiplet $\Phi$ and the $\cN{=}1$ vector multiplet $W_\alpha$, both $N\times N$ matrices. The hypermultiplets, in terms of $\cN{=}1$ chiral multiplets, can be written as \begin{equation}
Q_i^a, \quad \tilde Q^i_a, \qquad a=1,\ldots,N;\quad i=1,\ldots, N_f.
\end{equation}
One branch of the supersymmetric vacuum is given by the condition \begin{equation}
[\Phi,\Phi^\dagger]=0.
\end{equation} This means that $\Phi$ can be diagonalized. We denote it by \begin{equation}
\Phi=\diag(a_1,\ldots,a_N),\qquad \sum a_i=0.
\end{equation}

Let us consider a generic situation when $a_i\neq a_j$ for all $i\neq j$. Then the gauge group is broken from $\SU(N)$ to $\U(1)^{N-1}$. The W-boson mass is given by \begin{equation}
M_W=|a_i-a_j|
\end{equation} for the W-boson coming from the entry $(i,j)$ of the $N\times N$ matrix. 
As for the monopole, it is known that the 't Hooft-Polyakov monopole solution for the breaking from $\SU(2)$ to $\U(1)$ can be directly regarded as a solution for the breaking from $\SU(N)$ to $\U(1)$, by choosing $2\times 2$ submatrices of $N\times N$ matrices, given by picking the entries at the positions $(i,i)$, $(i,j)$, $(j,i)$ and $(j,j)$ for $i\neq j$. The masses are then \begin{equation}
M_\text{monopole}=|\tau_{UV}(a_i-a_j)|.
\end{equation}

The one-loop running is \begin{equation}
\LambdaRG \frac{d}{d\LambdaRG} \tau=\frac{i}{2\pi} (2N-N_f).
\end{equation} Then the theory is asymptotically free when $0\le N_f < 2N$. The dynamical scale is then \begin{equation}
\Lambda^{2N-N_f} := \Lambda^{2N-N_f}_{UV} e^{2\pi i \tau_{UV}}.
\end{equation} 
When $N_f=2N$, the theory is asymptotically conformal, and $\tau_{UV}$ is a dimensionless parameter in the quantum theory. 

When there are flavors, the $\cN{=}1$ superpotential in this vacua is \begin{multline}
\sum_i (Q_i\Phi\tilde Q^i -\mu_i Q_i \tilde Q^i)
=\\ 
\sum_i (Q_i^1,Q_i^2,\ldots, Q_i^N) 
\begin{pmatrix}
a_1 -\mu_i \\
& a_2-\mu_i \\
&&\ddots\\
&&& a_N-\mu_i
\end{pmatrix}
\begin{pmatrix}
\tilde Q^i_1\\
\tilde Q^i_2\\
\vdots\\
\tilde Q^i_N
\end{pmatrix}.
\end{multline}
Then we have one massless charged hypermultiplet component whenever we have $a_i-\mu_s=0$ for some $i$ and $s$.

In the strongly-coupled quantum theory, the definition of $a_i$ as the diagonal entries of the gauge-dependent quantity $\Phi$ does not make much sense. Instead, as we did in the case of $\SU(2)$ gauge theory, we define $a_i$ as the complex numbers entering in the BPS mass formula: \begin{equation}
M\ge | n^i a_i + m_i a_D^i + \sum_s f_s \mu_s |
\end{equation} where $(n^i,m_i)$ are the electric and the magnetic charges under $\U(1)^{N-1}$ infrared gauge group,
and $f_s$ are the flavor charges. 
We can also consider gauge-invariant combinations of $\Phi$ defined as \begin{equation}
x^N+ u_2 x^{N-2} + \cdots + u_{N-1} x + u_N := \vev{\det(x-\Phi)}
\end{equation} where $x$ is a dummy variable. 
For $N=2$, we had $\Phi\sim \diag(a,-a)$ and therefore $u_2=-a^2$ up to quantum corrections. 
Similarly, for general $N$, $u_k$ is the degree $k$ elementary symmetric polynomials of the variables $a_1$, \ldots,  $a_N$ up to quantum corrections. Our task then is to determine the mapping between $(u_2,\ldots, u_N)$ and $(a_1,\ldots, a_N)$ including the quantum corrections.

\subsection{Pure $\SU(N)$ theory}\label{sec:pureSUN}
\subsubsection{The curve}
Without further ado, let us introduce the \SeibergWitten\ curves. 
First, the \SeibergWitten\ curve for the pure $\SU(N)$ theory is given by \begin{equation}
\Sigma: \qquad \frac{\Lambda^N}z +\Lambda^N z = x^N+u_2 x^{N-2}+\cdots + u_N\label{puresuNcurve}
\end{equation} with the differential $\lambda=xdz/z$ as always. 
The \Gaiotto\ curve $C$ is just a sphere with the complex coordinate $z$.
At each point on the \Gaiotto\ curve $z$, we have $N$ solutions to the equation above.
Therefore, $\Sigma$ is an $N$-sheeted cover of $C$.

\begin{figure}[h]
\[
\inc{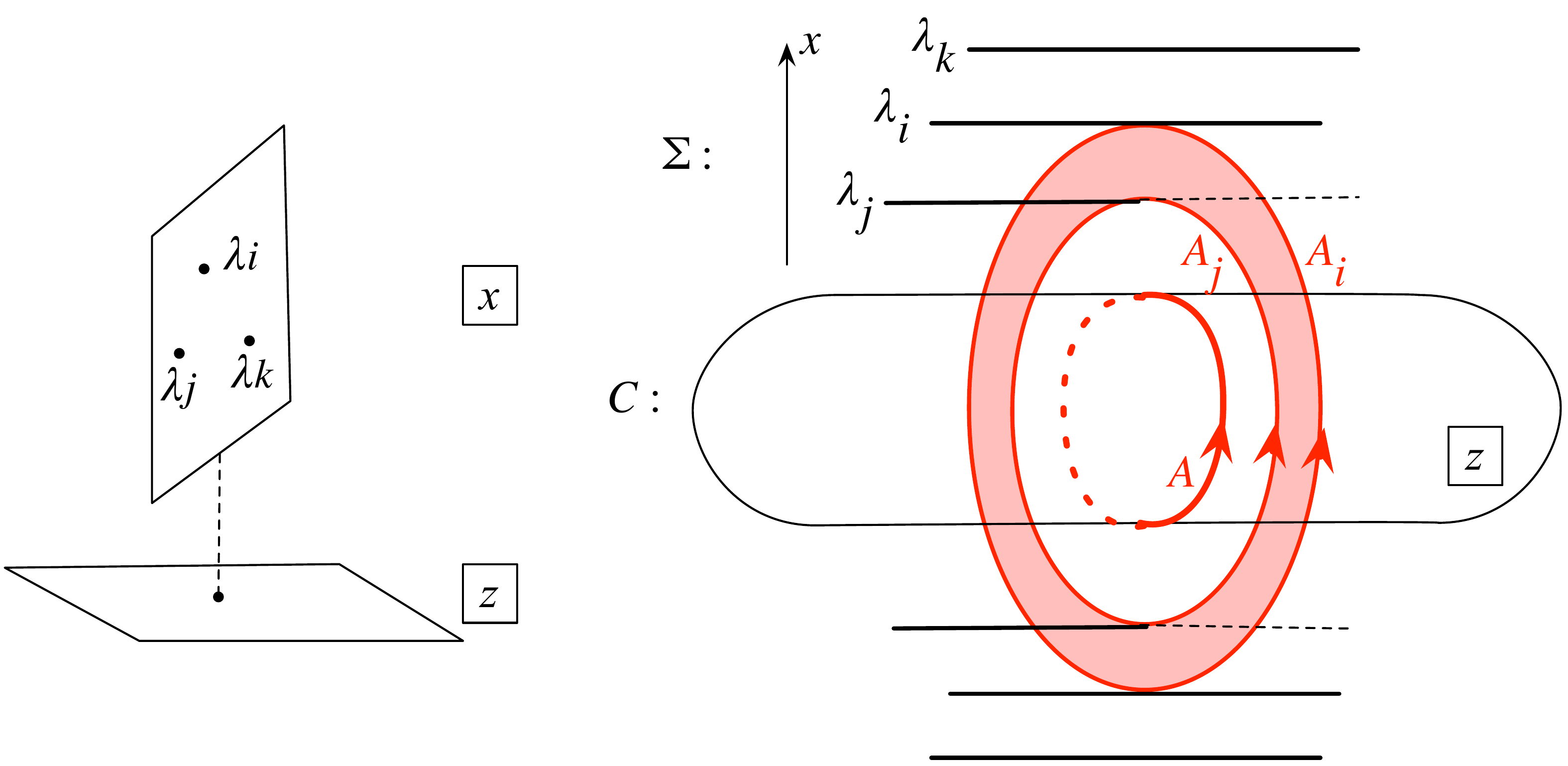}
\]
\caption{W-boson of the $\SU(N)$ theory\label{fig:SUN-W}}
\end{figure}
Let us check that this curve reproduces the semiclassical behavior. 
We introduce variables $\underline{a}_i$ via \begin{equation}
x^N+u_2 x^{N-2}+\cdots + u_N=\prod_i (x-\underline{a}_i).
\end{equation}

We declare the A-cycle on the \Gaiotto\ curve to be the unit circle $|z|=1$.
As the \SeibergWitten\ curve is an $N$-sheeted cover, we can lift this curve to each sheet, which we call the cycle $A_i$.
Assume we are in the regime $|\underline{a}_i| \sim \LambdaRG  $ independent of $i$, and  $\LambdaRG  \gg \Lambda$.
Then, we can approximately solve \eqref{puresuNcurve} by \begin{equation}
x_i(z)=\underline{a}_i + O(\Lambda).
\end{equation} It is more convenient to regard $\lambda=xdz/z$ itself to be the coordinate of the sheets. Then we have \begin{equation}
\lambda_i=\underline{a}_i \frac{dz}z +O(\Lambda).
\end{equation} The situation is shown in Fig.~\ref{fig:SUN-W}. 
The integral of $\lambda$ on the cycle $A_i$ is easy to evaluate: \begin{equation}
a_i:=\frac{1}{2\pi i}\oint_{A_i} \lambda=\underline{a}_i + O(\Lambda).\label{suNpureA}
\end{equation}

Now we can suspend a ring-shaped membrane suspended between the $i$-th sheet and the $j$-th sheet. The mass of this object is \begin{equation}
|\frac{1}{2\pi i}\oint_{A_i}\lambda-
\frac{1}{2\pi i}\oint_{A_j}\lambda|
=|\frac{1}{2\pi i}\oint_{A}(\lambda_i-
\lambda_j)|=|a_i-a_j| .
\end{equation} This reproduces the mass of the W-boson. 

\begin{figure}[h]
\[
\inc{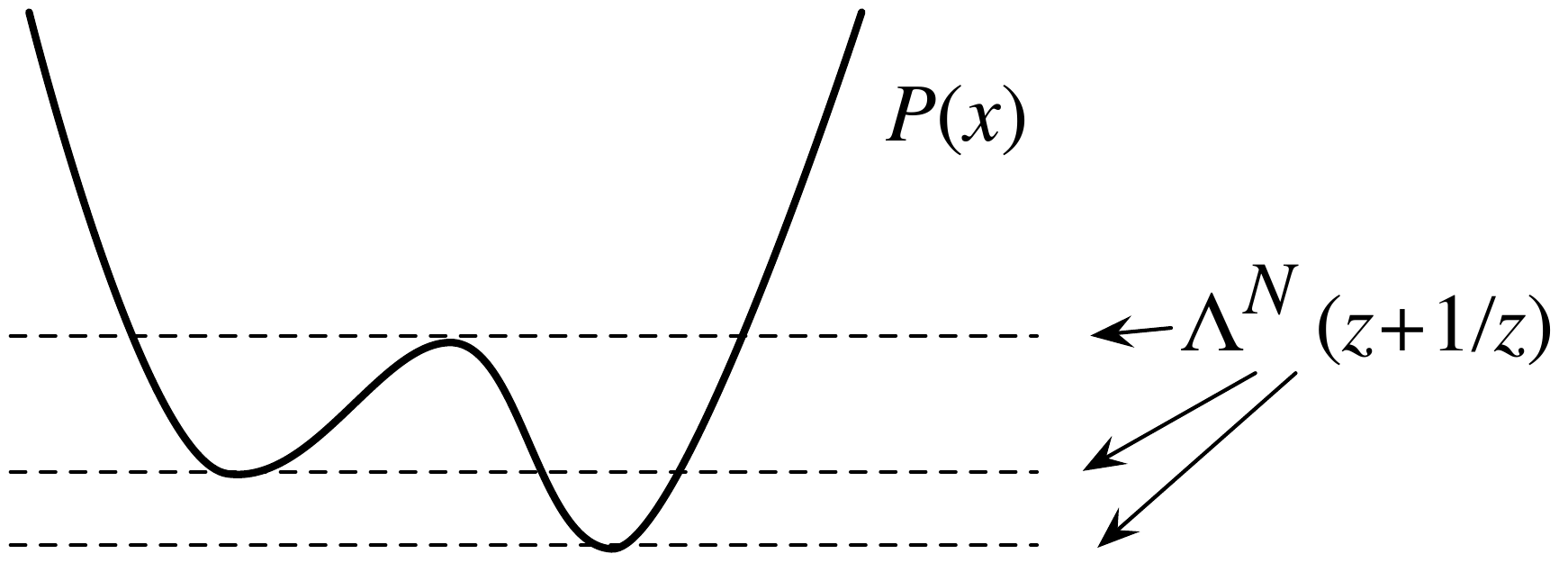}
\]
\caption{There are $(N-1)$ pairs of branch points in $\SU(N)$ pure theory\label{fig:suNbranching}}
\end{figure}

To see the monopoles, we need to understand the structure of the branching of the $N$-sheeted cover $\Sigma\to C$. 
It is convenient to regard the combination $y=\Lambda^N(z+1/z)$ as one coordinate. Then,
the equation \eqref{puresuNcurve} can be thought of determining the intersections of the graph of the polynomial \begin{equation}
y=P(x)=x^N+u_2 x^{N-2} +\cdots + u_N
\end{equation} and a horizontal line \begin{equation}
y=\Lambda^N(z+\frac 1z)
\end{equation} as shown in Fig.~\ref{fig:suNbranching}. Of course the figure needs to be complexified, but the reader should be able to get the idea. 

\begin{figure}[h]
\[
\inc{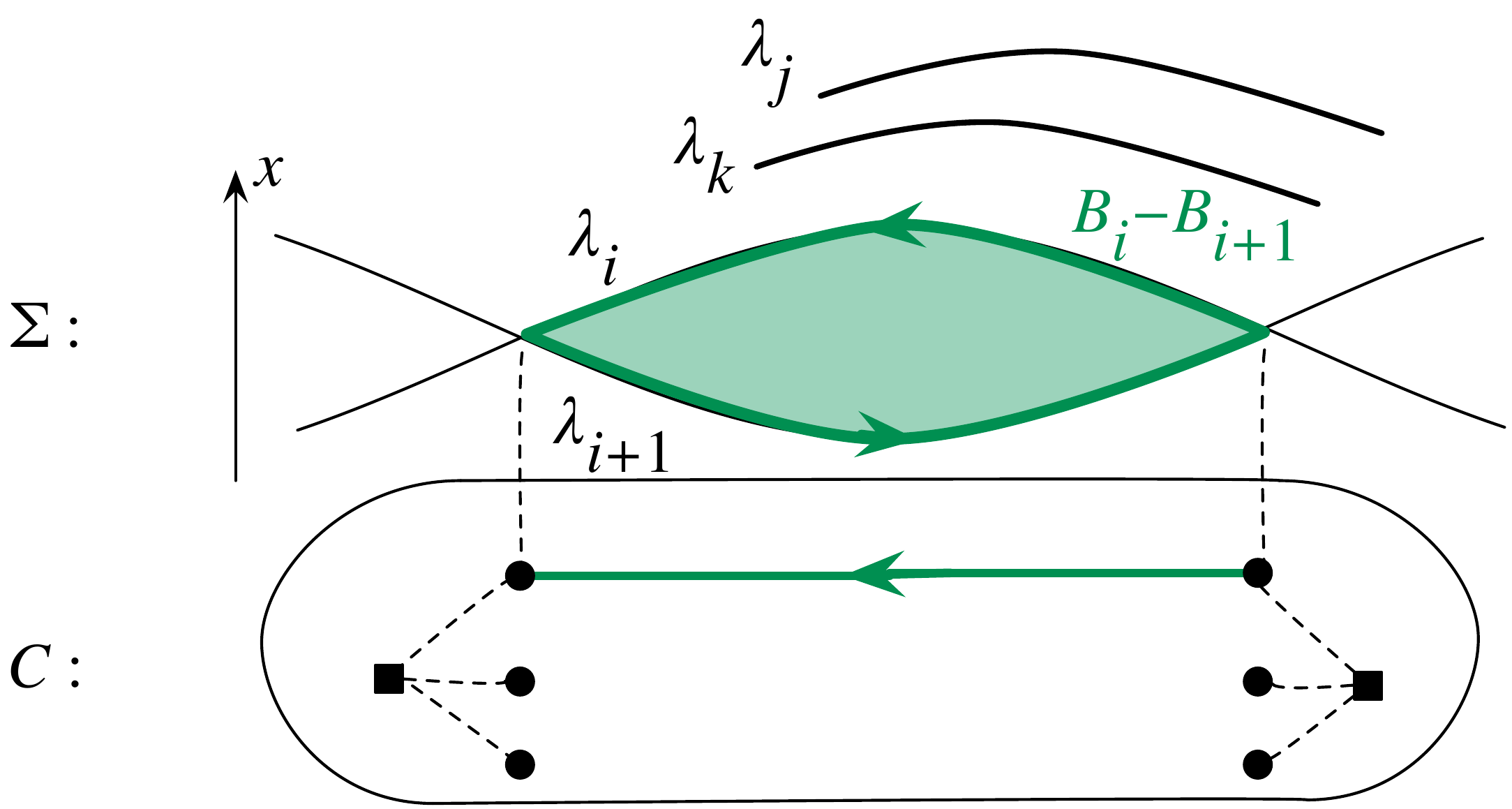}
\]
\caption{Monopoles of $\SU(N)$ pure theory\label{fig:SUNmono}}
\end{figure}

As is apparent,  two out of $N$ sheets meet at $(N-1)$ values of $y=\Lambda^N(z+1/z)$,
each of which becomes a pair $z_i^\pm$ of branch points on the $z$-sphere with $z_i^+z_i^-=1$. 
Note that the $i$-th sheet and the $(i+1)$-st sheet meet at this pair of branch points. 
Then we can suspend a disk-shaped membrane between this pair of branch points, as shown in Fig.~\ref{fig:SUNmono}.

In the semiclassical regime when $|\underline{a}_i| \sim |\LambdaRG | \gg |\Lambda|$, 
we have \begin{equation}
|z_i^+| \sim \frac{\LambdaRG ^N}{\Lambda^N},\qquad
|z_i^-| \sim \frac{\Lambda^N}{\LambdaRG ^N}.
\end{equation}
We call the path connecting $z_i^+$ and $z_i^-$ as $B_i$. Then \begin{align}
M_\text{monopole}&%=|\frac{1}{2\pi i}\int_{B_i} \lambda|
=|\frac{1}{2\pi i}\int_{B_i} (\lambda_i-\lambda_{i+1})| \\
& \sim |(a_i-a_{i+1}) \frac{1}{2\pi i} \int_{\Lambda^N/\LambdaRG ^N}^{\LambdaRG ^N/\Lambda^N} \frac{dz}z| \\
& \sim |(a_i-a_{i+1}) \frac{2N}{2\pi i} \log \frac{\LambdaRG }{\Lambda}|.
\end{align} This reproduces the mass of the monopole, by identifying \begin{equation}
\tau(\LambdaRG )=\frac{2N}{2\pi i} \log \frac{\LambdaRG }{\Lambda}.
\end{equation} This correctly reproduces the one-loop running of the pure $\SU(N)$ theory. 

%Note that we found W-bosons for any pair $(i,j)$ with $i\neq j$, but we found monopoles only for $(N-1)$ specific pairs $(i,i+1)$. This agrees with the known semi-classical analysis of the monopoles: 't Hooft-Polyakov monopoles associated to other pairs are in fact unstable \cite{Hollowood:1996zd}. 

\subsubsection{Infrared gauge coupling matrix}
\begin{figure}[h]
\[
\inc{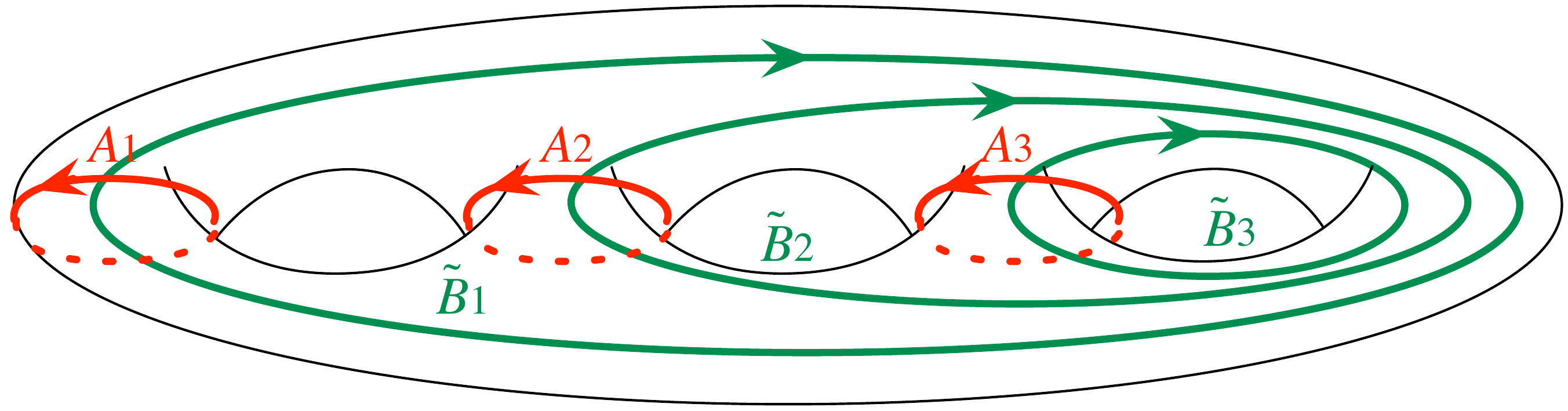}
\]
\caption{Cycles $A_i$ and $\tilde B_i$ on the \SeibergWitten\ curve of the pure $\SU(N)$ theory.\label{fig:he}}
\end{figure}

Let us check that our curve satisfies the condition that the coupling matrices of the low-energy $\U(1)^{N-1}$ theory
is positive definite. For this purpose we need to understand the geometry of the \SeibergWitten\ curve $\Sigma$ better. 
This is an $N$-sheeted cover of $C$ with $2N-2$ branch points $z_i^\pm$ of order 2
and 2 branch points $z=0,\infty$ of order $N$.  The genus $g$ of $\Sigma$ is then determined by the Riemann-Hurwitz formula\footnote{This is not hard to derive. 
Let us say we triangulate the curve $C$ with $V$ vertices, $E$ edges and  $F$ triangles so that the branch points are all at the vertices. We have $\chi(C)=V-E+F$.  We can just lift the edges and triangles to $\Sigma$: we have $NE$ edges and $NF$ triangles. The vertices are however less than $NV$.  At each vertex $p_i$ let the degree of the branching be $\deg p_i$.
Then there are $NV-\sum (\deg p_i-1)$ vertices in the triangulation of $\Sigma$. We end up $\chi(\Sigma)=N\chi(C)-\sum_i(\deg p_i -1)$. }:  \begin{equation}
\chi(\Sigma)=N \chi(C) - (2N-2)-2(N-1)
\end{equation} where $\chi(\Sigma)=2-2g$ and $\chi(C)=2$  are the Euler number of the respective surfaces.
We find $g=N-1$.  The basis of the 1-cycles consists of $(2N-2)$ cycles $A_i$ and $\tilde B^i$, $i=1,\ldots,N-1$, where the intersections are given by  \begin{equation}
A_i \cdot A_j =0=\tilde B^i \cdot \tilde B^j,\qquad
A_i \cdot \tilde B^j =\delta_i^j.
\end{equation} Here the dot product counts the number of intersections (including signs) of two one-cycles.  
The resulting set of cycles is shown in Fig.~\ref{fig:he}. 

The figures \ref{fig:SUNmono} and \ref{fig:he} are drawn in a rather different manner. 
The cycles from $A_1$ to $A_{N-1}$ can be directly identified. We have \begin{equation}
A_N=-A_1-A_2\cdots -A_{N-1}
\end{equation} as far as the line integral of holomorphic forms are concerned. 
Correspondingly, the variables $a_i$ as defined in \eqref{suNpureA} are not linearly independent,
and we have \begin{equation}
a_N=-a_1-\cdots - a_{N-1}.
\end{equation}

The combination $B_i-B_{i+1}$ in Fig.~\ref{fig:SUNmono} intersects with $A_i$ positively and with $A_{i+1}$ negatively. 
Then, we see \begin{equation}
B_i-B_{i+1}=\tilde B^i-\tilde B^{i+1}.
\end{equation} Equivalently, $\tilde B^i$ is a closed one-cycle completing the open path $B_i$ in a way independent of $i$.
 Then we define \begin{equation}
a_D^i:=\frac{1}{2\pi i}\oint_{\tilde B^i} \lambda
\end{equation} on the curve. Let us consider \begin{equation}
\tau^{ij}:=\frac{\partial a_D^i}{\partial a_j}
=X_D^{ik} (X^{-1})^{j}_k\qquad
\end{equation}
where \begin{equation}
X_i^k:=\frac{\partial a_i}{\partial u_k},\qquad
X_D^{jk}:=\frac{\partial a^j_D}{\partial u_k}.
\end{equation} Defining \begin{equation}
\omega_k= \frac{\partial}{\partial u_k} \lambda\Big|_\text{constant $z$},
\end{equation} we find \begin{equation}
\tau^{ij}=X_D^{ik} (X^{-1})^{j}_k\qquad\text{where}\quad
X_{i}^{k}=\oint_{A_i} \omega_k,\quad
X_D^{jk}=\oint_{\tilde B_j} \omega_k.\label{periodmatrix}
\end{equation}
It can be checked that $\omega_{2,3,\ldots,N}$ form a basis of holomorphic non-singular one-forms on $\Sigma$. The matrix $\tau^{ij}$ formed this way is known mathematically as the period matrix of $\Sigma$, and is known to satisfy \begin{equation}
\tau^{ij}=\tau^{ji}, \qquad \Im\tau^{ij}\ \text{is positive definite}.\label{wzw}
\end{equation}  From the first condition, we see that there is locally a function $F(a_i)$ such that \begin{equation}
a_D^i=\frac{\partial F}{\partial a_i},\qquad
\tau^{ij}=\frac{\partial^2 F}{\partial a_i\partial a_j}.
\end{equation}  
This justifies that we identify $a_i$, $a^i_D$ defined this way with the $a_i$ appearing in the low-energy description of $\U(1)^{N-1}$ gauge theory. The inverse gauge coupling matrix is given by $\Im\tau^{ij}$, whose positive definiteness is guaranteed by the mathematical relation \eqref{wzw}.

\subsection{$\SU(N)$ theory with fundamental flavors}
\subsubsection{$N_f=1$}
Next, consider the $\SU(N)$ theory with one flavor $(Q,\tilde Q)$ of bare mass $\mu$. 
The curve is given by \begin{equation}
\Sigma: \qquad \frac{\Lambda^{N-1}(x-\mu)}z +\Lambda^N z = x^N+u_2 x^{N-2}+\cdots + u_N.
\end{equation}  
Recall that in the semiclassical analysis we saw that a light charged hypermultiplet arises when $a_i \sim \mu$. Let us check that the curve written above reproduces this behavior. 

First, we introduce $\underline{a}_i$ as before, and consider the semiclassical regime when all $|\underline{a}_i|$ is far larger than $|\Lambda|$. The A-cycle on the \Gaiotto\ curve was $|z|=1$ as before. Then we find $a_i \sim \underline{a_i}+O(\Lambda)$ just as was in the case of the pure theory.  

To see additional singularities in the weakly-coupled region, define $\tilde z=z/\Lambda^{N-1}$. The curve is then \begin{equation}
\frac{x-\mu}{\tilde z} +\Lambda^{2N-1} \tilde z = x^N+u_2 x^{N-2}+\cdots + u_N,
\end{equation}  which can be approximated by \begin{equation}
\frac{x-\mu}{\tilde z}  = x^N+u_2 x^{N-2}+\cdots + u_N = \prod (x-\underline{a}_i)
\end{equation} in the extremely weakly coupled limit. 
 The equation factorizes 
and  the curve separates into two when $\underline{a}_i=\mu$; otherwise the curve is a smooth degree-$N$ covering of the $z$ sphere.  This shows that when $\underline{a}_i=\mu$, a one-cycle on the \SeibergWitten\ curve shrinks, and the membrane suspended there produces a massless hypermultiplet, see Fig.~\ref{fig:connection}.

The one-loop running can also be checked. The branch points $z_i^+$ in the large $z$ region is unchanged, as the structure of the $N_f=1$ curve in the large $z$ region itself is unchanged from the pure curve. 
Then 
\begin{equation}
z_i^+ \sim (\LambdaRG /\Lambda)^N.
\end{equation}
 In the small $z$ region, the branch points are around where $\Lambda^{N-1} x /z$ and $P(x)$ are of the same order. Assuming $|x|\sim |\underline{a}_i| \sim |\LambdaRG |$, we see 
\begin{equation} 
z_i^- \sim (\Lambda/\LambdaRG )^{N-1}.
\end{equation} Then the monopole has the mass \begin{align}
M_\text{monopole}&=|\frac{1}{2\pi i}\int_{B_i} \lambda| \\
& \sim |(a_i-a_{i+1}) \frac{1}{2\pi i} \int_{\Lambda^{N-1}/\LambdaRG ^{N-1}}^{\LambdaRG ^N/\Lambda^N} \frac{dz}z| \\
& \sim |(a_i-a_{i+1}) \frac{2N-1}{2\pi i} \log \frac{\LambdaRG }{\Lambda}|.
\end{align} This gives \begin{equation}
\tau(\LambdaRG )=\frac{2N-1}{2\pi i} \log \frac{\LambdaRG }{\Lambda}
\end{equation} as it should be.

\subsubsection{General number of flavors}
More generally, we can consider the curve given by \begin{multline}
\Sigma:\qquad
\frac{\Lambda^{N-N_L}\prod_{i=1}^{N_L}(x-\mu_i)}z
+ z\Lambda^{N-N_R}\prod_{i={1}}^{N_R}(x-\mu'_i) \\
= x^N+u_2 x^{N-2}+\cdots+u_{N-1}x+u_N\label{suNNf}
\end{multline}  where $N_L,N_R\le N$. 
When $N_L=N_R=N$, we need to introduce complex numbers $f$, $f'$ as in the curve of $\SU(2)$ with four flavors, \eqref{naive_nf_4}: \begin{multline}
\Sigma:\qquad
f\cdot \frac{\prod_{i=1}^{N}(x-\mu_i)}z
+ f'\cdot z\prod_{i={1}}^{N}(x-\mu'_i) \\
= x^N+u_2 x^{N-2}+\cdots+u_{N-1}x+u_N.\label{suN-saturated}
\end{multline} We also need to distinguish the mass parameters in the curve and the mass parameters in the BPS mass formula, carefully studied in Sec.~\ref{sec:ident} for $\SU(2)$ with four flavors. 
In the following we mainly discuss the case with less than $2N-1$ flavors. 

Consider the case when $\mu_i$ and $\mu'_i$ are all small.
Further, consider the regime where $|\underline{a}_i| \gg \Lambda$.
As always we find $a_i =\underline{a}_i + O(\Lambda)$.
The branch points are at \begin{equation}
|z_i^+| \sim \frac{\LambdaRG ^{N-N_R}}{\Lambda^{N-N_R}},\qquad
|z_i^-| \sim \frac{\Lambda^{N-N_L}}{\LambdaRG ^{N-N_L}}.
\end{equation} Then we find \begin{align}
M_\text{monopole}
& \sim |(a_i-a_{i+1}) \frac{1}{2\pi i} \int_{\Lambda^{N-N_L}/\LambdaRG ^{N-N_L}}^{\LambdaRG ^{N-N_R}/\Lambda^{N-N_R}} \frac{dz}z| \\
& \sim |(a_i-a_{i+1}) \frac{2N-(N_L+N_R)}{2\pi i} \log \frac{\LambdaRG }{\Lambda}|,
\end{align} and therefore the one-loop running is \begin{equation}
\tau(\LambdaRG )=-\frac{2N-(N_L+N_R)}{2\pi i} \log \frac{\LambdaRG }{\Lambda}.
\end{equation}

\begin{figure}[h]
\[
\inc{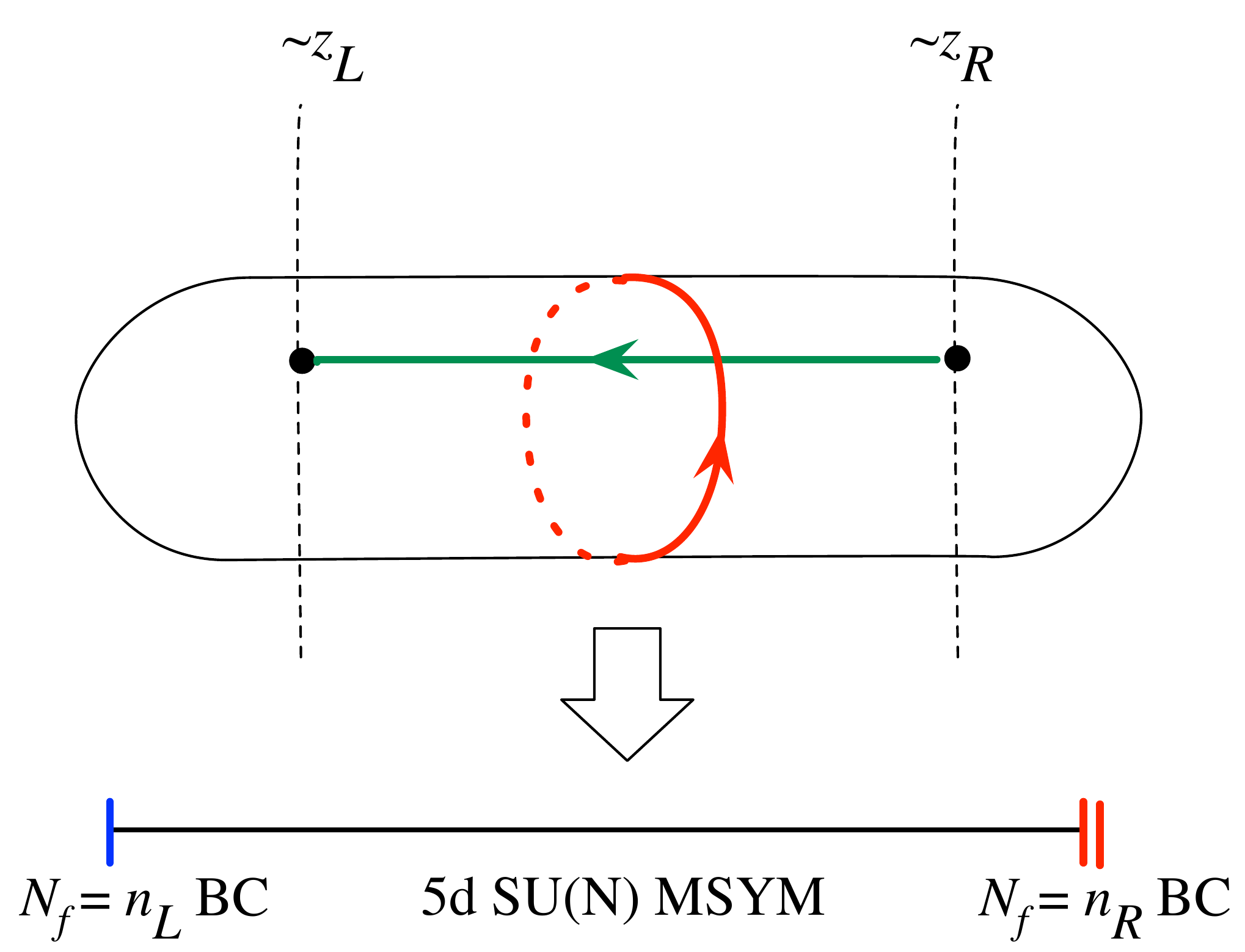}
\]
\caption{$\SU(N)$ theory with flavors\label{fig:suNgeneral}}
\end{figure}

In the other regime when $|\mu_i|, |\underline{a}_i| \gg \Lambda$, 
we can use the redefining trick to find singularities on the Coulomb branch. 
For example, defining $\tilde z=z/\Lambda^{N-N_L}$, the curve is  \begin{equation}
\frac{\prod_{i=1}^{N_L}(x-\mu_i)}{\tilde z}
+ \tilde z\Lambda^{2N-N_R-N_L}\prod_{i={1}}^{N_R}(x-\mu'_i)
= x^N+u_2 x^{N-2}+\cdots+u_{N-1}x+u_N.
\end{equation}  Then the limit $\Lambda\to 0$ can be taken, which gives \begin{equation}
\frac{\prod_{i=1}^{N_L}(x-\mu_i)}{\tilde z}
=\prod_{i=1}^N (x-\underline{a}_i).
\end{equation} This means that whenever $\underline{a}_i=\mu_s$ for some $i$ and $s=1,\ldots,N_L$, the curve splits into two, because the equation can be factorized. 
The same can be done for the variable $w=1/z$. Then we also find singularities when $\underline{a}_i=\mu'_s$ for some $i$ and $s=1,\ldots,N_R$. 
In total, these reproduce the semiclassical, weakly-coupled physics of $\SU(N)$ theory with $N_f=N_R+N_L$ hypermultiplets in the fundamental representation.  The situation is summarized in Fig.~\ref{fig:suNgeneral}.

We have a sphere $C$ described by the coordinate $z$. 
The curve $\Sigma$ is an $N$-sheeted cover of $C$.
We have one M5-brane wrapping $\Sigma$.
We call the 6d theory living on $C$ the $\cN{=}(2,0)$ theory of type $\SU(N)$. 
Roughly speaking, it arises from $N$ coincident M5-branes. 

Consider $\Arg z$ as the sixth direction $x_6$, and $\log |z|$ as the fifth direction $x_5$.
Reducing along the $x_6$ direction, we have a 5d theory on a segment.
The 5d theory is the maximally supersymmetric Yang-Mills theory with gauge group $\SU(N)$. 
The term \begin{equation}
\frac{\Lambda^{N-N_L}\prod_{i=1}^{N_L}(x-\mu_i)}z\label{Lb}
\end{equation} in the curve can be thought of defining a certain boundary condition on the left side of the fifth direction. 
We regard it as giving $N_L$ hypermultiplets in the $\SU(N)$ fundamental representation there.
Similarly, the term \begin{equation}
z\Lambda^{N-N_R}\prod_{i=1}^{N_R}(x-\mu'_i)\label{Rb}
\end{equation} is regarded as the boundary condition such that $N_R$ fundamental hypermultiplets there. 
By further reducing the theory along the fifth direction, we have $\SU(N)$ gauge theory with $N_f=N_L+N_R$ fundamental hypermultiplets in total.
We saw that the effect of the boundary conditions becomes noticeable around when \begin{equation}
\log |z_R| \sim (N-N_R) \log \frac{\LambdaRG }{\Lambda} < 0,\qquad
\log |z_L| \sim (N-N_L) \log \frac{\Lambda}{\LambdaRG } >0.
\end{equation} In the five dimensional Yang-Mills, we have monopole strings,
which have ends around $|z_R|$ and $|z_L|$. From the four-dimensional point of view, $\log |z_L|/|z_R|$ then controlled the mass of the monopoles, which then gave the one-loop running of the theory. 

Note that from the four-dimensional point of view, the split of $N_f$ into $N_R$ and $N_L$ is rather arbitrary. In fact, by redefining $z$, we can easily come to the form of the curve given by \begin{equation}
\frac{\Lambda^{2N-N_f}\prod_i^{N_f}(x-\mu_i)}z 
+z = x^N+ u_2 x^{N-2} + \cdots u_N
\end{equation} where we defined $\mu_{N_L+i}:=\mu'_i$. In this form the symmetry exchanging all $N_f$ mass parameters is manifest. 

From the higher-dimensional perspective, it is however sometimes convenient to stick to the situation where the equation of the \SeibergWitten\ curve $\Sigma$ is of degree $N$ regarded as a polynomial in $x$. 
This guarantees that $\Sigma$ is always an $N$-sheeted cover of the \Gaiotto\ curve $C$. Numerically, 
this condition means that the boundary condition such as \eqref{Lb} and \eqref{Rb} also has degrees less than or equal to $N$.
 This imposes the constraint $N\ge N_{L,R}$, and therefore $2N\ge N_f$. This is the condition that the theory is asymptotically free or asymptotically conformal.

\subsection{$\SO(2N)$ theories}\label{sec:SO}
Now let us quickly discuss the $\SO(2N)$ gauge theories. 
\subsubsection{Semi-classical analysis}
The vector multiplet scalar $\Phi$ is an $2N\times 2N$ antisymmetric matrix. 
Let us denote the hypermultiplets by $(Q_i^a,\tilde Q^i_a)$ where $a=1,\ldots,2N$ and 
$i=1,\ldots, N_f$. 
We consider the branch of the supersymmetric vacuum given by \begin{equation}
[\Phi,\Phi^\dagger]=0.
\end{equation} As $\Phi$ is antisymmetric, the outcome of the diagonalization is \begin{equation}
\Phi=\diag(a_1,\ldots,a_N,-a_1,\ldots,-a_N). 
\end{equation}  In general the gauge group is broken to $\U(1)^N$. 
The gauge invariant combination is given by \begin{equation}
x^{2N}+u_2 x^{2N-2} + u_4 x^{2N-4} + \cdots + u_{2N}
= \det (x+\Phi)
\end{equation} 
where $x$ is a dummy variable.
Note that the odd powers automatically vanish due to the antisymmetry.
In fact $\tilde u_N$ defined by the condition \begin{equation}
u_{2N}=\tilde u_N{}^2 ,\qquad \tilde u_N=a_1a_2\ldots a_N
\end{equation} is also invariant under $\SO(2N)$ but not under $\OO(2N)$.

The W-bosons have masses \begin{equation}
|{}\pm a_i \pm a_j|
\end{equation} for $i\neq j$. Similarly, the monopole has the mass \begin{equation}
|\tau (\pm a_i \pm a_j)|.
\end{equation}

By expanding the superpotential \begin{equation}
\sum_i (Q_i\Phi\tilde Q^i + \mu_i Q_i \tilde Q^i) ,
\end{equation} classically  we find that there is a massless hypermultiplet charged under one of $\U(1)$ gauge fields when $\mu_s=\pm a_i$ for some $i$ and $s$. 

The one-loop running is given by  \begin{equation}
\Lambda\frac{d}{d\Lambda} \tau = -\frac{1}{2\pi i}(2(2N-2)-2N_f).
\end{equation} Therefore the theory is asymptotically free for $N_f<2N-2$,
and is asymptotically conformal when $N_f=2N-2$. 

\subsubsection{Pure $\SO(2N)$ theory}\label{sec:pureSON}

The \SeibergWitten\ curve of the pure theory is given by \begin{equation}
x^2(\frac{\Lambda^{2N-2}}z + \Lambda^{2N-2} z)
=x^{2N}+u_2 x^{2N-2} + u_4 x^{2N-4} + \cdots + u_{2N}\label{pureSOcurve}
\end{equation} with the differential $\lambda=xdz/z$.
This is a $2N$-sheeted cover of the \Gaiotto\ curve $C$, which is just a sphere with the complex coordinate $z$.
By solving the equation, one finds $2N$ local solutions $\pm x_i(z)$. 
Correspondingly, we define $\pm\lambda_i=\pm x_i(z) dz/z$. 

Let us study the weakly-coupled regime. 
We introduce $\underline{a}_i$ by \begin{equation}
x^{2N}+u_2 x^{2N-2} + u_4 x^{2N-4} + \cdots + u_{2N}
= \prod_{i=1}^N(x^2-\underline{a}_i{}^2).
\end{equation}  The regime we are interested in is when $|\underline{a}_i| \gg |\Lambda|$.

\begin{figure}[h]
\[
\inc{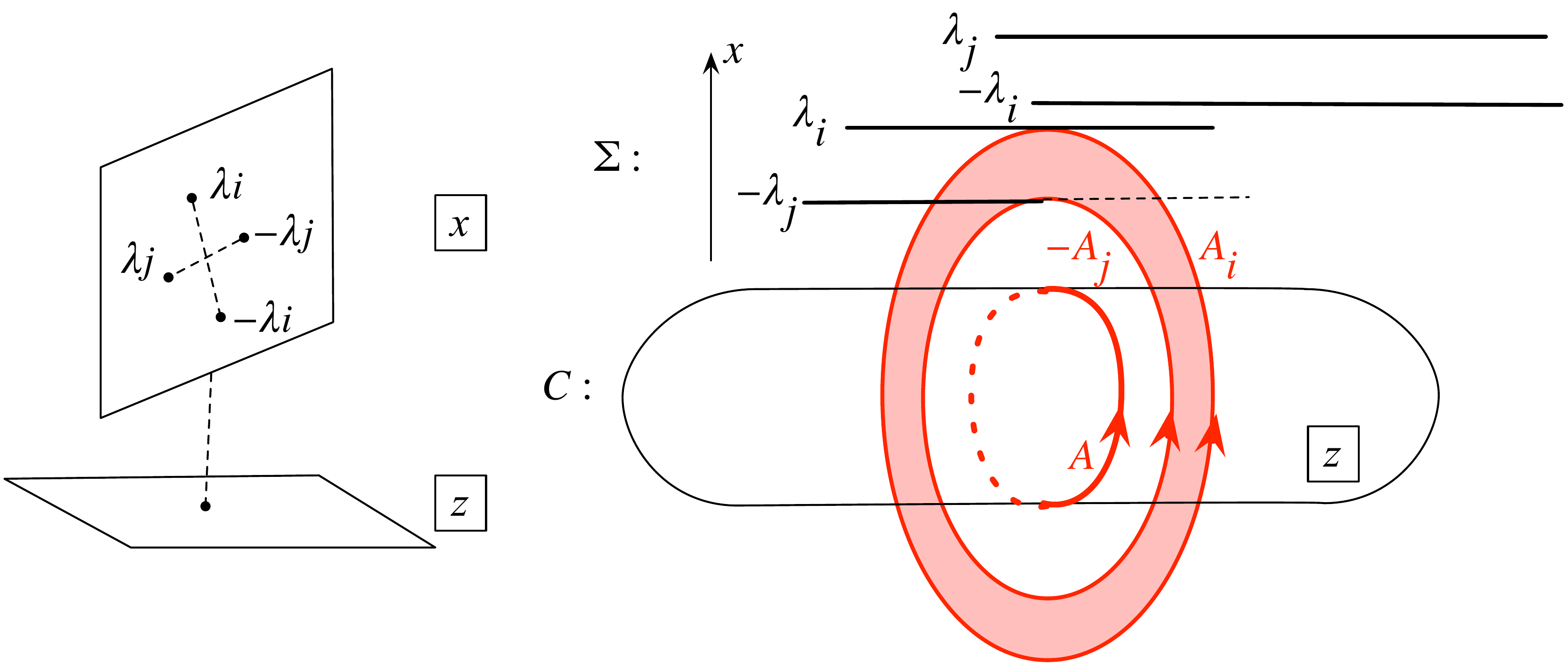}
\]
\caption{W-boson of the $\SO(2N)$ theory\label{fig:soN}}
\end{figure}

We draw the A-cycle on the \Gaiotto\ curve at $|z|=1$, see Fig.~\ref{fig:soN}.
On the A-cycle, the equation \eqref{pureSOcurve} can be solved approximately to give \begin{equation}
x_i(z) = \underline{a}_i + O(\Lambda).
\end{equation}
We lift the A-cycle on $C$ to the sheets of $\Sigma$. We have $N$ pairs of cycles $\pm A_{i}$.
Then \begin{equation}
a_i=\frac{1}{2\pi i} \oint_{A_i} \lambda = \frac{1}{2\pi i} \oint \lambda_i = \underline{a}_i+ O(\Lambda).
\end{equation}
We can now suspend ring-like membranes between sheets. They clearly have masses \begin{equation}
|{\pm a_i\pm a_j}|.
\end{equation} We find that we need to impose the constraint that M2-brane cannot be suspended between the $i$-th sheet and the $(-i)$-th sheet, to forbid the W-boson with mass $|{\pm 2a_i}|$. 
As for the monopoles, the branch points are at around \begin{equation}
z^+ \sim \left(\frac{\LambdaRG }{\Lambda}\right)^{2N-2},\quad
z^- \sim \left(\frac{\Lambda}{\LambdaRG }\right)^{2N-2}.
\end{equation}
Then the monopole mass can be approximately computed as in the case of $\SU(N)$ gauge theory: we find \begin{equation}
\sim | (a_i-a_j) \frac{1}{2\pi i}\int_{z^-}^{z^+} \frac{dz}z |
= |  (a_i-a_j) \frac{2(2N-2)}{2\pi i} \log  \frac{\LambdaRG }{\Lambda} |.
\end{equation} From this we see that the running coupling is \begin{equation}
\tau(\LambdaRG )=
\frac{2(2N-2)}{2\pi i} \log  \frac{\LambdaRG }{\Lambda} ,
\end{equation} correctly reproducing the one-loop analysis.

Let us study the low-energy coupling matrix $\tau^{ij}$. 
The branch points are at $z=0,\infty$ together with $N$ pairs on generic places of the $z$-sphere.
At $z=\infty$, there are $N-2$ solutions behaving as $x\sim z^{1/(2N-2)}$
and two solutions behaving as $x\sim z^{-1/2}$. Therefore it counts as a  branch point of degree $2N-2$ and another of degree 2. The structure of the branching at $z=0$ is the same. 
Next, consider one of $N$ pairs of branch points  of these latter type.  When the sheets $i$ and $j$ meet there, the sheets $-i$ and $-j$ meet at the same time. Slightly moving them apart, we find that there are $4N$ branch points of degree 2 in total.  Using the Riemann-Hurwitz theorem, we see \begin{equation}
\chi(\Sigma)=2N\chi(C)-2(2N-3)-2-4N.
\end{equation} Therefore the genus of the \SeibergWitten\ curve is $g=2N-1$.
Therefore, the independent 1-cycles on $\Sigma$ can be labeled as $\tilde A_1$, \ldots, $\tilde A_{2N-1}$ and $\tilde B^1$,\ldots, $\tilde B^{2N-1}$ with the intersection \begin{equation}
\tilde A_i\cdot \tilde A_j=0=\tilde B^i\cdot \tilde B^j,\qquad
\tilde A_i \cdot \tilde B^j=\delta ^j_i.
\end{equation}
Note that the curve $\Sigma$ has the symmetry $\bZ_2$ acting by $x\to -x$. 
Under this symmetry, the differential is odd: $\lambda\to -\lambda$. 
Correspondingly, only the 1-cycles $L$ odd under this $\bZ_2$ action can have $\oint_L \lambda\neq 0$. 
The cycles $A_i$ for $i=1,\ldots,N$ obtained by lifting the A-cycle on the \Gaiotto\ curve $C$ to $\Sigma$ are indeed odd.
The period matrix $\tau^{ij}$ computed as in \eqref{periodmatrix}  is an $(2N-1)\times (2N-1)$ matrix, which is symmetric and whose imaginary part is positive definite. 
By restricting to the subspace odd under $\bZ_2$ action, we end up having $N\times N$ matrix, which is again symmetric and whose imaginary part is positive definite. 

\subsubsection{$\SO(2N)$ theory with flavors in the vector representation}

The curve of the $\SO(2N)$ theory with one hypermultiplet in the $2N$-dimensional representation is \begin{equation}
x^2(\frac{\Lambda^{2N-4}(x^2-\mu^2)}z + \Lambda^{2N-2} z)
=x^{2N}+u_2 x^{2N-2} + u_4 x^{2N-4} + \cdots + u_{2N}.
\end{equation}
Let us just see that there is a singularity in the Coulomb branch when $a_i=\pm \mu$ for some $i$. 
As always, we assume $|\underline{a}_i|, |\mu|\gg |\Lambda|$ ,
  make the redefinition $\tilde z=z/\Lambda^{2N-4}$ and take the limit of the curve: \begin{equation}
x^2\frac{(x^2-\mu^2)}z 
=x^{2N}+u_2 x^{2N-2} + u_4 x^{2N-4} + \cdots + u_{2N}.
\end{equation} This equation is factorized when $\pm \underline{a}_i=\mu$ or $\underline{a}_i=0$ for some $i$. 
The latter choice does not fit the assumption that $|\underline{a}_i| \gg |\Lambda|$.
Then we find the singularities when $\pm\underline{a}_i \sim \mu $ in the weakly-coupled region. 

In general, the curve of the $\SO(2N)$ with $N_f=N_R+N_L$ hypermultiplets in the vector representation is given by \begin{multline}
x^2(\frac{\Lambda^{2(N-N_R-1)}\prod_{i=1}^{N_R}(x^2-\mu_i^2)}z + \Lambda^{2(N-N_L-1)} z\prod_{i=1}^{N_L}(x^2-\mu'_i{}^2))\\
=x^{2N}+u_2 x^{2N-2} + u_4 x^{2N-4} + \cdots + u_{2N}.\label{so2NNf}
\end{multline} Strictly speaking, this is only for $N_L+N_R < 2N-2$. When $N_L=N_R=N-1$, we need to put two complex numbers $f$ and $f'$ instead of the powers of $\Lambda$, much as in \eqref{suN-saturated} for the case of the $\SU(N)$ theory with $2N$ flavors.

Let us check the one-loop running when $\mu_i=\mu_i'=0$. 
Assume $|\underline{a}_i| \gg |\Lambda|$. As always we find $a_i =\underline{a}_i+O(\Lambda)$.
The branch points on the \Gaiotto\ curve are at around\begin{equation}
z^+ \sim \left(\frac{\LambdaRG }{\Lambda}\right)^{2N-2-2N_L},\quad
z^- \sim \left(\frac{\Lambda}{\LambdaRG }\right)^{2N-2-2N_R}.
\end{equation}
Then the monopole mass can be approximately computed as in the case of $\SU(N)$ gauge theory: we find \begin{equation}
\sim \left| (a_i-a_j) \frac{1}{2\pi i}\int_{z^-}^{z^+} \frac{dz}z \right|
= \left|  (a_i-a_j) \frac{2(2N-2-2(N_L+N_R))}{2\pi i} \log  \frac{\LambdaRG }{\Lambda} \right|.
\end{equation} From this we see that the running coupling is \begin{equation}
\tau(\LambdaRG )=
\frac{2(2N-2)-2(N_L+N_R)}{2\pi i} \log  \frac{\LambdaRG }{\Lambda} ,
\end{equation} correctly reproducing the one-loop analysis.
Again, the condition that the theory is asymptotically free or conformal is related to the fact that the left hand side of the equation of the curve has lower degree than or equal degree to the right hand side. 

\subsection{Argyres-Douglas CFTs}\label{sec:ADAD}
Let us study the most singular point in the Coulomb branches of  the theories we analyzed in this section. \subsubsection{Pure $\SU(N)$ theory}
\begin{figure}[h]
\[
\inc{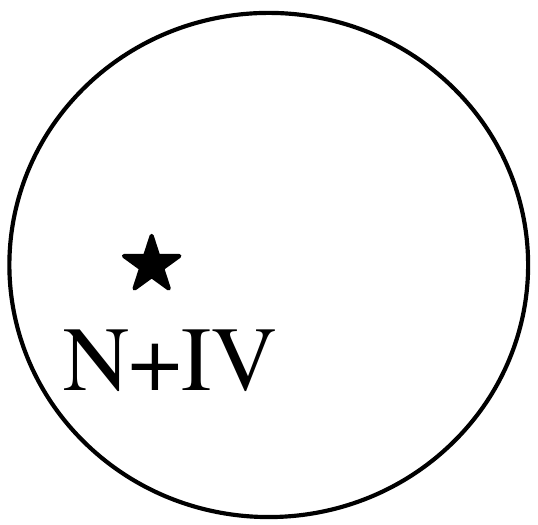}
\]
\caption{The most singular point of pure $\SU(N)$ theory\label{fig:pureSUAD}}
\end{figure}

First, take the curve of the pure $\SU(N)$ theory: \begin{equation}
\frac{\Lambda^N}{z}+\Lambda^N z = x^N+ \cdots + u_N
\end{equation} with the differential $\lambda=x dz/z$.
We set $z=1+\delta z$, $u_N=2\Lambda^N + \delta u_N$ and take the limit where both $\delta z$ and $\delta u_N$ are very small. We find \begin{equation}
c\,\delta z^2 = x^N+u_2  x^{N-2}+\cdots + \delta u_N
\end{equation} where $c$ is an unimportant constant.  
The differential is now given by $\lambda=x d \delta z\sim \delta z dx$.  
Introducing $\tilde z= 1/x$, we find that the curve in this limit can be written as \begin{equation}
c\,\lambda^2 = \frac{1+ u_2 \tilde z^2 + u_3 \tilde z^3 + \cdots +  (\delta u_N) \tilde z^N}{\tilde z^{N+4}} d\tilde z^2.
\end{equation}   Note that it has the same form as the curve we saw in Sec.~\ref{sec:generalAD},
which arose from considering the curve \begin{equation}
\lambda^2 = \phi(\tilde z)
\end{equation} where $\phi(\tilde z)$ is a quadratic differential with one pole of order $N+4$, see Fig.~\ref{fig:pureSUAD}. This is the same as the theory $Y_{N+4}$ introduced in Fig.~\ref{fig:XNYN}.
We have \begin{equation}
AD_{N_f=0}(\SU(N))=Y_{N+4}.\label{@@@}
\end{equation}

Demanding that $\lambda$ has scaling dimension 1, we find that \begin{equation}
[u_k]=\frac{2k}{N+2}. 
\end{equation} 
Note that we have $[u_k]+[u_{N+2-k}]=2$.
At this point it is instructive to recall our discussions  around \eqref{generalCFTfeatures}.
We consider the prepotential deformation \begin{equation}
\int d^4\theta u_k u_{N+2-k}
\end{equation} where $d^4\theta$ is the chiral $\cN{=}2$ superspace integral.
As $[u_k]\le 1 \le [u_{N+2-k}]$ when $k\le N+2-k$, 
we consider $u_k$ is the deformation parameter for the physical operator $u_{N+2-k}$.

Take the simplest case $N=3$. We have the theory with one operator with $[u_3]=6/5$ and a corresponding parameter with $[u_2]=4/5$. These are the same as those of the Argyres-Douglas CFT which arose from $\SU(2)$ with one flavor; in fact the curve and the differential are completely the same: \begin{equation}
AD_{N_f=0}(\SU(3))=Y_{7}=AD_{N_f=1}(\SU(2)).
\end{equation}

\subsubsection{$\SU(N)$ theory with two flavors}

Next, consider $\SU(N)$ theory with two flavors. The curve is \begin{equation}
(x-\mu_1)\frac{\Lambda^{N-1}}{z}+(x-\mu_2)\Lambda^{N-1} z = x^N+ \cdots + u_N.
\end{equation} We already studied the case $N=2$, so let us set $N>3$. Then we expand as \begin{equation}
u_{N-1}=2\Lambda^{N-1}+\delta u_{N-1},\quad
u_N=(2x-(\mu_1+\mu_2))\Lambda^{N-1}+\delta u_N,\quad
z=1+\delta z
\end{equation}  and take the limit where $\delta z$, $\delta u_{N-1}$, $\delta u_N$ and $\mu_{1,2}$ are all small. The curve is \begin{equation}
\Lambda^{N-1}\,(x - \mu_1)\delta z^2 +\Lambda^{N-1}\, (\mu_1-\mu_2) \delta z=x^{N} +\cdots + u_{N-2} x^2 + \delta u_{N-1}x +\delta {u_N}
\end{equation} with the differential $\lambda = x d \delta z\sim \delta z dx$. 
Here $c$ and $c'$ are unimportant constants.

We now define $x'$ by
$x'=x-\mu_1$,  shift $\delta z$ by 
$\delta z\to \delta z-(\mu_1-\mu_2)/(2x')$,  
and introduce $\tilde z=1/x'$. The curve is now \begin{equation}
\lambda^2 =\frac{  1+ \tilde u_1 \tilde z+ \tilde  u_{2} \tilde z^2 + \cdots + \tilde  u_N \tilde z^N + (\tilde\mu_1-\tilde\mu_2)^2 \tilde z^{N+1}   }{\tilde z^{N+3}} d\tilde z^2.
\end{equation} 
Here we absorbed various unimportant numerical constants into the definition of variables with tildes. 

\begin{figure}[h]
\[
\inc{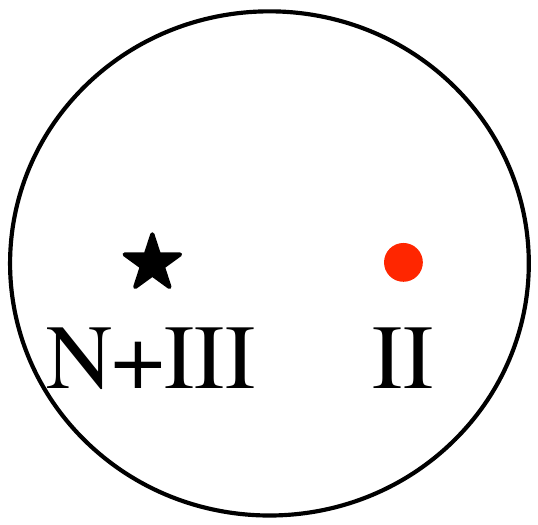}
\]
\caption{The most singular point of $\SU(N)$ theory with two flavors\label{fig:suNADwithF}}.
\end{figure}

This is the curve $\lambda^2=\phi(z)$ with $\phi$ having one pole of order $N+3$ and another of order 2, see Fig.~\ref{fig:suNADwithF}.
This is the  theory $X_{N+3}$ introduced in Fig.~\ref{fig:XNYN}.
We have \begin{equation}
AD_{N_f=2}(\SU(N))=X_{N+3}.\label{===}
\end{equation}
The most singular point of $\SU(N)$ theory with odd number of flavors gives an $\cN{=}2$ CFT, analyzed in \cite{Cecotti:2013lda,Giacomelli:2013tia}.  The most singular point of $\SU(N)$ theory with even number of flavors $N_f\ge 4$ does \emph{not} give an $\cN{=}2$ CFT, as we will see in Sec.~\ref{sec:applicationL}.

\subsubsection{Pure $\SO(2N)$ theory}

\begin{figure}[h]
\[
\inc{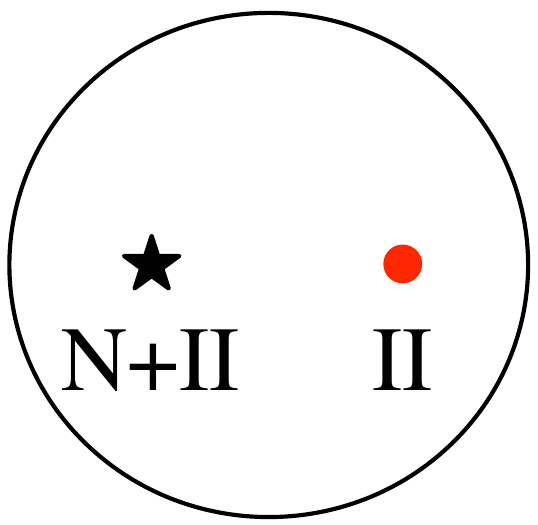}
\]
\caption{The most singular point of $\SO(2N)$ theory\label{fig:so2NAD}}
\end{figure}

Next, take the pure $\SO(2N)$ theory \begin{equation}
x^2(\frac{\Lambda^{2N-2}}z+\Lambda^{2N-2}z) = x^{2N} + u_2 x^{2N-2} + \cdots + u_{2N}.
\end{equation}
Take \begin{equation}
u_{2N-2}=2\Lambda^{2N-2}+\delta u_{2N-2},\qquad z=1+\delta z
\end{equation} and go to the limit where $\delta u_{2N-2}$, $\delta z$ are both small. The curve is \begin{equation}
c\,\delta z^2=x^{2N-2} + \cdots + \delta u_{2N-2}+\frac{u_{2N}}{x^2}
\end{equation}where $c$ is an unimportant constant. 
The differential is given by  $\lambda=x d\delta z\sim \delta z dx$. In terms of $\tilde z=1/x^2$, the curve is \begin{equation}
c\,\lambda^2 =\frac{  1+ \tilde u_2 \tilde z + \cdots + \tilde  u_{2N-2} \tilde z^{N-1} + u_{2N} \tilde z^{N}   }{\tilde z^{N+2}} d\tilde z^2.
\end{equation} This is again the curve $\lambda^2=\phi(z)$ with $\phi$ having one pole of rather high order $N+2$ and another of order 2, see Fig.~\ref{fig:so2NAD}. Therefore we find \begin{equation}
AD_{N_f=0}(\SO(2N))=X_{N+2}.\label{qqq}
\end{equation}

Now, $\SU(4)$ and $\SO(6)$ have the same Lie algebra. Using \eqref{@@@} and \eqref{qqq}, we find \begin{equation}
Y_8= AD_{N_f=0}(\SU(4))=AD_{N_f=0}(\SO(6))=X_5.
\end{equation} Using \eqref{===}, we find that these are also equivalent to $AD_{N_f=2}(\SU(2))$.
This set of equivalences explains what we saw in \eqref{accidents}.

\subsubsection{Argyres-Douglas CFTs and the Higgs branch}
 The $\SU(2)$ theory with $N_f=2$ flavors has a Higgs branch of the form $\bC^2/\bZ_2$,
but the pure $\SU(4)$ theory does not have it in the ultraviolet. We just claimed \begin{equation}
AD_{N_f=0}(\SU(4))=AD_{N_f=2}(\SU(2)).
\end{equation}
How is this compatible? The discussion below summarizes the content of \cite{Argyres:2012fu}.

Note that the limiting Argyres-Douglas theory has an operator $u$ of scaling dimension 4/3, a corresponding parameter $m$
of scaling dimension $2/3$ and an additional mass parameter $\mu_1-\mu_2$ of scaling dimension 1. 
When we realize it as a limit of the $\SU(2)$ theory with $N_f=2$ flavors, clearly the low energy theory has just one $\U(1)$ multiplet and $\mu_1-\mu_2$ is an external parameter. 

When we realize the same theory as a limit of the pure $\SU(4)$ theory, however, originally the low energy theory has $\U(1)^3$ vector multiplet, and three Coulomb branch parameters $u_2$, $u_3$ and $u_4$. We saw that $\delta u_2$ has scaling dimension 2/3, $\delta u_4$ scaling dimension 4/3, and $\delta u_3$  is of scaling dimension 1. 
Therefore, we see that the mass parameter $\mu_1-\mu_2$ of the limiting Argyres-Douglas theory 
is now promoted to the vev $\delta u_3$ of a $\U(1)$ multiplet in this realization. 
Equivalently, the $\U(1)$ subgroup of the $\SU(2)$ flavor symmetry of the limiting theory is weakly dynamically gauged, thus removing the Higgs branch. 

Similarly, we saw here that the pure $\SO(8)$ theory and the $\SU(3)$ theory with $N_f=2$ flavors both give rise to the CFT $X_6$.
In Sec.~\ref{sec:generalAD},  we also learned that $\SU(2)$ theory with $N_f=3$ flavors also has a point on the Coulomb branch where the low energy limit is described by the same theory $X_6$, see \eqref{accidents}.
The situation concerning their  Higgs branches can also be studied similarly as above. 
The limiting theory itself has an operator $u$ of scaling dimension 3/2, a corresponding deformation parameter $m$ of dimension 1/2, and two mass parameters $\mu_1 -\mu_2$ and $\mu_1-\mu_3$ for the $\SU(3)$ flavor symmetry.
This is most clearly seen in the description as a point on the Coulomb branch of the $\SU(2)$ theory with $N_f=3$ flavors. 

In terms of $\SU(3)$ theory with $N_f=2$ flavors, we have two  Coulomb branch operators $u_2$, $u_3$, the mass parameter $m$ for the $\U(1)$ part of the flavor symmetry, and the mass parameter for the $\SU(2)$ part $\mu_1-\mu_2$.
We see that $u_2$ and $u_3$ has scaling dimensions $1$ and $3/2$ respectively, $m$ has scaling dimension $1/2$,
and $\mu_1-\mu_2$ has dimension $1$.  Then we see that $\U(1)$ subgroup of the flavor symmetry $\SU(3)$ is weakly gauged. The vev of this weakly-gauging $\U(1)$ vector multiplet is $u_2$.  

In terms of pure $\SO(8)$ theory, we  have four Coulomb branch operators  $u_2$, $u_4$, $u_6$ and $u_8$,
but as we discussed above, $u_8=\tilde u_4^2$. 
Close to the Argyres-Douglas point, we see that $u_2$, $u_4$, $u_3$ and $\tilde u_4$ has scaling dimensions $1/2$, $1$, $3/2$ and $1$ respectively.
We see that $\U(1)^2$ subgroup of the flavor symmetry $\SU(3)$ is weakly gauged by the two $\U(1)$ vector multiplets with scalar components $u_4$ and $\tilde u_4$. The action of the outer automorphism $S_3$ of $\SO(8)$  on the dimension-1 operators $u_4$ and $\tilde u_4$ are generated by the parity operation $\tilde u_4\to -\tilde u_4$ and a $120^\circ$ rotation acting on the $u_4$-$\tilde u_4$ plane. This is exactly how the Weyl group of the flavor symmetry $\SU(3)$ acts on the two mass parameters $\mu_1$, $\mu_2$, $\mu_3$ with $\mu_1+\mu_2+\mu_3=0$.  
Therefore we see that the outer-automorphism symmetry of $\SO(8)$ can be identified with the Weyl group of the $\SU(3)$ flavor symmetry. 

\subsection{Seiberg-Witten solutions for various other simple gauge groups}\label{sec:others}
Let us close this section by very briefly mentioning the Seiberg-Witten solution of various other gauge theories.  First, let us copy the solutions for $\SU(N)$ and $\SO(2N)$ with $N_f$ flavors.
The \SeibergWitten\ curve for $\SU(N)$ with $N_f$ flavors was 
\begin{equation}
\frac{\Lambda^{N}}z
+ z\Lambda^{N-N_f}\prod_{i={1}}^{N_f}(x-\mu_i) 
= x^N+u_2 x^{N-2}+\cdots+u_{N-1}x+u_N
\end{equation}
and the \SeibergWitten\ curve for $\SO(2N)$ with $N_f$ flavors was 
\begin{equation}
x^2\left[\frac{\Lambda^{2N-2}}z + \Lambda^{2N-2-2N_f} z\prod_{i=1}^{N_f}(x^2-\mu_i{}^2)\right]
=x^{2N}+u_2 x^{2N-2} + u_4 x^{2N-4} + \cdots + u_{2N}.
\end{equation} 
Here, for simplicity, we dropped the flavor terms multiplying $1/z$  on the left hand sides. For the full expressions, see \eqref{suNNf} and \eqref{so2NNf}, respectively. 

Without derivations, we present here  the \SeibergWitten\ curves for other classical gauge groups. 
The \SeibergWitten\ curve for $\SO(2N+1)$ with $N_f$ flavors is \begin{equation}
x\left[\frac{\Lambda^{2N-1}}{z^{1/2}} + \Lambda^{2N-1-2N_f} {z^{1/2}}\prod_{i=1}^{N_f}(x^2-\mu_i{}^2)\right]
=x^{2N}+u_2 x^{2N-2} + u_4 x^{2N-4} + \cdots + u_{2N},
\end{equation} and the \SeibergWitten\ curve for $\Sp(N)$ with $N_f$ flavors is \begin{multline}
\frac{\Lambda^{2N+2}}{z^{1/2}} + 2c + \Lambda^{2N+2-2N_f} z^{1/2}\prod_{i=1}^{N_f}(x^2-\mu_i{}^2)\\
=x^2(x^{2N}+u_2 x^{2N-2} + u_4 x^{2N-4} + \cdots + u_{2N})
\end{multline} where $c^2=\Lambda^{4N+4-2N_f} \prod_{i=1}^{N_f} (-\mu_i^2)$.
The differential is always just $\lambda=xdz/z$. We again dropped the flavor terms multiplying $1/z$ on the left hand sides. 

The curves so far can be always written as \begin{equation}
\frac{F(x)}z + z \tilde F(x) = P(x)\label{FP}
\end{equation} for some polynomials $F(x)$, $\tilde F(x)$ and $P(x)$.
In the older literature, it is more common to find the curve and the differential in the form \begin{equation}
y^2=P(x)^2 - 4F(x)\tilde F(x),\qquad \lambda=\frac{x}2 d\log\frac{P(x)-y}{P(x)+y}.\label{yFP}
\end{equation}   To relate \eqref{FP} and \eqref{yFP}, note that the equation \eqref{yFP} implies that the combination\begin{equation}
\zeta_\pm=\frac12(P(x)\pm y)
\end{equation}satisfies  \begin{equation}
\zeta_++\zeta_-=P(x),\qquad \zeta_+\zeta_-= F(x)\tilde F(x).
\end{equation} Comparing with \eqref{FP}, we find \begin{equation}
\zeta_+= \frac{F(x)}z,\qquad \zeta_-=z\tilde F(x).
\end{equation}  This also explains the differential given in \eqref{yFP}. 

This older form is mathematically easier to deal with in certain situations, because the branch cut of the function $y(x)$ is at most of order 2. Mathematically, such Riemann surfaces are called hyperelliptic, and have  a few special properties compared to more general Riemann surfaces. That said,  the forms we use in the rest of the lecture note is much more physical and usually more useful. 

A good summary of the curves for classical gauge groups listed above can be found e.g.~in \cite{Nekrasov:2004vw}. For exceptional gauge groups, the situation is more complicated. Although one can write the \SeibergWitten\ curve, it is more natural to study the \SeibergWitten\ geometry,  which is a complex 3-dimensional space, fibered over the \Gaiotto curve $C$.  
A very nice presentation for the pure theory can be found in \cite{Hollowood:1997pp}. With matter hypermultiplets, a modern reference is \cite{Tachikawa:2011yr}.

\section{Argyres-Seiberg-Gaiotto duality for $\SU(N)$ theory}\label{sec:duality}
\subsection{S-dual of $\SU(N)$ with $N_f=2N$ flavors, part I}
\subsubsection{Rewriting of the curve}
We learned in the last section that the curve of $\SU(N)$ theory with $2N$ flavors is given by: \begin{equation}
\frac{\prod_{i=1}^N(\tilde x-\tilde\mu_i)}{\tilde z}+
f{\prod_{i=1}^N(\tilde x-\tilde\mu'_i)}{\tilde z} = \tilde x^N+\tilde u_2 \tilde x^{N-2}+\cdots +  u_N\label{step1}
\end{equation}
where $f$ is a complex number; the differential is $\tilde\lambda=\tilde x dz/z$. 
This theory is a superconformal theory deformed by mass terms, and $f$ is a function of the UV coupling constant $\tau_{UV}$.
We would like to understand the strong-coupling limits of this theory. 

As we did in Sec.~\ref{sec:gaiotto}, it is convenient to rewrite the curve in terms of the Seiberg-Witten differential $\lambda$, to the form \begin{equation}
\lambda^N+\phi_2(z)\lambda^{N-2}+\cdots + \phi_N(z)=0.\label{suN-gaiotto}
\end{equation}
We start from \eqref{step1}. First we gather terms with the same power of $\tilde x$: \begin{equation}
(1-\frac1{\tilde z}-f\tilde z)\tilde x^N + \heartsuit_1 \tilde x^{N-1}+\heartsuit_2 \tilde x^{N-2} + \cdots + \heartsuit_N=0
\end{equation} where \begin{equation}
\heartsuit_1=\frac{\sum\tilde\mu_i}{\tilde z} + f\tilde z\sum\tilde\mu'_i.
\end{equation}
We divide the whole equation by $(1-1/\tilde z-f\tilde z)$ and define $x=\tilde x+\heartsuit_1/(1-1/\tilde z-f\tilde z)/N$. We now have \begin{equation}
x^N + \clubsuit_2 x^{N-2}+\cdots + \clubsuit_N=0
\end{equation} where $\clubsuit_k$ has poles of order $k$ at two zeros $\tilde z_{1,2}$ of $1-1/\tilde z-f\tilde z=0$,
due to the shift from $\tilde x$ to $x$. We set $z=\tilde z/\tilde z_1$ so that one zero is now at 1, and another is at $q=\tilde z_2/\tilde z_1$. 

Introducing $\lambda=xdz/z$, we have an equation of the form \eqref{suN-gaiotto};
$\phi_k(z)$ has poles of order at most $k$ at $z=0$, $q$, $1$ and $\infty$.
Consider the case when all $\tilde\mu_i$ and $\tilde\mu'_i$ are generic, and assume $q\ll 1$. 
Then it is straightforward to determine how $\lambda$ behaves close to each of the singularity. 
As we are solving a degree-$N$ equation, we have $N$ residues at each singularity. 
They are given by \begin{equation}
\begin{array}{llllr@{\qquad}l@{}llll}
  \mu_1,&\mu_2,&\ldots,&\mu_{N-1},&\mu_N,&   z&\sim 0,&\\
  \mu,&\mu,&\ldots,&\mu,&(1-N)\mu,&   z&\sim q,&\\
  \mu',&\mu',&\ldots,&\mu',&(1-N)\mu',&   z&\sim 1,&\\
  \mu'_1,&\mu'_2,&\ldots,&\mu'_{N-1},&\mu'_N,&  z&\sim \infty.
\end{array}
\label{simple-full-massive-condition}
\end{equation}
Here  \begin{align}
\mu_i &= \tilde \mu_i - \frac{1}{N}\sum_i\tilde \mu_i + O(q) , &
\mu_i' &= \tilde \mu_i' - \frac{1}{N}\sum_i\tilde \mu_i' + O(q) , \\
\mu &=\frac{1}{N}\sum_i \tilde\mu_i + O(q),&
\mu' &=-\frac{1}{N}\sum_i \tilde\mu_i' + O(q).
\end{align} Here $\sum \mu_i=\sum \mu_i' = 0$.
Note that $\mu_i$ and $\mu$ are the mass parameters which enter the BPS mass formula.
We found that they are related to the parameters $\tilde \mu_i$ via a finite renormalization.

When $N=2$, the structure of the residues at all four punctures were of the same type, as they are all given by  $(m,-m)$.
For $N>2$, we see that the structure of the residues at $z=0,\infty$ and the structure at $z=q,1$ are different. The former is of the form $(m_1,\ldots,m_N)$ with $\sum m_i=0$,
and the latter is of the form $m(1,1,\ldots,1-N)$. 

It is also instructive to consider the completely massless case, when we have $\tilde\mu_i=\tilde\mu_i'=0$ for all $i$. The original curve is just \begin{equation}
\frac{x^N}{z}+
f{x^N}{ z} =  x^N+ u_2 x^{N-2} +\cdots+  u_N.
\end{equation} After the same manipulation as above, we find \begin{equation}
\phi_k(z)=\frac{\hat u_k}{(z-q)(z-1)}\frac{dz^k}{z^{k-1}}.
\end{equation} Therefore, 
\begin{equation}
\begin{array}{rll}
\text{$\phi_k(z) $ has poles of order} & k-1 & \text{when $z=0,\infty$}, \\
\text{$\phi_k(z) $ has poles of order} & 1 & \text{when $z=q,1$}. 
\end{array}\label{simple-full-condition}
\end{equation}
We observe here again that the behavior of the poles are all the same when $N=2$, while the behavior at $z=0,\infty$
and the behavior at $z=1,q$ are distinct when $N>2$.

\begin{figure}[h]
\[
\inc{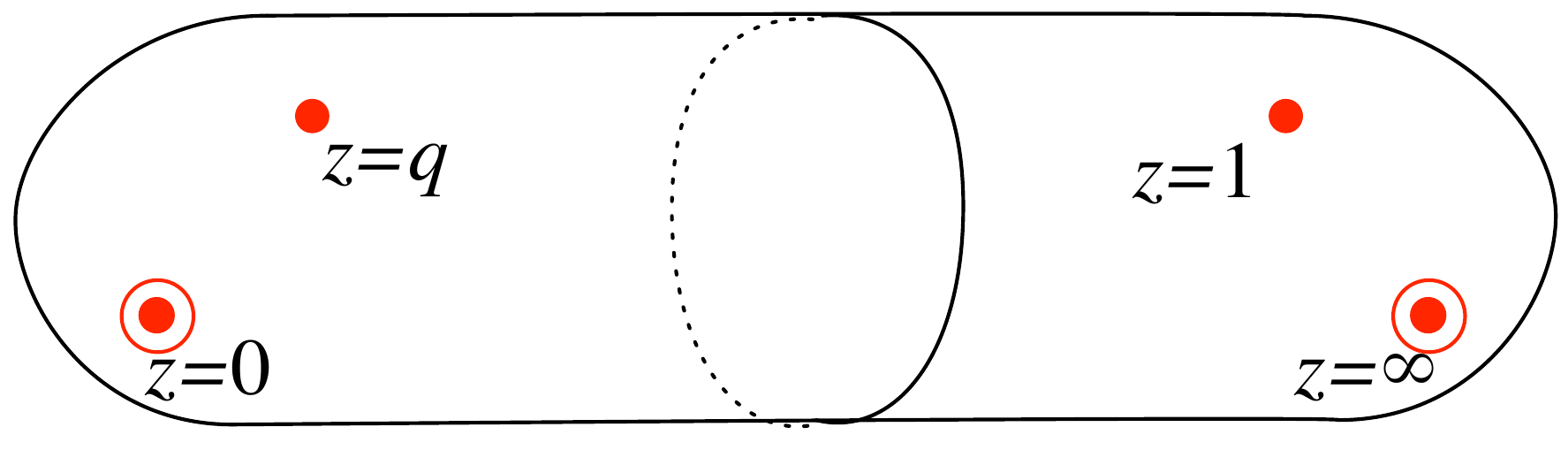}
\]
\caption{The \Gaiotto\ curve of $\SU(N)$ theory with $2N$ flavors\label{fig:suNgaiotto}}
\end{figure}

We have $2N$ mass terms in the system. First of all we split them into $N$ mass terms encoded in the region $z\sim 0$,
and another $N$ mass terms in the region $z\sim\infty$.
Correspondingly, we started from the flavor symmetry $\U(2N)$ and decomposed it into $\U(N)\times \U(N)$. We further decompose each of $\U(N)$ into $\SU(N)$ and $\U(1)$.
Combined, we use the decomposition of the flavor symmetry of the form
 \begin{equation}
\U(2N) \supset \U(N)\times \U(N)
\simeq [\SU(N)_A \times \U(1)_B] \times [\U(1)_C\times \SU(N)_D].
\end{equation}
The residues of $\lambda$ at the puncture $A$ at $z=0$ and 
those at the puncture $D$ at $z=\infty$ encode the mass terms for $\SU(N)_{A,D}$ respectively,
whereas those at the puncture $B$ at $z=q$ and 
those at the puncture $C$ at $z=1$  encode the mass terms for  $\U(1)_{B,C}$; compare \eqref{simple-full-massive-condition}.

We then say that the singularity at $z=0$ carry the $\SU(N)$ symmetry,
the one at $z=q$ carry the $\U(1)$ symmetry,
and similarly for those at $z=1$, $=\infty$. We can visualize the situation as in Fig.~\ref{fig:suNgaiotto}.
We call the punctures at $z=0,\infty$ the full punctures,
and those at $z=q,1$ the simple punctures. 
In the 6d viewpoint, these are four-dimensional defect objects extending along the Minkowski $\bR^{3,1}$, and they carry respective flavor symmetries on them. 

When $N=2$, the original symmetry is not just $\U(2)$ but $\SO(4)$. Accordingly, the split 
$\U(2)\simeq \SU(2)\times \U(1)$ is enhanced to the following structure \begin{equation}
\begin{array}{ccccc}
\SO(4) &\supset& \SU(2)&\times& \SU(2)\\
\cup && \rotatebox{90}{$=$} && \cup\\
\U(2) &\supset& \SU(2)&\times& \U(1)
\end{array} 
\end{equation} and therefore the distinction of the types of punctures is gone.

\subsubsection{Weak-coupling limit}

\begin{figure}[h]
\[
\inc{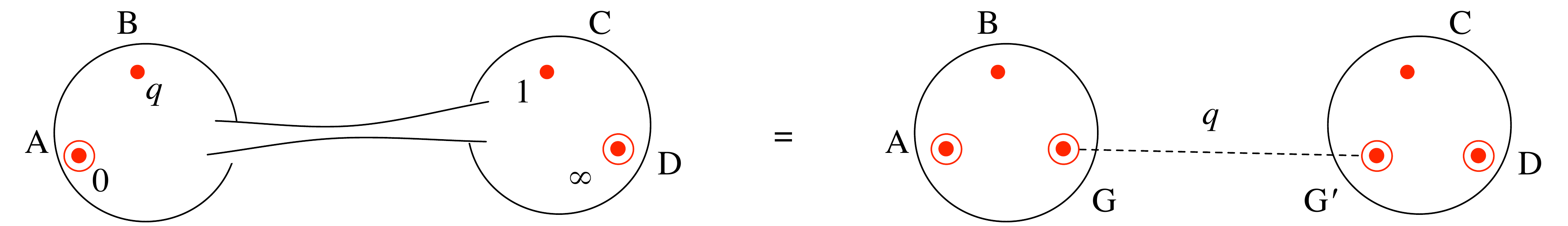}
\]
\caption{Weakly coupled limit \label{fig:weakSUN}}
\end{figure}
Clearly $f\sim q\sim e^{2\pi i \tau_{UV}}$ in the weak coupling region, see Fig.~\ref{fig:weakSUN}.
When the coupling is extremely weak, we can think that the four-punctured sphere on the left 
is composed of two three-punctured spheres. 
In the tube region connecting the two, the behavior of $\lambda$ is essentially given just by 
\begin{equation}
\phi_k(z)\sim u_k\frac{dz^k}{z^k}.
\end{equation} Writing \begin{equation}
\prod(x-a_i) = x^N + u_2 x^{N-2} + \cdots +u_N,
\end{equation} we find that the residues of $\lambda$ in the tube region is given by $a_1,\ldots,a_N$.
Therefore, we find full punctures after we split off two spheres. 

The resulting three-punctured sphere has one simple puncture and two full punctures.
Therefore it should carry $\U(1)\times \SU(N)\times \SU(N)$ symmetry. 
The four-punctured sphere represents the $\SU(N)$ theory with $2N$ flavors.
The tube region carries the $\SU(N)$ vector multiplet.
Then each three-punctured sphere just represents $N$ flavors, i.e.~hypermultiplets $(Q_i^a,\tilde Q^i_a)$
where $a,i=1,\ldots,N$. 
Then two $\SU(N)$ symmetries can be identified with those acting on the index $a$ and $i$ respectively,
and the $\U(1)$ symmetry is such that $Q$ has charge $+1$ while $\tilde Q$ has charge $-1$.

The \Gaiotto\ curve of the $\SU(N)$ theory with $2N$ flavors, shown in Fig.~\ref{fig:suNgaiotto},
is composed of two copies of this three-punctured sphere.
The $2N$ hypermultiplets  are split into $N$ hypermultiplets $(Q_i^a,\tilde Q^i_a)$ charged under 
$\SU(N)_A$ and $\U(1)_B$,
and another $N$ hypermultiplets $(Q'{}_i^a,\tilde Q'{}^i_a)$ charged under 
$\SU(N)_D$ and $\U(1)_C$.

\begin{figure}[h]
\[
\inc{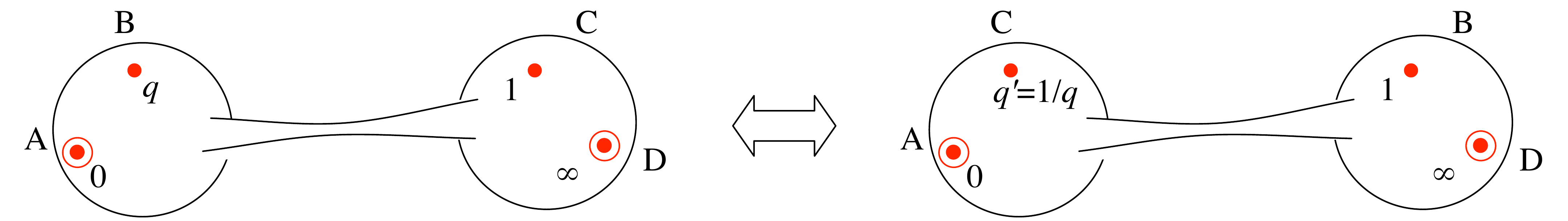}
\]
\caption{S-duality of $\SU(N)$ $2N$ flavors\label{fig:SUN-S}}
\end{figure}

\subsubsection{A strong-coupling limit}

Let us consider what happens when $q\to \infty$. As shown in Fig.~\ref{fig:SUN-S}, it just ends up exchanging the puncture $B$ and $C$, at the same time redefining the coupling $q$ via $q'=1/q$.  This means that this strongly-coupled limit turns out to be another weakly-coupled $\SU(N)$ gauge theory with $2N$ flavors. 
This time,  the $2N$ hypermultiplets  are split into $N$ hypermultiplets $(q_i^a,\tilde q^i_a)$
and another $N$ hypermultiplets $(q'{}_i^a,\tilde q'{}^i_a)$, but notice that the first $N$ are charged under $\SU(N)_A$ and $\U(1)_C$ while the second $N$ are charged under $\SU(N)_D$ and $\U(1)_B$.
As we learned for the case of the $\SU(2)$ theory with four flavors in Sec.~\ref{sec:su2-4-strong}, the new quarks are magnetic from the point of view of the original theory. 

\begin{figure}[h]
\[
\inc{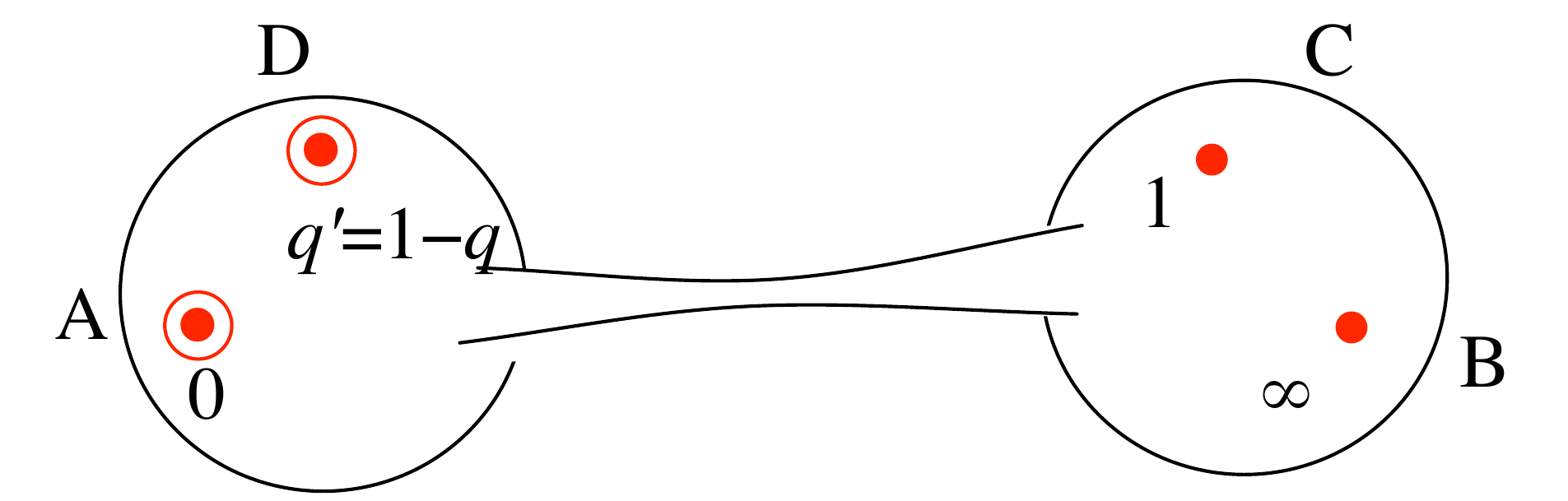}
\]
\caption{Another limit of $\SU(N)$ $2N$ flavors:\label{fig:Argyres-Seiberg}}
\end{figure}
We would like to understand the limit $q\to 1$ too. We need to split the four-punctured sphere as shown in Fig.~\ref{fig:Argyres-Seiberg}. But the configuration of punctures are not what we already know: we have two full punctures on one side, and two simple punctures on the other side.
We need to study more about the 6d construction before answering what happens in the limit.

\subsection{$\SU(N)$ quiver theories and tame punctures}\label{sec:tame}

\subsubsection{Quiver gauge theories}
To this aim, we introduce a new diagrammatic notation for $\cN{=}2$ gauge theories. This notation is related to  but distinct from the trivalent one introduced in Sec.~\ref{sec:trivalent}.

A diagram is composed of squares and circles with integers written in them,
and edges connecting squares and circles. 
A square with $N$ stands for a $\U(N)$ flavor symmetry,
and a circle with $N$ an $\SU(N)$ gauge symmetry. 
An edge connecting two objects with $N$ and $M$  written within them represents a hypermultiplet $(Q_i^a,\tilde Q^i_a)$
where $i=1,\ldots,N$ and $j=1,\ldots,M$. They are in the tensor product of the fundamental representation of $\SU(N)$ and $\SU(M)$, and is called the bifundamental hypermultiplet. 
Such a diagram specifies an $\cN{=}2$ gauge theory. This class of theories is often called quiver gauge theories.

\begin{figure}[h]
\[
\inc{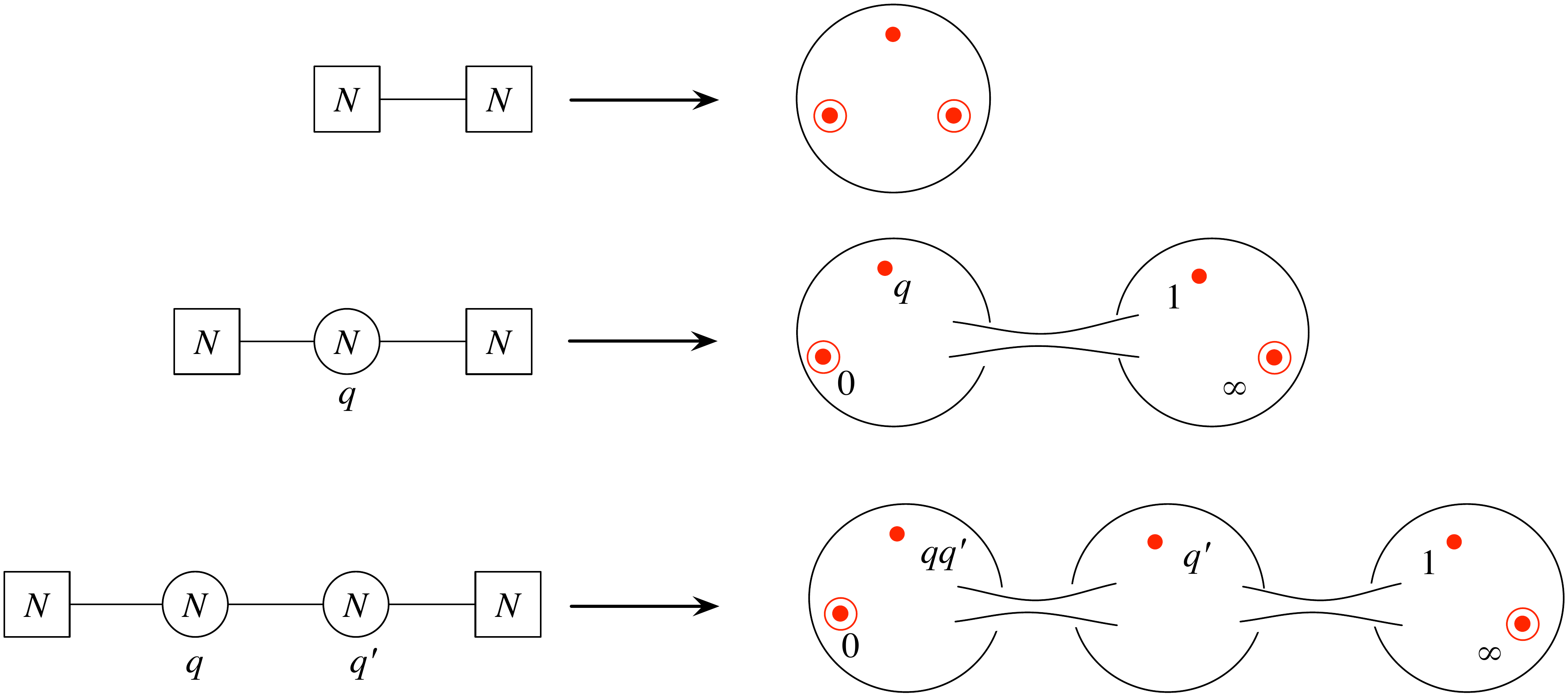}
\]
\caption{$\SU(N)$ quiver theory\label{fig:SUN-quiver}}
\end{figure} 

The simplest cases are when all the squares and circles have the same number $N$ written in them, see Fig.~\ref{fig:SUN-quiver}.  The first one in the figure is just a bifundamental hypermultiplet.  The second one is the $\SU(N)$ theory with $2N$ flavors.
The last one is an $\SU(N)_1\times \SU(N)_2$ theory, so that
\begin{itemize}
\item  there is a bifundamental hypermultiplet for $\SU(N)_1\times \SU(N)_2$, and
\item there are $N$ fundamental hypermultiplets for $\SU(N)_1$, and 
\item there are $N$ fundamental hypermultiplets for $\SU(N)_2.$
\end{itemize} Note that both $\SU(N)_{1}$ and $\SU(N)_2$ have zero beta function.

Their Seiberg-Witten solutions can be obtained by combining the knowledge we acquired so far. Namely,
each edge corresponds to the bifundamental hypermultiplet of $\SU(N)\times \SU(N)$, which we know to come from
a three punctured sphere of 6d theory of type $\SU(N)$,
with two full punctures and one simple puncture. 
All we have to do then is to prepare one such sphere for each edge, and connect pairs of full punctures by tubes. 
For example,  the Seiberg-Witten solution for the third theory in Fig.~\ref{fig:SUN-quiver} is given by \begin{equation}
\lambda^N+\phi_2(z)\lambda^{N-2}+\cdots + \phi_N(z)=0\label{qwerty}
\end{equation} where $\phi_k(z)$ has five singularities, such that two at $z=0$, $=\infty$ are full and the other three at $z=1,$ $q$ and $qq'$  are simple. 

For simplicity, let us assume that all the mass parameters are zero. Then, from the condition of the order of the poles of the singularities given in \eqref{simple-full-condition}, the fields $\phi_k(z)$ are uniquely fixed to be \begin{equation}
\phi_k(z)=\frac{u_k^{(1)}z+u_k^{(2)}}{(z-1)(z-q)(z-qq')} \frac{dz^k}{z^{k-1}}. 
\end{equation} The reader should check that it has the correct behavior at $z=\infty$.
This theory is superconformal, as both $\SU(1)_1$ and $\SU(2)_2$ have zero one-loop beta function. 
This is reflected by the fact that the variables appearing in the \SeibergWitten\ curve \eqref{qwerty} can be assigned scaling dimensions in a natural way. The differential $\lambda$ should have scaling dimension one, since its integral gives the BPS mass formula: $[\lambda]=1$.  We then set $[z]=0$ and $[\phi_k]=k$. 
This means that $u_k^{(i=1,2)}$ should be two Coulomb branch operators with scaling dimension $k$.
Indeed, we are dealing with an $\SU(N)^2$ gauge theory which is superconformal, and there are exactly one Coulomb branch operator of scaling dimension $k$ for $k=2,\ldots,N$. 

\subsubsection{$\cN{=}2^*$ theory}
\begin{figure}[h]
\[
\inc{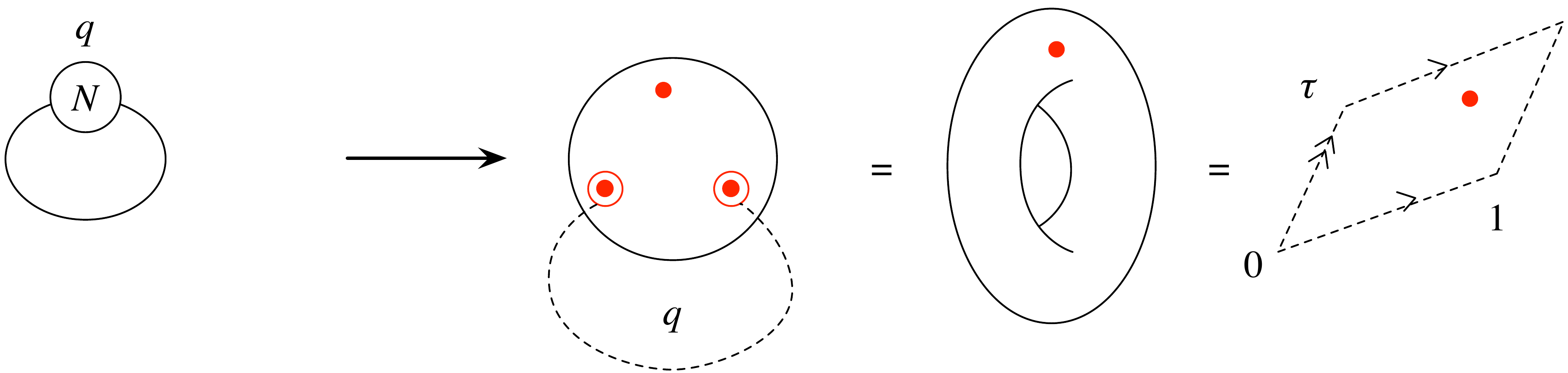}
\]
\caption{$\SU(N)$ plus adjoint: the $\cN{=}2^*$ theory. \label{fig:N2*}}
\end{figure}

A rather degenerate situation arises when we take just one bifundamental hypermultiplet $(Q_i^a,\tilde Q^a_i)$ and couple one $\SU(N)$ gauge multiplet to both indices, see Fig.~\ref{fig:N2*}.
The $N\times N$ hypermultiplet components now behave as an adjoint representation plus a singlet. 
The singlet part is completely decoupled, and therefore  the theory is essentially the $\SU(N)$ gauge theory with an adjoint hypermultiplet. When massless this is the $\cN{=}4$ super Yang-Mills, whereas it is called $\cN{=}2^*$ theory when massive. 
The Seiberg-Witten solution can then be obtained by taking a three-punctured sphere and connecting the two full punctures.  
We end up having a torus with one simple puncture.
This solution was first found in \cite{Donagi:1995cf}, to which the readers should refer for details. 

%  More explicitly, we have \begin{equation}
%\lambda^N+\phi_2(z)\lambda^{N-2}+\cdots + \phi_N(z)=0
%\end{equation} with $z\sim z+1\sim z+\tau$. We put the simple puncture at $z=0$ without loss of generality.
%At the simple puncture, $\lambda$ should have $N$ residues given by \begin{equation}
%\mu, \mu, \ldots,  \mu, (1-N)\mu.
%\end{equation} This condition uniquely fixes the residues of the elliptic functions $\phi_k(z)$. 
%Pick one such elliptic function and call it $\phi_k^{(0)}(z)$. Then the difference is a holomorphic function without any residue on the torus, which is a constant: \begin{equation}
%\phi_k(z)=\phi_k^{(0)}(z)+u_k dz^k.
%\end{equation} We can identify $u_k$ with the dimension-$k$ Coulomb branch operator of $\SU(N)$ gauge multiplet.  

\subsubsection{Linear quiver theories}

So far we learned how to solve gauge theories  shown in Fig.~\ref{fig:SUN-quiver}. 
They have the gauge group \begin{equation}
\SU(N)\times \cdots \times \SU(N)\times \cdots \times \SU(N)
\end{equation} with bifundamentals between adjacent $\SU(N)$ groups,
and additional $N$ flavors each for the first and the last $\SU(N)$ groups. All $\SU(N)$ groups have zero beta function.

Let us consider a slight generalization of this class of theories. The gauge group is of the following form \begin{equation}
\SU(N)\times\cdots\times \SU(N)\times \SU(N_k) \times \SU(N_{k-1}) \times  \cdots \SU(N_2) \times \SU(N_1).\label{quivergauge}
\end{equation} We put the bifundamental hypermultiplets between adjacent $\SU(N)$ and $\SU(N')$.
Such gauge theories are often called linear quiver gauge theories, since the gauge factors are arranged in a linear fashion. 

Here, we introduce additional flavors for every $\SU$ group, so that they all have zero beta functions. Define $N_0=0$ and $N_{k+1}=N$.
Then the condition we need to impose is \begin{equation}
N_{i-1}+N_{i+1}+n_i=2N_i,  \qquad i=1,\ldots,k
\end{equation}  where  $n_i$ is the number of additional fundamental hypermultiplet for $\SU(N_i)$. 
Since $n_i\ge 0$, we have $s_i \ge s_{i+1}$ where $s_i=N_{i}-N_{i-1}$.
Clearly $\sum_{i=1}^{k+1} s_i=N$. 

\begin{figure}[h]
\[
\inc{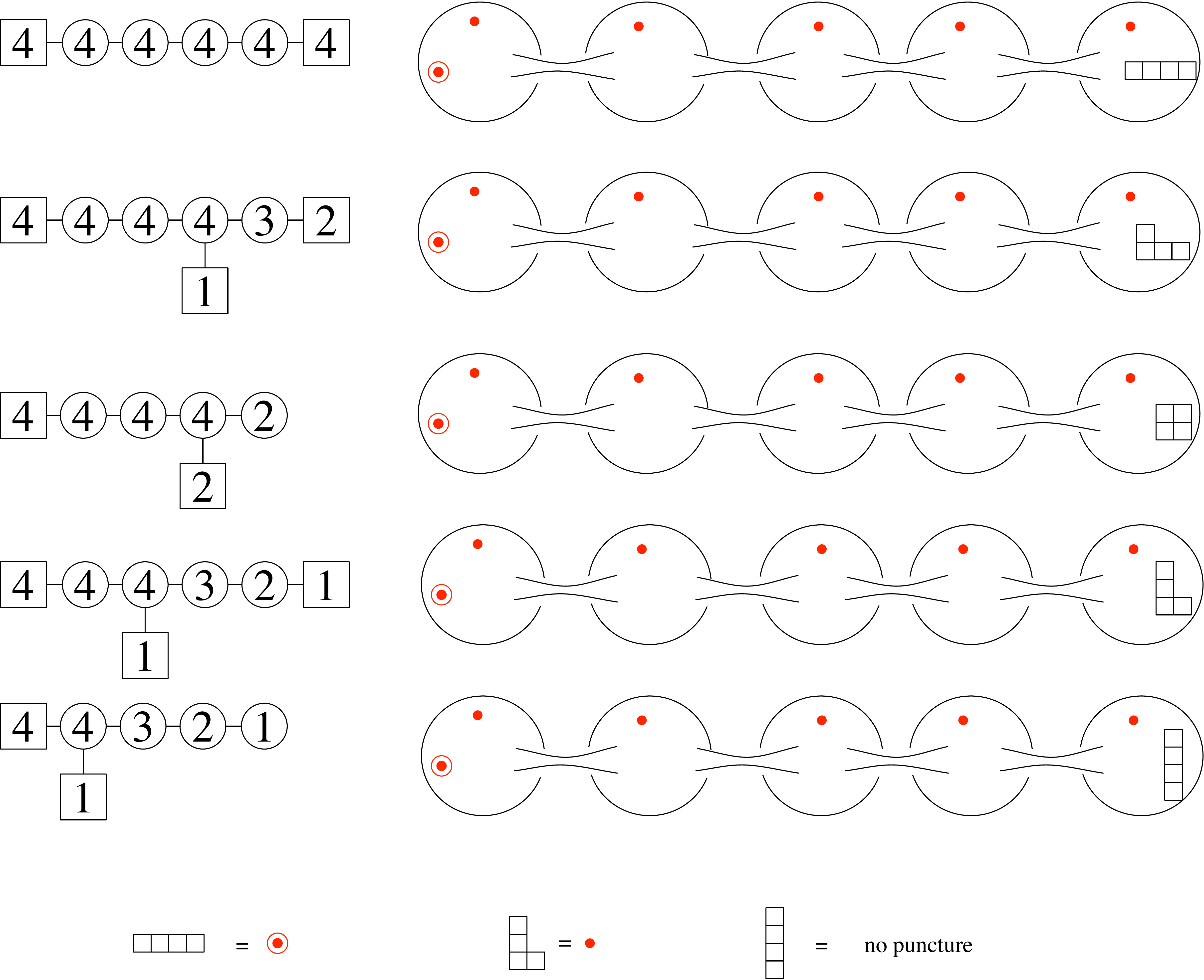}
\]
\caption{$\SU(N)$ tame punctures\label{fig:SUNtame}}
\end{figure}

A decreasing sequence of integers $s_{1} \ge s_2 \ge \cdots \ge s_{k+1}$ whose sum is $N$ is called a partition of $N$. 
Then we can phrase our finding here by saying that this type of gauge theory can be characterized by a partition of $N$.
A partition can be graphically represented by a Young diagram. Here we draw it by arranging boxes so that the widths of the rows are given by $s_i$. Examples are shown for $N=4$ on the left hand side of Fig.~\ref{fig:SUNtame}.
There, additional $n_i$ flavors are shown by connecting a box $n_i$ to a circle $N_i$.

What is  the Seiberg-Witten solutions of this class of theories?
There are a few independent methods to arrive at the solutions. 
Originally they are obtained using a configuration of branes in type IIA string theory and lifting it to M-theory \cite{Witten:1997sc}.
We now also have a field theoretical derivation in terms of instanton computation \cite{Nekrasov:2012xe}.
In this subsection, we just state the results, and give a few justification.
We will come back to this point in more details in Sec.~\ref{sec:higgsing}.

The Seiberg-Witten solution is obtained by the following procedure.
First, consider a sphere of 6d theory of type $\SU(N)$, realizing the theory where all $N_i$ is equal to $N$.
It was given by  the \SeibergWitten\ curve of the form \begin{equation}
\lambda^N+\phi_2(z)\lambda^{N-2}+\cdots + \phi_N(z)=0.\label{mnb}
\end{equation} 
 As explained above,
we have two full punctures and a number of simple punctures. 
We then replace one full puncture at $z=\infty$ with a new type of puncture labeled by the Young diagram, see the right hand side of Fig.~\ref{fig:SUNtame}.  These new types of punctures, together with the simple and the full punctures introduced already, are called tame $\SU(N)$ punctures.

The change of the type of the puncture is the change of the structure of the singularities of the fields $\phi_k(z)$.  We can also write the curve \eqref{mnb} as \begin{equation}
\det(\lambda-\varphi(z))=0\label{vcx}
\end{equation} where $\varphi(z)$ is a meromorphic one-form which is a traceless $N\times N$ matrix,
as we did for the $\SU(2)$ case in Sec.~\ref{sec:hitchin}. Then $\phi_k(z)$ is given by an elementary symmetric degree-$k$ polynomial combination of the eigenvalues of $\varphi(z)$.  Then the structure of the singularities of $\phi_k(z)$ 
can be described also by the structure of the residue of $\varphi(z)$. 

\subsubsection{Tame punctures}
We already saw a full puncture carries the flavor symmetry $\SU(N)$, and a simple puncture $\U(1)$. 
To correctly reproduce the flavor symmetry of the total theory, the singularity at $z=\infty$ labeled by the  Young diagram $s_1\ge s_2 \ge \cdots \ge s_{k+1}$ 
needs to be associated to the flavor symmetry \begin{equation}
\mathrm{S}[\U(n_1)\times \U(n_2)\times \ldots \U(n_k)]\label{young-flavor}
\end{equation} where the $\mathrm{S}[\cdots]$ means that we remove the diagonal $\U(1)$ of the following unitary gauge groups.

\begin{figure}[h]
\[
\inc{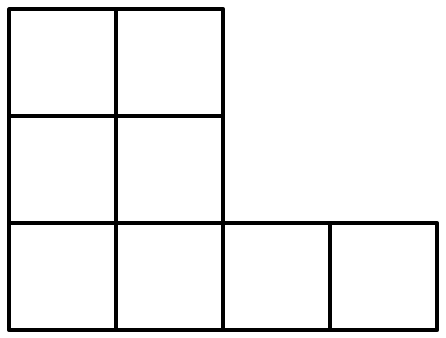}
\]
\caption{The Young diagram shown here has $(s_i)=(4,2,2)$,
$(\nu_k)=(1,1,1,1,2,2,3,3)$, $(p_k)_{k=1}^8=(0,1,2,3,3,4,4,5)$ and  $(t_i)=(3,3,1,1)$.
The standard convention is to use the column heights $(t_i)$  to label punctures. 
 \label{fig:young}}
\end{figure}

The description becomes complete once we describe how the fields $\phi_k(z)$ behave at this new puncture.  
When the hypermultiplets are all massless, the rule is given as follows.
Given a Young diagram with row widths $s_1\ge s_2 \ge \cdots $,  define
$p_k=k-\nu_k$ where \begin{equation}
(\nu_1,\nu_2,\ldots,\nu_N)=
(\underbrace{1,\ldots,1}_{s_1},\underbrace{2,\ldots,2}_{s_2},\ldots,)\label{pole-structure}
\end{equation}
Then $\phi_k(z)$ should have a pole of order $p_k$ at the puncture.  For an example, see Fig.~\ref{fig:young}.  

In terms of the $N\times N$ matrix-valued one-form $\varphi(z)$ the statement is somewhat simpler. 
Namely, the residue of $\varphi(z)$ at the puncture should be given by \begin{equation}
\Res \varphi(z) \sim J_{s_1}\oplus J_{s_2} \oplus \cdots \oplus J_{s_k}\label{ooo}
\end{equation} where $J_{s}$ is an $s\times s$ Jordan block: \begin{equation}
J_s= \underbrace{\begin{pmatrix}
0&1 \\
&0&1 \\
&& 0 & 1\\
&&&\ddots &\ddots \\
&&& &0 
\end{pmatrix}}_s .
\end{equation}
It is a good exercise to check that the pole orders $p_k$ of $\phi_k(z)$ 
can be reproduced by plugging in \eqref{ooo} into \eqref{vcx} and comparing it with \eqref{mnb}.

When the hypermultiplets are massive, the rule goes instead as follows.
Take the same Young diagram, but describe it with column heights $t_1\ge t_2\ge \cdots t_x$
where $x$ is the number of columns. 
 Then $\lambda$ should have $N$ residues with following structure:\begin{equation}
(\underbrace{\mu_1,\ldots,\mu_1}_{t_1},\underbrace{\mu_2,\ldots,\mu_2}_{t_2},\ldots,)\label{mass-structure}
\end{equation} where we need to impose \begin{equation}
\sum t_i \mu_i=0.
\end{equation}
This is equivalent to say that the residue of the matrix-valued one-form $\varphi(z)$ 
should be conjugate to a diagonal matrix with entries given by \eqref{mass-structure}.

We identify these residues with the mass parameters associated to the flavor symmetry \eqref{young-flavor}.
There are $n_i$ mass parameters $\mu^{(i)}_{a}$, $a=1,\ldots,n_i$ for each $\U(n_i)$. 
We then make the identification 
 \begin{equation}
(\mu^{(1)}_1,\ldots,\mu^{(1)}_{n_1} ;\mu^{(2)}_1,\ldots,\mu^{(2)}_{n_2} ;\cdots;
\mu^{(k)}_1,\ldots,\mu^{(k)}_{n_k} )
= (\mu_1,\mu_2,\ldots,\mu_x). 
\end{equation}  Note that $\sum n_i$ equals the number of columns $x$.
The individual $n_i$ corresponds to the number of columns of a certain given height, say $h$,
then there is an index $a$ such that \begin{equation}
t_a=t_{a+1}=\cdots=t_{a+n_i-1} = h.
\end{equation} Then the Weyl group of the $\U(n_i)$ flavor symmetry 
can be identified with the permutation of the columns of height $h$. 

It is conventional in the $\cN{=}2$ literature to label the punctures using column heights $(t_i)$. 
The full puncture is then associated to the Young diagram $(1,1,\ldots,1)$,
and the simple puncture has the Young diagram $(N-1,1)$.
We can indeed check that the general formulas \eqref{pole-structure} and \eqref{mass-structure}
reproduce \eqref{simple-full-massive-condition} and \eqref{simple-full-condition}.
Note also that the puncture of type $(N)$ does not have poles at all. This corresponds to an absence of the puncture.

\begin{figure}[h]
\[
\inc{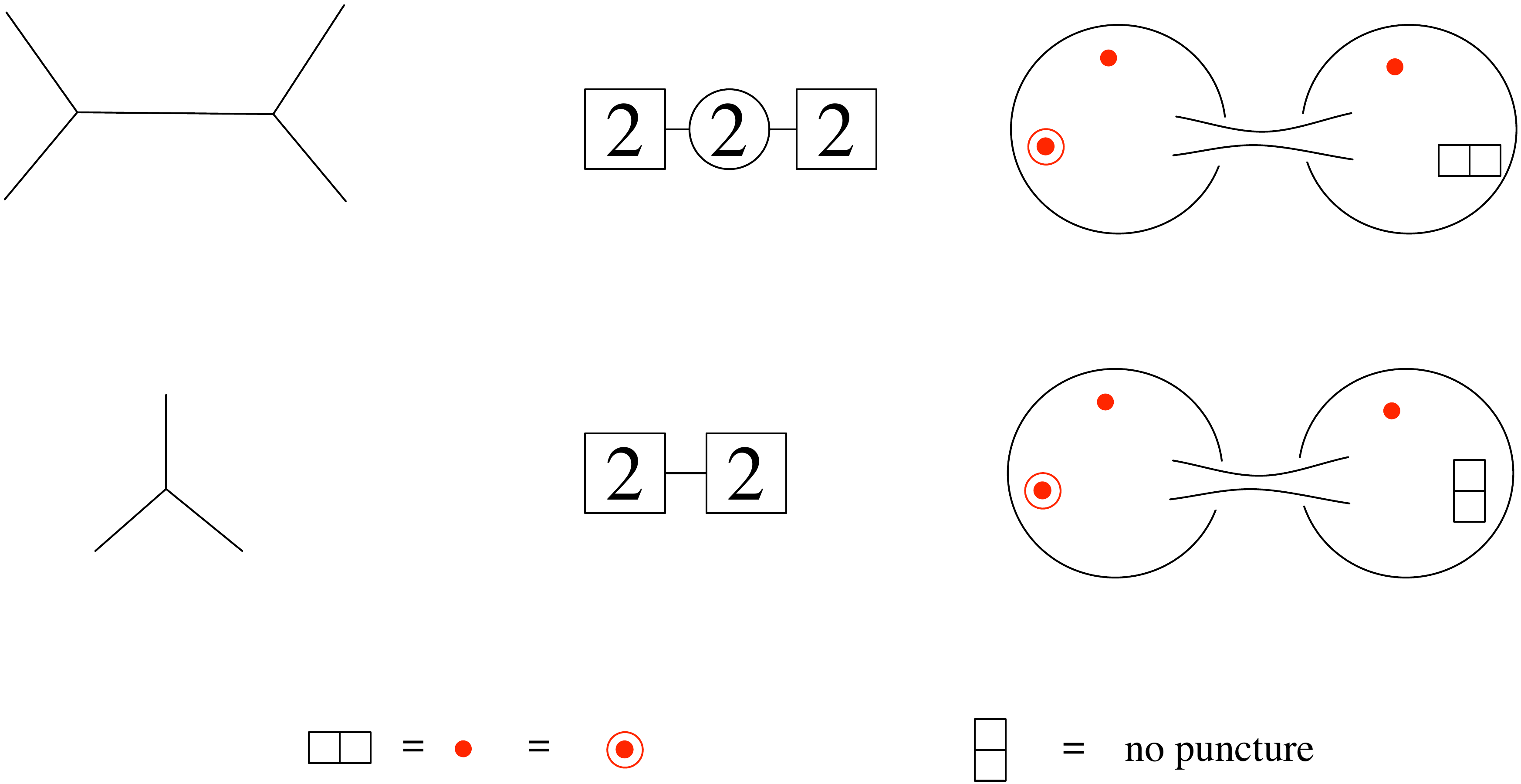}
\]
\caption{$\SU(2)$ tame punctures\label{fig:SU2-tame}}
\end{figure}
Let us apply this general discussion to the particular case  $N=2$ which we discussed extensively in Sec.~\ref{sec:gaiotto}.
There, we introduced a different diagrammatic notation using trivalent vertices, reflecting special properties of $\SU(2)$, see Fig.~\ref{fig:SU2-tame}.
In the current approach, we see that both the full puncture and the simple puncture for $N=2$ have the Young diagram $(1,1)$, thus losing the distinction.  The only other type of puncture is $(2)$, which corresponds to the absence of puncture in the first place.
Therefore the construction in this section does not give anything new for $N=2$.

\subsubsection{Tame punctures and the number of Coulomb branch operators}
Let us check  that the prescription described above  reproduces the expected number of Coulomb branch operators.
Compare, for example, the first and the fourth rows of Fig.~\ref{fig:SUNtame}.
The Seiberg-Witten solutions are both given by \begin{equation}
\lambda^4+\phi_2(z)\lambda^2+\phi_3(z)\lambda+\phi_4(z)=0.
\end{equation} In both cases, $\phi_k(z)$ has one full puncture at $z=0$ and five simple punctures at $z=z_i$.
The puncture at $z=\infty$ changes types. For the theory at the first row, the puncture at $z=\infty$ is a full puncture,
where $\phi_k(z)$ has poles of order $k-1$. This determines the fields $\phi_k(z)$ to be given by \begin{equation}
\phi_k(z)=\frac{u_k^{(1)}+u_k^{(2)}z+u_k^{(3)}z^2+u_k^{(4)}z^3}{\prod_i^5(z-z_i)}\frac{dz^k}{z^{k-1}}.\label{phifull}
\end{equation} Note that the degree of the polynomial in the numerator is fixed by the order of the pole at $z=\infty$. 
We identify $u_k^{(i)}$ as the dimension-$k$ Coulomb branch operator of the $i$-th $\SU(4)$  gauge group. 

Now change the type of the puncture at $z=\infty$. The allowed order of the pole there is reduced by $\nu_k$ as given in \eqref{pole-structure}.
In this particular case, the orders of the poles for $\phi_2(z)$, $\phi_3(z)$, $\phi_4(z)$ are reduced by $0$, $1$, $2$  respectively.
This reduces the degree of the polynomials in the numerator of \eqref{phifull} by $0$, $1$, $2$ respectively, resulting in \begin{align}
\phi_2(z)&=\frac{u_2^{(1)}+u_2^{(2)}z+u_2^{(3)}z^2+u_2^{(4)}z^3}{\prod_i^5(z-z_i)}\frac{dz^2}{z}\\
\phi_3(z)&=\frac{u_3^{(1)}+u_3^{(2)}z+u_3^{(3)}z^2}{\prod_i^5(z-z_i)}\frac{dz^3}{z^2}\\
\phi_4(z)&=\frac{u_4^{(1)}+u_4^{(2)}z}{\prod_i^5(z-z_i)}\frac{dz^4}{z^3}.
\end{align} 
We identify $u_k^{(i)}$ as a dimension $k$ Coulomb branch operator for the $i$-th gauge group.
We see that the third gauge group now has the Coulomb branch operators of dimension 2 and of dimension 3,
and that the fourth gauge group only has the Coulomb branch operator of dimension 2. 
This agrees with our claim that the gauge group is now $\SU(4)\times \SU(4)\times \SU(3)\times \SU(2)$.

This analysis of the number of the Coulomb branch operators can be extended to arbitrary $N$ and to arbitrary Young diagram.
By a straightforward but somewhat cumbersome combinatorial computation we see that the pole structure \eqref{pole-structure}
reproduces the structure of the gauge group as given in \eqref{quivergauge}.

\subsubsection{Tame punctures and the decoupling}

Now let us study what happens when we make the coupling of the last gauge group in \eqref{quivergauge} very weak. 
When we completely turn off the coupling, we lose the last gauge group $\SU(N_k)$.
The new last gauge group is $\SU(N_{k-1})$, which is now coupled to $N_k+n_{k-1}$ hypermultiplets in the fundamental representation.
Note that $N_k$ of them originally came from the bifundamental hypermultiplet for $\SU(N_{k-1})\times \SU(N_k)$.

\begin{figure}[h]
\[
\inc{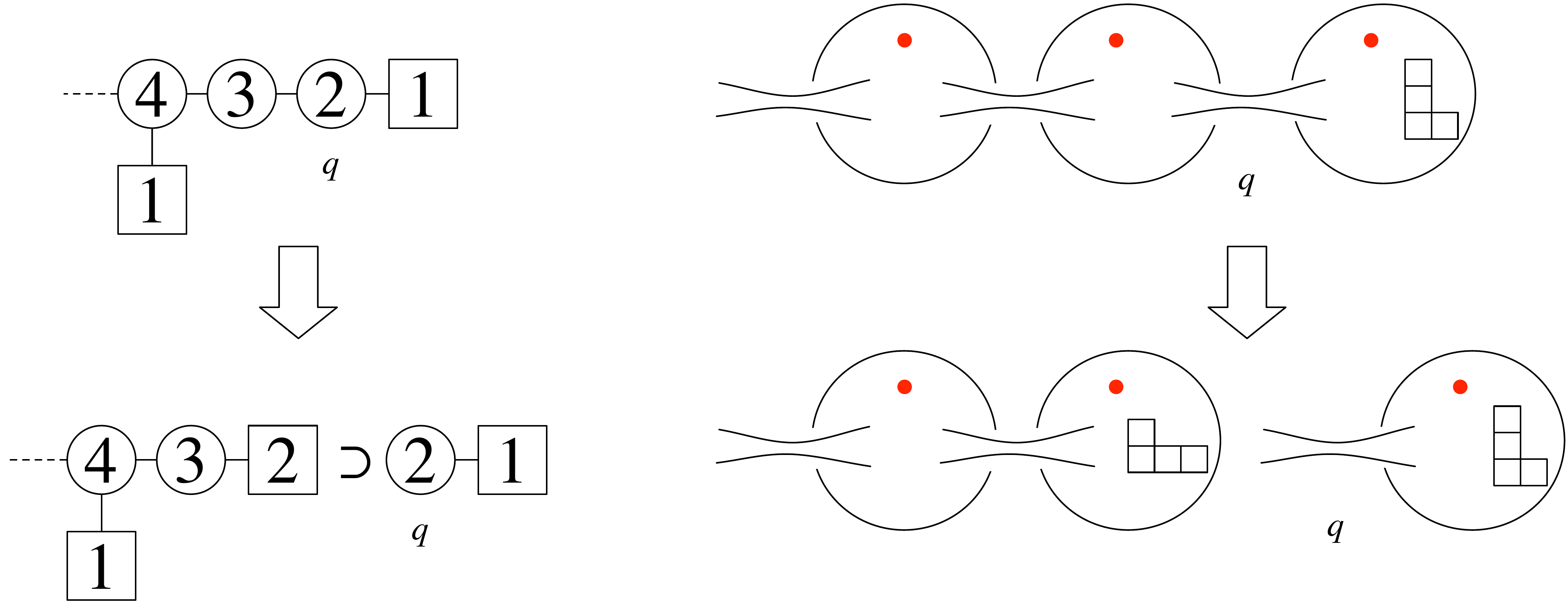}
\]
\caption{Decoupling one.\label{fig:decoupling1}}
\end{figure}

This process for the quiver tail characterized by the Young diagram $(3,1)$ is shown on the right hand side of Fig.~\ref{fig:decoupling1}.
In terms of the \Gaiotto\ curve, turning off the coupling of the last gauge group corresponds to splitting off the last two punctures. 
When we completely decouple the gauge group, we find a new puncture emerging. The type of this new puncture can be determined by the rule explained above, from the resulting gauge theory with one less gauge group. In this case, the newly appearing puncture on the left has the Young diagram $(2,1,1)$. 
The decoupled three-punctured sphere on the right hand side represents one hypermultiplet in the doublet  representation of $\SU(2)$.
 We intentionally do not discuss the  new puncture arising on this decoupled three-punctured sphere on the right; for more details, see \cite{Chacaltana:2010ks,Gaiotto:2011xs}.

\begin{figure}[h]
\[
\inc{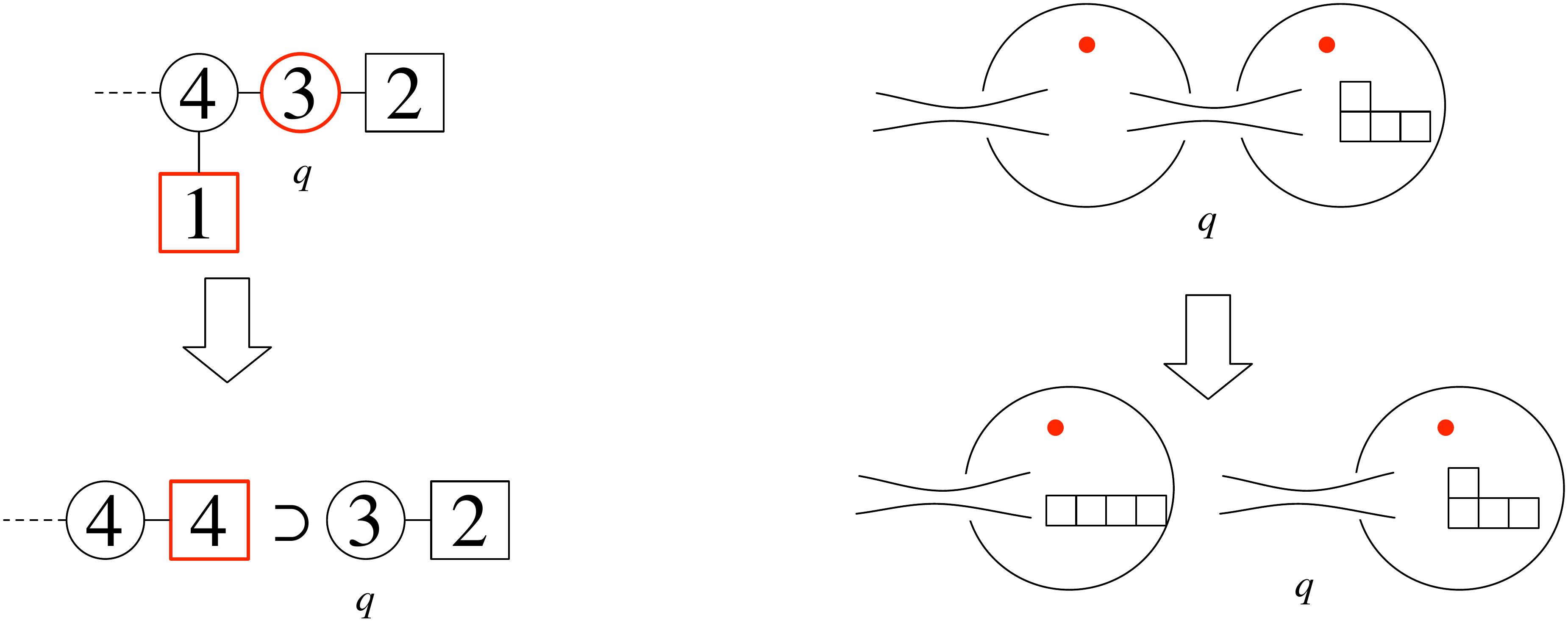}
\]
\caption{Decoupling the next.\label{fig:decoupling2}}
\end{figure}
We can continue the process. Decoupling the next gauge group, the Young diagram becomes $(1,1,1,1)$, i.e.~the full puncture. 
The situation is shown in Fig.~\ref{fig:decoupling2}.
The decoupled three-punctured sphere on the right hand side represents two hypermultiplets in the triplet   representation of $\SU(3)$.

Note that $\SU(3)$ gauge group before the complete decoupling can be thought of as gauging the $\SU(3)$ subgroup of the $\SU(4)$ flavor symmetry of the full puncture, as shown in the second row of the figure. 
This splits four fundamental flavors coupled to $\SU(4)$ into a set of three  flavors and an additional one flavor. 
The $\SU(3)$ gauge group makes the first three into the bifundamental hypermultiplet of $\SU(4)\times \SU(3)$,
and one flavor remains to couple just to $\SU(4)$ on the upper row. 

\begin{figure}[h]
\[
\inc{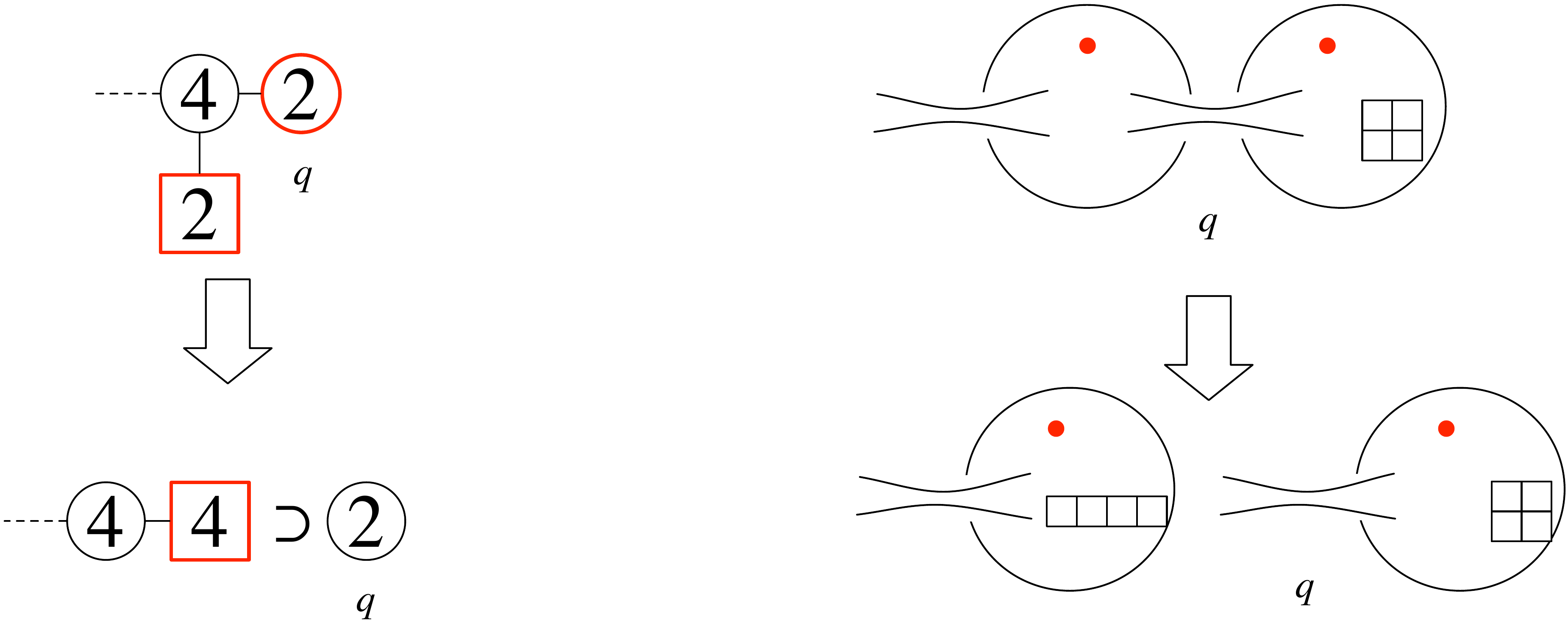}
\]
\caption{Another example of decoupling.\label{fig:decoupling3}}
\end{figure}

Another example of decoupling process for the puncture of type $(2,2)$ is shown in Fig.~\ref{fig:decoupling3}. The decoupled three-punctured sphere on the right hand side represents an empty theory.

\subsection{S-dual of $\SU(N)$ with $N_f=2N$ flavors, part II}\label{sec:partII}
\subsubsection{For general $N$}
\begin{figure}[h]
\[
\inc{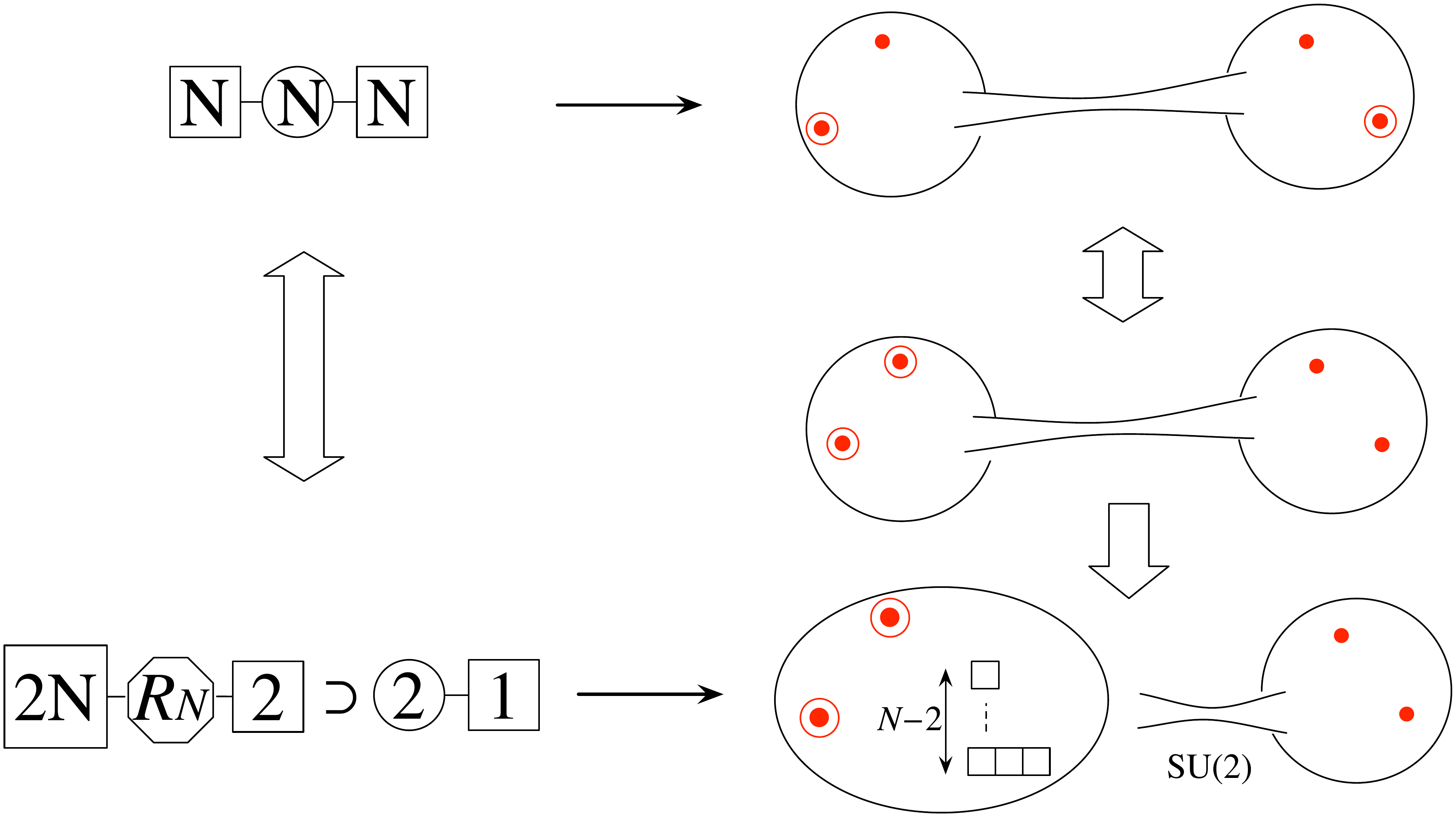}
\]
\caption{S-dual of $\SU(N)$ with $2N$ flavors, explained.\label{fig:SUN-Sdual}}
\end{figure}
Now we have learned enough techniques to understand the S-dual of $\SU(N)$ theory with $2N$ flavors, see the first row of Fig.~\ref{fig:SUN-Sdual}.
Originally, we have a sphere with four punctures: two at $z=0$, $\infty$ are full punctures,
and two at $z=q$, $1$ are simple punctures.
We would like to understand the limit $q\to 1$.  We end up decoupling two simple punctures from the other two. 
We already learned what happens in this decoupling process.

The simple puncture is a puncture of type $(N-1,1)$. 
Decoupling two of them, we generate a puncture of type $(N-2,1,1)$.
This puncture has a flavor symmetry $\SU(2)\times \U(1)$ when $N>3$, and $\SU(3)$ when $N=3$. 
The behavior of the duality when $N=3$ is somewhat more peculiar than the other cases. 
In any case, there is an $\SU(2)$ symmetry exchanging the last two columns of height 2,
and a weakly-coupled dynamical $\SU(2)$ group gauges this $\SU(2)$ symmetry.
There is in addition  one flavor in the doublet representation for this $\SU(2)$ gauge group coming from the almost decoupled sphere on the right, see the last row of Fig.~\ref{fig:SUN-Sdual}.

\begin{figure}[h]
\[
\inc{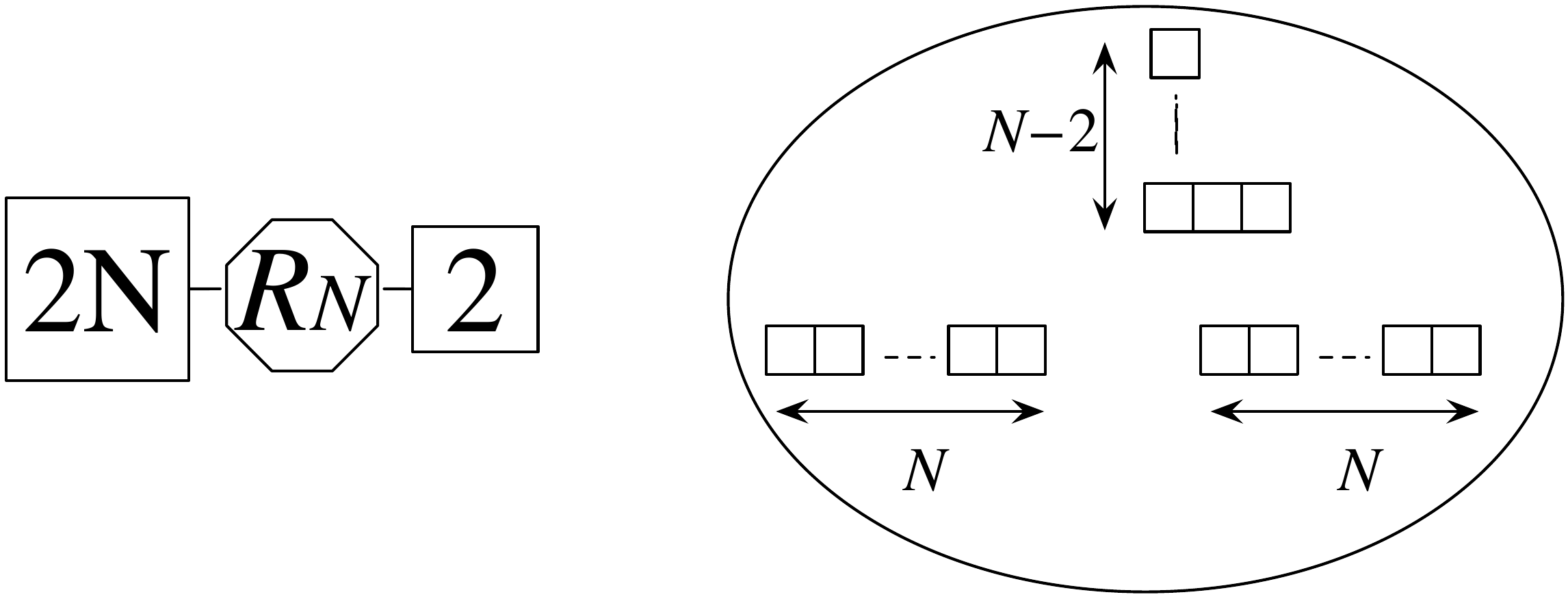}
\]
\caption{Strange theory of Chacaltana-Distler, $R_N$.\label{fig:RN}}
\end{figure}

The question is the nature of the sphere on the left hand side. It has three punctures: two are full punctures, and one is of type $(N-2,1,1)$.
Assuming all the mass parameters are zero, we can determine the behavior of fields $\phi_k(z)$ easily,
as the pole structure at $z=\infty$ is $(p_2,p_3,\ldots,p_N)=(1,2,\ldots,2)$. We see that \begin{equation}
\phi_2(z)=0,\qquad
\phi_k(z)=\frac{u_k}{(z-1)^{k-1}z^{k-1}}dz^k.\label{phi-RN}
\end{equation} This theory has one dimension-$k$ operator for each $k=3,4,\ldots, N$. 
The flavor symmetry is at least $\SU(N)\times \SU(N)$ associated to the full punctures,
and $\SU(2)\times \U(1)$ associated to the puncture of type $(N-2,1,1)$. 
Call this funny conformal field theory $R_N$, for which we introduce a graphical notation  as in Fig.~\ref{fig:RN}.
In the original theory, the symmetry $\SU(N)\times\SU(N)\times \U(1)$ was part of the flavor symmetry $\SU(2N)$ rotating the whole $2N$ hypermultiplets in the fundamental representation.
We then need to demand that this theory $R_N$ has a larger flavor symmetry \begin{equation}
\SU(2N)\times \SU(2) \supset [\SU(N)\times \SU(N)\times\U(1) ] \times \SU(2).\label{enhancement}
\end{equation} 

\begin{figure}[h]
\[
\inc{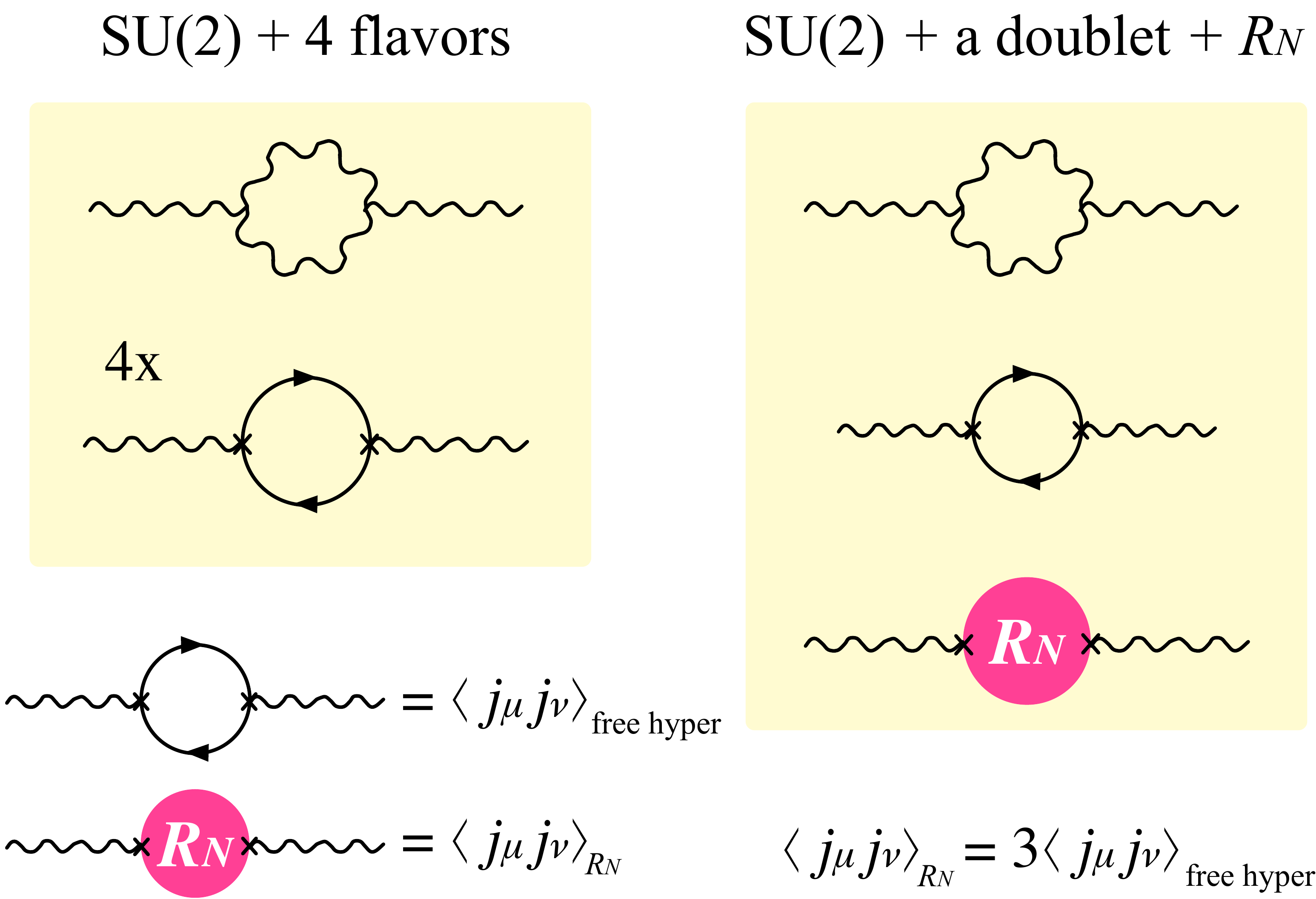}
\]
\caption{The $\SU(2)$ flavor symmetry current of the $R_N$ theory.\label{fig:dd} }
\end{figure}

We finally have the S-duality statement: \begin{equation}
\begin{array}{c}
\text{$\SU(N)$ theory with $2N$ flavors at the strong coupling $q\to 1$} \\
\rotatebox{90}{$\Leftrightarrow$}\\
\text{weakly-coupled $\SU(2)$ gauge multiplet coupled to one doublet and to the $R_N$ theory.}
\end{array}
\end{equation}
This general statement was found by Chacaltana and Distler in \cite{Chacaltana:2010ks}.
We know that the dual $\SU(2)$ gauge coupling has zero beta function. Applying the analysis as in Sec.~\ref{sec:generalAD}, we find that  the $\SU(2)$ flavor symmetry of the $R_N$ theory contributes to the running of the $\SU(2)$ coupling as if it has effectively three hypermultiplets in the doublet. Equivalently, we have \begin{equation}
\vev{j_\mu j_\nu}_{R_N}=3 \vev{j_\mu j_\nu}_\text{free hyper in a doublet of $\SU(2)$}\label{k-of-su2}
\end{equation} where $j_\mu$ is the $\SU(2)$ flavor symmetry current.
See Fig.~\ref{fig:dd}.

\subsubsection{$N=3$: Argyres-Seiberg duality}\label{sec:AS}
\begin{figure}[h]
\[
\inc{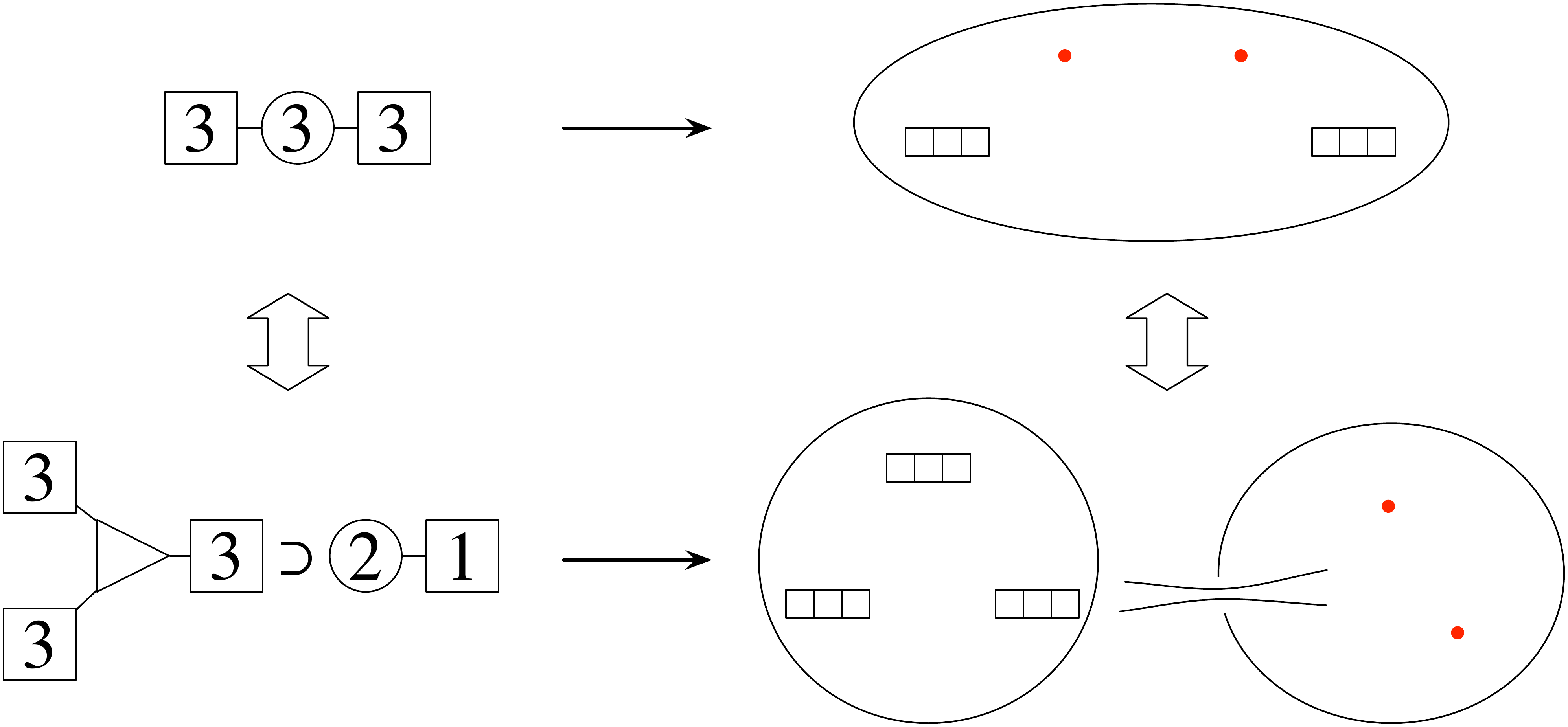}
\]
\caption{S-duality of $\SU(3)$ with 6 flavors involves  the theory $MN(E_6)$.\label{fig:Argyres-Seiberg-E6}}
\end{figure}
When $N=3$ we can say a little more about this duality.
This was originally found by Argyres and Seiberg in \cite{Argyres:2007cn}; the presentation here follows that given by Gaiotto in \cite{Gaiotto:2009we}.

 Now the puncture of type $(N-2,1,1)=(1,1,1)$  is a full puncture. Therefore the theory $R_3$ is given by a sphere with three full punctures, see Fig.~\ref{fig:Argyres-Seiberg-E6}.
The structure of $\phi_k(z)$ is already given in \eqref{phi-RN}. Therefore, this theory has just one Coulomb branch operator, of dimension 3.

\begin{figure}[h]
\[
\inc{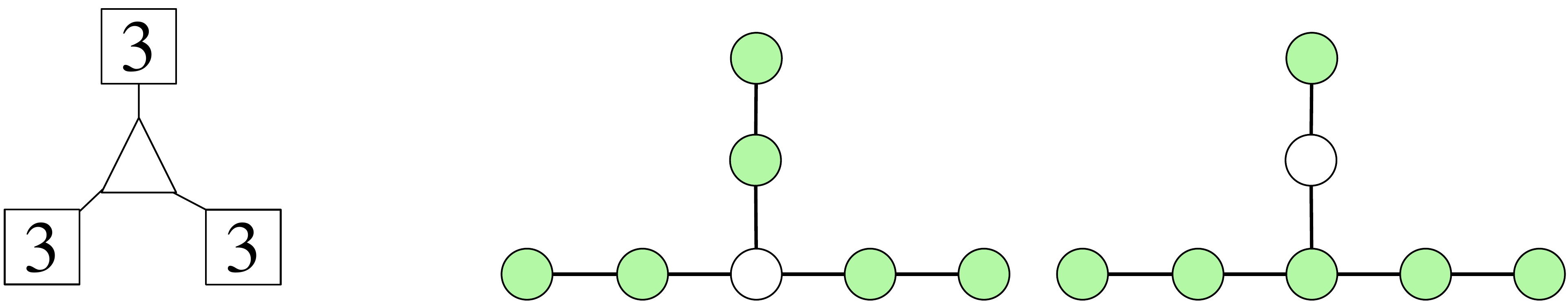}
\]
\caption{The theory $MN(E_6)=R_3=T_3$ \label{fig:E6}}
\end{figure}

We know that there is an enhancement of the flavor symmetry $\SU(3)\times \SU(3)$ associated to two full punctures to $\SU(6)$, as in \eqref{enhancement}.
We have three full punctures. Therefore, it should be that the flavor symmetry $F$ of this theory should be such that we have the following diagram \begin{equation}
\begin{array}{cccc}
F &\supset& \SU(6)&\times \SU(2) \\
\cup & & \cup \\
\SU(3)\times \SU(3)\times \SU(3) & \supset & \SU(3) \times \SU(3)\times \U(1) & \times \SU(2)
\end{array}
\end{equation} for any choice of two out of three $\SU(3)$s. Fortunately, there is unique such $F$, that is $E_6$, see Fig.~\ref{fig:E6}.
There, on the left, we introduce a diagrammatic notation for this theory.
On the center and on the right, we have the extended Dynkin diagram of $E_6$ with one node removed.\footnote{There is a general theorem for any $G$ stating that there is always a maximal subgroup whose Dynkin diagram is given by the extended Dynkin diagram of $G$ minus one node.}
We clearly see subgroups $\SU(3)^3$ and $\SU(6)\times \SU(2)$.
We already saw above that this theory has only one Coulomb branch operator, and its dimension is three.  This nicely fits the feature of a rank-1 superconformal theory announced to exist in Sec.~\ref{sec:rank-1SCFT}. This is equivalent to Minahan-Nemeschansky's theory $MN(E_6)$.

We conclude that we have the following duality:
\begin{equation}
\begin{array}{c}
\text{$\SU(3)$ theory with $6$ flavors at the strong coupling $q\to 1$} \\
\rotatebox{90}{$\Leftrightarrow$}\\
\text{weakly-coupled $\SU(2)$ gauge multiplet coupled to one doublet}\\
\text{and to the theory $MN(E_6)$  of Minahan-Nemeschansky.}
\end{array}
\end{equation}

We can give a few more checks to this duality. The first one concerns the current two-point functions. 
Firstly, we computed the current two-point function for the $\SU(2)$ flavor symmetry in \eqref{k-of-su2}. 
Then the whole $E_6$ flavor currents, which include the $\SU(2)$ ones, should have the same coefficient in front of the two-point function. Note that $\SU(6)$ flavor symmetry of the $\SU(3)$ gauge theory with six flavors is also a subgroup of this $E_6$ flavor symmetry. Therefore, we should have \begin{equation}
\vev{j_\mu^{\SU(6)} j_\nu^{\SU(6)}}_{\SU(3),\ N_f=6}=3 \vev{j_\mu j_\nu}_\text{free hyper in the fundamental of $\SU(6)$}.
\end{equation} This is indeed the case, since the left hand side can be computed in the extreme weakly-coupled regime, where they just come from three hypermultiplets in the fundamental representation of $\SU(6)$. 

The second check is about the Higgs branch. The $\SU(3)$ theory with six flavors has a Higgs branch of quaternionic dimension \begin{equation}
3\cdot 6-\dim \SU(3)=10.
\end{equation} Let us perform the computation in the dual side. The theory $MN(E_6)$ has a Higgs branch of quaternionic dimension 11, as we tabulated in Table~\ref{tab:rank1sing}. We have a doublet of $\SU(2)$ in addition, and we perform the hyperk\"ahler quotient with respect to $\SU(2)$ gauge group. Therefore the quaternionic dimension is  \begin{equation}
11+2-\dim \SU(2)=10,
\end{equation} which agrees with what we found above in the original gauge theory side.  Here we only compared the dimensions, but they can be shown to be equivalent as hyperk\"ahler manifolds, see \cite{Gaiotto:2008nz}.

\subsection{Applications}\label{sec:application}
\subsubsection{$T_N$} 
\begin{figure}[h]
\[
\inc{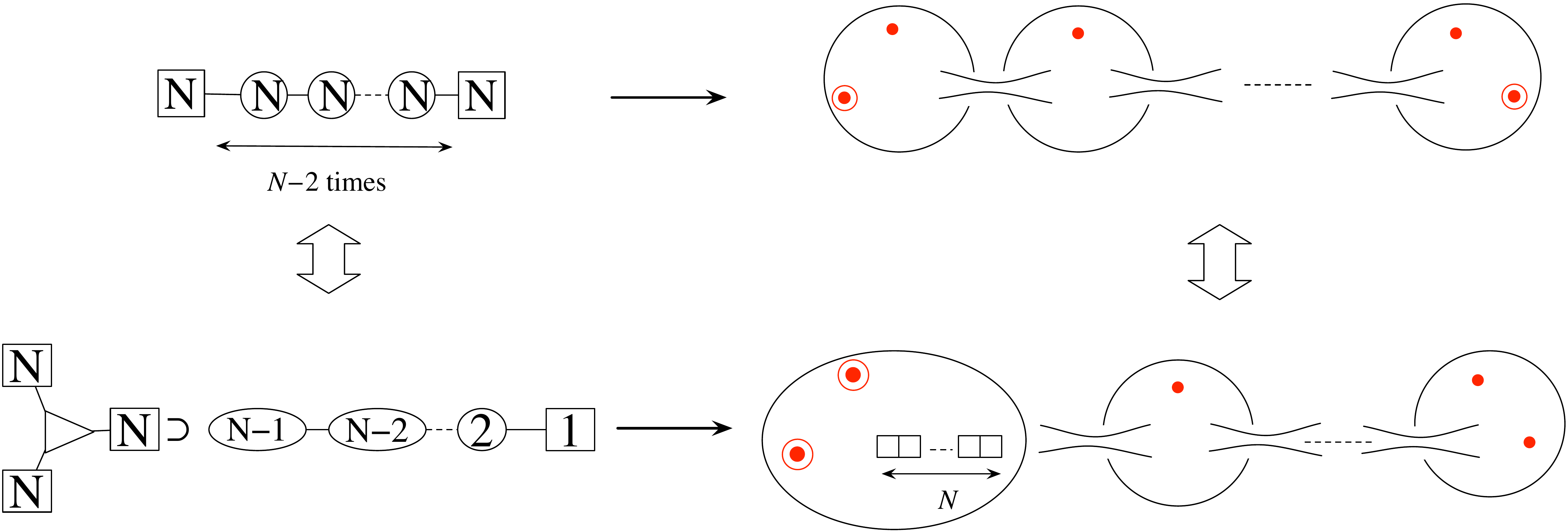}
\]
\caption{Duality producing $T_N$ theory\label{fig:constructingTN}}
\end{figure}
We can now have some fun manipulating punctures. 
For example, consider a gauge theory with gauge group $\SU(N)^{N-2}$, with bifundamental hypermultiplets between consecutive groups, together with $N$ additional fundamental hypermultiplets for the first and the last one, see the first row of Fig.~\ref{fig:constructingTN}.
The Seiberg-Witten solution is easily given: it is given by a sphere of type $\SU(N)$ theory,
with two full punctures and $N-1$ simple punctures. 
We go to a duality frame where we decouple all of these $N-1$ simple punctures. Applying the decoupling procedure we learned in Sec.~\ref{sec:tame}, we find that we generate a quiver tail with gauge group  \begin{equation}
\SU(N-1)\times \SU(N-2)\times \cdots \SU(2),
\end{equation} with bifundamental hypermultiplets between two consecutive groups and one doublet for the last $\SU(2)$.
The first $\SU(N-1)$ gauges an $\SU(N-1)$ subgroup of the flavor symmetry $\SU(N)$ of the puncture of type $(1,1,\ldots,1)$, i.e.~the full puncture. 

\begin{figure}[h]
\[
\inc{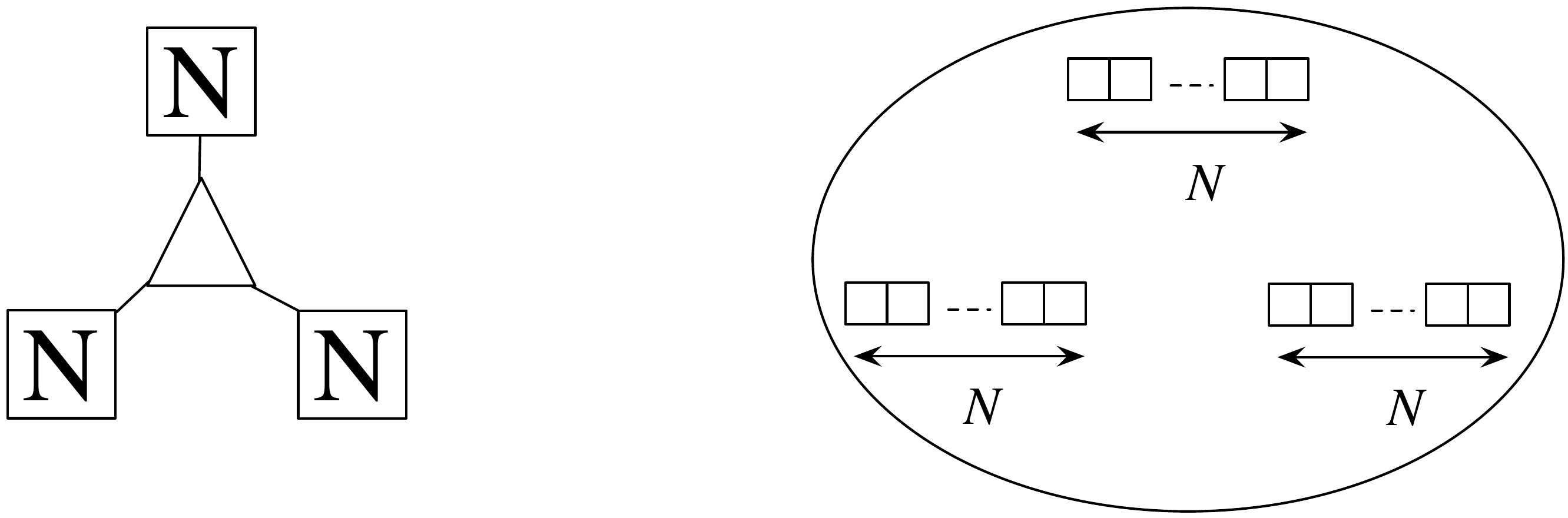}
\]
\caption{The  $T_N$ theory\label{fig:TN}}
\end{figure}
In this way, we can construct a theory described by a sphere with three full punctures. This is called the $T_N$ theory, see Fig.~\ref{fig:TN}. 
Note that $R_3=T_3$. 
As we have three full punctures, the flavor symmetry is at least $\SU(N)^3$. When $N=3$, we saw above that this flavor symmetry enhances to $E_6$.
When $N\ge 4$, there are more than one gauge group in the original gauge theory. Therefore, we do not have an enhancement from $\SU(N)\times \SU(N)$ to any other group. This matches with the fact that there is no group containing $\SU(N)^3$ such that $\SU(N)^2$ enhances to $\SU(2N)$ when $N\ge 4$. Putting the punctures at $z=0,1,\infty$, we see that $\phi_k$ has the form \begin{equation}
\phi_k=\frac{ u_k^{(1)} + \cdots + u_k^{(k-2)}z^{k-3} }{z^{k-1}(z-1)^{k-1}}dz^k.
\end{equation} Therefore this theory has one Coulomb branch operator of dimension $3$,
two Coulomb branch operators of dimension $4$, \ldots, and $N-2$ Coulomb branch operators of dimension $N$.

\begin{figure}[h]
\[
\begin{array}{c}
\inc{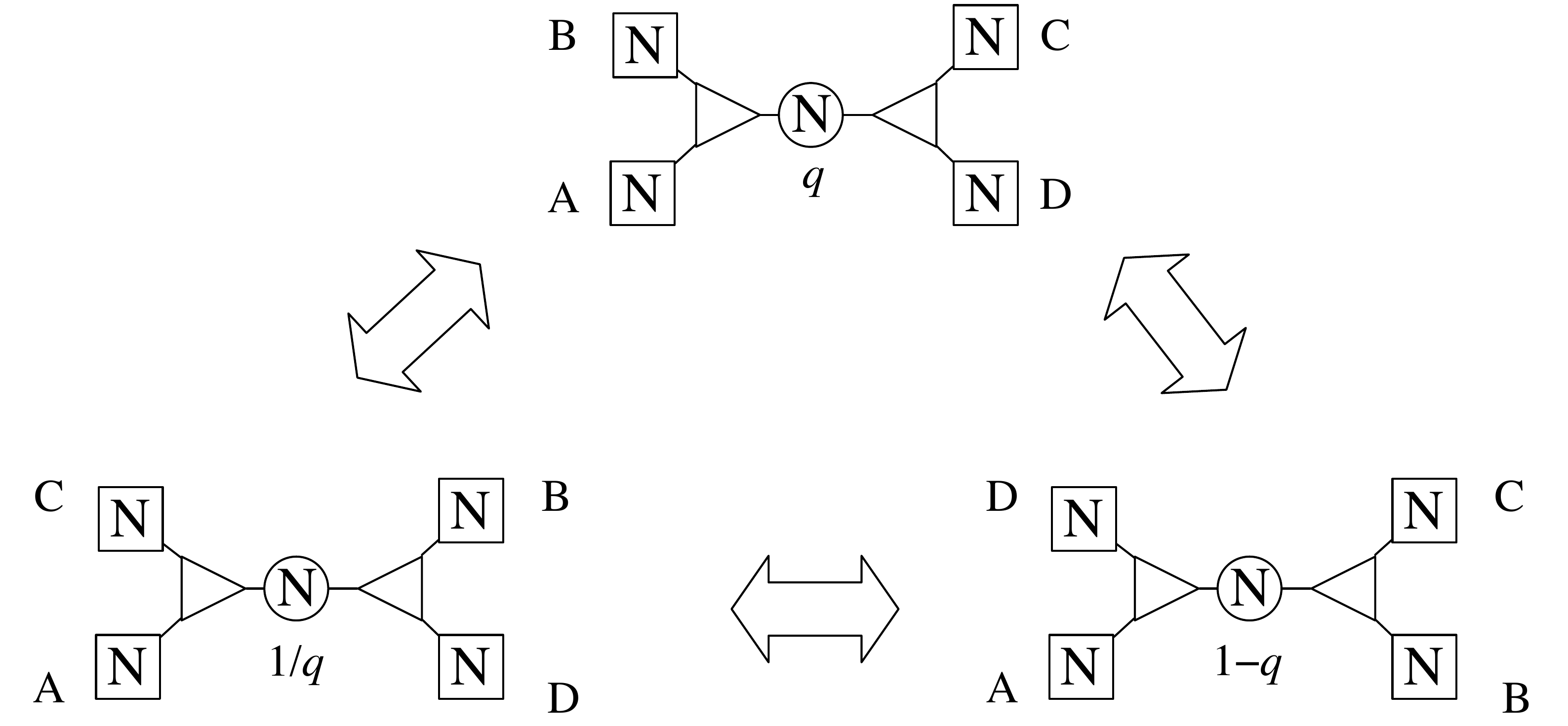}\\
\inc{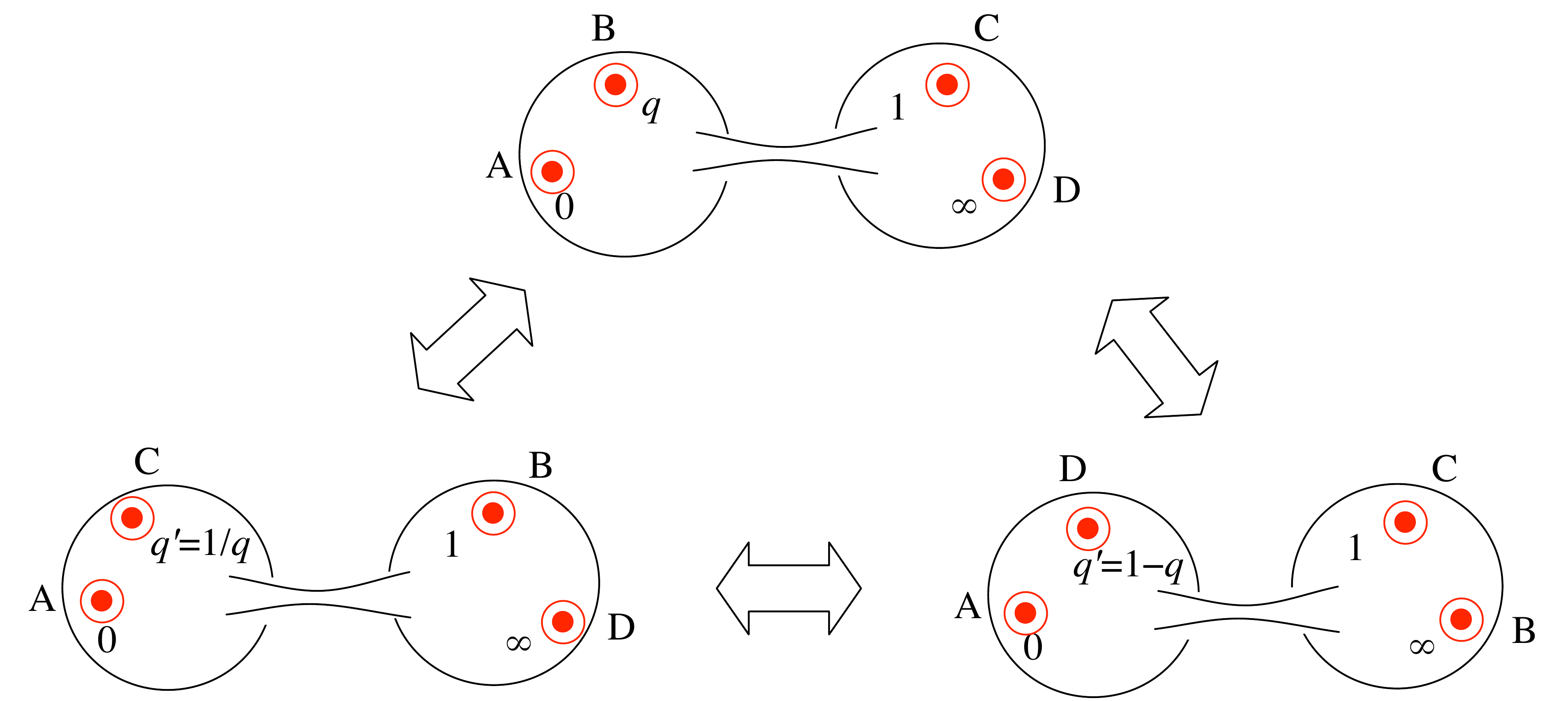}
\end{array}
\]
\caption{S-duality of coupled copies of  $T_N$ theory \label{fig:TNTN}}
\end{figure}
Now we can take two copies of this $T_N$ theory and couple them by an $\SU(N)$ gauge multiplet. In the 6d construction, we just have four full punctures on the sphere. Therefore, we have the S-duality structure exactly as in $\SU(2)$ theory with four flavors, exchanging all four punctures. 
In fact, $T_2$ theory is just the trifundamental hypermultiplet $Q_{ijk}$.

\subsubsection{$MN(E_7)$}

\begin{figure}[h]
\[
\inc{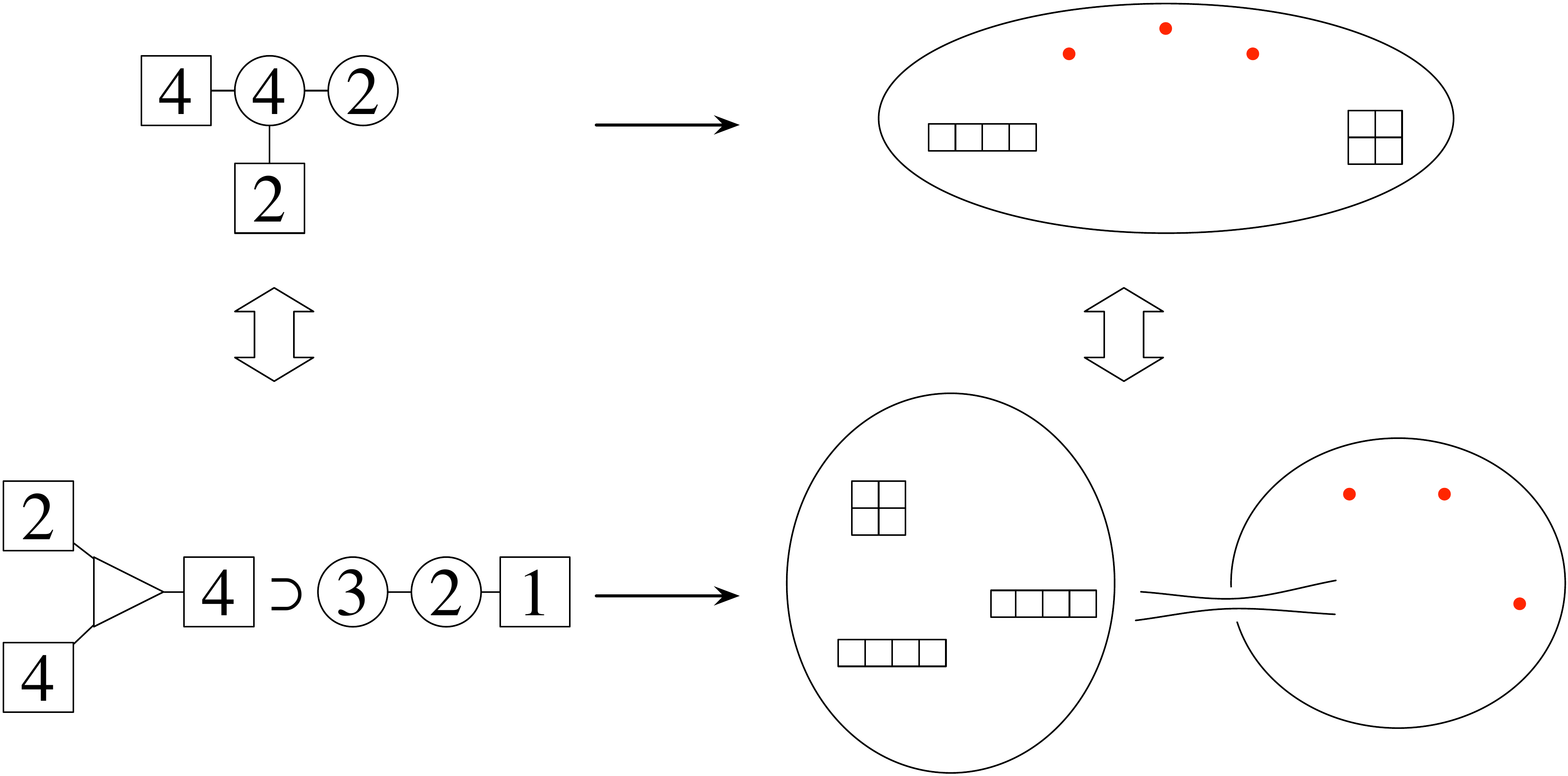}
\]
\caption{Duality producing the $MN(E_7)$  \label{fig:constructingE7}}
\end{figure}
Next, consider the duality shown in Fig.~\ref{fig:constructingE7}. We end up with a three-punctured sphere with two full puncture and one puncture of type $(2,2)$. 
In the original gauge theory, we have six fundamental flavors coupling to the $\SU(4)$ gauge multiplet with $\SU(6)$ flavor symmetry.
To construct the \Gaiotto\ curve, we split these six flavors into four flavors and two flavors, and applied the rule shown in the third row of Fig.~\ref{fig:SUNtame}.
Therefore, we see that the theory represented by the three-punctured sphere have a flavor symmetry $F$ of the form
\begin{equation}
\begin{array}{cccc}
F &\supset& \SU(6)&\times \SU(3) \\
\cup & & \cup \\
\SU(4)\times \SU(2)\times \SU(4) & \supset & \SU(4) \times \SU(2)\times \U(1) & \times \SU(3)
\end{array}.
\end{equation} 

\begin{figure}[h]
\[
\inc{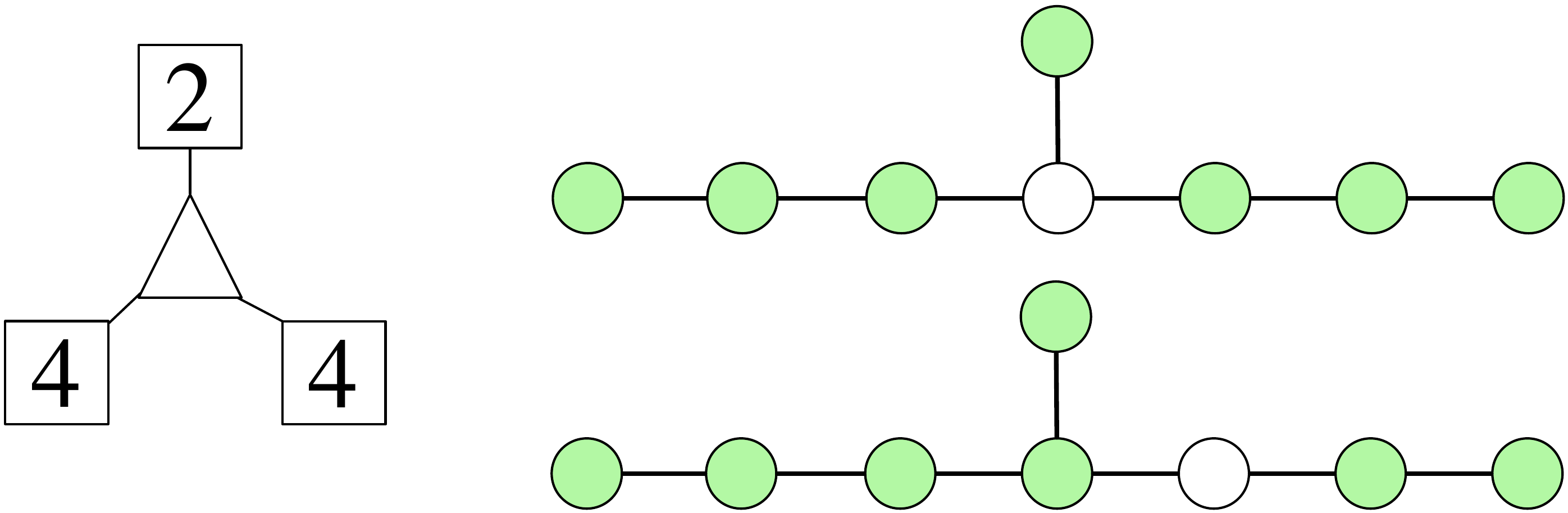}
\]
\caption{The theory $MN(E_7)$.\label{fig:E7}}
\end{figure}
Thankfully, there is a unique such group $F$, that is $E_7$, see Fig.~\ref{fig:E7}. We can of course compute the number of Coulomb branch operators this theory has, by studying $\phi_k(z)$. Here, let us try a different procedure. 
Originally, we had the gauge group $\SU(4)\times \SU(2)$. Therefore, the numbers of the Coulomb branch operators of dimension 2,3,4 were respectively $2,1,1$. 
On the dual side, the quiver tail contains $\SU(3)\times \SU(2)$, which has two operators of dimension 2 and one operator of dimension 1. 
The theory represented by the three-punctured sphere should account for the difference. Therefore there is just one Coulomb branch operator, of dimension 4. 
This again fits  the feature of a rank-1 superconformal theory announced to exist in Sec.~\ref{sec:rank-1SCFT}. This is equivalent to Minahan-Nemeschansky's theory $MN(E_7)$. We can also check the agreement of the current two-point functions and the dimensions of the Higgs branch, as we did at the end of Sec.~\ref{sec:AS}.

\subsubsection{$MN(E_8)$} 

\begin{figure}[h]
\[
\inc{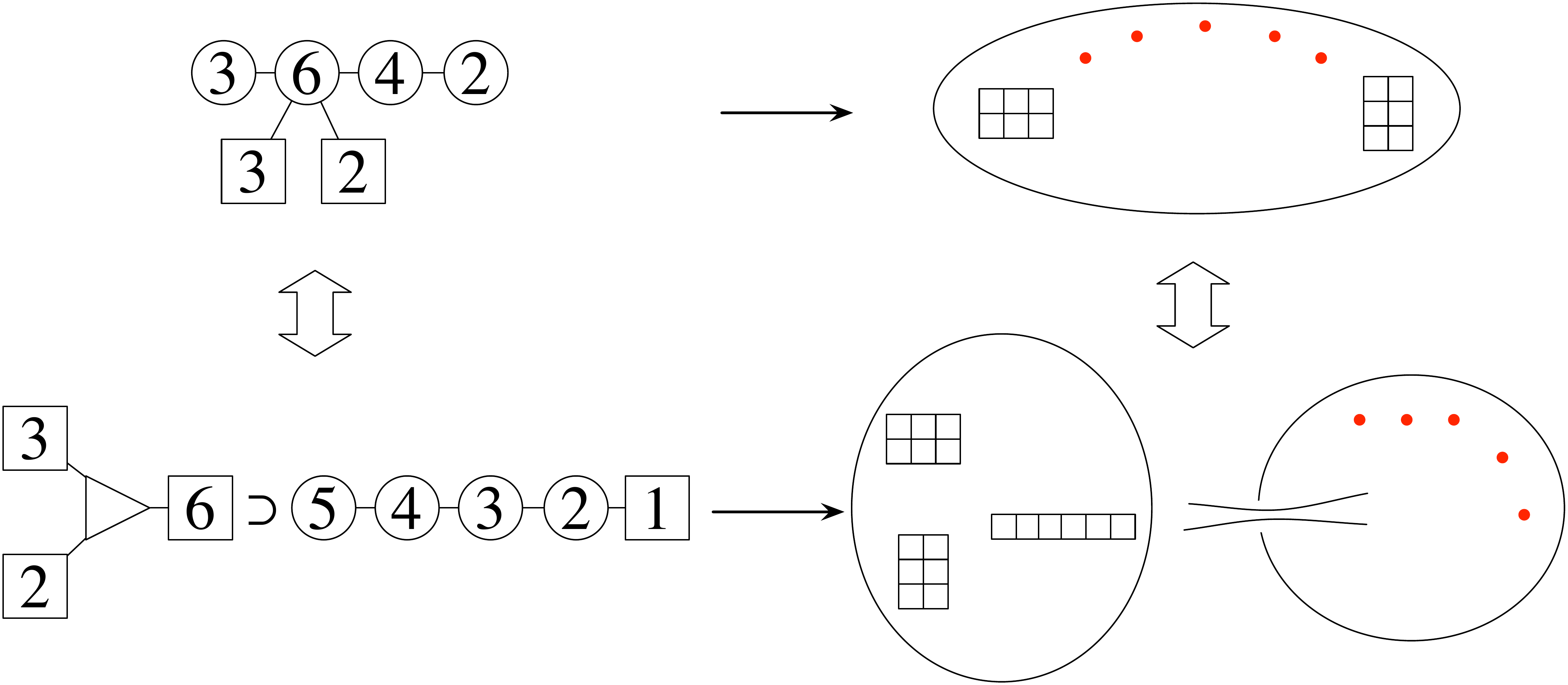}
\]
\caption{Duality producing the $MN(E_8)$ theory\label{fig:constructingE8}}
\end{figure}
\begin{figure}[h]
\[
\inc{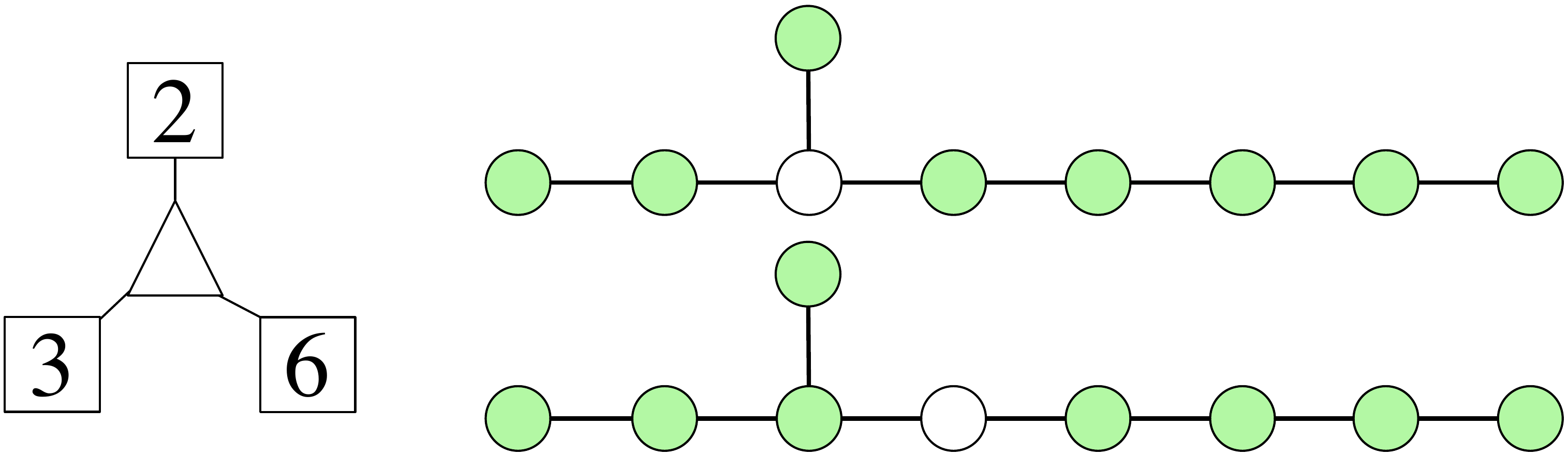}
\]
\caption{$E_8$ theory\label{fig:E8}}
\end{figure}

Generalizing this to the $E_8$ symmetry is by now rather straightforward. We perform the duality as shown in Fig.~\ref{fig:constructingE8}. 
In the dual side, we have a three-punctured sphere with one full puncture, another of type $(2,2,2)$, and of type $(3,3)$. 
We see that the flavor symmetry $F$ of the theory should satisfy \begin{equation}
\begin{array}{cccc}
F &\supset& \SU(5)&\times \SU(5) \\
\cup & & \cup \\
\SU(2)\times \SU(3)\times \SU(6) & \supset & \SU(2) \times \SU(3)\times \U(1) & \times \SU(5)
\end{array}.
\end{equation}
This nicely fits the structure of Minahan-Nemeschansky's  theory $MN(E_8)$, see Fig.~\ref{fig:E8}. Checks of various properties are left as an exercise to the reader. 

\subsubsection{The singular limit of $\SU(N)$ with even number of flavors} \label{sec:applicationL}

Finally, let us study a non-conformal example. Consider $\SU(N)$ theory with $N_f=2n$ flavors, with $N>n$. 
%For simplicity we take all masses to be the same. Then 
The curve is \begin{equation}
\frac{\Lambda^{N-n} \prod_{i=1}^n (x+\mu+\mu_i)}z
+\Lambda^{N-n} \prod_{i=1}^n (x+\mu+\tilde\mu_i)^n z = x^N+u_2 x^{N-2}+\cdots+ u_N\label{Nn}
\end{equation} with the differential $\lambda=xdz/z$.  Here we demanded $\sum_i\mu_i+\tilde \mu_i=0$
and split the $\U(1)$ mass term as $\mu$.
Clearly something happens when $u_{N-n}=2\Lambda^{N-n}$ around $z\sim 1$. 
This point was first considered in \cite{Eguchi:1996vu}. The correct physics was first discussed in \cite{Gaiotto:2010jf}.
We will see below  that the low-energy limit is an infrared-free $\SU(2)$ gauge theory coupled to the theories $R_n$ and $X_{N-n+4}$.

To study the infrared behavior, we let \begin{equation}
u_{N-n,\text{old}}=2\Lambda^{N-n}+u_{N-n,\text{new}},\qquad z=1+\delta z
\end{equation} and
assume the scaling \begin{equation}
\mu_i\sim \epsilon,\qquad 
u_N \sim \epsilon^n,\  
u_{N-1} \sim \epsilon^{n-1},\  \label{JJJ}
\ldots,\ 
u_{N-n+2}\sim \epsilon^2,\ 
\end{equation} and \begin{equation}
u_2 \sim \epsilon'{}^2,\ 
u_3 \sim \epsilon'{}^3,\ \ldots,\ 
u_{N-n+2}\sim \epsilon'{}^{N-n+2}.
\end{equation} We then need to assume
\begin{equation}
\epsilon'{}^{N-n+2}\sim \epsilon^{2}. \label{ee}
\end{equation} In particular  we have \begin{equation}
\epsilon \ll \epsilon' \ll 1.
\end{equation}

In the region $x\sim \epsilon$, we can approximate the curve \eqref{Nn} as \begin{multline}
\frac{\Lambda^{N-n} \prod_{i=1}^n (x+\mu+\mu_i)}z + \Lambda^{N-n}\prod_{i=1}^n (x+\mu+\tilde\mu_i) z\\
=(2\Lambda^{N-n}+u_{N-n})x^n + u_{N-n+2} x^{n-2} + \cdots + u_N 
\end{multline} with the scaling \eqref{JJJ}. 
When this is written as a degree-$n$ equation for $x$, the  coefficient of the $x^n$ term is given by  \begin{equation}
\frac{\Lambda^{N-n}}z + \Lambda^{N-n}z -2\Lambda^{N-n}-u_{N-n} \label{***}
\end{equation}
In the limit $\epsilon\to 0$, two zeros of \eqref{***}   collide at $z=1$. This is exactly the situation we studied in Sec.~\ref{sec:partII} for $\SU(n)$ theory with $2n$ flavors
 in the $q\to 1$ limit. 
We see that we generate the $R_n$ theory coupled to $\SU(2)$ gauge group; the operator $u_{N-n+2}$ is now regarded as the Coulomb branch vev of this $\SU(2)$.  The  parameters $\mu_i$ and $\tilde \mu_i$ are now the mass parameters for the $\SU(2n)$ symmetry of the $R_n$ theory. 

In the region $x\sim \epsilon'$, the curve \eqref{Nn} can be approximated as \begin{equation}
c\,\delta z^2=(x^{N-n} + u_{2} x^{N-n-2} + \cdots + \frac{u_{N-n+1}}x+\frac{u_{N-n+2}}{x^2}),
\end{equation} where the differential $\lambda=xd \delta z$ and $c$ is an unimportnat constant.
We already encountered this in Sec.~\ref{sec:ADAD}; this is the curve describing the Argyres-Douglas point of $\SU(N-n+1)$ theory with 2 flavors. Equivalently, we called this theory $X_{N-n+4}$ in Sec.~\ref{sec:generalAD}.

\begin{figure}[h]
\[
\inc{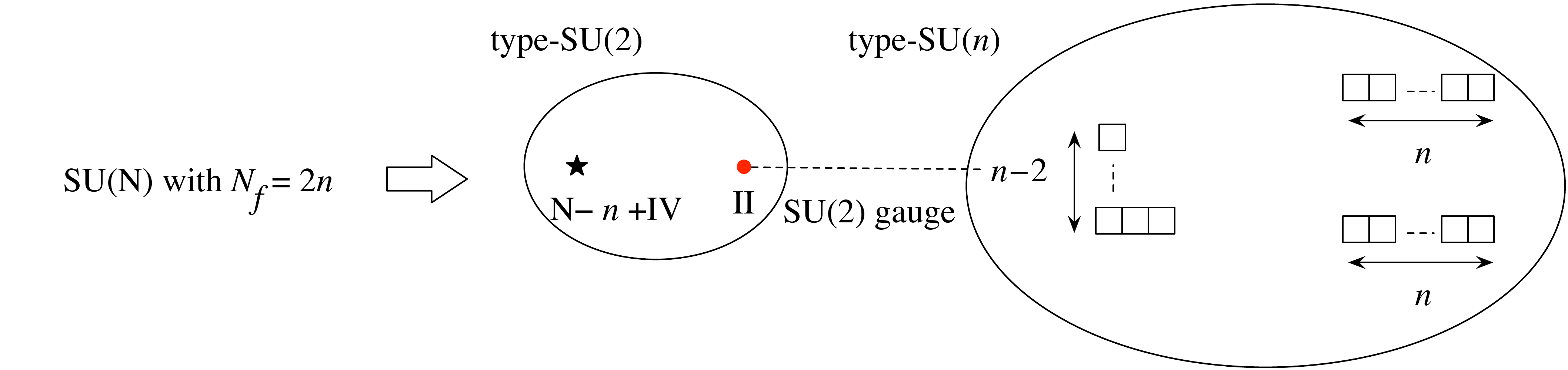}
\]
\caption{The most singular point of $\SU(N)$ with $N_f=2n$ flavors\label{fig:EHIY}}
\end{figure}
Summarizing, we see that the limiting theory has the structure given in Fig.~\ref{fig:EHIY}.  Namely, there is a weakly-coupled $\SU(2)$ gauge group, connecting the region $x\sim \epsilon$ given by a sphere of 6d theory of type $\SU(n)$, representing the $R_n$  theory,
to the region $x\sim \epsilon'$, given by a sphere of 6d theory of type $\SU(2)$, representing the theory $X_{N-n+4}$.

In the intermediate region $\epsilon'\gg x \gg \epsilon$, the curve is just \begin{equation}
\delta z^2 \sim \frac{u_{N-n+2}}{x^2}
\end{equation} with $\lambda=\delta z dx \sim \sqrt{u_{N-n+2}} dx/x$.
We see that there is an $\SU(2)$ gauge group, with \begin{equation}
a\sim \frac{1}{2\pi i}\oint \delta z\frac{dx}{x} \sim \sqrt{u_{N-n+2}}.
\end{equation} The dual coordinate $a_D$ is then given roughly by \begin{equation}
a_D\sim \frac{2}{2\pi i}\int^{x\sim \epsilon}_{x\sim \epsilon'} \sqrt{u_{N-n+2}}\frac{dx}x
\sim \frac{2}{2\pi i} a\log \frac{\epsilon}{\epsilon'}.
\end{equation} 
Using $a\sim \epsilon$ and the relation \eqref{ee}, we see \begin{equation}
a_D= \frac{2}{2\pi i}\frac{N-n}{N-n+2} a\log a +\cdots.
\end{equation}
Recall that the running is given by \begin{equation}
a_D \sim \frac{2}{2\pi i}(4-N_f)a\log a + \cdots
\end{equation} for $\SU(2)$ theory with $N_f$ flavors. This system then effectively has \begin{equation}
N_f=\frac{5N-5n+8}{N-n+2} >4.
\end{equation}
The $\SU(2)$ is now infrared free.
Note that this is correctly the sum of the effective number of flavors of the $R_N$ theory 
and the $X_{N-n+4}$ theory, as computed already. 
Indeed, it is $3$ for the $R_N$ theory,
and $2(N-n+1)/{(N-n+2)}$ for the $X_{N-n+4}$ theory, see  \eqref{k-of-su2} and  \eqref{GOO}, respectively. 

\subsection{Tame punctures and Higgsing}\label{sec:higgsing}

\begin{figure}[h]
\[
\inc{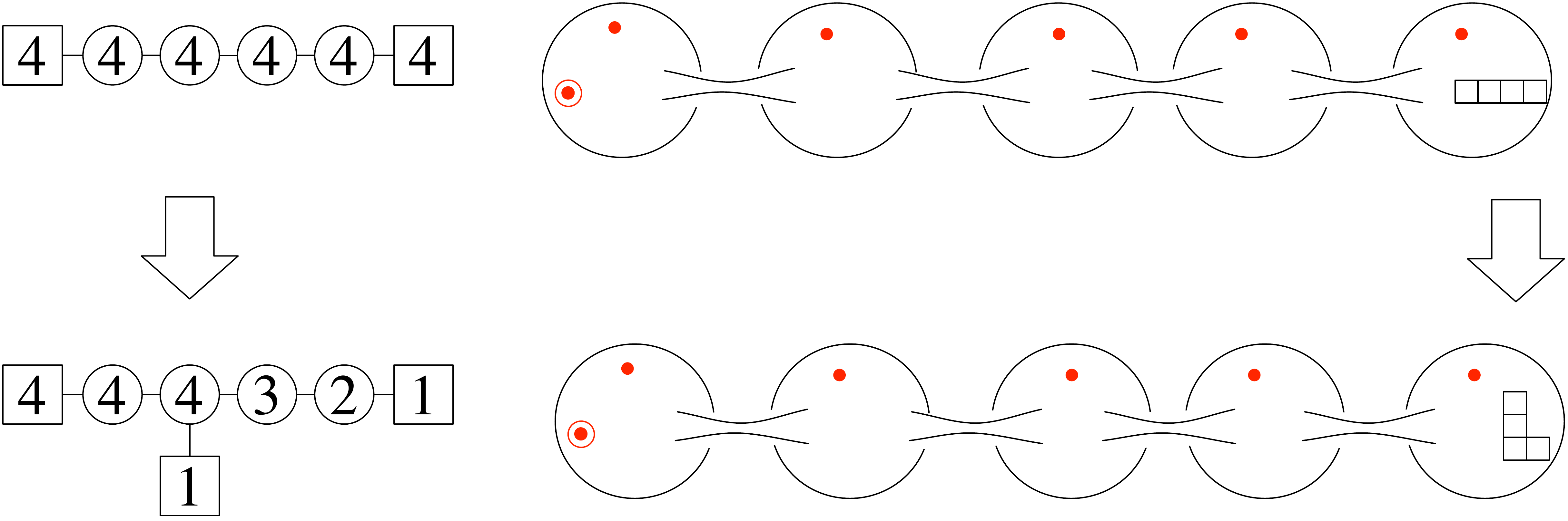}
\]
\caption{Change of the type of the puncture.\label{fig:partial-closing}}
\end{figure}

In Sec.~\ref{sec:tame}, we introduced punctures on the \Gaiotto\ curve labeled by Young diagrams in a rather ad hoc manner. 
Examples for $\SU(4)$ case were shown in Fig.~\ref{fig:SUNtame}. In this last subsection of the note,
we would like to study the meaning of the Young diagram in slightly more detail. For example, how should we understand the process of changing the full puncture to the simple puncture, i.e.~the puncture of type $(3,1)$, shown in Fig~\ref{fig:partial-closing}? 
We will use this particular example of changing the full puncture $(1,1,1,1)$ to the simple puncture $(3,1)$ as a concrete example throughout this section. The extension to the general punctures should be left as an exercise to the reader. 
The content of this section is based on an unpublished work with Francesco Benini, done sometime between 2009 and 2010.

The \SeibergWitten\ curves are both given by 
\begin{equation}
\lambda^4+\phi_2(z)\lambda^2+\phi_3(z)\lambda+\phi_4(z)=0.
\end{equation} In both cases, $\phi_k(z)$ has one full puncture at $z=0$ and  five simples punctures at $z=z_i$.
For the first, the puncture at $z=\infty$ was full
and for the second, it is a simple puncture, of type $(3,1,1)$.

For the first,  the fields $\phi_k(z)$ are given by \begin{equation}
\phi_k(z)=\frac{u_k^{(1)}+u_k^{(2)}z+u_k^{(3)}z^2+u_k^{(4)}z^3}{\prod_i^5(z-z_i)}\frac{dz^k}{z^{k-1}} .\label{former}
\end{equation} For the second, they are given by 
\begin{equation}
\begin{aligned}
\phi_2(z)&=\frac{u_2^{(1)}+u_2^{(2)}z+u_2^{(3)}z^2+u_2^{(4)}z^3}{\prod_i^5(z-z_i)}\frac{dz^2}{z},\\
\phi_3(z)&=\frac{u_3^{(1)}+u_3^{(2)}z+u_3^{(3)}z^2}{\prod_i^5(z-z_i)}\frac{dz^3}{z^2},\\
\phi_4(z)&=\frac{u_4^{(1)}+u_4^{(2)}z}{\prod_i^5(z-z_i)}\frac{dz^4}{z^3}.
\end{aligned} \label{latter}
\end{equation} 
Here, $u_k^{(i)}$ is the dimension-$k$ Coulomb branch operator of the $i$-th gauge group, and 
the way to determine them from the pole structure was described around \eqref{phifull}. 

It is clear that $\phi_k(z)$ in \eqref{latter} is obtained by setting $u_4^{(3,4)}=u_3^{(4)}=0$ in \eqref{former}. 
We will explain below that we can start from the first theory, set the Coulomb branch parameters to this subspace,
and then move to the Higgs branch, realizing the second theory. 

\begin{figure}[h]
\[
\inc{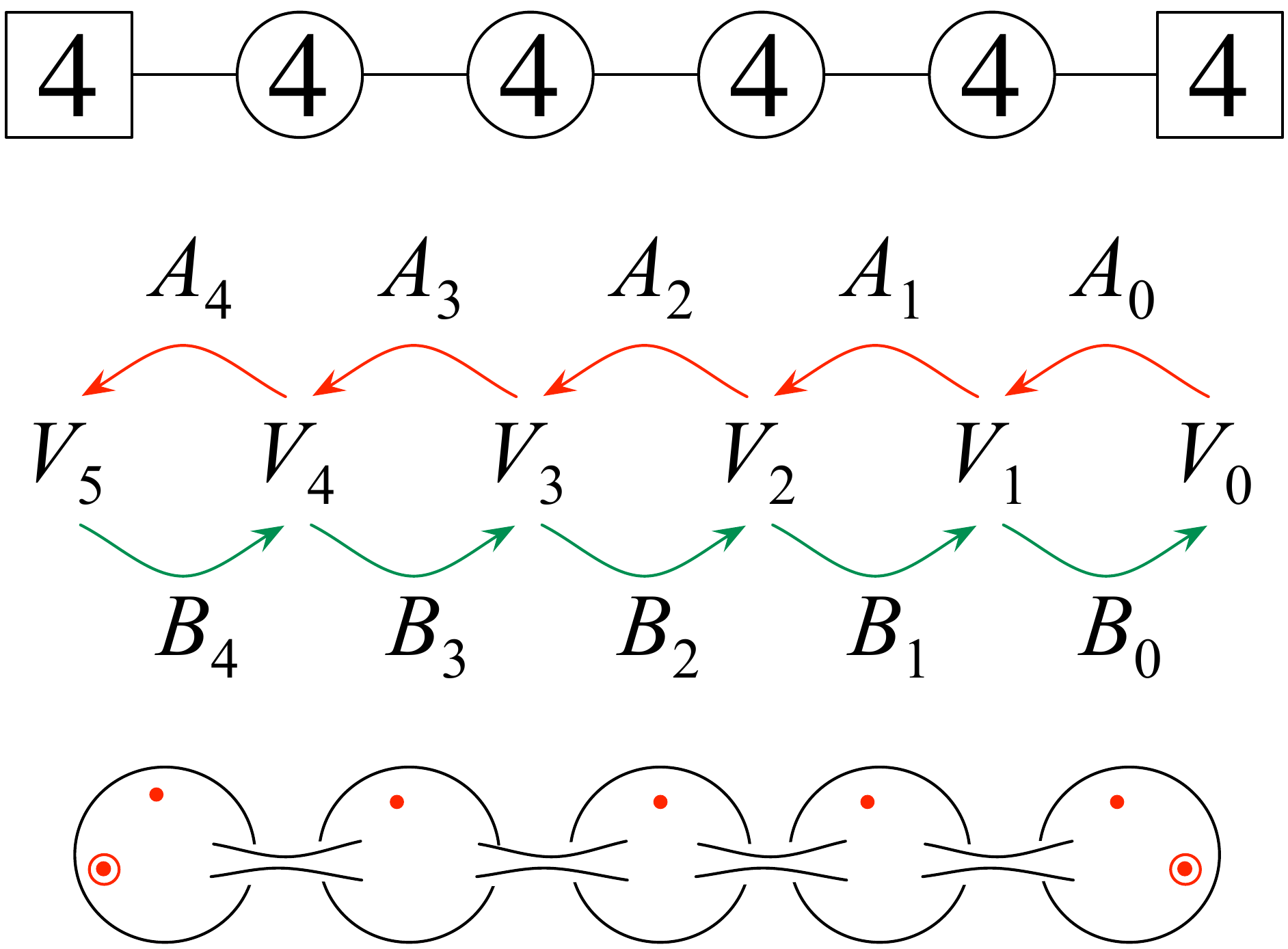}
\]
\caption{Assignment of new names to the fields.\label{fig:renaming-fields}}
\end{figure}
To facilitate the analysis of the Higgs branch, we introduce new names to the bifundamentals, see Fig~\ref{fig:renaming-fields}.
We name the rightmost $\SU(N)$ flavor symmetry $\SU(N)_0$, and the gauge groups $\SU(N)_{i=1,2,3,\ldots}$ from the right to the left. 
Introduce an auxiliary $N$-dimensional complex space $V_i$ for each of them.
For each consecutive pair  $\SU(N)_{i+1}\times \SU(N)_i$, we have a bifundamental hypermultiplet $(Q_b^a,\tilde Q^b_a)$
where $a=1,\ldots,N$ and $b=1,\ldots,N$ are the indices for $\SU(N)_{i+1}$, $\SU(N)_i$ respectively. 
We regard $Q_b^a$  as a linear map $A_i:V_i\to V_{i+1}$ and $\tilde Q^b_a$ as  a map in the reverse direction  $B_i:V_{i+1}\to V_i$.
Note that each pair $(A_i,B_i)$ comes from one of the several three-punctured spheres comprising the \Gaiotto\ curve, as shown in the figure.  Let us say that there are $k$ three-punctured spheres in total.

Let us introduce the notation \begin{equation}
M_i':=B_iA_i,\qquad \primeM _i:=A_i B_i.
\end{equation} 
We will use the trivial identity \begin{equation}
\tr M_i'{}^n = \tr B_iA_i \cdots B_i A_i = \tr A_i B_i \cdots A_i B_i = \tr {}\primeM _i{}^n
\end{equation} repeatedly below.

 Note that $\tr M_i:=\tr M_i'=\tr \primeM _i$ is the mass term for the $i$-th $\U(1)$ flavor symmetry, which can be naturally associated to the simple puncture of the $i$-th three-punctured sphere.
We also have two other gauge invariant combinations, namely \begin{equation}
M_0'|_\text{traceless}:= M_0'-\frac1N \tr M_0, \qquad
\primeM _k|_\text{traceless}:= \primeM _k-\frac1N \tr M_k.
\end{equation}
$M_0'|_\text{traceless}$ is an adjoint of $\SU(N)$ flavor symmetry associated to the full puncture of the rightmost sphere, at $z=\infty$.
Similarly,  $\primeM _k|_\text{traceless}$ is an adjoint of the $\SU(N)$ flavor symmetry at the puncture $z=0$.

%\begin{figure}[h]
%\[
%\ic{f-term-relation}
%\]
%\caption{Mesons and F-term relations.\label{fig:f-term-relation}}
%\end{figure}

%Fig~\ref{fig:f-term-relation}

Now, we would like to make a local modification at the puncture $z=\infty$,
by giving a non-zero vev to the adjoint field $M_0'|_\text{traceless}$.
Other gauge-invariant combinations $\tr M_i$ for $i=1,\ldots,k$ and $\primeM _k|_\text{traceless}$ are `localized' at other punctures. So we choose to keep them zero.

The F-term relation from the adjoint scalar in the gauge multiplet for  $\SU(N)_i$ is \begin{equation}
M_{i+1}'|_\text{traceless}={}\primeM _i|_\text{traceless}.
\end{equation} As we are imposing the condition $\tr M_i=0$, we can drop the tracelessness condition 
and just say \begin{equation}
M_{i+1}'={}\primeM _i.
\end{equation} 
Then we have the following relations: \begin{equation}
\tr M_0'{}^n = \tr\primeM _0{}^n = \tr M_1'{}^n = \tr \primeM _1{}^n= \cdots = \tr M_k'{}^n = \tr \primeM _k{}^n=0
\end{equation} for arbitrary $n$. 

This means that the gauge-invariant combination $M_0'$, transforming in the adjoint of the $\SU(N)$ flavor symmetry, is a nilpotent matrix. They can be put in the Jordan normal form by a complexified $\SU(N)$ rotation: \begin{equation}
M_0' = J_{t_1}\oplus J_{t_2} \oplus \cdots ,\qquad \sum_i t_i =N\label{nilpotent-vev}
\end{equation} where $J_{t}$ is the Jordan cell of size $t$,\begin{equation}
J_t= \underbrace{\begin{pmatrix}
0&1 \\
&0&1 \\
&& 0 & 1\\
&&&\ddots &\ddots \\
&&& &0 
\end{pmatrix}}_t .
\end{equation}

We again found a partition $(t_i)$ of $N$. We argue below that this partition $(t_i)$ is exactly the Young diagram labeling the punctures introduced in Sec.~\ref{sec:tame}.
To study the effect of the vev \eqref{nilpotent-vev}, we need to find a choice of hypermultiplet fields $(A_i,B_i)$ solving the F-term and the D-term relations. 

\begin{figure}[h]
\[
\inc{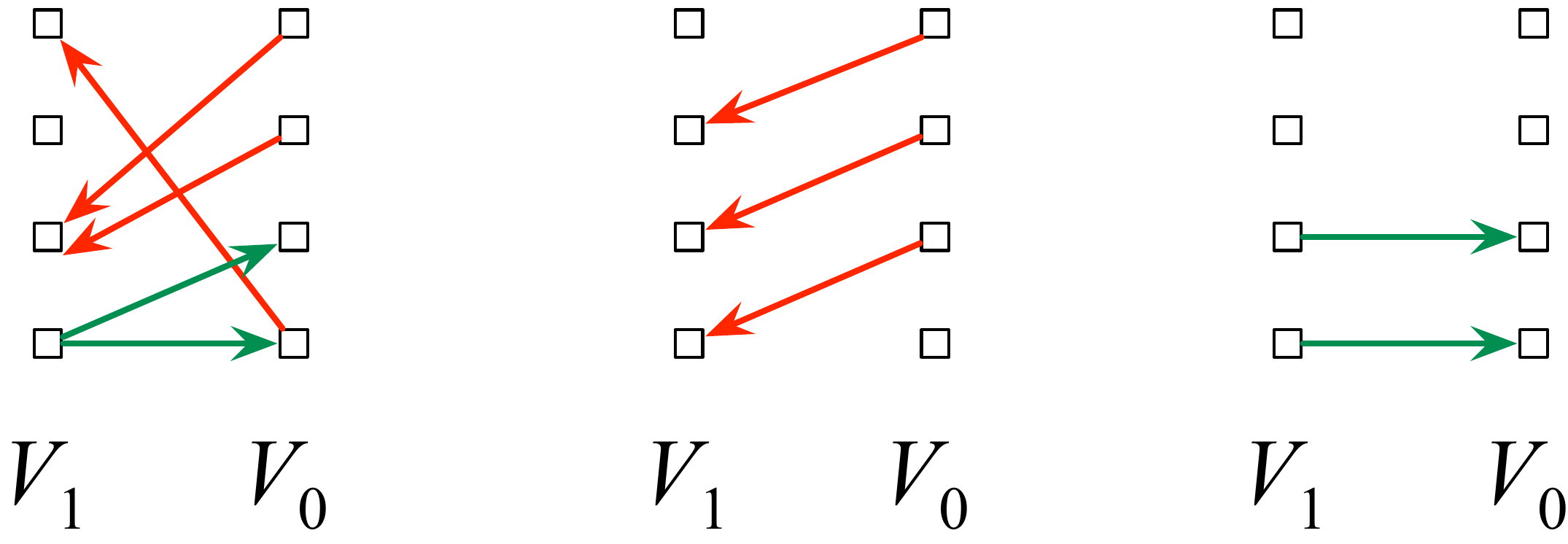}
\]
\caption{A graphical notation for matrices.\label{fig:more-notation}}
\end{figure}
To write down such a choice, it is useful to introduce a further diagrammatic notation, see Fig~\ref{fig:more-notation}. An $N$-dimensional vector space $V$ has $N$ basis vectors.  Let us denote them by a column of $N$ dots. A matrix whose entries are 0 or 1, from $V$ to $V'$
can be represented by a set of arrows connecting the $a$-th dot for $V$ to the $b$-th dot for $V'$ if and only if the $(a,b)$-th entry of the matrix is 1.  In the center of  Fig~\ref{fig:more-notation} we denoted a Jordan block $J_4$ of size $4$. The rightmost diagram of the same figure is for a projector to the last two basis vectors.

\begin{figure}[h]
\[
\inc{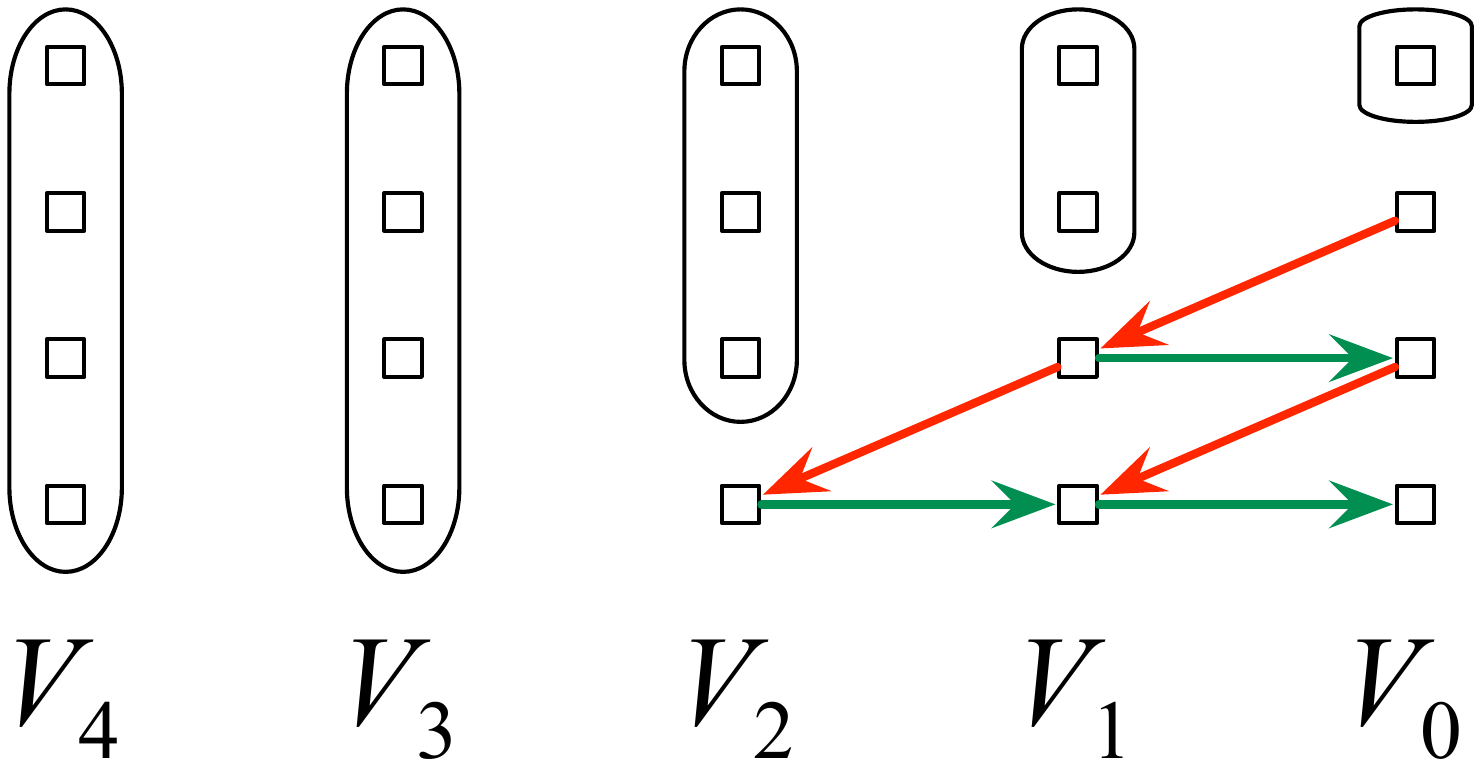}
\]
\caption{A particular point on the Higgs branch.\label{fig:higgsing}}
\end{figure}
For concreteness, let $N=4$, and give a nilpotent vev to $M_0'$ of  type $(3,1)$, namely it is given by $J_3\oplus J_1$. A solution to the F-term  relations are given in Fig.~\ref{fig:higgsing}.
There, we see that the unbroken gauge group is now $\SU(4)\times \SU(4)\times \SU(3)\times \SU(2)$.

In general, a solution to the F-term relations can be constructed as follows. Let us say we would like to set $M_0'=X$, where $X$ is in a Jordan normal form. We identify the vector spaces $V_0=V_1=V_2= \cdots$.
Let us introduce the notation $W_i = \Im X^i$
and denote the projector to $W_i$ by $P_{W_i}$.
We then  set \begin{equation}
A_0=X,\quad A_1=XP_{W_1},\quad A_2=XP_{W_2},\ldots
\end{equation} and take \begin{equation}
B_0=P_{W_1},\quad B_1=P_{W_2},\quad B_2=P_{W_3},\ldots.
\end{equation} 
Clearly, the remaining gauge group is of the form \begin{equation}
\cdots \times\SU(N_3)\times \SU(N_2)\times \SU(N_1) 
\end{equation} where \begin{equation}
N_i=N-\dim W_{i} = N-\rank X^{i}.
\end{equation} 
Define $s_i=N_i-N_{i-1}$. 
A short combinatorial computation shows that when $X$ has the type described by a Young diagram
whose $i$-th column from the left has height $t_i$,
the sequence $(s_1,s_2,\ldots)$ is such that $s_i$ is the width of the $i$-th row from the bottom.
This is exactly the rule we already introduced in Sec.~\ref{sec:tame} for the gauge group.
Now let us determine the massless matter content of the resulting theory. 

An indirect but fast way to determine the matter content is as follows. 
We started from a superconformal theory without any parameters.
After the Higgsing, the only parameter with mass dimensions is the vev of the hypermultiplet fields.
By the general decoupling of the hypermultiplet and the vector multiplet side of the Lagrangian, which we discussed in Sec.~\ref{sec:hypergeneral},
we see that there cannot be any mass terms or dynamical scales in the low-energy theory after the Higgsing.
Therefore, the resulting theory is also superconformal. We already determined $N_i$,
and we can only have bifundamental fields or fundamental fields. 
This shows that $\SU(N_i)$ should have exactly \begin{equation}
n_i=2N_i-N_{i+1}-N_{i-1}\label{additional-funds}
\end{equation} fundamental hypermultiplets in addition. 

\begin{figure}[h]
\[
\inc{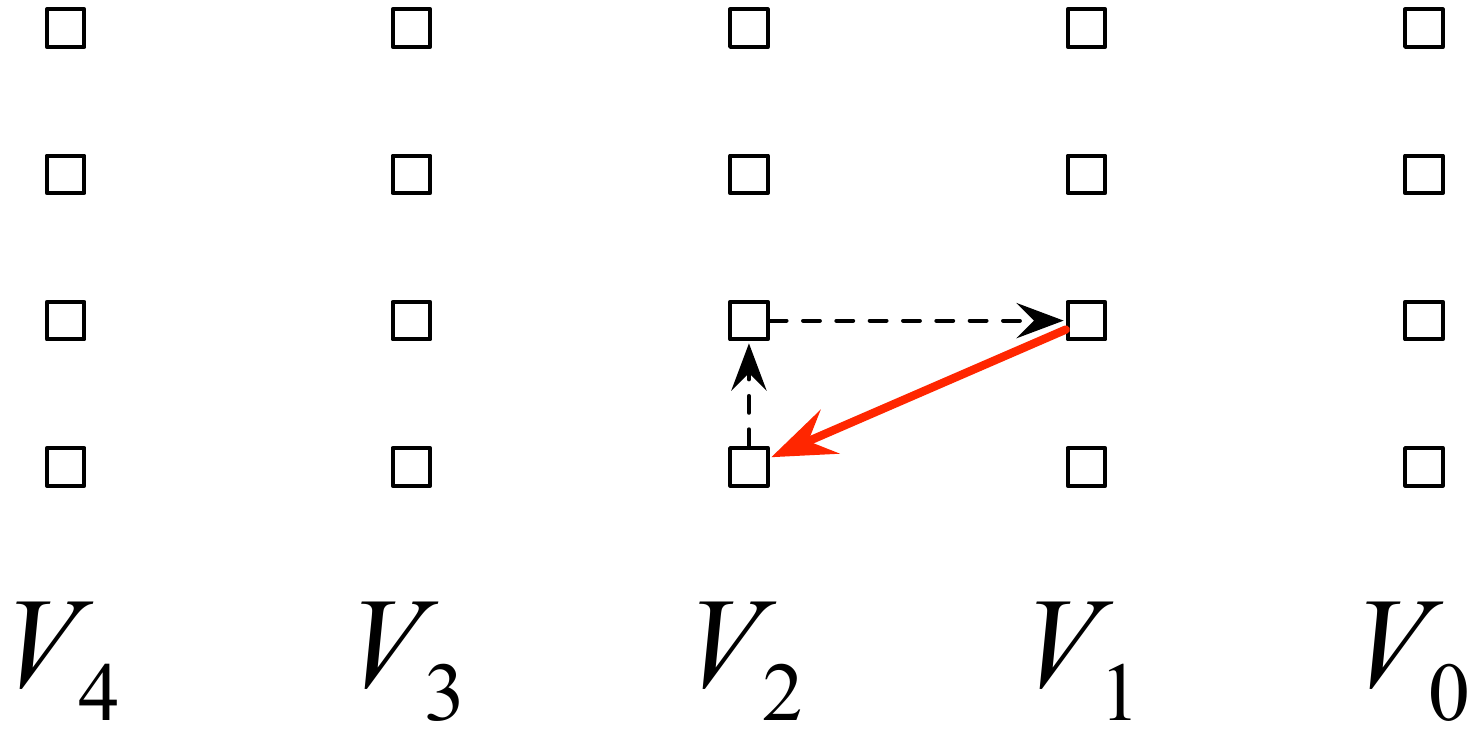}
\]
\caption{Mass terms generated for scalar fields.\label{fig:mass-term}}
\end{figure}
Of course this result can also be obtained by a direct computation of the mass terms of the various fields in the system.
Note that originally, there is an $\cN{=}1$ superpotential $\tr A_i\Phi_i B_i$ and $\tr B_i\Phi_{i+1}A_i$ where $\Phi_i$ is the adjoint scalar of the $\SU(N)_i$ vector multiplet. 
As we gave vevs to some components to $A_i$ and $B_i$, we see that certain components of hypermultiplets scalars and vector multiplet scalars pair up, due to the three-point couplings. One example is  shown in Fig~\ref{fig:mass-term}. There, the vev of $A_1$ represented by a  down-left arrow gives a mass term of a component of the vector multiplet scalar of the gauge group for $V_2$ and a component of $B_1$.

\begin{figure}[h]
\[
\inc{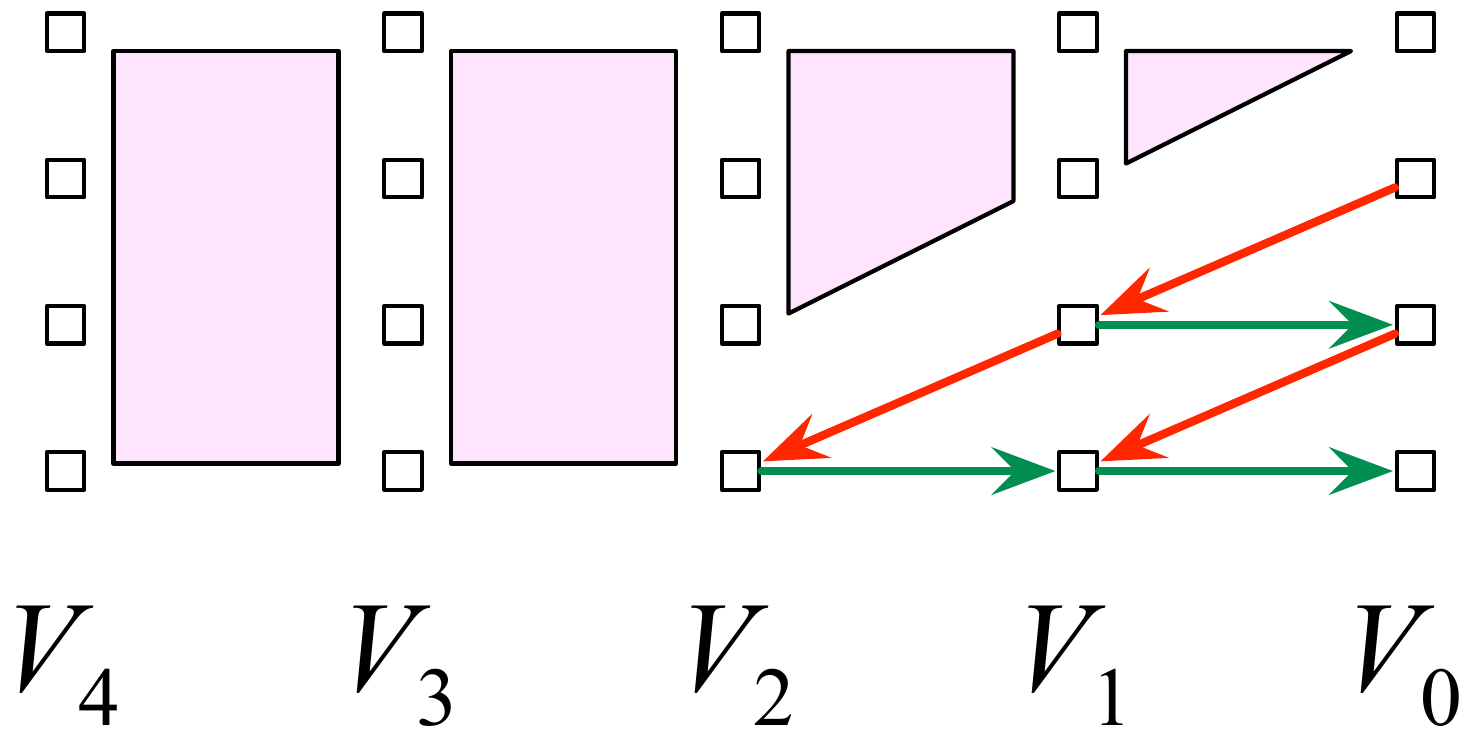}
\]
\caption{Remaining fields after the Higgsing.\label{fig:remaining-fields}}
\end{figure}
We see that always a bifundamental in $\SU(N_{i+1})\times \SU(N_i)$ remains massless.
But  from a careful analysis of the mass terms, we see that sometimes more charged hypermultiplets
remain massless. For example,  as shown in  Fig~\ref{fig:remaining-fields},
the whole bifundamental between $V_3$ and $V_2$ remains massless.
At  $V_2$, $\SU(4)$ is broken to $\SU(3)$. Therefore, from the point of view of the unbroken $\SU(4)$ at $V_3$, we see there are an $\SU(4)\times \SU(3)$ bifundamental together with a fundamental of $\SU(4)$. 
This can be generalized to see that the number of additional fundamental hypermultiplets of $\SU(N_i)$ is given by \eqref{additional-funds}.

\begin{figure}[h]
\[
\inc{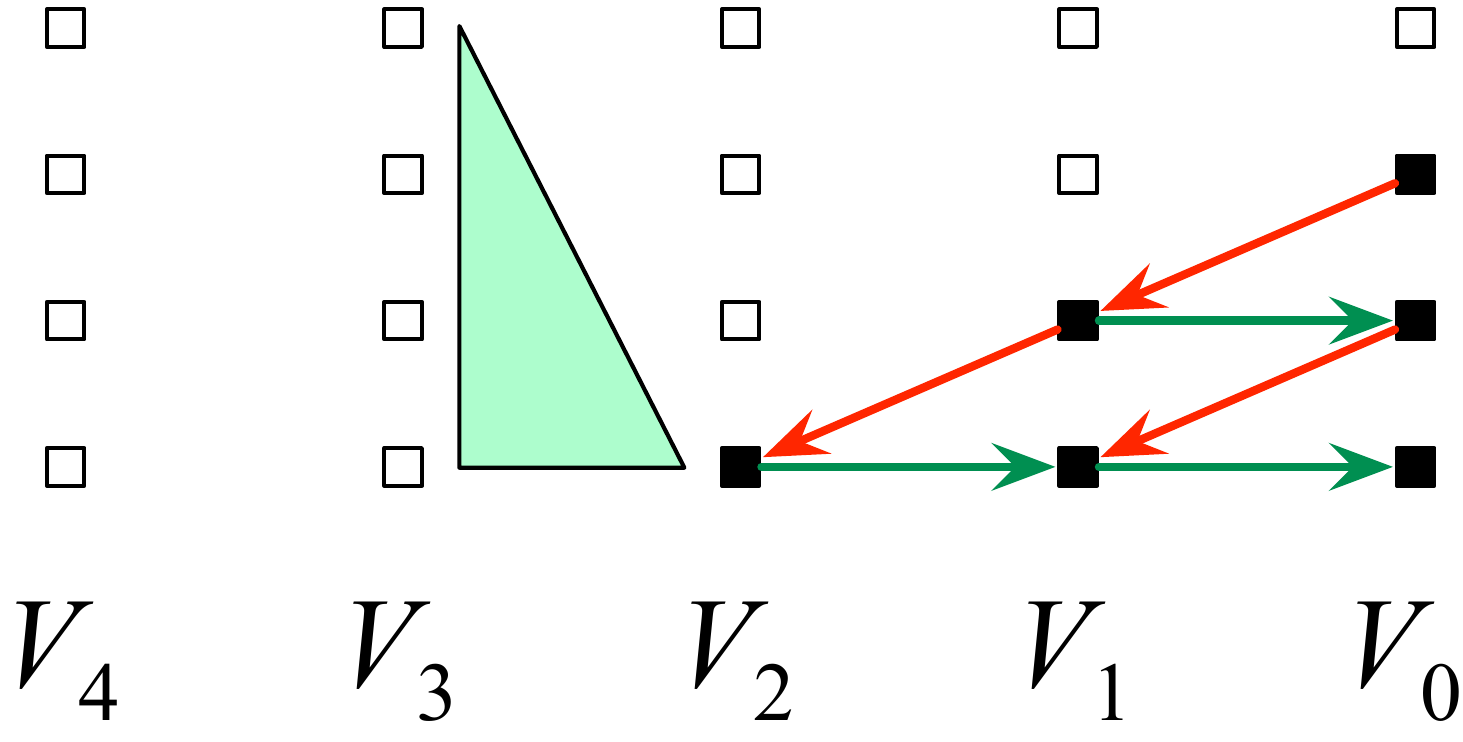}
\]
\caption{Flavor symmetry assignment.\label{fig:flavor-telekinesis}}
\end{figure}

In Sec.~\ref{sec:tame}, we said that the puncture at $z=\infty$ carries all the flavor symmetry associated to the additional $n_i$ fundamental hypermultiplets attached to $\SU(N_i)$. This sounded somewhat counter-intuitive, since the flavor symmetry $\SU(n_i)$ looks more associated to the $i$-th node. 
Now we understand the physical mechanism operating here. 
Let us take the puncture of type $(3,1)$ again for concreteness, see Fig~\ref{fig:flavor-telekinesis}.
The vev $X=M_0'$, which is from our rule is given by $X=J_3\oplus J_1$,
is invariant under the $\U(1)$ rotation acting on the three basis vectors,
as denoted by black dots in the figure. 
This symmetry, if unaccompanied by the gauge rotation, does not fix the Higgs vevs $\vev{A_i}$ and $\vev{B_i}$.  To make the symmetry compatible with the Higgs vev, we need to rotate at the same time all the other basis vectors connected from the original black dots by the arrows representing $A_i$ and $B_i$.

We see that the Higgs vevs identify the $\U(1)$ flavor symmetry rotating three basis vectors of $V_0$
and the $\U(1)$ flavor symmetry rotating the last basis vector of $V_3$. 
After the Higgsing, this latter $\U(1)$ symmetry is exactly the flavor symmetry carried by the additional one fundamental hypermultiplet of $\SU(4)$ at $V_3$, denoted by the filled triangle in the figure. 
This analysis can be generalized to arbitrary types of punctures.

Summarizing, we found a new interpretation of the punctures introduced in Sec.~\ref{sec:tame}.
Such a puncture can always be obtained from the full puncture, by first choosing the Coulomb branch vevs to the right subspace, and then giving a nilpotent vev to the hypermultiplet combination $M_0'$ which transforms in the adjoint of the flavor $\SU(N)$ associated to the full puncture. 
The vev given to $M_0'$ causes some of the other hypermultiplet fields $A_i$, $B_i$ for $i>0$ to have non-zero vevs, breaking the original gauge group $\cdots \times\SU(N)\times \SU(N)\times \SU(N)$
to $\cdots \times \SU(N_3)\times \SU(N_2)\times \SU(N_1)$. 

\section{Conclusions and further directions}\label{sec:conclusions}

In this lecture note, we first discussed the Lagrangian of $\cN{=}2$ supersymmetric gauge theory,
and then studied the Coulomb and Higgs branches of $\SU(2)$ gauge theories with various number of flavors. 
Two related concepts, the \SeibergWitten\ curve and the \Gaiotto\ curve played very important roles along the way. 
We then analyzed what happens when Coulomb branch vevs or exactly-marginal coupling parameters are finely tuned. Sometimes the limit was described by a dual weakly-coupled gauge theory, as was the case with $\SU(2)$ theory with four flavors. Most often, however, we saw that  we end up with new superconformal field theories, of Argyres-Douglas-type or of Gaiotto-type. 

For example, we saw the theories $AD_{N_f=1,2,3}(\SU(2))$ and $MN(E_{6,7,8})$ in Sec.~\ref{sec:rank-1SCFT}, the theories $X_N$ and $Y_N$ in Sec.~\ref{sec:generalAD}, $R_N$ in Sec.~\ref{sec:partII} and $T_N$ in Sec.~\ref{sec:application}. More and more $\cN{=}2$ superconformal theories are being discovered, see e.g.~\cite{Cecotti:2010fi,Cecotti:2012jx,Cecotti:2013lda}. 
 This means that, to fully understand the interrelations of $\cN{=}2$ supersymmetric systems, we cannot restrict our attention to just $\cN{=}2$ theories composed of vector multiplets and hypermultiplets. 

 The topics we covered in this lecture note are only a  tip of a huge iceberg that is the study of $\cN{=}2$ dynamics, and there are many other further directions of research. Let us list some of them.\footnote{The author did not try to be exhaustive and comprehensive here, and just cited a few recent ones. He is happy to add as many citations upon request.} 
  First, we can put an $\cN{=}2$ theory on a nontrival manifold: \begin{itemize}
\item Using the topological twisting, it can be put on an arbitrary manifold \cite{Witten:1988ze}. When the manifold is compact, the partition function is equivalent to what is known as the Donaldson invariant to mathematicians. Applying the Seiberg-Witten solution in the case of pure $\SU(2)$ theory, Witten introduced a new mathematical invariant, now called the Seiberg-Witten invariant \cite{Witten:1994cg}, which revolutionized four-dimensional differential geometry twenty years ago. 
\item We can put it on $S^1$.  Then the theory is effectively three-dimensional. 
As was first analyzed in \cite{Seiberg:1996nz}, the Coulomb branch as a three-dimensional theory is naturally a fibration over the Coulomb branch as a four-dimensional theory. The 3d Coulomb branch is hyperk\"ahler, and has the structure of a classical integrable system with finite degrees of freedom. This integrable system   was originally introduced in \cite{Donagi:1995cf}. For modern developments,  see e.g.~\cite{Gaiotto:2008cd}.
\item On the so-called $\Omega$ background. Very roughly speaking, it involves a forced rotation of the entire Euclidean system on $\bR^4$ around the origin. The spacetime is effectively compact and we can define the partition function, which is usually called Nekrasov's partition function. For a recent comprehensive discussion, see e.g.~\cite{Nekrasov:2012xe}. In a certain limiting case, it is found in \cite{Nekrasov:2009rc} that it gives rise to a quantized integrable system which is a quantized version of the Donagi-Witten integrable system.
\item On a round or deformed $S^4$.  The spacetime is compact and the partition function can be computed exactly, see e.g.~\cite{Pestun:2007rz,Hama:2012bg,Nosaka:2013cpa}. The partition function is also known to be related to 2d conformal field theories on the \Gaiotto\ curve, see  e.g.~\cite{Alday:2009aq,Wyllard:2009hg}.
\item On $S^1\times S^3$.  The partition function is called the superconformal index, and gives rise to 2d topological field theories on the \Gaiotto\ curve. 
It also has a deep relation to various important orthogonal polynomials, see e.g.~\cite{Gadde:2011uv,Gaiotto:2012xa}.
\item Other backgrounds can also be considered. See \cite{Luo:2013nxa} for $S^2\times S^1\times \bR$. A study of  $\cN{=}1$ theories on  $T^2\times S^2$ can be found in \cite{Closset:2013sxa}, and surely $\cN{=}2$ systems can be similarly considered there. 
\end{itemize}
Second, we can study dynamical excitations and externally-introduced operators of these theories:
\begin{itemize}
\item We have seen how we can read off the number of BPS-saturated particle types from the 6d construction.  The number is an integer and therefore it cannot usually change, but it does jump at certain loci in the Coulomb branch. This is called the wall-crossing and is an intensively-studied area, see e.g.~\cite{Gaiotto:2009hg}. The resulting spectrum can often be summarized using a diagram, called the BPS quiver.
This point of view was originally introduced in the context of $\cN{=}2$ supergravity in \cite{Denef:2002ru}. For more recent developments, see e.g.~\cite{Denef:2007vg,Cecotti:2011rv,Manschot:2012rx}.
% maybe I should split the sugra version and the susy version. Hmm.
\item Instead of dynamical particles, we can introduce worldlines of external objects. These are called line operators. See e.g.~\cite{Drukker:2009tz,Gaiotto:2010be}.
\item Once we allow the introduction of external line operators, there is no reason not to introduce higher-dimensional external objects. When they have two spacetime dimensions, they are called surface operators.  A \SeibergWitten\ curve can be defined intrinsically as the infra-red moduli space of a surface operator \cite{Gaiotto:2009fs}. Another interesting recent paper worth studying is \cite{Gaiotto:2013sma}.
\item We can then consider objects with three spacetime dimensions. This is an external domain-wall. A recent study can be found e.g.~in \cite{Dimofte:2013lba}.
\end{itemize}
On these topics, the review \cite{MooreReview} is a great source of information, although the review itself is meant for mathematicians.

Third, the method described in this lecture note is not yet powerful enough to solve arbitrary $\cN{=}2$ gauge theories. Many 4d $\cN{=}2$ theories do come from the 6d $\cN{=}(2,0)$ theory, but there are also many which presently do not. Therefore we should also study alternative approaches.\begin{itemize}
\item  The 6d construction itself needs to be developed further. For tame punctures, further discussions can be found in e.g.~\cite{Chacaltana:2010ks,Chacaltana:2012ch,Chacaltana:2013oka} and for wild punctures, more can be found in \cite{Xie:2012hs,Xie:2013jc}. 
\item  A 6d construction of 4d $\cN{=}2$ theory can always be uplifted to Type IIB string theory on a non-compact Calabi-Yau manifold, which is a fibration over the \Gaiotto\ curve. Even when the non-compact Calabi-Yau is not a fibration over a curve, Type IIB string theory on it often realizes a 4d $\cN{=}2$ field theory, and this gives an alternative to find the solution to the $\cN{=}2$ systems, see e.g.~\cite{Tachikawa:2011yr,Cecotti:2012sf}.
\item The $\cN{=}2^*$ theories, i.e.~$\cN{=}4$ super Yang-Mills deformed by a mass term for the adjoint hypermultiplet, have been long solved for general gauge group $G$ \cite{D'Hoker:1999ft}. Somewhat surprisingly, when $G\neq \SU(N)$, there is no known explicit string theory or M-theory construction of these solutions.  This clearly shows how primitive our current understanding is. 
\end{itemize}

Fourth, there are many properties of $\cN{=}2$ theories which are satisfied by all known examples, but we do not currently have any way to derive them. It would be fruitful to devise new methods to study these properties.  Let us list a few questions in this direction.  \begin{itemize}
\item The chiral operators on the Coulomb branch of the $\cN{=}2$ gauge theories are clearly always freely generated. For example, in an $\SU(N)$ gauge theory, it is generated by $\tr \phi^k$, ($k=2,\ldots,N$), which have no nontrivial relations. Experimentally, all the non-Lagrangian theories obtained from the 6d construction still satisfy this property: the Coulomb branch operators are freely generated.  The author conjectures this is in fact a theorem applicable to every $\cN{=}2$ supersymmetric systems. 
\item In \cite{Buican:2013ica}, it was argued that there is a non-zero lower bound in the change in the central charge $a$ along the RG flow between two $\cN{=}2$ superconformal field theories. Is there are more rigorous derivation of this fact? 
\item Is it possible to characterize the whole zoo of $\cN{=}2$ theories itself? As an analogy, consider all the representation of $\SU(2)$.  If we allow only the direct sum, we need all irreducible representations  to construct all possible representations. If we also allow the tensor product and the extraction of an irreducible summand, we only need the two-dimensional irreducible representation to generate all others. 

We can pose a similar question for $\cN{=}2$ theories. If we allow only weak gauging, what kind of generalized matter contents, i.e.~hypermultiplets and other `irreducible' strongly-coupled theories, are needed to generate all the $\cN{=}2$ theories? If we also allow the strongly-coupled limit, S-duality, and decomposition into the constituent parts, how much do we need?  What `percentage' of the theories can be obtained via 6d, string or M-theory constructions?

$\cN{=}2$ theories that are complete (in a certain technical sense) were classified in \cite{Cecotti:2011rv}, and $\cN{=}2$ weakly-coupled gauge theories were classified in \cite{Bhardwaj:2013qia}. These are however but two tiny steps into the vast space of all possible $\cN{=}2$ theories.  The theories we saw in this lecture note are summarized in the Appendix~\ref{sec:zoo}.
\end{itemize}

Finally, the author would like to emphasize that even such innocent looking gauge theories  as \begin{itemize}
\item $\cN{=}2$ supersymmetric $\SU(7)$ gauge theory with a hypermultiplet in the three-index anti-symmetric tensor representation, or 
\item $\cN{=}2$ supersymmetric $\SU(2)^3$ gauge theory with a massive full hypermultiplet in the trifundamental, $(Q_{aiu},\tilde Q^{aiu})$
\end{itemize} have not been solved yet. He would be happy to offer a dinner at the Sushi restaurant in the Kashiwa campus to the first person who finds the solution to either of the two theories. There are many other $\cN{=}2$ gauge theories without known solutions, as listed in \cite{Bhardwaj:2013qia}. So this field should be considered still wide-open.

Hopefully, those readers who came to this point should be at least moderately equipped to tackle these and other recent articles on $\cN{=}2$ supersymmetric theories. It would be a pleasure for the author if they would continue the study and contribute to extend the frontier of the research.

\appendix 
\section{Prepotential and the instanton computation}\label{sec:instcount}
In Sec.~\ref{sec:purecurve}, we learned how to obtain $a$, $a_D$ of the pure $\SU(2)$ theory in terms of the \SeibergWitten\ curve. By integrating $a_D$ twice, we can then determine the prepotential $F(a)$ concretely. 
In this appendix, we first perform this computation, and then explain very briefly how the same prepotential can be obtained from a microscopic path-integral calculation. 

\subsection{Prepotential from the curve}
The curve of the pure $\SU(2)$ theory was given in \eqref{curve_pure}, which we reproduce here: \begin{equation}
\Lambda^2 z + \frac{\Lambda^2}z=x^2-u,
\end{equation} with the differential $\lambda=xdz/z$. 
Then $a$ and $a_D$ are determined via 
\begin{equation}
a=\frac{1}{2\pi i}\oint_A \lambda,\qquad
a_D=\frac{1}{2\pi i}\oint_B \lambda. \label{aad}
\end{equation} From this, we should determine $F(a)$ such that $a_D=\partial F/\partial a$. The low-energy coupling is $\tau(a)=\partial a_D/\partial a$.
In \eqref{tauexpansion}, we obtained the form of the weak-coupling expansion of $\tau(a)$. Integrating twice, we see that $F(a)$ has the expansion \begin{equation}
2\pi iF(a)=-4a^2\log\frac{a}{\Lambda}+ \sum_{k=1}^\infty d_k \frac{\Lambda^{4k}}{a^{4k-2}}\label{Fweak}
\end{equation} where $d_k$ are purely numerical numbers. Let us determine them explicitly, following  \cite{Chan:1999gj}.  

For this, we use the so-called renormalization group relation \begin{equation}
2\pi i \Lambda\frac{\partial}{\partial\Lambda} F(a,\Lambda)= 4u\label{matone}
\end{equation}  derived originally in \cite{Matone:1995rx}.  This relation can be proved in various ways. One immediate way is to see that \begin{equation}
2\pi i \Lambda\frac{\partial}{\partial\Lambda}= 4\frac{\partial}{\partial \tau_{UV}}
\end{equation} due to \eqref{lambdaSU2pure}, where we keep the UV regulator $\Lambda_{UV}$.
Recall that the prepotential at the ultraviolet scale is $F(\phi)=(\tau_{UV}/2) \tr \phi^2$, see the end of Sec.~\ref{sec:lel}. Then, the relation \eqref{matone} is just the statement \begin{equation}
\frac{\partial}{\partial \tau_{UV}}\vev{F} = \frac12\vev{\tr\phi^2} = u.
\end{equation} 

If one prefers, we can show the relation \eqref{matone} from \eqref{aad} purely mathematically. Let us start from the obvious relation \begin{equation}
(\Lambda\frac{\partial}{\partial\Lambda} + a\frac{\partial}{\partial a} )F= 2F. 
\end{equation} Using this, we find \begin{align}
\frac{\partial}{\partial u} \Lambda\frac{\partial}{\partial\Lambda} F 
&= 
\frac{\partial}{\partial u} (2F-a\frac{\partial}{\partial a} F )
=
a_D \frac{\partial a}{\partial u} - a \frac{\partial a_D}{\partial u}   \\
&=\frac1{(2\pi i)^2} \left(\oint_B \lambda \oint_A \frac{\partial}{\partial u}\lambda
-\oint_A \lambda \oint_B \frac{\partial}{\partial u}\lambda\right).
\end{align} Now $\lambda=xdz/z$ is a differential of the second kind (i.e.~meromorphic with zero residue),
and $\partial\lambda/\partial u=dz/(2xz)$ is a differential of the first kind (i.e.~holomorphic). Then we can use the reciprocity law of Riemann to compute it, see e.g.~Sec.~2.3 of \cite{GH}. Namely, the second-order pole of $\lambda$ is at $z=0$ and $\infty$. At both points, $w:=1/x$ is a good local coordinate, and we have the expansion \begin{equation}
\lambda=x\frac{dz}z=-2 \frac{dw}{w^2}+\cdots, \qquad 
\frac{\partial}{\partial u}\lambda=\frac{dz}{2xz}=-dw +\cdots.
\end{equation} Therefore we have \begin{equation}
\frac{\partial}{\partial u} \Lambda\frac{\partial}{\partial\Lambda} F =\frac{4}{2\pi i},
\end{equation} proving \eqref{matone}.

Let us now come back to the computation of the coefficients $d_k$ in \eqref{Fweak}. 
Let $u=\underline{a}^2$. Then \begin{align}
a(\underline a)&=\frac1{2\pi i}\oint_A \lambda=\frac1{2\pi i}\oint_A (\sum_{l=0}^\infty 
(z+\frac1z)^l \frac{(-1)^l(2l)!}{(1-2l) (l!)^2 2^{2l}}
\frac{\Lambda^{2l}}{\underline{a}^{2l-1}}) \frac{dz}z \\
&=\underline{a}\sum_{k=0}^\infty \frac{\Lambda^{4k}}{\underline{a}^{4k}} \frac{1}{(1-4k)2^{4k}}\frac{(4k)!}{(2k)!k!k!}. \label{113}
\end{align} Now, plug in the expansion \eqref{Fweak} into the relation \eqref{matone}. We have the relation \begin{equation}
a^2 (1+ \sum_{k=1}^\infty k d_k \frac{\Lambda^{4k} }{a^{4k}}) = \underline{a}^2.\label{114}
\end{equation}
The relations \eqref{113} and \eqref{114} together are sufficient to determine $d_k$ recursively. The first few terms are \begin{equation}
d_1=\frac12, \ 
d_2=\frac5{64}, \ 
d_3=\frac3{64}, \ 
d_4=\frac{1469}{32768}, \ \ldots.\label{results-from-curve}
\end{equation}

\subsection{Prepotential from the instanton computation}
Let us present here, another computation of the same prepotential, from a more microscopic point of view. This computation can be phrased entirely in terms of the path integral of the pure $\SU(2)$ gauge theory \cite{Nekrasov:2002qd}. Essentially the same calculation can be phrased more physically using a closely-related five-dimensional field theory. This is the approach we take here. The presentation here will be very brief; for more details, readers are referred to \cite{Tachikawa:2014dja}.

\subsubsection{The $\Omega$ background}
We start from the fact that given a four-dimensional $\cN{=}2$ supersymmetric gauge theory with gauge group $G$ and hypermultiplets in the representation $R$ of $G$, we can also consider a five-dimensional supersymmetric gauge theory with the same gauge group $G$ and hypermultiplets in the same representation $R$, such that when we compactify this theory on $S^1$ with radius $\beta$ and then take the zero-radius limit $\beta\to 0$, we recover the original four-dimensional theory.

\begin{figure}[h]
\[
\inc{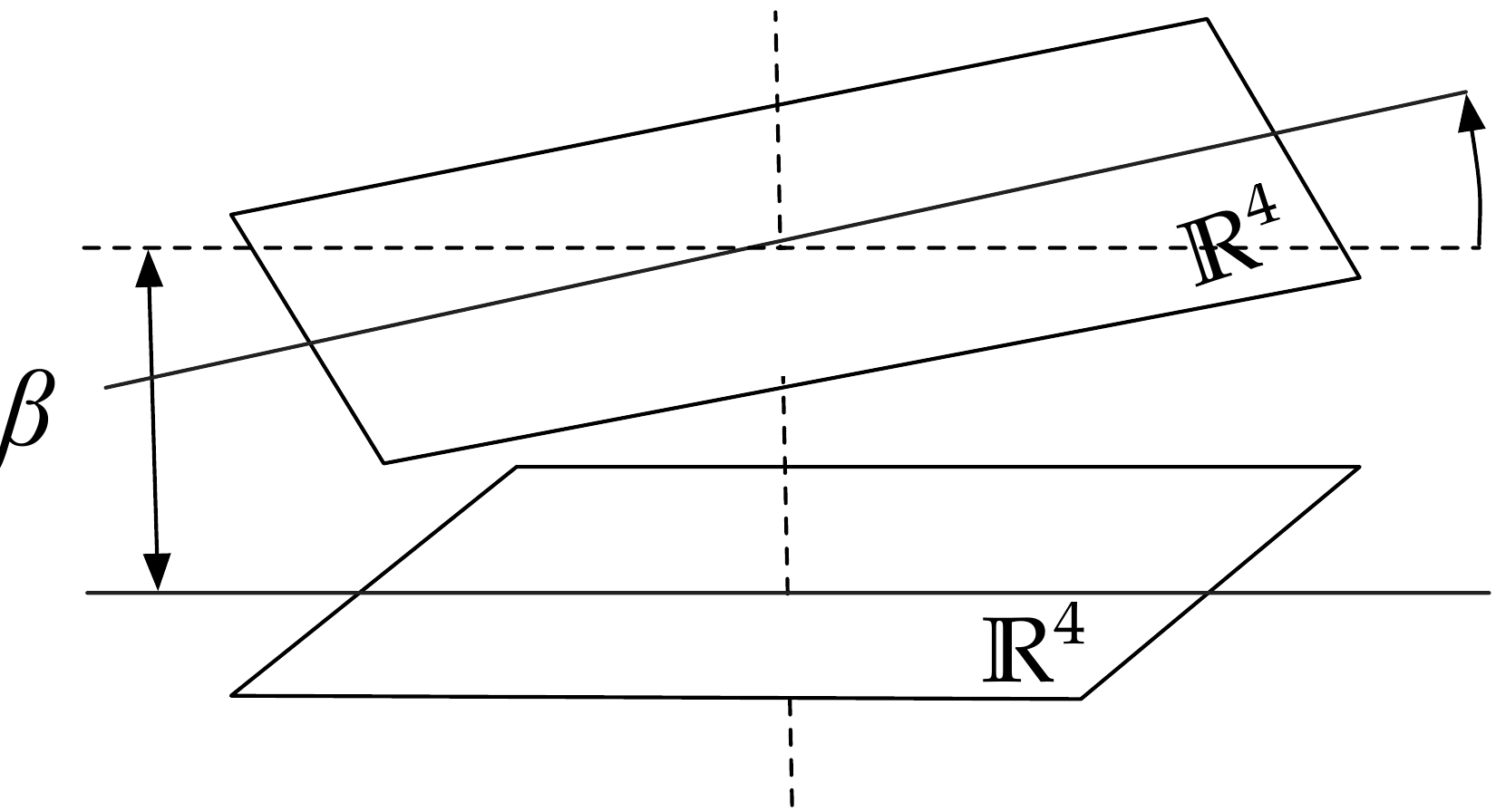}
\]
\caption{Schematic drawing of the five-dimensional $\Omega$ background. \label{fig:omega}}
\end{figure}

We consider this five-dimensional theory on the so-called $\Omega$ background, which is obtained as follows.  We first regard $\bR^5$ as $\bC^2 \times \bR$, with coordinates $(z_1,z_2,x_5)$ where $z_{1,2}$ are complex and $x_5$ is real. We then make an identification \begin{equation}
(z_1,z_2,x_5)\sim (z_1 e^{i\beta\epsilon_1},z_2e^{i\beta\epsilon_2},x_5+\beta).\label{omegaident}
\end{equation}  To preserve supersymmetry, we need to accompany this identification \eqref{omegaident} with an $\SU(2)_R$ rotation \begin{equation}
\diag(e^{i\beta(\epsilon_1+\epsilon_2)/2},e^{-i\beta(\epsilon_1+\epsilon_2)/2}) \in \SU(2)_R.
\end{equation} We can also choose to perform a gauge rotation by \begin{equation}
e^{i\beta \vec a} \in G
\end{equation} where $\vec a$ is an element in the Lie algebra of $G$, when we make the identification \eqref{omegaident}.
See Fig.~\ref{fig:omega} for a schematic drawing of this system.

The fifth direction of this spacetime is $\beta$, and this comes with a rotation of the $\bR^4$ side by the angle $\beta\epsilon_{1,2}$.  When we regard $x_5$ as the time direction, the system are forced to rotate, and any excitations far from the origin of $\bC^2$ cost a lot of energy. This makes the system effectively of finite volume, of size $\sim 1/(\epsilon_1\epsilon_2)$. Therefore the system has a finite partition function $Z(\epsilon_{1,2},\tau_{UV};\vec a)$ and its logarithm behaves as \begin{equation}
\log Z(\epsilon_{1,2},\tau_{UV};\vec a) = \frac{1}{\epsilon_1\epsilon_2} \cF(\tau_{UV};\vec a) + \text{terms less singular in $\epsilon_{1,2}$}.
\end{equation} where $\cF$ is the free energy per volume of the system. 
Here, we define $\tau_{UV}$ using the five-dimensional coupling constant $g_{UV5}$, which we define 
 so that the gauge kinetic term is of the form \begin{equation}
\int dx_5 \int d^4 x \frac{1}{2g_{UV5}{}^2}\tr F_{\mu\nu} F_{\mu\nu}  . \label{5dgaugekin}
\end{equation} By first integrating over $x_5$ which produces a factor of $\beta$, we find \begin{equation}
\tau_{UV}=\beta\frac{4\pi i}{g_{UV5}{}^2}.
\end{equation}

A basic fact is that this free energy $\cF$  equals the prepotential: \begin{equation}
2\pi i F(\tau_{UV};\vec a)
= \cF(\tau_{UV};\vec a) 
= \lim_{\epsilon_{1,2}\to 0} \epsilon_1\epsilon_2 \log Z(\tau_{UV};\vec a).\label{main-instanton-formula}
\end{equation} By taking a controlled limit $\beta\to 0$, we can then extract the prepotential of a purely four-dimensional theory.  This universal fact about any $\cN{=}2$ theory can be derived in various ways, see \cite{Nekrasov:2002qd} or the Appendix of \cite{Tachikawa:2004ur}.

\subsubsection{Reduction to supersymmetric quantum mechanics}

This reformulation might sound cumbersome, but the nice thing is that the computation of the prepotential is now reduced to that of the partition function $Z$ of the system in the $\Omega$ background \eqref{omegaident}, which can be done in much the same way by reducing the system to supersymmetric quantum mechanics, as we did in Sec.~\ref{sec:box} when we analyzed pure $\cN{=}1$ gauge theories in a box. Note that we regard the extra dimension $x^5$ as the time direction $t$ here.

The total partition function $Z$ is a product of the perturbative part $Z_\text{pert}$ and the non-perturbative part $Z_\text{non-pert}$. The first part just reproduces the leading logarithmic term in \eqref{Fweak}, and the way it is reproduced is itself quite interesting, but we do not reproduce it here. Instead, let us concentrate on the non-perturbative part, $Z_\text{non-pert}$.  To compute this, one can restrict to the lowest-energy states in each topological sector, as in Sec.~\ref{sec:box}.  From \eqref{5dgaugekin}, we see that the energy of a gauge configuration has the contribution \begin{equation}
\int d^4 x \frac{1}{2g_{UV5}{}^2}\tr F_{\mu\nu} F_{\mu\nu} ,\label{5dprefactor}
\end{equation}  which is bound from below by $8\pi^2|k|/g_{UV5}{}^2 $ as we saw in \eqref{bound}, where $k$ is the instanton number. 

Therefore, the only classical field configuration that contributes to the computation of $Z$ is such that at every constant time surface $t=\text{const}$, we have an instanton configuration. As the instanton number $k$ is an integer, this cannot change as we move along time. However, there is a continuous family of instanton with fixed $k$, and therefore the shape of the instanton can vary.  Denote by $M_k$ the parameter space of this continuous family.  This is known as the instanton moduli space.
Then, given a point $p\in M_k$, we have an instanton solution $
F_{\mu\nu}(x_{1,2,3,4};p)
$ for $\mu,\nu=1,2,3,4$.  As the time $t=x_5$ changes, the shape of the instanton changes. In other words, we have a path $p(t)$ in $M_k$. Then the five-dimensional action \eqref{5dgaugekin} 
becomes approximately  \begin{equation}
\int dt \int d^4x \frac{1}{2g_{UV5}{}^2}\tr F_{\mu\nu} F_{\mu\nu} 
= \int dt (\frac{8\pi^2}{2g_{UV5}{}^2} + G_{IJ}(p) \partial_t p^I(t) \partial_t p^J(t))\label{sqm}
\end{equation} where the first term and the second term on the right hand side come from the terms on the left hand side with spacetime indices $\mu,\nu$ without $5$ or with $5$, respectively. 
Here, the index $I=1,\ldots,\dim M_k$ is  for the local coordinates of $M_k$, and $G_{IJ}(p)$ is the metric on $M_k$.

For $G=\SU(2)$, the moduli space $M_k$ is known to be a $4|k|$ dimensional space.  Then, the system for a given $k$ reduces to a supersymmetric quantum mechanics of a point particle moving in $M_k$; the action of the bosonic part is given in \eqref{sqm}.
This is essentially a Born-Oppenheimer approximation; a significant fact is that in a supersymmetric background, this approximation is in fact exact. 

Around \eqref{bound}, we explained that an instanton with $k\ge 0$ and $k\le 0$ preserves a different supercharge. Due to this, only instantons with $k\ge 0$ contribute to the computation of $Z$; those with $k\le 0$ contribute to $\bar Z$. From these considerations, we have the following formula: \begin{equation}
Z_\text{non-pert}=\sum_{k=0}^\infty Z_k 
\end{equation} Here, $Z_k$ is the partition function of the supersymmetric quantum mechanics on $M_k$ whose bosonic part has the action \eqref{sqm}.  As the first term of \eqref{sqm} just gives a constant shift of energy, it is useful to separate it by writing \begin{equation}
Z_k=e^{2\pi i \tau_{UV} k} \tilde Z_k\label{constantshift}
\end{equation}  where $\tilde Z_k$ is the partition function of the supersymmetric quantum mechanics on $M_k$ where the lowest energy state has zero energy. 

\subsubsection{Concrete computations}

Let us compute $Z_k$. We start from the simplest case $k=0$. In this case, the moduli space $M_0$ is just a single point,  and a point particle moving in a point cannot do much. There is just one-dimensional Hilbert space with zero Hamiltonian, and therefore we have \begin{equation}
Z_0= 1. 
\end{equation}

Next, let us consider $k=1$. In this case, the instanton moduli space is $\bC^2\times (\bC^2/\bZ_2)$. 
Here, the first $\bC^2$ specifies the center of mass of the instanton on $\bC^2$. 
An $\SU(2)$ instanton with $k=1$ is further specified by a size $\rho$, which is a non-negative  real number, and  a gauge direction in the group manifold. As gauge fields are valued in the adjoint representation, the gauge direction takes values in the $\SO(3)$ group manifold, which is $S^3/\bZ_2$. Combined with the size $\rho$, they parameterize $\bC^2/\bZ_2$. 

Introduce coordinates $(z_1,z_2)$ for $\bC^2$ and $(u,v)$ for $\bC^2/\bZ_2$, with an identification $(u,v)\simeq -(u,v)$. Under the angular rotation by $\beta\epsilon_{1,2}$, they are affected as \begin{equation}
(z_1,z_2,u,v)\mapsto  (e^{i\beta \epsilon_1} z_1,\epsilon_{i\beta\epsilon_2}z_2, 
e^{i\beta(\epsilon_1+\epsilon_2)/2} u,
e^{i\beta(\epsilon_1+\epsilon_2)/2} v).
\end{equation} The actions on $z_1$ and on $z_2$  come directly from the definition; the actions on $u$ and $v$ are more subtle. Under the gauge rotation \begin{equation}
\diag(e^{i\beta a},e^{-i\beta a}) \in \SU(2),
\end{equation} $z_{1,2}$ are unchanged and \begin{equation}
(u,v)\mapsto (e^{i\beta a}u,e^{-i\beta a}v).
\end{equation}

These data are sufficient to compute $Z_2$. First, let us consider the contribution from the degrees of freedom on the motion described by $z_1\in \bC$. The supersymmetric wavefunctions are holomorphic functions of $z_1$ (times a non-holomorphic Gaussian factor which are irrelevant in our analysis). The state whose wavefunction is $z_1^n$ gets multiplied by $e^{in\beta\epsilon_1}$ by the spatial rotation. Therefore, the trace of this element over this part of the Hilbert space is \begin{equation}
\sum_{n=0}^\infty e^{in\beta\epsilon_1}=\frac{1}{1-e^{i\beta\epsilon_1}}.
\end{equation} Similarly, the motion described by $z_2$ gives a factor $1/(1-e^{i\beta\epsilon_2})$. 
The motion along $u,v$ would give, similarly, a factor \begin{equation}
\frac{1}{1-e^{i\beta((\epsilon_1+\epsilon_2)/2+a)}}
\frac{1}{1-e^{i\beta((\epsilon_1+\epsilon_2)/2-a)}}
\end{equation} if it were not for the identification $(u,v)\simeq (-u,-v)$. 
Among the wavefunctions $u^n v^m$ for $n,m$ non-negative integers, only those with even $n+m$ are compatible with this identification. Therefore, the correct contribution from the motion along $u$ and $v$ is \begin{equation}
\sum_{n\ge 0,\ m\ge 0,\ n+m:\text{even}} e^{in\beta((\epsilon_1+\epsilon_2)/2+a)}
e^{im\beta((\epsilon_1+\epsilon_2)/2-a)}
=\frac{1+e^{i\beta(\epsilon_1+\epsilon_2)}}{ (1-e^{i\beta(\epsilon_1+\epsilon_2+2a)})
 (1-e^{i\beta(\epsilon_1+\epsilon_2-2a)}) }. 
\end{equation} 
Finally, we should not forget that there is a contribution from the constant shift of the energy as in \eqref{constantshift}. 

Combined, we find that the one-instanton contribution is \begin{equation}
 Z_1=e^{2\pi i \tau_{UV} } 
\frac{1}{1-e^{i\beta\epsilon_1}}\frac{1}{1-e^{i\beta\epsilon_2}}
\frac{1+e^{i\beta(\epsilon_1+\epsilon_2)}}{ (1-e^{i\beta(\epsilon_1+\epsilon_2+2a)})
 (1-e^{i\beta(\epsilon_1+\epsilon_2-2a)}) }.\label{inst-z1}
\end{equation}

Let us consider how we can take the four-dimensional limit $\beta\to 0$. 
The denominator of \eqref{inst-z1} is of order $\beta^4$; therefore, to take a sensible limit of $Z_1$, we must assume that \begin{equation}
e^{2\pi i\tau_{UV}} \sim \beta^{4}. 
\end{equation}  This is compatible with the one-loop running \eqref{lambdaSU2pure} of $\tau_{UV}$, by identifying $\beta^{-1}=\Lambda_{UV}$.  Indeed, we should consider the five-dimensional description goes to an effectively four-dimensional one around the energy scale inverse to the radius of the $x_5$ direction. 
Let us decide to take the $\beta\to 0$ limit fixing \begin{equation}
\underline\Lambda^4:=\beta^{-4} e^{2\pi i\tau_{UV}}.
\end{equation} Then we find  \begin{equation}
Z_1 \to \underline\Lambda^4 \frac12 \frac{1}{\epsilon_1}\frac{1}{\epsilon_2} \frac{1}{(\epsilon_1+\epsilon_2)/2 +a} \frac{1}{(\epsilon_1+\epsilon_2)/2 -a}.
\end{equation}
Plugging it into \eqref{main-instanton-formula}, we find the instanton contribution to the prepotential to this order: \begin{equation}
2\pi i F_\text{non-pert}=
\lim_{\epsilon_{1,2}\to 0}\epsilon_1\epsilon_2 \log (1+e^{2\pi i\tau_{UV}} Z_1+\cdots)
= \frac12\frac{\underline\Lambda^4}{a^2}+\cdots.
\end{equation} We find that we reproduce the same coefficient $d_1=1/2$ as in \eqref{results-from-curve}, but we cannot claim victory here, as in this casual derivation, there is no way to tell a priori the instanton factor $\Lambda^4$ on the curve side and the corresponding factor $\underline\Lambda^4$ on the instanton computation side are the same. Instead, this serves as a confirmation that $\Lambda^4=\underline\Lambda^4$.

\begin{figure}[h]
\[
\inc{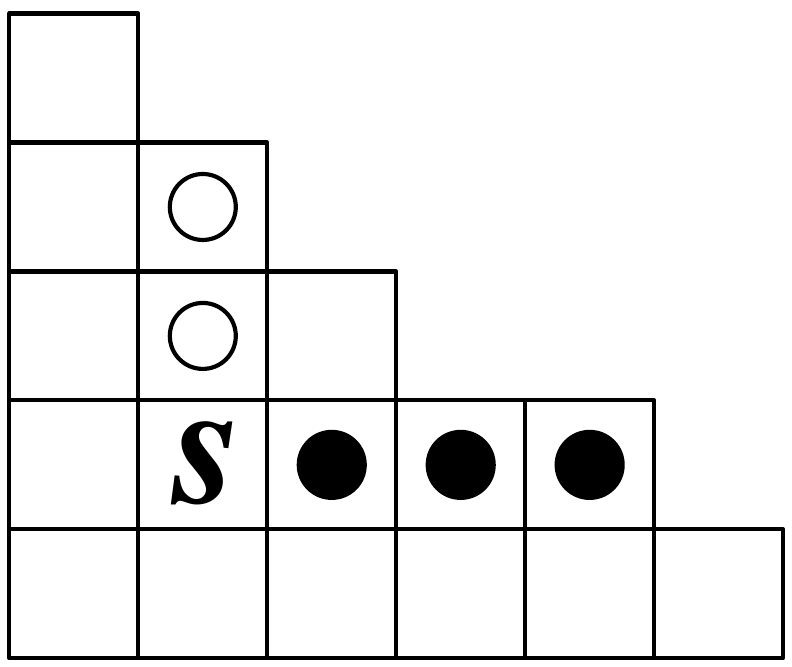}
\]
\caption{Definition of the arm length and the leg length, given a box $s$ in a Young diagram. 
$A_Y(s)$  and $L_Y(s)$ are the numbers of white and black circles, respectively. \label{fig:armleg}}
\end{figure}

Therefore, to really see the agreement, we need to go further.  Much more pages are needed to explain how to perform this computation, but the result can be written down in a combinatorial formula given below: \begin{multline}
\tilde Z_k=  \sum_{Y_1,Y_2} \prod_{n,m=1}^2
\prod_{s\in Y_n} \frac{1}{1-e^{i\beta(-L_{Y_m}(s)\epsilon_1+(A_{Y_n}(s)+1)\epsilon_2 + a_m-a_n)}} \\
\prod_{t\in Y_m}\frac{1}{1-e^{i\beta((L_{Y_n}(t)+1)\epsilon_1-A_{Y_m}(s)\epsilon_2 + a_m-a_n)}}.\label{inst-general}
\end{multline} Here, the sum  runs over pairs $(Y_1,Y_2)$ of Young diagrams with the number of total boxes being $k$, $s\in Y$ denotes that $s$ is a box in a Young diagram $Y$, and finally 
the functions $A_{Y}(s)$ and $L_{Y}(s)$ are the arm length and the leg length of the box $s$ in a Young diagram $Y$, as given in Fig.~\ref{fig:armleg}.

It is a good exercise to reproduce $Z_1$ given in \eqref{inst-z1} from this general formula; the contributing pairs of diagrams are $(\Young{1},0)$ and $(0,\Young{1})$. To compute $Z_2$, one needs to sum over five pairs $(\Young{2},0)$, 
$(\Young{11},0)$, $(\Young{1},\Young{1})$, 
$(0,\Young{2})$, $(0,\Young{11})$; you will find that \begin{equation}
Z_2\to  \Lambda^8 \frac{( 8(\epsilon_1+\epsilon_2)^2 +\epsilon_1\epsilon_2 - 8a^2)}
{\epsilon_1^2\epsilon_2^2 ((\epsilon_1+\epsilon_2)^2-4a^2)
((2\epsilon_1+\epsilon_2)^2-4a^2) ((\epsilon_1+2\epsilon_2)^2-4a^2) }
\end{equation} in the limit $\beta\to 0$. 

Plugging it again into \eqref{main-instanton-formula}, we find the instanton contribution to the prepotential to this order: \begin{align}
2\pi i F_\text{non-pert}&=
\lim_{\epsilon_{1,2}\to 0}\epsilon_1\epsilon_2 \log (1+ Z_1+ Z_2+\cdots) \\
&= \lim_{\epsilon_{1,2}\to 0} \epsilon_1\epsilon_2 \left( Z_1 
+  (Z_2 - \frac12 Z_1{}^2)  + \cdots \right) \label{225}\\
&= \frac12\frac{\Lambda^4}{a^2}+\frac5{64}\frac{\Lambda^8}{a^6}+\cdots, \label{226}
\end{align}  where we also took the limit $\beta\to 0$ to go from \eqref{225} to \eqref{226}.
We indeed reproduce $d_2=5/64$ we already saw in \eqref{results-from-curve}.
Using any computer algebra system one likes, it is straightforward to check that the coefficients $d_k$ 
as determined from the curve \eqref{results-from-curve} and those as determined from the instanton formula \eqref{inst-general} always agree.  A general mathematical proof was given e.g.~in \cite{Nekrasov:2003rj}.

\section{The zoo of $\cN{=}2$ theories}\label{sec:zoo}
In this lecture note, we have seen many four-dimensional $\cN{=}2$ theories. The aim of this appendix is to summarize them in a single place, since many of them are gradually introduced along the way and appear in scattered places.\footnote{The author thanks Simone Giacomelli for helping him prepare this section.}

Unfortunately, we cannot present a coherent classification of all the theories involved, as our understanding of the totality of $\cN{=}2$ theories is yet quite inadequate. However, we will already be able to see some pattern: there are many standard infinite series, together with some `exceptional' ones. Furthermore, some of the first members of the infinite series accidentally become equal to each other.  
This is just as in the classification of the simple Lie algebras: there are four infinite series $A_k=\SU(k{+}1)$, $B_k=\SO(2k{+}1)$, $C_k=\Sp(k)$ and $D_k=\SO(2k)$, together with exceptionals $E_{6,7,8}$, $F_4$ and $G_2$. There are accidental coincidences such as $A_1=B_1=C_1$, $B_2=C_2$, $A_3=D_3$, and some of the exceptionals can be naturally continued below, so that $E_5=D_5$ and $E_4=A_4$.

\subsection{Gauge theories}\label{sec:zoogauge}
Let us first discuss  Lagrangian gauge theories. As we discussed extensively in Sec.~\ref{sec:lagrangian}, a Lagrangian $\cN{=}2$ supersymmetric theory is basically specified by \begin{itemize}
\item a gauge group $G$, which can be a product of many simple factors
\item and the (half)-hypermultiplet in a representation $R$ of $G$. 
\end{itemize} 
In addition, we have one complexified gauge coupling for each simple factor of the gauge group $G$,
and we can add mass terms for the hypermultiplets.  

In order for the theory to be UV-complete, i.e.~to be defined at every energy scale, the one-loop beta function as defined in \eqref{n2running} needs to be non-negative. This puts a constraint on the possible  hypermultiplet representation $R$ given $G$.  For simple $G$, the list of all such allowed representations are known for quite some time \cite{Koh:1983ir,Dong:1984vt,Derendinger:1984bu}. The classification for semi-simple $G$ was done in \cite{Bhardwaj:2013qia}. 

A typical gauge theory with semi-simple $G$ is obtained as follows. Take a graph, consisting of vertices and edges. For each vertex $v$, introduce a gauge multiplet $\SU(N_v)$, and for each edge $e$ connecting vertices $v$ and $v'$, introduce a bifundamental hypermultiplet of $\SU(N_v)\times \SU(N_{v'})$. 
Given a graph, one can find an assignment of the number of colors $N_v$ such that 
\begin{itemize}
\item the theory is asymptotically free if and only if the graph is a finite Dynkin diagram $\Gamma$, and 
\item the theory is superconformal if and only if the graph is an affine Dynkin diagram $\hat\Gamma$. 
\end{itemize} These are often called quiver gauge theories, of type $\Gamma$ or $\hat\Gamma$, respectively, and were systematically analyzed first in \cite{Katz:1997eq}.

More generally, we can start from a possibly non-Lagrangian theory $T$ with a flavor symmetry $F$. We can then pick a subgroup $G\subset F$, and try to couple a vector multiplet with gauge group $G$ to the theory $T$.  As long as the coefficient of the two-point function of the current of $G$, defined as in \eqref{GOO}, is not too big, the combined system should make sense at every energy scale. Such combined system can be called a semi-Lagrangian system, in the sense that we have both a part $T$ for which we do not have a Lagrangian and the vector multiplet part for which we do have a Lagrangian.   Below, we do not explicitly list such semi-Lagrangian systems.

\subsection{6d constructions}\label{sec:app6d}
In this lecture note, the Seiberg-Witten solutions of Lagrangian gauge theories are almost always phrased in terms of a certain 6d $\cN{=}(2,0)$ supersymmetric theory compactified on the \Gaiotto\ curve, as explained in Sec.~\ref{sec:6d}. We mainly focused on the 6d theory obtained from $N$ M5-branes: this is the 6d $\cN{=}2$ theory of type $\SU(N)=A_{N-1}$.   In addition, it is known that there are 6d $\cN{=}(2,0)$ theories  of type $\SO(2N)=D_N$ and $E_{6}$, $E_7$ and $E_8$. 

\subsubsection{Types of 6d $\cN{=}(2,0)$ theories}
For theories of type $A_1$, we saw in Sec.~\ref{sec:gaiotto} that a field $\phi(z)$ on the 6d worldvolume, so that the \SeibergWitten\ curve is given by $\lambda^2=\phi(z)$.  We also saw there that we can introduce various punctures on the \Gaiotto\ curve, giving poles of various orders to the field $\phi(z)$. 
For theories of type $A_{N-1}$, we saw in Sec.~\ref{sec:duality} that there are fields $\phi_2(z)$, \ldots, $\phi_N(z)$.  Again, we discussed there various punctures on the \Gaiotto\ curve, that introduce poles of various orders to the field $\phi_{2,\ldots,N}(z)$.   Note that the field $\phi(z)$ for type $A_1$ corresponds to $\phi_2(z)$ in a more general notation.

We discussed 6d theories of type $D_N$ implicitly in Sec.~\ref{sec:SO}. There, we used the worldvolume fields $\phi_2(z)$, $\phi_4(z)$, \ldots, $\phi_{2N-2}(z)$ and $\phi_{2N}(z)$. But it is known that the last one should be viewed as a square of another field: $\phi_{2N}(z)=\tilde \phi_N(z)^2$. 
For theories of type $E_{n}$, we have worldvolume fields $\phi_i(z)$ where $i$ runs over the degrees of invariants of the Weyl group of type $E_n$. In particular, the one with the highest $i$ has $i=12$, $18$, $30$ for $E_6$, $E_7$, $E_8$, respectively. 
Note that for  6d theory of any type $G$,  the worldvolume field $\phi_i(z)$ with highest $i$ has $i=h^\vee(G)$, where $h^\vee$ is the dual Coxeter number. 

\subsubsection{Punctures of the 6d theory}
A tame puncture is a puncture such that the pole of $\phi_i(z)$ is of order less than or equal to $i$.  A wild puncture is one such that at least one of the pole of $\phi_i(z)$ is of order larger than $i$. 
Tame/wild punctures are also called regular/irregular punctures.
Possible types of tame punctures for any type are classified, see \cite{Chacaltana:2012zy}. 
As discussed in Sec.~\ref{sec:tame}, we saw that for type $A_{N-1}$, a tame puncture is labeled by a partition of $N$.  Regard a partition $(t_1,\ldots,t_k)$ with a nilpotent matrix with the Jordan block decomposition \eqref{nilpotent-vev} specified by this partition.   Then, for 6d theory of an arbitrary type $G$, a tame puncture is labeled by a nilpotent element of the complex Lie algebra of $G$, up to conjugation by the complexified adjoint action of $G$.  One particular nilpotent element which always exists for any $G$ is the zero element itself. The corresponding puncture is called the full puncture, and has an associated flavor symmetry $G$. For $G=A_{N-1}$, this was explained in detail in Sec.~\ref{sec:tame}.

Possible types of wild punctures are not studied as well, although there is a work on type $A_k$ theories \cite{Xie:2012hs}.  One special type of wild puncture is known to exist for any type, and is specified by the order $N$ of the pole for $\phi_{i}(z)$ with $i=h^\vee(G)$. 
Call this a standard wild puncture of order $N$. By definition, $N>h^\vee(G)$.
For $G=A_1$, this is the type of punctures we discussed rather extensively in Sec.~\ref{sec:generalAD}. In the figures in that section, we denoted those punctures by labeling it by the pole order in a roman numeral. 

\subsubsection{Basic building blocks}
With this preparation, we can construct a rather large class of 4d $\cN{=}2$ theories, by taking the following steps: \begin{itemize}
\item Choose the type $G=A_N,$ $D_N$, $E_{6,7,8}$ of the 6d theory.
\item Pick a \Gaiotto\ curve $C$ on which to put this 6d theory.
\item Pick a collection of tame and wild punctures, labeled by various data, on $C$.
\item Go to the low-energy limit to decouple the higher-dimensional ingredients. 
\end{itemize}
Four-dimensional theories obtained in this manner are called \emph{theories of class S} in the literature. 

When $C$ has of high genus or there are many punctures, we can chop the \Gaiotto\ curve $C$. In general, this process produces two full punctures at the place we cut the curve $C$.  A reverse process is to couple a vector multiplet $G$ to the diagonal $G$ subgroup of the two $G$ flavor symmetries at the two full punctures. For $G=A_{N-1}$ this process was described at length in Sec.~\ref{sec:duality}.

Repeating this process, we find that any class S theories can be obtained by coupling the following basic ingredients by gauge multiplets: \begin{enumerate}
\item A sphere with three tame punctures, or
\item A sphere with one tame and one wild puncture, or
\item A sphere with one wild puncture. 
\end{enumerate} 
The theories in the first class are called tinkertoys and extensively studied in \cite{Chacaltana:2010ks,Chacaltana:2011ze,Chacaltana:2012ch,Chacaltana:2013oka}. 
For type $A_{N-1}$,  we discussed some examples of them in Sec.~\ref{sec:duality}. 
When three punctures are all full punctures, the theory is often denoted by $T_G$, and when $G=A_{N-1}$, we also call it the $T_N$ theory, following \cite{Gaiotto:2009we}. All the other tinkertoys are obtained by giving an appropriate Higgs vev to the $T_G$ theory, as we discussed in Sec.~\ref{sec:higgsing}.

The theories in the second and the third classes are not studied as well. For type $A_1$, we discussed them in Sec.~\ref{sec:generalAD} under the names $X_N$ and $Y_N$. A natural generalization of $X_N$ and $Y_N$ exists for any type: \begin{itemize}
\item the theory $X_N(G)$, obtained from a sphere of 6d theory of type $G$, with a full puncture and a standard wild puncture of order $N$,
\item the theory $Y_N(G)$, obtained from a sphere of 6d theory of type $G$, with  a single standard wild puncture of order $N$.
\end{itemize}
These notations are not at all standard in the literature.
In the notation of \cite{Cecotti:2013lda}, we have 
\begin{equation}
X_N(G)=D_{N-h^\vee(G)}(G). \label{lda} %=D(G,N{-}h^\vee(G){-}1).
\end{equation}

In Sec.~\ref{sec:pureSUN} and Sec.~\ref{sec:pureSON}, we saw the Seiberg-Witten solution of pure $G$ gauge theory for $G=\SU(N)$ and $\SO(2N)$. In general, when $G=A_{N-1}$, $D_N$ or $E_{6,7,8}$, the solution of the pure $G$ gauge theory was found in \cite{Martinec:1995by}, and their results can be phrased in our language in the following way. Namely, the theory $X_{h^\vee(G)+1}(G)$ is an empty theory with $G$ symmetry. Take two copies of them, and couple a $G$ vector multiplet to their full punctures.  We have pure $G$ gauge theory, and its 6d realization is a sphere with two standard wild punctures of order $h^\vee(G)+1$.
The Argyres-Douglas points of these pure theories can be easily obtained by colliding two wild singularities, as we did in Sec.~\ref{sec:generalAD} for type $A_1$.   Two standard wild punctures merge into a single standard wild puncture of order $2h^\vee(G)+2$. Therefore, we find that \begin{equation}
AD_\text{pure}(G)=Y_{2h^\vee(G)+2}(G).\label{ady}
\end{equation}
In Sec.~\ref{sec:ADAD}, we learned that \begin{equation}
AD_\text{pure}(\SU(N))=Y_{N+4}(A_1),\qquad
AD_\text{pure}(\SO(2N))=X_{N+2}(A_1).\label{ADDD}
\end{equation} Therefore we see that there are coincidences of $X_N(G)$, $Y_N(G)$ with different $N$ and $G$.

It is known that the quiver theories of type $A$, $\hat A$, $D$ and $\hat D$  can be obtained in this 6d construction.  It is currently not known if the quiver theories of type $E$ and $\hat E$ can be obtained in this manner. 

\subsection{Other stringy constructions}
There are two additional classes of constructions of $\cN{=}2$ theories, that utilize string theory. One is the F-theory construction briefly explained in Sec.~\ref{sec:F}; another is to use Type IIB string on a singular Calabi-Yau manifold.

\subsubsection{F-theoretic construction}
F-theory is a version of Type IIB string theory where the axiodilaton $\tau$ is allowed to have nontrivial $\SL(2,\bZ)$ monodromy, generated by 7-branes. 
$N$ D7-branes carry $A_{N-1}$ gauge symmetry, and $N$ D7-branes on top of an O7-plane carry $D_N$ symmetry. In addition, there are 7-branes which carry flavor symmetry $A_{0,1,2}$ and $E_{6,7,8}$. 
This additional type of 7-branes with flavor symmetry $A_{0,1,2}$ is often called 7-branes of type $H_{0,1,2}$, to distinguish it from the usual D7-branes. 

Now, we take the spacetime to be \begin{equation}
\bR^{1,3}\times \bC\times \bC^2,\label{Fgeom}
\end{equation} and pick a 7-brane of type $G=H_{0,1,2}$ or $E_{6,7,8}$ and put them at
\begin{equation}
\bR^{1,3}\times \{0\} \times \bC^2.\label{Fgeomm}
\end{equation} Finally we introduce  $k$ D3-branes at \begin{equation}
\bR^{1,3}\times \{0\} \times \{0\}.
\end{equation}
Let us denote the resulting $\cN{=}2$ theory by $F(G,k)$. As we already explained in Sec.~\ref{sec:rank-1SCFT}, \begin{equation}
AD_{N_f=n}(\SU(2))=F(H_{n-1},1)
\end{equation} for $n=1,2,3$, and \begin{equation}
MN(E_{n})=F(E_n,1)
\end{equation} for $n=6,7,8$.  The theories $F(G,k)$ with $k>1$ are higher-rank versions of these theories. 
There are $k$ Coulomb branch operators in the theory $F(G,k)$, representing the motion of $k$ D3-branes along the direction $\bC$ in \eqref{Fgeom}.  The Higgs branch of $F(G,k)$ describes the process where D3-branes become instantons along $\bC^2$ in \eqref{Fgeom} on the 7-brane, and therefore is given by the $k$-instanton moduli space of the group $G$.

This construction can be further generalized by replacing the direction $\bC^2$ in \eqref{Fgeom}, \eqref{Fgeom} by an ALE orbifold $\bC^/\Gamma$, the possible type of which was already listed in Table~\ref{tab:ADE}. These theories are not studied yet. 

In Sec.~\ref{sec:duality} and in particular in Sec.~\ref{sec:application}, we already discussed that $MN(E_6)=T_3$ and that $MN(E_{7,8})$ can also be represented as a three-punctured sphere. 
The realization of $F(E_{6,7,8},k)$ for general $k$ as a tinkertoy was given in \cite{Benini:2009gi}.

\subsubsection{Type IIB on a singular Calabi-Yau}
The final construction we explain in this note goes as follows.  Type IIB string theory on a compact Calabi-Yau manifold with complex dimension 3 gives rise to four-dimensional $\cN{=}2$ supergravity system. This setup was studied in great detail in relation to the celebrated mirror symmetry.  Here we try to make the Calabi-Yau non-compact, to let the gravity decouple.  To have something non-trivial in the end, we want to make the Calabi-Yau space singular. An easy way to describe a singular Calabi-Yau space of complex dimension 3 is to write down an equation with four complex variables, e.g.~ \begin{equation}
x^2+y^2+z^2+w^2=0,\label{conifold}
\end{equation}  which is called the conifold.  Type IIB string theory on this conifold geometry is known to produce just one free hypermultiplet. 

There are many types of such singular Calabi-Yau geometry. For example, it can be used to describe quiver gauge theories associated to arbitrary Dynkin diagrams \cite{Katz:1997eq}. Instead, let us discuss here the Calabi-Yau geometries which would give non-Lagrangian theories. 

%\paragraph{First class of examples} 
For this purpose, 
let us start our discussion from singular Calabi-Yau geometries of complex dimension \emph{two} instead of three. Such geometries are known to be exhausted by the ALE orbifold $\bC^2/\Gamma$, listed in Table~\ref{tab:ADE}.  It is known that all these singularities $\bC^2/\Gamma$ can be written as an equation in three variables, shown in Table~\ref{tab:ADEeq}. It always have the form $x^2+P_\Gamma(y,z)=0$.

\begin{table}
\[
\begin{array}{c|l@{}l}
G_\Gamma & x^2+ P(y,z)&=0 \\
\hline 
 \SU(n) & x^2 + y^2 + z^n &=0\\
 \SO(2n) &  x^2 + y^2z + z^{n-1} &=0\\
 E_6 & x^2 + y^3 + z^4 &=0\\
 E_7 &  x^2+y^3+ yx^3 &=0\\
 E_8 &x^2+y^3+z^5 &=0
\end{array}
\]
\caption{Simply-laced Lie groups $G_\Gamma$ and the corresponding singularities $\bC^2/\Gamma$ as  equations with three variables. \label{tab:ADEeq}}
\end{table}

Now, define Calabi-Yau geometries of complex dimension three labeled by a pair $(G_\Gamma,G_{\Gamma'})$ by \begin{equation}
P_\Gamma(x,y)+P_{\Gamma'}(z,w)=0.\label{geomGG}
\end{equation} For example, the conifold geometry \eqref{conifold} corresponds to a pair $(A_1,A_1)$.
By abusing the notation, we denote the $\cN{=}2$ theory obtained by putting Type IIB string theory on the geometry labeled by $(G,G')$ by the same symbol $(G,G')$. This class of theories was introduced and studied in \cite{Cecotti:2010fi}. Obviously, we have \begin{equation}
(G,G')=(G',G).
\end{equation} In addition, the equalities \begin{equation}
(D_4,A_3)=(E_6,A_2),\qquad (E_8,A_3)=(E_6,A_4)
\end{equation} are manifest, just by writing down the geometries \eqref{geomGG}.

Note that the 6d $\cN{=}(2,0)$ theory of type $G_\Gamma$ can also be obtained from putting Type IIB theory on $\bC^2/\Gamma$. Therefore, the class of models discussed in Sec.~\ref{sec:app6d} can also be obtained by putting Type IIB theory on a Calabi-Yau geometry of complex dimension three, which is a fibration of deformed $\bC^2/\Gamma$ singularities over the \Gaiotto\ curve $C$.  The theories $(G,G')$ introduced here instead uses a pair, but when $G'=A_k$, they can be related to the 6d construction of type $G$: \begin{equation}
(G,A_k)=Y_{2h^\vee+k+1}(G).
\end{equation} Combined with \eqref{ady}, we see \begin{equation}
AD_\text{pure}(G)=Y_{2h^\vee+2}(G)=(G,A_1).\label{ADG}
\end{equation} When we further set $G=A_{N-1}$,  the equality becomes \begin{equation}
AD_\text{pure}(A_{N-1})=Y_{2N+2}(A_{N-1})=(A_{N-1},A_1)=(A_1,A_{N-1})=Y_{N+4}(A_1).
\end{equation}
Further letting $G=A_1$, we see that $(A_1,A_1)$ corresponds to the most singular point of the pure $\SU(2)$ theory, i.e.~the monopole point representing a single hypermultiplet.  This is consistent with what we said below \eqref{conifold}.

When we set $G=D_N$ in \eqref{ADG}, we have \begin{equation}
AD_\text{pure}(D_N)=Y_{4N-2}(D_N)=(D_N,A_1)
\end{equation} but we saw in \eqref{ADDD} that this is also equal to $X_{N+2}(A_1)$, which is denoted by $D_N(A_1)$ in \cite{Cecotti:2013lda}, as we mentioned in \eqref{lda}. 

There are a few other accidental equivalences of the theories introduced so far in this section. But the full discussion of such coincidences should probably  be better done after we have a clearer idea of the entirety of the zoo of $\cN{=}2$ theories in general. 

\if0
\paragraph{Second class of examples} 
The second class of examples based on the Calabi-Yau geometry is for the quiver gauge theories we discussed in Sec.~\ref{sec:zoogauge}.  As an example, take the quiver gauge theory of type $\hat E_6$. 
This has a gauge group \begin{equation}
(\SU(N)\times \SU(2N))^3 \times \SU(3N),
\end{equation} and the bifundamental hypermultiplets between three pairs of $\SU(N)\times \SU(2N)$ and between each of $\SU(2N)$ and $\SU(3N)$.  This gauge theory is clearly superconformal, as can be seen by computing the one-loop beta function for the seven gauge coupling constants. 

This is also given by the compactification of the type IIB theory on a Calabi-Yau, obtained in the following manner.  The two-dimensional singularities corresponding to finite Dynkin diagrams are tabulated in Table~\ref{tab:ADEeq}. There are also those for affine Dynkin diagrams. For example, For $\hat E_6$, it is given by \begin{equation}
x^3+y^3+z^3=0. 
\end{equation} To obtain the Calabi-Yau geometry for the $\hat E_6$ quiver gauge theory, we simply introduce deformations of this geometry by another coordinate $w$: \begin{multline}
x^3+y^3+z^3 + \mu x y z
+  P_N(w) x^2 + P_{2N}(w) x   \\ 
+  Q_N(w) y^2+ Q_{2N}(w) y   + R_N(w) z^2 + R_{2N}(w) z
 + S_{3N}(w)=0
\end{multline}
where $P_k(w)$, $Q_k(w)$, $R_k(w)$ and $S_k(w)$ are polynomials of degree $k$.  
\fi

\newpage

\bibliographystyle{ytamsalpha}
%\small\baselineskip=.93\baselineskip
%\let\bbb\bibitem\def\bibitem{\itemsep1pt\bbb}
\bibliography{ref}

\end{document}